\documentclass[twocolumn,
amsmath,amssymb,
aps,
longbibliography, superscriptaddress
]{revtex4-2}
\usepackage[utf8]{inputenc}
\usepackage{amsmath}
\usepackage{amsfonts}
\usepackage{amssymb}
\usepackage{braket}
\usepackage{graphicx}
\usepackage{float}
\usepackage{overpic}
\usepackage{tikz}
\usepackage{bm}
\usetikzlibrary{arrows,decorations.pathmorphing,backgrounds,positioning,fit,petri,decorations.pathreplacing,decorations.markings}
\usepackage{xcolor}
\definecolor{mygreen1}{rgb}{0, 0.4, 0}
\usepackage[margin=0.7in]{geometry}
\newcommand{\mbeq}{\overset{!}{=}}
\usepackage{pifont}
\usepackage{hyperref}
\hypersetup{
	colorlinks=true,
	linkcolor=blue,
	filecolor=magenta,
	urlcolor=cyan,
	citecolor=blue
}

\tikzset{
	on each segment/.style={
		decorate,
		decoration={
			show path construction,
			moveto code={},
			lineto code={
				\path [#1]
				(\tikzinputsegmentfirst) -- (\tikzinputsegmentlast);
			},
			curveto code={
				\path [#1] (\tikzinputsegmentfirst)
				.. controls
				(\tikzinputsegmentsupporta) and (\tikzinputsegmentsupportb)
				..
				(\tikzinputsegmentlast);
			},
			closepath code={
				\path [#1]
				(\tikzinputsegmentfirst) -- (\tikzinputsegmentlast);
			},
		},
	},
	mid arrow/.style={postaction={decorate,decoration={
				markings,
				mark=at position .5 with {\arrow[#1]{stealth}}
	}}},
}

\definecolor{myblue1}{rgb}{0.9, 0.9, 1}
\begin{document}
	
	\title{Excitations in the higher lattice gauge theory model for topological phases II: The (2+1)-dimensional case}
	\author{Joe Huxford}
\affiliation{Rudolf Peierls Centre for Theoretical Physics, Clarendon Laboratory, Oxford OX1 3PU, UK}
\affiliation{Department of Physics, University of Toronto, Ontario M5S 1A7, Canada}
\author{Steven H. Simon}
\affiliation{Rudolf Peierls Centre for Theoretical Physics, Clarendon Laboratory, Oxford OX1 3PU, UK}
	
	\begin{abstract}
	
	In this work, the second paper of this series, we study the 2+1d version of a Hamiltonian model for topological phases based on higher lattice gauge theory. We construct the ribbon operators that produce the point-like excitations. These ribbon operators are used to find the braiding properties and topological charge carried by the point-like excitations. The model also hosts loop-like excitations, which are produced by membrane operators. By considering a change of basis, we show that, in certain cases, some loop-like excitations represent domain walls between patches corresponding to different symmetry-related ground states, and we find this symmetry. We also map the higher lattice gauge theory Hamiltonian to the symmetry enriched string-net model for symmetry enriched topological phases (SETs) described by Heinrich, Burnell, Fidkowski and Levin [\textit{Phys. Rev. B}, \textbf{94}, 235136 (2016)], again in a subset of cases.

	\end{abstract}

		\maketitle
	\tableofcontents

		\section{Introduction}
		\label{Section_Introduction_2}
	
		In this series of papers, we are examining the commuting projector model for topological phases from Ref. \cite{Bullivant2017}, based on higher lattice gauge theory (see also Refs. \cite{Bullivant2018, Bullivant2020,Bullivant2020b, Delcamp2018, Delcamp2019, Koppen2021} for further work on these phases, as well as Refs. \cite{Gukov2013, Kapustin2014, Kapustin2017} for work on similar models). In the previous paper in this series, we discussed the higher lattice gauge theory model in 3+1d, but the model can also be defined in 2+1d, as discussed in Ref. \cite{Bullivant2017} (with a generalization of this 2+1d model also studied in Ref. \cite{Koppen2021}). This lower dimensional case exhibits interesting properties in its own right. Despite the spatial lattice being 2d, the model hosts loop-like excitations. Furthermore, the ground-state on a sphere is generally degenerate \cite{Bullivant2017}. These features are both somewhat unusual, and are not found in generic models for 2+1d topological phases. We will find that, in certain cases, both of these features can be explained by an additional, spontaneously broken, symmetry. The loop-like excitations (or at least a subset of them) form domain walls between patches corresponding to the different ground states. In these cases, the model is therefore an example of a so-called ``symmetry enriched topological phase", or SET phase.

		 In an SET phase, in addition to long-range entanglement, there is an enforced symmetry \cite{Mesaros2013, Chen2013}. This can be contrasted with a symmetry protected topological phase (SPT phase), which is a short-range entangled phase where the degrees of freedom can be ``disentangled" smoothly into a product state with a suitable local unitary evolution, but only by breaking the symmetry \cite{Chen2013, Chen2010}. When considering additional enforced symmetries, the space of phases becomes even richer, and so there has been considerable interest in SET phases, and Hamiltonian models describing them \cite{Wen2002, Barkeshli2019, Mesaros2013, Heinrich2016, Lu2016, Maciejko2010, Swingle2011, Levin2012, Essin2013, Teo2015, Tarantino2016, Chang2015, Hermele2014, Lan2017, Wen2017, Hung2013a, Hung2013b, Wen2002}. One Hamiltonian model of particular interest is the symmetry enriched string-net model of Ref. \cite{Heinrich2016}, which generalizes the Levin Wen string-net model \cite{Levin2005} to allow for an additional symmetry. In this symmetry enriched string-net model, which we describe in more detail and relate to the higher lattice gauge theory model in Section \ref{Section_Mapping_SN_SET}, there are additional degrees of freedom on the plaquettes of the 2d lattice, with the symmetry affecting these degrees of freedom. In some cases this symmetry may permute the anyons of the model, converting one anyon type into another (although we do not find this feature for the higher lattice gauge theory model).

		Another interesting feature exhibited by the higher lattice gauge theory model is condensation/confinement. This refers to the situation where some topological charges join the topological vacuum (trivial charge) \cite{Bais2009, Burnell2018, Neupert2016, Bais2003, Eliens2014}. During this process some charges become confined, so that it costs energy to separate an anyon from its associated anti-anyon. The theory of anyon condensation in 2+1d is well developed \cite{Bais2009, Burnell2018, Neupert2016, Bais2003, Eliens2014}, and commuting projector models that exhibit this property include the family examined in Ref. \cite{Bombin2008}, which is a generalization of Kitaev's Quantum Double model \cite{Kitaev2003}. Indeed, we will draw on the methods developed in Ref. \cite{Bombin2008} to explicitly show which topological sectors are confined. We do this by constructing the topological charge measurement operators and demonstrating that certain measurement operators are only compatible with confined excitations.

		There are other advantages to considering the higher lattice gauge theory model in 2+1d, despite the existence of the 3+1d model. The first advantage is theoretical: there already exists a well understood structure for considering many topological phases in 2+1d. Bosonic topological phases are believed to be characterised by modular tensor categories, along with the chiral central charge and other objects relating to any symmetry present \cite{Wen2017, Barkeshli2019, Lan2017}. In addition, many features of topological phases in 2+1d, including the previously mentioned condensation and confinement, have been extensively studied compared to their 3+1d counterparts. This gives us a framework with which to understand the features of any particular topological phase, but it also means that any physics found outside the accepted structure would be an interesting challenge to existing knowledge. The other advantage is more pragmatic: there are by now many examples of experimentally observed 2+1d topological phases, including fractional quantum Hall systems \cite{Tsui1982, Wen1990, Stern2008, Chakraborty1995, Sarma1997} as well as small fabricated systems \cite{Satzinger2021ReducedScience}. It is therefore more realistic to suppose that a 2+1d phase may be observed in experiment than a 3+1d one (at least for the near future).

		\subsection{Structure of this paper}
		
		Here we briefly outline the layout of the rest of this paper. We begin in Section \ref{Section_Recap_Paper_2} by reminding the reader of some concepts and mathematical ideas discussed in Ref. \cite{HuxfordPaper1}. Then in Sections \ref{Section_RO_2D_Tri_Trivial} and \ref{Section_2D_RO_Fake_Flat} we start describing our results by presenting the \textit{ribbon operators} in 2+1d, for two of the special cases described in Ref. \cite{HuxfordPaper1} and more concisely here in Section \ref{Section_Recap_Paper_2} (Cases 1 and 3 from Table \ref{Table_Cases_2d}). Ribbon operators create, annihilate and move the quasiparticles of our theory, so these are the key to many of our other results. By obtaining these operators, we can work out the \textit{fusion rules} of our theory, that is how we can bring multiple particles together and consider them as a single object. The ribbon operators can also be used to obtain the \textit{braiding relations} of our theory, which describe what happens when we move the excitations around each-other. We consider these braiding relations in Section \ref{Section_2D_Braiding}. Having looked at these ideas in fairly general cases, in Section \ref{Section_2D_Particular_examples} we take two specific examples that highlight some of the important features of this model. In Section \ref{Section_2D_irrep_basis}, we interpret various features of the 2+1d model when a particular operator called $\rhd$ (which we define in Section \ref{Section_Recap_Paper_2}) is trivial, by changing basis from group elements to irreps, and expose a symmetry in the model. We then use this basis to construct a mapping between some of the higher lattice gauge theory models and the symmetry enriched string-net construction from Ref. \cite{Heinrich2016}. Finally, in Section \ref{Section_2D_topological_Charge} we consider the \textit{topological sectors} of the models. The topological sectors classify the conserved charges that the excitations carry. These sectors therefore determine what distinct species of anyon are present in the model. We examine how these sectors change when the model undergoes a \textit{condensation-confinement} transition, where some charges join the topological vacuum and become trivial and other charges become confined.

		The proofs for many of our results are relegated to the Supplemental Material. In Section \ref{Section_2D_Ribbon_Operators_Appendix}, we calculate the commutation relations between the ribbon and membrane operators and the energy terms, therefore showing which energy terms are excited. Then, in Section \ref{Section_topological_membrane_operators}, we prove that the non-confined ribbon operators are topological, meaning that the ribbon can be smoothly deformed without changing the action of the ribbon operator. Next, in Section \ref{Section_Condensation_Magnetic_2D}, we demonstrate the pattern of condensation for the magnetic excitations (i.e., which magnetic excitations are topologically trivial and can be produced by local operators). An explicit calculation of the braiding of the various excitations is presented in Section \ref{Section_braiding_supplemental}, while the measurement operators for topological charge are constructed in Section \ref{Section_topological_charge_supplemental}. Then, in Section \ref{Section_2D_irrep_basis_Appendix}, we give some miscellaneous algebraic proofs that support Section \ref{Section_2D_irrep_basis} of the main text. Finally, in Section \ref{Section_rhd_non-trivial_condense_confine}, we show that, in certain cases, it is possible to construct the magnetic excitations of the model even when $\rhd$ is non-trivial. In these cases, the magnetic and electric excitations have an interesting pattern of condensation and confinement that depends on which degenerate ground state we create the excitations from.

		\section{Summary of the model}
		\label{Section_Recap_Paper_2}	
			
		Before describing our results, we would briefly like to review some ideas we covered in the previous paper in this series, Ref. \cite{HuxfordPaper1}, to prevent readers from needing to refer back to that work (or Ref. \cite{Bullivant2017}, where the model was originally defined) for definitions. The higher lattice gauge theory model is based on a generalisation of ordinary lattice gauge theory, where in addition to the ordinary (1-)gauge field, there is a second 2-gauge field which describes a gauge symmetry on the first gauge field itself. That is, while an ordinary gauge field describes a transformation of point-like objects (matter) the higher gauge field describes a transformation of extended objects, such as the Wilson lines.

		We work with a 2d lattice, where the (directed) edges are labelled by elements of a group $G$ (the 1-gauge field) and the plaquettes (which have both an orientation and a distinguished vertex called the base-point) are labelled by elements of a second group $E$ (the 2-gauge field). These groups are part of a ``crossed module" and each crossed module defines a different lattice model. A crossed module $(G,E, \partial, \rhd )$ consists of the two previously mentioned groups, together with two maps $\partial:E \rightarrow G$ and $\rhd:G \rightarrow \text{Aut}(E)$. Here $\partial$ is a group homomorphism from $E$ to $G$ which controls the action of the 2-gauge transforms on the 1-gauge field, while $\rhd$ is another homomorphism that maps elements of $G$ to automorphisms on $E$ (isomorphisms from $E$ to itself) and controls the action of the 1-gauge transforms on the 2-gauge field. We write the automorphisms in the form $g \rhd$ (so each $g \rhd$ is a particular isomorphism), and we denote this map acting on an element $e$ of $E$ by $g \rhd e$. The fact that $\rhd$ is an homomorphism means that these maps have a group structure: $g_1 \rhd (g_2 \rhd e) = (g_1 g_2) \rhd e$ for any elements $g_1, g_2 \in G$ and $e \in E$. In addition to restrictions put upon the maps $\rhd$ and $\partial$ by them being homomorphisms, the two maps must also satisfy additional constraints, called the Peiffer conditions, in order for the parallel transport processes described by the gauge fields to be consistent \cite{Baez2002, Pfeiffer2003}. These Peiffer conditions are
			\begin{align}
				\partial( g \rhd e)&= g \partial(e) g^{-1} \label{Peiffer_1}\\
				\partial(e) \rhd f &= efe^{-1}. \label{Peiffer_2}
			\end{align} 
			
				\begin{center}
				\noindent\fcolorbox{black}{myblue1}{%
					\parbox{0.9\linewidth}{%
						\textbf{Definition 1:}
						A crossed module is a collection $(G, E, \partial, \rhd)$, where $G$ and $E$ are groups, and $\partial:E \rightarrow G$ and $\rhd:G \rightarrow \text{Aut}(E)$ are group homomorphisms satisfying the Peiffer conditions Equations \ref{Peiffer_1} and \ref{Peiffer_2}.}}
			\end{center}
			
		While the spatial dimensions are represented by a lattice, the time dimension is continuous with evolution controlled by a Hamiltonian. This Hamiltonian, as introduced in Ref. \cite{Bullivant2017}, is a sum of commuting projector terms:
			\begin{equation}
			H = - \hspace{-0.1cm} \sum_{\text{vertices, }v} \hspace{-0.4cm} A_v \: - \sum_{\text{edges, } i} \hspace{-0.2cm} \mathcal{A}_i \: -\hspace{-0.3cm}\sum_{\text{plaquettes, }p} \hspace{-0.5cm} B_p, \label{Equation_Hamiltonian}
			\end{equation}
		where, unlike the version we used in Ref. \cite{HuxfordPaper1}, there are no blob (3-cell) terms, because the 2d lattice has no 3-cells. Here the vertex term $A_v$ is an average over the 1-gauge transforms at vertex $v$ labelled by the different elements of $G$:
			\begin{equation}
			A_v = \frac{1}{|G|} \sum_{g \in G} A_v^g, \label{Equation_vertex_energy_term}
			\end{equation}
		where $A_v^g$ fluctuates the states of the surrounding edges $i$ and also changes the state of any plaquette $p$ for which $v$ is the base-point:
			\begin{align}
			A_v^g: g_{i} &\rightarrow \begin{cases} gg_{i} &\text{ if $v$ is the start of $i$}\\
			g_ig^{-1} &\text{ if $v$ is the end of $i$}\\
			g_{i} &\text{ otherwise} \end{cases} \notag\\
			A_v^g : e_p &\rightarrow \begin{cases} g \rhd e_p &\text{if $v$ is the base-point of $p$}\\
			e_p &\text{otherwise.} \label{Equation_vertex_transform_definition}\end{cases}
			\end{align}

	From this definition, we see that $\rhd$ describes how the (1-gauge) vertex transforms act on the surface labels that are based at that vertex. Geometrically, the vertex transform acts like parallel transport of that vertex across an edge, and so $\rhd$ describes how the 2-gauge field changes under such parallel transport. Equivalently, it describes how the label of the field changes when we alter the base-point of a plaquette or surface \cite{Bullivant2017}.

	The individual vertex transforms satisfy the algebra $A_v^g A_v^h =A_v^{gh}$, which further leads to the result $A_v^g A_v = A_v$ (i.e., the vertex transforms can be absorbed into the corresponding vertex term). Because of this, the vertex term is a projector which projects onto states which are 1-gauge invariant at that vertex. This also means that the ground states (which are eigenstates of each vertex term with eigenvalue 1) can absorb vertex transforms (i.e., the ground-state is 1-gauge invariant):
			$$A_v^g \ket{GS} =A_v^g A_v \ket{GS} =A_v \ket{GS}= \ket{GS},$$
	for all vertices $v$ and elements $g \in G$. In fact we can absorb the vertex transform $A_v^g$ into any state for which the vertex $v$ is unexcited, not just the ground states.

\begin{figure}[h]
	\begin{center}
		\includegraphics[width=\linewidth]{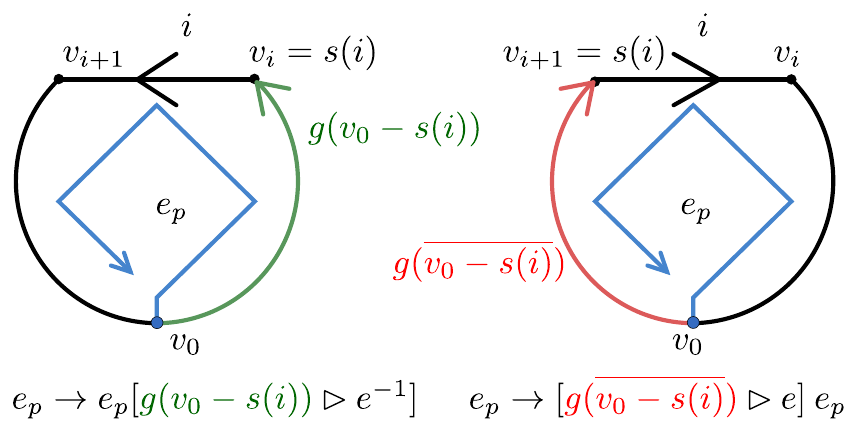}

		\caption{(Copy of Figure 27 from Ref. \cite{HuxfordPaper1}) The path involved in the effect of the 2-gauge transform $\mathcal{A}_i^e$ on a plaquette $p$ depends on whether the edge $i$ is aligned with the $p$ (as in the left case) or anti-aligned (as in the right case). If the edge is aligned with the plaquette, then the path $(v_0-s(i))$ in the transformation of the plaquette label is aligned with $p$, whereas if $i$ is anti-aligned with $p$ then the path $(\overline{v_0-s(i)})$ is anti-aligned with $p$. Either way, the path is aligned with the edge $i$.}
		\label{pathsonplaquettepaper2}
	\end{center}
\end{figure}

		Similarly, the edge term $\mathcal{A}_i$ is an average of the 2-gauge transforms at edge $i$ labelled by the different elements of the group $E$:
			\begin{equation}
			\mathcal{A}_i = \frac{1}{|E|} \sum_{e \in E}\mathcal{ A}_i^e. \label{Equation_edge_energy_term}
			\end{equation}
		Here $\mathcal{ A}_i^e$ fluctuates the labels of the plaquettes adjacent to the edge $i$, and multiplies the label of edge $i$ itself by $\partial(e)$:
			\begin{align}
			\mathcal{A}_i^e: g_{i'} &\rightarrow \begin{cases} \partial(e) g_{i'} &\text{ if $i=i'$}\\
			g_{i'} &\text{ otherwise} \end{cases} \notag \\
			\mathcal{A}_i^e : e_p &\rightarrow \begin{cases} e_p (g(v_0(p) - s(i)) \rhd e^{-1}) &\text{if $i$ is on $p$ and}\\& \text{aligned with $p$}\\
			(g(\overline{v_0(p) - s(i)}) \rhd e) e_p &\text{if $i$ is on $p$ and}\\& \text{points against $p$}\\
			e_p &\text{otherwise.} \end{cases} \label{Equation_edge_transform_definition}
			\end{align}
		Here $v_0(p)$ is the base-point of plaquette $p$ and $s(i)$ is the source of edge $i$ (an edge points from its source vertex to its target vertex). $v_0(p) - s(i)$ is the path around the plaquette, following the circulation of the plaquette, from the base-point of $p$ to the source of $i$. On the other hand, $\overline{v_0(p) - s(i)}$ is the path from the base-point to the source of $i$ travelling against the circulation of the plaquette, as illustrated in Figure \ref{pathsonplaquettepaper2}.

		These edge transforms satisfy an algebra analogous to the vertex transforms. That is, we have $\mathcal{A}_i^e \mathcal{A}_i^f =\mathcal{A}_i^{ef}$. This again allows the individual transforms to be absorbed into the corresponding energy term: $\mathcal{A}_i^e \mathcal{A}_i = \mathcal{A}_i$. Because of this, the edge terms project to states which are 2-gauge invariant at that edge and in particular the ground state satisfies $\mathcal{A}_i^e \ket{GS} = \ket{GS}$ for all edges $i$ and elements $e \in E$. 
			
The two types of gauge transform satisfy the mutual algebra \cite{Bullivant2017}
\begin{equation*}
A_v^g \mathcal{A}_i^e = \begin{cases}
	\mathcal{A}_i^{g \rhd e} A_v^g & \text{ if $v$ is $s(i)$} \\
	\mathcal{A}_i^e A_v^g & \text{otherwise.}
\end{cases}
\end{equation*}
	This means that the 2-gauge symmetry is not entirely 1-gauge symmetric, with the discrepancy described by $\rhd$. Nonetheless, the overall energy terms $\mathcal{A}_i$ and $A_v$ do commute, because the $\rhd$ action can be absorbed into the sum over elements of $E$ in Equation \ref{Equation_edge_energy_term} for $\mathcal{A}_i$.
			
	Finally, the plaquette term $B_p$ enforces so-called fake-flatness on the plaquette $p$ \cite{Bullivant2017}. A plaquette satisfies fake-flatness if its boundary path element $g_p$ is related to its plaquette label $e_p$ by $\partial(e_p)g_p=1_G$. $B_p$ then projects onto the states for which fake-flatness of the plaquette $p$ is satisfied. The plaquette term enters the Hamiltonian with a minus sign, so this leads to a reduction in energy for states satisfying fake-flatness. This energy term commutes with both the 1-gauge transforms and 2-gauge transforms, so the Hamiltonian is indeed a commuting projector Hamiltonian. For readers familiar with Kitaev's Quantum Double model \cite{Kitaev2003}, which is based on lattice gauge theory, this plaquette term (and the vertex term from earlier) may seem familiar, except for the factor of $\partial(e_p)$. Whereas in the Quantum Double model, the plaquette term ensures that contractible closed loops have trivial label, in the higher lattice gauge theory model such closed loops can have any label in the subgroup $\partial(E) \subset G$, provided that this label matches the surface label contained within the closed loop. Indeed, the edge transforms described by Equation \ref{Equation_edge_transform_definition} ensure that the ground state contains closed loops with these non-trivial labels. As we will discuss in more detail later, this leads to some of the magnetic vortices associated to $G$ becoming condensed, meaning they proliferate within the ground state.

	This similarity to Kitaev's Quantum Double model extends beyond the plaquette term. Indeed, if we take the group $E$ to be trivial, then the higher lattice gauge theory model reduces to Kitaev's model \cite{Kitaev2003}, with the vertex and plaquette terms becoming the analogous terms from Kitaev's model and the edge terms becoming trivial. This is because the group $E$ describes a higher gauge field that acts on the original gauge field and so taking it to be trivial recovers the ordinary lattice gauge theory described by Kitaev's model. For this reason, it will be useful to compare and contrast the features of the higher lattice gauge theory model to those of the Kitaev Quantum Double model. As we go on, we will find many similarities with the Quantum Double model, but will also find interesting properties not present in Kitaev's model, such as loop-like excitations, confinement and a ground-state degeneracy on the sphere.

	The final thing that we would like to discuss before getting to our results is the special cases of the model. As described in Ref. \cite{HuxfordPaper1}, there are some obstructions that prevent us from considering a general crossed module without restricting to fake-flatness. In particular, in the most general case the edge energy terms are not invariant under changes to the branching structure of the lattice and moving the base-point of a plaquette all the way around the plaquette changes the plaquette label. In 2+1d, we are only able to consider two special cases fully (rather than the three in 3+1d, although we consider some additional cases in Section \ref{Section_Example_Z_2_Z_3} and Section \ref{Section_rhd_non-trivial_condense_confine} in the Supplemental Material). These are the case when $\rhd$ is trivial, but we do not restrict the Hilbert space, and the case of a general crossed module, but where we restrict the Hilbert space to only include fake-flat configurations (where configurations are basis states for which each edge and plaquette is labelled by an appropriate group element). These two special cases are shown in Table \ref{Table_Cases_2d}. In the 3+1d case we were also able to deal with the case where $E$ is Abelian and $\partial$ maps to the centre of $G$, but $\rhd$ is general and fake-flatness is not enforced on the level of the Hilbert space (Case 2 in Table \ref{Table_Cases_2d}). However in the 2+1d model we were not able to construct the magnetic (fake-flatness violating) excitations in this case (at least not generally, although see Section \ref{Section_Z2_Z3_Magnetic} and Section \ref{Section_confined_magnetic} in the Supplemental Material for a construction in some cases). We note that restricting to the case where $\rhd$ is trivial also simplifies the group $E$ and the map $\partial$. The first Peiffer condition $\partial(g \rhd e)= g \partial(e)g^{-1}$ (for all $g \in G$ and $e \in E$) becomes $\partial(e)= g \partial(e) g^{-1}$, which implies that all elements $\partial(e)$ are in the centre of $G$. Furthermore, the second Peiffer condition $\partial(e)\rhd f = efe^{-1}$ (for each pair $e, \ f \in E$) becomes $f=efe^{-1}$, which implies that the group $E$ is Abelian.

			\begin{table}[h]
				\begin{center}
					\begin{tabular}{ |c|c|c|c|c| } 
						\hline
						& & & & Full\\
						Case & $E$ & $\rhd$ & $\partial(E)$ & Hilbert \\ 
						& & & &Space\\
						\hline
						1 & Abelian & Trivial & $\subset$ centre($G$) & Yes\\ 
						2 & Abelian & General & $\subset$ centre($G$) & Yes\\ 
						3 & General & General & General & No \\
						\hline
					\end{tabular}
					
					\caption{A summary of the special cases of the model in 2+1d. Note that Case 2, which is a generalization of Case 1 and was examined closely for the 3+1d case in Ref. \cite{HuxfordPaper1}, will not be considered here for the 2+1d case.}
					\label{Table_Cases_2d}
				\end{center}
				
			\end{table}

		\section{Ribbon operators when $\rhd$ is trivial}
		\label{Section_RO_2D_Tri_Trivial}
		
		The creation and movement of the topological excitations in the 2+1d model are governed by ribbon operators. The name ribbon operators \cite{Kitaev2003} originates from the fact that, in order to produce a pair of excitations, we typically have to act along a ribbon whose end-points lie at the positions of the excitations. We must act along a ribbon, rather than just locally at the location of our excitations, because the excitations may carry a conserved charge, called \textit{topological charge} \cite{Kitaev2006}. Because this charge is conserved, we cannot create a single charge-carrying excitation, and can instead only transport charge along a path by producing two excitations with opposite charge at the ends of the path. To do so we must act along the entire path, hence the requirement for an operator with linearly extended support. In addition to being linearly extended, the width of the creation operator may also be small but finite, hence the name ribbon operators, rather than string operators. These ribbon operators underpin most of our other results, and so we will start by constructing them and looking at how they interact with the energy terms.

		 As we mentioned in Section \ref{Section_Introduction_2}, due to the consistency issues described in Ref. \cite{HuxfordPaper1} (in Section I F), we restrict to the special cases described in Table \ref{Table_Cases_2d}. The first case that we look at is the one where $\rhd$ is trivial (that is $g \rhd e = e \: \: \forall g \in G, \: e \in E$). In this case we find two simple types of point-like excitations: excitations that violate the vertex terms (electric excitations) and excitations that violate the plaquette terms (magnetic excitations). These excitations are analogous to the electric and magnetic excitations of the Kitaev Quantum Double model \cite{Kitaev2003}, as we will see in the next sections. We can also fuse these two types of excitation to produce dyonic excitations, which excite both the vertices and plaquettes at the ends of the ribbon.

		\subsection{Electric excitations}
		\label{Section_2D_electric}
		
		The first excitations that we look at are the electric excitations, corresponding to violations of the vertex energy terms. In Section \ref{Section_Introduction_2} we explained that the vertex terms enforced 1-gauge symmetry on their lower energy eigenstates. Therefore, the electric excitations correspond to violations of this gauge symmetry.

\begin{figure}[h]
			\begin{center}
			\includegraphics[width=0.97\linewidth]{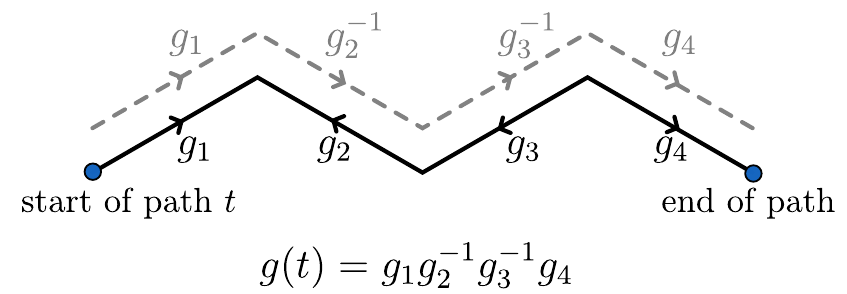}

				\caption{(Copy of Figure 41 from Ref. \cite{HuxfordPaper1}.) An electric ribbon operator measures the value of a path and assigns a weight to each possibility, creating excitations at the two ends of the path. In order to find the group element associated to the path, we must first find the contribution of each edge to the path. In this example, the edges along the path are shown in black. Some of the edges are anti-aligned with the path and so we must invert the elements associated to these edges to find their contribution to the path. This is represented by the grey dashed lines, which are labelled with the contribution of each edge to the path.}
				\label{path_image_2d}
			\end{center}
		\end{figure}

		To study these excitations, we first construct their associated ribbon operators, which produce the excitations at the ends of a ribbon. The electric ribbon operator acts along a path made of connected edges from the lattice. This ribbon operator measures the group element associated to that path and then assigns a weight to each possible value. We define an operator $\hat{g}(t)$ which measures the group element of a path $t$, where the path element is the product of the elements associated to the edges along the path (or the inverse of the edge element if an edge is anti-aligned with the path, as shown in Figure \ref{path_image_2d}). Then the electric ribbon operators applied on a path $t$ are given by
		\begin{equation}
		\hat{S}^{\vec{\alpha}}(t)= \sum_{g \in G} \alpha_{g} \delta( \hat{g}(t), g),
		\label{Equation_electric_ribbon_definition_rhd_trivial}
		\end{equation}
		where $\alpha_g$ are coefficients which describe the ribbon operator. These operators commute with all of the plaquette energy terms, because both the ribbon operators and the plaquette terms are diagonal in the basis described by group elements (where the basis states are those for which each edge is labelled by an element of $G$ and each plaquette is labelled by an element of $E$). These ribbon operators also commute with all of the vertex terms, except for the two at the ends of the path. This is because the vertex transform only affects the labels of paths starting or ending at that vertex, not paths that just pass through the vertex. As we show in Section \ref{Section_Electric_Ribbon_Operator_Proof} in the Supplemental Material, if a path passes through a vertex, then the vertex transform $A_v^g$ gives a factor of $g^{-1}$ to the contribution to the path label from the edge on which the path enters the vertex, and a factor of $g$ to the contribution from the edge on which the path leaves the vertex, which cancel. This means that the only vertex transforms that affect the path element are the ones applied on the two ends of the path, and so only the vertex energy terms at these ends may fail to commute with the ribbon operator.
		
		 If our ribbon operator commutes with particular energy terms, it means that the ribbon operator does not excite those energy terms. Therefore, the operator may excite the two vertices at the ends of the path, but not any other vertices or plaquettes. Notice that we have not yet mentioned the edge terms. We will see later in this section that the edge terms are significant and provide a mechanism for the confinement of some of the electric particles.

		Looking at the expression for our ribbon operators, given in Equation \ref{Equation_electric_ribbon_definition_rhd_trivial}, we see that there is a $|G|$-dimensional space of electric operators, because we have a coefficient for each group element. An appropriate basis for this space has basis vectors labelled by the irreducible representations of the group $G$. For these basis vectors, the coefficients are given by matrix elements of the irreps:
		\begin{equation}
		S^{R,a,b}(t) = \sum_{g \in G} [D^{R}(g)]_{ab} \delta(\hat{g}(t), g),
		\label{electric_basis}
		\end{equation}
		where $[D^{R}(g)]$ is the matrix representation of $g$ in representation $R$, while $a$ and $b$ are the indices of the matrix. The fact that the operators labelled by the possible $R$, $a$ and $b$ in this way do indeed form a basis for the space of electric ribbon operators follows from the Grand Orthogonality Theorem of representation theory (a proof that such a change of basis from group elements to irreps is allowed is given in Ref. \cite{Buerschaper2009} for example). We expect (and show in Section \ref{Section_2D_charge_electric_excitation} in the Supplemental Material) that the \textit{topological sectors} of pure electric excitations are labelled by these irreps. A topological sector is a group of particle types that carry the same conserved charge, called topological charge. Any degrees of freedom within the sector (such as the matrix indices $a$ and $b$ here) describe non-conserved, local quantities. These local quantities may be changed by local operators, but to change the topological charge of an excitation we must transport that topological charge away using a ribbon operator. A particularly important idea related to topological charge is that there must be a so-called vacuum topological charge, which is carried by the ground state. We can see this reflected in the electric ribbon operators. When $R$ is the trivial irrep, given by $D^{1_R}(g)=1 \ \forall g \in G$, the ribbon operator is given by
		$$\sum_{g \in G} \delta( \hat{g}(t), g) =1,$$
		which is just the identity operator. Naturally, the identity operator does not produce any excitations, and so will not excite the two vertices at the end-points of the path. In fact, of the electric ribbon operators labelled by the irreps of $G$, only the electric ribbon labelled by the trivial irrep fails to excite the two vertices. This suggests that the irrep labels the topological charge and the trivial irrep corresponds to the vacuum charge. This is true for pure electric excitations, but the picture is more complicated when we also allow for magnetic excitations. Nonetheless, the trivial irrep labels the case where the electric part of the topological charge is trivial and labels the vacuum if the magnetic part is also trivial.

		Those familiar with Kitaev's Quantum Double model \cite{Kitaev2003} may notice that these ribbon operators are exactly the same as the ribbon operators corresponding to the electric excitations in the Kitaev Quantum Double model. However in the higher lattice gauge theory case, these electric excitations have an additional feature: They may be \textit{confined}. By confined, we mean that when we create a pair of excitations from the vacuum, there is an energy cost for pulling the pair apart. This energy grows with the length of the ribbon operator we apply to move them. In this model, the mechanism for this confinement is through the edge terms of the Hamiltonian. Unlike in the Kitaev Quantum Double model, there are energy terms associated to the edges of our lattice that allow the edges to be excited. We find that some of the electric operators excite every edge along the path that they act on, so that the energy cost of the ribbon operator has a component proportional to the length of the ribbon.

		 In order to determine which electric excitations are confined, we need to consider how the edge terms interact with the path element measured by the ribbon operators. Recall from Section \ref{Section_Introduction_2} that an edge transform $\mathcal{A}_i^e$ changes the label of the edge $i$ on which the operator is applied by the element $\partial(e)$. Because a ribbon operator measures the group element of a path on the lattice, the ribbon operator can be sensitive to this change in path element and so will not commute with the edge transform in general. However the element $\partial(e)$ is not a general element of $G$, but instead lies in the image of $\partial$. This means that the potential non-commutativity of the ribbon operators and edge transforms is controlled by the map $\partial$. For example, if $\partial$ mapped only to the identity element of $G$, then the factor $\partial(e)$ would be trivial and so the edge transform would not affect the path element, meaning that all of the ribbon operators would commute with the edge transforms. Even when $\partial$ is non-trivial, not every electric ribbon operator is sensitive to factors in the image of $\partial$. Considering the general electric ribbon operator given in Equation \ref{Equation_electric_ribbon_definition_rhd_trivial}, we can divide the ribbon operators into a class of confined ribbon operators and a class of unconfined ones. As we prove in Section \ref{Section_Electric_Ribbon_Operator_Proof} of the Supplemental Material, a ribbon operator is unconfined if the coefficients $\alpha_g$ satisfy 
		 \begin{equation}
		 \alpha_g = \alpha_{g\partial(e)} \text { for all $g \in G$ and $e \in E$} .
		 \end{equation}
		That is, the ribbon operator is unconfined if $\alpha_g$ is only a function of the coset of $\partial(E)$ in $G$ and not of elements within the coset. On the other hand a ribbon operator is confined in orthogonal cases, where the coefficients within each coset sum to zero:
		 \begin{equation}
		 \sum_{e \in E} \alpha_{g \partial(e)}= 0 \ \forall g \in G.
		 \end{equation}
		 A general ribbon operator will have coefficients that can be split into a sum of two parts, the first being a function of coset only and the second summing to zero over a coset. This pattern of confinement is the same as that for the related 4d field theory discussed in Ref. \cite{Gukov2013}, where the electric operators are confined if they can detect factors in a subgroup $\pi_1(H)$ (equivalent to $\partial(E)$ here)

		 Another way to look at which ribbon operators are confined is to use the irrep basis described in Eq. \ref{electric_basis}, for which it is easy to separate the confined and unconfined excitations. Because the image of $\partial$ is a (normal) subgroup of $G$, each irreducible representation of $G$ gives us a representation of $\partial(E)$, by restricting the irrep to the subgroup. That is, for each irrep $R$ of $G$ we have a representation of $\partial(E)$, $R_{\partial}$, defined by $R_{\partial}(h)=R(h)$ for $h \in \partial(E)$. If this representation is trivial, then the electric ribbon operator is not sensitive to changes to the path element in $\partial(E)$ and so is not sensitive to the edge transform, therefore commuting with the transform (and the edge energy term). On the other hand, if the ribbon operator carries some non-trivial irrep of the image of $\partial$, then applying the edge term (which is an average of the edge transforms with each label in $E$) applies factors to the path element which average over the elements in the image of $\partial$. Because averaging over a non-trivial irrep gives zero by the orthogonality theorem, this leads to the edge energy term annihilating the state produced by the ribbon operator. This means that the edge energy terms along the ribbon are excited by the ribbon operator, and so the corresponding excitations are confined.

		For example, consider the quaternion group $\mathbb{Q}_8$. This group has elements $1$, $-1$, $i$, $-i$, $j$, $-j$, $k$ and $-k$. The group has a two-dimensional irrep with corresponding matrices given by $\pm \mathbf{1}_2$ and $\pm i \sigma_l$ (for $l$=1, 2, 3), where $\sigma_l$ are the three Pauli matrices. $\mathbb{Q}_8$ has a $\mathbb{Z}_2$ subgroup $\set{1,-1}$. By restricting the two-dimensional irrep to these elements, we get the matrices $\pm \mathbf{1}_2$, which form a (reducible) representation of the subgroup $\mathbb{Z}_2$.	As we see in this $\mathbb{Q}_8$ example, the restricted representation $R_{\partial}$ need not be an \textit{irreducible} representation of $\partial(E)$. However, when $\rhd$ is trivial, $\partial(E)$ is in the centre of $G$ (and so the subgroup is Abelian). This means that, from Clifford's theorem \cite{Clifford1937}, the restricted representation $R_{\partial}$ branches into copies of one particular (1D) irrep of $\partial(E)$, $R_{\partial}^{\text{irr.}}$. Because $\partial(E)$ is in the centre of $G$, this can also be understood from Schur's Lemma, which enforces that the matrices in the restricted irrep must be scalar multiples of the identity matrix. As an example of this branching, in the $\mathbb{Q}_8$ example above, the 2D irrep gives us two copies of the non-trivial irrep of $\mathbb{Z}_2$, which has phases $\set{1,-1}$ representing the two elements of $\mathbb{Z}_2$. If the irrep $R_{\partial}^{\text{irr.}}$ is trivial, so that $R_{\partial}^{\text{irr.}}(g)=1$ for all $g \in \partial(E)$, then the excitations labelled by the irrep $R$ are unconfined. On the other hand, if the irrep $R_{\partial}^{\text{irr.}}$ is non-trivial (as in the example above) then the excitations labelled by $R$ are confined. A similar argument holds when $\rhd$ is not trivial (so that $\partial(E)$ is not in the centre of $G$), though in that case $R_{\partial}^{\text{irr.}}$ need not be 1D, and rather than copies of $R_{\partial}^{\text{irr.}}$ we will have different irreps related to $R_{\partial}^{\text{irr.}}$ by conjugation, as we can deduce from Clifford's theorem \cite{Clifford1937}.

		Having obtained the ribbon operators that produce the electric excitations, we can make use of them to examine the properties of these excitations. One important property is the set of \textit{fusion} rules of the excitations. To understand fusion, recall that our excitations carry some conserved charge. Then given two such excitations, we can bring them close together and ask what their total charge is. We say that the two charges fuse to the total charge. To find the fusion properties we consider applying two electric operators in sequence on the same path. This corresponds to the process where we produce a pair of excitations and separate them using a ribbon operator, before producing another pair of excitations and moving them to the locations of the first pair, i.e., to the ends of the same ribbon. Then we want to ask what total charge is located at each end of the ribbon. Applying these ribbon operators, we have that
		\begin{align*}
		\sum_{g \in G}& [D^{R_1}(g)]_{ab} \delta(g,\hat{g}(t))\sum_{k \in G} [D^{R_2}(k)]_{cd} \delta(k,\hat{g}(t))\\
		&=\sum_{g \in G} \sum_{k \in G} [D^{R_1}(g)]_{ab} [D^{R_2}(k)]_{cd} \delta(g,\hat{g}(t)) \delta(k,\hat{g}(t))\\
		&=\sum_{g \in G} \sum_{k \in G}[D^{R_1}(g)]_{ab} [D^{R_2}(k)]_{cd} \delta(g,k) \delta(k,\hat{g}(t))\\
		&=\sum_{g \in G} [D^{R_1}(g)]_{ab} [D^{R_2}(g)]_{cd} \delta(g,\hat{g}(t)).
		\end{align*}
		As described previously in this section, the irreps $R_1$ and $R_2$ label conserved charges, while the matrix indices describe internal degrees of freedom. By allowing $a$, $b$, $c$ and $d$ to vary here (which moves us through the internal spaces of the individual sectors), we see all possible results from fusing the charges labelled by $R_1$ and $R_2$. The result is the tensor product of the irreps $R_1 \otimes R_2$, with four matrix indices. This tensor product is a representation but not necessarily an irreducible one. However we can then decompose this representation into irreps. The resulting irreps will describe the possible charges resulting from fusion. The fusion of the electric excitations is therefore described by the decomposition of products of irreps of the group $G$ (i.e., the electric excitations are described by the fusion category Rep($G$), although the non-confined excitations form a subcategory).

		For fusion it is not strictly necessary to have the ribbon operators acting on the same path. It is sufficient to bring the excitations close together. However when the paths do not align, we cannot simply write the product of ribbon operators as a single ribbon operator, and so we must use another method to find the charge of the two excitations. We consider how to find the charge of a general set of excitations in a different way in Section \ref{Section_2D_topological_Charge}. Having said that, there are cases where we can combine operators on different paths simply. This occurs when the path $s$ has the same start and end-points as $t$, and $s$ can be deformed smoothly into $t$ without crossing any excitations. In this case, due to the plaquette constraints satisfied by unexcited regions of the lattice, the group elements assigned to the paths $s$ and $t$ only differ by an element of $\partial(E)$. For the unconfined excitations, this has no effect, because such excitations are described by trivial representations of $\partial(E)$. This illustrates a general point: the ribbon operators associated to non-confined excitations in our model are topological, in the sense that deforming the ribbons over a region with no excitations does not affect the result of acting with the operator. This is proven for each ribbon and membrane operator in the Supplemental Material (see Section \ref{Section_topological_membrane_operators}).

		\subsection{Magnetic excitations}
		\label{Section_2D_Magnetic}
\begin{figure}[h]
			\begin{center}
				\includegraphics[width=\linewidth]{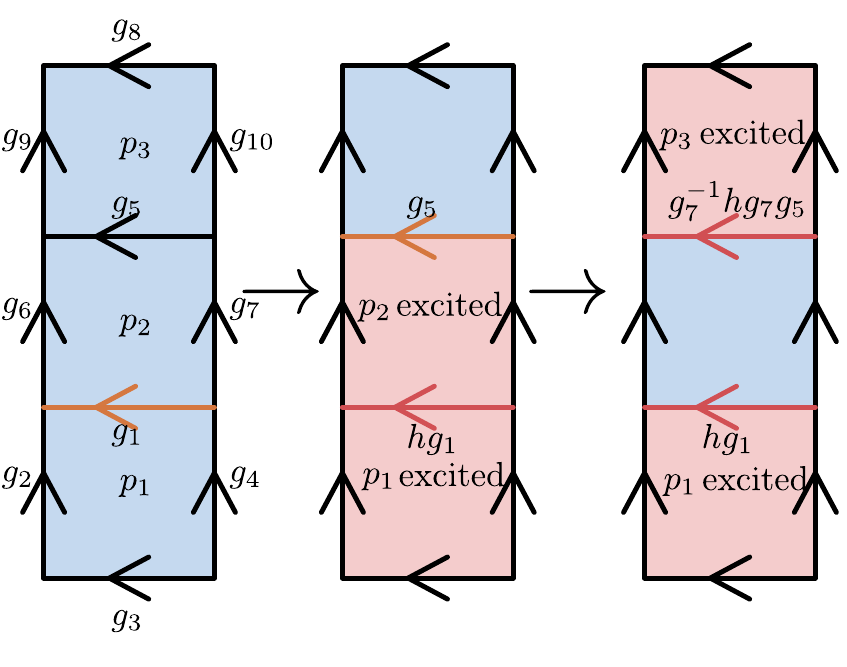}
				
				\caption{If we consider trying to produce a single plaquette excitation by changing an edge label (here $g_1$) we instead excite two plaquettes, as shown in the second diagram. Changing further edges (here the edge labelled by $g_5$ in the second diagram) just pushes these excitations apart, as shown in the third diagram.}
				\label{plaquette_2D_step}
			\end{center}
			
		\end{figure}

		The next type of excitation that we examine is called the magnetic excitation, due to the similarity with the magnetic excitation in the Kitaev Quantum Double model \cite{Kitaev2003}. The magnetic excitations are ``flux" excitations, corresponding to excited plaquette terms. They are called magnetic excitations because they are associated to closed loops with non-trivial flux (1-holonomy), just like the magnetic field in the Aharanov-Bohm effect. Just as with the electric excitations, the ribbon operators that produce the magnetic excitations are the same as those in the Kitaev Quantum Double model \cite{Kitaev2003}. To find the ribbon operators for the magnetic excitations, we consider trying to excite a plaquette term. Recall from Section \ref{Section_Introduction_2} that the plaquette term measures the path around the plaquette and checks if it is compatible with the surface element of the plaquette. Denoting the path around plaquette $p$, starting at the base-point of $p$ and matching the orientation of the plaquette, by $r(p)$, the plaquette term is $\delta(\partial(e_p )g(r(p)),1_G)$. Starting in the ground state, to excite such a term we can either change the plaquette label or the edge labels of the edges that make up the path $r(p)$. The former case is considered in Section \ref{Section_Single_Plaquette_1}. Here we consider the latter approach. We consider trying to change a single edge on a plaquette, by multiplying it by some group element in $G$. However each edge is shared by two plaquettes, so performing this operation will excite two plaquettes rather than one. For example, consider the situation in Figure \ref{plaquette_2D_step}. In the left image we have one of the configurations in the ground state, satisfying the plaquette constraints. Then we multiply the edge label $g_1$ by an element $h$ (which will label our magnetic ribbon operator) to give $hg_1$. Looking at the bottom plaquette ($p_1$), initially $$\partial(e_{p_1} )g(r(p_1))=\partial(e_{p_1})g_1g_2^{-1}g_3^{-1}g_4=1_G.$$ Then after our operator acts, we have 
		\begin{align*}
		\partial(e_{p_1} )g(r(p_1))&= \partial(e_{p_1})hg_1g_2^{-1}g_3^{-1}g_4\\
		&=h \partial(e_{p_1}) g_1g_2^{-1}g_3^{-1}g_4\\
		&=h 1_G\\
		&=h, 
		\end{align*}
		where we used the fact that $\partial(e_{p_1})$ is in the centre of $G$ due to $\rhd$ being trivial (see Section \ref{Section_Recap_Paper_2}). Looking at Figure \ref{plaquette_2D_step}, we see that changing $g_1$ will also excite the plaquette $p_2$, as shown in the middle diagram, because the corresponding edge is shared by the two plaquettes $p_1$ and $p_2$. Therefore, we consider trying to correct the plaquette holonomy of $p_2$, to leave $p_2$ unexcited, by changing another edge label. To do this, we try multiplying the edge label $g_5$ by some element $x$. The plaquette holonomy of $p_2$ was originally given by $$\partial(e_{p_2} )g(r(p_2))= \partial(e_{p_2})g_5g_6^{-1}g_1^{-1}g_7=1_G.$$ After changing the two edges, we have instead that
		\begin{align}
		\partial(e_{p_2} )g(r(p_2))&= \partial(e_{p_2})xg_5g_6^{-1}g_1^{-1}h^{-1}g_7 \notag \\
		&=x \partial(e_{p_2}) g_5g_6^{-1}g_1^{-1}(g_7 g_7^{-1})h^{-1}g_7 \notag \\
		&=x (\partial(e_{p_2}) g_5g_6^{-1}g_1^{-1} g_7) g_7^{-1}h^{-1}g_7 \notag\\
		&=x 1_G g_7^{-1}h^{-1}g_7 \notag\\
		&=x g_7^{-1}h^{-1}g_7. \label{Equation_magnetic_try}
		\end{align}
		We see that for this to give $1_G$ (and so to leave the plaquette unexcited), we need to choose $x=g_7^{-1}hg_7$. We note that the value of $x$ needed depends on the edge label $g_7$ and therefore depends on the state that the ribbon operator is acting on. This action reduces to multiplying both edges by $h$ when $G$ is Abelian, but in the non-Abelian case we need to keep track of additional edge labels.

		While this action leaves the plaquete $p_2$ unexcited, because the second edge that we changed is also shared by two plaquettes, a third plaquette ($p_3$ in the figure) must now be excited. Therefore, we have simply separated the two plaquette excitations that we initially produced and cannot remove the second excitation (at least not without moving it back to the first excitation), indicating that the magnetic excitations are produced in pairs.

\begin{figure}[h]
			
			\begin{center}
				\includegraphics[width=0.95\linewidth]{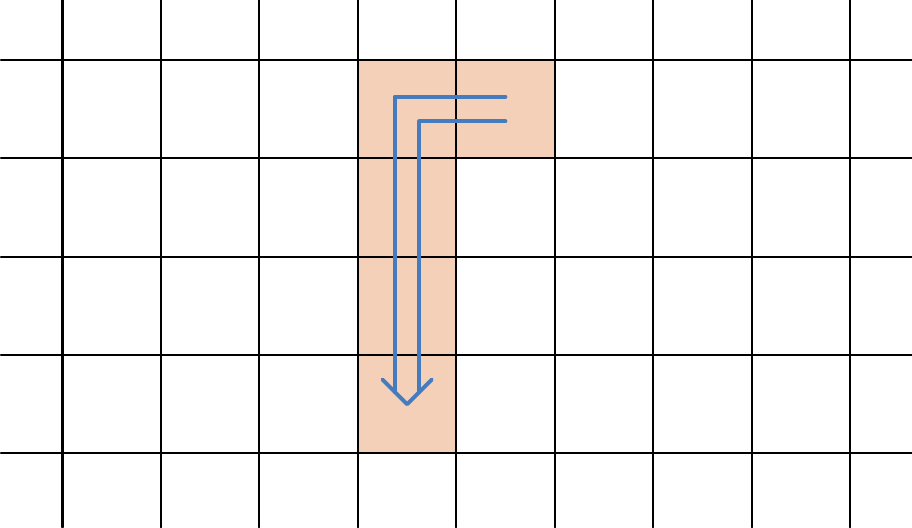}

				\caption{An example of a dual path on a lattice. This dual path includes the shaded plaquettes, which take the role usually held by vertices on a path. Note that the path passes through, rather than along, the edges of the lattice.}
				
				\label{DualPath}
			\end{center}
			
		\end{figure}

		Looking at Figure \ref{plaquette_2D_step}, we note that the edges that we changed (the ones labelled by $g_5$ and $g_1$) are bisected by a path between the two excited plaquettes. Indeed, generally to create two plaquette excitations we must act on all edges that are bisected by a path that connects those two plaquettes. This path exists on the dual lattice: it travels between the centres of plaquettes and bisects the edges of the lattice. This is in contrast with a path on the direct lattice, which passes from vertex to vertex along the edges of the lattice. Therefore, we call this path the dual path of the ribbon operator. A simple example of such a dual path is shown in Figure \ref{DualPath}.

\begin{figure}[h]
			
			\begin{center}
				\includegraphics[width=\linewidth]{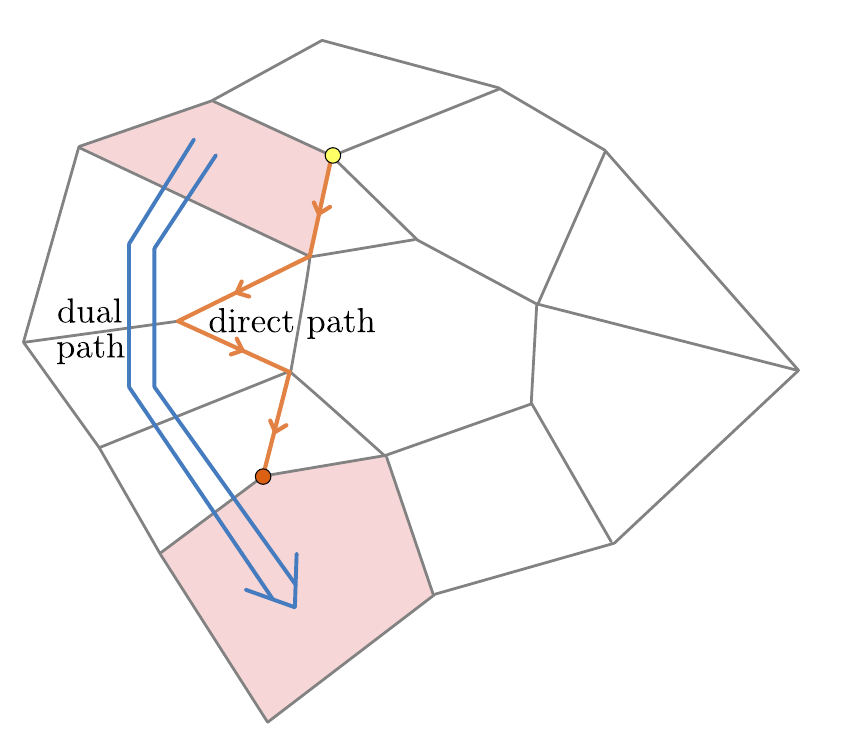}
				
				\caption{An example of the dual and direct paths. The dual path is the (blue) double arrow between the shaded plaquettes and the direct path is the (orange) solid line between the two vertices shown by circles. The edges cut by the dual path of the magnetic ribbon operator have their labels changed by the operator.}
				
				\label{PathAndDualPath}
			\end{center}
			
		\end{figure}

		We already mentioned that the factor $x=g_7^{-1} h g_7$ by which we must change the second edge in our simple example (see Figure \ref{plaquette_2D_step}) is not a constant, but instead depends on the edge label $g_7$. The edge labelled by $g_7$ lies on a path between the two excited plaquettes, but on the direct lattice rather than the dual lattice. Generally, to define our magnetic ribbon operator we must define both a direct path, which passes from vertex to vertex along the edges of the lattice, and a dual path, which passes between plaquettes and so cuts through edges on the lattice. For example, in Figure \ref{plaquette_2D_step} the direct path is the edge labelled by $g_7$, while the dual path passes from plaquette $p_1$ to plaquette $p_2$ and cuts through the edges labelled by $g_1$ and $g_5$. The space between the dual and direct paths forms a ribbon, which is the origin of the name ribbon operator \cite{Kitaev2003}. We give an example of a ribbon, along with its corresponding direct and dual paths, in Figure \ref{PathAndDualPath}.

		Having given a rough motivation and description of the magnetic ribbon operator, we will now be more specific about its action. As we just described, when defining our ribbon operator we must specify a dual path, which connects the two plaquettes to be excited. We must also give a direct path, which runs from a specified start-point (which we usually take to lie on the first excited plaquette) and a vertex on the second excited plaquette. Which edges are affected by the multiplication from the ribbon operator depends on the dual path, but precisely which factor they are multiplied by depends on the label of an appropriate section of the direct path. A magnetic ribbon operator $C^h(t)$, labelled by an element $h$ of $G$ and acting on a ribbon $t$, affects the label $g_i$ of an edge $i$ cut by the dual ribbon according to
		\begin{equation}
		C^h(t):g_i= \begin{cases} g(\tilde{t}_i)^{-1}h g(\tilde{t}_i)g_i &\mbox{if edge } i \mbox{ points away}\\ &\mbox{ from the direct path}\\ g_i g(\tilde{t}_i)^{-1}h^{-1} g(\tilde{t}_i)&\mbox{if } i \mbox{ points towards}\\ &\mbox{ the direct path,} \end{cases}
		\label{Equation_Magnetic_Action_C}
		\end{equation}
		given that $\tilde{t}_i$ is the path from the start-point of the direct path up to the edge $i$.

		\begin{figure}[h]
			\begin{center}
				\includegraphics[width=\linewidth]{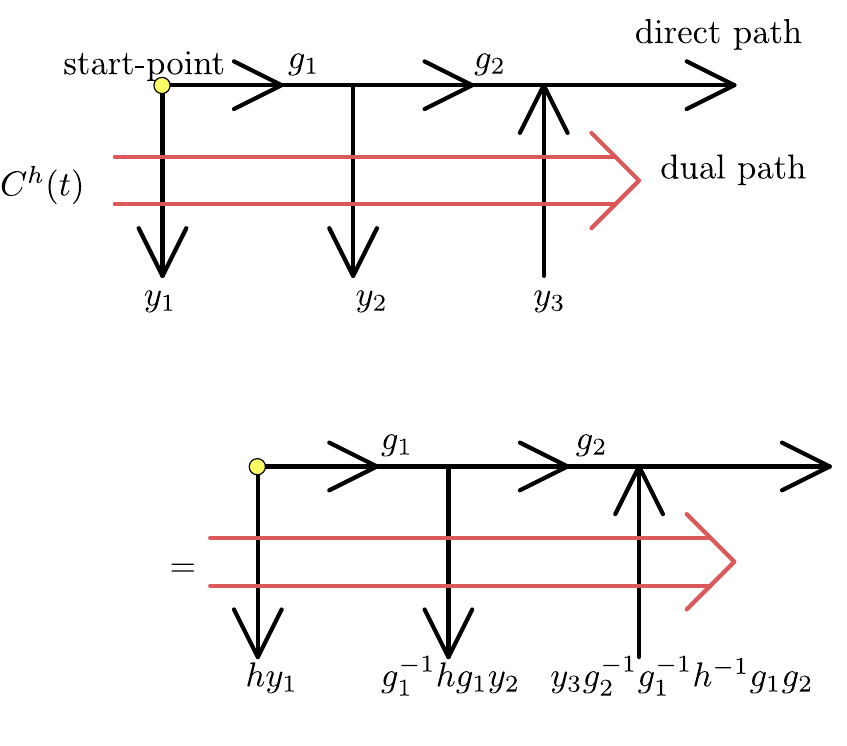}
					
				\caption{An explicit example of the action of the magnetic ribbon operator on the edges. The action on the edges cut by the dual path (the vertical edges) depends on the labels of the edges on the direct path (horizontal line) and on the orientation of the affected edges.}
				\label{magnetic_ribbon_1}
			\end{center}
		\end{figure}

		As an example of the action of the magnetic ribbon operator, consider Figure \ref{magnetic_ribbon_1}. We see that the cut edges are either left-multiplied by some element in the conjugacy class of $h$, or right multiplied by some element in the conjugacy class of $h^{-1}$, depending on whether the edge points into or out of the ribbon. The element within the conjugacy class is determined by the direct path from the start of the ribbon to the edge being cut. The reason that this dependence on the direct path is necessary is that it ensures that plaquettes along the ribbon (apart from at the ends of our ribbon) are not excited, as we saw in our earlier example.

		As we have discussed already, the ribbon operator has the effect of exciting the two plaquettes at the end of the ribbon (the shaded ones in Figure \ref{PathAndDualPath}). However this is not the only excitation that may be produced by the ribbon operator. The first vertex on the direct path, which we call the start-point of the ribbon operator (the yellow circle in Figure \ref{PathAndDualPath}), may be excited. The reason that this start-point may be excited is the dependence of the action of the magnetic ribbon operator on the direct path. As shown in Figure \ref{magnetic_ribbon_1} and Equation \ref{Equation_Magnetic_Action_C}, the ribbon operator multiplies the element of each edge cut by the dual path by an element $g(\tilde{t}_i)^{-1}h g(\tilde{t}_i)$ or its inverse, where $\tilde{t}_i$ is the path from the start-point to the edge being changed, and $g(\tilde{t}_i)$ is the group element assigned to that path. If we apply a vertex transform at the start-point, then we will change this path element, and so the vertex transform will not commute with the ribbon operator. Specifically, the action of the vertex transform $A_v^g$ takes the path element $g(\tilde{t}_i)$ to $g g(\tilde{t}_i)$, which means that the factor $g(\tilde{t}_i)^{-1}h g(\tilde{t}_i)$ in the action of the ribbon operator becomes $g(\tilde{t}_i)^{-1}g^{-1}h g g(\tilde{t}_i)$. We can think of this as replacing the label $h$ of the ribbon operator by $g^{-1}hg$ and so our ribbon operator $C^h(t)$ becomes $C^{g^{-1}hg}(t)$ instead. In terms of a commutation relation, we have 
		\begin{equation}
		C^h(t)A_v^g=A_v^g C^{g^{-1}hg}(t),
		\label{Equation_magnetic_ribbon_vertex_commutation_1}
		\end{equation}
		where $v$ is the start-point of $t$, and so generally the vertex transform at the start-point does not commute with the ribbon operator. To determine whether the vertex is excited, we need to look at the vertex energy term, which is an average over the vertex transforms: $A_v = \frac{1}{|G|} \sum_{g \in G} A_v^g$. Applying this to our commutation relation Equation \ref{Equation_magnetic_ribbon_vertex_commutation_1}, we have 
		\begin{equation}
		A_v C^h(t) = \frac{1}{|G|} \sum_{g \in G} C^{ghg^{-1}}(t)A_v^g 	\label{Equation_magnetic_ribbon_vertex_commutation_2}.
		\end{equation}
	
		 Now we need to use this to evaluate the state produced by acting with the ribbon operator on the ground state, to see if the vertex is excited. Because $A_v$ is a projector, with eigenvalue one for the lower energy state and eigenvalue zero for the higher energy state, the expression $A_v C^h(t)\ket{GS}$ will give zero if the vertex is excited and $C^h(t)\ket{GS}$ if it is not excited. To evaluate this expression we can first use the fact that the ground state is an eigenstate of the vertex term to pull out a factor of $A_v$, to obtain
		\begin{align*}
		A_v C^h(t)\ket{GS} &= 	A_v C^h(t)A_v\ket{GS}.
		\end{align*}
		
		We can then use the commutation relation between the ribbon operator and vertex term given in Equation \ref{Equation_magnetic_ribbon_vertex_commutation_2}, to find that
		\begin{align*}
		A_v C^h(t)\ket{GS}&=\frac{1}{|G|} \sum_{g \in G} C^{ghg^{-1}}(t)A_v^g A_v\ket{GS} .
	\end{align*}
	Then, using the fact that $A_v^g A_v =A_v$ (as described in Section \ref{Section_Recap_Paper_2}) we have
		\begin{align}
		A_v C^h(t)\ket{GS}&=\frac{1}{|G|} \sum_{g \in G} C^{ghg^{-1}}(t) A_v\ket{GS} \notag\\
		&=\frac{1}{|G|} \sum_{g \in G} C^{ghg^{-1}}(t) \ket{GS}, \label{Equation_commutation_magnetic_ribbon_start-point_term}
		\end{align}
		where we used the relation $A_v\ket{GS} =\ket{GS}$ in the final step. We therefore obtain an average over the ribbon operators that have label in the conjugacy class of $h$. We see that the magnetic ribbon operator does not generally produce an eigenstate of $A_v$ when acting on the ground-state, and so the start-point vertex is not generally in a excited or unexcited state. In order to construct ribbon operators that leave the vertex in a definite energy state, we must construct a linear combination of magnetic ribbon operators with different labels $h$. A general magnetic operator on the path $t$ is given by $ C_{\vec{\alpha}}(t) = \sum_{h \in G} \alpha_h C^h(t)$, where $\vec{\alpha}$ is a set of coefficients. Using Equation \ref{Equation_commutation_magnetic_ribbon_start-point_term} with this linear combination, we obtain
		\begin{align*}
			A_v C_{\vec{\alpha}}(t)\ket{GS}&= A_v \sum_{h \in G} \alpha_h C^h(t) \ket{GS}\\
			&= \frac{1}{|G|} \sum_{g \in G} \sum_{h \in G} \alpha_h C^{ghg^{-1}}(t) \ket{GS}.
			\end{align*}
		
		Replacing the dummy index $h$ with $h' =ghg^{-1}$, we have
		
		\begin{align}
			A_v C_{\vec{\alpha}}(t)\ket{GS}&= \frac{1}{|G|} \sum_{g \in G} \sum_{h' \in G} \alpha_{g^{-1}h'g} C^{h'}(t) \ket{GS} \notag\\
		&= \sum_{h' \in G} \big(\frac{1}{|G|} \sum_{g \in G} \alpha_{g^{-1}h'g}\big) C^{h'}(t) \ket{GS}. \label{Equation_commutation_magnetic-ribbon_start-point_term_2}
	\end{align}
		Then the coefficients $\vec{\alpha}$ determine whether or not the start-point is excited. If the coefficients $\alpha_h$ are a function of conjugacy class, so that $\alpha_h=\alpha_{xhx^{-1}} \: \forall h,x \in G$, then the start-point vertex is not excited. For example, we could have the operator $C_{[h]}(t)= \sum_{x \in G} C^{xhx^{-1}}(t)$, which is an equal sum over all elements of the conjugacy class of $h$. In this case the expression $\frac{1}{|G|} \sum_{g \in G} \alpha_{g^{-1}h'g}$ in Equation \ref{Equation_commutation_magnetic-ribbon_start-point_term_2}, which averages over elements in the conjugacy class of $h'$, just gives us $\alpha_{h'}$, because the coefficients are already a function of conjugacy class (specifically $\alpha_{h'}$ is one if $h'$ is in the same class as $h$ and zero otherwise). Therefore, the vertex term commutes with such a linear combination of magnetic ribbon operators and so the start-point vertex would be unexcited. On the other hand, if the coefficients for the elements of each conjugacy class sum to zero, then the vertex is excited. That is, the vertex is definitely excited if $\sum_{g \in G} \alpha_{g^{-1}h'g^{-1}}=0$ for each $h'$ in $G$, as we can see from Equation \ref{Equation_commutation_magnetic-ribbon_start-point_term_2}. For example, consider the case where we have a conjugacy class with two elements, $x$ and $y$, and we have a superposition over only these two elements: $aC^x(t) +b C^y(t)$. If $a+b=0$, then the vertex is excited.

		Generally any linear combination of basis magnetic ribbon operators can be written as a sum of two parts, the first of which is a function of conjugacy class and the second of which has coefficients which sum to zero within each conjugacy class. This is because, given coefficients $a_g$ for the elements within a conjugacy class $[g_1]$, we can extract the average coefficient to obtain the part which is a function of conjugacy class, while the remainder will sum to zero. That is, we write 
		\begin{equation}
		\sum_{g \in [g_1]} a_g C^g(t)= \langle a \rangle \sum_{g \in [g_1]} C^g(t) + \sum_{g \in [g_1]}(a_g - \langle a \rangle)C^g(t),
		\label{magnetic_coefficient_split}
		\end{equation}
		where $\langle a \rangle $ is the mean coefficient for that conjugacy class. The first term has the same coefficient for each operator in the conjugacy class, and the coefficients in the second term sum to zero by the definition of the mean. Therefore, our space of ribbon operators can be decomposed in terms of ribbon operators that excite the start-point and ones that do not.

		From considering the start-point excitations, we can see that conjugacy classes of $G$ are important when considering the magnetic ribbon operators. These conjugacy classes are also important for considering the conserved topological charge of the excitations, along with their braiding. We can think of the magnetic excitation as being labelled by a conjugacy class and some internal degrees of freedom which describe the coefficients within the conjugacy class. The topological sectors will be unions of these classes, as we will discuss in Section \ref{Section_2D_Condensation_Confinement}, while the internal degrees of freedom contribute to the braiding of the magnetic excitations around each-other together with the conjugacy class.

		Now that we have found the ribbon operators for the magnetic excitations, we can consider fusion of the magnetic excitations. If we apply two magnetic ribbon operators, labelled by $g$ and $h$, on the same ribbon in sequence, we get $gh$ or $hg$ depending on the order in which we apply them (note that these are in the same conjugacy class and so the resulting excitations are in the same sector). That is, we have
		\begin{equation}
		C^g(t) C^h(t) =C^{gh}(t), \label{Equation_magnetic_ribbon_fusion}
		\end{equation}
		when combining two ribbons of the form shown in Figure \ref{magnetic_ribbon_1}.

		In addition to considering ribbon operators that are purely magnetic or purely electric, we can consider ribbons made by applying both a magnetic and electric ribbon on the same space (that is, the electric ribbon is placed on the direct path of the magnetic one). We write this as $F^{h,g}(t)=\delta(g,g(t))C^h(t)$, following the notation used by Kitaev in Ref. \cite{Kitaev2003}. Note that, while the magnetic part acts on a ribbon $t$, the electric operator only acts on the direct path of that ribbon. Nonetheless, following Kitaev's notation, we will use $t$ as the argument for the electric part as well, and hope that it is clear which part of the ribbon is relevant for the electric operator. These more general ribbon operators obey the fusion rules
		\begin{equation}
		F^{h_1,g_1}(t) F^{h_2,g_2}(t)= \delta(g_1,g_2) F^{h_1 h_2, g_1}(t). \label{Equation_fusion_EM_ribbons_1}
		\end{equation}
		Note that the electric part is not labelled by an irrep, so to recover the familiar fusion of electric excitations from Section \ref{Section_2D_electric} we must take appropriate linear combinations of these ribbon operators.
		
		\subsection{Single plaquette multiplication operators}
		\label{Section_Single_Plaquette_1}
		
		In Section \ref{Section_2D_Magnetic}, we mentioned that there are two ways to produce excitations of the plaquette energy terms. In addition to exciting the plaquettes by changing the edges around them (as we did to produce the magnetic excitations), we can change the plaquette label itself. This can be done simply by multiplying a single plaquette label $e_p$ by a group element $e$ (note that when $E$ is Abelian we do not need to worry about the order of multiplication). We denote the operator that does this to a plaquette $p$ by $M^e(p)$ and refer to such operators as ``single plaquette multiplication operators". If $e$ is in the kernel of $\partial$, applying the operator $M^e(p)$ does not change $\partial(e_p)$. Because the plaquette condition $\partial(e_p)g(\text{boundary})=1_G$ only depends on $e_p$ through $\partial(e_p)$, this means that no plaquette excitation is formed. On the other hand, if $e$ is not in the kernel, this operator causes a fake-flatness violation and so we have an excitation of a single plaquette $p$ (no other plaquettes are affected). Because this excitation is formed on its own by a local operator, the excitation does not correspond to a non-trivial anyon (it cannot carry non-trivial topological charge).

		From the fact that the single plaquette multiplication operators, much like the magnetic ribbon operators, produce plaquette excitations, we may suppose that there is a connection between the two types of operator. Indeed, as we show in Section \ref{Section_Condensation_Magnetic_2D} in the Supplemental Material, some of the magnetic ribbon operators act on the ground state in the same way as a pair of single plaquette multiplication operators applied on the two plaquettes at the ends of the ribbon. Specifically, this is true for magnetic ribbon operators whose labels lie in $\partial(E)$. The single plaquette multiplication operators act on the plaquettes, whereas the ribbon operator acts on the edges of the lattice, but for these particular ribbon operators the difference between the two types of action is a series of edge transforms that act trivially on the ground state. This means that the action of such a magnetic ribbon operator on the ground state is equivalent to that of local operators (the single plaquette multiplication operators) at the ends of the ribbon. This indicates that the excitations produced by such a ribbon operator are equivalent to those produced by the (local) single plaquette multiplication operators and so are not topologically protected (the excitations must carry trivial topological charge). Such an excitation is called \textit{condensed} (we will discuss the concept of condensation further in Section \ref{Section_2D_Condensation_Confinement}). Recalling from Section \ref{Section_2D_Magnetic} (see Equation \ref{Equation_magnetic_ribbon_fusion}) that the fusion rule for magnetic ribbons is $C^g(t) C^h(t) = C^{gh}(t)$, we have that $C^g(t) C^{\partial(e)}(t)=C^{g\partial(e)}(t)$. Then because $C^{\partial(e)}(t)$ is equivalent to the action of local operators when acting on the ground state, we see that $C^{g}(t)$ and $C^{g\partial(e)}(t)$ are equivalent up to some local operators (again, when acting on the ground state). This indicates that the magnetic excitations labelled by $g$ and $g\partial(e)$ belong to the same topological sector. The sectors are therefore not just given by conjugacy classes of $G$, unlike in Kitaev's Quantum Double model \cite{Kitaev2003}, but instead are given by unions of conjugacy classes that are related by multiplication by elements of $\partial(E)$. We can think of these unions as conjugacy classes of cosets of $\partial(E)$ in $G$. This is because a class that includes an element $g$ includes all elements of each coset $xgx^{-1} \partial(E)$ for each $x \in G$ (i.e., the elements of cosets corresponding to elements of the conjugacy class of $g$ in $G$). Because $\partial(E)$ is a normal subgroup, these cosets can also be written as $xg \partial(E)x^{-1}$.

		\subsection{Loop excitations}
		\label{Section_2D_Loop}
\begin{figure}[h]
		\begin{center}
			\includegraphics{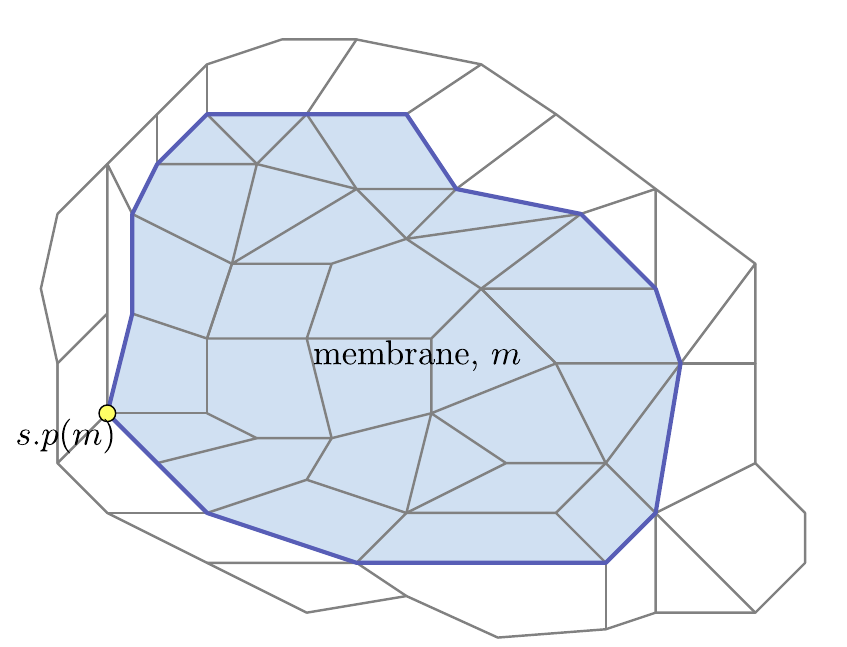}
			
			\caption{Consider an $E$-valued membrane operator applied on a fragment of the lattice. We measure the surface label of the membrane (shaded area) and apply a weight to each possibility. This excites the edges on the boundary of the membrane (solid blue edges). If $\rhd$ is non-trivial, we must choose a base-point for the surface, which we call the start-point of the membrane (here represented by the yellow dot).}
			\label{E_valued_membrane_example_2D}
		\end{center}
	\end{figure}

		In addition to the new features of confinement and condensation (which can be seen in other extensions to the Kitaev Quantum Double model, such as the models studied in Ref. \cite{Bombin2008}), there are additional loop-like excitations, despite the 2+1d nature of the model. These loop excitations are formed by excited edges, with these edges connecting together to give us a loop. To construct the excitations, we must act with an operator that has support across an extended surface, rather than just on a ribbon. Such an operator is called a \textit{membrane operator}, and produces excitations along the boundary of the membrane, as shown in Figure \ref{E_valued_membrane_example_2D}.

		The membrane operators for our loop excitations, which we will call $E$-valued membrane operators, act by measuring the total group element assigned to the surface (calculated using the rules for composing surfaces given in Ref. \cite{Bullivant2017}) and then applying a weight based on the group element measured. The operator that measures the total surface element of the membrane $m$ is denoted by $\hat{e}(m)$ and can be expressed as
		$$\hat{e}(m) = \prod_{\text{plaquettes }p \in m} g(s.p(m) -v_0(p)) \rhd e_p^{\sigma_p},$$
	where $v_0(p)$ is the base-point of plaquette $p$, $s.p(m)$ is the start-point of the membrane (which is the base-point with respect to which we measure the surface label) and $\sigma_p$ is $+1$ or $-1$ to account for the relative orientations of $p$ and $m$ (1 if $p$ and $m$ are aligned, $-1$ otherwise). For a general crossed module, the order of multiplication and the paths $(s.p(m)-v_0(p))$ must be determined using the rules for composing surfaces. In the present case, where $\rhd$ is trivial (Case 1 in Table \ref{Table_Cases_2d}), the expression for the surface label simplifies to
		$$\hat{e}(m) = \prod_{\text{plaquettes }p \in m} e_p^{\sigma_p},$$
		and because $E$ is Abelian when $\rhd$ is trivial, the order of multiplication is arbitrary. The membrane operator is then
		\begin{equation}
		L^{\vec{\gamma}}(e)= \sum_{e \in E} \gamma_e \delta(e,\hat{e}(m)),
		\end{equation}
		where $\gamma_e$ is some set of coefficients. Allowing these coefficients to vary gives us a space of possible membrane operators. This space is (much like the space of electric ribbon operators) conveniently spanned by irreducible representations, this time of $E$. In the case where $\rhd$ is trivial, the group $E$ must be Abelian and so these irreps are 1D. We therefore do not need to include matrix indices for our basis vectors, as we did for the electric operators in Eq. \ref{electric_basis}. Then the basis operator labelled by irrep $\mu$ of $E$ is
		\begin{equation}
		L^{\mu}(m)= \sum_{e \in E} \mu(e) \delta(e,\hat{e}(m)).
		\label{Loop_Excitation_Basis_Abelian}
		\end{equation}
 		Note that we use Greek letters to represent irreps of $E$ throughout this paper, in order to differentiate them from the irreps of $G$, which we represent with upper case Roman letters (typically $R$).

		In a similar way to the fusion of particles, we can fuse two loops by applying their membrane operators on the same surface. The membrane operators obey a similar algebra to the electric ribbon operators. We have 
		\begin{align}
		L^{\mu}(m) L^{\nu}(m) &= \sum_{e \in E} \mu(e) \delta(e,\hat{e}(m)) \sum_f \nu(f) \delta(f,\hat{e}(m)) \notag \\
		&= \sum_{e \in E} \mu(e) \delta(e,\hat{e}(m)) \sum_f \nu(f) \delta(f,e) \notag \\
		& = \sum_{e \in E} \mu(e) \nu(e) \delta(e,\hat{e}(m)) \notag\\
		& = L^{\mu \cdot \nu}(m), \label{Equation_E_valued_loop_fusion}
		\end{align}
		from which we see that two excitations labelled by irreps fuse by multiplication of those irreps. Because the irreps are 1D, the result of this fusion is an irrep (rather than generically being a reducible representation as for the general case). This means that there is only one possible fusion channel (unlike for the electric excitations where we had branching into multiple irreps). Such fusion is called \textit{Abelian} \cite{Kitaev2006}.

\begin{figure}[h]
			\begin{center}
				\includegraphics{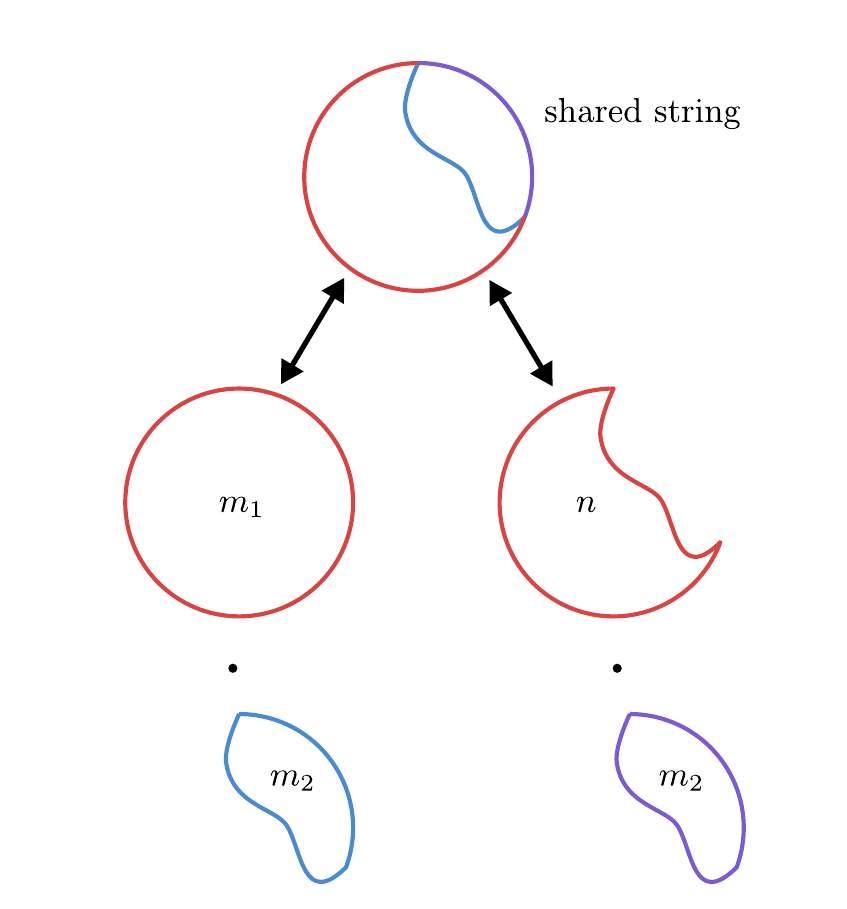}
				
				\caption{Consider applying two $E$-valued membrane operators on two membranes $m_1$ and $m_2$ (enclosed within the red and blue loops on the left side of the image), where one membrane ($m_2$) is contained within the other ($m_1$) and the two membranes share part of their boundary. Then applying both of these membrane operators results in the situation shown in the top of the image, where the common boundary is represented by the thicker purple line at the top-right. We can decompose this product of membrane operators as two membrane operators acting on the membranes $n$ and $m_2$ shown on the right of the image, which do not share any area. Then the only operator acting near the previously shared boundary is the membrane operator acting on the region $m_2$ enclosed by the (purple) string in the bottom right of the image, and the label of this membrane operator can be obtained from the labels of the original membrane operators by using the fusion rules from Equation \ref{Equation_E_valued_loop_fusion}.}
				\label{E_valued_membrane_partial_fuse}
			\end{center}
		\end{figure}

		Unlike point-particles, loop-like excitations are extended, and do not need to be fused along their entire length. That is, we can consider a case where parts of the loops fuse into a combined string and then split further along the loops' boundary, as shown in Figure \ref{E_valued_membrane_partial_fuse}. For example, this can occur if the two membrane operators that produce the loops share part, but not all, of the membrane with each-other. Consider the case where we have two membrane operators, $L^{\mu}(m_1)$ and $L^{\nu}(m_2)$ applied on membranes $m_1$ and $m_2$, for which $m_2$ is entirely contained within $m_1$. Further, in order to fuse the strings produced along the boundaries of the membranes, we consider the case where the boundary of $m_2$ includes some of the boundary of $m_1$, as shown in Figure \ref{E_valued_membrane_partial_fuse}. Then some of the excited edges, those on the shared boundary of the two membranes, are excited due to the combined action of the two membrane operators, and so should correspond to a fused string. We therefore expect this part of the string to correspond to an irrep $\mu \cdot \nu$, just as for the case of complete fusion discussed above. This can be shown directly by explicitly considering the action of the two membrane operators on the shared parts of the membrane and the parts exclusively in $m_1$ (i.e., not in $m_2$). We will refer to this latter part of the membrane as $n$. Then we can write the surface element of $m_1$ in terms of the elements of the shared and exclusive parts as $\hat{e}(m_1)=\hat{e}(m_2) \hat{e}(n)$. This allows us to rewrite the product of membrane operators as
		\begin{align*}
		L^{\mu}(m_1)& L^{\nu}(m_2)\\
		&= \sum_{e \in E} \mu(e) \delta(e,\hat{e}(m_1)) \sum_f \nu(f) \delta(f,\hat{e}(m_2))\\
		&=\sum_{e \in E} \mu(e) \delta(e,\hat{e}(m_2) \hat{e}(n)) \sum_f \nu(f) \delta(f,\hat{e}(m_2))\\
		&=\sum_{e \in E} \mu(e) \delta(\hat{e}(m_2)^{-1} e,\hat{e}(n)) \sum_f \nu(f) \delta(f,\hat{e}(m_2)).
		\end{align*}
		
		Then we can use the fact that we are applying $\delta(f,\hat{e}(m_2))$ to replace the element $\hat{e}(m_2)^{-1}$ in the other Kronecker delta with $f^{-1}$. This gives us
		\begin{align*}
		L^{\mu}(m_1)& L^{\nu}(m_2)\\
		&=\sum_{e \in E} \mu(e) \delta(f^{-1} e,\hat{e}(n)) \sum_f \nu(f) \delta(f,\hat{e}(m_2))\\
		&= \sum_{e' = f^{-1} e} \mu(f e') \delta(e',\hat{e}(n)) \sum_f \nu(f) \delta(f,\hat{e}(m_2))\\
		&= \sum_{e' \in E} \mu(e') \delta(e',\hat{e}(n)) \sum_f \mu(f) \nu(f) \delta(f,\hat{e}(m_2)),
		\end{align*}
		where we used the fact that $\mu$ is an irrep of $E$ to write $\mu(e' f) = \mu(e') \mu(f)$. Then we note that $\mu(f) \nu(f) = (\mu \cdot \nu)(f)$ and so
		\begin{align*}
		L^{\mu}(m_1)& L^{\nu}(m_2)\\
		&=\sum_{e' \in E} \mu(e') \delta(e',\hat{e}(n)) \sum_f (\mu \cdot \nu)(f) \delta(f,\hat{e}(m_2))\\
		&= L^{\mu}(n) L^{ \mu \cdot \nu}(m_2).
		\end{align*}
		
		Now only the operator $L^{ \mu \cdot \nu}(m_2)$ acts on the part of the membrane near the boundary shared by $m_1$ and $m_2$ (i.e., near the fused string), and so we see that, as we expect from the fusion rules given in Equation \ref{Equation_E_valued_loop_fusion}, this part of the boundary corresponds to the irrep $\mu \cdot \nu$.
		
		While it is convenient to label the loop-like excitations with irreps, some of these irreps label condensed excitations. The reason that some of the loop excitations are condensed can be seen fairly clearly without detailed calculation. Consider a closed electric ribbon operator placed on the path around the boundary of some membrane, $m$. This operator is sensitive to the path element around the boundary. However due to fake-flatness, when acting on the ground-state this path element is related to the surface element by $\partial(e(m))g(\text{boundary})=1_G$. That is, the path element can resolve information about $e(m)$ up to elements of the kernel of $\partial$. This means that any membrane operators that cannot resolve elements within the kernel (i.e., excitations labelled by irreps that trivially restrict to the kernel) are equivalent to electric ribbon operators placed on the boundary of the membrane. However the boundary operator is \textit{local} to the excitation: it only acts near the excited loop. We contrast this with the case we have considered so far, where to create a large loop we need to act on an entire membrane bounded by it. This is local in a slightly different sense to the usual meaning, because rather than the operators having no linear extent, operating on only a few degrees of freedom, the operators instead act on a region extended in one dimension (a loop). However the boundary operators are local in the sense that they only act in a region that is close to the loop excitation (the region has potentially large extension only in the same direction as the loop). This fact is particularly important in the 3+1d case, which we will study in a future paper (Ref. \cite{HuxfordPaper3}), where the loop-like excitations are truly topological excitations and so the fact that the excitations can be produced by an operator on the boundary of a membrane means that the membrane operator cannot move topological charge across the membrane. Any charge must only be moved along the support of the operator, and the fact that the membrane operator is equivalent to an operator on the boundary (when acting on the ground state) means that charge can only be moved along the boundary of the membrane. This means that the excitations cannot carry loop-like charge and so are condensed. The picture in 2+1d is a little different however. In Section \ref{Section_2D_irrep_basis} we will see that the loop excitations in 2+1d, at least when $\rhd$ is trivial, are related to a symmetry of the model. This symmetry is spontaneously broken when $\ker(\partial)$ is a non-trivial group and the uncondensed excitations then represent domain walls between different ground states of the model (with the ground states labelled by different irreps of $\ker(\partial)$). On the other hand, if $\ker(\partial)$ is trivial then all of the $E$-valued loops are labelled by irreps with trivial representations of the kernel and so all of the $E$-valued loop excitations are condensed.

		\subsection{Condensation and confinement}
		\label{Section_2D_Condensation_Confinement}
		
		When discussing our excitations in Section \ref{Section_2D_electric}, we noted that some of the electric excitations are confined. Whereas unconfined excitations cost no energy to separate, the confined excitations have an energy cost proportional to the length of the ribbon operator producing them. In this model, this energy cost is due to the edges along the ribbon being excited. We found in Section \ref{Section_2D_electric} that the confined excitations are the electric excitations whose labelling representation has non-trivial restriction to the subgroup $\partial(E)$. We also discussed the idea that some of our magnetic excitations were equivalent to local operators in Section \ref{Section_Single_Plaquette_1}, and we described such excitations as condensed. This is a general feature exhibited by certain topological models \cite{Bais2009, Burnell2018} and is particularly important when we consider transitions between topological phases.

		Given a topological phase, we can cause a transition by allowing some of the (bosonic) topological charges to join the ground state. That is, some of the topologically non-trivial excitations become trivial, with their conserved charge joining the vacuum sector, just as we already discussed for some of our excitations. This process is known as condensation (hence our use of the term condensed excitations). When this happens, any excitations that originally braid non-trivially with the condensing charges must become confined \cite{Bais2009, Burnell2018} in the condensed phase. The ribbon operators of such confined excitations can no longer be topological, because their non-trivial interaction with the vacuum means that we cannot freely deform the ribbons through space. This restriction manifests in the confinement of the excitations.

		We can see this process in the context of the higher lattice gauge theory models. In the case of the higher lattice gauge theory model with $\rhd$ trivial, there can be several models with the same groups $E$ and $G$ (but different $\partial$) and so with the same Hilbert space on a given lattice. We can therefore consider transitions between these models. In particular, we have the model $(G,E, \partial \rightarrow 1_G, \rhd \rightarrow \text{id})$, for which $\partial(e)=1_G \ \forall e \in E$. This means that the image of $\partial$ only contains the identity element, which in turn means that all irreps of this image are trivial. This means that there is no confinement of the electric excitations and no condensation of the magnetic excitations. We refer to this model as the uncondensed model. However we can then consider ``switching $\partial$ on", by changing $\partial$ to map onto some larger subgroup of $G$ and considering this new higher lattice gauge theory model. In doing this, we condense the magnetic excitations with label in the image of the new $\partial$. We will see in Section \ref{Section_2D_Braiding_Tri_Trivial} that these excitations braid non-trivially with the electric excitations that carry non-trivial representations of this subgroup, which results in these electric excitations being confined.

		We also saw that some of our loop excitations were condensed. When we go from our uncondensed $(G, E, \partial \rightarrow 1_G, \rhd \rightarrow \text{id})$ case to a more general $\partial$, we condense the loop excitations whose irreps restrict to trivial representations of the kernel of $\partial$. That is, any irreps that have $\mu(e_K)=1$ for every $e_K$ in the kernel of $\partial$ are associated to condensed excitations. As we will see in Section \ref{Section_2D_irrep_basis}, the uncondensed loop excitations are associated to domain walls related to a symmetry of the model, so it is more appropriate to consider the condensation in that context rather than as condensation of a topological charge.

		\section{Ribbon operators when $\rhd$ is non-trivial}
		
		\label{Section_2D_RO_Fake_Flat}
		
		So far we have dealt with the case where $\rhd$ is trivial. We also consider the case where $\rhd$ is non-trivial but we restrict to fake-flat configurations (configurations that satisfy the plaquette constraints). In this case, many of the excitations are largely the same as in the $\rhd$ trivial case discussed in Section \ref{Section_RO_2D_Tri_Trivial}, and so we will not repeat results from that section. Instead we will describe the differences between the two cases. The biggest difference is that we do not allow any magnetic excitations in the fake-flat case, because the magnetic excitations violate fake-flatness. The other difference is that the loop excitations gain an extra feature. Consider the $E$-valued membrane operator for a loop excitation, which measures the total surface label of the membrane. As discussed in Ref. \cite{Bullivant2017}, the label of a surface depends on its base-point when $\rhd$ is non-trivial. Therefore, when we specify the membrane operator we must specify the base-point of the surface that we measure. We call this base-point the start-point of the membrane operator. In addition to the edge excitations, the membrane operator may excite this start-point vertex. Recall that the vertex transform $A_v^g$ acting on the base-point of a plaquette takes that plaquette label from $e_p$ to $ g \rhd e_p$. This is also true for a general surface made up of multiple plaquettes (as we show in Section \ref{Section_E_Membrane_Commutation_Proof} in the Supplemental Material). This means that the vertex operator at the start-point of our membrane will affect the surface element of the membrane, and so may not commute with our membrane operator. Denoting our membrane operator (acting on a membrane $m$) by $\sum_{e \in E} \gamma_e \delta(\hat{e}(m),e)$ and the start-point of this operator by $s.p$, we have that 
		\begin{align*}
		\sum_{e \in E} \gamma_e \delta(\hat{e}(m),e)A_{s.p}^g&=A_{s.p}^g \sum_{e \in E} \gamma_e \delta(g \rhd \hat{e}(m),e)\\
		&=A_{s.p}^g \sum_{e \in E} \gamma_e \delta(\hat{e}(m),g^{-1} \rhd e)\\
		&= A_{s.p}^g \sum_{e'= g^{-1} \rhd e} \gamma_{g \rhd e'} \delta(\hat{e}(m), e').
		\end{align*}
		
		This indicates that the $E$-valued membrane operator only commutes with the vertex transform $A_{s.p}^g$ at the start-point when the set of coefficients $\gamma_e$ satisfy $\gamma_{g \rhd e}=\gamma_e$ for all $e \in E$. Then the start-point vertex is not excited when this condition is satisfied for all $g \in G$ and it is excited in orthogonal cases. Explicitly, after the action of a membrane operator $\sum_{e \in E} \gamma_e \delta(e, \hat{e}(m))$ the start-point vertex is not excited (indicating that the operator is gauge invariant) if the coefficients for the membrane operator satisfy
		\begin{equation}
		\gamma_e=\gamma_{g \rhd e} \quad \forall g \in G, \label{Equation_2D_E_Loop_Start_Point_Unexcited}
		\end{equation}
		and it is excited if the coefficients satisfy
		\begin{equation}
		\sum_{g \in G} \gamma_{g \rhd e} = 0 \quad\forall e \in E.
		\label{Equation_2D_E_Loop_Start_Point_Excited}
		\end{equation}

		All possible sets of coefficients can be written as a sum of two sets of coefficients, one set satisfying Equation \ref{Equation_2D_E_Loop_Start_Point_Unexcited} and one set satisfying Equation \ref{Equation_2D_E_Loop_Start_Point_Excited}, in a similar way to how the coefficients for magnetic ribbon operators could be split into parts that were functions of conjugacy class and parts that gave zero when summed over a conjugacy class (as shown in Equation \ref{magnetic_coefficient_split}). This is simply because, given an arbitrary set of coefficients, we can separate the part which is invariant under $\rhd$ action from the rest, analogous to separating symmetric and antisymmetric parts of a matrix. Explicitly, for an arbitrary set of coefficients $\gamma_e$ we have
		$$\gamma_e =\frac{1}{|G|}\sum_{g \in G} \gamma_{g \rhd e} + (\gamma_e -\frac{1}{|G|}\sum_{g \in G} \gamma_{g \rhd e}),$$
		where the first term satisfies Equation \ref{Equation_2D_E_Loop_Start_Point_Unexcited} and the second term satisfies Equation \ref{Equation_2D_E_Loop_Start_Point_Excited}. This means that we can construct a basis where each basis element corresponds to either the excited case or the unexcited case.

		This excitation of the start-point of the $E$-valued membrane operator is, as already mentioned, similar to how the magnetic ribbon operator can excite its start-point, depending on what linear combination of magnetic ribbon operators is taken. In the case of the magnetic ribbon operators, it is the conjugacy classes that are the significant objects for determining when the start-point is excited. For the $E$-valued membrane operators, instead of conjugation it is the action of $\rhd$ that matters. We can define an equivalence relation by $e \sim f$ if and only if there exists an element $g \in G$ such that $e= g \rhd f$ (reflexivity, symmetry and transitivity all follow from the group properties of the map $\rhd$), so much like how $G$ is partitioned by conjugacy classes, there is a partition of $E$ by ``$\rhd$-classes". The vertex at the start-point is not excited if the coefficients of the membrane operators are functions of $\rhd$-class (that is, group elements in a particular $\rhd$-class all have the same coefficient). This is similar to how the magnetic excitations have no vertex excitations if their operator has coefficients that are a function of conjugacy class.

		To better understand these $\rhd$-classes, it may be useful to go through an example. Consider the crossed module $(G=\mathbb{Z}_2, E=\mathbb{Z}_3, \partial \rightarrow 1_G, \rhd)$, where, denoting the elements of $G$ by $1$ and $-1$, the map $\rhd$ is defined by $-1 \rhd e = e^{-1}$ and $1 \rhd e = e$ for all $e \in E$. Denoting the elements of $E=\mathbb{Z}_3$ by $1_E$, $\omega_E$ and $\omega_E^2$, we can write the action of $-1 \: \rhd$ explicitly as $-1 \rhd 1_E=1_E$, $-1 \rhd \omega_E=\omega^2_E$ and $-1 \rhd \omega^2_E = \omega_E$. We therefore see that $\omega_E$ and $\omega^2_E$ are in the same $\rhd$-class, because they are related by the action of $-1 \: \rhd$. On the other hand, $1_E$ is in a $\rhd$-class of its own, because it is left invariant by any $\rhd$ action. Therefore, there are two $\rhd$-classes. In this case the condition for the start-point vertex of an $E$-valued membrane operator not to be excited is that the coefficients of $\omega_E$ and $\omega^2_E$ in the membrane operator are the same, while the conditions for the vertex to be definitely excited are that the coefficients of $\omega_E$ and $\omega^2_E$ must sum to zero and the coefficient of $1_E$ must be zero.

		So far, we have been considering the $E$-valued membrane operators in a basis labelled by elements of $E$. However, in a similar way to the $\rhd$ trivial case, it is useful to use irreducible representations to form a basis for our membrane operators. Unlike the $\rhd$ trivial case however, $E$ need not be Abelian and so these irreps do not have to be 1D. Therefore, we also need to include the matrix indices for the representation in our basis membrane operators. We define 
		\begin{equation}
		L^{\mu,a,b}(m) = \sum_{e \in E} [D^{\mu}(e)]_{ab} \delta(e,\hat{e}(m)), \label{Equation_E_Loops_Irrep_Labels}
		\end{equation}
		where $D^{\mu}(e)$ is the matrix representation of $e$ for irrep $\mu$ and $a$ and $b$ are its indices. Under an edge transform $\mathcal{A}_i^f$, the surface label $e(m)$ transforms in the same way as individual plaquettes (see Figure \ref{pathsonplaquettepaper2}) and so transforms as $e(m) \rightarrow [g(t) \rhd f] e(m)$ or $e(m) \rightarrow e(m) [g(t) \rhd f^{-1}]$ for some path element $g(t)$ (as we consider in more detail in Section \ref{Section_E_Membrane_Commutation_Proof} in the Supplemental Material). This means that we expect our individual edge transforms $\mathcal{A}_i^f$ to affect our membrane operator as
		\begin{align*}
		\sum_{e \in E} [D^{\mu}(e)]_{ab} &\delta(e,\hat{e}(m)) \\
		&\rightarrow \sum_{e \in E} [D^{\mu}([g \rhd f]e)]_{ab} \delta(e,\hat{e}(m))\\
		&= \sum_{e \in E}\sum_{c=1}^{|\mu|} [D^{\mu}(g \rhd f)]_{ac} [D^{\mu}(e)]_{cb} \delta(e,\hat{e}(m)),
		\end{align*}
		where some details of the transformation (such as the label of $g$, and whether $g \rhd f$ or its inverse appears) may depend on the branching structure of the lattice and details of the membrane operator. Regardless of the precise form of the transformation, we see that the edge transforms mix labels within a representation (it mixes operators labelled by different indices $c$ within the representation). These edge transforms are local operators. From the fact that the local operator $\mathcal{A}_i^f$ can change the matrix indices, we see that, as we may expect, the matrix elements do not label distinct topological charges, but instead describe local degrees of freedom. In addition, there is some mixing of different labels due to the vertex transforms. While edge transforms only mix excitations with different indices but the same irreps, vertex transforms mix different representations within a class of representations that we call a ``$\rhd$-Rep class" of representations. To explain what we mean by this, consider the object
		$$D^{g \rhd \mu}(e):=D^{\mu}(g \rhd e).$$
		This defines a representation $g \rhd \mu$ because
		\begin{align*}
		D^{g \rhd \mu}(ef)&=D^{\mu}(g \rhd (ef))=D^{\mu}([g \rhd e] \: [g \rhd f])\\
		&=D^{\mu}(g \rhd e)D^{\mu}(g \rhd f) = D^{g \rhd \mu}(e) D^{g \rhd \mu}(f).
		\end{align*}

		We say that two representations $\mu$ and $\alpha$ are in the same $\rhd$-Rep class if $g \rhd \mu = \alpha$ for some $g \in G$. This is an equivalence relation because of the group properties of the $\rhd$ map. We can define this equivalence relation simply in the case of 1D reps, but for higher dimensional reps we need to be careful because each irrep is related to a set of equivalent irreps which we can generate by conjugating the matrix representation by a constant matrix. Then it could be possible for the $g \rhd$ action on an irrep to give an irrep that is equivalent to one of our irreps, but in a different form. In most cases, we will either be dealing with the case where $E$ is Abelian or will only care about subgroups in the centre of $E$ (specifically the kernel of $\partial$). In this case conjugation is trivial and the irreps are unaffected. However, to be concrete, in the non-Abelian case we should talk about classes of characters, or define the $\rhd$-Rep classes to account for conjugation.

		Having discussed these $\rhd$-Rep classes, we can now see how they relate to the action of the vertex transform. The action of the vertex transform on an $E$-valued membrane operator is to take $L^{\mu,a,b}(m) = \sum_{e \in E} [D^{\mu}(e)]_{ab} \delta(e,\hat{e}(m))$ to 
		\begin{align*}
		\sum_{e \in E} [D^{\mu}(e)]_{ab}& \delta(e,g^{-1} \rhd \hat{e}(m))\\
		&= \sum_{e' = g \rhd e} [D^{\mu}(g^{-1} \rhd e')]_{ab} \delta(e',\hat{e}(m))\\
		& =\sum_{e' \in E} [D^{g^{-1} \rhd \mu}( e')]_{ab} \delta(e',\hat{e}(m))\\
		& = L^{g^{-1} \rhd \mu,a,b}(m),
		\end{align*}
		which is labelled by another irrep in the same $\rhd$-Rep class as $\mu$. The fact that vertex operators link excitations labelled by irreps within these classes suggests that we should group the membrane operators into $\rhd$-Rep classes of representations of $E$. However, just as we discussed in the $\rhd$ trivial case in Section \ref{Section_2D_Loop}, some of the $E$-valued loops are condensed, which means that some membrane operators labelled by different irreps are in fact equivalent up to operators on the boundary of the membrane. Just as in the $\rhd$ trivial case, the condensed $E$-valued loops are those labelled by irreps that have trivial restriction to the kernel of $\partial$. This follows from the same reasoning as for the $\rhd$ trivial case: an electric ribbon around the boundary of a surface can measure the surface element up to an element of the kernel when that surface satisfies fake-flatness. This condensation of the excitations corresponding to irreps with trivial restriction to the kernel means that we should group the loop-like excitations into $\rhd$-Rep classes of irreps of the kernel of $\partial$. These groupings will fuse in a non-Abelian way, because two $\rhd$-Rep classes of irreps can potentially fuse to more than one $\rhd$-Rep class. This is true even if the irreps themselves are 1D (in fact, irreps of the kernel are always 1D because the kernel is Abelian), so that an Abelian group $E$ does not necessarily give Abelian fusion.

		 As an example, consider the crossed module $(\mathbb{Z}_2, \mathbb{Z}_3, \partial \rightarrow 1_G, \rhd)$ that we used earlier in this section. Here the map $\rhd$ is defined by $-1_G \rhd e =e^{-1}$ for any $e \in \mathbb{Z}_3$ (while $1_G \rhd$ is the identity map). The group $\mathbb{Z}_3$ has three irreps. The trivial one, which we call $1_R$, is invariant under the $\rhd$ action (because $1_R(g \rhd e)=1$ regardless of $g$), and so is in a $\rhd$-Rep class of its own. The other two irreps, which we denote by $\alpha_R$ and $\alpha^2_R$, are defined by $\alpha_R(\omega_E)=e^{\frac{2 \pi i}{3}}$, $\alpha_R( \omega_E^2)=e^{\frac{4 \pi i}{3}}$ and $\alpha_R^2(e)=\alpha_R(e^{-1})$ (where $\omega_E$ and $\omega_E^2$ are the two non-trivial elements of $E$). We therefore see that 
		 $$-1_G \rhd \alpha_R(e)= \alpha_R(-1_G \rhd e) = \alpha_R(e^{-1})=\alpha_R^2(e),$$ 
		 and so $\alpha_R$ and $\alpha_R^2$ belong to the same $\rhd$-Rep class. Now consider an $E$-valued membrane operator made from a combination of the $\alpha_R$ and $\alpha_R^2$ irreps:
		$$L^{\vec{A}}(m)= A_1L^{\alpha_R}(m)+ A_2 L^{\alpha_R^2}(m),$$
		where $A_1$ and $A_2$ are non-zero coefficients. If we fuse two copies of this membrane operator, we obtain
		
		\begin{align*}
		&L^{\vec{A}}(m) L^{\vec{A}}(m)\\
		&= \big( A_1L^{\alpha_R}(m)+ A_2 L^{\alpha_R^2}(m)\big) \big( A_1L^{\alpha_R}(m)+ A_2 L^{\alpha_R^2}(m)\big)\\
		&= A_1^2 L^{\alpha_R}(m) L^{\alpha_R}(m) + A_2^2 L^{\alpha_R^2}(m)L^{\alpha_R^2}(m)\\
		& \hspace{0.5cm} + 2A_1A_2 L^{\alpha_R}(m) L^{\alpha_R^2}(m). 
		\end{align*}
		
		Using the fusion rules given in Equation \ref{Equation_E_valued_loop_fusion}, we have
		\begin{align*}
		L^{\alpha_R}(m) L^{\alpha_R}(m)&= L^{\alpha_R \cdot \alpha_R}(m)= L^{\alpha_R^2}(m),\\
		L^{\alpha_R^2}(m) L^{\alpha_R^2}(m)&= L^{\alpha_R^2 \cdot \alpha_R^2}(m)= L^{\alpha_R}(m), \intertext{and}\\
		L^{\alpha_R}(m) L^{\alpha_R^2}(m)&= L^{\alpha_R \cdot \alpha_R^2}(m)= L^{1_R}(m).
		\end{align*}
	Therefore,
			\begin{align*}
			L^{\vec{A}}(m) &L^{\vec{A}}(m)\\
			&= A_1^2 L^{\alpha_R^2}(m) + A_2^2 L^{\alpha_R}(m)+ 2A_1A_2 L^{1_R}(m). 
			\end{align*}
		We see that the fusion product includes contributions from the $\rhd$-Rep class containing $\alpha_R$ and $\alpha_R^2$, but also a term corresponding to the trivial $\rhd$-Rep class, indicating that the fusion is non-Abelian. 
		
		\section{Braiding}
		\label{Section_2D_Braiding}
		
\begin{figure}[h]
			\begin{center}
				\includegraphics{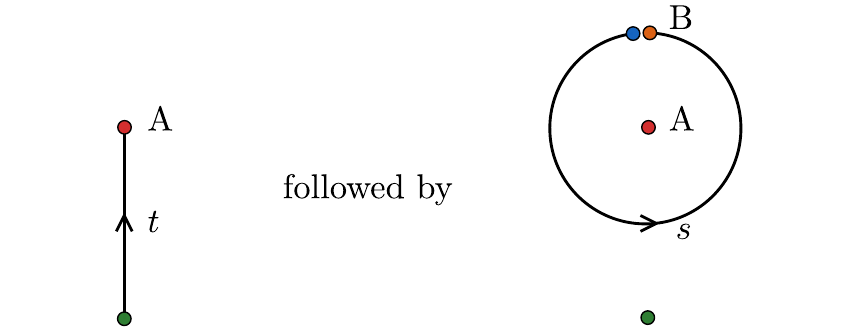}
				\caption{We can create the situation where a particle B braids around A by first producing A and then moving B around it with ribbon operators.}
				\label{2D_braiding_1}
			\end{center}
		\end{figure}

		In 2+1d, one major feature of topological phases of matter is that the excitations may support \textit{braiding statistics} that generalize the familiar bosonic and fermionic exchange phases \cite{Kitaev2006, Leinaas1977, Wilczek1982, Rao1992}. When one point particle is taken on a path all the way around another (which would give a trivial transformation for either fermions or bosons) it can result in a transformation that is a general phase, or even some more complicated transformation. Before we discuss the braiding results for this model, we should first explain how we find the braiding relations. We mentioned earlier how the ribbon operators produce and move excitations. This lets us recast braiding in terms of commutation relations of the ribbon operators. Consider the following situation, as shown in Figure \ref{2D_braiding_1}. We create a pair of excitations and move one of them (A) far away from the other, using a ribbon operator on a path $t$. Then we produce another pair of excitations and move one of these (B) around A, by acting with a ribbon operator on a new path $s$.
		
\begin{figure}[h]
			\begin{center}
			\includegraphics{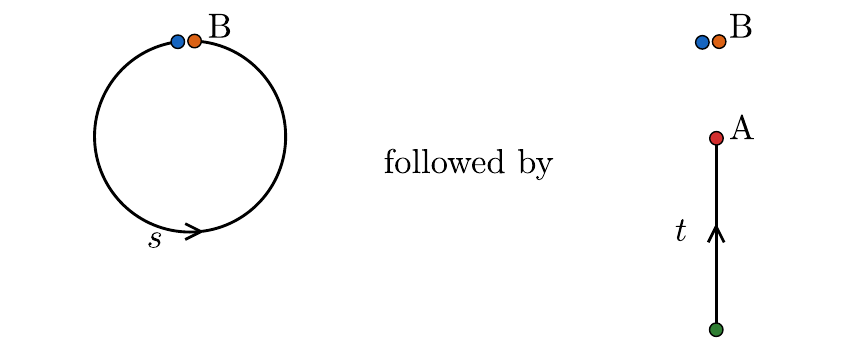}
				\caption{By applying the ribbon operators in the other order, we move particle B through empty space rather than around A.}
				\label{2D_braiding_2}
			\end{center}
		\end{figure}

		Now consider applying the operators in the other order, acting on $s$ first, as shown in Figure \ref{2D_braiding_2}. This moves B around empty space and then creates the other pair and moves A into the region of interest. Therefore, no braiding has occurred, because no particle moved around another.
		
		In the first case (shown in Figure \ref{2D_braiding_1}), B braided with A but in this second case (Figure \ref{2D_braiding_2}) it has not. However in the two cases the particles end up in the same places and have moved through the same space. Comparing the two (by working out the commutation relation of the two ribbon operators used to perform the movement) therefore isolates the effect of braiding B around A from any other details and gives the braiding relation.
		
		\subsection{The $\rhd$ trivial case}
		\label{Section_2D_Braiding_Tri_Trivial}
		We first look at braiding in the case where $\rhd$ is trivial. In this case, the non-trivial braiding involves only the electric and magnetic excitations, and not the loop-like excitations. This is because the braiding relations can be calculated from commutation relations, and the electric and magnetic operators both commute with the $E$-valued membrane operators. We can see this from the fact that the electric and magnetic ribbons act only on the edge labels, whereas the membrane operators only act on the surface labels.

		We also have trivial braiding between two electric excitations. The braiding is trivial because the electric ribbons are diagonal in the configuration basis (i.e., the basis where each edge is labelled by a group element in $G$ and each plaquette by a group element in $E$), so that they all commute with each-other. On the other hand, braiding an electric excitation with a magnetic excitation is not trivial. When the start-points of the operators are the same, electric operators transform under braiding with magnetic excitations according to the representation of the electric excitation and the group element of the magnetic one. For braiding of an electric excitation labelled by irrep and matrix indices $\set{R,a,b}$ with a magnetic excitation labelled by $h$, we have (as we show in Section \ref{Section_2D_braiding_electric_magnetic} of the Supplemental Material):

					\begin{center}
			\noindent\fcolorbox{black}{myblue1}{%
				\parbox{0.9\linewidth}{% 
					\centering \textbf{Electric-magnetic braiding:}
				
		\begin{align}
		S^{R,a,b}(t)&C^h(r) \ket{GS} \notag \\
		 &=C^h(r) \sum_{c=1}^{|R|} [D^{R}(h)]_{ac} S^{R,c,b}(t)\ket{GS}. \label{Equation_electric_magnetic_braiding_summary}
		\end{align}}}
	\end{center}
		
		We can understand this relation by considering that the magnetic excitation sets up a flux labelled by $h$. When the electric excitation crosses the magnetic ribbon, it is acted on by this flux, so that the path measured by the electric excitation gains an additional factor of $h$. When we consider the irrep basis, this leads to the factor of $[D^{R}(h)]_{ac}$. We see that different indices within the representation of the electric excitation are mixed by this braiding. Note that the precise formula depends on the orientation of the braiding procedure, so there may be an inverse on $h$ compared to the result in Equation \ref{Equation_electric_magnetic_braiding_summary}. This braiding between the electric and magnetic excitations is the same as in the Kitaev Quantum Double model \cite{Kitaev2003}, because the ribbon operators have the same commutation relations as the corresponding operators in that model. In Ref. \cite{Kitaev2003}, the commutation relations are calculated by using the group element basis for the electric part of the ribbon, rather than the irrep basis. Then to obtain the relation given above we simply need to take appropriate linear combinations of the operators used in Ref. \cite{Kitaev2003}.

		One interesting case of the braiding relation given in Equation \ref{Equation_electric_magnetic_braiding_summary} is where we take $h$ to be an element of $\partial(E)$. In this case, as we discussed in Section \ref{Section_Single_Plaquette_1}, the magnetic excitation is condensed. Taking $h= \partial(e)$, the braiding relation becomes
		\begin{align*}
			S^{R,a,b}(t)&C^{\partial(e)}(r) \ket{GS} \notag \\ & = \sum_{c=1}^{|R|} [D^{R}(\partial(e))]_{ac} C^{\partial(e)}(r) S^{R,c,b}(t) \ket{GS}. 
		\end{align*}
		
		When $\rhd$ is trivial, $\partial(E)$ is in the centre of $G$ (see Section \ref{Section_Recap_Paper_2}), and so $[D^{R}(\partial(e))]$ must be a scalar multiple of the identity from Schur's Lemma. The braiding relation therefore simplifies to 
		\begin{align}
		S^{R,a,b}(t)&C^{\partial(e)}(r) \ket{GS} \notag \\ & = \sum_{c=1}^{|R|} [D^{R}(\partial(e))]_{11} \delta_{ac} C^{\partial(e)}(r) S^{R,c,b}(t) \ket{GS} \notag\\
		&= [D^{R}(\partial(e))]_{11} C^{\partial(e)}(r) S^{R,a,b}(t) \ket{GS}.
		\end{align}
		
		The diagonal element $[D^{R}(\partial(e))]_{11}$ can be considered as an irrep of $\partial(E)$ (the irrep branching from the restricted representation we described in Section \ref{Section_2D_electric}), and the braiding is only non-trivial if this irrep is non-trivial. As we described in Section \ref{Section_2D_electric}, this irrep of $\partial(E)$ being non-trivial is also the condition for an electric excitation to be confined. We therefore see that the only electric excitations that braid non-trivially with the condensed magnetic excitations are the confined excitations, as we expect.

		The final non-trivial braiding is between two magnetic excitations, with this braiding only being non-trivial when the group $G$ is non-Abelian. Under braiding, two magnetic ribbons with the same start-point transform by conjugation. When the end of one ribbon (the end is a quasiparticle) is taken all the way around one of the other quasiparticles (not enclosing any other quasiparticle), the label of each is conjugated by the product of the labels of each quasiparticle. If we then consider ribbon operators that are linear combinations of operators labelled by elements in a certain conjugacy class, then the braiding results in mixing within this conjugacy class. In particular, note that if one of the magnetic ribbons is an equal superposition of all fluxes in a conjugacy class (meaning it carries no start-point excitation, as discussed in Section \ref{Section_2D_Magnetic}), conjugation has no effect because it just permutes these fluxes within their conjugacy class. If the ribbon is not an equal superposition then it transforms in some other, more complex, way. This means that the degrees of freedom within a conjugacy class of the flux are non-conserved internal degrees of freedom that describe how they braid with other fluxes, as we mentioned in Section \ref{Section_2D_Magnetic}. Again, this braiding is the same as the braiding of the equivalent excitations in Kitaev's Quantum Double model \cite{Kitaev2003}, so we will not go into too much detail here (although the detailed results are discussed in Section \ref{Section_2D_braiding_two_magnetic_appendix} in the Supplemental Material).

		The non-trivial braiding relations that we have discussed so far hold only when the ribbon operators have the same start-point. This is because the theory is generally non-Abelian. In a non-Abelian anyon theory, the braiding between two anyons depends on what fusion channel they are in \cite{Kitaev2006}. That is, the braiding operator $R^{ab}_c$ for two anyons $a$ and $b$ depends on their total anyonic charge $c$. The excitations are generally only in a definite fusion channel (as opposed to some operator superposition) when the ribbons used to create them share a start-point. Therefore, they only have a definite braiding transformation under the same circumstances.

		The notion of definite braiding only occurring when our operators have a ``start-point" at the same location also has an interpretation in terms of the gauge theory picture. In non-Abelian gauge theory, fluxes are associated with some closed loop \cite{Alford1992}. This closed loop has a definite start-point and moving this start-point should be accompanied by some change of basis. Therefore, to compare two fluxes (such as when we want to fuse or braid them), they need to be defined with the same start-point. If we start with two fluxes that have different start-points, we therefore need to write one of these fluxes in terms of a flux based at the other start-point (i.e., work out its label with respect to a different start-point). In our lattice theory, the flux label corresponding to a new start-point is the original flux label conjugated by a path element operator (not just a fixed group element), so the flux does not have definite label with respect to that new start-point.

		\subsection{The fake-flat case}
		\label{Section_braiding_2D_fake_flat}

		When we restrict to fake-flat configurations, but allow $\rhd$ to be non-trivial (Case 3 in Table \ref{Table_Cases_2d} in Section \ref{Section_Recap_Paper_2}), we do not have the magnetic excitations. In addition the $E$-valued loops and the electric ribbons still braid trivially, due to the fact that both are diagonal in the configuration basis. Therefore, there is no non-trivial braiding in this case. There is non-trivial commutation between the $E$-valued membrane operators and the single plaquette multiplication operators (which is also the case in the $\rhd$ trivial case). However, this does not correspond to braiding, because the single plaquette operators are local operators and have no interpretation in terms of moving excitations.

		While restricting to fake-flatness rules out any magnetic excitations, the signatures of these missing excitations are still present in the model. If we consider a manifold with non-contractible cycles, such as a torus, then some ground states contain closed loops with labels outside of $\partial(E)$ (just like we would expect around a non-condensed magnetic excitation). If we apply an electric ribbon operator around such a cycle $t$, then this is equivalent to producing a pair of electric excitations and passing one of them around the cycle. We can use this to find the transformation from moving the excitation around the handle. Consider comparing an electric ribbon operator that produces and separates a pair of excitations along a small path $s$, to one that produces and separates the pair along $s$, but then further moves the excitation at the end of $s$ around the non-contractible cycle $t$. That is, we compare a ribbon operator applied on $s$ to one applied on the composite path $s \cdot t$. If the ribbon operators are labelled by irrep $R$ of $G$, with matrix indices $a$ and $b$, then the latter ribbon operator is given by
		$$S^{R,a,b}(s \cdot t) = \sum_{g \in G} [D^R(g)]_{ab} \delta(g, \hat{g}(s \cdot t)).$$
		
		We can then separate the path element $\hat{g}(s \cdot t)$ into two parts, corresponding to $s$ and $t$: $\hat{g}(s \cdot t) = \hat{g}(s) \cdot \hat{g}(t)$. This tells us that
		\begin{align*}
		S^{R,a,b}(s \cdot t) &= \sum_{g \in G} [D^R(g)]_{ab} \delta(g, \hat{g}(s) \cdot \hat{g}(t))\\
		&= \sum_{g \in G} [D^R(g)]_{ab} \delta(g\hat{g}(t)^{-1}, \hat{g}(s) )\\
		&= \sum_{ g' = g\hat{g}(t)^{-1} \in G} [D^R(g' \hat{g}(t))]_{ab} \delta(g', \hat{g}(s)).
		\end{align*}
		
		We can then split the matrix element $[D^R(g' \hat{g}(t))]_{ab}$ into contributions from $g'$ and $\hat{g}(t)$, to give
		\begin{align*}
		S^{R,a,b}(s \cdot t) &= \sum_{ g' \in G} \sum_{c=1}^{|R|} [D^R(g')]_{ac} [D^R(\hat{g}(t))]_{cb} \delta(g, \hat{g}(s)).
		\end{align*}
		
		Recognising $\sum_{ g' \in G} [D^R(g')]_{ac}\delta(g, \hat{g}(s))$ as the electric ribbon operator $S^{R,a,c}(s)$ on the ribbon $s$, we have
		\begin{align*}
		S^{R,a,b}(s \cdot t) &= \sum_{c=1}^{|R|} [D^R(\hat{g}(t))]_{cb} S^{R,a,c}(s),
		\end{align*}
		so that moving the excitation around $t$ induces a transformation that mixes the ribbon operators labelled by different matrix indices for the irrep $R$, by (right-) matrix multiplication by $D^R(\hat{g}(t))$ (note the similarity to the transformation we predicted in Section \ref{Section_Recap_Paper_2} in Ref. \cite{HuxfordPaper1}, except that right-multiplication is required to ensure correct composition, rather than left-multiplication). Here $\hat{g}(t)$ is an operator, and cycles (including non-contractible ones) in the ground state are not generally in an eigenstate of this path measurement operator (that is, the cycles are in a linear combination of states with different labels), so this transformation is not usually a simple one. In Section \ref{Section_2D_braid_fake_flat_non_trivial_cycle} of the Supplemental Material, we discuss in more detail the circumstances where we can simplify this transformation. We find that, for ground-states where $\hat{g}(t)$ is minimally mixed, and where we annihilate the two excitations at the end of the ribbon operator, the transformation can be described by the character of the irrep $R$.

		\section{Particular examples}
		\label{Section_2D_Particular_examples}

		It may be useful to consider some simple examples in more detail, to see the interesting features of the higher lattice gauge theory model. Therefore, we will look at two examples that each highlight a particular feature of the model.

		\subsection{$\mathbb{Z}_2$, $\mathbb{Z}_3$ model}
		\label{Section_Example_Z_2_Z_3}
		
		The first features to highlight are the loop excitations and how these excitations are related to the ground state degeneracy. For simplicity we first look at these in the absence of condensation and confinement. A convenient model with which to examine these features is described by the crossed module $(\mathbb{Z}_2,\mathbb{Z}_3,\partial \rightarrow 1_G, \rhd)$ (which we used as an example in Section \ref{Section_2D_RO_Fake_Flat}, but define again here for reference). To distinguish the elements of the two groups, we will write the elements of $G=\mathbb{Z}_2$ as $\pm1$ and those of $E=\mathbb{Z}_3$ as $1_E$, $\omega_E$ and $\omega^2_E$. Then $\rhd$ is defined by
		\begin{align*}
			1 \rhd e= e \hspace{0.1cm} \forall e \in E,& \hspace{0.2cm} -1 \rhd 1_E=1_E,
			\\ -1 \rhd \omega_E=\omega_E^2,& \hspace{0.2cm} -1 \rhd \omega_E^2= \omega_E.
		\end{align*}

		This can be summarized as $1 \rhd e = e, -1 \rhd e = e^{-1}$. One simple way to generalize this crossed module slightly is to replace the group $\mathbb{Z}_3$ with $\mathbb{Z}_n$, where $n$ is odd, and define $\rhd$ by $-1 \rhd e =e^{-1}$. However, the features of such a model are very similar to the $\mathbb{Z}_3$ case, so we only present the $\mathbb{Z}_3$ case here. The properties of this crossed module and the notation used in this section are summarized in Tables \ref{properties_Z_2_Z_3} and \ref{properties_Z_2_Z_3_maps} to refer back to as necessary.
		
		\begin{table}[h]	
			\begin{center}
				\begin{tabular}{|c|c |c|} 
					\hline
					& $\bm{G}$ & $\bm{E}$ \\ 
					\hline
					Group & $\mathbb{Z}_2$ & $\mathbb{Z}_3$ \\
					\hline
					\rule{0pt}{12pt}
					Elements & $\set{1, -1}$ & $\set{ 1_E, \omega_E, \omega_E^2}$ \\ 
					\hline
					\rule{0pt}{12pt}
					Irreps & $\set{1^G_R,-1_R^G}$ & $\set{1_R, \alpha_R, \alpha_R^2}$ \\ 
					\hline
					
				\end{tabular}
				\caption{The notation used for the two groups $\mathbb{Z}_2$ and $\mathbb{Z}_3$ in this section.}
				\label{properties_Z_2_Z_3}
			\end{center}
		\end{table} 
		
		\begin{table}[h]	
			\begin{center}
				\begin{tabular}{|c|c |c|c|} 
					\hline
					$\bm{e}$ & $\bm{\partial(e)}$ & $\bm{1 \rhd e}$ & $\bm{-1 \rhd e}$\\ 
					\hline
					$1_E$ & $1_G$ & $1_E$ & $1_E$\\ 
					\hline
					\rule{0pt}{12pt}
					$\omega_E$ & $1_G$ & $\omega_E$ & $\omega_E^2$\\ 
					\hline
					\rule{0pt}{12pt}
					$\omega_E^2$ & $1_G$ & $\omega_E^2$ & $\omega_E$ \\ 
					\hline
					\rule{0pt}{12pt}
					General $e \in E$ & $1_G$ & $e$ & $e^{-1}$ \\ 
					\hline
				\end{tabular}
				\caption{The result of each map acting on the elements of $E$. The final line shows the general action of each map.}
				\label{properties_Z_2_Z_3_maps}
			\end{center}
		\end{table}

\begin{figure}[h]
		 	\begin{center}
		 	\includegraphics[width=0.8\linewidth]{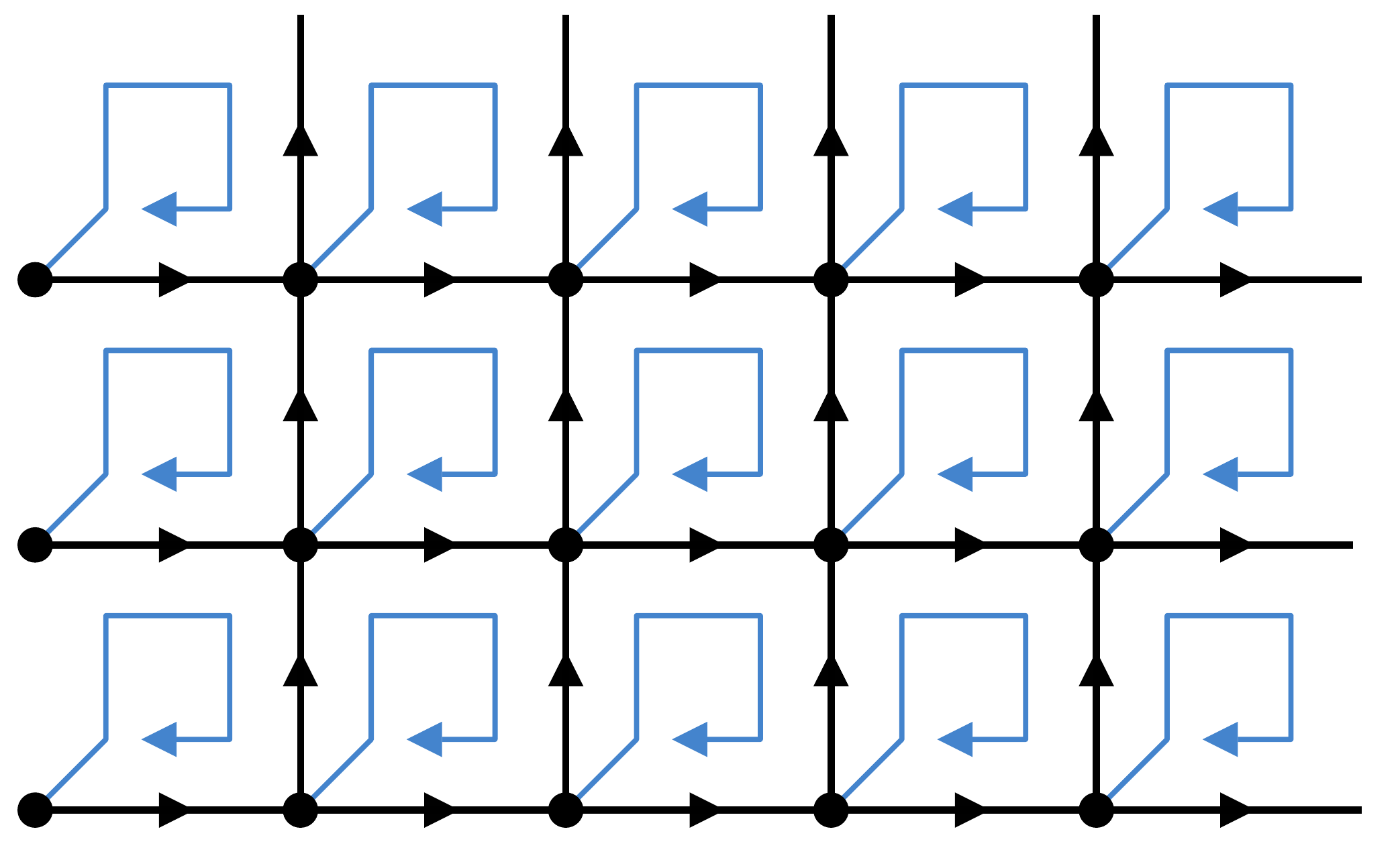}

		 		\caption{This image shows a fragment of the lattice used for our example model. The (blue) circulating arrows represent the plaquettes, which are based at the vertices (shown as small black circles) that are attached to the arrows.}
		 		\label{2Dlattice}
		 	\end{center}
		 \end{figure}

\begin{figure}[h]
			\begin{center}
				
				\includegraphics{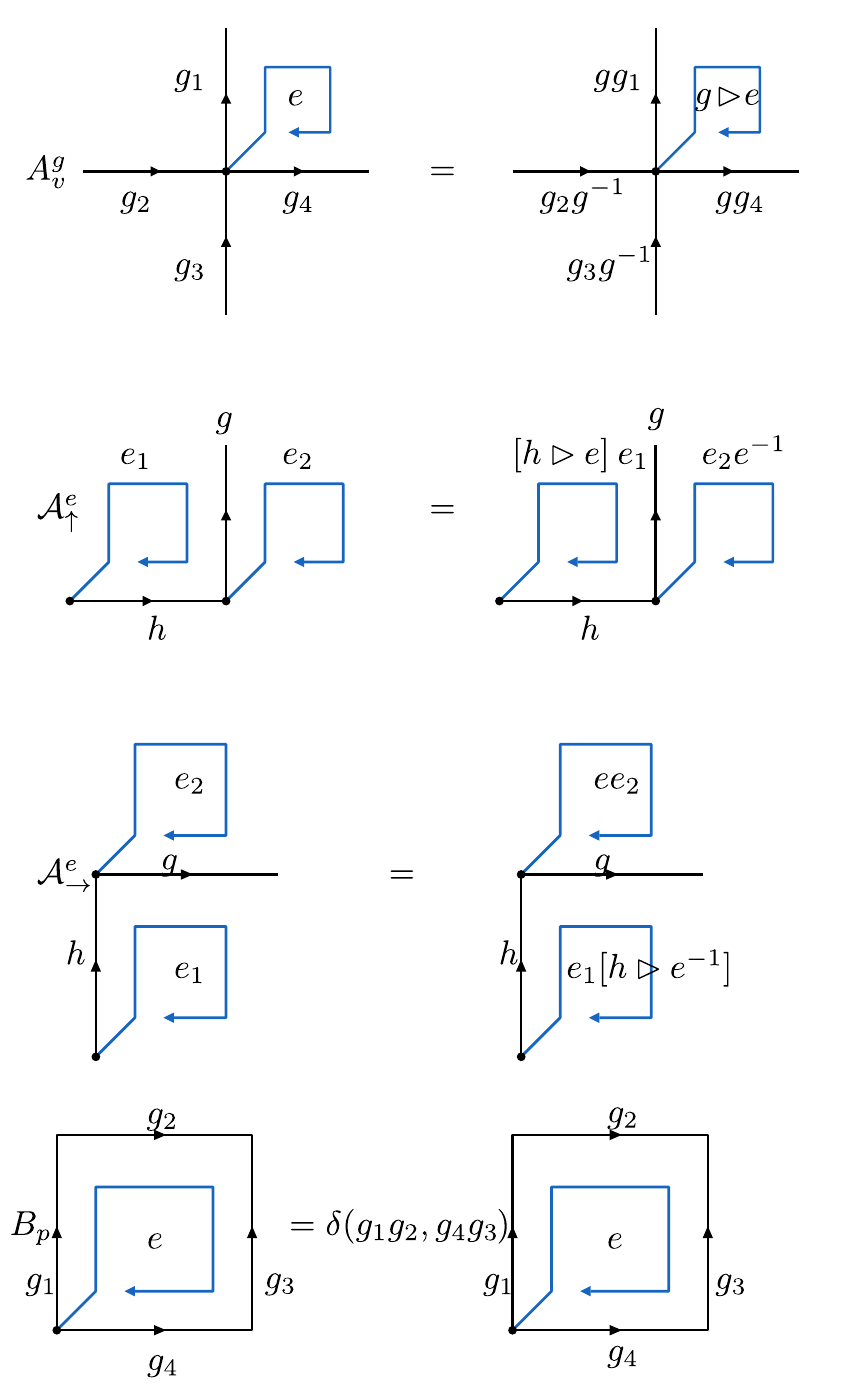}
				\caption{Here we indicate the action of the vertex transforms, edge transforms and plaquette term of the example model.}
				\label{2Dtransforms}
			\end{center}
		\end{figure}

		 In addition to choosing this particular crossed module, we also choose a specific lattice, a fragment of which is shown in Figure \ref{2Dlattice}. In this lattice we have one kind of vertex, two types of edge (horizontal and vertical) and one kind of plaquette. The actions of the corresponding operators are shown in Figure \ref{2Dtransforms}.

\begin{figure}[h]
			\begin{center}
			\includegraphics{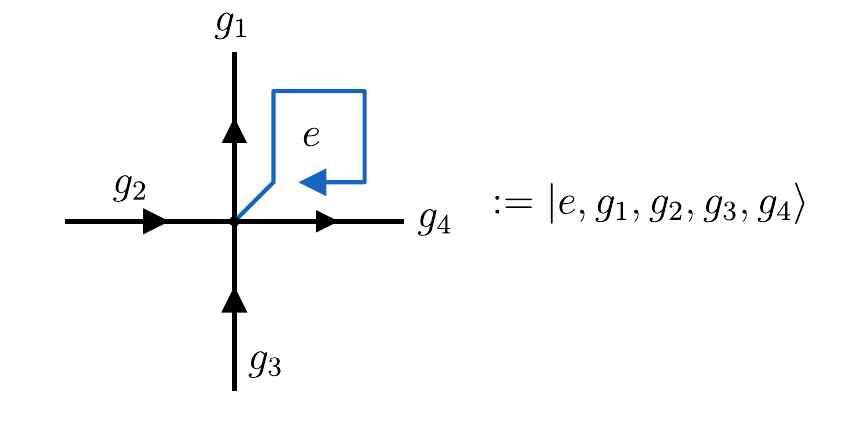}
				\caption{We wish to consider the effect of a vertex transform on the degrees of freedom around the vertex as we change basis. It is therefore useful to have a simple notation for the state of the degrees of freedom affected by the vertex transform in the original basis, which we define here.}
				\label{support_vertex_1}
			\end{center}
		\end{figure}

		It will be convenient to consider a change of basis, to simplify the action of the operators. For the plaquette elements we change variables from group elements to irreps. We denote the three irreps of $\mathbb{Z}_3$ by $1_R$, $\alpha_R$ and $\alpha_R^2$. The 1D irreps form a group under the multiplication defined by $(\mu_1\cdot \mu_2)(e)=\mu_1(e) \cdot \mu_2(e)$ for irreps $\mu_1$ and $\mu_2$. $1_R$ is the identity element for this group, while $\alpha_R$ and $\alpha_R^2$ are inverses to each-other. We can then define states for the plaquette degrees of freedom using these irreps:
		\begin{align*}
		\ket{1_R}&= \frac{1}{\sqrt{3}}(\ket{1_E}+\ket{\omega_E}+\ket{\omega^2_E})\\
		\ket{\alpha_R}&= \frac{1}{\sqrt{3}}(\ket{1_E}+\omega \ket{\omega_E}+\omega^2\ket{\omega^2_E})\\
		\ket{\alpha^2_R}&= \frac{1}{\sqrt{3}}(\ket{1_E}+\omega^2 \ket{\omega_E}+\omega\ket{\omega^2_E}),
		\end{align*}
		where $\omega$ is $e^{2 \pi i /3}$.	We now wish to see how the various energy terms act in this new basis. First we consider the vertex transforms, as defined in Equation \ref{Equation_vertex_transform_definition} in Section \ref{Section_Recap_Paper_2}. We consider the state of the degrees of freedom around a particular vertex, using the notation $\ket{e,g_1,g_2,g_3,g_4}$ defined in Figure \ref{support_vertex_1}. We suppress any labels corresponding to the degrees of freedom in the rest of our lattice (because these other degrees of freedom are unaffected by the vertex transform). Then the corresponding state where the plaquette is labelled by an irrep $\mu$ of $E$ is given by
		$$ \ket{\mu,g_1,g_2,g_3,g_4}= \frac{1}{\sqrt{3}} \sum_{e \in \mathbb{Z}_3} \mu(e)\ket{e,g_1,g_2,g_3,g_4}.$$
		
		From the definition of the vertex transforms (Equation \ref{Equation_vertex_transform_definition}), we know that the vertex transform labelled by the identity element of $G$ is the identity operator. On the other hand the vertex transform $A_v^{-1}$ acts on this new basis as
		\begin{align}
		&A_v^{-1} \ket{\mu,g_1,g_2,g_3,g_4} \notag \\
		&=A_v^{-1} \frac{1}{\sqrt{3}} \sum_{e \in \mathbb{Z}_3} \mu(e)\ket{e,g_1,g_2,g_3,g_4} \notag \\
			&= \frac{1}{\sqrt{3}} \sum_{e \in \mathbb{Z}_3} \mu(e) \ket{-1 \rhd e,-1 \cdot g_1, -1 \cdot g_2, -1 \cdot g_3, -1 \cdot g_4} \notag \\
		&= \frac{1}{\sqrt{3}} \sum_{e \in \mathbb{Z}_3} \mu(e) \ket{e^{-1},-1 \cdot g_1, -1 \cdot g_2, -1 \cdot g_3, -1 \cdot g_4} \notag \\
		&= \frac{1}{\sqrt{3}} \sum_{e'=e^{-1} } \mu(e'^{-1}) \ket{e',-1 \cdot g_1, -1 \cdot g_2, -1 \cdot g_3, -1 \cdot g_4} \notag \\
		&= \frac{1}{\sqrt{3}} \sum_{e' \in \mathbb{Z}_3} \mu(e')^{-1} \ket{e',-1 \cdot g_1, -1 \cdot g_2, -1 \cdot g_3, -1 \cdot g_4} \notag \\
		&= \frac{1}{\sqrt{3}} \sum_{e' \in \mathbb{Z}_3} \mu^{-1}(e') \ket{e',-1 \cdot g_1, -1 \cdot g_2, -1 \cdot g_3, -1 \cdot g_4} \notag \\
		&= \ket{\mu^{-1},-1 \cdot g_1, -1 \cdot g_2, -1 \cdot g_3, -1 \cdot g_4}. \label{Equation_Z2_Z3_vertex_transform}
		\end{align}
		
\begin{figure}[h]
			\begin{center}
			\includegraphics{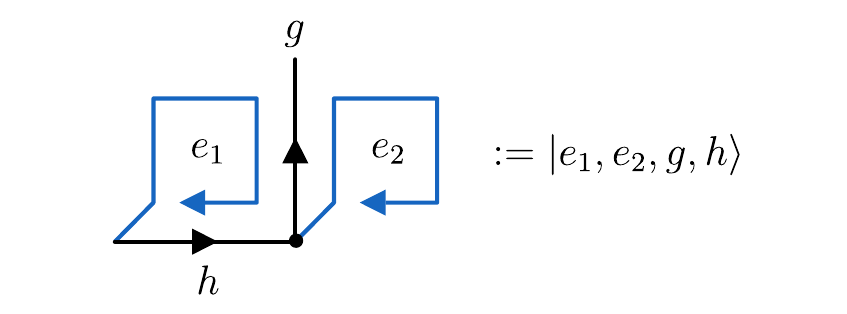}
				\caption{We introduce notation to represent the state of the degrees of freedom around the vertical edges of our lattice (here the one labelled by $g$).}
				\label{vert_edge_op}
			\end{center}
		\end{figure}

		Next we consider the edge terms, starting by considering the edge terms for the vertical edges, $\mathcal{A}_{\uparrow}$. In Figure \ref{vert_edge_op} we define the state $\ket{e_1,e_2,g,h}$ for the degrees of freedom near the edge. Again, for simplicity we suppress labels corresponding to the degrees of freedom on the rest of our lattice, which are unaffected by the edge term.

		As described in Equation \ref{Equation_edge_energy_term} in Section \ref{Section_Recap_Paper_2}, the edge energy terms are given by averages of different edge transforms, where the edge transforms are defined in Equation \ref{Equation_edge_transform_definition}. That is, the edge energy term is given by $\mathcal{A}_{\uparrow} = \frac{1}{|E|} \sum_{f \in E} A_{\uparrow}^f$. Then the action of the edge term in the new basis is
		\begin{align}
		\mathcal{A}_{\uparrow}& \ket{\mu_1,\mu_2,g,h} \notag\\
		&= \frac{1}{3}\sum_{e_1,e_2 \in \mathbb{Z}_3}\mathcal{A}_{\uparrow} \mu_1(e_1)\mu_2(e_2)\ket{e_1,e_2,g,h} \notag\\
		&= \frac{1}{3}\sum_{e_1,e_2\in \mathbb{Z}_3} (\frac{1}{3} \sum_{f \in E} \mathcal{A}_{\uparrow}^f) \mu_1(e_1)\mu_2(e_2)\ket{e_1,e_2,g,h} \notag \\
		&=\frac{1}{9} \sum_{e_1,e_2\in \mathbb{Z}_3} \mu_1(e_1)\mu_2(e_2) \sum_{f \in E} \ket{[h \rhd f]e_1, e_2f^{-1},g,h} \notag\\
		&= \frac{1}{9} \sum_{f,e_1',e_2' \in \mathbb{Z}_3} \mu_1([h \rhd f^{-1}]e_1')\mu_2( e_2'f) \ket{e_1',e_2',g,h},
		\label{Equation_Z2_Z3_irrep_basis_vertical_edge_transform_1.5}
		\end{align}
		where we replaced the dummy indices $e_1$ and $e_2$ with $e_1'= [h \rhd f] e_1$ and $e_2'=e_2f^{-1}$ in the last step. We can then use the fact that $\mu_1$ and $\mu_2$ are irreps to split off the contribution from $f$ from the contributions of $e_1'$ and $e_2'$ in Equation \ref{Equation_Z2_Z3_irrep_basis_vertical_edge_transform_1.5}, to obtain
		\begin{align}
		\mathcal{A}_{\uparrow}& \ket{\mu_1,\mu_2,g,h}\notag\\ 
		&= \frac{1}{9} \sum_{f,e_1',e_2' \in \mathbb{Z}_3} \mu_1(h \rhd f^{-1}) \mu_2(f)\mu_1(e_1') \mu_2(e_2')\notag\\ &\hspace{2cm} \times \ket{e_1', e_2',g,h}\notag\\
		&= \frac{1}{3} \sum_{f \in \mathbb{Z}_3}\mu_1(h \rhd f^{-1}) \mu_2(f) \ket{\mu_1,\mu_2,g,h}. \label{Equation_Z2_Z3_irrep_basis_vertical_edge_transform_2}
		\end{align}
		
		The above equation includes the quantity $\mu_1(h \rhd f^{-1})$. Because we are working with irreps, rather than group elements, it is useful to consider the $h \rhd$ as acting on the irrep rather than the group, just as we did in Section \ref{Section_2D_RO_Fake_Flat} for the $E$-valued membrane operators. Consider a general expression $\mu(g \rhd e)$, where $\mu$ is an irrep of $E$ and $g$ is a general group element of $G$. We see that this will give us a new irrep of $E$ acting on $e$, $\mu'(e)$, with the resulting irrep depending on $g$:
		\begin{align*}
		1_R(g \rhd e)&=1=1_R(e)\\
		\alpha_R(1 \rhd e)&=\alpha_R(e)\\
		\alpha^2_R(1 \rhd e) &= \alpha^2_R(e)\\
		\alpha_R(-1 \rhd e)&= \alpha_R(e^{-1})=\alpha_R^{-1}(e)=\alpha_R^2(e)\\
		\alpha_R^2(-1 \rhd e)&= \alpha_R^2(e^{-1})=(\alpha_R^2)^{-1}(e)=\alpha_R(e).
		\end{align*}
		
		Therefore, we can write $\mu(g \rhd e)$ as a group element acting on an irrep, by defining the irrep $g \rhd \mu$ according to
		$$\mu(g \rhd e) = [g \rhd \mu](e).$$
		We say that irreps that are related by this $\rhd$ action, for some value of $g \in G$, are in the same \textit{$\rhd$-Rep class}. For this crossed module, $\alpha_R$ and $\alpha_R^2$ are in the same $\rhd$-Rep class (they are related by the action $-1 \rhd$), while the trivial irrep is in a class of its own. We note that the composition rule for this action on the irreps is generally reversed:
		\begin{align}
		[(gh) \rhd \mu](e) &= \mu((gh) \rhd e)= \mu (g \rhd (h \rhd e)) \notag\\
		&= [g \rhd \mu](h \rhd e) = [h \rhd (g \rhd \mu)](e), \label{rhd_irrep_composition}
		\end{align}
		although this does not matter in this case because $G$ is Abelian. Under this definition of the $\rhd$ action on the irreps, for an arbitrary irrep $\mu$ of $E$, we have $1\rhd \mu=\mu$ and $-1 \rhd \mu=\mu^{-1}$.

		Returning to our calculation of the action of the edge terms, we can insert the relation $$\mu_1(h \rhd f^{-1}) = [h \rhd \mu_1](f^{-1})$$ into our expression for the action of $\mathcal{A}_{\uparrow}$ from Equation \ref{Equation_Z2_Z3_irrep_basis_vertical_edge_transform_2}, to obtain
		\begin{align}
		\mathcal{A}_{\uparrow} \ket{\mu_1,\mu_2,g,h} &= \frac{1}{3} \sum_{f \in \mathbb{Z}_3} [h \rhd \mu_1(f^{-1})] \mu_2(f) \ket{\mu_1,\mu_2,g,h} \notag \\
		&=\delta(\mu_2,h \rhd \mu_1) \ket{\mu_1,\mu_2,g,h}, \label{Equation_Z2_Z3_irrep_basis_vertical_edge_transform_3} 
		\end{align}
		where the last equality follows from standard orthogonality relations for irreps of groups. We see that the edge transform enforces that the irreps labelling the plaquettes separated by the edge must be related by the action of $h \rhd$ in the low energy state, where $h$ labels the edge separating the base-points of the two plaquettes.

\begin{figure}[h]
		\begin{center}
			\includegraphics{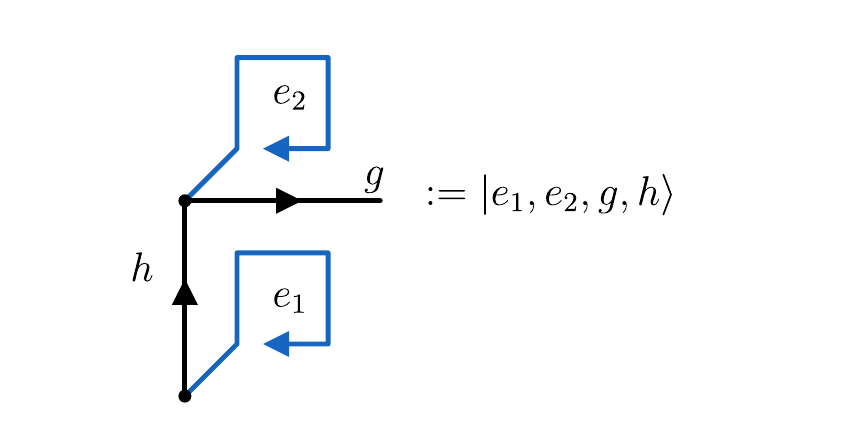}
			\caption{We introduce notation to describe the state of the degrees of freedom affected by a horizontal edge transform.}
			\label{2D_horizontal_edge_transform_1}
		\end{center}
	\end{figure}

		Next we consider the other type of edge energy term, $\mathcal{A}_{\rightarrow}$, corresponding to the horizontal edges of our lattice, for which we will find a similar result. As with the vertex transform and other edge transform, we use a simplified notation to describe the degrees of freedom in the support of this energy term, as shown in Figure \ref{2D_horizontal_edge_transform_1}. Then, following the same procedure as for the vertical edge energy terms, we have the following action for this horizontal edge energy term (again using the definition of the edge transforms from Equation \ref{Equation_edge_transform_definition} and averaging them to obtain the edge term):
		\begin{align}
		\mathcal{A}_{\rightarrow}&\ket{\mu_1,\mu_2,g,h} \notag \\
		&= \frac{1}{9} \sum_{e_1,e_2,f \in \mathbb{Z}_3} \mu_1(e_1)\mu_2(e_2)\mathcal{A}_{\rightarrow}^f \ket{e_1,e_2,g,h} \notag \\
		&= \frac{1}{9} \sum_{e_1,e_2,f\in \mathbb{Z}_3} \mu_1(e_1)\mu_2(e_2) \ket{e_1[h \rhd f^{-1}],fe_2 ,g,h} \notag \\
		&=\frac{1}{9} \sum_{e_1',e_2',f\in \mathbb{Z}_3} \mu_1(e_1' [h \rhd f])\mu_2(e_2'f^{-1}) \ket{e_1',e_2',g,h} \notag \\
		&= \frac{1}{3} \sum_{f \in \mathbb{Z}_3} \mu_1(h \rhd f) \mu_2(f^{-1}) \ket{\mu_1,\mu_2,g,h} \notag \\
		&= \delta(\mu_2,h \rhd \mu_1) \ket{\mu_1,\mu_2,g,h}. \label{Equation_Z2_Z3_irrep_basis_horizontal_edge_transform_1} 
		\end{align}
	
		Again we see that the irreps labelling the plaquettes separated by the edge must be related by the $\rhd$ action of the label of the edge separating their base-points. This suggests that the plaquette labels must be different to account for the effect of choosing different base-points (recall that moving the base-point of a plaquette has a similar $\rhd$ action on the plaquette label). That is, if we had chosen the plaquettes to have the same base-points, the irreps labelling them would be the same in the low energy space. Indeed, we shall see that this is true when we consider this in more detail in Section \ref{Section_Z4_Z4_Alternate_Lattice}.

		The last energy term to consider is the plaquette term, which is described in the group element basis in the bottom of Figure \ref{2Dtransforms}. In a general higher lattice gauge theory model, the plaquette term $B_p$ depends on the plaquette label $e_p$ only through $\partial(e_p)$. This means that in the example model under consideration, where $\partial$ maps to the identity element $1_G$, the plaquette term does not act on the plaquette labels, and so it acts exactly the same in the basis where the plaquettes are labelled by irreps as in our original basis where they were labelled by group elements. That is, the plaquette term is satisfied if the boundary of the plaquette has total label $1_G$, and gives zero otherwise. 
		
		\subsubsection{Ground states}
		\label{Section_Z2_Z3_Ground_States}
		Having considered the various energy terms of this example model, we wish to use them to examine the ground states. The number of ground states will generally depend on the topology of the manifold. Our fragment of lattice can be completed to a variety of manifolds, depending on the boundary conditions of our lattice. We want to keep the discussion general, but it may be useful to have the sphere in mind. Typically, topological models on the sphere have a unique ground-state, in the absence of symmetry. In order to close the lattice we have been considering into a sphere, we may need to include additional vertices that do not look like the ones in the fragment of lattice shown in Figure \ref{2Dlattice}. However the general features of the different energy terms will still be preserved when we include such vertices.

		Now consider the restrictions for states in the ground state sector. Flatness constraints (the plaquette terms) restrict our choice of configuration, by restricting the allowed edge labels. Then given a set of edge labels, the plaquette labels (in representation basis) are restricted by the edge transforms. We can see from Equations \ref{Equation_Z2_Z3_irrep_basis_vertical_edge_transform_3} and \ref{Equation_Z2_Z3_irrep_basis_horizontal_edge_transform_1} that the edge transforms fix the labels $\mu_1$ and $\mu_2$ of neighbouring plaquettes to be related by $\mu_2=h \rhd \mu_1$, where $h$ labels the edge separating the base-points of the two plaquettes. To see how this determines the allowed plaquette labels, consider taking one of these plaquettes and choosing a particular irrep $\mu_1$ to label it. If we choose $\mu_1$ to be the trivial irrep $1_R$, then the label $\mu_2$ of an adjacent plaquette must satisfy $\mu_2=1_R$, because $h \rhd$ acts trivially on the trivial irrep for any group label $h \in G$. Then by iteration, this will be true for any plaquette connected to the first by a path on the dual lattice. Therefore, for a path connected manifold, once we choose one plaquette to be labelled by $1_R$, all of the other plaquette labels must also take that value.

	\begin{figure}[h]
	\begin{center}
		\includegraphics{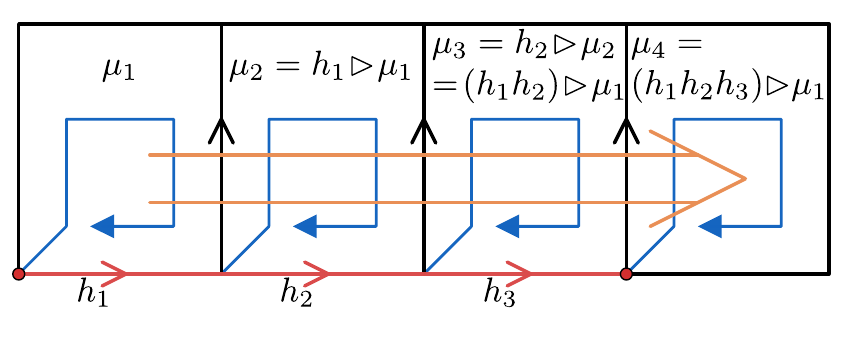}
		\caption{The edge terms relate the labels of neighbouring plaquettes, so that the labels $\mu_1$ and $\mu_2$ of plaquettes 1 and 2 in the figure are related by $\mu_2 = h_1 \rhd \mu_1$. By iterating this relation, we can relate the labels of distant plaquettes, such as plaquettes 1 and 4 in the figure, as long as they are connected by a ribbon (here the direct path of the ribbon is the thicker red edges at the bottom and the dual path is the large orange arrow in the middle).}
		\label{Z2_Z3_related_plaquettes}
	\end{center}
\end{figure}

		Next we look at the case where we choose the irrep label $\mu_1$ of the first plaquette to be $\mu_1=\alpha_R$ or $\alpha_R^2$ instead of $1_R$. Then the label $\mu_2$ of an adjacent plaquette could be either $\alpha_R$ or $\alpha^2_R$, but which one is fixed by the edge configuration (the labels of the two plaquettes match if the edge connecting their base-points is labelled by $1$ and are opposite if the edge is labelled by $-1$). Again, all path-connected plaquettes will be fixed in this way. By iterating the relationship that two adjacent plaquettes 1 and 2 labelled $\mu_1$ and $\mu_2$ must satisfy $\mu_2 = h \rhd \mu_1$, where $h$ is the label of the edge connecting the base-points of plaquette 1 and plaquette 2, we can find the relationship between any two plaquettes connected by a path (see Figure \ref{Z2_Z3_related_plaquettes} for an example). For a plaquette 1 with base-point at the start of path $t$ and plaquette n with base-point at the end of the path, we must have $\mu_n =g(t) \rhd \mu_1$, where $g(t)$ is the label of path $t$. If there are multiple paths between two plaquettes, we must ask if these different paths always give consistent conditions on the two irreps. In order to answer this question, consider two such paths $t_1$ and $t_2$ between the two plaquettes. Because these two paths share the same start and end-point, we can construct a closed path $s=t_1 t_2^{-1}$ by concatenating one with the inverse of the other. If the two paths $t_1$ and $t_2$ can be smoothly deformed into one-another, then the closed path $s$ is contractible and so must satisfy fake-flatness in the ground state. For a general crossed module, fake-flatness enforces that the label of the closed path is related to the label $e_s$ of the surface enclosed by that path by $\partial(e_s)g(s)=1_G$. In the case of this specific $\mathbb{Z}_2$, $\mathbb{Z}_3$ model however, $\partial$ maps to the identity element of $G$ and so $\partial(e_s)$ is always $1_G$. Therefore, $g(s)=g(t_1)g(t_2)^{-1}=1_G$, meaning that the labels $g(t_1)$ and $g(t_2)$ of the two paths are the same, and so the two paths give consistent conditions $\mu_n =g(t_1) \rhd \mu_1 = g(t_2) \rhd \mu_1$. Of course, this relies on the two paths being related by a smooth deformation, which means that the closed path $s$ is contractible. However if this is not so and the closed path is non-contractible then the closed path $s$ can have either label $\pm 1$. If it has the label $-1$ then we have $g(t_1) = -1 \cdot g(t_2)$, which leads to the two conditions
		$$\mu_n = g(t_2) \rhd \mu_1$$
		and $$\mu_n = (-1 \cdot g(t_2)) \rhd \mu_1,$$
		which together imply that $\mu_1 = -1 \rhd \mu_1$ in order for the two conditions to be consistent and all of the edge terms to be satisfied. This can only be true if $\mu_1= 1_R$, not if $\mu_1 = \alpha_R$ or $\alpha_R^2$. We therefore see that, if there are non-contractible paths, not all edge configurations are compatible with choosing the first plaquette label to be $\alpha_R$ or $\alpha_R^2$ (specifically, a configuration is incompatible with this choice if there are closed paths with non-trivial label).

		From this procedure of fixing the edge configurations, then making a particular choice for the label of one plaquette, we might think that on the sphere we have three choices of plaquette configurations per edge configuration, one for each irrep of $E$. However this is not quite true, because some of the resulting sets of plaquette labels are connected by the energy terms. If we apply $\prod_{\text{all }v} A_v^{-1}$, the product of vertex transform $A_v^{-1}$ on every vertex of the lattice, the edge labels are unchanged because we act on both ends of the edge, so that $g \rightarrow -1 \cdot g \cdot -1 =g$. On the other hand, we apply a transform on the base-point of each of the plaquettes precisely once, so all of the plaquette labels are affected by the $\rhd$ action from the vertex transform. This means that the label $\mu_p$ of each plaquette $p$ transforms as $\mu_p \rightarrow -1 \rhd \mu_p$ (from Equation \ref {Equation_Z2_Z3_vertex_transform}). This leaves the identity irrep unchanged, and so does not affect the state arising from an initial choice of $\mu_1 =1_R$. On the other hand, if we took a non-trivial irrep as the label of the first plaquette, then this plaquette swaps label from $\alpha_R$ to $\alpha_R^2$ or vice-versa. In addition, all of the other plaquettes swap their label between these two irreps. The state resulting from this is the same state we would have if we had chosen the other label for the first plaquette and then used that to determine the label of all of the other plaquettes. Then, because each ground state is invariant under the vertex transforms, the ground state satisfies $\prod_{\text{all } v} A_v^{-1} \ket{GS}=\ket{GS}$. This means that the two plaquette configurations arising from choosing the label of the first plaquette to be $\mu_1 = \alpha_R$ or $\alpha_R^2$ for a given edge configuration must appear with equal amplitude in the ground state (the configurations are connected by the vertex terms). Therefore, the ground states produced from the two initial choices of plaquette label are not distinct.

		This means that, for each edge configuration, we have a choice of only two distinct plaquette configurations, corresponding to the two $\rhd$-Rep classes of the crossed module. From these configurations, we can generate a ground state by acting with a product of all of the vertex energy terms. Given that the plaquette labels are fixed by our initial choice (to be $1_R$ or an entangled combination of $\alpha_R$ and $\alpha_R^2$) and the vertex transforms commute with this fixing, we can forget about the plaquettes after making that choice. Then the problem reduces to finding the ground state of Kitaev's Quantum Double model. This gives us two parts to our ground state sector, one for the normal Kitaev Quantum Double model in a tensor product with $1_R$ at each plaquette and one for the Quantum Double model entangled with a more complicated plaquette set. In the case where our manifold is the sphere, this gives us a ground-state degeneracy of two, rather than giving a ground state degeneracy of one as we may expect for a generic topological phase on a sphere. However the different ground states can be detected locally (by measuring the $\rhd$-Rep class of the plaquette label in a single location), so the degeneracy on the sphere is not topologically protected. Note that the result for the ground state degeneracy agrees with a more general calculation given in Ref. \cite{Bullivant2017}, as expected. 
		
		\subsubsection{Excitations}
		\label{Section_Z2_Z3_Excitations}
\begin{figure}[h]
			\begin{center}
			\includegraphics{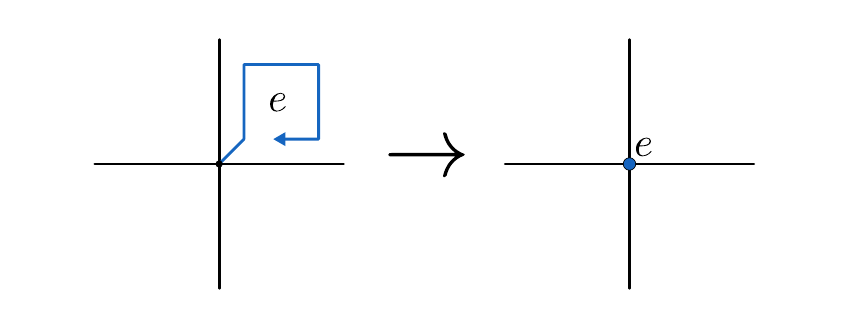}
				\caption{We imagine the label of a plaquette to be placed at the base-point of that plaquette}
				\label{plaquette_label_at_vertex}
			\end{center}
		\end{figure}

		In this model, we have the usual electric excitations of the Kitaev Quantum Double model, and these particles behave exactly as in the Quantum Double model \cite{Kitaev2003}. On the other hand, we find loop excitations which are not present in Kitaev's Quantum Double model. To understand these loop excitations it is convenient to change how we label the plaquettes. Instead of considering the plaquette labels to belong to the plaquette itself, we put them on the vertex that the plaquette is based at, as shown in Figure \ref{plaquette_label_at_vertex}.

\begin{figure}[h]
			\begin{center}
			\includegraphics{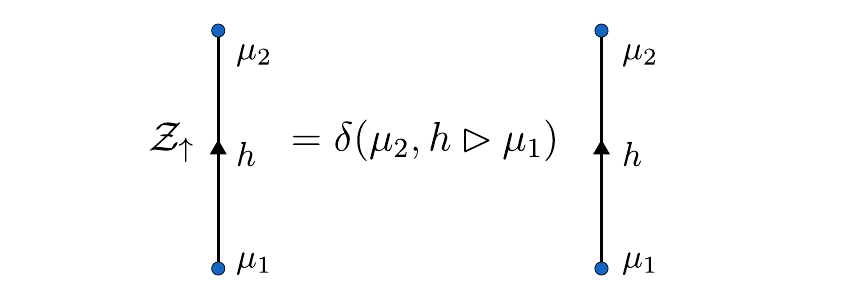}
				\caption{Our new vertical edge operator}
				\label{New_vertical_edge_transform_1}
			\end{center}
		\end{figure}

\begin{figure}[h]
			\begin{center}
			\includegraphics{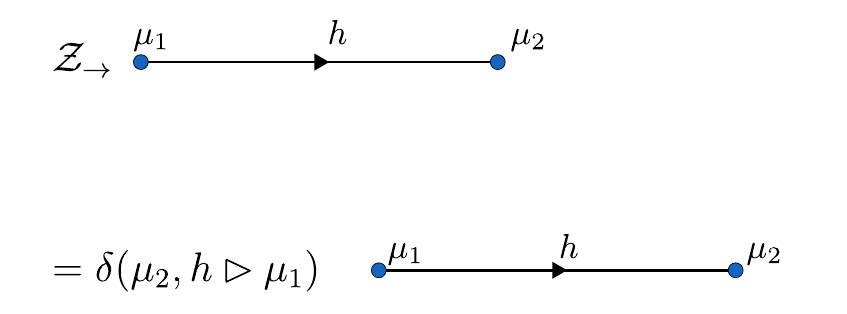}
				\caption{Our new horizontal edge operator}
				\label{New_horizontal_edge_transform_1}
			\end{center}
		\end{figure}

		Next we note that the edge transform corresponding to an edge $i$ does not change the label of that edge, because $\partial(e)=1_G$ for all $e \in E$ in this example model. However the action of the edge transform on the plaquettes does depend on a nearby edge, as indicated by Equations \ref{Equation_Z2_Z3_irrep_basis_vertical_edge_transform_3} and \ref{Equation_Z2_Z3_irrep_basis_horizontal_edge_transform_1}. Specifically, the action of the edge transform depends on the edge connecting the base-points of the plaquettes affected by the edge transform. It is therefore convenient to relabel our operators to reflect the edge that matters for its action. Then the edge term $\mathcal{A}_{\rightarrow}$ becomes $\mathcal{Z}_{\uparrow}$ (applied on the edge down and to the left of the original horizontal edge) and $\mathcal{A}_{\uparrow}$ becomes $\mathcal{Z}_{\rightarrow}$ (again applied on the edge down and to the left of the vertical edge), and we now refer to the new labelled edges as being excited or not. The action of these operators can be written more compactly now that we no longer keep track of the extra edge. The action of these new edge transforms are defined in Figures \ref{New_vertical_edge_transform_1} and \ref{New_horizontal_edge_transform_1}.

\begin{figure}[h]
			\begin{center}
				\includegraphics[width=0.4\linewidth]{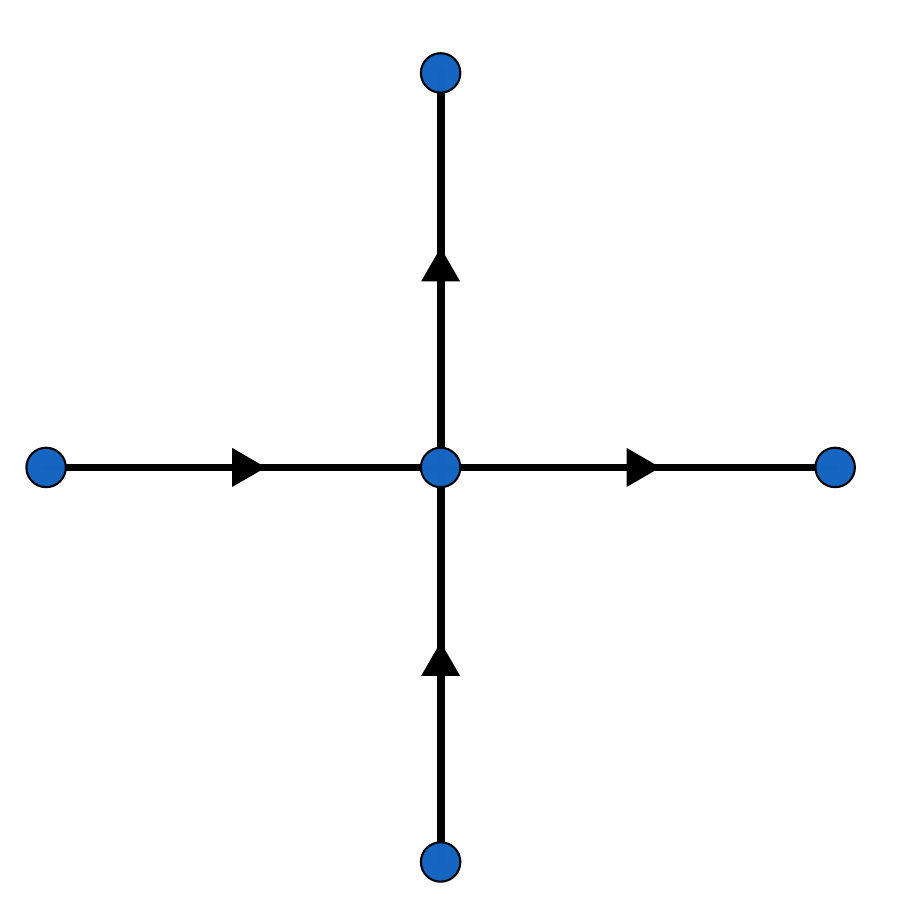}

				\caption{A vertex and its neighbours, connected by edges}
				\label{2Dfivevertices}
			\end{center}
		\end{figure}

\begin{figure}[h]
			\begin{center}
				\includegraphics[width=0.4\linewidth]{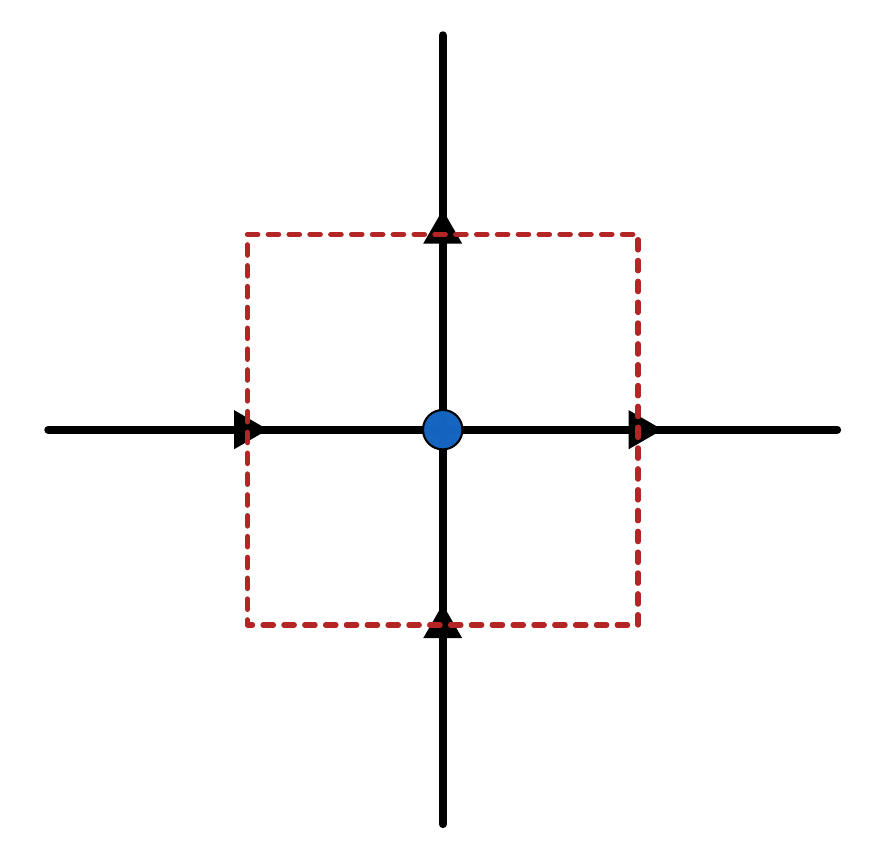}

				\caption{Changing the label of a vertex in the ground state breaks its bonds with the neighbouring vertices, resulting in the edges connecting the vertex to its neighbours becoming excited.}
				\label{2Ddomainwall1}
			\end{center}
		\end{figure}

\begin{figure}[h]
			\begin{center}
				\includegraphics[width=0.5\linewidth]{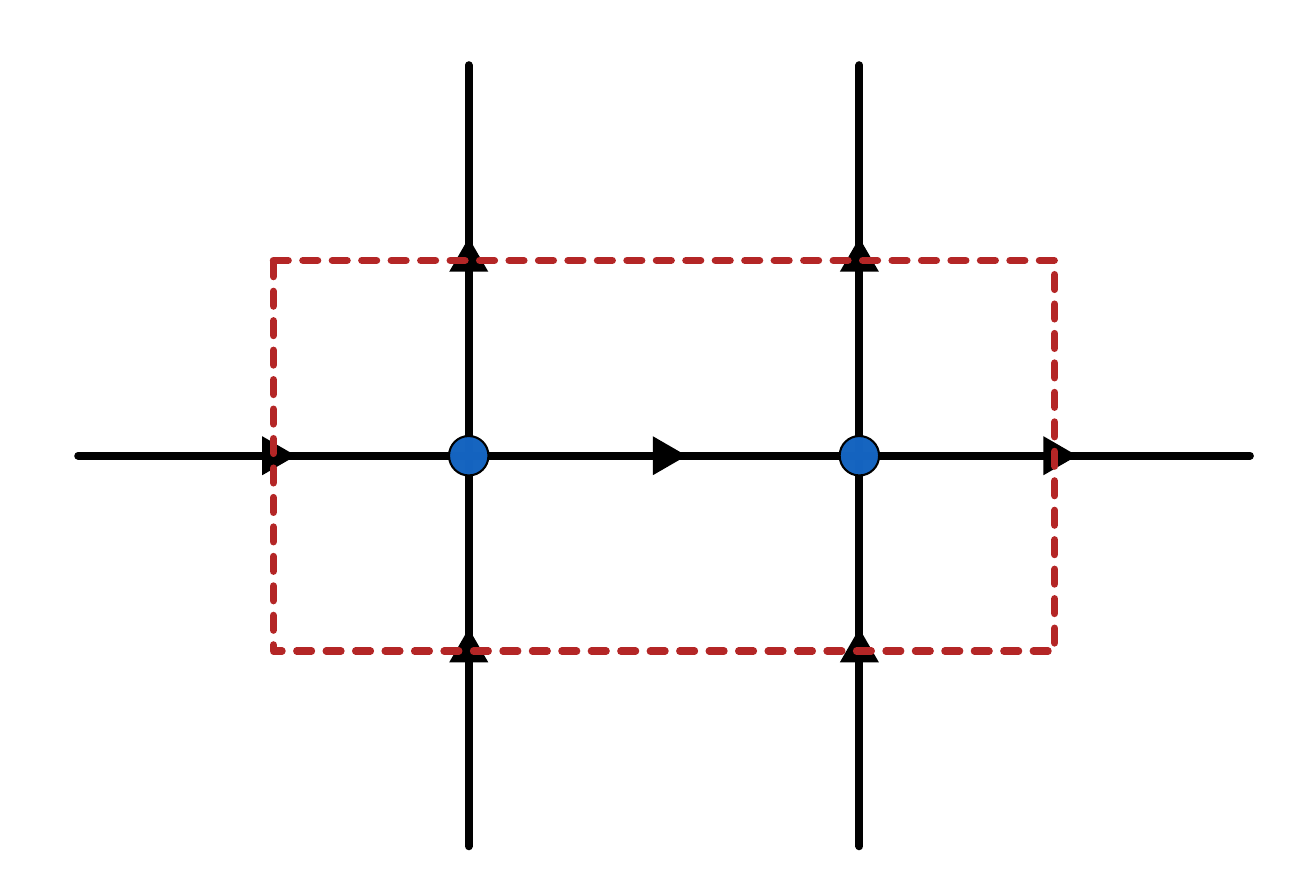}

				\caption{Attempting to fix a broken bond may lead to more excited edges, thereby expanding or changing the shape of the loop-like excitation}
				\label{2Ddomainwall2}
			\end{center}
		\end{figure}

		Now that we have relabelled our edge transforms, we see that they enforce that the (now vertex) labels at either end of the edge that we act on are related by the $\rhd$ action of the edge label. We can think of adjacent vertices satisfying this relation as being linked. Then violating an edge transform can be thought of as ``breaking" the link between vertices. Given a vertex and its neighbours (as shown in Figure \ref{2Dfivevertices}), we can excite the edges connecting the vertex to its neighbours by changing the label of that vertex. These excited edges are cut by a loop in the dual lattice, shown in Figure \ref{2Ddomainwall1} as a red dashed line, indicating that we have produced a loop-like excitation. Note that in our original construction (where we use the original edge transforms $\mathcal{A}_i$), we would have placed the excited loop on the direct lattice, up and to the right of where we place it here. We can fix one of the broken bonds between the vertex and its neighbours, by changing the label of a neighbouring vertex so that the edge condition between the two vertices is satisfied, restoring the link between them. However changing the label of this neighbouring vertex will break more links on the neighbouring vertex, as shown in Figure \ref{2Ddomainwall2}. This simply changes the shape of the excited loop.
		
		Recall that when we looked at the ground states, after fixing the edge labels we built the plaquette labels by starting at one plaquette and making the other plaquettes agree with that first one (in the sense that all of the edges connecting the plaquettes are satisfied). Similarly, to build the loop excitation, we start by changing one plaquette label, now living at a vertex, and then make the other labels within the loop agree with that vertex by ensuring that all links within this region are satisfied. That is, the region within the excited loop satisfies the same conditions as the ground state (although the bonds crossing the boundary of the loop are not satisfied). This means that we can think of the loop as a domain wall between two regions representing different ground states or ground state-like configurations. This is most clear when we start in the ground state where all of the vertex labels are $1_{R}$. Then to create the loop excitation we multiply one vertex label by $\alpha_R$ or $\alpha_R^2$ and make all of the other vertices within the loop agree with that label (so that the edge terms within the loop are satisfied). However, recalling the general case (see Section \ref{Section_2D_Loop}), the loop excitation may have an excited vertex term at the start-point of the membrane operator unless the membrane operator assigns equal weight to all surface elements within a particular class of elements. In this particular model, to avoid a vertex excitation we must take an equal superposition of the cases where we multiply the first label by $\alpha_R$ (then make all of the others agree) and the case where we multiply the first label by $\alpha^2_R$ (then make all of the others agree). The loop then corresponds to a domain wall between the $1_R$ ground state and the ground state made of a combination of $\alpha_R$ and $\alpha_R^2$, because the vertex labels within the loop are those that we would expect in the state made from $\alpha_R$ and $\alpha_R^2$ and those outside are those we expect in the $1_R$ state. By extending this membrane operator over the entire lattice, we can move from the $1_R$ ground state to the other ground state. On the other hand, the privileged vertex will be excited if we take an orthogonal superposition of the two states built from $\alpha_R$ and $\alpha_R^2$ (i.e., we take the state built from starting with the first vertex in the state $\alpha_R$ and making the other vertices agree and then make a linear combination by subtracting the state built from starting with that vertex in the state $\alpha_R^2$).
		
		\subsubsection{Magnetic excitations}
		\label{Section_Z2_Z3_Magnetic}
		
\begin{figure}[h]
			\begin{center}
				\includegraphics{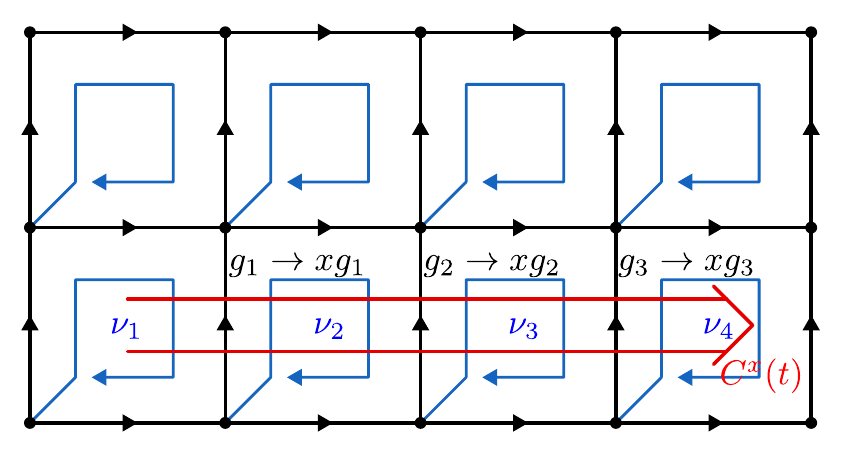}
				\caption{We consider the action of the usual magnetic ribbon operator, in the basis where the plaquettes are labelled by irreps of $E$ and the edges by elements of $G$. The ribbon operator acts only on the edges of the lattice, and because $G$ is Abelian this action is simple (without the dependence on the path element of sections of the direct path seen in the more general case, as described by Equation \ref{Equation_Magnetic_Action_C}). In this example, all of the edges cut by the ribbon point in the same direction (upwards), so the action of the ribbon operator $C^x(t)$ is to multiply each edge label by $x$ (if an edge pointed the opposite way with respect the ribbon, it would have its label multiplied by $x^{-1}$ instead).}
				\label{Z2_Z3_magnetic_1}
			\end{center}
		\end{figure}

		As we described in Section \ref{Section_Recap_Paper_2}, we are not typically able to construct the magnetic excitations in 2+1d when $\rhd$ is non-trivial. For this simple model however (and some generalisations, as we describe in Section \ref{Section_confined_magnetic} in the Supplemental Material), and with this fixed branching structure, we are able to find these excitations. In fact, the ribbon operator that produces the magnetic excitation in the $(\mathbb{Z}_2, \mathbb{Z}_3)$ model has exactly the same form as the ones used in Section \ref{Section_2D_Magnetic} for the $\rhd$ trivial case (and indeed for Kitaev's Quantum Double model \cite{Kitaev2003}). However, unlike in the $\rhd$ trivial case, the properties of the magnetic excitations depend on which ground state we create them from, as we shall see shortly. In order to reveal these properties, we consider applying the magnetic ribbon operator on a basis state in the basis where the plaquettes are labelled by irreps of $E$ (but the edges are still labelled by group elements of $G$), as shown in Figure \ref{Z2_Z3_magnetic_1}.

\begin{figure}[h]
			\begin{center}
				\includegraphics[width=\linewidth]{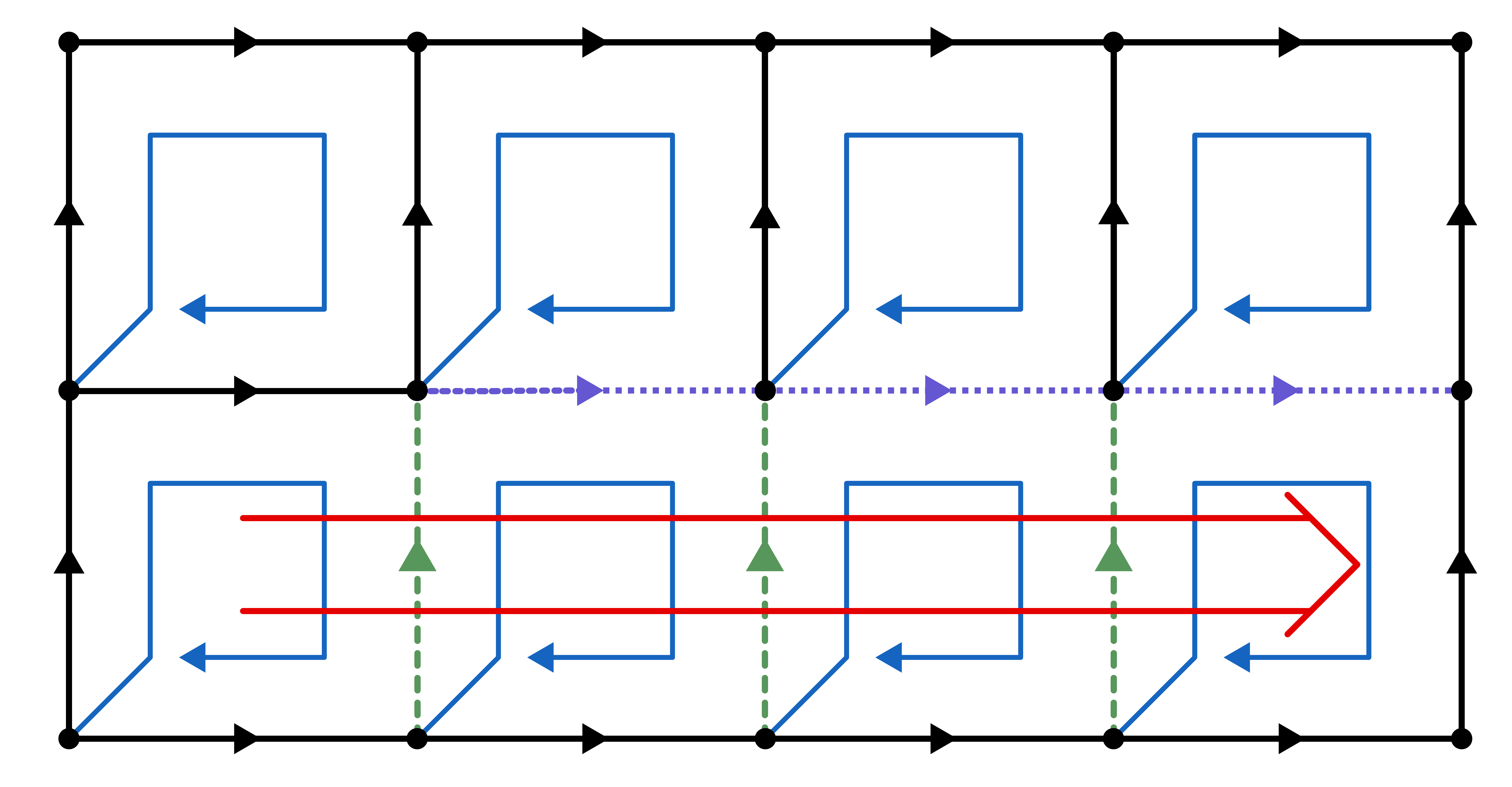}
					
				\caption{The edge terms that may be excited are those whose action on the plaquette labels depends on the edges cut by the ribbon operator (the green dashed edges). That means that the edge terms on the purple dotted edges may be excited if we are using the original $A_i$ edge terms, or the edge terms on the dashed green edges themselves may be excited if we are using the altered $\mathcal{Z}_{\uparrow}$ edge terms.}
				\label{Z2_Z3_magnetic_2}
			\end{center}
		\end{figure}

		We want to consider the commutation relations between the ribbon operator and the different energy terms. The magnetic ribbon operator only acts on the edges of the lattice, and it does so in the same way as in the case where $\rhd$ is trivial. This means that it has the usual commutation relations with the vertex terms (which commute with it because $G=\mathbb{Z}_2$ is Abelian) and the plaquette terms (which commute with it, except at the two end plaquettes), both of which act on the edges of the lattice in the same way as in the $\rhd$ trivial case. On the other hand, the action of the edge term on the plaquettes depends on the edge labels (see Equation \ref{Equation_Z2_Z3_irrep_basis_vertical_edge_transform_3}), unlike in the $\rhd$ trivial case, and so there is some potential non-commutativity here. Looking at Equation \ref{Equation_Z2_Z3_irrep_basis_vertical_edge_transform_3}, we see that the edge transforms that may fail to commute with the magnetic ribbon operator are those whose action on the plaquette labels depends on an edge label of an edge that is cut by the ribbon operator (e.g., if the edge label $h$ in Figure \ref{vert_edge_op} is changed by the ribbon operator, then that edge transform may fail to commute). If we are using the $\mathcal{Z}_{\rightarrow}$ and $\mathcal{Z}_{\uparrow}$ edge terms, then this means that edge transforms applied on the edges cut by the ribbon operator may fail to commute, whereas for the original $\mathcal{A}_i$ edge operators, it is the edge transforms on the edges up and to the right of the cut edges that may fail to commute (recall from Section \ref{Section_Z2_Z3_Excitations}, and in particular Figures \ref{New_vertical_edge_transform_1} and \ref{New_horizontal_edge_transform_1}, that the $\mathcal{Z}$ edge term on an edge is equivalent to the original $\mathcal{A}$ term on the edge above and to the right). This is illustrated in Figure \ref{Z2_Z3_magnetic_2}.
		
		\begin{figure}[h]
		\begin{center}
		\includegraphics{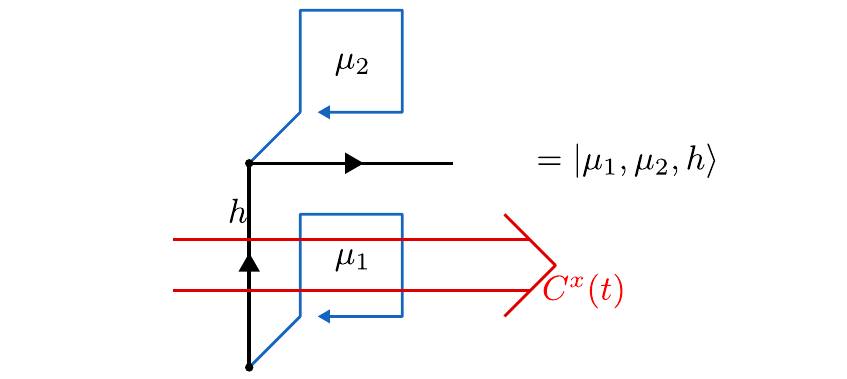}
			\caption{We wish to consider the commutation relation between the edge transforms and magnetic ribbon operator. To do so we consider the degrees of freedom affected by the edge transforms indicated in Figure \ref{Z2_Z3_magnetic_2}, and provide a shorthand for the associated state. }
			\label{Z2_Z3_magnetic_shorthand}
		\end{center}
	\end{figure}

		Having determined which edge transforms may fail to commute, we now look explicitly at the commutation relations. To do so we must consider the degrees of freedom around one of the edges cut by the magnetic ribbon, and so we will use the shorthand illustrated in Figure \ref{Z2_Z3_magnetic_shorthand}. Then the action of the edge term $\mathcal{Z}_{\uparrow}$ on this state is (from Figure \ref{New_vertical_edge_transform_1})
		$$\mathcal{Z}_{\uparrow} \ket{\mu_1, \mu_2 , h} = \delta( \mu_2, h \rhd \mu_1)\ket{\mu_1, \mu_2 , h},$$
		whereas the action of the magnetic ribbon operator $C^x(t)$ on these degrees of freedom is
		
		$$C^x(t)\ket{\mu_1, \mu_2 , h} = \ket{\mu_1, \mu_2 , xh}.$$
		
		This means that applying the edge term and then the magnetic ribbon operator gives
			$$C^x(t)\mathcal{Z}_{\uparrow}\ket{\mu_1, \mu_2 , h} = \delta( \mu_2, h \rhd \mu_1)\ket{\mu_1, \mu_2 , xh},$$
		whereas applying the magnetic ribbon operator and then the edge term gives
		$$\mathcal{Z}_{\uparrow}C^x(t)\ket{\mu_1, \mu_2 , h} = \delta( \mu_2, (xh) \rhd \mu_1)\ket{\mu_1, \mu_2 , xh}.$$
		
		In order to determine whether the ribbon operator excites the edge, we consider the product 
		$$\mathcal{Z}_{\uparrow} C^x(t) \mathcal{Z}_{\uparrow},$$
		which first projects to the case where the edge is initially unexcited, then acts with the ribbon operator and then projects again. If the ribbon operator excites the edge, then this product will give zero. On the other hand, if the ribbon operator leaves the edge unexcited, then we will obtain $C^x(t)\mathcal{Z}_{\uparrow}$ (because the second projection is trivial). We find that
		\begin{align}
		\mathcal{Z}_{\uparrow} C^x(t)& \mathcal{Z}_{\uparrow}\ket{\mu_1, \mu_2 , h} \notag \\
		 &= \mathcal{Z}_{\uparrow} \delta( \mu_2, h \rhd \mu_1)\ket{\mu_1, \mu_2 , xh} \notag \\
		&= \delta( \mu_2, (xh) \rhd \mu_1) \delta( \mu_2, h \rhd \mu_1)\ket{\mu_1, \mu_2 , xh}\notag \\
		&= \delta( h \rhd \mu_1, (xh) \rhd \mu_1)\delta( \mu_2, h \rhd \mu_1)\ket{\mu_1, \mu_2 , xh} \notag\\
		&=\delta( \mu_1, x \rhd \mu_1)\delta( \mu_2, h \rhd \mu_1)\ket{\mu_1, \mu_2 , xh} \notag\\
		&= \delta( x \rhd \mu_1, \mu_1) C^x(t) \mathcal{Z}_{\uparrow}\ket{\mu_1, \mu_2 , h}. \label{Equation_Z2_Z3_magnetic_edge_excited}
		\end{align}
		
		This indicates that whether the edge is excited or not depends on the term $\delta( x \rhd \mu_1, \mu_1)$. Taking $x$ to be the only non-trivial element $-1$ of $G=\mathbb{Z}_2$, and using the rule $ -1 \rhd \mu(e) = \mu( -1 \rhd e)$ with the definition of $\rhd$ given in Table \ref{properties_Z_2_Z_3_maps} ($-1 \rhd e = e^{-1}$), we see that
		$$-1\rhd \mu(e) = \mu (e^{-1})= \mu^{-1}(e).$$
		
		Therefore, 
		\begin{align*}
		\delta( -1 \rhd \mu_1, \mu_1)&= \delta( \mu_1^{-1}, \mu_1).
		\end{align*}
		This Kronecker delta is satisfied (and so the edge is not excited) if $\mu_1=1_R$, and is not satisfied (implying an excited edge) if $\mu_1$ is $\alpha_R$ or $\alpha^2_R$. A general state will not be an eigenstate of this operator and so does not give us a well defined energy for the edge. However, the ground states that we constructed in Section \ref{Section_Z2_Z3_Ground_States} (which form a basis for the ground state sector) will. To see this, recall that we found two unique ground states on the sphere. In the first, every plaquette was labelled by the trivial irrep $1_R$. In this case, the plaquette label will always satisfy the Kronecker delta in Equation \ref{Equation_Z2_Z3_magnetic_edge_excited}, and so the edge will never be excited by the ribbon operator. This is true for each edge, and so the magnetic excitation is not confined. On the other hand, in the second ground state every irrep is in some combination of the states $\alpha_R$ and $\alpha_R^2$ (entangled with the states of each other irrep). Regardless of which of these states the plaquette is in (or how it is entangled), the plaquette will therefore not satisfy $\delta( \mu_1^{-1}, \mu_1)$ and so the edge will be excited. This will be true for every edge cut by the ribbon operator (or the edges above and to the right if we use the original edge terms $\mathcal{A}_i$). This gives an energy cost that grows with the length of the ribbon, and so we see that the magnetic ribbon operator is confined. However, unlike the confinement of the electric excitations that we have seen previously, this confinement depends on which ground state we create the magnetic excitation from. While we have only been considering a specific crossed module here, this feature is more generic, as we explain in Section \ref{Section_confined_magnetic} in the Supplemental Material.

		\subsubsection{Condensation}
		\label{Section_Z2_Z3_Condensation}
\begin{figure}[h]
			\begin{center}
		\includegraphics{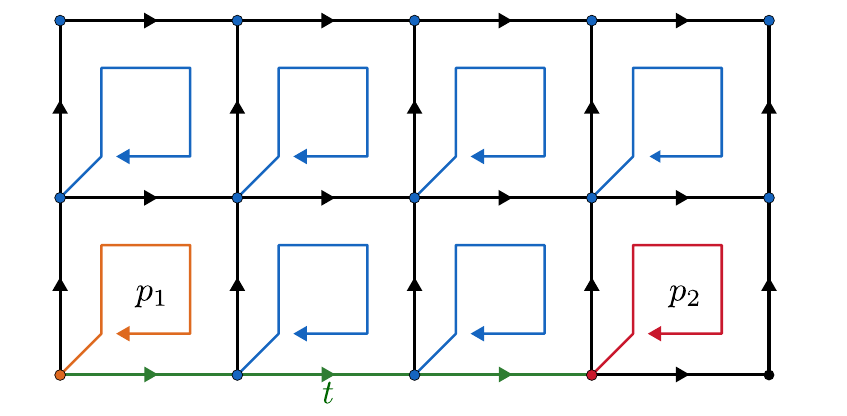}
				\caption{Given an electric ribbon operator applied on a path $t$, the two ends of the ribbon are the base-points for some plaquettes $p_1$ and $p_2$. In the ground state, the labels $\mu_1$ and $\mu_2$ of the plaquettes are related by $\mu_2 = g(t) \rhd \mu_1$, where $g(t)$ is the path element of path $t$. Locally measuring the two plaquette labels therefore gives information about the path element $g(t)$, and so can reproduce some of the supposedly non-local electric ribbon operators on $t$, indicating that the corresponding electric excitations are condensed.}
				\label{Z2_Z3_electric}
			\end{center}
		\end{figure}

		Given that we have a confined magnetic excitation in one of the ground states, we also expect to see some condensation in that ground state, in the form of a condensed electric excitation (because confined excitations arise due to non-trivial braiding with condensed excitations). Recall from Section \ref{Section_2D_Condensation_Confinement} that we use the term condensed excitation to refer to an excitation which carries trivial topological charge (usually after some condensation-confinement transition). If we can find some local operator that reproduces the action of the electric ribbon operator on the ground state, then we know that the corresponding electric excitation is condensed. Consider an electric ribbon operator applied on a path $t$ in the lattice. Because of how we have chosen our lattice, the two end-points of the path will be the base-points for two plaquettes, $p_1$ and $p_2$, as illustrated in Figure \ref{Z2_Z3_electric}. As we showed in Section \ref{Section_Z2_Z3_Ground_States}, in the ground state the labels of these two plaquettes in the irrep basis are related by the action of $g(t) \rhd$. That is, if the labels of $p_1$ and $p_2$ are $\mu_1$ and $\mu_2$ respectively, then $\mu_2 = g(t) \rhd \mu_1$. Now consider the bi-local operator $\delta(\hat{\mu}_1, \hat{\mu}_2)$, where $\hat{\mu}_i$ measures the value of plaquette $i$. In the ground state labelled by the trivial irrep, this operator acts as the identity because both plaquettes are labelled by the trivial irrep. However in the $\alpha_R/ \alpha^2_R$ ground state this is not the case. Instead, this Kronecker delta is one when the path element $g(t)$ separating the plaquettes is $1_G$ and zero when the path element is $-1_G$ (because $-1_G \rhd \alpha_R = \alpha^2_R$). Therefore, when acting on the ground state $\ket{\alpha_R/ \alpha^2_R}$, we have
		$$\delta(\hat{\mu}_1, \hat{\mu}_2)\ket{\alpha_R/ \alpha^2_R} = \delta(\hat{g}(t),1_G)\ket{\alpha_R/ \alpha^2_R},$$
		which is just an electric ribbon operator acting on the ground state. We can also construct
		$$\delta(\hat{\mu}_1, \hat{\mu}_2^{-1})\ket{\alpha_R/ \alpha^2_R} = \delta(\hat{g}(t), -1_G)\ket{\alpha_R/ \alpha^2_R}.$$
		
	This means that a general electric ribbon operator
	 $$a\delta(\hat{g}(t),1_G) + b \delta(\hat{g}(t),-1_G)$$
	(where $a$ and $b$ are arbitrary coefficients) can be reproduced (when acting on this particular ground state) by the bi-local operator
	 $$a\delta(\hat{\mu}_1, \hat{\mu}_2) + b\delta(\hat{\mu}_1, \hat{\mu}_2^{-1}).$$
	 
	 This indicates that the electric excitations are condensed (although when $G=\mathbb{Z}_2$ there is only one independent electric excitation in the first place, because there are two orthogonal electric ribbon operators, of which one is the trivial operator). In the $\alpha_R/\alpha^2_R$ ground state, where the non-trivial magnetic excitation is confined, the only non-trivial electric excitation is condensed (as we may expect from the idea that the confined excitation should braid non-trivially with the condensed one). On the other hand, there is no condensation or confinement in the $1_R$ ground state. Just like the confinement of magnetic excitations, this ground state dependent condensation occurs in the 2+1d theory for other crossed modules, as we show in Section \ref{Section_condensed_magnetic_electric} in the Supplemental Material.
		
		\subsubsection{Alternate lattice}
		\label{Section_Z4_Z4_Alternate_Lattice}
		
		Returning to the discussion of the ground states, we can make a further simplification to the model that will make the structure of the ground states more obvious. Instead of basing every plaquette at the bottom left corner of the plaquette, we can base every plaquette at a single special vertex $v_0$ (in order to do this in a consistent way, we must have no non-contractible loops in the lattice). Because the choice of base-point is analogous to a choice of gauge, this is equivalent to choosing all of the surface elements to be in the same gauge. This should therefore simplify certain expressions. Apart from the choice of base-point, we proceed as before, replacing the plaquette labels with representations using a change of basis and placing the plaquette labels at the vertex down and to the left of the plaquette (although we note that the plaquette labels are no longer based at that vertex).

\begin{figure}[h]
			\begin{center}
			\includegraphics{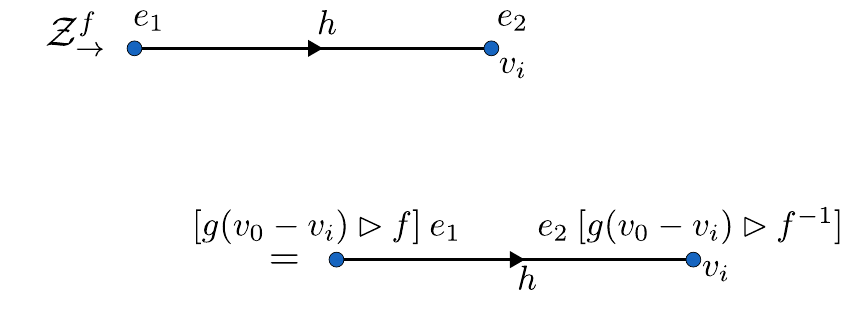}
				\caption{The action of the horizontal edge operator $\mathcal{Z}_{\rightarrow}^f$ in the group-valued basis. Here $g(v_0-v_i)$ is the path element from a privileged vertex in our lattice, $v_0$, to $v_i$, the vertex at the end of the edge that is being transformed.}
				\label{2D_horizontal_transform_Z}
			\end{center}
		\end{figure}

		We start by considering the edge transforms on this new lattice. Just as before, the edge transforms do not affect the edge on which they are applied, and so it is convenient to instead associate them with the edge joining the two plaquette labels that they affect. That is, we define new operators $\mathcal{Z}^f_{\rightarrow}=A_{\uparrow}^f$ and $\mathcal{Z}^f_{\uparrow}=A^f_{\rightarrow}$, similar to the operators $\mathcal{Z}_{\uparrow}$ and $\mathcal{Z}_{\rightarrow}$ defined in Figures \ref{New_vertical_edge_transform_1} and \ref{New_horizontal_edge_transform_1}, except for individual edge transforms rather than edge energy terms. The action of $\mathcal{Z}^f_{\rightarrow}$ on the group element basis is illustrated in Figure \ref{2D_horizontal_transform_Z} (and we can obtain the action of $\mathcal{Z}^f_{\uparrow}$ by rotating the figure ninety degrees anticlockwise). The associated energy terms $\mathcal{Z}_{\rightarrow}$ and $\mathcal{Z}_{\uparrow}$ are obtained by averaging these transforms over all elements $f \in E$ as usual. For instance
		$$\mathcal{Z}_{\rightarrow} = \frac{1}{|E|} \sum_{f \in E} \mathcal{Z}^f_{\rightarrow}.$$

		Then for the action in the representation basis, we have
		\begin{align*}
		&\mathcal{Z}_{\rightarrow} \ket{\mu_1,\mu_2}\\
		&= \big(\frac{1}{|E|} \sum_{f \in E} \mathcal{Z}_{\rightarrow}^f\big) \big(\frac{1}{|E|} \sum_{e_1,e_2 \in E} \mu_1(e_1)\mu_2(e_2)\ket{e_1,e_2}\big)\\
		&= \frac{1}{9} \sum_{e_1,e_2,f \in \mathbb{Z}_3} \mu_1(e_1)\mu_2(e_2)\\
		&\hspace{0.5cm} \ket{ [g(v_0-v_i) \rhd f] e_1, [g(v_0-v_i) \rhd f^{-1}] e_2} \\
		&= \frac{1}{9} \sum_{e_1,e_2,f'=g(v_0-v_i) \rhd f}\hspace{-0.8cm}\mu_1(e_1)\mu_2(e_2)\ket{f^{'}e_1,f^{'-1}e_2}\\
		&= \frac{1}{9} \sum_{f' \in \mathbb{Z}_3} \sum_{e_1' =f'e_1} \sum_{e_2'=f^{' -1}e_2}\mu_1(f^{'-1}e_1')\mu_2(f'e_2')\ket{e_1',e_2'}\\
		&= \frac{1}{3} \sum_{f' \in \mathbb{Z}_3} \mu_1(f^{'-1}) \mu_2(f') \frac{1}{3} \sum_{e_1', e_2' \in \mathbb{Z}_3}\mu_1(e_1')\mu_2(e_2')\ket{e_1',e_2'}\\
		&= \frac{1}{3} \sum_{f' \in \mathbb{Z}_3}\mu_1(f^{'-1})\mu_2(f')\ket{\mu_1,\mu_2}\\
		&=\delta(\mu_1,\mu_2)\ket{\mu_1,\mu_2}.
		\end{align*}
		
		Note that the edge label $h$ is not present in this expression. The two irrep labels on either end of the edge are forced to be the same in the low energy state, rather than being related by the action $h \rhd$. In order to obtain this result, we implicitly used flatness to make sure that all paths with the same end points have the same label (the same rather than the same up to an element of $\partial(E)$ because $\partial(E)$ is trivial), provided that all such paths can be deformed into one-another (the manifold is simply connected). The vertical edge transform gives something similar, enforcing that the irrep labels at the two ends of the vertical edge are the same. By fixing each plaquette to have the same base-point, we have removed the impact of the edge labels on the edge transforms. The vertex transforms are also simplified, because now only the vertex transform at the privileged point $v_0$ affects the surface labels, with the other vertex transforms only affecting the edge labels. This is because vertex transforms only affect plaquettes if that vertex is the base-point of the plaquette, and we have chosen all of the plaquettes to have the same base-point. The vertex transform $A_{v_0}^g$ at the privileged base-point now takes all of the (group-valued) plaquette labels $e_p$ to $g \rhd e_p$. On the other hand, the vertex transforms at the other vertices do not change the surface labels, so they act exactly like the vertex transforms in Kitaev's Quantum Double model \cite{Kitaev2003}. The other energy terms to consider are the plaquette terms. These energy terms are unaffected by changing the base-point. As shown in Figure \ref{2Dtransforms}, the plaquette term in the example model checks that the boundary of the plaquette is labelled by the identity element of $G$. If we move the base-point of the plaquette along an edge labelled by $g$, then the label of the boundary of the plaquette is conjugated by $g$. However this has no effect because the identity is preserved by conjugation (and besides the group $G$ is Abelian in this case).

		We can consider the ground-state degeneracy in the same way that we did before we changed the base-points of the plaquettes, provided that the manifold is simply connected. That is, the plaquette terms again determine the allowed edge configurations and the edge terms then determine the allowed plaquette configurations for each edge configuration. Then we fluctuate the configurations by applying the vertex terms. However when all of the plaquettes have the same base-point it is easier to interpret the different ground states. This is because the edge terms now enforce that all of the plaquettes have the same irrep label in the ground state. In addition, only the vertex transform at the privileged vertex fluctuates these irreps and it simply flips between the labels $\alpha_R$ and $\alpha^2_R$ for every plaquette. This means that there are always two choices for the plaquettes in the ground state (again assuming that there are no non-contractible paths). Either every plaquette is labelled by $1_R$, or the plaquettes are in a linear combination of a product state where all plaquettes are in the $\mu=\alpha_R$ state and a product state where all plaquettes are in the $\mu=\alpha^2_R$ state.
		
		\subsubsection{Topological phase}
		
		Having considered the properties of the $\mathbb{Z}_2$, $\mathbb{Z}_3$ model, we can try to identify its topological phase. We found that the ground states of this model on the sphere can be split into two sectors, which can be distinguished locally. Namely, one sector has all plaquettes labelled by the trivial irrep of $\mathbb{Z}_3$, while the other involves a combination of the two non-trivial irreps. In the former sector the ground state is equivalent to that of the toric code, and the excitations also have the same properties as the toric code. We therefore identify the topological phase in this sector as that of the toric code. For the other sector, we found that the magnetic excitation is confined, while the electric excitation is condensed. This implies that the topological phase for this sector is trivial.

		From the ground state labelled by the trivial irrep, we can obtain the other ground state by applying a membrane operator over the entire lattice. However we do not regard this operator as a symmetry because it only commutes with the Hamiltonian in the ground state space (which is why the properties of the excitations are different in the two sectors). It therefore appears that the model contains two separate, but related, phases that happen to have degenerate ground states that are not protected by a true symmetry. We expect that this is a general feature of the models with $\rhd$ non-trivial, although the inability to construct the magnetic excitations for general models makes it difficult to classify these phases.

		\subsection{$\mathbb{Z}_4,\mathbb{Z}_4$ model}
		\label{Section_Z2_Z4}

		The next example is the model based on the crossed module $(G=\mathbb{Z}_4, \ E=\mathbb{Z}_4, \ \partial \rightarrow \mathbb{Z}_2, \ \rhd \text{ trivial})$. This example is intended to highlight the condensation/confinement aspect of the model. We denote the elements of $\mathbb{Z}_4$ by $1$, $i$, $-1$ and $-i$. Then we define the map $\partial$ as follows: $\partial(1)=\partial(-1)=1$ and $\partial(i)=\partial(-i)=-1$. This map preserves the group product, as we can see from the following relations:
		\begin{align*}
		\partial(\pm 1 \cdot \pm 1)&=1=1 \cdot 1=\partial( \pm 1) \cdot \partial(\pm 1)\\
		\partial(\pm 1 \cdot \pm i)&=-1= 1 \cdot -1 = \partial(\pm 1) \cdot \partial(\pm i)\\
		\partial(\pm i \cdot \pm i)&=1= -1 \cdot -1 = \partial(\pm i) \cdot \partial(\pm i).
		\end{align*}
		
		Therefore, $\partial$ is indeed a group homomorphism. We can see that $\partial$ maps $\mathbb{Z}_4$ onto the $\mathbb{Z}_2$ subgroup of $\mathbb{Z}_4$: ($1, -1$). Having chosen $\partial$ in this way, we now choose $\rhd$ to be trivial. The crossed module then satisfies the Peiffer conditions, as we now verify directly. The first Peiffer condition, Equation \ref{Peiffer_1} in Section \ref{Section_Recap_Paper_2}, requires that $\partial(g \rhd e) =g \partial(e)g^{-1}$ for all $g \in G$ and $e \in E$. This is true because $\rhd$ is trivial and $G$ is Abelian: 
		\begin{align*}
		\partial(e)\overset{\rhd \text{ trivial}}{=}&\partial(g \rhd e), \: \partial(e) \overset{G\text{ Abelian}}{=}g \partial(e) g^{-1}\\
		& \implies \partial(g \rhd e) = g \partial(e) g^{-1}.
		\end{align*}
		
		The second Peiffer condition, Equation \ref{Peiffer_2} in Section \ref{Section_Recap_Paper_2}, states that $\partial(e) \rhd f = efe^{-1}$ for all $e, \: f \in E$. This is true for this example model because $\rhd$ is trivial and $E$ is Abelian:
		\begin{align*}
		f\overset{\rhd \text{ trivial}}{=}& \partial(e) \rhd f, \: f \overset{E \text{ Abelian}}{=} efe^{-1}\\
		& \implies \partial(e) \rhd f =efe^{-1}.
		\end{align*}
		
		Because the maps $\partial$ and $\rhd$ satisfy the homomorphism conditions and the Peiffer conditions, the collection of objects $(\mathbb{Z}_4,\mathbb{Z}_4, \partial, \rhd)$ is a valid crossed module and so this crossed module defines a higher lattice gauge theory model. We note that there is a related model with the same groups $G$ and $E$ and where $\rhd$ is again trivial, but where $\partial:E \rightarrow {1_G}$, as discussed in Section \ref{Section_2D_Condensation_Confinement}. In this case, because $\rhd$ and $\partial$ are both trivial, the edge and plaquette labels completely decouple in the Hamiltonian. Furthermore, the excitations in that model are neither condensed nor confined. We will therefore refer to that model as the unconfined model and the model with $\partial \rightarrow \mathbb{Z}_2$ as the confined model in this section. We can think of the confinement and condensation in the confined model as arising from a condensation-confinement transition between the unconfined and confined models.

\begin{figure}[h]
			\begin{center}
				\includegraphics[width=0.9\linewidth]{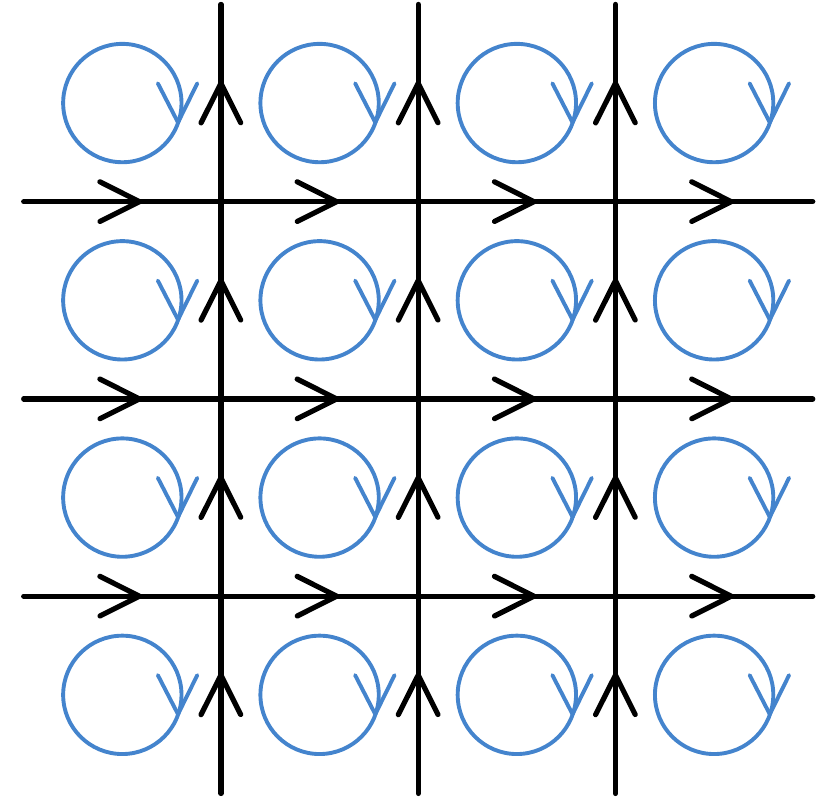}

				\caption{A square lattice, with oriented edges and plaquettes}
				\label{2Dlattice1}
			\end{center}
		\end{figure}

		Returning to the confined model where $\partial$ maps onto the $\mathbb{Z}_2$ subgroup, we now need to choose a lattice. We take a square lattice, as we did in the previous example. Because $\rhd$ is trivial, we don't need to specify the base-point of each plaquette, so we simply draw the plaquettes as oriented loops with no other feature. A fragment of this lattice is shown in Figure \ref{2Dlattice1}. There are two kinds of edges in this lattice, vertical edges and horizontal edges. Each vertical edge has the same environment (i.e., the same objects surrounding it) and each horizontal edge has the same environment. Similarly, there is only one kind of plaquette, because we have chosen each plaquette to have the same orientation, and one kind of vertex. Because of this, in later figures we may not draw the circulation of the plaquette at all and simply put a label in the middle of the plaquette. For now, we will not worry about the boundary of this lattice and simply assume that there is a sensible way to close it into a sphere or other desired manifold, although this may necessitate introducing vertices or plaquettes with different environments.

\begin{figure}[h]
			\begin{center}
				\includegraphics{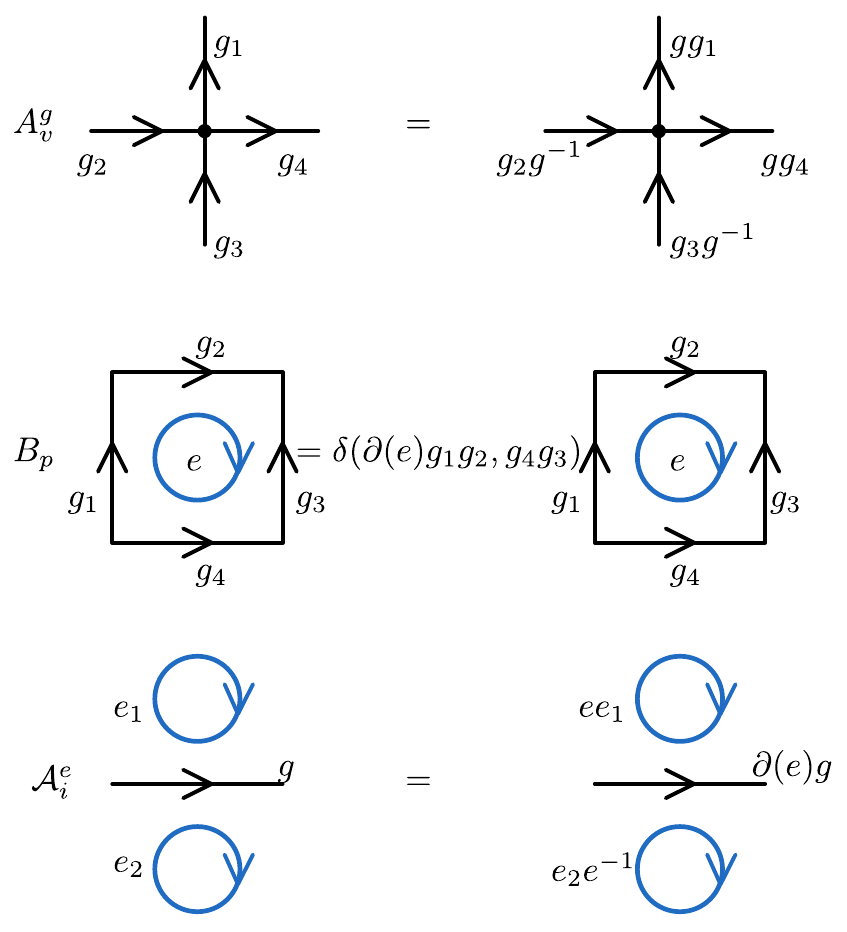}
				\caption{The gauge transforms and plaquette terms in this example model. We can obtain the action of the vertical edge transform by rotating the bottom diagrams by ninety degrees anticlockwise.}
				\label{2D_energy_terms_1}
			\end{center}
		\end{figure}

		With the lattice fixed, we next consider the operators that make up the Hamiltonian: vertex transforms, edge transforms and plaquette terms. Because of the simple lattice and crossed module, the action of these operators can be expressed relatively concisely, as shown in Figure \ref{2D_energy_terms_1}. 
		
	\subsubsection{Excitations}
		
		We can now consider the excitations of this example model, starting with the purely electric ones. To examine the excitations, we construct the ribbon operators that produce them. As we discussed in Section \ref{Section_2D_electric}, these can be labelled by irreps of the group $G$. Because $\mathbb{Z}_4$ is Abelian, the irreps (shown in Table \ref{irreps_Z_4}) are 1D, which means that we do not need to give matrix indices in order to specify an electric ribbon operator. Then the basis operators for the space of electric ribbon operators are given by
		$$\hat{S}^{R}(t) = \sum_{g \in G} R(g) \delta(g, \hat{g}(t)),$$
		where $R$ is an (1D) irrep of $G$, so that $R(g)$ is the representation of the element $g$, and $t$ is the path on which we apply the operator. We discussed this construction for more general models in Section \ref{Section_2D_electric}.

		\begin{table}[h]	
			\begin{center}
				\begin{tabular}{|c|c |c| c| c|} 
					\hline
					& $\bm{1}$ & $\bm{i}$ & $\bm{-1}$ &$\bm{-i}$\\ 
					\hline
					$\bm{1_{\text{Rep}}}$ & $1$ & $1$ & $1$ & $1$ \\ 
					\hline
					$\bm{-1_{\text{Rep}}}$ & $1$ & $-1$ & $1$ & $-1$ \\ 
					\hline
					$\bm{i_{\text{Rep}}}$ & $1$ & $i$ & $-1$ & $-i$ \\ 
					\hline
					$\bm{-i_{\text{Rep}}}$ & $1$ & $-i$ & $-1$ & $i$ \\ 
					\hline
					
				\end{tabular}
				\caption{The elements of the unitary irreps of $\mathbb{Z}_4$}
				\label{irreps_Z_4}
			\end{center}
		\end{table}

		These ribbon operators commute with the plaquette terms, because both types of operator are diagonal in the basis labelled by group elements on each edge. The ribbon operators also have the same commutation relations with the vertex operators as the equivalent operators in Kitaev's Quantum Double model \cite{Kitaev2003}. Namely, the ribbon operators commute with all vertex operators except those at the two end points of $t$, which are excited by the ribbon operator (except for the ribbon operator labelled by the trivial irrep, which is just the identity operator). To see this, we first note that, as discussed in Section \ref{Section_2D_electric}, the vertex transform only affects the labels of paths that start or end at that vertex, not those that just pass through it. This is because under the action of a gauge transform $A_v^g$ the path obtains a factor $g^{-1}$ from entering the vertex and a factor $g$ from leaving the vertex, which cancel. Therefore, the ribbon operator will commute with all vertex transforms except those at the start and end of the path, just as we discussed for a more general case in Section \ref{Section_2D_electric}. Then for the starting vertex operator, we have the commutation relation
		$$\hat{g}(t) A_{s.p}^g = A_{s.p}^g \: g \hat{g}(t),$$ 
		because the path picks up an additional $g$ if the path is acted on by $A_{s.p}^g$ before we measure it. Then if we wish to reverse this commutation relation, we have 
		$$ A_{s.p}^g \hat{g}(t) = g^{-1} \hat{g}(t) A_{s.p}^g.$$
		
		Then we can check whether the ribbon operator excites the vertex ${s.p}$ when acting on the ground state by applying the energy term $A_{s.p}$ to the left of the ribbon operator. We see that
		\begin{align*}
		A_{s.p} \hat{S}^{R}(t) &\ket{GS} =A_{s.p} \sum_{g \in G} R(g) \delta(g, \hat{g}(t)) \ket{GS}\\
		&= \frac{1}{|G|} \sum_{x \in G} A_{s.p}^x \sum_{g \in G} R(g) \delta(g, \hat{g}(t)) \ket{GS}\\
		&= \frac{1}{|G|} \sum_{x \in G} \sum_{g \in G} R(g) \delta(g, x^{-1}\hat{g}(t)) A_{s.p}^x \ket{GS}\\
		&= \frac{1}{|G|} \sum_{x \in G} \sum_{g \in G} R(g) \delta(xg,\hat{g}(t)) A_{s.p}^x \ket{GS}.
		\end{align*}
		
		Then we can use the fact that $A_{s.p}^x \ket{GS} =\ket{GS}$ as discussed in Section \ref{Section_Recap_Paper_2}, to write this as
		\begin{align*}
		A_{s.p} \hat{S}^{R}(t) &\ket{GS} =\frac{1}{|G|} \sum_{x \in G} \sum_{g \in G} R(g) \delta(xg,\hat{g}(t))\ket{GS}\\
		&= \frac{1}{|G|} \sum_{x \in G} \sum_{g'=xg} R(x^{-1}g') \delta(g',\hat{g}(t))\ket{GS}\\
		&= \big(\frac{1}{|G|} \sum_{x \in G} R(x^{-1}) \big) \sum_{g' \in G} R(g') \delta(g',\hat{g}(t))\ket{GS}\\
		&=\big(\frac{1}{|G|} \sum_{x \in G} R(x^{-1}) \big) \hat{S}^{R}(t) \ket{GS},
		\end{align*}
		where we used the fact that $R$ is an irrep to split the contributions from $x$ and $g'$. If $R$ is a non-trivial irrep, this is zero by orthogonality of characters, and if $R$ is the trivial irrep it is just $\hat{S}^{R}(t) \ket{GS}$. If $A_{s.p} \hat{S}^{R}(t) \ket{GS}=0$, the vertex is excited, so a non-trivial irrep gives an excited vertex at the start of the path. A similar calculation holds for the vertex transform at the end of the path and so that vertex is also excited if the irrep is non-trivial.

		Next we consider the commutation relations of the ribbon operator with the edge transforms. The ribbon operator commutes with edge transforms for edges that are not on the path, but those transforms for edges on the path change the path label $\hat{g}(t)$. This is because, in addition to changing the plaquette labels, the edge transform $\mathcal{A}_i^e$ multiplies the label of the edge it acts on by $\partial (e)$. Then we have that 
		$$ \hat{g}(t) \mathcal{A}_i^e = \mathcal{A}_i^e \partial(e)^{(\pm 1)} \hat{g}(t) ,$$
		where the plus or minus depends on the direction of the edge relative to the direction of $t$ ($+$ if they are aligned, $-$ if they are anti-aligned). This means that 
		$$ \mathcal{A}_i^e \hat{g}(t) = \partial(e)^{(\mp 1)} \hat{g}(t) \mathcal{A}_i^e.$$
		
		Therefore, if we act with the edge energy term $\mathcal{A}_i$ to the left of the ribbon operator, we have
		\begin{align*}
		\mathcal{A}_i \hat{S}^{R}(t)&\ket{GS}\\
		& = \frac{1}{|E|} \sum_{e \in E} \mathcal{A}_i^e \sum_{g \in G} R(g) \delta(g, \hat{g}(t))\ket{GS}\\
		&= \frac{1}{|E|} \sum_{e \in E} \sum_{g \in G} R(g) \delta(g, \partial(e)^{(\mp 1)} \hat{g}(t)) A_i^e \ket{GS}.
		\end{align*}
		
		As discussed in Section \ref{Section_Recap_Paper_2}, the ground states satisfy $\mathcal{A}_i^e \ket{GS} = \ket{GS}$, and so we can write this commutation relation as
		\begin{align*}
		\mathcal{A}_i &\hat{S}^{R}(t)\ket{GS} =\frac{1}{|E|} \sum_{e \in E} \sum_{g \in G} R(g) \delta(g, \partial(e)^{(\mp 1)} \hat{g}(t)) \ket{GS}\\
		& = \frac{1}{|E|} \sum_{e \in E} \sum_{g \in G} R(g) \delta( \partial(e)^{(\pm 1)}g, \hat{g}(t)) \ket{GS}\\
		&=\frac{1}{|E|} \sum_{e \in E} \sum_{g' = \partial(e)^{(\pm 1)}g} R(\partial(e)^{(\mp 1)} g') \delta(g', \hat{g}(t)) \ket{GS}\\
		&= \big (\frac{1}{|E|} \sum_{e \in E} R(\partial(e)^{(\mp 1)}) \big) \sum_{g' \in G} R( g') \delta(g', \hat{g}(t)) \ket{GS}\\
		&= \big (\frac{1}{|E|} \sum_{e \in E} R(\partial(e)^{(\mp 1)}) \big) \hat{S}^{R}(t)\ket{GS},
		\end{align*}
		where we used the fact that $R$ is an irrep of $G$ to separate the contribution from $g'$ and $\partial(e)$. From this commutation relation, we see that whether the edge energy term is excited or not (i.e., whether the above expression is zero or not) depends on the expression $\big (\frac{1}{|E|} \sum_{e \in E} R(\partial(e)^{(\mp 1)}) \big)$. This expression is zero if the irrep $R$ is non-trivial in the subgroup $\partial(E) \subset G$. To see this, let us consider the irreps of $G$ in this model. So far, our discussion of the electric ribbon operator has been valid for any crossed module with trivial $\rhd$ and Abelian $G$. However we now consider the specific groups $G=E=\mathbb{Z}_4$ and the map $\partial \rightarrow \mathbb{Z}_2$. For $\sum_{e \in E} R(\partial(e))$, we have 
		\begin{align*}
		\sum_{e \in E} R&(\partial(e))\\
		&= R( \partial(1)) + R(\partial(-1))+R(\partial(i))+R(\partial(-i))\\
		&= R(1) + R(1)+ R(-1) + R(-1)\\
		&= 2 R(1) + 2 R(-1).
		\end{align*}
		
		If $R = \pm i_{\text{Rep}}$ (i.e., if $R$ is one of the irreps sensitive to factors in $\partial(E)$), then $R(1)=1$ and $R(-1)=-1$, implying that
		$$\sum_{e \in E} R(\partial(e))= 2 \cdot 1 + 2 \cdot -1 =0.$$ 
		This means that, in the case where $R = \pm i_{\text{Rep}}$, we have $$\mathcal{A}_i \hat{S}^{R}(t) \ket{GS}=0.$$ This indicates that $\hat{S}^{R}(t) \ket{GS}$ is an eigenstate of $\mathcal{A}_i$ where the edge $i$ is excited for all edges $i$ on the path $t$. The string operator therefore costs one unit of energy per edge and so the corresponding excitations are confined.

		On the other hand, the other two irreps satisfy $R(\partial(e))=1$. This means that for $R= \pm 1_{\text{Rep}}$, 
		$$\sum_{e \in E} R(\partial(e))= 2 \cdot 1 + 2 \cdot 1 =4,$$
		 so that 
		 $$\frac{1}{|E|}\sum_{e \in E} R(\partial(e))=1, $$ 
		 and so
		 $$\mathcal{A}_i \hat{S}^{R}(t)\ket{GS} =\hat{S}^{R}(t)\ket{GS}.$$
		Therefore, the electric ribbon operators labelled by these two irreps commute with the edge terms. This means that the edges are not excited by the ribbon operators, so the corresponding excitations are not confined. We can see that the ribbon operators labelled by irreps with trivial restriction to the image of $\partial$ (the irreps $\pm 1_{\text{Rep}}$ ) are not confined and those with non-trivial restriction ($\pm i_{{\text{Rep}}}$) are confined, similar to the general claim made in Section \ref{Section_2D_electric}. Note that in the unconfined model, where $\partial(E)=\set{1_G}$, every irrep of the subgroup $\partial(E) \subset G$ is trivial, so that we have no confined excitations.

\begin{figure}[h]
			\begin{center}
			\includegraphics{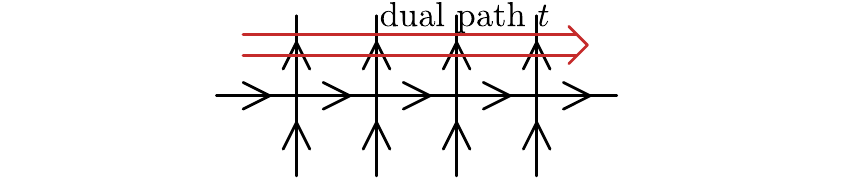}
				\caption{An example of a dual path on our lattice. A magnetic ribbon operator acting on this dual path will produce excitations at the plaquettes at the two ends of the path.}
				\label{dual_path_example}
			\end{center}
		\end{figure}

\begin{figure}[h]
			\begin{center}

		\includegraphics{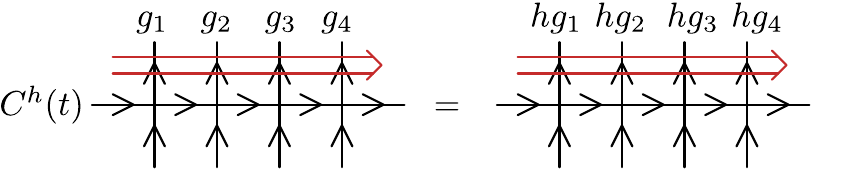}
				\caption{An example of the action of the magnetic operator}
				\label{magnetic_example_1}
			\end{center}
		\end{figure}

		Next we consider the magnetic excitations (which we can find because $\rhd$ is trivial). In order to construct a ribbon operator to produce these excitations, we first choose a path on the dual lattice, such as the one shown in Figure \ref{dual_path_example}. The magnetic ribbon operator $C^h(t)$ multiplies the edges cut by the dual path by either $h$ or $h^{-1}$ depending on the orientation of the edges. In the example shown in the figure, all of the cut edges point in the same direction, and so we multiply the label of these edges by the same element $h$, as shown in Figure \ref{magnetic_example_1}. Note that this particular form of the magnetic ribbon operator depends on the Abelian nature of $G=\mathbb{Z}_4$ and a more general case is given in Section \ref{Section_2D_Magnetic} (because $G$ is Abelian we do not need to define a direct path for the ribbon, although we still need a convention for whether edges are multiplied by $h$ or $h^{-1}$). Then we consider how this affects the plaquette terms near the ribbon. The internal plaquettes, which are those passed through by the dual path (i.e., those on the ribbon but not at the ends of the dual path) have two edges cut by the dual path. Each of these edges will gain a factor of $h$. However one of these edges is aligned with the boundary of the plaquette, while the other is anti-aligned with the boundary. This means that the boundary of the plaquettes gain one factor of $h$ and one factor of $h^{-1}$, which cancel. Therefore, the magnetic operator commutes with those internal plaquette operators. However the end plaquettes gain only a factor of $h$ or $h^{-1}$, so they are excited just as in Kitaev's Quantum Double model \cite{Kitaev2003}.
		
\begin{figure}[h]
			\begin{center}
				\includegraphics{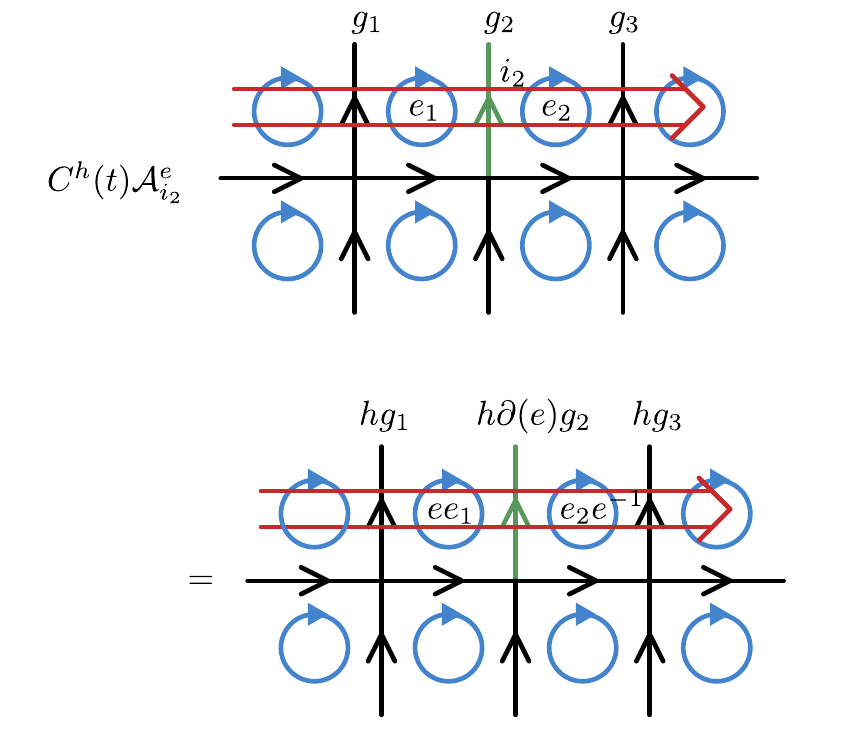}
				\caption{An example of applying an edge transform followed by a magnetic ribbon operator. We apply the edge transform on the edge $i_2$ (green), initially labelled by the element $g_2$. The final label of $i_2$ is $h \partial(e)g_2$. If we applied the operators in the opposite order, the label would be $\partial(e)hg_2$, which is the same for this model.}
				\label{2D_magnetic_edge_commute}
			\end{center}
		\end{figure}

		Taking the group $G$ to be Abelian means that the magnetic ribbon operator commutes with all of the vertex transforms (again, just like in Kitaev's Quantum Double model \cite{Kitaev2003}). This is because both the vertex and ribbon operators multiply a set of edges by fixed group elements (there is no configuration dependence, unlike in the non-Abelian case), and the Abelian nature of the group $G$ means that the order in which this multiplication is done does not matter. The ribbon operator also commutes with the edge transforms for the same reason. We give an example of this commutation in Figure \ref{2D_magnetic_edge_commute}, where we first apply an edge transform $\mathcal{A}_i^e$ on the edge labelled by $g_2$, then apply a magnetic ribbon operator $C^h(t)$. Under these operations, the edge label $g_2$ becomes $ h \partial(e)g_2$, whereas applying the operators in the opposite order would give $\partial(e)h g_2$. However $h \partial(e)g_2= \partial(e)hg_2$ because $G$ is Abelian, so the operators commute.

		Having discussed all of the energy terms, we see that the magnetic ribbons only excite the two end plaquettes and commute with all of the other energy terms. In Section \ref{Section_2D_Magnetic}, we saw that for a general crossed module model the start-point vertex of the ribbon could be excited by the ribbon operator, but for this particular model the vertex is not excited because $G$ is Abelian. Either way, just as in the more general $\rhd$ trivial case, the magnetic excitations are not confined because excitations are only produced at the ends of the ribbon rather than along the length of the ribbon.

\begin{figure}[h]
			\begin{center}
				\includegraphics{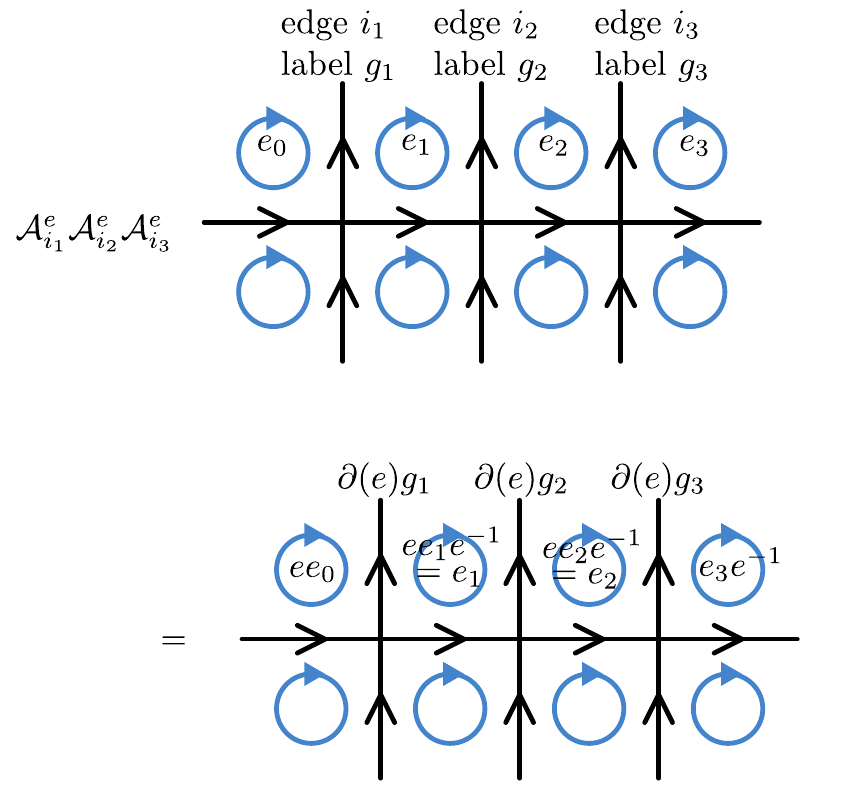}
				\caption{Performing a series of edge transforms on the edges cut by the dual ribbon has a similar effect to the magnetic ribbon operator, changing the edge labels and leaving the internal plaquette labels (here $e_1$ and $e_2$) invariant. However, the two plaquettes at the ends of the ribbon (here initially labelled by $e_0$ and $e_3$) are affected by the edge transforms, whereas they are not affected by the magnetic ribbon operator. This difference between the action of the edge transforms and the action of the magnetic ribbon operator is equivalent to the action of local operators at the two ends of the ribbon.}
				\label{series_edge_transforms}
			\end{center}
		\end{figure}

		While none of the magnetic excitations are confined, some of them are ``condensed". As we explained in Section \ref{Section_Single_Plaquette_1}, by condensed we mean that some of the excitations can be produced by local operators, acting only near the excitations themselves. This means that they cannot carry a non-trivial topological charge. To see which excitations are condensed, consider a ground state. Any ground state is an eigenstate of the edge terms with eigenvalue one and is unchanged by applying any edge transform $\mathcal{A}_i^e$. Therefore, we are free to apply these transforms without changing the state. The action of some of the magnetic ribbon operators is equivalent to a series of edge transforms (which act trivially on the ground state) and some local operators. To see this, consider applying an edge transform $\mathcal{A}_i^e$ on each edge cut by the dual path of the magnetic ribbon operator, as shown in Figure \ref{series_edge_transforms}. 	Apart from the action on the two end plaquettes, the action shown in Figure \ref{series_edge_transforms} looks exactly like the action of a magnetic ribbon with label $\partial(e)$. In fact, we can turn this action into the action of the magnetic ribbon simply by applying \textit{local} operators on the two end plaquettes to correct their labels. That is,
		\begin{equation}
			C^{\partial(e)}(t)=M^{e^{-1}}(p_0)M^{e}(p_3) \mathcal{A}_{i_1}^e \mathcal{A}_{i_2}^e \mathcal{A}_{i_3}^e, \label{Equation_Z4_Z4_condensed_1}
		\end{equation}
		where $M^e(p)$ multiplies the single plaquette $p$ by $e$ and is therefore local to the plaquette, and $p_x$ refers to the plaquette initially labelled by $e_x$ in Figure \ref{series_edge_transforms}. The operators $M^{e^{-1}}(p_0)$ and $M^e(p_3)$ are therefore local to the excitations produced at the two ends of the ribbon.

		Then acting with both sides of Equation \ref{Equation_Z4_Z4_condensed_1} on a ground-state gives us the equality
		\begin{align*}
		C^{\partial(e)}(t)\ket{GS}&=M^{e^{-1}}(p_0)M^{e}(p_3) \mathcal{A}_{i_1}^e \mathcal{A}_{i_2}^e \mathcal{A}_{i_3}^e \ket{GS}\\
		&= M^{e^{-1}}(p_0)M^{e}(p_3) \ket{GS},
		\end{align*}
		where the second equality results from absorbing the $\mathcal{A}_i^e$ into the ground state as explained earlier. This equality means that the excitations produced by the magnetic operator can be reproduced by local operators acting on the ground state, so those excitations are not topological. That is, they are condensed. We see that the magnetic excitations produced by all magnetic operators with label in the image of $\partial$ are condensed. Furthermore, the fusion rule $C^g(t) C^h(t)=C^{gh}(t)$ means that $C^{g \partial(e)}(t) \ket{GS}=C^g(t) M^{e^{-1}}(p_1)M^{e}(p_4) \ket{GS}$, so that the $g$ and $g \partial(e)$ excitations differ only by the application of local operators, and so must be in the same topological sector. Therefore, the magnetic topological sectors are not labelled by group elements, but instead by cosets $g\partial(E)$. A similar argument holds for any crossed module model where $\rhd$ is trivial (although if $G$ is non-Abelian the sectors are also widened by conjugation, as described in Section \ref{Section_2D_Magnetic}). For the particular crossed module considered in this section, we see that the elements $1$ and $-1$ of $G$ are in the image of $\partial$. Therefore, the magnetic excitation labelled by $-1$ is condensed ($1$ corresponds to no excitation, so we do not say that it is condensed) and the excitations labelled by $i$ and $-i$ are in the same sector, but are not condensed.

\begin{figure}[h]
			\begin{center}
			\includegraphics{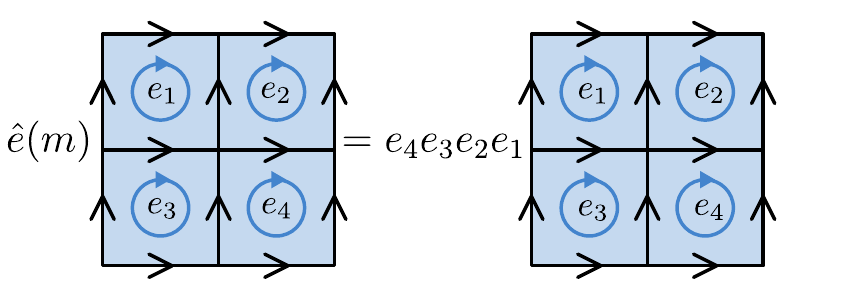}
				\caption{The surface measurement operator $\hat{e}(m)$ is simplified when $\rhd$ is trivial and the plaquettes all have the same orientation. For a surface $m$ (shaded) formed from the combination of four plaquettes, the surface label is the product of the labels of each of these plaquettes, and the order of the product is irrelevant. }
				\label{combined_surface_example_1}
			\end{center}
		\end{figure}

		Having considered the point-like excitations of this example model, we next examine the loop-like excitations. Thanks to $\rhd$ being trivial and all of the plaquettes in the lattice having the same orientation, the membrane operators that produce these loop excitations are relatively simple. As discussed in Section \ref{Section_2D_Loop}, the membrane operators measure the total surface label of the membrane and assign a weight to each possible value. Because $\rhd$ is trivial, when combining the plaquette labels into a total surface label we do not need to worry about defining base-points and moving them around to ensure that these match. Combined with the fact that we have chosen all plaquettes to have the same orientation, this means that the total surface element of a surface made from multiple plaquettes becomes a simple product of the operators for the individual plaquettes. The expression $\delta(\hat{e}(m),e)$ from the membrane operator in the general case then becomes $\delta(\prod_{p \in m} \hat{e}_p,e)$, where $\hat{e}_p$ is the operator that measures the label of plaquette $p$. Furthermore, because the group $E$ is Abelian we do not need to worry about the order of multiplication. As an example, consider a membrane made of four square plaquettes, shown in Figure \ref{combined_surface_example_1}. The total surface label of this membrane is $e_4 e_3 e_2 e_1$, where $e_p$ is the plaquette label of plaquette $p$ in the membrane and $p$ runs from 1 to 4.

\begin{figure}[h]
			\begin{center}
				\includegraphics{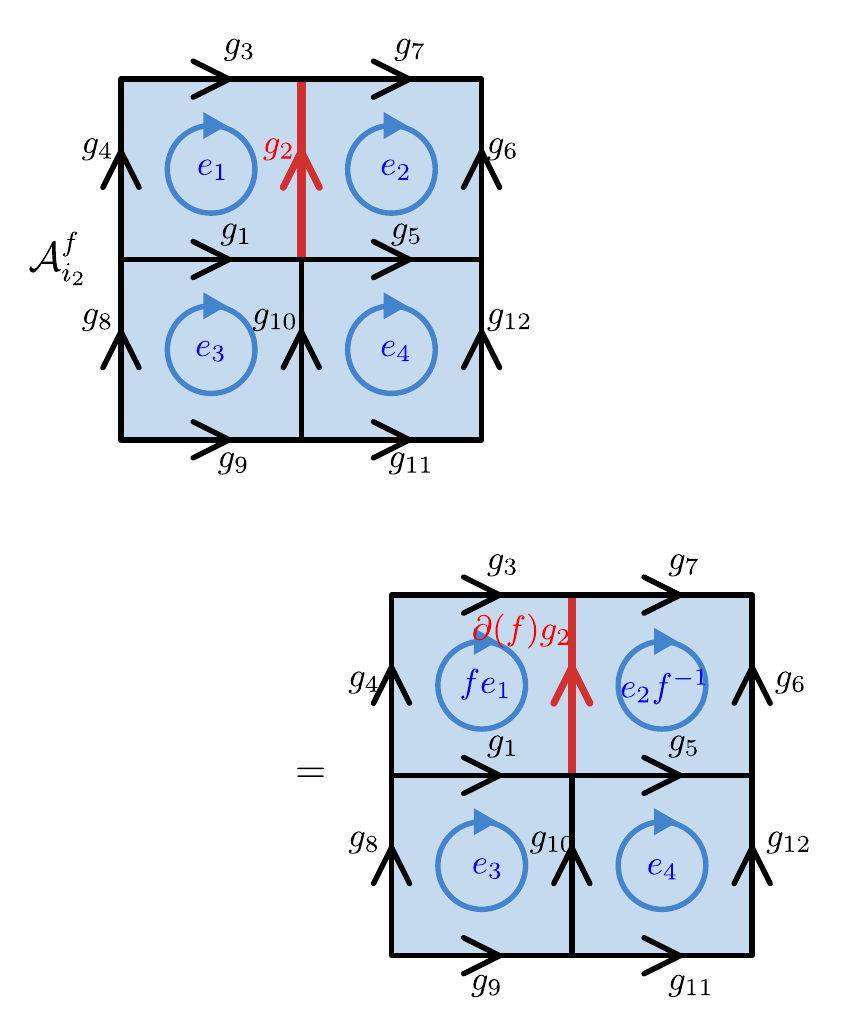}
				\caption{An example of the action of an ``internal" edge operator (applied on the thicker red edge with initial label $g_2$) acting on our combined surface. Note that one plaquette label gains a factor of $f$ and another gains a factor of $f^{-1}$. These factors will cancel when we consider the total label of the surface.}
				\label{edge_operator_example_1}
			\end{center}
		\end{figure}

		Now consider the commutation relations between the membrane operator and the energy terms. The operator $\delta(e, \hat{e}(m))$ clearly commutes with every vertex transform $A_v^g$, because the vertex transform does not affect any plaquette labels when $\rhd$ is trivial. Similarly, the operator $\delta(e, \hat{e}(m))$ commutes with the plaquettes terms $B_p$, because both $B_p$ and the surface measurement operator are diagonal in the element basis. The membrane operator also commutes with ``internal" edge transforms. An internal edge is an edge where both plaquettes adjacent to that edge are within the membrane. For example, consider Figure \ref{edge_operator_example_1}. In the figure, the red vertical edge (initially labelled by $g_2$) between the plaquettes labelled by $e_1$ and $e_2$ is an example of an internal edge, because both adjacent plaquettes are within the surface $m$. In the case shown in the figure, we see that under the edge transform $\mathcal{A}_{i_2}^f$ the surface label $e(m)$ transforms as 
		$$e(m)=e_4e_3e_2e_1 \rightarrow e_4 e_3 e_2 f^{-1} f e_1 =e_4e_3e_2e_1.$$
		
		For any surface, the edge transform $\mathcal{A}_i^f$ on an edge wholly within that surface will generate a factor of $f$ (from one of the plaquettes adjacent to that edge) and a factor of $f^{-1}$ (from the other plaquette), which will cancel, leaving the surface label unchanged.

		On the other hand, consider transforms on edges that are on the boundary of $m$ (such as the one labelled by $g_7$ in Figure \ref{edge_operator_example_1}). Of the two plaquettes adjacent to such an edge, only one plaquette is in $m$, so the surface label $e(m)$ gains a factor of $f$ or $f^{-1}$ from that plaquette (depending on the orientation of the edge) from an edge transform labelled by $f$, without the compensating factor from the other plaquette. This means that the edge transform does not generally commute with the membrane operator, and so the membrane operator may create edge excitations. When we construct basis membrane operators labelled by irreps of $E$, of the form
		$$L^{\mu}(m) = \sum_{e \in E} \mu(e) \delta(\hat{e}(m),e)$$
		for an irrep $\mu$, the membrane operator then excites the boundary edges if the irrep $\mu$ is non-trivial. If the irrep $\mu$ is trivial, then the membrane operator is just the identity operator.

		In Section \ref{Section_2D_Loop}, we explained that some of the loop excitations are condensed in the general case. We now wish to consider how this arises in our example model in more detail. Consider acting with a surface measurement operator $\hat{e}(m)$ on the (or a) ground state. The ground state satisfies fake-flatness, which means that we can relate the plaquette labels to the labels of the path around the the boundary of the plaquettes. This holds not only for individual plaquettes, but also for surfaces made up of multiple plaquettes. For example, considering Figure \ref{edge_operator_example_1}, from fake-flatness each plaquette label must satisfy $\partial(e_p) = g_{dp}^{-1}$, where $g_{dp}$ is the label of the boundary of the plaquette. This means that
		\begin{align*}
		\partial(e_1)&=g_2g_3^{-1}g_4^{-1}g_1\\
		\partial(e_2)&=g_5g_6g_7^{-1}g_2^{-1}\\
		\implies& \partial(e_2e_1)=g_5g_6g_7^{-1}g_2^{-1} g_2g_3^{-1}g_4^{-1}g_1 \\
		&\quad \quad \quad \: = g_5g_6g_7^{-1}g_3^{-1}g_4^{-1}g_1.
		\end{align*}
		
\begin{figure}[h]
			\begin{center}
			\includegraphics{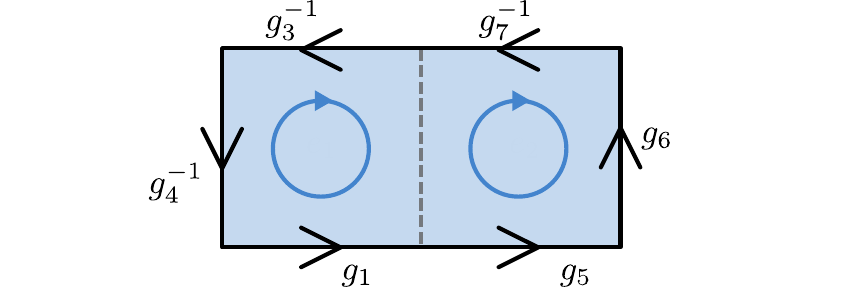}
				\caption{The combination of the boundary of two plaquettes gives the boundary of the combined surface}
				\label{combined_surface_boundary_1}
			\end{center}
		\end{figure}

		We can recognise the right-hand side of the final line as the label of the boundary of the combined surface of the two plaquettes (in reverse), as shown in Figure \ref{combined_surface_boundary_1}.

\begin{figure}[h]
			\begin{center}
			
				\includegraphics{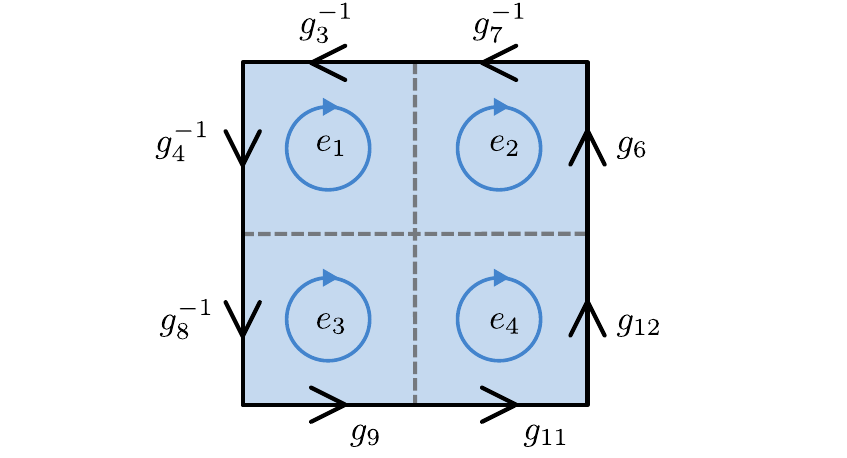}
				\caption{The boundary of four combined plaquettes}
				\label{combined_surface_boundary_2}
			\end{center}
		\end{figure}

		Then we wish to combine this with the other plaquettes from Figure \ref{edge_operator_example_1}. The labels of these plaquettes satisfy
		\begin{align*}
		\partial(e_3)&=g_9g_{10}g_1^{-1}g_8^{-1}\\
		\partial(e_4)&=g_{11}g_{12}g_5^{-1}g_{10}^{-1},
		\end{align*}
		so that
		\begin{align*}
		\partial(e_4e_3e_2e_1)&=g_{11}g_{12}g_5^{-1}g_{10}^{-1} g_9g_{10}g_1^{-1}g_8^{-1}\\
		& \hspace{1cm} \times g_5g_6g_7^{-1}g_3^{-1}g_4^{-1}g_1 \\
		&=g_{11}g_{12}g_6g_7^{-1}g_3^{-1}g_4^{-1}g_8^{-1}g_9,
		\end{align*}
		where we used the fact that $G$ is Abelian to cancel group elements and their inverses (a similar result holds in the general case, but we have to make more careful use of the Peiffer conditions if $G$ is non-Abelian). This is the label of the combined boundary of all four plaquettes (in reverse), indicating that the relationship between the boundary and surface label holds. The boundary for this combined surface is indicated in Figure \ref{combined_surface_boundary_2}. 
		
		Now with this information in hand, consider the four membrane operators labelled by the irreps of the group $E=\mathbb{Z}_4$. The membrane operator $L^{\alpha}(m)$ labelled by an irrep $\alpha$ and applied on a membrane $m$ is given by $L^{\alpha}(m)=\sum_{e \in E} \alpha(e) \delta(e, \hat{e}(m))$. Then we have
		\begin{align*}
		L^{1_{\text{Rep}}}(m)&= \delta(\hat{e}(m),1)+\delta(\hat{e}(m),-1)+\delta(\hat{e}(m),i)\\
		&\hspace{0.5cm}+\delta(\hat{e}(m),-i)\\
		&=1\\
		L^{-1_{\text{Rep}}}(m)&= \delta(\hat{e}(m),1)+\delta(\hat{e}(m),-1)-\delta(\hat{e}(m),i)\\
		&\hspace{0.5cm}-\delta(\hat{e}(m),-i)\\
		L^{i_{\text{Rep}}}(m)&=\delta(\hat{e}(m),1)-\delta(\hat{e}(m),-1)+i\delta(\hat{e}(m),i)\\
		&\hspace{0.5cm}-i\delta(\hat{e}(m),-i)\\
		L^{-i_{\text{Rep}}}(m)&= \delta(\hat{e}(m),1)+\delta(\hat{e}(m),-1)-i\delta(\hat{e}(m),i)\\
		&\hspace{0.5cm}+i\delta(\hat{e}(m),-i).\\
		\end{align*}
		
		We see that the membrane operator labelled by the trivial irrep is just the identity operator, because summing over all of the Kronecker delta just gives one. The membrane operator labelled by $-1_{\text{Rep}}$ can also be written in a simpler form by combining different Kronecker deltas. Noting that $\partial(1)=\partial(-1)=1$ and $\partial(i)= \partial(-i)=-1$, we can see that 
		$$\delta(\hat{e}(m),1)+ \delta(\hat{e}(m),-1)=\delta(\partial(\hat{e}(m)),1)$$
		and
		$$\delta(\hat{e}(m),i)+ \delta(\hat{e}(m),-i)=\delta(\partial(\hat{e}(m)),-1).$$
		Therefore
		$$L^{-1_{\text{Rep}}}(m)= \delta(\partial(\hat{e}(m)),1) -\delta(\partial(\hat{e}(m)),-1).$$
		
		That is, we can write this membrane operator entirely in terms of $\partial(\hat{e}(m))$, which is not sensitive to elements in the kernel of $\partial$. In the ground state, the surface label is related to the label of the boundary of that surface, as we have seen previously:
		$$\partial(\hat{e}(m)) \ket{GS} = \hat{g}_{dm}^{-1} \ket{GS},$$
			where we label the boundary element for membrane $m$ by $g_{dm}$. This means that, when acting on the ground state, the membrane operator labelled by $-1_{\text{Rep}}$ gives
		$$L^{-1_{\text{Rep}}}(m) \ket{GS} = (\delta(\hat{g}_{dm}^{-1},1)- \delta(\hat{g}_{dm}^{-1},-1)) \ket{GS}.$$
	 However the operator 
		$$(\delta(\hat{g}_{dm}^{-1},1)- \delta(\hat{g}_{dm}^{-1},-1))$$
		is simply an electric ribbon operator applied around the closed loop that forms the boundary of $m$ (specifically a confined electric operator, as we require from the fact that the boundary edges are excited). This indicates that we can replicate the action of the \textit{surface} operator by a \textit{path} operator applied on the boundary of the surface. This path is local to the excitation (which lies on the boundary of the surface), and so the excitation is condensed and cannot be a domain wall.

		On the other hand, consider the last two membrane operators, $L^{i_{\text{Rep}}}(m)$ and $L^{-i_{\text{Rep}}}(m)$. The action of these operators on the ground state cannot be written in terms of the path operator $\hat{g}_{dm}$, so these membrane operators are not condensed. We note that restricting the irreps that label these operators to the kernel of $\partial$, we have $\pm i_{\text{Rep}}(1)=1$ and $\pm i_{\text{Rep}}(-1)=-1$. We therefore see that the non-condensed excitations are labelled by irreps with non-trivial restriction to the kernel, as we expect from our more general claim in Section \ref{Section_2D_Condensation_Confinement}.

		 This pattern of condensation is reflected in the fusion rules of the loop-like excitations. These excitations satisfy a similar fusion rule to the magnetic excitations, as we showed more generally in Section \ref{Section_2D_Loop}. Given two membrane operators labelled by irreps $\alpha_1$ and $\alpha_2$ of $E$, the fusion rule is
		\begin{align*}
		L^{\alpha_1}(m)&L^{\alpha_2}(m)\\
		&=\sum_{e_1,e_2 \in E} \alpha_1(e_1) \delta(e_1, \hat{e}(m)) \alpha_2(e_2) \delta(e_2, \hat{e}(m))\\
		&=\sum_{e_1,e_2 \in E} \alpha_1(e_1) \alpha_2 (e_2) \delta(e_1,e_2) \delta(e_1, \hat{e}(m))\\
		&=\sum_{e_1 \in E} \alpha_1(e_1) \alpha_2(e_1) \delta(e_1, \hat{e}(m)),
		\end{align*}
		so that two membrane operators labelled by irreps $\alpha_1$ and $\alpha_2$ fuse to one operator labelled by the irrep $\alpha_1 \cdot \alpha_2$, defined by $\alpha_1 \cdot \alpha_2 (e) = \alpha_1(e)\alpha_2(e)$. From this rule, along with the definition of the irreps given in Table \ref{irreps_Z_4}, we can see that the condensed membrane operator (labelled by $-1_{\text{Rep}}$), together with the trivial operator (labelled by $1_{\text{Rep}}$), form a closed subset of the membrane operators under fusion, as we may expect. We can also see that the membrane operator labelled by $i_{\text{Rep}}$ is related to the membrane operator labelled by $-i_{\text{Rep}}$ by fusion with the condensed membrane operator $-1_{\text{Rep}}$, and so can be regarded as belonging to the same sector of excitations. We therefore have two sectors of loop-like excitation, with all of the condensed excitations belonging to the same sector and all of the non-condensed excitations belonging to another sector in this case.

		\subsubsection{Ground states in the irrep basis}
		\label{Section_Z4_Z4_irrep_basis}
		
\begin{figure}[h]
			\begin{center}
			\includegraphics{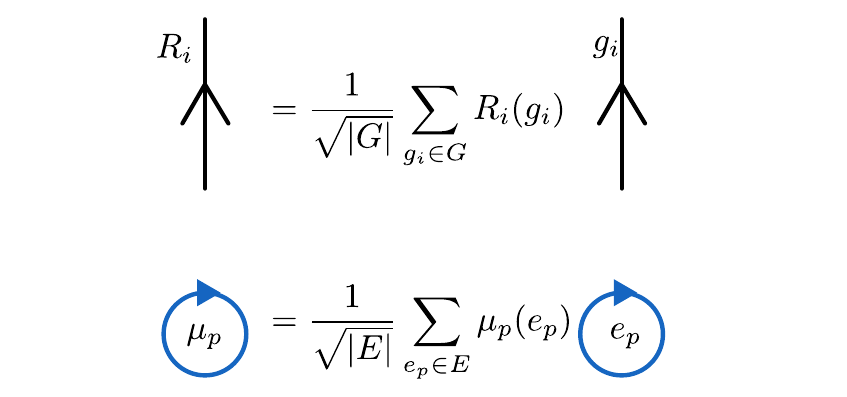}
				\caption{In order to examine the ground states, we change basis from states where the edges and plaquettes are labelled by elements of $G$ and $E$ to states where they are labelled by irreps of those same groups.}
				\label{changebasis1}
			\end{center}
		\end{figure}

		Having considered each of the excitations, we now consider the ground states and ground state degeneracy. To do this, it will be simplest to change basis from elements of the groups $G$ and $E$ to irreps of the groups. This change of basis for the edge elements and surface elements is shown in Figure \ref{changebasis1}.
		
		\begin{figure}[h]
			\begin{center}
			\includegraphics{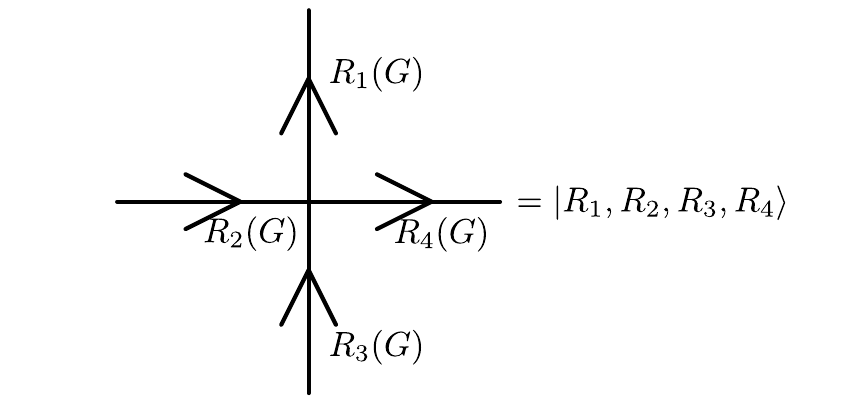}
				\caption{It is convenient to introduce a shorthand for the state of the edges around a vertex}
				\label{vertex_support_1}
			\end{center}
		\end{figure}

		We then consider how the energy terms act in this new basis, starting with the vertex terms. We use the shorthand notation shown in Figure \ref{vertex_support_1} for a state with given edge elements around the vertex in question.	After some simple algebra performed (for a more general case) in Section \ref{Section_2D_irrep_basis_Appendix} in the Supplemental Material, we find that the vertex terms act in the irrep basis according to
		\begin{align}
		A_v& \ket{R_1,R_2,R_3,R_4} \notag \\ &= \delta\big((R_1^{-1} \cdot R_2 \cdot R_3 \cdot R_4^{-1}), 1_{\text{Rep}}\big) \ket{R_1,R_2,R_3,R_4}. \label{Equation_Z4_Z4_irrep_vertex}
		\end{align}
		
		We can see that the vertex term ensures that the irreps labelling the surrounding edges fuse to the identity irrep at the vertex in the ground state (with inverses on some of the irreps to account for the orientation of the edges). Equivalently, the irreps labelling the edges entering the vertex fuse to the same channel as the irreps of the edges leaving the vertex.

\begin{figure}[h]
			\begin{center}
			\includegraphics{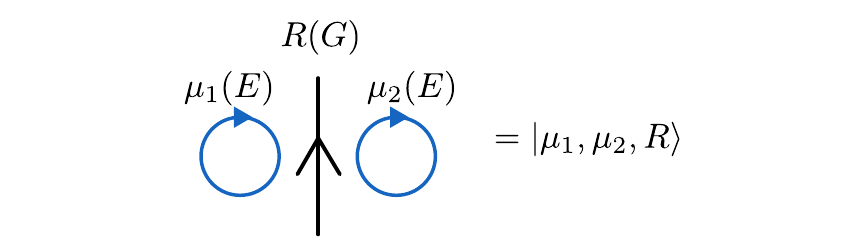}
				\caption{We introduce a shorthand for the labels around an edge. $\mu_1$ and $\mu_2$ here label the plaquettes adjacent to the edge.}
				\label{edge_support_1}
			\end{center}
		\end{figure}

		Next consider the edge terms in this irrep basis. Again we define a shorthand for the state of the labels around an edge, shown in Figure \ref{edge_support_1}.
		
\begin{figure}[h]
			\begin{center}
			\includegraphics{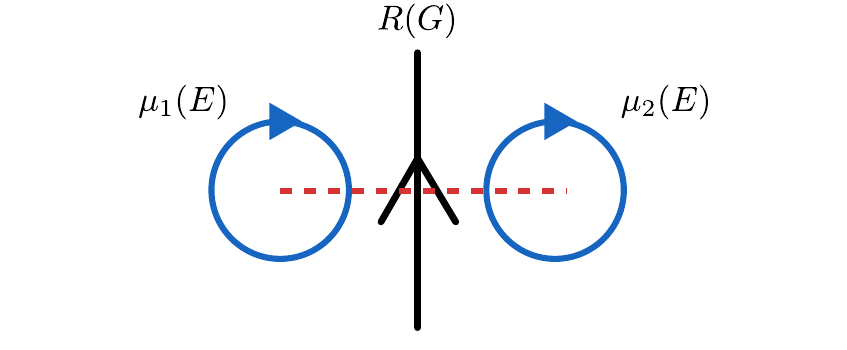}
				\caption{We can consider the plaquettes on either side of an edge as being linked by the edge (just as adjacent vertices are linked by an edge). Here the link is represented by the dashed line, which can be obtained by rotating the edge by ninety degrees.}
				\label{link1}
			\end{center}
		\end{figure}

		Then (again omitting algebraic steps described in Section \ref{Section_2D_irrep_basis_Appendix} in the Supplemental Material) the edge term acts on the edge and surrounding plaquettes as
		\begin{align}
		&\mathcal{A}_i \ket{\mu_1,\mu_2,R} \notag \\
		&= \delta \big((\mu_1^{-1} \cdot \mu_2)(E) , \mu^R(E)\big) \ket{\mu_1,\mu_2,R}, \label{Equation_Z4_Z4_irrep_edge}
		\end{align}
		where $\mu^R$ is defined by 
		\begin{equation}
		\mu^R(e)=R(\partial(e)), \label{Equation_defect_label}
		\end{equation}
		for all $e\in E$ and is itself an irrep of $E$. The representation $\mu^R$ derived from the label of the separating edge is called the defect label for the representation $R$ of that edge. In the particular case considered in this section, the irreps $R= \pm 1_{\text{Rep}}$ have a defect label given by the identity irrep and the irreps $R =\pm i_{\text{Rep}}$ have $-1_{\text{Rep}}$ as the defect label. From Equation \ref{Equation_Z4_Z4_irrep_edge} we see that the edge energy terms enforce a fusion rule, that two neighbouring plaquettes fuse with the edge separating them (or rather, with the defect label derived from the edge) to give the identity representation. This is also true for the horizontal edges, which act in an equivalent way to the vertical edge transforms (just consider rotating Figure \ref{edge_support_1} ninety degrees clockwise and applying the same mathematics). We say that two neighbouring plaquettes separated by an edge are linked, with the dual path between the two plaquettes called the link. In the ground state, two linked plaquettes must be related by the fusion rule derived in Equation \ref{Equation_Z4_Z4_irrep_edge}. For example, given the link shown in Figure \ref{link1}, the edge term enforces that, in the ground state
		\begin{equation}
		\mu_2 \mbeq \mu_1 \cdot \mu^R. \label{Equation_Z4_Z4_irrep_edge_link}
		\end{equation}

		\begin{figure}[h]
			\begin{center}
				
			\includegraphics{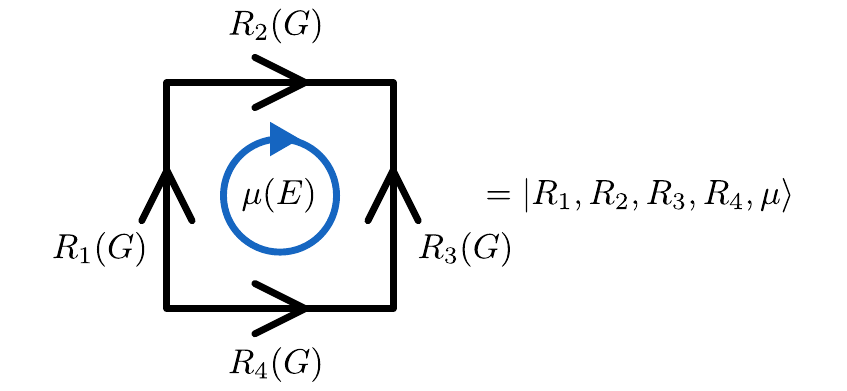}
				\caption{The final energy term to consider is the plaquette term. We use a shorthand for the state of labels near a plaquette which are acted on by the plaquette energy term.}
				\label{plaquette_support_1}
			\end{center}
		\end{figure}

		Finally, consider the plaquette term in this new basis. Once again, we define a shorthand for the labels of the degrees of freedom near the plaquette, as shown in Figure \ref{plaquette_support_1}. After some algebra, we find that
		\begin{align}
		&B_p \ket{R_1,R_2,R_3,R_4,\mu} \notag\\
		&=\frac{1}{|G|} \sum_{\substack{\text{irreps }R \\ \text{ of } G}} \ket{ R \cdot R_1, R \cdot R_2, R^{-1} \cdot R_3, R^{-1} \cdot R_4, \mu^R \cdot \mu}, \label{Equation_Z4_Z4_irrep_plaquette}
		\end{align}
		where (as described earlier in the section), $\mu^R$ is the irrep of $E$ defined by $\mu^R(e)=R(\partial(e))$ for irrep $R$ of $G$. We can see from Equation \ref{Equation_Z4_Z4_irrep_plaquette} that we can split the plaquette energy term into a sum of terms, one per irrep $R$, each of which fluctuates the labels of the edges around the plaquette by $R$ and the label of the plaquette itself by $\mu^R$. We call these individual terms plaquette transforms and write the transform corresponding to irrep $R$ as $B_p^{R}$. Then $B_p = \frac{1}{|G|} \sum_{\text{irreps }R}B_p^{R}$. Because of the group structure of the 1D irreps, these plaquette transforms satisfy a similar algebra to the vertex transforms:
		$$ B_p^{R_1} B_p^{R_2} = B_p^{ R_1 \cdot R_2},$$
		as can be directly verified from Equation \ref{Equation_Z4_Z4_irrep_plaquette}. In particular, this algebra means that a plaquette transform can be absorbed into the corresponding plaquette term (just as for vertex transforms):
		$$B_p^{R}B_p=B_p.$$
		This implies that plaquette transforms have no effect on the ground state, because the ground state is an eigenstate of each plaquette term with eigenvalue one, and so 
		$$B_p^{R} \ket{GS} = B_p^{R}B_p \ket{GS} =B_p \ket{GS}=\ket{GS}.$$

		Having considered the energy terms of the model in this irrep basis, we can now examine the ground states of the model. To construct a ground state, we first find a configuration in the irrep basis that satisfies all of the vertex and edge constraints, then we apply the plaquette terms to produce a suitable superposition of these. We first pick an edge configuration (in the irrep basis) that satisfies the vertex terms. These vertex terms are the same here as in Kitaev's Quantum Double model (when that model is treated in the irrep basis), and so any irrep configurations which satisfy the vertex terms of Kitaev's quantum double model will also satisfy the vertex conditions of this model.

\begin{figure}[h]
			\begin{center}
				
			\includegraphics{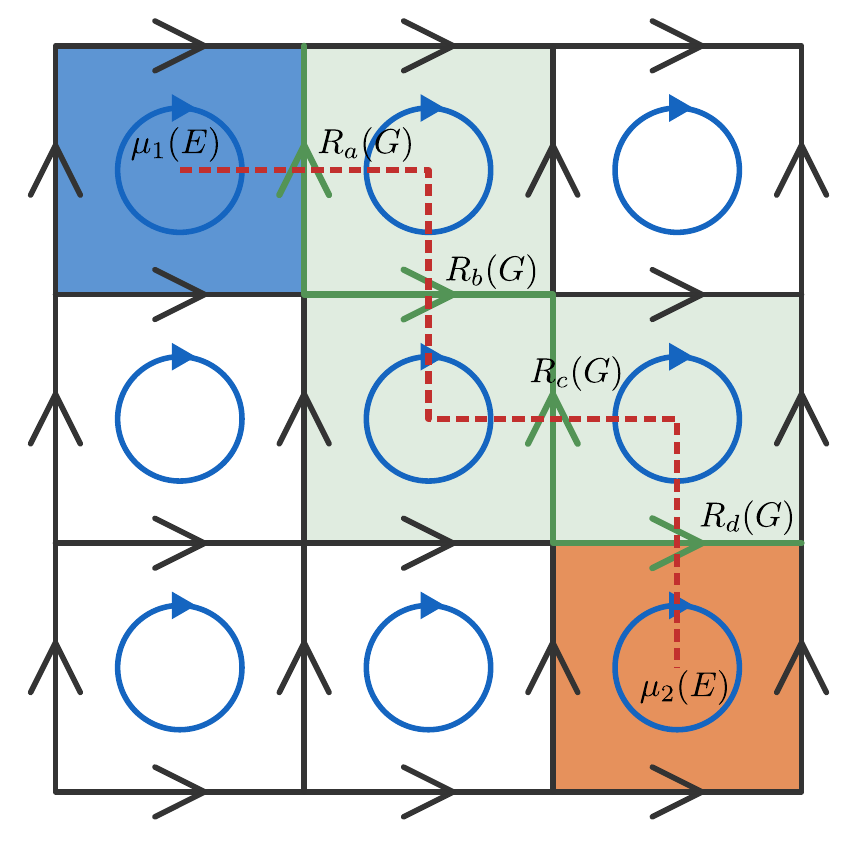}
				\caption{Given two plaquettes that can be connected by a path in the dual lattice, the labels of the two plaquettes in the ground state are related due to the conditions imposed by the edge energy terms. For example, here the heavily shaded (blue and orange) plaquettes are connected by a dual path through the lightly shaded (green) plaquettes, represented by the (red) dashed line. The edge terms on the green edges that intersect with this line then enforce that the irrep $\mu_2$ of the orange plaquette must be related to the irrep $\mu_1$ of the blue plaquette by $\mu_2= \mu_1 \cdot \mu^{R_a} \cdot \mu^{R_b}\cdot \mu^{R_c}\cdot\mu^{R_d}$ in the ground state, where the $R_x$ are the irreps labelling the edges intersected by the dual path.}
				\label{Z4_Z4_plaquettes_connected}
			\end{center}
		\end{figure}

		The next step in finding the ground states is to choose plaquette labels that satisfy the edge terms, given the edge labels. In a path-connected manifold (where each pair of plaquettes is connected by a dual path made of a series of links) the choice of any single plaquette label (along with the previously determined edge labels) fixes the labels of the other plaquettes. To see how this occurs, consider two plaquettes on the lattice, such as the heavily shaded blue and orange plaquettes in Figure \ref{Z4_Z4_plaquettes_connected}. Because the lattice is path-connected, we can construct a path on the dual lattice between the two plaquettes (through the lightly shaded green plaquettes in the example in Figure \ref{Z4_Z4_plaquettes_connected}). By rotating the edges cut by the dual path by ninety degrees in order to form links, we obtain a path of links that connect the two plaquettes (the red dashed lines in Figure \ref{Z4_Z4_plaquettes_connected}). The edge term (as given in Equation \ref{Equation_Z4_Z4_irrep_edge}) forces the label of two adjacent plaquettes to be related by fusion of the defect label of the link connecting them (i.e., of the edge separating the plaquettes). This means that choosing the label of the first plaquette in the dual path fixes the label of the next in the path and so on, until we reach the end of the path. This means that once we have set the label of one plaquette, the label of any other plaquette connected to that one by a dual path is also fixed.

		For example, in Figure \ref{Z4_Z4_plaquettes_connected} the blue plaquette, with label $\mu_1$, is separated from the next plaquette along the path by an edge with label $R_a$. Considering the action of the edge term applied on that edge, which enforces the relation shown in Figure \ref{link1} in the ground state, the label of the next plaquette on the path must be $\mu_1 \cdot \mu^{R_a}$, where $\mu^{R_a}$ is defined in terms of $R_a$ by Equation \ref{Equation_defect_label}. We can repeat this for the next plaquette in the path, and the next, until we reach the final plaquette. This tells us that the label of the orange plaquette, the last plaquette on the path, is $\mu_2 = \mu_1 \cdot \mu^{R_a} \cdot \mu^{R_b}\cdot \mu^{R_c}\cdot\mu^{R_d}$. This is equivalent to $\mu_2 = \mu_1 \cdot \mu^{R_a \cdot R_b \cdot R_c \cdot R_d}$. We can imagine taking the edges intersected by the dual path and rotating them by ninety degrees (clockwise), to form a path connecting the blue and orange plaquettes (shown as a red dashed line). Then $R_a \cdot R_b \cdot R_c \cdot R_d$ is the irrep that we would obtain from fusing the labels of the rotated edges along that path. We see that the label of the orange plaquette is obtained by fusing the label of the blue plaquette with the defect label of the dual path connecting the two plaquettes (where $\mu^R$ is the defect label corresponding to a label of $R$). In this example, the direction of each rotated edge matches the direction of the path. If a rotated edge pointed against the dual path, that edge would instead contribute the inverse irrep to the dual path (in a similar way to how edges on the lattice contribute to a direct path), which matches the inverse that would appear in the condition imposed by the corresponding edge term on the adjacent plaquette labels due to the orientation of the edge being reversed.

\begin{figure}[h]
			\begin{center}
				
				\includegraphics[width=\linewidth]{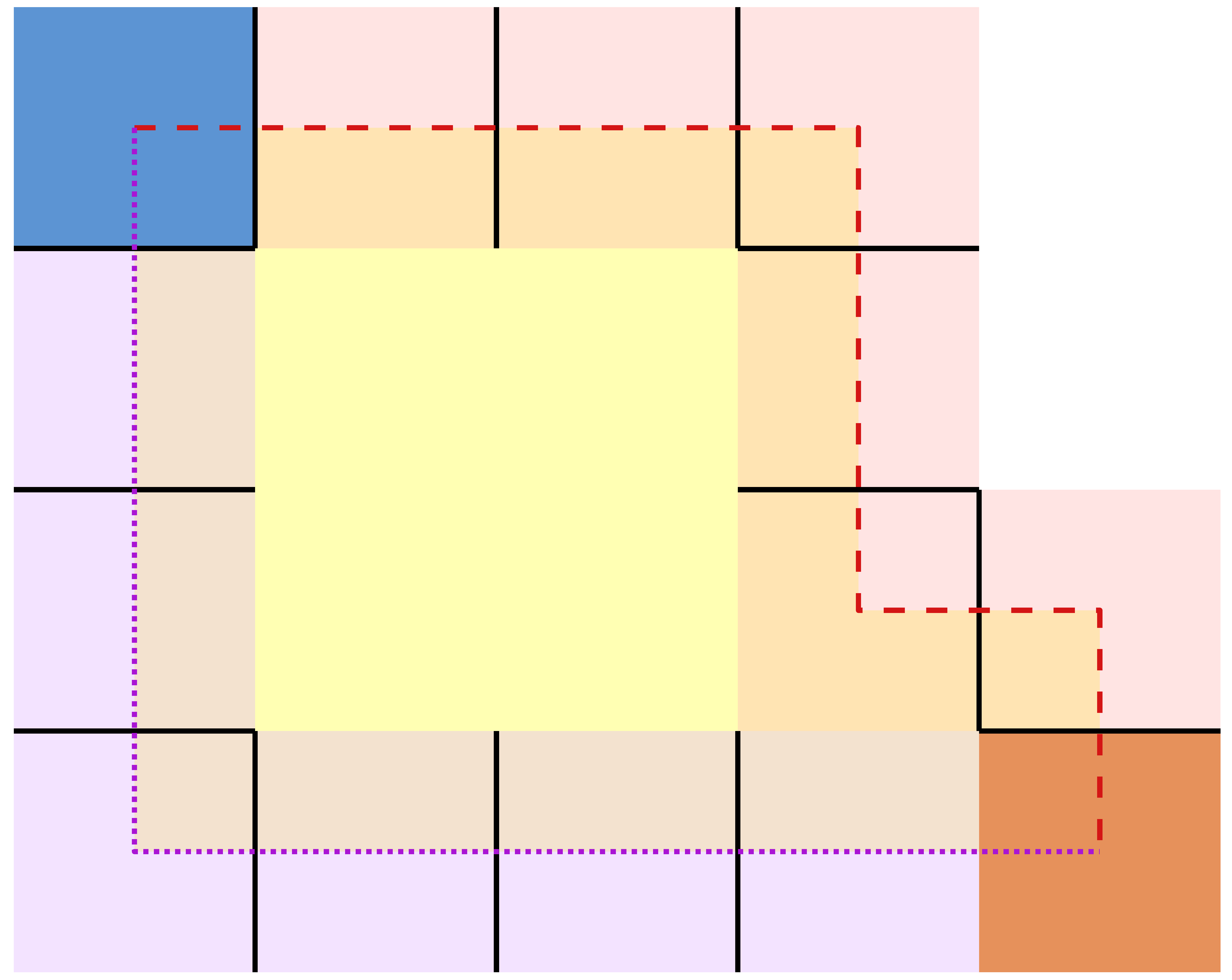}
		
				\caption{There may be many dual paths that connect any two plaquettes. For example, both the red dashed line and purple dotted line connect the two heavily shaded (blue and orange) plaquettes and so both of these paths give constraints on the labels of the two plaquettes (as described in Figure \ref{Z4_Z4_plaquettes_connected}). If the two paths can be deformed into one-another, then they enclose a contractible surface (the yellow region in this figure). The vertex terms in this region enforce fusion rules, so that the total charge carried in or out of the region by the black edges is trivial, and this ensures that the two paths agree (carry the same defect label) in the ground state. This guarantees that both paths give consistent relations between the two plaquettes in the ground state.}
				\label{Z4_Z4_two_dual_paths}
			\end{center}
		\end{figure}

		While on a generic path-connected manifold any two plaquettes can be connected by links in this way, this does not guarantee that we can assign the second plaquette a label in a consistent way. Two plaquettes may be connected by many different such dual paths, and it is not guaranteed that these different paths will give a consistent label for the second plaquette. However, if we consider a simply connected manifold, the various different paths can all be deformed into one-another, which means that they differ only by a closed contractible path on the dual lattice. This closed dual path will enclose some region of the lattice, as shown in the example in Figure \ref{Z4_Z4_two_dual_paths}, and all of the edges that point into or out of this region (the black edges in Figure \ref{Z4_Z4_two_dual_paths}) will be cut by the dual path. However the vertex terms in that region enforce that the edges coming into or out of the region must fuse to the identity (individual vertex terms enforce that the edges coming in and out of the vertex fuse to the identity, as indicated by Equation \ref{Equation_Z4_Z4_irrep_vertex}). This means that the path label (in the irrep basis) associated to the closed dual path must be trivial. In turn, this means that the two possible dual paths between the two plaquettes (which together form the closed path) have the same labels, and so guarantees that the different paths will give consistent answers for the label of the second plaquette. 
		
\begin{figure}[h]
			\begin{center}
				
			\includegraphics{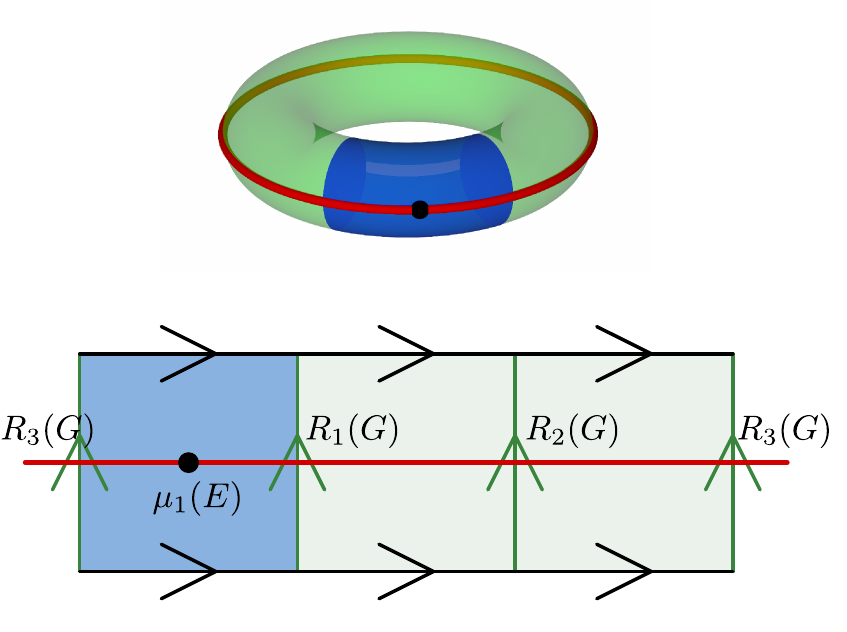}
				\caption{We consider an example of a manifold that is not simply connected, namely the torus in the upper part of the figure. The torus can be represented by a rectangle with its opposite sides identified, as in the lower part of the figure. As we described in Figure \ref{Z4_Z4_plaquettes_connected}, any two plaquettes connected by a path in the dual lattice must have their labels related by fusion with the defect label of the dual path, if the edges crossed by the dual path are unexcited. This means that the blue (dark gray in grayscale) plaquette in this figure must have a label that is related to itself by fusion with the defect label of the closed dual path shown here. In the case shown in the lower part of the figure, this means that $\mu_1(E) = \mu_1(E) \cdot \mu^{R_1(G) \cdot R_2(G) \cdot R_3(G)}$. More generally if the closed dual path is $t$ and its defect label is $\mu^{R(t)}$, then we must have $\mu_1 = \mu_1 \cdot \mu^{R(t)}$. This means that the label of the closed dual path around the cycle of the torus must satisfy $\mu^{R(t)}=1_{\text{Rep}}$ in the ground state.}
				\label{Z4_Z4_torus}
			\end{center}
		\end{figure}

		On the other hand, if the manifold is not simply-connected, then we cannot guarantee this consistency, and so some sets of edge labels may not generate a ground state (i.e., edge excitations may be inevitable for some otherwise valid sets of edge labels, which is only revealed when trying to find appropriate plaquette labels). For example, consider the case of a torus, such as the one shown in Figure \ref{Z4_Z4_torus}. Suppose that the label of one of the plaquettes is given by $\mu_1(E)$. Then, just as we did in Figure \ref{Z4_Z4_plaquettes_connected}, we can see what labels any other plaquette should have (if the edges are all unexcited) by fusing this plaquette label with the defect label of the dual path connecting the two plaquettes. If the path is $t$, with defect label $\mu^{R(t)}$ (the general version of $\mu^{R_a} \cdot \mu^{R_b}\cdot \mu^{R_c}\cdot\mu^{R_d}$ from Figure \ref{Z4_Z4_plaquettes_connected}), then the label $\mu_2$ of the second plaquette must satisfy $\mu_2= \mu_1 \cdot \mu^{R(t)}$. Now consider a non-contractible closed cycle connecting the first plaquette to itself, such as the cycle of the torus shown in Figure \ref{Z4_Z4_torus} (the red cycle), so that $\mu_2 = \mu_1$. This gives us the condition $\mu_1 = \mu_1 \cdot \mu^{R(t)}$. That is, in order to be consistent with the edge energy terms, the label of the plaquette must be the same as the label obtained by fusing the plaquette label with the defect label of the dual path. If it is not, then there is a contradiction and so the edge energy terms cannot all be satisfied. This means that the edge labels are only consistent with the edge energy terms when $\mu^{R(t)}= 1_{\text{Rep}}$ (irrespective of the label of the plaquette itself), which is true when the dual path is labelled by $R(t)= \pm 1_{\text{Rep}}$. This tells us that the ground state (which must satisfy the edge energy terms) can only have certain quantum numbers wrapping the cycles of the torus (it can only have irreps that have trivial defect labels).
		
		Once we have chosen consistent plaquette and edge configurations in the irrep basis (i.e., a set of irrep labels that satisfy the vertex and edge energy terms), a ground state is acquired by applying the plaquette energy terms to fluctuate the configurations. Knowing that all of the different allowed edge configurations for a spherical manifold are connected by the fluctuations from the plaquette term (because Kitaev's Quantum Double model \cite{Kitaev2003} has a unique ground state on the sphere), we would expect the ground state degeneracy to arise from the different choices for the label of the first plaquette. This would suggest a ground state degeneracy of four on the sphere, with one state per irrep of $E$ that we can take as our choice for the first plaquette. However this value for the ground state degeneracy is not quite correct. If we perform a plaquette transform $B_p^{R}$ on each plaquette, it affects each edge twice in a way that cancels out. This is because each edge is aligned with one adjacent plaquette and so the edge label gains a factor of $R$ from the transform associated to that plaquette, but the edge is anti-aligned with another plaquette and so the edge label gains a factor of $R^{-1}$ from that plaquette, which cancels with the first factor. On the other hand, each plaquette label is only affected by one transform and so the series of plaquette transforms changes the label of each plaquette by $\mu^R$. Because these plaquette transforms must leave the ground state invariant (analogous to the behaviour of vertex and edge transforms on the ground state), the state resulting from applying this series of plaquette transforms must appear with equal weight as the original state in the ground state.

		This means that an initial choice for the first plaquette label of irrep $\beta$ is not an independent choice from choosing $ \mu^R\cdot \beta$ instead. The initial choice of irrep should only be considered modulo the possible $ \mu^R$. In this example model, the only non-trivial irrep that can be expressed as $\mu^R(e)=R(\partial(e))$ for some irrep $R$ of $G$ is the $-1_{\text{Rep}}$ representation. This means that our initial choice of plaquette label is defined only up to multiplication by the $-1_{\text{Rep}}$ irrep, i.e., up to a minus sign. Therefore, the only independent choices are $1_{\text{Rep}}$ and $i_{\text{Rep}}$. This implies that the ground state degeneracy should be two on the sphere, which matches the general formula given in Ref. \cite{Bullivant2017}. Note that this degeneracy is the same as the number of sectors of loop excitations, and the non-condensed loop excitations will form domain walls between different regions that look like the different ground states. As with the ($\mathbb{Z}_2$, $\mathbb{Z}_3$) model considered in Section \ref{Section_Example_Z_2_Z_3}, these ground states are locally distinguishable rather than being topologically protected. Unlike for the ($\mathbb{Z}_2$, $\mathbb{Z}_3$) model, these ground states are related by a spontaneously broken symmetry. We mentioned that multiplying each plaquette irrep by $-1_{\text{Rep}}$ must leave the ground state invariant. This is a global symmetry, although it is not spontaneously broken. On the other hand we could also multiply each plaquette irrep by $\pm i_{\text{Rep}}$. This is also a symmetry, although it is spontaneously broken in the ground state space. In other words, there is a $\mathbb{Z}_4$ symmetry that is spontaneously broken down to $\mathbb{Z}_2$, with the broken symmetry resulting in a non-topological degeneracy and the unbroken part giving a symmetry enriched topological phase (as we discuss further in Section \ref{Section_Mapping_SN_SET}). We will discuss this idea in more detail in Section \ref{Section_2D_irrep_basis}, where we consider the more general $\rhd$ trivial case.

		\section{The $\rhd$ trivial case in the irrep basis}
		\label{Section_2D_irrep_basis}

		In Section \ref{Section_Z4_Z4_irrep_basis} we saw that, at least for a particular example of a higher-lattice gauge theory model, changing the basis for our Hilbert space allowed us to reveal information about the ground-state structure of the model. Specifically, it was useful to change from a basis where the edges and plaquettes were labelled by group elements of $G$ and $E$ to one where they were labelled by irreducible representations of those groups. In this section we will take the more general class of higher lattice gauge theory models in 2+1d for which $\rhd$ is trivial and apply the same change of basis. We will then examine the various energy terms in this basis and discuss the ground states in particular. This change of basis is motivated by work from Ref. \cite{Buerschaper2009}, which mapped Kitaev's Quantum Double model \cite{Kitaev2003} to another set of models for 2+1d topological phases, called string-net models \cite{Levin2005}. In Section \ref{Section_Mapping_SN_SET}, we will show that this change of basis similarly allows us to map some of the higher-lattice gauge theory models to a construction for symmetry enriched topological phases described in Ref. \cite{Heinrich2016}.

		We start by defining our change of basis, which is a generalization of the one used in Section \ref{Section_Z4_Z4_irrep_basis}. Consider a single edge of our lattice. In our usual basis, we describe this edge with a group label. For example, if the edge is labelled by $g \in G$, we could denote the state of the edge by $\ket{g}$. However we can construct an alternate basis, where the edge is instead labelled by an irrep of $G$ and the matrix indices for that irrep. For an irrep $R$ and matrix indices $a$ and $b$, we define the state \cite{Buerschaper2009}
		\begin{equation}
		\ket{R,a,b} = \sqrt{\frac{|R|}{|G|}} \sum_{g \in G} [D^R(g)]_{ab} \ket{g},
		\end{equation}
		where $|R|$ is the dimension of the irrep $R$ and $[D^R(g)]$ is the matrix representing $g$ in the irrep $R$. This is the same change of basis we used in Section \ref{Section_Z4_Z4_irrep_basis}, except that because $G$ may be non-Abelian we must include matrix indices for the representations in the change of basis. We can apply a similar change of basis to the plaquette labels, but because $E$ is Abelian when $\rhd$ is trivial, all of the irreps are one-dimensional. Therefore, for an irrep $\mu$ of $E$ we can write the corresponding plaquette state as
		\begin{equation}
		\ket{\mu}= \sqrt{\frac{1}{|E|}} \sum_{e \in E} \mu(e) \ket{e},
		\label{Equation_E_irrep_basis}
		\end{equation}
		where $\ket{e}$ is the state in which the plaquette is labelled by the group element $e \in E$ and $\mu(e)$ is the phase representing $e$ in the 1D irrep (or equivalently is the character of the irrep).

		Having defined our new basis for each degree of freedom, with the basis for the entire lattice made from tensor products of these individual basis elements, we can see how the energy terms of our model act in this new basis. This will again be similar to the results that we found for the example model in Section \ref{Section_Z4_Z4_irrep_basis}, except for features arising from the possibility of $G$ being non-Abelian. We will also not restrict to a particular lattice, such as the square lattice used in Section \ref{Section_Z2_Z4}. We start by considering the vertex energy term. As described in Section \ref{Section_Recap_Paper_2}, a vertex transform $A_v^x$ acts on each edge attached to that vertex (and when $\rhd$ is trivial the transform acts on no other degrees of freedom). For an edge $i$ with label $g_i$ attached to $v$, this action is
		\begin{equation}
		A_v^x:g_i = \begin{cases} xg_i & \text{ if $i$ points away from $v$} \\
		g_ix^{-1} & \text{ if $i$ points towards $v$.}
		\end{cases}
		\end{equation}
		
\begin{figure}[h]
			\begin{center}
			\includegraphics{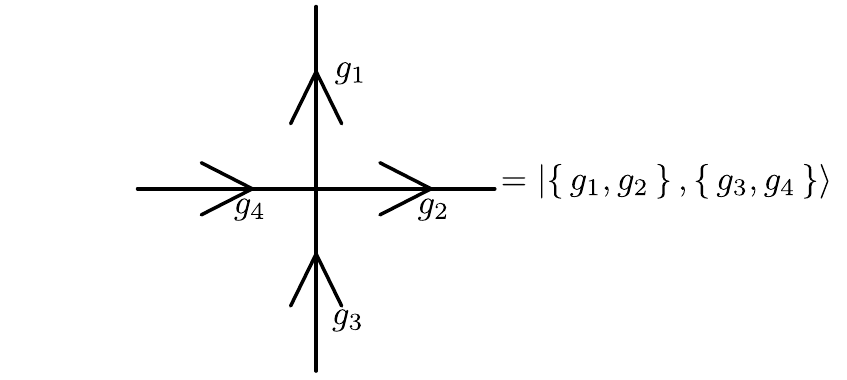}
				\caption{The vertex transform acts on the edges adjacent to that vertex in a way that depends on whether those edges point towards or away from the vertex. We therefore need to differentiate between the inwards and outwards edges in the state, which we do by collecting their labels in separate sets when we write the state.}
				\label{vertex_support_rhd_trivial}
			\end{center}
		\end{figure}

		In order to describe the action of the vertex transform on all of the surrounding edges simultaneously, we denote the state of the adjacent edges by $\ket{\set{g_i}_{\text{out}}, \set{g_j}_{\text{in}}}$, where the subscript ``in" denotes the inwards pointing edges and the subscript ``out" refers to the outwards pointing ones. For example, in Figure \ref{vertex_support_rhd_trivial} we would denote the state of the edges around the vertex by $\ket{\set{g_1,g_2}, \set{g_3,g_4}}$. The action of the vertex transform $A_v^x$ on the state $\ket{\set{g_i}_{\text{out}}, \set{g_j}_{\text{in}}}$ can then be written as
		\begin{equation}
			A_v^x\ket{\set{g_i}_{\text{out}}, \set{g_j}_{\text{in}}} = \ket{\set{xg_i}_{\text{out}}, \set{g_jx^{-1}}_{\text{in}}}.
		\end{equation}

		The vertex energy term $A_v$ is then given by $A_v = \frac{1}{|G|}\sum_{x \in G} A_v^x$. In order to apply this term in our new basis, we define a state
		\begin{align*}
		&\ket{ \set{R_i,a_i,b_i}_{\text{out}}, \set{R_j,a_j,b_j}_{\text{in}}}\\
		& = \sum_{\set{g_i}_{\text{out}}} \sum_{\set{g_j}_{\text{in}}} \bigg(\prod_{\substack{\text{outgoing } \\ \text{edges }i}} \sqrt{\frac{|R|}{|G|}} [D^{R_i}(g_i)]_{a_i b_i} \bigg) \\
		& \hspace{0.5cm}\bigg(\prod_{\substack{\text{incoming } \\ \text{edges }j}} \sqrt{\frac{|R|}{|G|}} [D^{R_j}(g_j)]_{a_j b_j} \bigg)\ket{\set{g_i}_{\text{out}}, \set{g_j}_{\text{in}}},
		\end{align*}
		where $\sum_{\set{g_i}_{\text{out}}}$ sums over each element of $G$ for each outgoing edge and $\sum_{\set{g_j}_{\text{in}}}$ does the same for each incoming edge. Then we can apply the vertex energy term to this state. By applying some algebraic manipulations, given in Section \ref{Section_2D_irrep_basis_Appendix} of the Supplemental Material, we find that

		\begin{align}
		A_v &\ket{ \set{R_i,a_i,b_i}_{\text{out}}, \set{R_j,a_j,b_j}_{\text{in}}} \notag \\
		&= \frac{1}{|G|} \sum_{x \in G} \sum_{\set{c_i}} \sum_{\set{c_j}} \bigg( \prod_{\substack{\text{outgoing } \\ \text{edges }i}} [D^{\overline{R}_i}(x)]_{c_i a_i} \bigg) \notag \\
		& \hspace{0.5cm} \bigg( \prod_{\substack{\text{incoming } \\ \text{edges }j}} [D^{R_j}(x)]_{c_j b_j} \bigg) \notag\\ & \hspace{0.5cm} \ket{\set{R_i,c_i,b_i}_{\text{out}}, \set{R_j,a_j,c_j}_{\text{in}}}, \label{Equation_irrep_basis_vertex_term_1}
		\end{align}
		where $\overline{R}$ is the representation of $G$ conjugate to $R$, which satisfies
		$$[D^{\overline{R}_i}(x)]_{c_i a_i}= [D^{R_i}(x)]_{c_i a_i}^*.$$
		
		The product
		$$\bigg( \prod_{\substack{\text{outgoing } \\ \text{edges }i}} [D^{\overline{R}_i}(x)]_{c_i a_i} \bigg) \bigg(\prod_{\substack{\text{incoming } \\ \text{edges }j}} [D^{R_j}(x)]_{c_j b_j} \bigg)$$ 
		is itself a matrix element of a (generally reducible) representation of $G$ formed from the combination of the individual representations. Because of the Grand Orthogonality Theorem for representations, summing this over the elements $x \in G$ projects to the identity representation. The vertex energy term therefore energetically penalizes states for which the irreps around the vertex do not fuse to the identity (with the direction of the edge with respect the vertex accounted for by using the original irrep or the conjugate irrep). For example, in the Abelian case, where the irreps are all one-dimensional and so we do not need to consider matrix indices, the action of the vertex term on a basis state becomes
		\begin{align}
		A_v &\ket{ \set{R_i}_{\text{out}}, \set{R_j}_{\text{in}}} \notag \\
		&= \delta\bigg( \big(\prod_{\substack{\text{outgoing } \\ \text{edges }i}} \overline{R}_i \big) \big(\prod_{\substack{\text{incoming } \\ \text{edges }j}} R_j \big),1_{\text{Rep}}\bigg) \notag \\
		& \hspace{0.5cm} \ket{\set{R_i}_{\text{out}}, \set{R_j}_{\text{in}}}. \label{Equation_vertex_transform_irrep_basis_Abelian}
		\end{align}

		Next we consider the edge term. An edge transform $\mathcal{A}_i^e$ acts on the edge $i$ itself and the two adjacent plaquettes. If the edge is initially labelled by $g_i$, then the edge transform acts on the edges as $\mathcal{A}_i^e:g_i = \partial(e)g_i$. For an adjacent plaquette $p$, with initial label $e_p$, the action of the transform is
		$$\mathcal{A}_i^e: e_p = \begin{cases} e_p e^{-1} & \text{if the circulation of $p$ matches}\\
		& \text{ the orientation of $i$} \\ 
		e e_p & \text{if the circulation of $p$ opposes}\\
		& \text{ the orientation of $i$.}
		\end{cases}$$
		Because $E$ is Abelian when $\rhd$ is trivial, we can simplify the notation here by neglecting the order of multiplication and writing
		$$\mathcal{A}_i^e: e_p = e_p e^{\sigma_p},$$
		where $\sigma_p$ is $-1$ if the circulation of $p$ aligns with the orientation of $i$ and is $1$ otherwise.
		
		\begin{figure}[h]
			\begin{center}
			\includegraphics{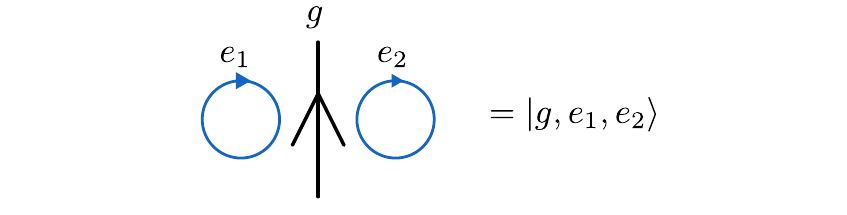}
				\caption{The edge transform affects the edge itself and the two neighbouring plaquettes. We wish to introduce a labelling scheme for the state of the degrees of freedom associated to the edge and these plaquettes. If the label of the edge is $g$, the label of the plaquette to the left of the edge (with respect to the orientation of the edge) is $e_1$ and the label of the right plaquette is $e_2$, then we denote the state by $\ket{g,e_1,e_2}$. }
				\label{edge_support_rhd_trivial}
			\end{center}
		\end{figure}

		Now we wish to consider the action of the edge transform in the irrep basis. As illustrated in Figure \ref{edge_support_rhd_trivial}, we denote the initial state of the three degrees of freedom (the edge and the two adjacent plaquettes) in the group element basis by $\ket{g,e_1,e_2}$, where $e_1$ labels the plaquette to the left of the edge, while $e_2$ labels the plaquette to the right of the edge. Then to implement the change of basis to the irrep basis, we define
		\begin{align*}
		\ket{ \set{R,a,b}, \mu_1, \mu_2} = \sqrt{\frac{|R|}{|G|}}& \sum_{g \in G} \frac{1}{|E|} \sum_{e_1,e_2 \in E} [D^R(g)]_{ab}\\
		& \mu_1(e_1) \mu_2(e_2) \ket{g,e_1,e_2},
		\end{align*}
		where we have used Greek letters to represent the irreps of $E$ to distinguish them from the irreps of $G$. Then in the new basis, the action of the edge transform is given by
		\begin{align}
		\mathcal{A}_i^e &\ket{ \set{R,a,b}, \mu_1, \mu_2} \notag \\
		&= \sum_{c=1}^{|R|} [D^R(\partial(e)^{-1})]_{ac} \mu_1(e^{- \sigma_1}) \mu_2(e^{- \sigma_2}) \notag \\
		& \hspace{1cm} \ket{\set{R,c,b}, \mu_1, \mu_2}, \label{Equation_edge_transform_irrep_basis_2}
		\end{align}
		as proven in Section \ref{Section_2D_irrep_basis_Appendix} in the Supplemental Material (where $\sigma_p$ is 1 if the plaquette $p$ is anti-aligned with the edge and is $-1$ otherwise). We now wish to consider the object $[D^R(\partial(e)^{-1})]_{ac}$ in closer detail. While $R$ is an irrep of $G$, the argument $\partial(e)$ is only ever in the subgroup $\partial(E)$ of $G$. Just as we discussed previously in Section \ref{Section_2D_electric} (in the context of electric ribbon operators), the fact that $\partial(E)$ is a subgroup in the centre of $G$ means that $R$ branches to multiple copies of the same irrep of $\partial(E)$, due to Clifford's theorem \cite{Clifford1937}. We can also see this from Schur's lemma, which tells us that the matrices of irrep $R$ that represent elements in $\partial(E)$ are scalar multiples of the identity, with the scalar being a 1D irrep of $\partial(E)$, which we will denote by $R_{\partial}^{\text{irr.}}$. Then we have 
		$$[D^R(\partial(e)^{-1})]_{ac} = \delta_{ac} R_{\partial}^{\text{irr.}}(\partial(e)^{-1}).$$
		
		Furthermore, we can use this irrep of $\partial(E)$ to define an irrep of $E$, which we call $\mu^R$, by $\mu^R(e)=R_{\partial}^{\text{irr.}}(\partial(e))$. This is a representation of $E$ due to the fact that $\partial$ is a group homomorphism, and it is an irrep because it is a 1D representation. Substituting these results into Equation \ref{Equation_edge_transform_irrep_basis_2} gives us
		\begin{align*}
		\mathcal{A}_i^e &\ket{ \set{R,a,b}, \mu_1, \mu_2}\\
		&= \mu^R(e^{-1}) \mu_1(e^{- \sigma_1}) \mu_2(e^{- \sigma_2}) \ket{\set{R,a,b}, \mu_1, \mu_2}.
		\end{align*}
		
		We now consider the energy term, which is an average over all of the edge transforms on $i$: $\mathcal{A}_i = \frac{1}{|E|}\sum_{e \in E} \mathcal{A}_i^e$. Defining $\mu^{-1}(e)=\mu^*(e)=\mu(e^{-1})$, we find from the Grand Orthogonality Theorem that
		\begin{align}
		\mathcal{A}_i&\ket{ \set{R,a,b}, \mu_1, \mu_2} \notag\\
		&= \delta( \mu^R \cdot \mu_1^{ \sigma_1} \cdot \mu_2^{ \sigma_2},1_{\text{Rep}(E)}) \ket{\set{R,a,b}, \mu_1, \mu_2}. \label{Equation_edge_transform_irrep}
		\end{align}
		
		We can see that this edge term checks that the irreps of the two plaquettes on either side of the edge are compatible with the irrep labelling the edge. This idea becomes more clear if we take a fixed orientation for the two plaquettes. For example, we can take the case in Figure \ref{edge_support_rhd_trivial}, where the two plaquettes have the same orientation (clockwise). Then with this orientation, the plaquette on the left is aligned against the direction of the edge, while the plaquette on the right is aligned with the edge. This means that the energy term becomes
		\begin{align}
		\mathcal{A}_i&\ket{ \set{R,a,b}, \mu_1, \mu_2} \notag\\
		&= \delta( \mu^R \cdot \mu_1 \cdot \mu_2^{ -1},1_{\text{Rep}(E)}) \ket{\set{R,a,b}, \mu_1, \mu_2} \notag \\
		&=\delta( \mu^R \cdot \mu_1, \mu_2) \ket{\set{R,a,b}, \mu_1, \mu_2}. \label{Equation_edge_term_irrep_basis}
		\end{align}
		
		Then it is clear that the edge term enforces that the irreps labelling the two plaquettes differ by the irrep $\mu^R$ derived from the irrep $R$ labelling the edge. We note that, because $\mu^R(e)=R_{\partial}^{\text{irr.}}(\partial(e))$, the irrep $\mu^R$ is trivial in the kernel of $\partial$. That is, for an element $e_k$ satisfying $\partial(e_k)
		=1_G$, we have $\mu^R(e_k)=R_{\partial}^{\text{irr.}}(1_G)=1$. This means that the Kronecker delta in Equation \ref{Equation_edge_term_irrep_basis} enforces that $\mu_1(e_k)=\mu_2(e_k)$, so the two plaquettes must carry the same irrep of the kernel of $\partial$.

		Finally, we consider the plaquette terms. The plaquette term $B^p$ acts in the group element basis according to $B^p \ket{\psi} = \delta(\partial(e_p)g_p,1_G)$, where $e_p$ is the label of the plaquette and $g_p$ is the path element corresponding to the boundary of the plaquette. This path element can be expressed as
		$$g_p = \prod_{\substack{\text{edge }i \text{ in } \\ \text{boundary}(p)}} g_i^{\sigma_i},$$
		where $\sigma_i$ is $1$ if the edge $i$ is oriented along the boundary and $-1$ if it is oriented against it. The product is taken in the order of the boundary, so that the first edge on the boundary is on the left of the product. We can denote the state of the plaquette and the edges on the boundary by $\ket{e_p, \set{g_i}}$. Then we can write the action of the plaquette operator on this state as
		\begin{align*}
		B_p\ket{e_p, \set{g_i}} & = \delta\big(\partial(e_p)\prod_{\substack{\text{edge }i \text{ in } \\ \text{boundary}(p)}} g_i^{\sigma_i}, 1_G \big)\ket{e_p, \set{g_i}}.
		\end{align*}

	Now we wish to consider the action of the plaquette term in the irrep basis. We define the irrep basis states by
		\begin{align*}
		\ket{ \mu , \set{R_i,a_i,b_i}}
		& = \sum_{\set{g_i}} \bigg( \prod_{i}\sqrt{\frac{|R_i|}{|G|}} [D^{R_i}(g_i)]_{a_i,b_i} \bigg) \\
		& \hspace{0.5cm}\sum_{e_p \in E} \sqrt{\frac{1}{|E|}} \mu(e_p)\ket{e_p, \set{g_i}}.
		\end{align*}
		
		We find that the action of the plaquette operator on this basis state is (as shown in Section \ref{Section_2D_irrep_basis_Appendix} of the Supplemental Material)
		\begin{align}
		B_p&\ket{ \mu, \set{R_i,a_i,b_i}} \notag\\
		&= \sum_{\text{irreps }R \text{ of }G} \sum_{\set{c_i}} \sum_{\set{g_i}} \frac{|R|}{|G|} \notag \\
		& \hspace{0.5cm} \bigg( \prod_{i} \sqrt{\frac{|R_i|}{|G|}} [D^{R_i}(g_i)]_{a_i,b_i} [D^R(g_i^{\sigma_i})]_{c_i c_{i+1}} \bigg) \notag \\
		& \hspace{0.5cm} \sum_{e_p \in E} \sqrt{\frac{1}{|E|}} \mu(e_p) \mu^R(e_p) \ket{e_p, \set{g_i}}. \label{Equation_plaquette_term_irrep_basis_3}
		\end{align}
		
		We can split the plaquette term into a sum of terms, where each term corresponds to one irrep $R$ on the right-hand side of Equation \ref{Equation_plaquette_term_irrep_basis_3}. That is, we write
		\begin{equation}
		B_p = \sum_{\text{irreps }R \text{ of }G} \frac{|R|}{|G|} B_p^R,
		\end{equation}
		where
		\begin{align*}
		B_p^R&\ket{ \mu, \set{R_i,a_i,b_i}}\\
		&= \sum_{\set{c_i}} \sum_{\set{g_i}} \bigg( \prod_{i} \sqrt{\frac{|R_i|}{|G|}} [D^{R_i}(g_i)]_{a_i,b_i} [D^R(g_i^{\sigma_i})]_{c_i c_{i+1}} \bigg)\\
		& \hspace{0.5cm} \sum_{e_p \in E} \sqrt{\frac{1}{|E|}} \mu(e_p) \mu^R(e_p) \ket{e_p, \set{g_i}}\\
		&=\sum_{\set{c_i}} \sum_{\set{g_i}} \bigg(\prod_{i} \sqrt{\frac{|R_i|}{|G|}} [D^{R_i}(g_i)]_{a_i,b_i} [D^R(g_i^{\sigma_i})]_{c_i c_{i+1}} \bigg)\\
		& \hspace{0.5cm} \ket{ \mu \cdot \mu^R, \set{g_i}}.
		\end{align*}
		
		We see that $B_p^R$ acts by fusing the irrep $R$ into the edges along the boundary of the plaquette (with an inverse if the edge is anti-aligned with the boundary) and fusing the irrep $\mu^R$ into the plaquette. For example, when $G$ is Abelian this becomes
		\begin{align*}
		B_p^R\ket{ \mu , \set{R_i}}&= \sum_{\set{g_i}} \bigg(\prod_{i} \sqrt{\frac{1}{|G|}} R_i(g_i) R(g_i^{\sigma_i}) \bigg) \\
		& \hspace{0.5cm} \sum_{e_p \in E} \sqrt{\frac{1}{|E|}} \mu(e_p) \mu^R(e_p) \ket{e_p, \set{g_i}}\\
		&= \sum_{\set{g_i}} \bigg(\prod_{i} \sqrt{\frac{1}{|G|}} (R_i\cdot R^{\sigma_i})(g_i) \bigg)\\ & \hspace{0.5cm}\sum_{e_p \in E} \sqrt{\frac{1}{|E|}} (\mu \cdot\mu^R)(e_p) \ket{e_p, \set{g_i}}\\
		&= \ket{\mu \cdot \mu^R, \set{R_i\cdot R^{\sigma_i}}},
		\end{align*}
		where $R^{\sigma_i}(g)=R(g^{\sigma_i})$. The total plaquette term $B_p$ then fluctuates the irreps of the edges and plaquette by averaging over the irreps fused into the edges and plaquette:
		\begin{align}
			B_p\ket{ \mu, \set{R_i}} &= \sum_{\text{irreps }R \text{ of }G} \frac{1}{|G|} \ket{\mu \cdot \mu^R, \set{R_i\cdot R^{\sigma_i}}}.
		\end{align}
	
		We note that the plaquette labels resulting from the action of the plaquette term have the form $\mu \cdot \mu^R$, where
		\begin{equation}
			\mu^R(e)=R_{\partial}^{\text{irr.}}(\partial(e)).
		\end{equation} 
		
		This means that when we restrict the irrep to the kernel of $\partial$, by considering elements $e_k$ in the kernel, we have 
		$$(\mu\cdot \mu^R) (e_k) = \mu(e_k) R_{\partial}^{\text{irr.}}(\partial(e_k))=\mu(e_k) R_{\partial}^{\text{irr.}}(1_G)=\mu(e_k),$$
		so that the resulting irrep that labels the plaquette has the same restriction to the kernel of $\partial$ as the original irrep label.

		Having considered the action of the different energy terms in the irrep basis in some detail, we now wish to consider what this means for the energy eigenstates and in particular the ground states. In the group element basis, the vertex term and the edge term both fluctuate the group elements around the edge or vertex, while the plaquette term puts constraints on what elements are allowed in the ground state. On the other hand, in the irrep basis this is reversed. The vertex and edge terms enforce fusion rules on the irreps in the ground states, while the plaquette term fluctuates the labels on and around the plaquette. In order to construct a ground state, we can first choose a set of edge irrep labels which satisfies the rules enforced by the vertex terms, which only act on the edge labels. Then we can pick a set of plaquette irrep labels that satisfy the edge terms, given our choice of edge labels. Finally, we apply the plaquette terms to fluctuate these degrees of freedom and produce the ground states, which are linear combinations of various irrep configurations satisfying the fusion rules. For those readers familiar with Kitaev's Quantum Double model, the first step is equivalent to how we would consider the ground states in that model in the irrep basis, because the vertex term is the same in both models. However it is interesting to consider the last two steps in more detail, because these involve the plaquette labels.

		 First, we want to choose plaquette labels that satisfy the edge terms. Recall that the edge term acts on an edge $i$ and the two adjacent plaquettes according to
		\begin{align*}
		\mathcal{A}_i&\ket{ \set{R_i,a,b}, \mu_1, \mu_2}\\
		&=\delta( \mu^{R_i} \cdot \mu_1, \mu_2) \ket{\set{R_i,a,b}, \mu_1, \mu_2},
		\end{align*}
		provided that the plaquettes have the same clockwise orientation, where $\mu_1$ is the initial label of the left plaquette and $\mu_2$ is the label of the right plaquette, as indicated in Figure \ref{edge_support_rhd_trivial}. This enforces that the two irreps of $E$ differ by multiplication by an irrep derived from the edge label $R_i$. This does not fix the label of either plaquette, but instead ensures that they are related by a factor. Once $\mu_1$ and $R_i$ are chosen, the irrep $\mu_2$ is set. As explained in Section \ref{Section_Z4_Z4_irrep_basis}, where we considered a particular example model, if the manifold represented by our lattice is simply connected, this means that choosing the label of one plaquette will fix the value of all other plaquettes on the lattice. In addition, because $\mu^{R_i}$ has a trivial restriction to the kernel of $\partial$, every plaquette label will have the same restriction to the kernel of $\partial$ (if we choose all of the plaquettes to have the same orientation, otherwise some will have the conjugate irrep of the kernel). Keeping this in mind, we now consider applying the plaquette terms. These fluctuate the values of the edge and plaquette labels, so that the ground state is not a simple product state. However, the plaquette term only fluctuates the plaquette labels by irreps $\mu^R$ which have trivial restriction to the kernel of $\partial$. This means that the restriction of the plaquette labels to the kernel of $\partial$ is unaffected by the application of the plaquette terms. Given that this restriction of the irrep labelling a plaquette to the kernel is also the same for each plaquette due to the edge terms, this means that this restriction of the plaquette label to the kernel of $\partial$ is a property of the overall ground state constructed in this way. This means (on a simply connected manifold) that we will have one ground state or ground state sector for each possible irrep of the kernel of $\partial$, and so there are multiple ground states even on the 2-sphere. This ground state degeneracy does not arise due to topological concerns, as we can tell from the fact that the ground-state can be determined locally by measuring the label of a single plaquette. Instead, for these models with $\rhd$ trivial, there is a symmetry that results in this ground state degeneracy.

	We claim that, if all of the plaquettes are oriented in the same way, this symmetry is given by multiplication of each plaquette's irrep label by the same irrep $\mu$ of $E$ (if a plaquette is oriented in the opposite way, we should instead multiply its label by the inverse irrep). We denote the operator that does this by $U^{\mu}$, where we have one such operator for each irrep $\mu$ of $E$. In order to be a symmetry of the model, this operator must commute with the energy terms. The symmetry operator clearly commutes with the vertex term, because the vertex term does not act on the plaquette labels. However we can also show that the symmetry operator commutes with the other energy terms. Recall that, when the plaquettes all have the same orientation, the edge energy term checks that the two plaquettes on either side of the edge have labels related by multiplication by an irrep associated to the edge. Because the 1D irreps of $E$ form an Abelian group, multiplying the irreps of both plaquettes by a common irrep $\mu$ does not change the relationship between the two irreps and so preserves this condition. That is, given an initial state $\ket{{R,a,b},\mu_1, \mu_2}$, acting first with the edge term and then the symmetry operator gives us
		\begin{align*}
		U^{\nu}\mathcal{A}_i&\ket{{R,a,b},\mu_1, \mu_2}\\
		&= U^{\nu} \delta( \mu^{R_i} \cdot \mu_1, \mu_2) \ket{\set{R_i,a,b}, \mu_1, \mu_2}\\
		&=\delta( \mu^{R_i} \cdot \mu_1, \mu_2) \ket{\set{R_i,a,b}, \nu \mu_1, \nu \mu_2}.
		\end{align*} 
		
		On the other hand, for the other order of operations we have
		\begin{align*}
		\mathcal{A}_iU^{\nu}&\ket{{R,a,b},\mu_1, \mu_2}\\
		&= \mathcal{A}_i \ket{\set{R_i,a,b}, \nu \mu_1, \nu \mu_2}\\
		&=\delta( \mu^{R_i} \cdot \nu \cdot \mu_1, \nu \cdot \mu_2) \ket{\set{R_i,a,b}, \nu \mu_1, \nu \mu_2}\\
		&=\delta( \mu^{R_i} \cdot \mu_1, \mu_2) \ket{\set{R_i,a,b}, \nu \mu_1, \nu \mu_2},
		\end{align*}
		where in the last line we used the fact that the irreps of $E$ form an Abelian group to remove the common factor of $\nu$ in each term of the Kronecker delta. This indicates that the symmetry operator does indeed commute with the edge terms. While we assumed that the plaquettes all have the same orientation, the same result holds generally (the symmetry operator then multiplies the plaquettes with the opposite orientation by the inverse irrep $\nu^{-1}$, but this inverse cancels with an inverse in the action of the edge term described by Equation \ref{Equation_edge_transform_irrep}).

		Lastly, we must consider the commutation between the plaquette terms and the symmetry. The plaquette term is made of a sum of operators $B_p^R$ which interact with the plaquette labels by multiplying them by an irrep $\mu^R$, but because the group of irreps of $E$ is Abelian this multiplication always commutes with the multiplication from the symmetry operator. Therefore, the symmetry operator commutes with the plaquette terms. Combining this with our previous results, we see that the symmetry operator commutes with all of the energy terms, as required. We have defined a symmetry operator $U^{\mu}$ for all irreps of $E$, but we mentioned earlier that the ground states were labelled by irreps of the kernel of $\partial$. The presence of degenerate ground states, even on a spherical manifold, suggests that the symmetry is spontaneously broken, but the fact that the ground states are labelled by irreps of the kernel of $\partial$ rather than irreps of the full group $E$ suggests that the symmetry is only partially broken, as we will see later in this section.

		Having constructed this symmetry operator by utilising the irrep basis for our model, it is enlightening to consider how this symmetry operator acts in our original basis. Consider a state $\ket{\set{\mu_p}}$, where $\set{\mu_p}$ denotes the irrep labels $\mu_p$ of each plaquette $p$ and we do not need to consider the edge labels. The action of the symmetry operator $U^{\nu}$ on this state is (assuming that all of the plaquettes have the same orientation)
		$$U^{\nu}\ket{\set{\mu_p}}= \ket{\set{\nu \mu_p}}.$$
		
		Now we can use our change of basis to see how this operator acts in the original basis. A state $\ket{\set{e_p}}$, where each plaquette is labelled by a group element $e_p$, is given in terms of the irrep basis states by
		$$ \ket{\set{e_p}} = \sum_{\set{\mu_p}} \bigg( \prod_p\sqrt{ \frac{1}{|E|}} \mu_p(e_p)^{-1} \bigg) \ket{\set{\mu_p}},$$
		where $\sum_{\set{\mu_p}}$ sums over all irreps $\mu_p$ of $E$ for each plaquette $p$ (this can directly be verified as the inverse transformation of Equation \ref{Equation_E_irrep_basis} for each plaquette). Then the symmetry operator $U^{\nu}$ acting on this state is
		\begin{align*}
		U^{\nu}\ket{\set{e_p}}&= \sum_{\set{\mu_p}} \bigg(\prod_p \sqrt{ \frac{1}{|E|}} \mu_p(e_p)^{-1} \bigg) U^{\nu}\ket{\set{\mu_p}}\\
		&= \sum_{\set{\mu_p}} \bigg(\prod_p \sqrt{ \frac{1}{|E|}} \mu_p(e_p)^{-1} \bigg) \ket{\set{\nu \mu_p}}.
		\end{align*}
		
		Now because the 1D irreps of $E$ form a group, we may replace the dummy index $\mu_p$ with $\mu'_p=\nu \mu_p$ in the sum for each $p$. This gives us
		\begin{align*}
		U^{\nu}&\ket{\set{e_p}}\\
		&=\sum_{\set{\mu'_p= \nu \mu_p}} \bigg(\prod_p \sqrt{ \frac{1}{|E|}} (\nu^{-1} \cdot \mu'_p)(e_p)^{-1} \bigg) \ket{\set{ \mu'_p}}\\
		&= \sum_{\set{\mu'_p}} \bigg(\prod_p \sqrt{ \frac{1}{|E|}} \nu(e_p) \mu'_p(e_p)^{-1} \bigg) \ket{\set{ \mu'_p}}\\
		&= \big(\prod_p\nu(e_p) \big) \sum_{\set{\mu'_p}} \bigg(\prod_p\sqrt{ \frac{1}{|E|}} \mu'_p(e_p)^{-1} \bigg) \ket{\set{ \mu'_p}}\\
		&=\big(\prod_p \nu(e_p) \big)\ket{\set{e_p}}\\
		&= \nu( \prod_p e_p) \ket{\set{e_p}}.
		\end{align*}
		
		Given that the plaquettes all have the same orientation, $\prod_p e_p$ is just the total surface element of the lattice. Writing this total surface element as $e(L)$, we see that the symmetry operator acts as
		\begin{align*}
		U^{\nu}\ket{\set{e_p}}&=\nu(e(L)) \ket{\set{e_p}}
		\end{align*}
		on a basis element. We can then write the operator acting on an arbitrary state $\ket{\psi}$ as
		\begin{align}
		U^{\nu}\ket{\psi}&=\sum_{e \in E} \nu(e) \delta(\hat{e}(L),e)\ket{\psi}, \label{Equation_symmetry_original_basis}
		\end{align}
		where $\hat{e}(L)$ is the operator that measures the total surface element. We note that, if $\ket{\psi}$ is a ground state, then fake-flatness implies that $\partial(\hat{e}(L))$ must be related to the path element $\hat{g}(L)$ corresponding to the path around the boundary of the lattice by $\partial(\hat{e}(L))\hat{g}(L)=1_G$. If our manifold is a sphere then the boundary is trivial, and so we have $\partial(\hat{e}(L))=1_G$. This implies that, for a ground state $\ket{GS}$ on the sphere, only the labels $e$ in the kernel of $\partial$ contribute, so that
		\begin{align*}
		U^{\nu}\ket{GS}&=\sum_{e \in E} \nu(e) \delta(\hat{e}(L),e)\ket{GS}\\
		&=\sum_{e \in \ker(\partial)} \nu(e) \delta(\hat{e}(L),e)\ket{GS}.
		\end{align*}
		
		This tells us that the action of the symmetry operator $U^{\nu}$ only depends on the restriction of $\nu$ to the kernel, and in particular is trivial if the restriction to the kernel is trivial. Therefore, the symmetries labelled by irreps with such trivial restrictions are unbroken in the ground states, which further explains (in the case of the 2-sphere) why the ground states are labelled by irreps of the kernel and not by general irreps of $E$.

		For a more general manifold, the picture is slightly more complicated. We can use the fake-flatness condition to write 
		$$\ket{GS} = \delta(\partial(\hat{e}(L)), \hat{g}(L)^{-1}) \ket{GS}.$$
		This allows us to write
		\begin{align*}
		&U^{\nu}\ket{GS}=\sum_{e \in E} \nu(e) \delta(\hat{e}(L),e)\delta(\partial(\hat{e}(L)), \hat{g}(L)^{-1})\ket{GS}.
		\end{align*}
		
		We can use $\delta(\hat{e}(L),e)$ to replace the operator $\hat{e}(L)$ in the fake-flatness condition with the variable $e$:
		\begin{align*}
		U^{\nu}\ket{GS}&=\sum_{e \in E} \nu(e) \delta(\hat{e}(L),e)\delta(\partial(e), \hat{g}(L)^{-1})\ket{GS}.
		\end{align*}

		We can represent an element $e \in E$ as a product $e_k q$, where $e_k$ is in the kernel of $\partial$ and $q$ is a representative of an element of the quotient group $E/ \ker(\partial)$ (so that each coset of $\ker(\partial)$ in $E$ is represented by a unique element $q$). Denoting the set of representatives of this quotient group by $Q_{\partial}$, we can write the sum over $e \in E$ as 
		$$\sum_{e \in E} = \sum_{e_k \in \ker(\partial)} \sum_{q \in Q_{\partial}}.$$
		
		This allows us to write the action of the symmetry operator as
		\begin{align*}
		U^{\nu}\ket{GS}&=\sum_{e \in E} \nu(e) \delta(\hat{e}(L),e)\delta(\partial(e), \hat{g}(L)^{-1})\ket{GS}\\
		&= \sum_{e_k \in \ker(\partial)} \sum_{q \in Q_{\partial}} \nu(e_k q) \delta(\hat{e}(L),e_k q)\\
		& \hspace{1cm} \delta(\partial(e_k q), \hat{g}(L)^{-1})\ket{GS}.
		\end{align*}
		
		We can then use a defining property of representations to write $\nu(e_kq) = \nu(e_k) \nu(q)$. Furthermore, $e_k$ is in the kernel of $\partial$, and so $\partial(e_k q)= \partial(q)$. This gives us
		\begin{align*}
		U^{\nu}\ket{GS}&= \sum_{e_k \in \ker(\partial)} \sum_{q \in Q_{\partial}} \nu(e_k) \nu(q) \delta(\hat{e}(L),e_k q)\\
		& \hspace{1cm}\delta(\partial(q), \hat{g}(L)^{-1})\ket{GS}.
		\end{align*}

		Because the $q \in Q_{\partial}$ are unique representatives of the cosets of $\ker(\partial)$ in $E$, there is a unique $q$ satisfying $\partial(q)= \hat{g}(L)^{-1}$. Denoting this value of $q$ (which is generally an operator) by $\hat{q}_L$, this allows us to remove the sum over $q\in Q_{\partial}$ and the Kronecker delta enforcing fake-flatness:
		\begin{align*}
		U^{\nu}\ket{GS}&= \sum_{e_k \in \ker(\partial)} \nu(e_k) \nu(\hat{q}_L) \delta(\hat{e}(L),e_k \hat{q_L})\ket{GS}.
		\end{align*}
		
		Then, even if $\nu$ is trivial in $\ker(\partial)$, so that $\nu(e_k)=1$ for all $e_k$, the action of the symmetry operator may still be non-trivial. When $\nu$ is trivial in the kernel, we have
		\begin{align*}
		U^{\nu}\ket{GS}&= \nu(\hat{q}_L) \sum_{e_k \in \ker(\partial)}\delta(\hat{e}(L),e_k \hat{q_L})\ket{GS}.
		\end{align*}
		
		Fake-flatness ensures that 
		$$\sum_{e_k \in \ker(\partial)}\delta(\hat{e}(L),e_k \hat{q_L})\ket{GS}= \ket{GS}$$ 
		(from the fact that $\partial(\hat{q}_L)= \hat{g}(L)^{-1}$), and so this action becomes
		\begin{align*}
		U^{\nu}\ket{GS}&= \nu(\hat{q}_L)\ket{GS},
		\end{align*} 
		which is a simple phase if $\ket{GS}$ is an eigenstate of $\hat{q}_L$ (i.e., has a well defined boundary label) but can be a more complicated transformation if $\ket{GS}$ is a linear combination of such states.

		The action of the symmetry operators given in Equation \ref{Equation_symmetry_original_basis} also reveals additional information about the model and in particular its excitations. We can see that the action of the symmetry operator is exactly the same as the action of an $E$-valued membrane operator
		 $$L^{\nu}(m)= \sum_{e \in E} \nu(e) \delta(\hat{e}(m),e),$$
		 applied over the entire lattice (see Section \ref{Section_2D_Loop}). This means that when we apply an $E$-valued membrane operator on part of the lattice it is the same as applying a symmetry operator over that section of the lattice. This means that we are changing which irrep that part of the lattice corresponds to, while leaving the rest of the lattice untouched. This creates a domain corresponding to a different ground state, and so we see that the loop-like excitation on the boundary of the membrane is in fact a domain wall between these different domains, just as we discussed for an example model in Section \ref{Section_Z4_Z4_irrep_basis}. This idea can also give us an alternative interpretation of the condensed loop excitations. The condensed $E$-valued membrane operators, which we found to be equivalent to a confined electric ribbon operator around the boundary of the membrane, are labelled by irreps with trivial restriction to the kernel of $\partial$. As we saw earlier, such irreps correspond to unbroken symmetries, which do not change the ground state when applied on the entire lattice (for a sphere). Therefore, it seems that these condensed membrane operators do not produce a domain corresponding to a different ground state compared to the rest of the lattice, which agrees with the idea that these operators act equivalently on the ground state to electric ribbon operators around the boundary of the membrane (because such local operators cannot make domains).

		Next we consider how to construct projectors to the various ground-state sectors. All of the plaquettes in a given basis ground state are labelled by irreps which have the same restriction to the kernel of $\partial$ (assuming the plaquettes all have the same orientation, otherwise plaquettes with the opposite orientation will restrict to the inverse irrep). This means that if we are in the space of ground states, we can determine which sector we are in by measuring this quantity for a single plaquette. Consider a plaquette $p$, with state $\ket{\mu_p}= \sqrt{\frac{1}{|E|}} \sum_{e_p \in E} \mu_p(e_p) \ket{e_p}$. Now consider applying a single plaquette multiplication operator $M^e(p)$ to this plaquette, which just multiplies the group label of the plaquette by $e$. Then we have
		\begin{align*}
		M^e(p) \ket{\mu_p}&= \sqrt{\frac{1}{|E|}} \sum_{e_p \in E} \mu_p(e_p) \ket{e e_p}\\
		&=\sqrt{\frac{1}{|E|}} \sum_{e_p'=ee_p \in E} \mu_p(e^{-1}e_p') \ket{e_p'}\\
		&= \sqrt{\frac{1}{|E|}} \sum_{e_p' \in E} \mu_p(e^{-1}) \mu_p(e_p') \ket{e_p'}\\
		&= \mu_p(e^{-1})\ket{\mu_p},
		\end{align*}
		so that the result is the accumulation of a phase $\mu_p(e^{-1})$. Then for each irrep $\nu$ of the kernel of $\partial$ we can construct an operator
		$$P^{\nu}= \frac{1}{|\ker(\partial)|}\sum_{e \in \text{ker}(\partial)} \nu(e) M^e(p).$$
		
		Applying this operator to the state $\ket{\mu_p}$ gives us
		\begin{align*}
		P^{\nu}\ket{\mu_p} &= \frac{1}{|\ker(\partial)|}\sum_{e \in \ker(\partial)} \nu(e) M^e(p) \ket{\mu_p}\\
		&=\frac{1}{|\ker(\partial)|} \sum_{e \in \text{ker}(\partial)} \nu(e) \mu_p(e^{-1})\ket{\mu_p}\\
		&=\frac{1}{|\ker(\partial)|}\sum_{e \in \text{ker}(\partial)} \nu(e) \mu_p(e)^*\ket{\mu_p}.
		\end{align*}
		
		From $\mu_p$ we can define an irrep $\mu_p^{\text{ker}}$ of ker$(\partial)$ by restricting $\mu_p$ to the kernel of $\partial$, so that $\mu_p^{\text{ker}}(e)= \mu_p(e)$ for $e$ in the kernel of $\partial$. We can therefore apply the Grand Orthogonality Theorem to write 
		\begin{align*}
		\frac{1}{|\ker(\partial)|}\sum_{e \in \text{ker}(\partial)} \nu(e) \mu_p(e)^* &= \delta(\nu,\mu_p^{\text{ker}}),
		\end{align*}
		so that
		\begin{align*}
		P^{\nu}\ket{\mu_p} & =\delta(\nu,\mu_p^{\text{ker}}) \ket{\mu_p}.
		\end{align*}

		That is, the operator $P^{\nu}$ projects onto states for which the plaquette $p$ is labelled by irreps of $E$ that restrict to the irrep $\nu$ of the kernel. We saw earlier that we can construct a basis for the ground states that is labelled by these irreps for the kernel, so that each plaquette in the ground state has the same restriction to the kernel. This means that, when applied in the space of ground states, the projection operator $P^{\nu}$ projects onto the ground state sector labelled by the irrep $\nu$ of the kernel of $\partial$.
		
		In this section we have demonstrated that the 2+1d higher lattice gauge theory model possesses a global symmetry when $\rhd$ is trivial, indicating that it describes some kind of symmetry enriched topological (SET) phase. It is therefore interesting to relate this model to other constructions for SET phases. In the following section we will do exactly this for one such construction, known as symmetry enriched string-nets.
		
		\section{Mapping the 2+1d higher lattice gauge theory model to the symmetry enriched string-net construction}
		\label{Section_Mapping_SN_SET}
		
		\subsection{Symmetry enriched string-nets}
		\label{Section_SN_SET_Description}
		The higher lattice gauge theory model in 2+1d can be considered as a generalization of Kitaev's Quantum Double model \cite{Kitaev2003}, as we mentioned in Ref. \cite{HuxfordPaper1}, in Section I E. It is known that Kitaev's Quantum Double model can be mapped to a subset of another class of models for topological phases, known as the string-net models \cite{Buerschaper2009}. In the same way, we can map the $\rhd$ trivial 2+1d higher lattice gauge theory model to a subset of a class of string-net-like models, known as symmetry enriched string-nets \cite{Heinrich2016}. In order to explain this connection, we will first give a summary of the symmetry enriched string-net models from Ref. \cite{Heinrich2016}.

		The symmetry enriched string-net model is considered on a honeycomb lattice (though it can be generalized to any trivalent lattice) \cite{Heinrich2016}. The plaquettes of the lattice are labelled by group elements in some group $H$. The (directed) edges of the lattice are labelled by objects in an \textit{$H$-extension} of a \textit{unitary fusion category} $\mathcal{C}$. For the purposes of this work, we can just consider a fusion category as a collection of objects with some fusion rules that describe how we can combine them. For example, given two objects $a$ and $b$, which can fuse to objects $c$ or $d$, we write $a \times b = c+d$. More generally, we write that $a \times b = \sum_c N^c_{ab} c$, where $N^c_{ab}$ are called fusion multiplicities. There must also be an identity (or vacuum) object which fuses trivially with the other objects. A familiar example of a category is the collection of representations of some finite group $X$, where the simple objects are the irreps of the group. These irreps can be fused (by considering a tensor product of the matrices from each representation) and the result of this fusion can be decomposed into a sum of irreps. There are other important components for a fusion category, such as the so-called F-matrices, which encode associativity relations on the fusion of multiple objects \cite{Heinrich2016}. These are important for the explicit construction of string-net models \cite{Levin2005}, but we will not need to consider them in detail here.

		Given a unitary fusion category $\mathcal{C}$, an $H$-extension of that category is another fusion category $\mathcal{D}$ which contains $\mathcal{C}$ and which has a property known as $H$-grading \cite{Heinrich2016}. Each object of the new category $\mathcal{D}$ is associated to a group element $h \in H$. This divides the category into parts $\mathcal{D}_h$ such that $\mathcal{D} = \oplus_{h \in H} \mathcal{D}_h$, with each part satisfying $\mathcal{D}_x \times \mathcal{D}_y \subset \mathcal{D}_{xy}$. That is, given an object in $\mathcal{D}_x$ and an object in $\mathcal{D}_y$, their fusion products must be in $\mathcal{D}_{xy}$. Furthermore, the part of $\mathcal{D}$ corresponding to the identity element of $H$ is $\mathcal{C}$, so that $\mathcal{D}_{1_H}=\mathcal{C}$ \cite{Heinrich2016}.

		Now that we have considered the objects labelling the directed edges (objects in the category $\mathcal{D}$) and plaquettes (elements of the symmetry group $H$) of the honeycomb lattice, we must consider the Hamiltonian. The Hamiltonian is a sum of commuting projector operators, where we have operators for the vertices, edges and plaquettes of the model, just as we do in the higher lattice gauge theory model. The Hamiltonian is then given by \cite{Heinrich2016}
		\begin{equation}
		H= - \sum_{\text{vertices } v} Q_v - \sum_{\text{edges } l} P_l - \sum_{\text{plaquettes }p} B_p,
		\end{equation}
		where we will now define each of these terms. The first operator, $Q_v$, is associated to a vertex $v$ of the lattice, and projects onto states where the three edges adjacent to that vertex obey the fusion rules (so if two edges go into the vertex and one comes out, the labels of the two incoming edges must fuse to the label of the outgoing edge) \cite{Heinrich2016}.

\begin{figure}[h]
			\begin{center}
			\includegraphics{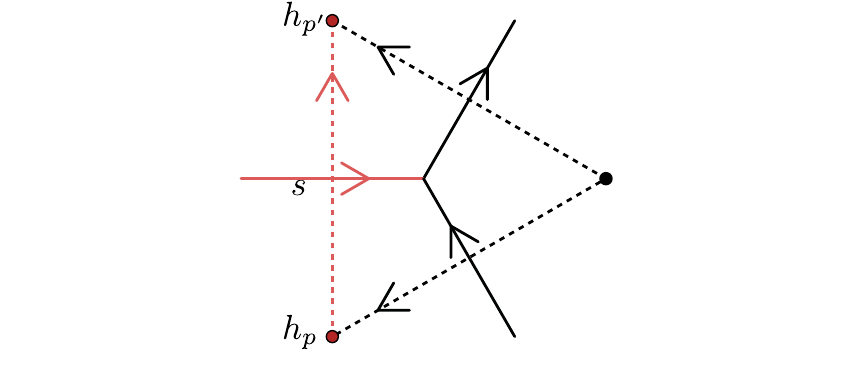}
				
				\caption{In this diagram, adapted from Figure 9 of Ref. \cite{Heinrich2016}, we consider the energy term associated to the edges of the lattice in the symmetry enriched string-net model. Each edge (solid arrows in this figure) separates two plaquettes (whose centres are represented by the dots). Then the edge can be associated to an edge in the dual lattice by rotating it ninety degrees anticlockwise (obtaining the dashed arrows which connect the plaquettes). The edge term then checks that the plaquette labels at the two ends of the dual edge differ by a label $h_s$, where $s$ is the label of the edge and $h_s$ is the group element associated to that label. For example, considering the red edge (labelled by $s$), the adjacent plaquettes $p$ and $p'$ are labelled by $h_p$ and $h_{p'}$ respectively. The dual edge corresponding to the red edge (the red dashed arrow) points from $p$ to $p'$ and so the edge term gives one if $h_{p'}^{-1}h_p =h_s$ and zero otherwise.}
				\label{SET_SN_link_term}
				
			\end{center}
		\end{figure}

		 The operator $P_l$, acting on an edge $l$, checks that the labels of the two plaquettes separated by that edge ``agree" with the label of the edge, in the following sense \cite{Heinrich2016}. The edge is labelled by an object $s$ in $\mathcal{D}$. This object is associated to a group element $h_s$ such that $s \in \mathcal{D}_{h_s}$. Then the operator $P_l$ projects onto the case where the group elements of the plaquettes on either side of the link differ by multiplication by this element $h_s$. For example, in the situation shown in Figure \ref{SET_SN_link_term}, where we have an edge $l$ bordered by plaquette $p$ and $p'$, the projector $P_l$ is one if $h_{p'}^{-1}h_p =h_s$ and zero otherwise. Which plaquette should play the role of $p'$ in this expression is determined by the orientation of the edge: when the edge is rotated by ninety degrees anticlockwise it should point from $p$ to $p'$.

		Finally, consider the plaquette term $B_p$. This is made of a sum of terms corresponding to the different object types in the category $\mathcal{D}$. We have
		$$B_p = \sum_{s \in \mathcal{D}} a_s B_p^s \tilde{U}_p^{h_s}.$$
		Here $B_p^s$ acts by fusing a closed string of object $s$ into the edges of plaquette $p$, while $\tilde{U}_p^{h_s}$ right-multiplies the plaquette label $h_p$ by $h_s$, where $h_s$ is the group label associated to object $s$. We have described $B_p^s$ qualitatively here, but it is the same as the string-net term $B_p^s$ from Ref. \cite{Levin2005}, for those familiar with the string-net model.
		
		While the input to the symmetry enriched string-net model is an $H$-extension of a category $\mathcal{C}$, the output topological phase is described by a braided $H$-crossed extension of $\mathcal{Z(C)}$ \cite{Heinrich2016}, where $\mathcal{Z(C)}$ is the Drinfeld center of $\mathcal(C)$ (i.e., the output from a standard string-net model). This category describes the fusion and braiding statistics of the topological excitations of the model, while the symmetry group is given by $H$.
		
		\subsection{Mapping from the higher lattice gauge theory model}

		We will now consider how we can map from the higher lattice gauge theory model in 2+1d (with $\rhd$ trivial) to these symmetry enriched string-net models. In Section \ref{Section_2D_irrep_basis}, we took the 2+1d higher lattice gauge theory model in the case where $\rhd$ is trivial, and we expressed each energy term in a new basis. In this new basis, each edge and plaquette is labelled by an irreducible representation (and matrix indices, for non-Abelian groups) of $G$ and $E$ respectively. In this section we will show that we can use this basis to relate the higher lattice gauge theory models (with $\rhd$ trivial) to the symmetry enriched string-net models. To do so, we consider placing the higher lattice gauge theory model on a honeycomb lattice, so that it is on the same lattice as the symmetry enriched string-net model. However there is still some difference in the decoration of the lattice in the two cases. In the higher-lattice gauge theory model, the plaquettes have an orientation, whereas they do not in the symmetry enriched string-net model. In order to address this, we simply choose all plaquettes to have the same orientation (clockwise) for the higher lattice gauge theory model.

		Having fixed the lattice, we now consider the labels of the edges and plaquettes in the two models. In the higher lattice gauge theory model, the edges and plaquettes are typically labelled by elements of the two groups $G$ and $E$ respectively. However, in the irrep basis introduced in Section \ref{Section_2D_irrep_basis}, the edges of this lattice are labelled by irreps of $G$ (and their matrix indices), while the plaquettes are labelled by irreps of $E$. Because $E$ is Abelian, the irreps of $E$ are 1D and form a group. We claim that this group of irreps corresponds to the symmetry group $H$ whose elements label the plaquettes in the symmetry enriched string-net model. On the other hand, the labels of the edges are given by irreps of $G$ and their matrix indices in our irrep basis for the higher lattice gauge theory model, while the edges in the symmetry enriched string-net model are labelled by objects in a fusion category. The irreps of a group form a fusion category and we claim that this is the category $\mathcal{D}$ whose objects label the edges in the symmetry enriched string-net model. We note that this identification is the same as the one used in Ref. \cite{Buerschaper2009}, where Kitaev's Quantum Double was mapped to the (non-enriched) string-nets. The mapping between the edge labels is the same whether we are considering a map from the Quantum Double model to the string-net model or a map from the higher lattice gauge theory model to the symmetry enriched string-net model. This is because when we omit the plaquette labels (by making the labels belong to the trivial group), the symmetry enriched string-net model becomes an ordinary string-net model and the higher lattice gauge theory model becomes Kitaev's Quantum Double model.

		One subtlety with the identification of the edge labels in the two models is that, in the irrep basis of the higher lattice gauge theory model, the edges are not only labelled by the irreps, which are the objects of $\mathcal{D}$, but also the matrix indices for those irreps (when $G$ is non-Abelian). As described in Ref. \cite{Buerschaper2009}, the matrix indices of the edges in Kitaev's Quantum Double model, and so in the higher lattice gauge theory model, can be associated to the vertices at the two ends of the edge. These additional degrees of freedom are fixed when the vertex terms are applied, with the indices being contracted to ensure appropriate fusion \cite{Buerschaper2009}. This means that, even if the two models have different Hilbert spaces when $G$ is non-Abelian, the ground states of the models are equivalent.

		So far, we have only considered the labels of the edges in the symmetry enriched string-net model as belonging to a fusion category. However, the category $\mathcal{D}$ whose objects label the edges in the symmetry enriched string-net model has additional structure, which is important for further discussion of the mapping. That is, $\mathcal{D}$ is an $H$-extension of another category, $\mathcal{C}$, where $H$ is the group whose elements label the plaquettes. This means that each object $s$ in $\mathcal{D}$ is associated with a group element $h_s$, and the objects in $\mathcal{D}$ that correspond to the identity of $H$ are the objects of $\mathcal{C}$, as discussed in Section \ref{Section_SN_SET_Description} (and in more detail in Ref. \cite{Heinrich2016}). How is this reflected in the higher-lattice gauge theory model? We have identified $\mathcal{D}$ as Rep($G$), the category formed from the irreps of $G$. Given an irrep $R$ of $G$, it is associated to a 1D irrep $R_{\partial}^{\text{irr.}}$ of the subgroup $\partial(E)$, where $R_{\partial}^{\text{irr.}}(\partial(e))=[D^R(\partial(e))]_{11}$, as we described in Section \ref{Section_2D_irrep_basis} when we first considered the irrep basis. This irrep $R_{\partial}^{\text{irr.}}$ is in turn associated to an irrep $\mu^R$ of $E$, satisfying $R_{\partial}^{\text{irr.}}(\partial(e))= \mu^R(e)$. This is the irrep of $E$ that we called the defect label of $R$ in Section \ref{Section_Z4_Z4_irrep_basis}. We claim that this irrep $\mu^R$ is the group element associated to the object $R$ (recall that we have identified the symmetry group $H$ from the symmetry enriched string-net model with the group formed by the irreps of $E$). That is, if $R$ is the object $s$ of $\mathcal{D}$, then $\mu^R$ is the group element $h_s$ of $H$ associated to that object. In particular, this means that the objects of $\mathcal{C}= \mathcal{D}_{1_H}$ are the irreps with trivial restriction to $\partial(E)$, which are equivalent to irreps of the quotient group $G/\partial(E)$ (note that $\partial(E)$ is central in $G$ for $\rhd$ trivial).

		There is a small subtletly in that, while $\mu^R$ is an irrep of $E$, it is not a generic irrep of $E$. This is because it is only sensitive to $\partial(e)$ and not to $e$ itself, which means that it acts trivially on the kernel of $\partial$. This means that some of the group elements $h \in H$ (which are irreps of $E$), specifically the irreps of $E$ with a non-trivial restriction to the kernel, cannot be written as $h=\mu^R$ for some irrep $R$ of $G$ and so have no objects associated to them. This means that some of the parts $\mathcal{D}_h$ of the graded category $\mathcal{D}$ are empty. Ref. \cite{Heinrich2016} specifically does not study cases with such empty parts of the graded category, and so for the sake of this mapping we will restrict to the case where the kernel of $\partial$ is trivial from here on (so that $E$ and $\partial(E)$ are isomorphic). From Section \ref{Section_2D_irrep_basis}, we know that the different ground states of the sphere are labelled by the irreps of the kernel, and so under this restriction we will not have degenerate ground states on the sphere (we do not have spontaneous symmetry breaking).

		Before we compare the Hamiltonians of the two models, we introduce a simple example to help describe the identification of the objects in the two models, which we will use throughout this section to explain features of the mapping. Consider the crossed module $(G=\mathbb{Z}_4, E=\mathbb{Z}_2, \partial \rightarrow \mathbb{Z}_2, \rhd \rightarrow \text{id})$, where $\partial$ maps the elements of $E=\mathbb{Z}_2$ to the $\mathbb{Z}_2$ subgroup of $\mathbb{Z}_4$. If we write the elements of $G$ as $1_G$, $i_G$, $-1_G$ and $-i_G$, and the elements of $E$ as $1_E$ and $-1_E$, then we have $\partial(1_E)=1_G$ and $\partial(-1_E)=-1_G$, which is a group homomorphism as required. This crossed module (as required) obeys the Peiffer condition Equation \ref{Peiffer_1}, because for any $g \in G$ and $e \in E$, $\partial(g \rhd e) = \partial(e)$ from $\rhd$ being trivial, and $\partial(e)=g\partial(e)g^{-1}$ because of the Abelian nature of $G$, so that $\partial(g \rhd e) = g \partial(e)g^{-1}$. It also obeys the second Peiffer condition, Equation \ref{Peiffer_2}, because for any pair $e, f \in E$ we have $\partial(e) \rhd f = f$ from $\rhd$ being trivial, and $f= efe^{-1}$ from $E$ being Abelian, so that $\partial(e) \rhd f =efe^{-1}$.

		 Having satisfied ourselves that $(G=\mathbb{Z}_4, E=\mathbb{Z}_2, \partial \rightarrow \mathbb{Z}_2, \rhd \rightarrow \text{id})$ is indeed a valid crossed module, we can find the corresponding symmetry enriched string-net model. The category $\mathcal{D}$ of this string-net model is given by Rep$(G=\mathbb{Z}_4)$, whose objects are the four (unitary) irreps of $\mathbb{Z}_4$, which we describe in Table \ref{irreps_Z_4_alt} below. Meanwhile, the symmetry group $H$ of the symmetry enriched string-net is the group of irreps of $E$, i.e., the irreps of $\mathbb{Z}_2$, where multiplication for this group is defined by $(\alpha_1 \cdot \alpha_2)(e)= \alpha_1(e) \alpha_2(e)$ for any two irreps $\alpha_1$, $\alpha_2$. The resulting group is itself a copy of $\mathbb{Z}_2$. As described in Section \ref{Section_SN_SET_Description}, in the symmetry enriched string-net model each object from the category is associated to an element of this symmetry group (the category has a $G$-grading). The group element associated to an irrep of $G$ (which is an object in the category) is the irrep of $E$ obtained by restricting that irrep of $G$ to the subgroup $\partial(E)$ of $G$ (note that for the mapping we restrict to cases where $\partial(E)$ is isomorphic to $E$). Considering the irreps of $\mathbb{Z}_4$, as defined in Table \ref{irreps_Z_4_alt}, we see that the irreps $1_R$ and $-1_R$ of $G=\mathbb{Z}_4$ restrict to the trivial irrep of $\mathbb{Z}_2 = \partial(E)$ (because $\pm1_R(1_G)=\pm1_R(-1_G)=1$). This means that these two irreps belong to $\mathcal{D}_{1_H}$, the part of $\mathcal{D}$ associated to the identity element of the symmetry group. On the other hand, the irreps $\pm i_R$ restrict to the non-trivial irrep of the $\mathbb{Z}_2$ subgroup (because $\pm i_R(1_G)=1$ and $\pm i_R(-1_G)=-1$), and so belong to the other part of the category, $\mathcal{D}_{-1_H}$, associated to the non-trivial symmetry element.
		
		\begin{table}[h]	
			\begin{center}
				\begin{tabular}{|c|c |c| c| c|} 
					\hline
					& $\bm{1_G}$ & $\bm{i_G}$ & $\bm{-1_G}$ &$\bm{-i_G}$\\ 
					\hline
					$\bm{1_{R}}$ & $1$ & $1$ & $1$ & $1$ \\ 
					\hline
					$\bm{-1_{R}}$ & $1$ & $-1$ & $1$ & $-1$ \\ 
					\hline
					$\bm{i_{R}}$ & $1$ & $i$ & $-1$ & $-i$ \\ 
					\hline
					$\bm{-i_{R}}$ & $1$ & $-i$ & $-1$ & $i$ \\ 
					\hline
					
				\end{tabular}
				\caption{The group $\mathbb{Z}_4$ has four irreps, which are one-dimensional. In each row, we give the phase representing each group element for a particular irrep (which is equivalently the character for that irrep).}
				\label{irreps_Z_4_alt}
			\end{center}
		\end{table}

		Next we wish to compare the Hamiltonians of the two models. We will show that the energy terms of the higher lattice gauge theory model are equivalent, in the irrep basis, to those of the symmetry enriched string-net model. First, we consider the vertex terms. In Section \ref{Section_2D_irrep_basis} we showed that the vertex term acts on neighbouring edges in the irrep basis by checking that the irreps fuse to the identity (if all of the edges point towards the vertex). This is equivalent to the term $Q_v$ from the symmetry enriched string-net model \cite{Heinrich2016}, which checks that the edges around the vertex obey fusion rules (so that if they all point inwards, they fuse to the identity). This is explained in more detail in Ref. \cite{Buerschaper2009} for the mapping between Kitaev's Quantum Double model and the string-net model (the vertex terms for the Quantum Double model are the same as those for higher lattice gauge theory when $\rhd$ is trivial, and similarly the vertex terms for the string-net model are the same as those for the symmetry enriched string-net model, so the mapping works the same in either case). In the case of the example model, where the group $G=\mathbb{Z}_4$ is Abelian, this term is simple, because fusion of two irreps $R_1$ and $R_2$ is just given by multiplication of the irreps defined by $(R_1 \cdot R_2)(g) =R_1(g) \cdot R_2(g)$ (note that the irreps form a group under this multiplication). Therefore, the vertex term just checks that the irreps (which are the natural labels for the edges in the symmetry enriched string-net model, and which become the labels of the edges in the higher lattice gauge theory model after a change of basis) multiply to give the identity irrep.

\begin{figure}[h]
			\begin{center}
			\includegraphics{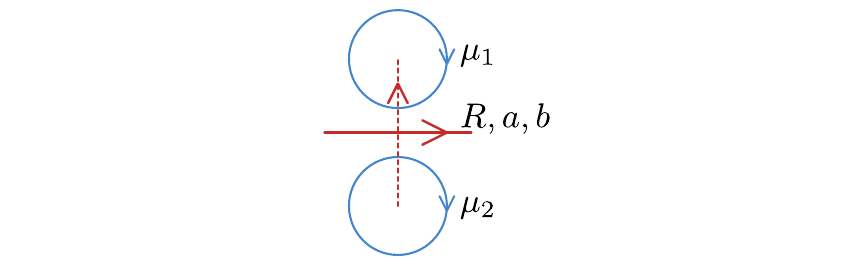}
				
				\caption{For the higher lattice gauge theory model in the irrep basis, the vertex term applied on the (red) edge enforces that the labels of the two adjacent plaquettes differ by an irrep $\mu^R$, where $R$ is the irrep of $G$ labelling the edge and $\mu^R$ is an irrep of $E$ derived from $R$. This has the same form as the edge term in the symmetry enriched string-net model, once we identify $R$ with an object of the category $\mathcal{D}$ and $\mu^R$ with the associated element of the symmetry group.}
				\label{CM_model_SET_SN_link_term_comparison}
				
			\end{center}
		\end{figure}

		The second energy term to consider is the edge term. In the symmetry enriched string-net model, the edge term checks that the two plaquettes separated by that edge have labels that agree with the edge label. If the two plaquettes, $p$ and $p'$, have labels $h_p$ and $h_{p'}$, while the edge $l$ has label $s_l$, then the energy term $P_l$ gives zero if $h_{p'}^{-1}h_p =h_{s_l}$ (where if we rotate the edge ninety degrees anticlockwise, it points from $p$ to $p'$ as shown in Figure \ref{SET_SN_link_term}). In Section \ref{Section_2D_irrep_basis} we saw that the edge term in the higher lattice gauge theory model in the irrep basis enforces that, for two plaquettes $1$ and $2$, labelled by irreps $\mu_1$ and $\mu_2$, that are separated by an edge $l$ with label $R_l$, the labels satisfy $ \mu^{R_l} \cdot \mu_1= \mu_2$. Here plaquette $1$ is the plaquette to the left of the edge and plaquette $2$ is to the right (with respect to the orientation of the edge, as shown in Figure \ref{CM_model_SET_SN_link_term_comparison}). This is equivalent to saying that if we rotate the edge by ninety degrees anticlockwise, it points from plaquette 2 to plaquette 1. That is, in the language of the string-net model, plaquette 1 is $p'$ and plaquette 2 is $p$. This means that the condition $\mu^{R_l} \cdot \mu_1= \mu_2$, which can also be written as $\mu^{R_l} = \mu_1^{-1} \mu_2 $ (using the fact that the group of 1D irreps is Abelian to swap the order of multiplication), is equivalent to the $h_{s_l} =h_{p'}^{-1} h_p$ condition from the symmetry enriched string-net model. $\mu^{R_l}$ is the irrep of $E$ associated to irrep $R_l$ of $G$ in the same way that $h_{s_l}$ is the group element of $H$ associated to object $s_l$ in the symmetry enriched string-net model. Therefore, the edge terms are the same in the two models.

		To understand this better, consider our example model. In this simple model, the plaquettes have two basis states, described by the identity irrep and the non-trivial irrep of $\mathbb{Z}_2$. Then we can distinguish between two types of edge: ones where the plaquette label changes across the edge (i.e., the two plaquettes on either side of the edge have different labels) and ones where it stays the same. The two situations allow different edge labels when the edge term is satisfied. If the plaquette label changes across the edge, then the edge term tells us that the edge label must correspond to the non-trivial element of the symmetry group, i.e., the non-trivial irrep of $\mathbb{Z}_2$, in order to satisfy $\mu^{R_l} \cdot \mu_1= \mu_2$. From our previous discussion, we know that the irreps of $G=\mathbb{Z}_4$ corresponding to this non-trivial irrep of $\mathbb{Z}_2$ are $\pm i_R$. Therefore, the edge labels that can separate two different plaquette labels are $\pm i_R$. On the other hand, if the plaquette label does not change across the edge, then the edge must correspond to a trivial irrep of $\mathbb{Z}_2$ ($\mu^{R_l}$ must be the trivial irrep), so the edge must be labelled by one of the irreps $\pm 1_R$ of $G=\mathbb{Z}_4$. In the general case, where the plaquette label can take more than two values, we have different kinds of transition of the plaquette label across the edge, and these must be consistent with the label of the edge, as described earlier.

		The final energy term to consider is the plaquette term. In the symmetry enriched string-net model, the plaquette term $B_p$ is given by
		$$B_p = \sum_{s \in \mathcal{D}} a_s \tilde{B}_p^s \tilde{U}_p^{h_s},$$
		where $\tilde{B}_p^s$ is the same as the plaquette term from the ordinary string-net model (it acts by fusing a string of label $s$ around the plaquette) and $\tilde{U}_p^{h_s}$ right-multiplies the plaquette label by $h_s$. Note that we have added a tilde to $\tilde{B}_p^s$ in order to differentiate it from a similar operator in the higher lattice gauge theory model. For the class of these symmetry enriched string-net models that we are mapping to, the objects are irreps of the group $G$. We can therefore write the plaquette term in these cases as
		$$B_p = \sum_{\text{irreps }R \text{ of }G} a_R \tilde{B}_p^R \tilde{U}_p^{h_R}.$$

		 As we saw in Section \ref{Section_2D_irrep_basis}, we can write the plaquette term for the higher lattice gauge theory model as
		$$B_p = \sum_{\text{irreps }R \text{ of }G} \frac{|R|}{|G|} B_p^R,$$
		where the effect of $ B_p^R$ is to fuse the irrep $R$ into the edges of the plaquette and also to multiply the plaquette label by $\mu^R$. This action on the edges is the same as the action of the plaquette term in Kitaev's Quantum Double model, and therefore the same as the action of the ordinary string-net plaquette term $\tilde{B}_p^s$ (as indicated by the mapping discussed in Ref. \cite{Buerschaper2009}). As noted in Ref. \cite{Buerschaper2009} for the mapping from the Quantum Double model to the string-net model, there is a subtlety in that the plaquette term in the string-net model is ill-defined if the fusion rules are not satisfied at adjacent vertices, and is typically chosen to be zero, whereas for the Quantum Double model the plaquette terms are always well-defined. However this does not affect the ground state because the fusion rules are satisfied in the ground state. The same thing is true for the mapping from the higher lattice gauge theory model to the symmetry enriched string-net models.

		In addition to the plaquette terms from the two models acting the same on the edges, the plaquette terms also have equivalent action on the plaquettes. The plaquette transform $B_p^R$ multiplies the plaquette term by $\mu^R$, while the $\tilde{U}_p^{h_R}$ term from the symmetry enriched string-net model multiplies the plaquette label by $h_R$. As we established earlier, the group element $h_R$ associated to object $R$ is $\mu^R$, so these are the same. When we include the action on both the edges and the plaquettes, we see that the operator $B_p^R$ from the higher lattice gauge theory model is equivalent to $\tilde{B}_p^R \tilde{U}_p^{h_R}$ from the symmetry enriched string-net model. Therefore, the plaquette term of the higher lattice gauge theory model maps onto the plaquette term of the symmetry enriched string-net model when we take $a_R=|R|/|G|$. While these coefficients $a_R$ can be chosen freely to define different string-net models \cite{Levin2005}, they are usually taken to be $a_R = d_R/D^2$, where $d_R$ is the so-called quantum dimension of $R$ and $D^2$ is the sum of the squared quantum dimensions of all of the objects: $D^2 = \sum_{s \in \mathcal{D}} d_s^2$. This choice of coefficients ensures that the string-net model is associated to a topological phase with a smooth continuum limit and that the energy terms are projectors \cite{Levin2005}. Indeed, these coefficients were used in Ref. \cite{Heinrich2016} for the symmetry enriched string-net construction. We should therefore check that these coefficients also agree between the symmetry enriched string-net model and the higher lattice gauge theory model. Identifying the quantum dimension of irrep $R$ as $|R|$ \cite{Buerschaper2009}, the total quantum dimension is given by $D^2 = \sum_{\text{irrep }R} |R|^2 = |G|$. This means that $d_R/D^2 = |R|/|G|$. Therefore, the coefficients $a_R$ for which the two plaquette terms match are the standard coefficients used for string-net models. Combining this with our results for the other energy terms, we see that all of the energy terms match between the two models.

		Now let us consider the plaquette term in our simple example model. In this case the plaquette term involves a sum over the four irreps of $G=\mathbb{Z}_4$. Because these irreps are all one-dimensional, the coefficient $a_R=|R|/|G|$ for these irreps is equal to $1/4$. Writing the plaquette term with the symmetry enriched string-net notation, we have
		$$B_p = \frac{1}{4} \sum_{\text{irreps }R \text{ of }G} \tilde{B}_p^R \tilde{U}_p^{h_R},$$
		where $h_R$ is the symmetry element associated to the irrep (or object, in the string-net language) $R$. Knowing that the irreps $\pm 1_R$ are associated to the identity element of the symmetry group, which we denote by $1_H$, while the irreps $\pm i_R$ are associated to the non-trivial element, which we denote by $-1_H$, the plaquette term can be written as
		$$B_p = \frac{1}{4} (\tilde{B}_p^{1_R} + \tilde{B}_p^{-1_R}) \tilde{U}_p^{1_H} + \frac{1}{4} (\tilde{B}_p^{i_R} + \tilde{B}_p^{-i_R}) \tilde{U}_p^{-1_H}.$$
		
		Noting that the action of $\tilde{U}_p^{h_R}$ is to multiply the plaquette label by $h_R$, we can see that $\tilde{U}_p^{1_H}$ is just the identity operator. Furthermore $\tilde{B}_p^{1_R}$ just fuses the vacuum string around the plaquette, so this is also the identity operator. This allows us to simplify our expression for the plaquette term to
		$$B_p = \frac{1}{4} (1 + \tilde{B}_p^{-1_R}) + \frac{1}{4} (\tilde{B}_p^{i_R} + \tilde{B}_p^{-i_R}) \tilde{U}_p^{-1_H}.$$
		
		By using the understanding of the ground state that we gained by examining the edge terms, we can see why the plaquette term requires the additional action of $\tilde{U}_p^{h_R}$ on the plaquette label itself when compared to the ordinary string-net model. The edge term enforces that a string of label $1_R$ or $-1_R$ must separate plaquettes with the same symmetry label, while a string of label $\pm i_R$ must separate plaquettes with different labels. If we were to fuse a string of label $\pm i_R$ around a plaquette without changing the plaquette label itself, this would violate that rule. For example, if the plaquette was originally surrounded by a single string of label $1_R$, its label must agree with all of the surrounding plaquettes' labels. But then when we fuse a string of label $i_R$ around the plaquette, the edge terms then require that the plaquette label must disagree with the surrounding plaquettes. In order to satisfy this rule, when we fuse the string around the plaquette we must also change the label of the plaquette.

		We have therefore shown that there is an equivalence between a subset of the higher lattice gauge theory models and a subset of the symmetry enriched string-net models. Specifically, we considered a higher lattice gauge theory model described by the crossed module $(G,E,\partial, \rhd)$, where $\rhd$ is trivial ($g \rhd e =e$ for all $g \in G$ and $e \in E$) and ker($\partial$) is just the identity element $1_E$. We saw that such a model is equivalent to the symmetry enriched string-net model labelled by a group $H=$ Rep($E$) and an $H$-graded fusion category Rep($G$). This means that the topological phase described by the two models is the same, allowing us to identify the SET phase realised by the higher lattice gauge theory model as described by a braided Rep($E$)-crossed extension of $\mathcal{Z(\text{Rep}(G/\partial(E)))}$. While we assumed that $E$ and $\partial(E)$ were isomorphic for this mapping, the fact that a non-trivial $\ker(\partial)$ introduces spontaneous symmetry breaking described by irreps of the kernel (as we discussed in Section \ref{Section_2D_irrep_basis}) suggests that a similar result holds for non-trivial $\ker(\partial)$, but with $H=\text{Rep}(E/ \text{ker}(\partial))$ as the unbroken part of the symmetry instead of the full symmetry (note that the properties of the electric and magnetic excitations are unaffected by changing $E$ while keeping $\partial(E)$ fixed).

		We can also use this correspondence between Hamiltonians to identify features in the two models. For example, in the higher lattice gauge theory model we know that we obtain loop-like excitations labelled by irreps of $E$. However, we also know that these loop-like excitations are equivalent to confined electric excitations (at least, in the absence of other excitations) if the irrep labelling them is not sensitive to the kernel of $E$. Because we took the kernel of $\partial$ to be trivial when we mapped to the symmetry enriched string-net model, this means that all of these loop-like excitations are of this type. We therefore expect the presence of confined excitations in the symmetry enriched string-net models, but not uncondensed loop excitations (which correspond to domain walls between different ground states). In the higher lattice gauge theory model, the confined electric excitations are labelled by irreps of $G$ which have non-trivial restriction to $\partial(E)$. In the language of the symmetry enriched string-net models, the confined electric excitations should be those whose label is associated to a non-trivial group element of $H$, meaning those labelled by objects that are not in the category $\mathcal{C}$ of which $\mathcal{D}$ is a group extension. We expect this to hold outside of the small subset of symmetry enriched string-nets that we can map to directly. This is because in Ref. \cite{Heinrich2016} it is stated that the anyonic excitations are the same as those produced by an ordinary string-net model with $\mathcal{C}$ as an input category, and so any extra excitations (such as the extra electric excitations we find in the higher lattice gauge theory model) must be either confined or condensed.

		As an example of the confined and condensed excitations, consider the simple model $(\mathbb{Z}_4,\mathbb{Z}_2,\partial, \rhd \rightarrow \text{id})$ we have been looking at throughout this section. From our general results for higher lattice gauge theory models (see Section \ref{Section_RO_2D_Tri_Trivial}), we know that there are four electric ribbon operators labelled by the different irreps of $G=\mathbb{Z}_4$. However, the excitations produced by the ribbon operators labelled by irreps with non-trivial restriction to the image of $\partial$, i.e., to the $\mathbb{Z}_2$ subgroup of $\mathbb{Z}_4$, are confined. Therefore, the excitations labelled by the irreps $\pm i_R$ are confined. These are the same irreps that are associated to the non-trivial symmetry group element (the non-trivial irrep of $E$), as we expect from the general discussion above. This just leaves the ribbon operators labelled by the irreps $1_R$ and $-1_R$, of which the one labelled by $1_R$ is the trivial operator. We also know that there are four purely magnetic ribbon operators, labelled by the group elements of $G=\mathbb{Z}_4$, but the operators labelled by elements in $\partial(E)$ correspond to condensed excitations. The ribbon operator labelled by the identity element is always trivial, but we also see that the excitation labelled by $-1_G$ is condensed (which also means that the two remaining magnetic excitations belong to the same topological sector). These confined electric and condensed magnetic excitations (and excitations built from fusion with them) would not be counted towards the excitations described in Ref. \cite{Heinrich2016}, because the authors of that work were only interested in distinct types of (unconfined) anyons. This leaves us with only one unconfined electric excitation and one distinct magnetic excitation, along with their fusion product and the trivial charge. These are the same excitations we would expect for the toric code. This matches the prediction from Ref. \cite{Heinrich2016}, because the category $\mathcal{C}$ of objects with trivial symmetry group label contains the irreps $1_R$ and $-1_R$, and so the excitations should be the same as those from the $C=$Rep($\mathbb{Z}_2$) string-net model (which is the toric code).

		Another important feature in both models is the presence of a symmetry. In Section \ref{Section_2D_irrep_basis}, we saw that the higher lattice gauge theory model (with $\rhd$ trivial) has a symmetry under multiplying each plaquette label (in the irrep basis) by an irrep $\nu$ of $E$. This is equivalent to the symmetry of the symmetry enriched string-net models of Ref. \cite{Heinrich2016} under multiplication of each plaquette label by a group element $h$ of $H$. This symmetry exists for all of the symmetry enriched string-net models, not just the ones we can map to from the higher lattice gauge theory model. In the symmetry enriched string-net model, this symmetry can permute the excitation types (for instance, an example is given in Ref. \cite{Heinrich2016} where the symmetry swaps between the electric and magnetic excitations of the toric code). In the case of the higher lattice gauge theory models however, the ribbon operators that produce the electric and magnetic excitations do not act on the plaquette labels and so these excitations are not affected by the symmetry.

		\section{Topological charge from closed ribbon operators when $\rhd$ is trivial}
		
		\label{Section_2D_topological_Charge}

 Now that we have looked at the excitations and their properties, it will be useful to discuss the conserved topological charge carried by the excitations in more detail, particularly to illustrate the ideas of confinement and condensation. The topological charge contained in a region of the lattice cannot be changed, except by moving some charge into or out of the region by using an operator that connects the interior and exterior of the region. That is, the charge cannot be changed by any local operator, where by local we here mean an operator entirely contained within the region. This means that any operator that changes the topological charge within a region will move charge through the boundary of that region and so can be detected by an operator on the boundary of the region. Indeed, the topological charge within a region can be measured by an operator on the boundary of the region. Because the topological charge is conserved in this sense, rather than being conserved by the action of the Hamiltonian, the charges are really properties of the Hilbert space. However, the Hamiltonian determines which particular set of charges are relevant for the particular model. The ground state acts as the topological vacuum for that set of charges, meaning that the ground state has trivial topological charge in all regions. This enforces additional properties for the topological charge realised by the topological model. Firstly, smoothly deforming the boundary within which we measure the charge should not change the value of charge measured, as long as the boundary does not cross any excitations in doing so (thereby including or excluding additional excitations in or from the region of interest). This is because deforming the surface will just include more or less of the vacuum, which has trivial topological charge. Secondly, the Hamiltonian does not produce or move excitations, so we require that the topological charge within a surface be conserved under the action of the Hamiltonian. In addition to these properties, we require some way to combine the charge in two regions to get a total topological charge, in order to reproduce the fusion rules of anyon theories.

		Given these properties, we look for measurement operators that would detect such a charge. Following Ref. \cite{Bombin2008}, we wish to construct these operators from the ribbon operators of our model. The ribbon operators (apart from the confined ones) have many of the properties that we require. They can freely be deformed, provided that we leave the end-points fixed. Furthermore, the ribbon operators form a complete set of operators, once we also include the $E$-valued membrane operators and single plaquette multiplication operators. To see this, note that we can apply each operator on a single edge or plaquette by choosing the associated ribbon or membrane appropriately. The electric ribbon operator then measures the value of an edge, while the magnetic operator allows us to multiply an edge by any element of $G$. Combining these therefore allows us to freely measure and control the state of an individual edge. Similarly the $E$-valued membrane operator allows us to measure the state of an individual plaquette and the single plaquette multiplication operator allows us to multiply the plaquette label by any element of $E$. To produce any operator, we can use the measurement operators on every edge and plaquette to fix which state in our Hilbert space we act on and then use the other operators to control the action on that state. Then summing over each possible state in our Hilbert space with a different action for each state will give us a general operator. We can therefore create any operator as a combination of ribbon and membrane operators (along with the single plaquette multiplication operators). However, the ribbon operators will generally produce excitations, with only closed ribbon operators potentially producing no excitations. Therefore, we must restrict to operators that can be produced by combining closed ribbon operators (and single plaquette multiplication operators or $E$-valued membrane operators, although these correspond to the symmetry content of the theory and we will only use these operators in a limited way, as we describe later). Some of the closed ribbon operators do still produce excitations where they join onto themselves (i.e., at the start of the ribbon, which is also the end), and so we must further restrict to closed ribbon operators that do not create such excitations. This is required to make operators that simply measure the topological charge, without moving said charge. These closed ribbon operators will then form an algebra, from which we can produce our charge measurement operators.

		In order to produce the charge measurement operators, we first choose a region whose topological charge we want to measure. We then consider applying a general electric and magnetic closed ribbon operator around the boundary of that surface. We choose the labels of these ribbon operators, and take linear combinations of them, to ensure that this closed ribbon operator does not produce any excitations. As we will see later, we may also need to apply a single plaquette multiplication operator to ensure that the magnetic ribbon operator does not produce an excitation at the start (which is also the end) of the ribbon. Apart from this, we do not use any single plaquette multiplication operators or $E$-valued membrane operators. This is because these operators either correspond to the symmetry content of the model, or are related to condensed versions of the ribbon operators. The single plaquette measurement operators in the kernel of $\partial$ distinguish between the symmetry-related ground states, while the $E$-valued membrane operators locally perform the symmetry transformation, as we discussed in Section \ref{Section_2D_irrep_basis}. Therefore, we regard two valid measurement operators that differ only by these operators as equivalent when considering the topological content of the model.

		For the higher lattice gauge theory models with $\rhd$ trivial, we can construct the charge measurement operators explicitly. These charge measurement operators will be useful for considering condensation and confinement. During condensation, a charge from the original uncondensed model joins the ground state of the new model (after condensation). Therefore, the expectation value of the charge measurement operator of the old model should be non-zero when applied to the ground-state of the new model \cite{Bombin2008}. Because of this fact, it is useful to first construct the charge measurement operators in the uncondensed higher lattice gauge theory models. We explained in Section \ref{Section_2D_Condensation_Confinement} that, given a model described by the crossed module $(G, E, \partial, \rhd \rightarrow \text{id})$, we can construct another crossed module $(G, E, \partial \rightarrow 1_G, \rhd \rightarrow \text{id})$, with $\partial(e)=1_G \: \: \forall e \in E$. Because the groups $G$ and $E$ are the same as in the original model, the Hilbert spaces of the models based on the two crossed modules are the same. However, in the new model the image of $\partial$ is $\partial(E)=\set{1_G}$ and the kernel is ker$(\partial)=E$. This indicates (from our analysis on the ribbons in Section \ref{Section_RO_2D_Tri_Trivial}) that there should be no condensation or confinement. In this unconfined model $E$ and $G$ completely decouple in the energy terms. The vertex term and plaquette terms become the equivalent terms from Kitaev's Quantum Double model \cite{Kitaev2003}. In addition, the edge transform now only affects the surface labels. We can therefore consider this model as a copy of Kitaev's Quantum Double model describing the edge elements combined with an independent model describing the surface elements. The surface elements are fluctuated by the edge transforms. Following the arguments in Section \ref{Section_Z4_Z4_irrep_basis}, but taking $\partial \rightarrow 1_G$, the edge terms project to states where the plaquettes are all labelled by the same irrep of $E$ in the irrep basis, with the choice of that irrep describing the various ground states on the sphere.

		Because of this decoupling of the surface and edge elements, the $G$-valued excitations (the electric and magnetic excitations) are exactly the same as in Kitaev's Quantum Double model \cite{Kitaev2003}, with no confinement. In particular, the ribbon algebra is the same and the ribbons only fail to commute with vertex and plaquette terms at the end sites of the ribbon. Then, following the methods of Bombin and Martin-Delgado \cite{Bombin2008}, we can find the measurement operators that detect the charges labelled by objects in $G$. As we show in Section \ref{Section_2D_Topological_Charge_Projectors} in the Supplemental Material, the relevant closed ribbon projectors are labelled by two quantities. The first label is a conjugacy class $C$ of $G$. Given such a conjugacy class, we choose a representative $r_C$. We can then construct the centralizer $N_C$ of that representative, which is the subgroup formed by elements that commute with $r_C$. The centralizers of different representatives of the conjugacy class are isomorphic, so the choice of representative is not significant. Then we can construct irreps of this centralizer. The irrep $R$ of the centralizer is the second label for our projector. The measurement operators on a closed ribbon $\sigma$ are then given by 
		\begin{equation}
		K_{\sigma}^{RC} := \frac{|R|}{|N_C|} \sum_{D \in (N_C)_{cj}} \overline{\chi}_R(D) K_{\sigma}^{DC},
		\end{equation}
		where
		\begin{equation}
		K_\sigma^{DC} := \sum_{q \in Q_C} \sum_{d \in D} F_{\sigma}^{qdq^{-1},qr_Cq^{-1}},
		\end{equation}
		just like the equivalent operators for the Quantum Double model found in Ref. \cite{Bombin2008}. Here $(N_C)_{cj}$ is the set of conjugacy classes of the centralizer $N_C$, $|N_C|$ is the size of $N_C$, $\chi_R$ is the character of irrep $R$, and $Q_C$ is a set of representatives of the quotient group $G/N_C$, so that for each $c_i \in C$ there is a unique $q_i \in Q_C$ such that $c_i=q_i r_Cq_i^{-1} $. $F^{h,g}_{\sigma}$ is the combined electric and magnetic ribbon that we defined at the end of Section \ref{Section_2D_Magnetic} (where the first argument in the superscript corresponds to the magnetic part and the second argument corresponds to the electric part). Then, substituting the expression for $K_\sigma^{DC}$ into the one for $K_{\sigma}^{RC}$, we see that
		\begin{equation}
		K_{\sigma}^{RC}= \frac{|R|}{|N_C|} \sum_{D \in (N_C)_{cj}} \overline{\chi}_R(D)\sum_{q \in Q_C} \sum_{d \in D} F_{\sigma}^{qdq^{-1},qr_Cq^{-1}}.
		\end{equation}
		
		The set of these operators over each conjugacy class and irrep of the centralizer of that conjugacy class (defined as the centralizer of the fixed representative) form a set of orthogonal projectors that make up a resolution of the identity, as described in Ref. \cite{Bombin2008}.

		From the fact that both the braiding (as described in Section \ref{Section_2D_Braiding_Tri_Trivial}) and the topological charge projectors match those from Kitaev's Quantum Double model in the uncondensed case when $\rhd$ is trivial, we infer that the phase described by the uncondensed model is the same as the Quantum Double model with group $G$, albeit with an additional symmetry described by the group $E$. 
		
		\subsection{Condensation}
		We have now constructed the topological charges for our uncondensed model. We can then use these charge measurement operators to see which of these charges condense in our original model (where $\partial$ need not be trivial, though we still require $\rhd$ to be trivial). Because the ground state is the topological vacuum of the model, any charges which are present in the ground state of the condensed model must have (at least partially) condensed. Therefore, we must calculate the expectation value of the charge measurement operators to see which charges have condensed. That is, we must work out
		$$ \braket{K_{\sigma}^{RC}} := \braket{GS|K_\sigma^{RC}|GS},$$
		where $K_\sigma^{RC}$ is a measurement operator in the uncondensed model and $\ket{GS}$ is a ground state of the condensed model. We have
		\begin{align*}
		&\braket{GS|K_\sigma^{RC}|GS}\\
		&= \bra{GS} \frac{|R|}{|N_C|} \sum_{D \in (N_C)_{cj}} \hspace{-0.4cm} \overline{\chi}_R(D)\sum_{q \in Q_C} \sum_{d \in D} F_{\sigma}^{qdq^{-1},qr_Cq^{-1}} \ket{GS}\\
		&= \bra{GS} \frac{|R|}{|N_C|} \sum_{D \in (N_C)_{cj}} \hspace{-0.4cm} \overline{\chi}_R(D)\sum_{q \in Q_C} \sum_{d \in D} C^{qdq^{-1}}(\sigma)\\
		 & \hspace{1cm} \delta(g(\sigma),qr_Cq^{-1})\ket{GS}. \intertext{Then we use the fact that $C^h(\sigma)\ket{GS}=\ket{GS}$ for a closed, contractible ribbon, as established in Section \ref{Section_Topological_Magnetic_Ribbons} of the Supplemental Material, to write}\\
		&\braket{GS|K_\sigma^{RC}|GS}\\
		&= \bra{GS} \frac{|R|}{|N_C|} \sum_{D \in (N_C)_{cj}} \hspace{-0.4cm} \overline{\chi}_R(D)\sum_{q \in Q_C} \sum_{d \in D} \delta(g(\sigma),qr_Cq^{-1})\\
		&\hspace{1cm}\ket{GS}\\
		&=\frac{|R|}{|N_C|} \sum_{D \in (N_C)_{cj}} \hspace{-0.4cm} \overline{\chi}_R(D)\sum_{q \in Q_C} \sum_{d \in D} \bra{GS} \delta(g(\sigma),qr_Cq^{-1})\\
		&\hspace{1cm}\ket{GS}.
		\end{align*}
		
		Because $\sigma$ is a contractible closed path, $g(\sigma)$ must be in $\partial(E)$ in the ground state due to fake-flatness. Indeed, because the edge terms fluctuate the path element by all of the elements in $\partial(E)$, $g(\sigma)$ is equally likely to be any element of this subgroup. Therefore, 
		$$\bra{GS} \delta(g(\sigma),qr_Cq^{-1})\ket{GS} = \frac{1}{|\partial(E)|}\delta(r_C \in \partial(E)).$$ 
		The expectation value of the measurement operator is then
		\begin{align*}
		&\braket{GS|K_\sigma^{RC}|GS}\\
		&=\frac{|R|}{|N_C|} \sum_{D \in (N_C)_{cj}} \hspace{-0.4cm} \overline{\chi}_R(D)\sum_{q \in Q_C} \sum_{d \in D} \frac{1}{|\partial(E)|}\delta(r_C \in \partial(E)).
		\end{align*}
	
	When $r_C$ is in $\partial(E)$, all elements of $G$ commute with it (because $\partial(E)$ is in the centre of $G$ when $\rhd$ is trivial) and so $N_C$ is just $G$. This also means that $Q_C$ is trivial and just includes the identity element. Therefore
		\begin{align*}
			&\braket{GS|K_\sigma^{RC}|GS}\\
		&=\frac{|R|}{|G|} \sum_{d \in G} \overline{\chi}_R(d)\frac{1}{|\partial(E)|}\delta(r_C \in \partial(E))\\
		&=\delta_{R,1_R} \frac{1}{|\partial(E)|}\delta(r_C \in \partial(E)),
		\end{align*}
		where in the last line we used the orthogonality of characters. The condensed charges are those for which this quantity is non-zero. Therefore, we see that the condensed charges are those where $R$ is trivial and $C$ is a subset of the image of $\partial$. If $C$ is in $\partial(E)$, then $C$ only contains a single element $\partial(e)$, because $\partial(E)$ is in the centre of $G$. This means that the charge measurement operators for the condensed charges can be written in a simpler way as $K^{1_{\text{Rep}}, \set{\partial(e)}}_{\sigma}=\frac{1}{|G|} \sum_{d \in G} F_{\sigma}^{d,\partial(e)}.$ Consider how this measurement operator acts when it encloses an excitation (produced by some open ribbon operator). It checks that the path of the measurement operator (the closed ribbon operator) has a path element of $\partial(e)$. This isolates the case where the excitation has a magnetic label in $\partial(E)$. The closed ribbon operator also acts with an equal superposition of every magnetic operator. This checks that the electric part of the excitation is trivial, so that the excitation is pure magnetic. Then the condensed magnetic excitations, those measured by the closed ribbon operators with non-zero expectation in the ground state, are the pure magnetic ones with label in the image of $\partial$. This agrees with our discussion in Section \ref{Section_Single_Plaquette_1}, where we came to the same conclusion based on the ribbon operators that produce the corresponding excitations.

		\subsection{Confinement}
		\label{Section_2D_Topological_Charge_Confinement}
		In the previous section, we put the measurement operators of the uncondensed model into the condensed model, to see which of the charges were condensed in the latter model. In this section, we look at the charges of the condensed model and see which of those are confined. To construct the topological charge measurement operators for the condensed model, we first throw out the ribbon operators that fail to commute with the edge terms. This is because our measurement operators must commute with all of the energy terms. We find that the remaining measurement operators are labelled by two objects, $C$ and $R$. When we discussed the charge measurement operators for the uncondensed model, the appropriate label $C$ was a conjugacy class of $G$. However for a general model, the appropriate label $C$ indicates a union of cosets, the ``conjugacy class of cosets" that was discussed earlier in Section \ref{Section_Single_Plaquette_1}. Specifically, $C$ labels a union of cosets $g\partial(E)$ for each $g$ within a conjugacy class of $G$. We can also define the classes $C$ with the following equivalence relation: $g \sim x\partial(e) g x^{-1} \ \forall x \in G, \: e \in E$. We choose for each such class a representative element $r_C \in G$. Then the second label of the measurement operator, $R$, is an irrep of the group $N'_C$, where $N'_C$ is the subgroup of $G$ made up of elements $g$ such that $gr_Cg^{-1}r_C^{-1} \in \partial(E)$. That is, $N_C'$ is the subgroup of elements that commute with $r_C$ up to an element of $\partial(E)$.

		The measurement operator also depends on the symmetry state of the plaquette at the start and end of the ribbon. In Section \ref{Section_2D_irrep_basis}, we discussed how there is a spontaneously broken symmetry, and the plaquettes in the different states correspond to different irreps of the kernel of $\partial$. If the plaquette is labelled by the trivial irrep then the projectors to definite topological charge are defined by
		\begin{equation}
		K_{\sigma}^{RC} := \frac{|R|}{|N_C'|} \sum_{D \in (N_C')_{\text{cj}}} \overline{\chi}_R (D) K_{\sigma}^{DC},
		\end{equation}
		where $(N_C')_{\text{cj}}$ is the set of conjugacy classes of $N_C'$ (so that each $D$ is a conjugacy class), $|R|$ is the dimension of irrep $R$, and $K_{\sigma}^{DC}$ is given by
		\begin{align}
		K_{\sigma}^{DC} :=& \sum_{q \in Q_C} \sum_{d \in D} \sum_{h \in \partial(E)} F_{\sigma}^{qdq^{-1},qr_Cq^{-1}h} M^{f(r_C,d)^{-1}}(p) \notag\\
		&\frac{1}{|\ker(\partial)|} \sum_{e_k \in \ker (\partial)}M^{e_k}(p).
		\end{align}
		Here $Q_C$ is a set of representatives of the quotient group $G/N_C'$, and conjugating $r_C$ by each element $q \in Q_C$ generates all elements of the equivalence class $C$ up to factors in $\partial(E)$. $M^{f(r_C,d)^{-1}}(p)$ is a single plaquette multiplication operator that multiplies the first plaquette on the ribbon $p$ (the start and end plaquette for the closed ribbon) by an element $f(r_C,d)^{-1}$ such that $\partial(f(r_C,d)) = r_C d r_C^{-1}d^{-1}$. This means that the topological charge projector labelled by union $C$ of cosets and irrep $R$ of $N'_C$ is given by:
			\begin{center}
			\noindent\fcolorbox{black}{myblue1}{%
				\parbox{0.9\linewidth}{%
					\centering \textbf{Topological charge projectors:}
		\begin{align}
		&K_{\sigma}^{RC} = \notag\\
		& \frac{|R|}{|N_C'|} \sum_{D \in (N_C')_{\text{cj}}} \overline{\chi}_R (D) \sum_{q \in Q_C} \sum_{d \in D} \sum_{h \in \partial(E)} F_{\sigma}^{qdq^{-1},qr_Cq^{-1}h} \notag \\
		& \hspace{0.2cm} M^{f(r_C,d)^{-1}}(p) \frac{1}{|\ker(\partial)|} \sum_{e_k \in \ker (\partial)}M^{e_k}(p).
		\end{align}}}
	\end{center}
		In Section \ref{Section_2D_Topological_Charge_Projectors} of the Supplemental Material, we prove that these are indeed orthogonal projectors. In addition, we construct the projectors corresponding to other symmetry states of the plaquette $p$ (by applying the symmetry operators $U^{\nu}$ from Section \ref{Section_2D_irrep_basis}).

\begin{figure}[h]
			\begin{center}
			\includegraphics{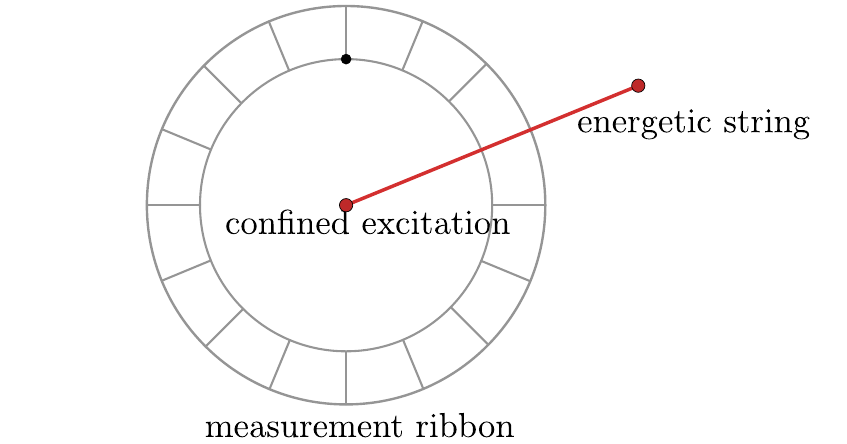}
				\caption{If our measurement operator encloses a confined excitation (or multiple such excitations), the energetic string dragged by this excitation will pass through our measurement ribbon and cause an edge to be excited.}
				\label{measure_confined}
			\end{center}
		\end{figure}

		Having found the projectors to definite topological charge, we want to consider which charges are confined. When we measure a confined charge, there will always be a violation of one or more energy terms on the measurement ribbon, due to the presence of the confining string. Given that we expect confinement to be due to the edge terms (from our discussion of the ribbon operators), we expect that at least one of the outward pointing edges on our ribbon should be violated in the measured state. An example situation is shown in Figure \ref{measure_confined}.
		
		We can therefore check whether a charge is confined by applying a product of all of the edge terms cut by the dual path of the ribbon $\sigma$. For a confined charge, at least one of the edges must be excited and so the state must be an eigenstate where one of the edge terms has eigenvalue zero. This means that the product of edge terms will give zero. That is, whenever the state being measured has definite topological charge labelled by $R$ and $C$, so that $K_{\sigma}^{RC} \ket{\psi}= \ket{\psi}$, if $R$ and $C$ correspond to a confined charge we must have $\prod_{i \in \sigma} \mathcal{A}_i \ket {\psi}=0$. This means that the product of operators $\prod_{i \in \sigma} \mathcal{A}_i K_{\sigma}^{RC}$ must give zero because $K_{\sigma}^{RC}$ projects to states for which the product of edge terms gives zero (and the edge terms commute with the measurement operator). On the other hand, if a charge is not confined, then a state $\ket{\psi}$ with that charge enclosed by the ribbon (and no other excitations near the ribbon) will satisfy
		$$\prod_{i \in \sigma} \mathcal{A}_i \ket{\psi}=\ket{\psi},$$
		because the edges will not be excited. Instead of considering a product of the edge terms, we can consider a product of edge transforms, which also leaves the state invariant when the edges are unexcited (because any edge transforms on unexcited edges leave the state invariant). We consider the product
		$$\prod_{i \in \sigma} \mathcal{A}_i^e,$$
	where we assume that all of the edges $i$ point outwards (we would simply replace $e$ with $e^{-1}$ for any edge that pointed inwards). Then the state $\ket{\psi}$ corresponding to an unconfined charge within $\sigma$ must satisfy
		$$\prod_{i \in \sigma} \mathcal{A}_i^e \ket{\psi}=\ket{\psi}.$$
		
		This implies that for an arbitrary state $\ket{\psi}$ we must have
		$$\prod_{i \in \sigma} \mathcal{A}_i^e K^{RC}_{\sigma} \ket{\psi} =K_{\sigma}^{RC} \ket{\psi},$$
		where $R$ and $C$ label an unconfined charge, because $K^{RC}_{\sigma}$ projects to states which correspond to that unconfined charge.

		We now wish to use this fact to identify the confined charges. As we prove in Section \ref{Section_Condensation_Magnetic_2D} in the Supplemental Material, any magnetic ribbon with label in $\partial(E)$ can be written as a product of edge transforms, multiplied by two operators at the ends of the ribbon that change the labels of the plaquettes at each end. In that section, we consider open ribbons. However for closed ribbons, the two end plaquettes are at the same place and the two single plaquette multiplication operators will cancel out. This means that any closed magnetic ribbon operator with label in $\partial(E)$ can be expressed simply as a product of edge transforms. The converse is also true, and a product of edge transforms along the dual path of a ribbon can be written as a condensed closed magnetic ribbon operator. That is, given a closed ribbon $\sigma$ where all of the edges cut by the dual path of the ribbon point outwards (away from the direct path of the ribbon), we have
		\begin{equation}
		C^{\partial(e)}(\sigma)=\prod_{i \in \sigma} \mathcal{A}_i^e.
		\label{Equation_closed_magnetic_edge_transforms}
		\end{equation}
		
		If some of the edges instead pointed inwards (towards the direct path), we would simply have to change the label $e$ to $e^{-1}$ for those edges in the right hand side of this equation. We can use this result to write the condition for the unconfined charges as
		$$C^{\partial(e)}(\sigma) K^{RC}_{\sigma} \ket{\psi} =K_{\sigma}^{RC} \ket{\psi}.$$

		Writing the condition in this way is useful because we can combine $C^{\partial(e)}(\sigma)$ with $K^{RC}_{\sigma}$, using our algebra of operators, to obtain a condition on $K^{RC}_{\sigma}$ itself. We have
		\begin{align*}
		&C^{\partial(e)}(\sigma)K^{RC}_{\sigma}\\
		&=\frac{|R|}{|N_C'|} \sum_{D \in (N_C')_{\text{cj}}} \overline{\chi}_R (D) \sum_{q \in Q_C} \sum_{d \in D} \sum_{h \in \partial(E)} C^{\partial(e) }(\sigma)\\
		& \hspace{1cm} F_{\sigma}^{qdq^{-1},qr_Cq^{-1}h} M^{f(r_C,d)^{-1}}(p)\\
		& \hspace{1cm} \frac{1}{|\ker(\partial)|} \sum_{e_k \in \ker (\partial)}M^{e_k}(p)\\
		&= \frac{|R|}{|N_C'|} \sum_{d \in N_C'} \overline{\chi}_R (d) \sum_{q \in Q_C} \sum_{h \in \partial(E)}F_{\sigma}^{\partial(e)qdq^{-1},qr_Cq^{-1}h} \\
		& \hspace{1cm}M^{f(r_C,d)^{-1}}(p) \frac{1}{|\ker(\partial)|} \sum_{e_k \in \ker (\partial)}M^{e_k}(p),
		\end{align*}
		where for convenience we replaced the sum over conjugacy classes $D$ of $N_C'$ and elements $d$ in that conjugacy class with a single sum over elements of $N'_C$. Because $\partial(E)$ is in the centre of $G$ (due to $\rhd$ being trivial), we can commute $\partial(e)$ next to $d$ in the label of $F_{\sigma}^{\partial(e)qdq^{-1},qr_Cq^{-1}h}$. Then we can replace the dummy variable $d$ with $d' = \partial(e)d$ (because $\partial(e)$ is in the centre of $G$, $d'$ will also be in $N'_C$), to obtain $F_{\sigma}^{qd'q^{-1},qr_Cq^{-1}h}$. The fact that $\partial(e)$ commutes with all elements of $G$ also means that $f(r_C,d)$ can be chosen to be the same as $f(r_C,d')$ (recall that $f(r_C,d)$ satisfies $\partial(f(r_C,d)) = r_C d r_C^{-1}d^{-1}$, which is the same as $ r_C d' r_C^{-1}d^{\prime -1}$). This gives us
		\begin{align*}
		&C^{\partial(e)}K^{RC}_{\sigma}\\
		&= \frac{|R|}{|N_C'|} \sum_{d'=\partial(e)d \in N_C'} \overline{\chi}_R (\partial(e)^{-1}d') \sum_{q \in Q_C} \sum_{h \in \partial(E)}\\ 
		& \hspace{1cm} F_{\sigma}^{qd'q^{-1},qr_Cq^{-1}h} M^{f(r_C,d')^{-1}}(p)\\
		& \hspace{1cm} \frac{1}{|\ker(\partial)|} \sum_{e_k \in \ker (\partial)}M^{e_k}(p).
		\end{align*}
	
	We can split the character $\overline{\chi}_R (\partial(e)^{-1}d')$ into the contributions from $\partial(e)$ and $d'$:
	\begin{align*}
	\overline{\chi}_R (\partial(e)^{-1}d') &= \sum_{c=1}^{|R|} [D^{R}(\partial(e)^{-1}d')]_{cc}^*\\
	&=\sum_{c=1}^{|R|} \sum_{a=1}^{|R|} [D^{R}(\partial(e)^{-1})]_{ca}^* [D^{R}(d')]_{ac}^*.
	\end{align*}

We can then use the fact that $\partial(E)$ is in the centre of $G$ (which also means that it is always in $N_C'$) to simplify this expression. From Schur's Lemma, 
$$ [D^{R}(\partial(e)^{-1})]_{ac}^* = \delta_{ac} [D^{R}(\partial(e)^{-1})]_{11}^*$$
 (because the matrix must be a scalar multiple of the identity). We can then define an irrep of $\partial(E)$, $R_{\partial}^{\text{irr.}}$, by
$$R_{\partial}^{\text{irr.}}(\partial(e))= [D^{R}(\partial(e))]_{11},$$
 as we have before in Section \ref{Section_2D_electric} when discussing the electric excitations. We then have
		\begin{align*}
		\overline{\chi}_R (\partial(e)^{-1}d') &= \sum_{c=1}^{|R|} \sum_{a=1}^{|R|} R_{\partial}^{\text{irr.}}(\partial(e^{-1}))^* \delta_{ac} [D^{R}(d')]_{ac}^*\\
		&= R_{\partial}^{\text{irr.}}(\partial(e^{-1}))^* \sum_{c=1}^{|R|} [D^{R}(d')]_{cc}^*\\
		&= R_{\partial}^{\text{irr.}}(\partial(e)) \overline{\chi}_R (d').
		\end{align*}
	
	This then implies that
		\begin{align}
			C^{\partial(e)}K^{RC}_{\sigma}
			&= \frac{|R|}{|N_C'|} \sum_{d'=\partial(e)d \in N_C'} \overline{\chi}_R (\partial(e)^{-1}d') \sum_{q \in Q_C} \sum_{h \in \partial(E)} \notag \\ 
			& \hspace{1cm} F_{\sigma}^{qd'q^{-1},qr_Cq^{-1}h} M^{f(r_C,d')^{-1}}(p) \notag\\
			& \hspace{1cm} \frac{1}{|\ker(\partial)|} \sum_{e_k \in \ker (\partial)}M^{e_k}(p) \notag\\
		&= R_{\partial}^{\text{irr.}} (\partial(e)) \frac{|R|}{|N_C'|} \sum_{d' \in N_C'} \overline{\chi}_R (d') \sum_{q \in Q_C} \sum_{h \in \partial(E)} \notag \\
		& \hspace{1cm} F_{\sigma}^{qd'q^{-1},qr_Cq^{-1}h} M^{f(r_C,d')^{-1}}(p) \notag \\ 
		& \hspace{1cm} \frac{1}{|\ker(\partial)|} \sum_{e_k \in \ker (\partial)}M^{e_k}(p) \notag \\
		&=R_{\partial}^{\text{irr.}}(\partial(e)) K_{\sigma}^{RC}. \label{Equation_charge_measurement_combine_magnetic}
		\end{align}

	We see that $K^{RC}_{\sigma}$ is left invariant, and so the charge labelled by the pair $R$ and $C$ is unconfined, when $R_{\partial}^{\text{irr.}}(\partial(e))=1$ for each $e \in E$, i.e., when the irrep $R_{\partial}^{\text{irr.}}$ is trivial. This gives us a simple way to check, from the label of a charge, that the charge is unconfined.

		From the above argument, we may expect that charges with irrep label $R$ corresponding to a non-trivial irrep $R_{\partial}^{\text{irr.}}$ of $\partial(E)$ are confined. We can check this directly, to make sure that the charges are always definitely confined or definitely unconfined (rather than confinement depending on some internal space of the charge) as we expect. As discussed previously, for a confined charge applying the edge terms on each of the edges surrounding the measurement region will give us zero, because at least one of the outwards edges must be excited. That is $\prod_{i \in \sigma} \mathcal{A}_i K_{\sigma}^{RC} =0$. We know that we can extract from the edge term $\mathcal{A}_i$ any edge transform: $\mathcal{A}_i= \mathcal{A}_i \mathcal{A}_i^e$. Therefore, our operator to check whether any of the edges of our ribbon are excited can be written as 
		$$\prod_{i \in \sigma} \mathcal{A}_i = \prod_{i \in \sigma} \mathcal{A}_i \prod_{i' \in \sigma} \mathcal{A}_{i'}^e = \frac{1}{|E|}\sum_{e \in E} \prod_{i \in \sigma} \mathcal{A}_i \prod_{i' \in \sigma} \mathcal{A}_{i'}^e.$$
		
		Then using this form and Equation \ref{Equation_closed_magnetic_edge_transforms}, we see that
		\begin{align*}
		\prod_{i \in \sigma} \mathcal{A}_i K_{\sigma}^{RC} &= \prod_{i \in \sigma} \mathcal{A}_i \frac{1}{|E|} \sum_{e \in E} \prod_{i' \in \sigma} \mathcal{A}_{i'}^e K_{\sigma}^{RC}\\
		&=\prod_{i \in \sigma} \mathcal{A}_i \frac{1}{|E|} \sum_{e \in E} C^{\partial(e)}(\sigma) K_{\sigma}^{RC}\\
		&=\prod_{i \in \sigma} \mathcal{A}_i \frac{1}{|E|} \frac{|E|}{|\partial(E)|} \sum_{h \in \partial(E)} C^h(\sigma) K_{\sigma}^{RC}.\\
	\end{align*}

Using Equation \ref{Equation_charge_measurement_combine_magnetic}, this implies that
		\begin{align*}
			\prod_{i \in \sigma} \mathcal{A}_i K_{\sigma}^{RC}	&= \prod_{i \in \sigma} \mathcal{A}_i \frac{1}{|\partial(E)|} \sum_{h \in \partial(E)} R_{\partial}^{\text{irr.}}(h^{-1}) K_{\sigma}^{RC}\\
		&=0 \text{ iff } R_{\partial}^{\text{irr.}} \text{ is not the trivial irrep.}
		\end{align*}
		
		This indicates that the charges with non-trivial irreps of $\partial(E)$ are confined, as we expect from our discussion of ribbon operators in Section \ref{Section_2D_electric}, where we saw that electric ribbon operators labelled by such irreps are confined.

		 So far we have not related the labels of the ribbon operators to the labels of the topological charge carried by the excitations they produce. However, in Section \ref{Section_2D_Topological_Charge_Examples} in the Supplemental Material, we use the charge measurement operators to verify our intuition about the charge carried by electric and magnetic excitations. For example, by applying the measurement operator around an excitation produced by a pure electric ribbon labelled by irrep $R_x$ of $G$, and matrix indices $a$ and $b$, we find that such an excitation has label $C$ given by the class containing the identity $1_G$ (this class is the image of $\partial$). We then find that the second label $R$ is $R_x$ or $\overline{R_x}$ depending on whether we measure the excitation at the start or the end of the ribbon. This matches our intuition that the irrep, and not the matrix indices $a$ and $b$, should label the topological charge.

\section{Conclusion}

The 2+1d higher lattice gauge theory model is interesting, in that it contains features outside those from the more heavily studied string-net and Kitaev Quantum Double models. Here we have been able to study many of these properties in two cases: one where $\rhd$ is trivial and one where we restrict to fake-flat configurations. By constructing the ribbon operators that produce the point-like excitations, we found that these particles are analogous to the electric and magnetic excitations from Kitaev's Quantum Double model \cite{Kitaev2003}. Unlike the excitations in the Quantum Double model however, some of the anyons in the higher lattice gauge theory model are confined, and others are condensed, in a similar way to the excitations from Ref. \cite{Bombin2008}. In the $\rhd$ trivial case, we were able to discuss this in terms of a condensation-confinement transition, just as we did for the 3+1d case in Ref. \cite{HuxfordPaper1}. We also used the ribbon operators to find the braiding relations of these point-like excitations, which are analogous to those for Kitaev's model \cite{Kitaev2003}, and to construct the projectors to definite topological charge (following the method used in Ref. \cite{Bombin2008}).

One intriguing feature of the 2+1d higher lattice gauge theory model is that, despite being in 2+1d, there are loop-like excitations that are produced by membrane operators. We found that, when $\rhd$ is trivial, these loop-like excitations can be interpreted in terms of two distinct phenomena. Firstly, some of the loop excitations correspond to confined versions of the electric excitations which are produced pairwise and separated, before being recombined, thereby dragging an energetic string which connects into a loop. More interestingly, some of the loop-like excitations can be interpreted as domain walls between patches of lattice corresponding to different ground states (with general loop-like excitations made from a combination of the two types). This revealed the presence of a symmetry, which leads to multiple ground states even on a spherical spatial manifold. Therefore, at least under certain circumstances, the lattice model represents a symmetry enriched topological phase (SET phase). We further confirmed this by mapping a subset of the higher lattice gauge theory models (where $\rhd$ is trivial and we further make $\partial$ injective) to a subset of the symmetry enriched string-net models from Ref. \cite{Heinrich2016}. However, unlike in the wider class of symmetry enriched string-net models, the symmetry in the higher lattice gauge theory models never permutes the anyon types (at least in the cases we were able to study in detail).

This leads us to an interesting avenue for potential further study: we have found a tidy interpretation of the higher lattice gauge theory model in 2+1d in terms of a SET phase only when $\rhd$ is trivial. It would be interesting to further study the case where $\rhd$ is non-trivial (beyond the examples considered in Section \ref{Section_Example_Z_2_Z_3}, as well as Section \ref{Section_rhd_non-trivial_condense_confine} of the Supplemental Material), if a method for dealing with the inconsistency under changes to the branching structure can be found. Furthermore, the fact that the symmetry in the cases where $\rhd$ is trivial does not permute the anyon types raises the question of whether there is an extension to this model which does allow the symmetry to permute anyon type.

\begin{acknowledgments}
	We would like to thank Jo\~{a}o Faria Martins and Alex Bullivant for informative discussions about the higher lattice gauge theory model. We are also grateful to Paul Fendley for advice on the preparation of this series of papers. We acknowledge support from EPSRC Grant EP/S020527/1. Statement of compliance with EPSRC policy framework on research data: This publication is theoretical work that does not require supporting research data.
	
\end{acknowledgments}

\bibliography{references2}{}

	\newpage
	
		\onecolumngrid		
\begin{center}
\textbf{\LARGE Supplemental Material}
\end{center}
\medskip

	\setcounter{equation}{0}
	\setcounter{figure}{0}
	\setcounter{table}{0}
	\makeatletter
	\renewcommand{\theequation}{S\arabic{equation}}
	\renewcommand{\thefigure}{S\arabic{figure}}
	\setcounter{section}{0}
	\renewcommand{\thesection}{S-\Roman{section}}
	
	\section{Energy of ribbon and membrane operators in 2+1d}
	\label{Section_2D_Ribbon_Operators_Appendix}
	In this section, we will derive the commutation relations of the ribbon and membrane operators with the energy terms in the 2+1d model, to demonstrate that they produce only the excitations described in Sections \ref{Section_RO_2D_Tri_Trivial} and \ref{Section_2D_RO_Fake_Flat} of the main text. 
	
	\subsection{Electric ribbon operators}
	\label{Section_Electric_Ribbon_Operator_Proof}
	As described in Section \ref{Section_2D_electric} of the main text, an electric ribbon operator has the form
	$$S^{\vec{\alpha}}(t)=\sum_{g \in G} \alpha_g \delta(g, \hat{g}(t)),$$
	where $t$ is the path on which we apply the operator and $\alpha_g$ are a set of coefficients. The operator $\hat{g}(t)$ is the product of edge element operators along the path, with inverses if the orientation of the edge is opposite to that of the path. This operator is diagonal in the basis of configurations, where the basis states are those for which each edge is labelled by an element of $G$ and each plaquette by an element of $E$ (rather than superpositions of different elements, or where the different degrees of freedom are entangled). The plaquette energy terms (and blob energy terms in 3+1d) are also diagonal in this basis, so they commute with the electric ribbon operators. Therefore, the electric ribbon operators do not excite these energy terms. This means that we just need to consider the vertex and edge transforms.

\begin{figure}[h]
		\begin{center}
			\begin{overpic}[width=0.75\linewidth]{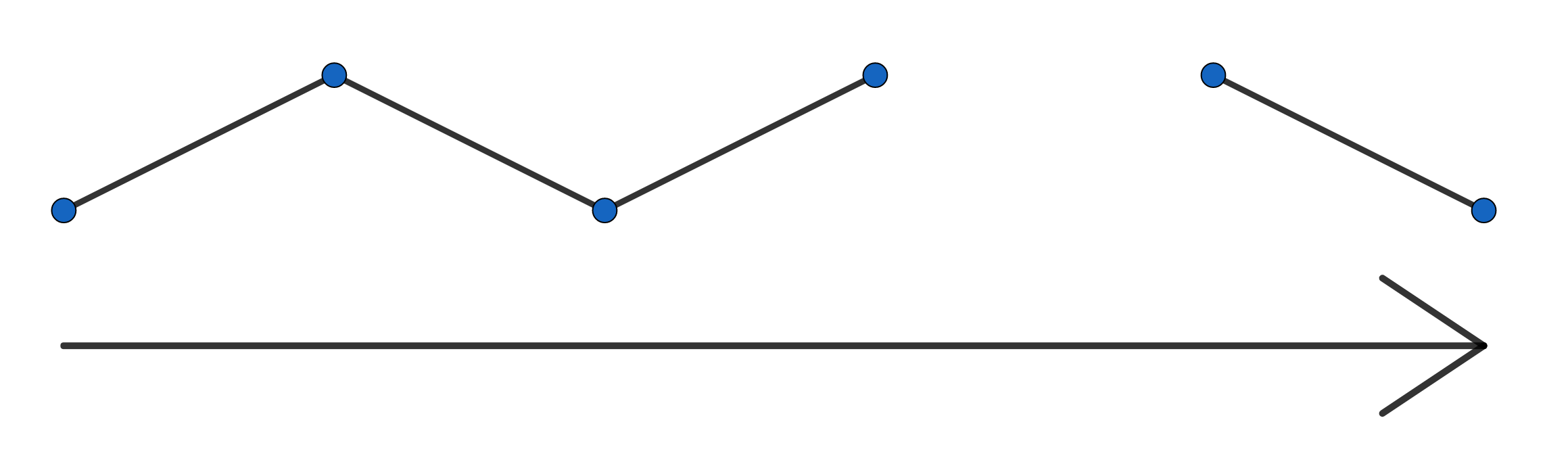}
				\put(0,13){$s.p=v_0$}
				\put(20,21){$v_1$}
				\put(38,13){$v_2$}
				\put(55,21){$v_3$}
				\put(76,21){$v_{n-1}$}
				\put(95,13){$v_n$}
				
				\put(64,25){\Huge ...}
				\put(70,2){path direction}
			\end{overpic}
			\caption{A path $t$ on the lattice passes from vertex to vertex along the edges of the lattice. The direction of the edges can be with or against the path.}
			\label{Path_vertices}
		\end{center}
	\end{figure}

	First we consider the vertex transforms. These will not generally commute with the electric ribbon operator if they change the path label $g(t)$ which is measured by the electric ribbon operator. The path $t$ starts at some vertex $s.p=v_0$ (the start-point of the path) and passes through a set of vertices $\set{v_i}$ before terminating on a vertex $v_n$, as shown in Figure \ref{Path_vertices}. We refer to all of these vertices as the vertices on the path $t$. The path is made up of a set of edges that pass between these vertices, and the path element is a product of these edge elements, in the order traversed along the path, with inverses if the edge points against the direction of the path. Therefore, operators that change the edge elements could fail to commute with the ribbon operator. An edge element is only affected by the vertex transforms at the two ends of the edge (and the edge transform on that edge, which we will examine later). The two vertices at the ends of an edge belonging to the path are vertices on the path. Therefore, we only need to consider vertex transforms for vertices on the path. We differentiate between three types of vertices on the path. These are the vertex at the start of the path, the one at the end of the path and the vertices in the middle of the path (any vertices which are not at the start or end of the path, but which the path passes through). As we discussed in Section \ref{Section_2D_electric} of the main text, the vertex transform does not change the path label of paths that pass through the vertex, only the label of paths that start or terminate on the vertex. Therefore, the transforms on the middle vertices should commute with the path element operator, as we will show explicitly here.

\begin{figure}[h]
		\begin{center}
			\begin{overpic}[width=0.6\linewidth]{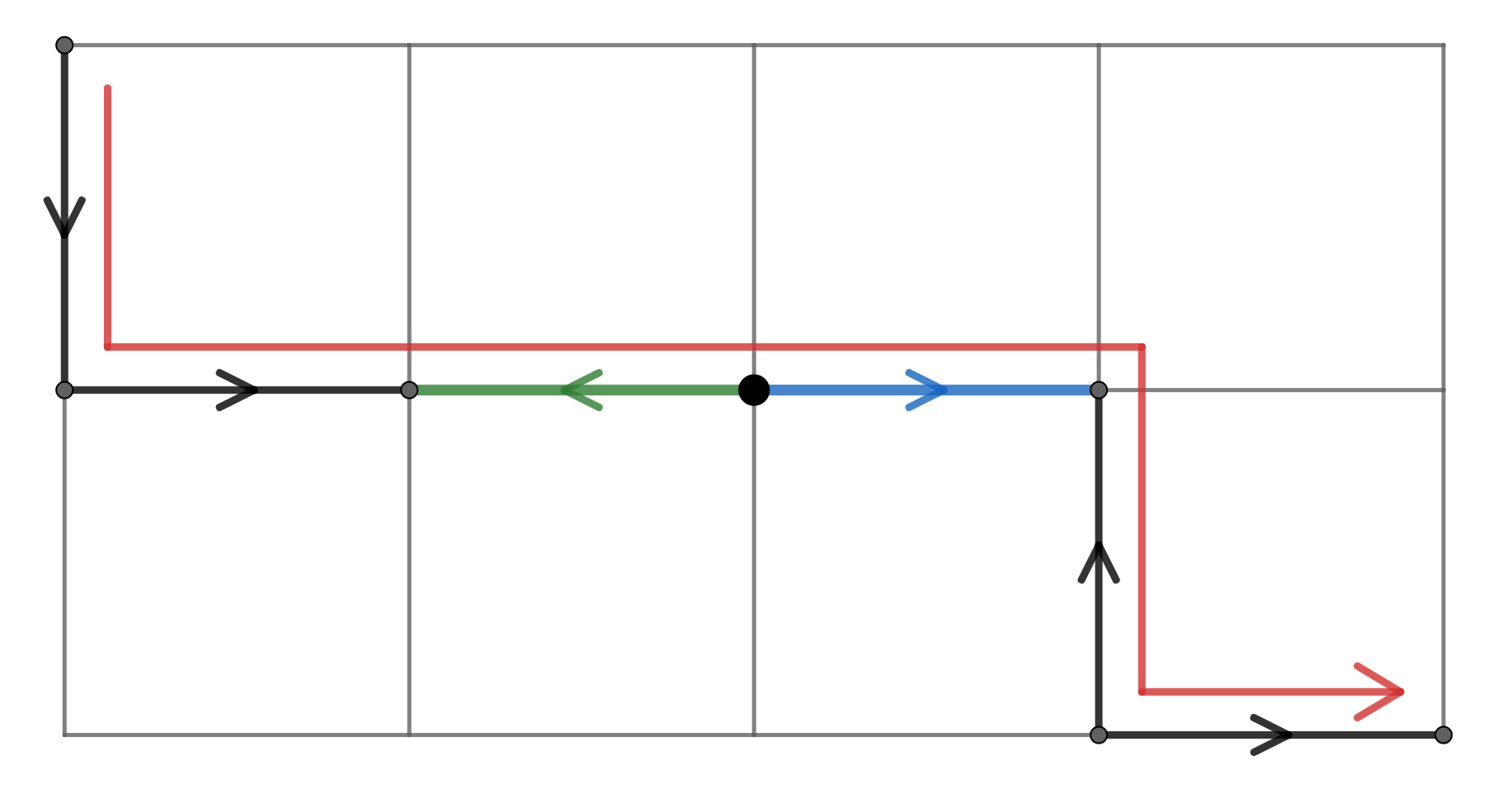}
				\put(25,22){\textcolor{mygreen1}{incoming edge}}
				\put(53,22){\textcolor{blue}{outgoing edge}}
				\put(45,32){vertex $v_i$}
				\put(10,40){\textcolor{red}{direction of path}}
			\end{overpic}
			\caption{A generic vertex $v_i$ on a path is both entered and exited by the path (whose direction is shown in red), and so is adjacent to two consecutive edges on the path, an incoming (green) and outgoing (blue) one. Here incoming and outgoing refer to the fact that the path (red) enters and exits the vertex along these edges respectively. The edges themselves can point in either direction. For example, here both edges point away from the vertex. The orientation of the edges with respect the path determines whether their labels contribute to the path element with an inverse or not. On the other hand, the orientation of the edges with respect the vertex (pointing towards or away from the vertex) determines how the edge labels transform under the vertex transform at $v_i$. The combination of these two effects ensures that the effect of the vertex transform on the contribution of the edge towards the path element does not depend on the actual orientation of the edge, only on whether the edge is incoming or outgoing in the sense explained earlier.}
			\label{middle_vertex_path}
		\end{center}
	\end{figure}

	The vertex transform on an internal vertex affects two edges that are placed consecutively on the path: an incoming and outgoing edge. This is indicated in Figure \ref{middle_vertex_path}. Here incoming and outgoing do not refer to the direction of the edge, but rather of the path. The path enters the vertex on the incoming edge and exits on the outgoing edge. The vertex transform therefore affects two of the labels in the product $g(t)=...g_i^{\sigma_i} g_{i+1}^{\sigma_{i+1}}...$, where $g_i$ is the incoming edge's label, $g_{i+1}$ is the label of the outgoing edge and each $\sigma$ depends on the orientation of the edge indicated in subscript. If the edge points along the direction of the path, then $\sigma$ is 1. On the other hand, if the edge points in the opposite direction, then $\sigma$ is $-1$. While the contribution of each edge to the path label depends on the orientation of the edge with respect to the path, the action of the vertex transform on the two edges depends instead on whether the edges point towards the vertex or away from it. For the incoming edge, towards the vertex is also along the path direction. Therefore, if the incoming edge points towards the vertex, then it appears as $g_i$ rather than the inverse in the path element. In addition, in this case the edge points towards the vertex, so under the action of the vertex transform $A_v^x$ the label $g_i$ becomes $g_i x^{-1}$. On the other hand, if the incoming edge points away from the vertex, the edge points against the path and so it appears as $g_i^{-1}$ in the expression for $g(t)$. However because the edge points away from the vertex, under the same vertex transform the label of the edge becomes $x g_i$. Therefore, $g_i^{-1}$ becomes $g_i^{-1} x^{-1}$. We therefore see that regardless of the orientation of the edge, the term $g_i^{\sigma_i}$ becomes $g_i^{\sigma_i} x^{-1}$. On the other hand, if the outgoing edge points along the path then it points away from the vertex. Therefore, the outgoing edge label $g_{i+1}$ becomes $xg_{i+1}$ if the edge points away from the vertex (along the path) and $g_i x^{-1}$ otherwise. In either case, $g^{\sigma_{i+1}}_{i+1}$ becomes $x g^{\sigma_{i+1}}_{i+1}$. Therefore, under the action of $A_v^x$, $g(t)=...g_i^{\sigma_i } g_{i+1}^{\sigma_{i+1}}...$ becomes $...g_i^{\sigma_i }x^{-1} x g_{i+1}^{\sigma_{i+1}}... =g(t)$. This means that the path element is unaffected by the vertex transform, and so these middle vertex transforms commute with the electric ribbon operator.
	
	On the other hand, consider the first vertex on the ribbon, the start-point of the path $t$. This vertex has an outgoing edge on the path, but no incoming edge. This means that under this vertex transform, the path element only gains the transformation from the outgoing edge. The outgoing edge label $g_1$ contributes a term $g_1^{\sigma_1}$ to $g(t) =g_1^{\sigma_1}...$ and, under the vertex transform $A_v^x$, $g_1^{\sigma_1}$ becomes $x g_1^{\sigma_1}$ as we discussed above. Therefore, for the vertex transform at the start of the path, we have that
	\begin{equation}
		A_v^x : g(t) = A_v^x: g_1^{\sigma_1}...= x g_1^{\sigma_1}... = x g(t). \label{Equation_vertex_transform_start_path_1}
	\end{equation}
	
	This means that 
	\begin{equation}
		\hat{g}(t) A_v^x = A_v^x x\hat{g}(t),
		\label{Equation_vertex_transform_start_path_operator}
	\end{equation}
	and so
	\begin{align*}
		\sum_{g \in G} \alpha_g \delta(\hat{g}(t),g) A_v^x&= A_v^x \sum_{g \in G} \alpha_g \delta(x \hat{g}(t),g)\\
		&= A_v^x \sum_{g \in G} \alpha_g \delta(\hat{g}(t), x^{-1}g)\\
		&= A_v^x \sum_{y=x^{-1}g} \alpha_{xy} \delta(\hat{g}(t), y).
	\end{align*}
	
	In a similar way, the last vertex on the path $t$ has an incoming edge, but no outgoing edge. Therefore, the vertex transform on this vertex acts on the path element as
	\begin{equation}
		A_v^x: g(t)=g(t)x^{-1}. \label{Equation_vertex_transform_end_path_1}
	\end{equation}
	
	This means that 
	\begin{equation}
		\hat{g}(t) A_v^x =A_v^x \hat{g}(t) x^{-1}, \label{Equation_vertex_transform_end_path_operator}
	\end{equation}
	and so
	\begin{align*}
		\sum_{g \in G} \alpha_g \delta(\hat{g}(t),g) A_v^x&= A_v^x \sum_{g \in G} \alpha_g \delta(\hat{g}(t)x^{-1},g)\\
		&= A_v^x \sum_{g \in G} \alpha_g \delta(\hat{g}(t), gx)\\
		&= A_v^x \sum_{y=gx} \alpha_{yx^{-1}} \delta(\hat{g}(t), y).
	\end{align*}
	
	So far, we have only considered the case where the path passes through each vertex on the path once. It is also possible that the path doubles back on itself and self-intersects, so that the path passes through a vertex multiple times. However in this case we can still pair up the incoming edges and outgoing edge for each visitation of the vertex by the path (except for where the path starts or terminates), and the effect of the vertex transform on the incoming and outgoing edges will cancel, just as described above.

\begin{figure}[h]
		\begin{center}
			\begin{overpic}[width=0.6\linewidth]{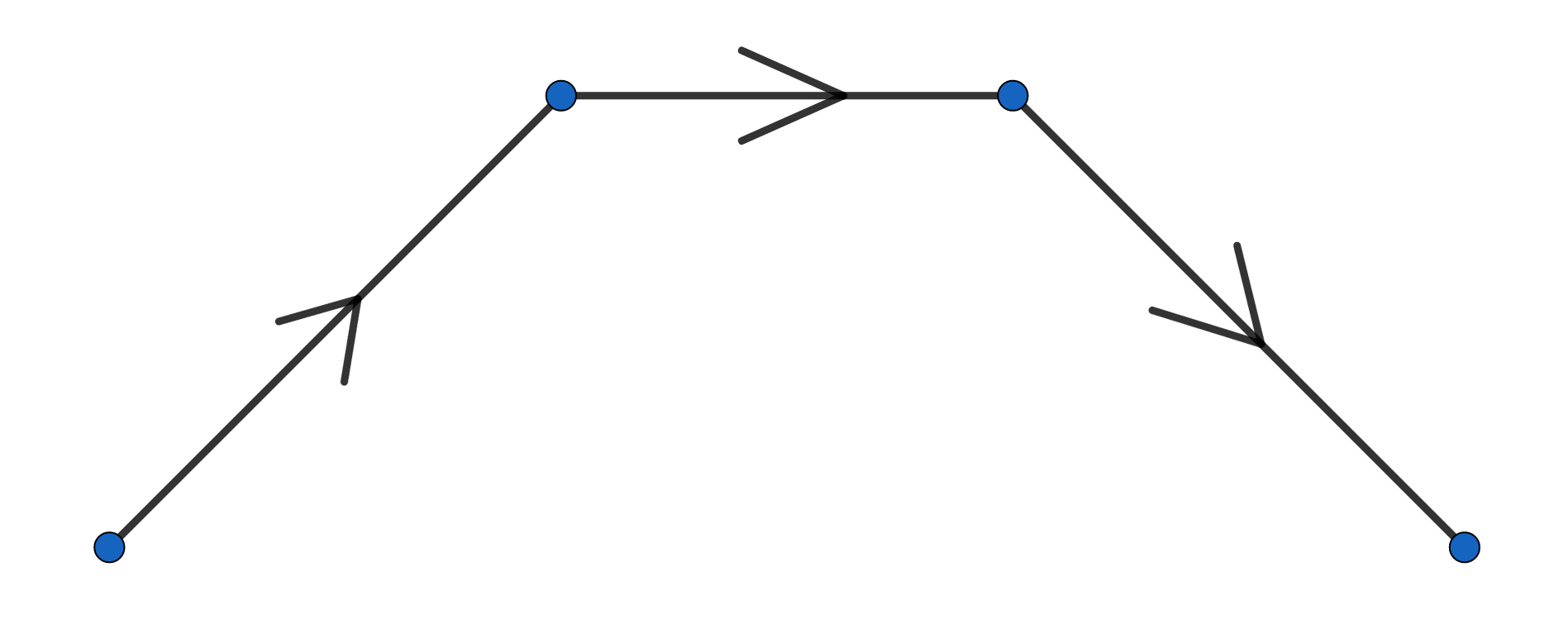}
				\put(5,7){$v_0$}
				\put(33,36){$v_i$}
				\put(61,36){$v_{i+1}$}
				\put(93,7){$v_n$}
				\put(13,27){$v_0-v_i$}
				\put(79,27){$v_{i+1}-v_n$}
				\put(44,40){edge $i$}
				\put(37,20){$\mathcal{A}_i^e: g_i = \partial(e)g_i$}
			\end{overpic}
			\caption{We consider the edge transform $\mathcal{A}_i^e$ acting on an edge $i$ on the path. To find the effect on the label of the overall path, we must split the path into sections before and after the edge, as well as the edge itself.}
			\label{edge_transform_path}
		\end{center}
	\end{figure}

	Finally we must consider the edge transforms. These can interact with the electric ribbon operator by changing the label of one of the edges on the path $t$, which we will denote by $i$. Let the edge in question be between consecutive vertices $v_i$ and $v_{i+1}$ on the path. We denote the path from the start of $t$ to $v_i$ by $v_0 - v_i$ and the path from $v_{i+1}$ to the end by $v_{i+1} - v_n$. This is shown in Figure \ref{edge_transform_path}. Then $g(t) = g(v_0 - v_i) g_i^{\sigma_i} g(v_{i+1} - v_n)$. In the case where the edge $i$ points along the path $t$, as shown in Figure \ref{edge_transform_path}, we have $\sigma_i=1$ and so
	\begin{align*}
		\mathcal{A}_i^e: g(t) &= g(v_0 - v_i) \partial(e)g_i g(v_{i+1} - v_n)\\
		&= g(v_0 -v_i)\partial(e) g(v_0 - v_i)^{-1} g(t)\\
		&= \partial( g(v_0-v_i) \rhd e)g(t),
	\end{align*}
	where in the last step we used the first Peiffer condition, Equation \ref{Peiffer_1} from the main text.
	
	Similarly, if the edge $i$ points against the path (i.e., if in Figure \ref{edge_transform_path}, the orientation of edge $i$ were flipped), the edge transform acts on the path as
	\begin{align*}
		\mathcal{A}_i^e: g(t) &= g(v_0 - v_i) g_i^{-1} \partial(e)^{-1} g(v_{i+1} - v_n)\\
		&= g(v_0 -v_{i+1})\partial(e)^{-1} g(v_0 - v_{i+1})^{-1} g(t)\\
		&= \partial( g(v_0-v_{i+1}) \rhd e^{-1})g(t).
	\end{align*}
	
	We can write the action of the edge transform in either case as $\mathcal{A}_i^e : g(t) = \partial(f_e)g(t)$, where $f_e$ is in one-to-one correspondence with $e$ (with the precise relation depending on the orientation of the edge and a path element $g(v_0-v_i)$ or $g(v_0-v_{i+1})$). Therefore,
	$$\hat{g}(t) \mathcal{A}_i^e =\mathcal{A}_i^e \partial(f_e) \hat{g}(t).$$
	This means that
	
	\begin{align}
		\sum_{g \in G} \alpha_g \delta(\hat{g}(t), g) \mathcal{A}_i^e &= \mathcal{A}_i^e \sum_{g \in G} \alpha_g \delta( \partial(f_e)\hat{g}(t), g) \notag\\
		&=\mathcal{A}_i^e \sum_{g \in G} \alpha_g \delta( \hat{g}(t), \partial(f_e)^{-1}g) \notag\\
		&= \mathcal{A}_i^e \sum_{g'=\partial(f_e)^{-1}g} \alpha_{\partial(f_e) g'} \delta( \hat{g}(t),g').
		\label{edge_transform_path_commutation}
	\end{align}
	
	This means that if 
	\begin{equation}
		\alpha_g = \alpha_{\partial(f)g} \label{Equation_electric_unconfined}
	\end{equation}
	for all $f \in E$, the edge transform commutes with the ribbon operator, and so the edge is not excited by the ribbon operator. However we also want to figure out what happens in the cases where $\alpha$ does not satisfy this condition. Now consider $\mathcal{A}_i \sum_{g \in G} \alpha_g \delta(\hat{g}(t), g) \ket{GS}$, where $\ket{GS}$ is one of the ground states of the model. This will check whether the edge is excited by the ribbon operator (in which case we get 0), not excited (in which case we get $\sum_{g \in G} \alpha_g \delta(\hat{g}(t), g) \ket{GS}$), or if $\sum_{g \in G} \alpha_g \delta(\hat{g}(t), g) \ket{GS}$ is not an eigenstate of the edge energy term at all. Therefore
	
	\begin{align*}
		\mathcal{A}_i \sum_{g \in G} \alpha_g \delta(\hat{g}(t), g) \ket{GS}&=\big(\sum_{e \in E} \frac{1}{|E|} \mathcal{A}_i^{e} \big) \sum_{g \in G} \alpha_g \delta(\hat{g}(t), g) \ket{GS} \\
		&= \sum_{e \in E} \frac{1}{|E|} \big[\mathcal{A}_i^{e} \sum_{g \in G} \alpha_g \delta(\hat{g}(t), g)\big] \ket{GS}. 
		\intertext{Then, using Equation \ref{edge_transform_path_commutation} to rewrite the term in square brackets, this means that}\\
		\mathcal{A}_i \sum_{g \in G} \alpha_g \delta(\hat{g}(t), g) \ket{GS}&= \sum_{e \in E} \frac{1}{|E|} \big[\sum_{g' \in G} \alpha_{\partial(f_e)^{-1}g'} \delta(\hat{g}(t), g') \mathcal{A}_i^{e}\big]\ket{GS}.
	\end{align*}
	
	As a ground state, $\ket{GS}$ will satisfy $\ket{GS} = \mathcal{A}_i^e \ket{GS}$ for all $e \in E$, as we discuss in Section \ref{Section_Recap_Paper_2} of the main text, so we can remove the $\mathcal{A}_i^{e}$ in the squared brackets (which acts directly on the ground state). Furthermore, because $e \rightarrow f_e$ is one-to-one and the label $e$ now only appears in $f_e$, we can replace the sum over $e$ with a sum over $f_e$. However, for convenience we define $f =f_e^{-1}$ and sum over that instead. This gives 
	\begin{align*}
		\mathcal{A}_i \sum_{g \in G} \alpha_g \delta(\hat{g}(t), g) \ket{GS}&= \frac{1}{|E|} \sum_{f \in E} \sum_{g' \in G} \alpha_{\partial(f)g'} \delta(\hat{g}(t), g') \ket{GS}\\
		&= \sum_{g' \in G} (\frac{1}{|E|} \sum_{f \in E} \alpha_{\partial(f)g'}) \delta(\hat{g}(t), g') \ket{GS}.
	\end{align*}
	
	If $\sum_f \alpha_{\partial(f)g'}=0$ for all $g'$ in $G$, then we get zero in the above equation, so the edge is excited. Because this condition on the coefficients is independent of which edge $i$ we consider on the path, if it holds for one edge it will hold for all edges on the path. This means that, if the condition is satisfied, there is an energetic string between the two end-points of the path. In this case, there is an energy cost associated to the length of the ribbon, and so the corresponding excitations are confined. On the other hand, as discussed previously, if $\alpha_{\partial(f)g'}=\alpha_{g'}$ for all $g' \in G$ and $f \in E$, then the edge will not be excited (again holding for each edge on the path). Given an arbitrary set of coefficients $\alpha_g$, we can write the coefficients as a sum of two sets of coefficients, one satisfying the first of these constraints and another satisfying the second:
	$$\alpha_g = \frac{1}{|E|} \sum_{e \in E} \alpha_{\partial(e)g} + \big(\alpha_g - \frac{1}{|E|} \sum_{e \in E} \alpha_{\partial(e)g}\big).$$
	This indicates that an arbitrary electric ribbon operator can be written as a sum of a confined and an unconfined ribbon operator.

	We have therefore proven that the electric ribbon operators commute with all of the energy terms except for the vertex transforms at the start and end of the path and possibly the edge transforms along the path. We have also established the commutation relations in the cases where the operators do not commute.
	
	\subsection{Magnetic ribbon operators when $\rhd$ is trivial}
	\label{Section_Magnetic_Ribbon_Proof}
	
	Next we want to demonstrate the commutation relations between the magnetic ribbon operators in 2+1d and the energy terms in the special case where $\rhd$ is trivial (Case 1 in Table \ref{Table_Cases_2d} in the main text). Taking this special case leads to some simplifications to the crossed module structure. When $\rhd$ is trivial, the group $E$ is Abelian and $\partial(E)$ is within the centre of $G$, as we described in Section \ref{Section_Recap_Paper_2} of the main text. This latter point will be particularly important here, because deforming a path over a fake-flat surface makes the path element pick up a factor of $\partial(e)$ for some $e \in E$, and this factor being in the centre of $G$ greatly simplifies some expressions that we will see later.

\begin{figure}[h]
		\begin{center}
			\begin{overpic}[width=0.6\linewidth]{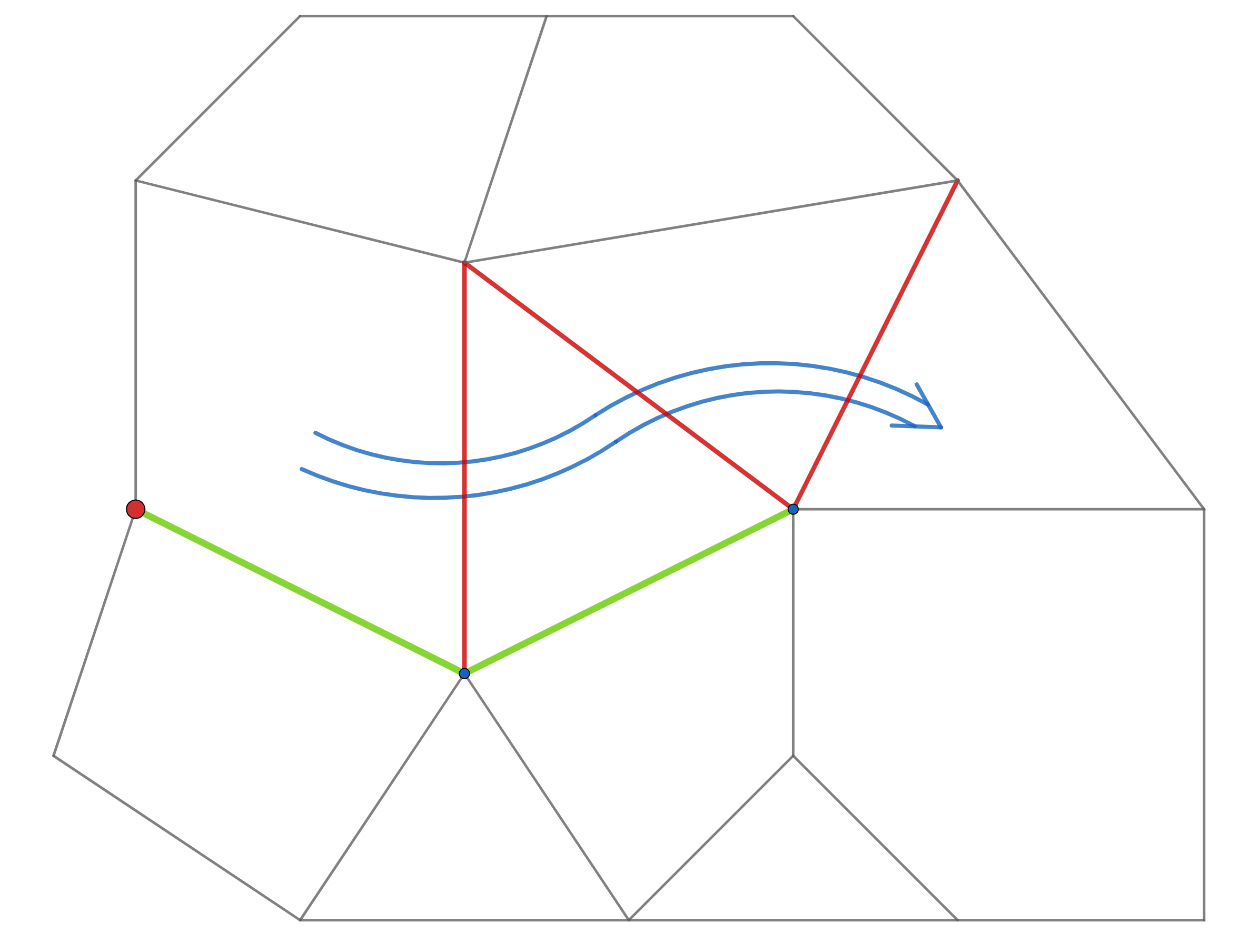}
				\put(-3,34){start-point}
				\put(15,22){direct path}
				\put(23,42){dual path}
				\put(38,30){$i$}
				\put(34,20){$v_i$}
				
			\end{overpic}
			\caption{When we define the magnetic ribbon operator, we must define a dual path (blue arrow) and a direct path (green). The ribbon operator acts on edges (red) that are cut by the dual path. The action on such an edge depends on the path element for the path from the start-point of the ribbon to the vertex on the direct path that is attached to the cut edge. For example, the action on the edge labelled $i$ depends on the path from the start-point to the vertex labelled $v_i$, along the direct path.}
			\label{magnetic_ribbon_example}
		\end{center}
	\end{figure}

	Let us start by briefly restating the action of the magnetic ribbon operator. Recall from Section \ref{Section_2D_Magnetic} of the main text that the ribbon upon which we apply our magnetic ribbon operator is defined by two paths: a direct and a dual path. The action of the magnetic ribbon operator $C^h(t)$ on an edge $i$ intersected by the dual path is
	\begin{equation}
		C^h(t): g_i = \begin{cases} g(s.p-v_i)^{-1}hg(s.p-v_i) g_i & i \text{ points away from direct path}\\ g_i g(s.p-v_i)^{-1}h^{-1}g(s.p-v_i) &i \text{ points towards direct path,} \end{cases} \label{Equation_magnetic_ribbon_appendix}
	\end{equation}
	where $s.p$ is the start-point of the ribbon and $v_i$ is the vertex on the direct path that is attached to edge $i$, as shown in Figure \ref{magnetic_ribbon_example}. Note that each vertex $v_i$ can be attached to multiple edges that are cut by the dual path (the red edges in Figure \ref{magnetic_ribbon_example}).

	First we will consider the commutation relation with the edge transforms. The magnetic ribbon operator only acts on the edges. When $\rhd$ is trivial, the action of the edge transform on the surface labels is independent of the value of the edge labels, with the surface labels simply being pre-multiplied by $e$ or post-multiplied by $e^{-1}$ under the action of an edge transform $\mathcal{A}_i^e$. Contrast this with the general case, where the action of the edge transform on a plaquette depends on the label of a path from the base-point of the plaquette to the edge. In this general case, the magnetic ribbon operator could affect the path label and therefore the action of the edge transform on the surface label would not commute with the magnetic ribbon operator. In the $\rhd$ trivial case, however, there is no dependence on any path labels, and the edge transform $\mathcal{A}_i^e$ only acts on the edge labels by multiplying the label of edge $i$ by $\partial(e)$. Therefore, we just need to check if this action commutes with the magnetic ribbon operator. There are two ways in which this action could cause the edge operators to fail to commute with the magnetic ribbon operator. Firstly, the edge transform could act on an edge that is also acted on by the magnetic ribbon operator (one of the edges cut by the dual path). These two actions on the same edge could fail to commute. However this is not the case, because the action of the two operators on the edge is
	\begin{align*}
		\mathcal{A}_i^e C^h(t): g_i &= \begin{cases}\partial(e) g(s.p-v_i)^{-1}hg(s.p-v_i) g_i & i \text{ points away from the direct path}\\ \partial(e)g_i g(s.p-v_i)^{-1}h^{-1}g(s.p-v_i) & i \text{ points towards the direct path} \end{cases}\\
		& = \begin{cases}g(s.p-v_i)^{-1}hg(s.p-v_i) \partial(e) g_i & i \text{ points away from the direct path}\\ \partial(e)g_i g(s.p-v_i)^{-1}h^{-1}g(s.p-v_i) & i \text{ points towards the direct path} \end{cases}\\
		&= C^h(t) \mathcal{A}_i^e : g_i,
	\end{align*}
	where we used the fact that $\partial(e)$ commutes with any element of $G$, because $\partial$ maps to the centre of $G$, to move the factor of $\partial(e)$ next to $g_i$ (if it wasn't already). Therefore, the edge transforms acting on the edges cut by the dual path commute with the ribbon operator.

	The second way in which the edge transform could fail to commute with the magnetic ribbon operator is by acting on an edge along the direct path, and therefore changing the path element $g(s.p-v_i)$ in the expression $g(s.p-v_i)^{-1}hg(s.p-v_i)$ which appears in the action of the ribbon operator given in Equation \ref{Equation_magnetic_ribbon_appendix}. Therefore, acting first with $\mathcal{A}_j^e$, where $j$ is an edge on the path $s.p-v_i$, could affect the action of the ribbon operator we apply afterwards. However, recall from the discussion of the electric ribbon operator in Section \ref{Section_Electric_Ribbon_Operator_Proof} that $\mathcal{A}_j^e: g(t)= \partial(f_e) g(t)$, where $f_e$ is some element of $E$ that depends on $e$. Because $\partial(f_e)$ commutes with all elements of $G$ (the image of $\partial$ is in the centre of $G$ when $\rhd$ is trivial), we see that 
	$$\mathcal{A}_j^e : g(s.p-v_i)^{-1}hg(s.p-v_i) = g(s.p-v_i)^{-1} \partial(f_e)^{-1}h \partial(f_e)g(s.p-v_i) = g(s.p-v_i)^{-1}hg(s.p-v_i).$$
	
	Therefore, the edge transform does not affect the action of the ribbon operator, and so these edge transforms also commute with the ribbon operator. Putting this together with our previous results, we see that all of the edge transforms commute with the magnetic ribbon operator.

\begin{figure}[h]
		\begin{center}
			\begin{overpic}[width=0.6\linewidth]{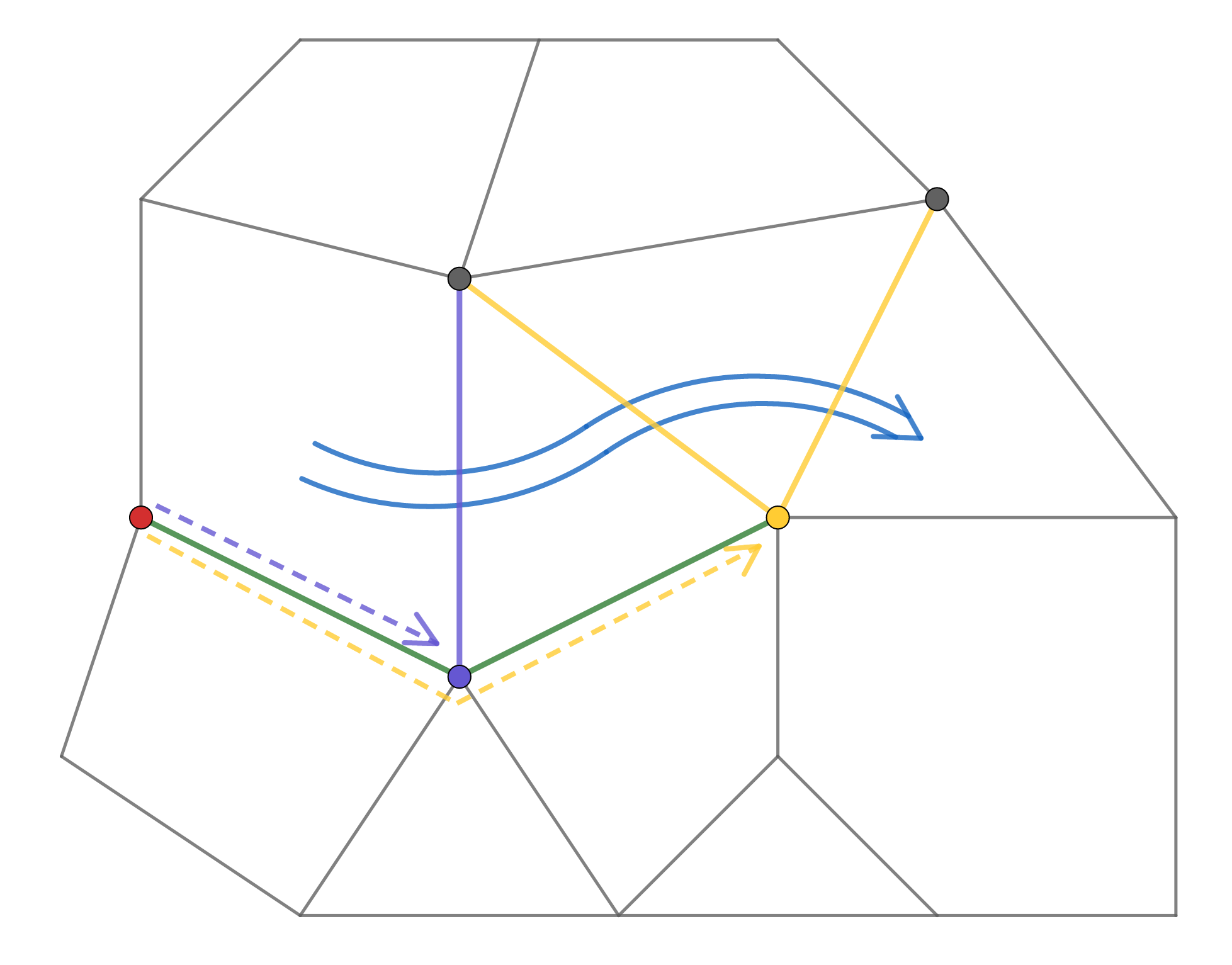}
				\put(-5,35){start-point}
				\put(23,44){dual path}
				\put(38,30){$i$}
				\put(35,20){$v_i$}
				\put(22,33){$s.p-v_i$}
				
			\end{overpic}
			\caption{We can group the edges affected by the magnetic ribbon operator (those cut by the dual path) according to the vertex on the direct path that they are attached to. In this example, we have split the edges into two groups. The purple (darker gray in grayscale) edge is attached to the purple vertex $v_i$. The action of the magnetic membrane operator on the purple edge depends on the path element for the purple dashed path from the start-point to the purple vertex. Similarly the yellow (lighter gray in grayscale) edges are both attached to the yellow vertex, so the action of the ribbon operator on each of these edges depends on the label of the yellow dashed path. The cut edges can be affected by the vertex transforms on either end of the edges. Therefore, we need to consider the vertex transforms at some vertices not on the direct path in addition to the vertex transforms on the direct path.}
			\label{vertex_on_magnetic_ribbon}
		\end{center}
	\end{figure}

	Next we consider the commutation relations with the vertex transforms. Just like with the edge transforms, there are two ways in which the vertex transform could fail to commute with the ribbon operator. The vertex transform could act on the same edges changed by the magnetic ribbon operator, or could affect the direct path label which determines the action of the ribbon operator. Consider the action of the magnetic ribbon on an edge, or set of edges, cut by the dual path and attached to a particular vertex $v_i$ on the direct path, as shown in Figure \ref{vertex_on_magnetic_ribbon}. Every edge attached to this vertex and cut by the dual path is acted on in a similar way (except for the dependence on the orientation of the edge). This is because, as shown in Equation \ref{Equation_magnetic_ribbon_appendix}, the action on an edge depends on the label of the section of the direct path up to the vertex, and edges attached to the same vertex will have the same path label $g(s.p-v_i)$ associated to them. We will start by considering which vertex transforms could change this path label $g(s.p-v_i)$. Recall from Section \ref{Section_Electric_Ribbon_Operator_Proof} that vertex transforms do not affect paths that pass through the vertex, only paths that start or end at that vertex. This narrows down which vertex transforms could fail to commute with the action on the edges. Only the transforms at the start-point of the ribbon ($s.p$) and the vertex $v_i$ can affect the label $g(s.p-v_i)$. In addition, as we mentioned earlier, we must consider vertex transforms that directly affect the labels of the edges in question (those attached to vertex $v_i$ and cut by the dual path). The vertex transforms that can do this are the vertex transforms at $v_i$ and at the other ends of the edges that are changed by the ribbon (as shown in Figure \ref{vertex_on_magnetic_ribbon}).

	Having found which vertex transforms may fail to commute with the action of the ribbon operator on these edges, we now wish to consider these transforms in more detail. First consider the vertex transforms applied on the vertices at the other ends of the edges cut by the dual path (i.e., the ends not on the direct path). Such a vertex transform $A_v^g$ left-multiplies the edge label of an edge attached to the vertex by $g$ if the edge points away from the vertex. However for a vertex transform away from the direct path, if the edge points away from the vertex then it points towards the direct path (because the other vertex attached to the cut edge is on the direct path). However in this same situation (where the edge points towards the direct path), the ribbon operator $C^h(t)$ right-multiplies the edge label by $g(s.p-v_i)^{-1}h^{-1}g(s.p-v_i)$. If the edge label is initially $g_i$, applying both of these factors gives us the expression $gg_i g(s.p-v_i)^{-1}h^{-1}g(s.p-v_i)$ regardless of which order the factors are applied in. This is because one of the factors is applied on the left and the other on the right and so it doesn't matter in which order these are applied (whereas if the vertex transform and ribbon operator both left-multiplied the edge label then they would not necessarily commute). So far we considered the case where the edge points away from the vertex and so towards the direct path. Similarly, if the edge points towards the vertex then the vertex transform right-multiplies the edge label by some group element, while the ribbon operator left-multiplies it by some element, and the fact that one of the operators applies the factor to the left while the other applies the factor to the right means that these effects commute. Therefore, these vertex transforms commute with the magnetic ribbon operator.

	Next we consider the vertex transform at the vertex $v_i$ itself (the vertex on the direct path that the edges we are considering are attached to). The vertex transform at this vertex both affects the path label $g(s.p-v_i)$ and acts directly on the edge labels changed by the magnetic ribbon operator. From Equation \ref{Equation_vertex_transform_end_path_1}, because $v_i$ is at the end of the path $s.p-v_i$, the transform acts on the label $g(s.p-v_i)$ as $A_{v_i}^x: g(s.p-v_i)= g(s.p-v_i)x^{-1}$. Therefore, when acting on one of the edges $g_i$ that is attached to $v_i$ and cut by the dual path, we have the following relation:
	\begin{align*}
		C^h(t)A_v^x :g_i &= \begin{cases} (g(s.p-v_i)x^{-1})^{-1}h(g(s.p-v_i)x^{-1})xg_i & i \text{ points away from the direct path}\\ g_i x^{-1} (g(s.p-v_i)x^{-1})^{-1}h^{-1}(g(s.p-v_i)x^{-1}) & i \text{ points towards the direct path} \end{cases}\\
		&= \begin{cases} x[g(s.p-v_i)^{-1}h g(s.p-v_i)]g_i & i \text{ points away from the direct path}\\ g_i [g(s.p-v_i)^{-1}h^{-1}g(s.p-v_i)]x^{-1} & i \text{ points towards the direct path} \end{cases}\\
		&= A_v^x C^h(t) :g_i.
	\end{align*}
	
	We have therefore shown that the actions of the magnetic ribbon operator and the vertex transform on $g_i$ commute. In addition, the actions of the two operators on the other edges attached to the vertex, but not cut by the ribbon, commute because the magnetic ribbon operator does not affect these edges. Therefore, this vertex transform commutes with the magnetic ribbon operator.

	This only leaves the vertex transform at the start-point of the ribbon. This transform affects all of the path elements $g(s.p-v_i)$ to each vertex $v_i$ (except the start-point itself). We have $A_{s.p}^x :g(s.p-v_i)=xg(s.p-v_i)$. Therefore, the action on edges away from the vertex $s.p$ is
	\begin{align*}
		C^h(t) A^x_{s.p} : g_i &= \begin{cases} (xg(s.p-v_i))^{-1}h(xg(s.p-v_i))g_i & i \text{ points away from the direct path}\\ g_i (xg(s.p-v_i))^{-1}h^{-1}(xg(s.p-v_i)) & i \text{ points towards the direct path} \end{cases}\\
		&= \begin{cases} g(s.p-v_i)^{-1}[x^{-1}hx]g(s.p-v_i)g_i & i \text{ points away from the direct path}\\ g_i g(s.p-v_i)^{-1}[x^{-1}h^{-1}x]g(s.p-v_i) & i \text{ points towards the direct path} \end{cases}\\
		&= A_{s.p}^x C^{x^{-1}hx}(t):g_i.
	\end{align*}
	
	This suggests that $C^h(t) A_{s.p}^x =A_{s.p}^x C^{x^{-1}hx}(t)$. However for this to be true it must hold for the action on every edge, including the edges attached to $s.p$ itself (assuming that these are cut by the dual path). For these edges
	\begin{align*}
		C^h(t) A_{s.p}^x :g_i &= \begin{cases} C^h(t) :x g_i &i \text{ points away from the direct path}\\ C^h :g_ix^{-1} & i \text{ points towards the direct path} \end{cases}\\
		&= \begin{cases} hx g_i &i \text{ points away from the direct path}\\ g_ix^{-1}h^{-1} & i \text{ points towards the direct path} \end{cases}\\
		&= \begin{cases} x (x^{-1}hx) g_i &i \text{ points away from the direct path}\\ g_i(x^{-1}h^{-1}x) x^{-1} & i \text{ points towards the direct path} \end{cases}\\
		&=A_{s.p}^x C^{x^{-1}hx}(t):g_i.
	\end{align*}
	Therefore, this relation holds for the action on each edge, so that the commutation relation for the vertex transform and the ribbon operator is
	\begin{equation}
		C^h(t) A_{s.p}^x =A_{s.p}^x C^{x^{-1}hx}(t). \label{Equation_magnetic_ribbon_vertex_commutation_appendix}
	\end{equation}
	
	We have now seen the commutation relations of the magnetic ribbon operator with each vertex transform. The ribbon operator commutes with each vertex transform except for transforms at the start-point of the ribbon.

	The final energy terms to consider are the plaquette terms. We should separately consider the plaquettes at the two ends of the ribbon (the first and final plaquette in the dual path) and the ``internal" plaquettes in the middle of the ribbon. In the same way as the direct path enters a vertex along an edge and leaves it along another, the dual path enters a plaquette by cutting through one edge and leaves it by cutting through another. This leads to an important distinction between the two types of plaquettes we just mentioned. The first kind are the plaquettes at the two ends of the ribbon. The dual path only enters or exits these plaquettes, not both (unless the ribbon is closed or self-intersects). The other type of plaquette are internal plaquettes: plaquettes which are both entered and exited by the dual path.

\begin{figure}[h]
		\begin{center}
			\begin{overpic}[width=0.6\linewidth]{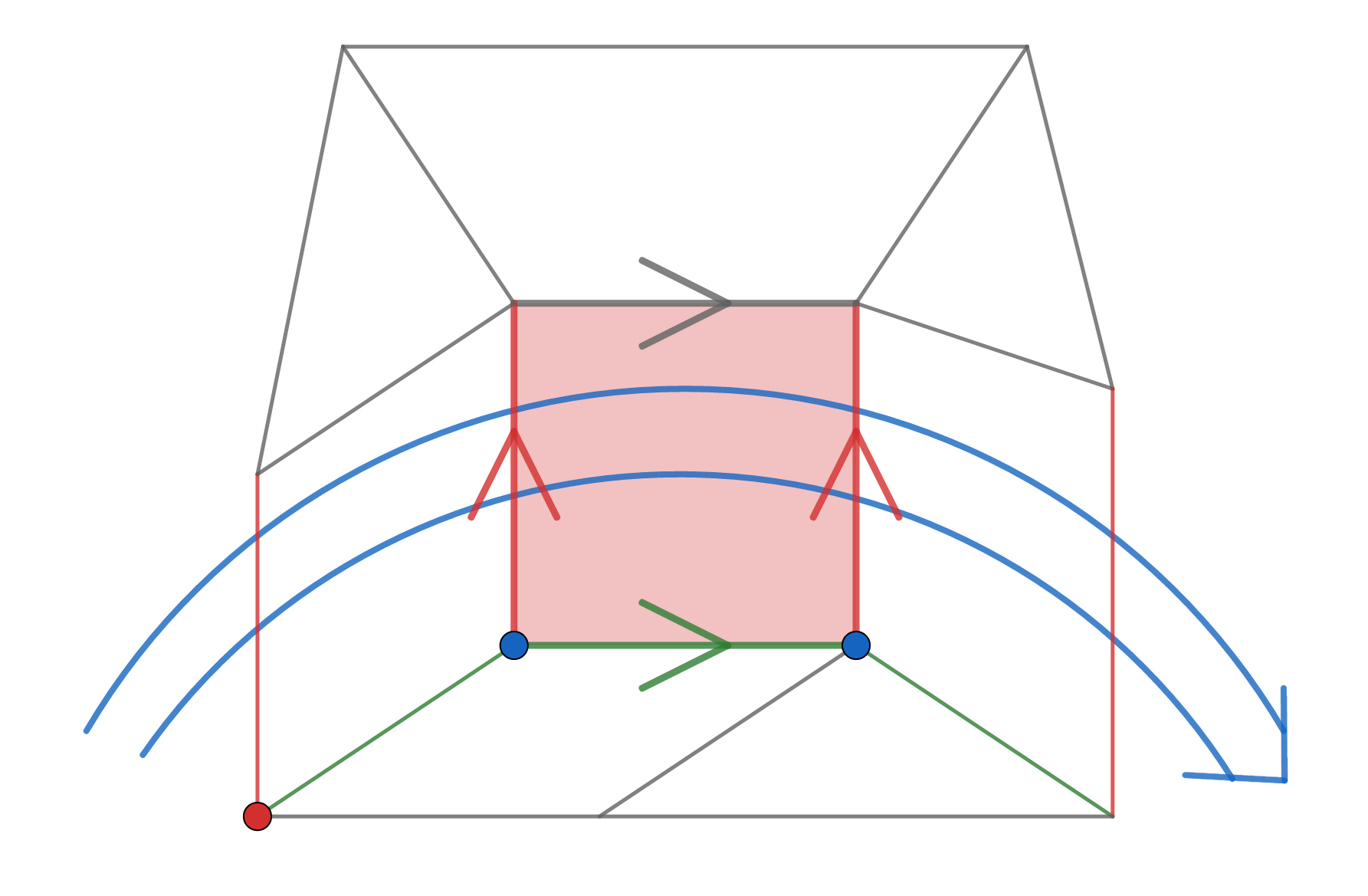}
				\put(5,2){start-point $s.p$}
				\put(37,15){$v_i$}
				\put(62,15){$v_{i+1}$}
				\put(0,21){dual path}
				\put(28,8){$g(s.p-v_i)$}
				\put(45,12){$g(1-2)$}
				\put(50,48){$g_x$}
				\put(32,30){$g_1$}
				\put(65,30){$g_2$}
			\end{overpic}
			\caption{The dual path passes through each internal plaquette, cutting through two of its edges. For example, consider the shaded plaquette. The dual path enters the plaquette by cutting through the edge labelled by $g_1$ and exits through the edge with label $g_2$. The other edges on the plaquette are included in the paths labelled by $g_x$ and $g(1-2)$, with $g(1-2)$ being part of the direct path of the ribbon. Note that the paths labelled by $g_x$ and $g(1-2)$ can contain any number of edges (including zero), so while we have drawn the plaquette as a square we do not need to assume that it has this shape. In addition, we can swap the orientations of the cut (red) edges, in which case we must invert their labels from $g_1$ and $g_2$ to $g_1^{-1}$ and $g_2^{-1}$.}
			\label{plaquette_on_magnetic_ribbon}
		\end{center}
	\end{figure}

\begin{figure}[h]
		\begin{center}
			\begin{overpic}[width=0.4\linewidth]{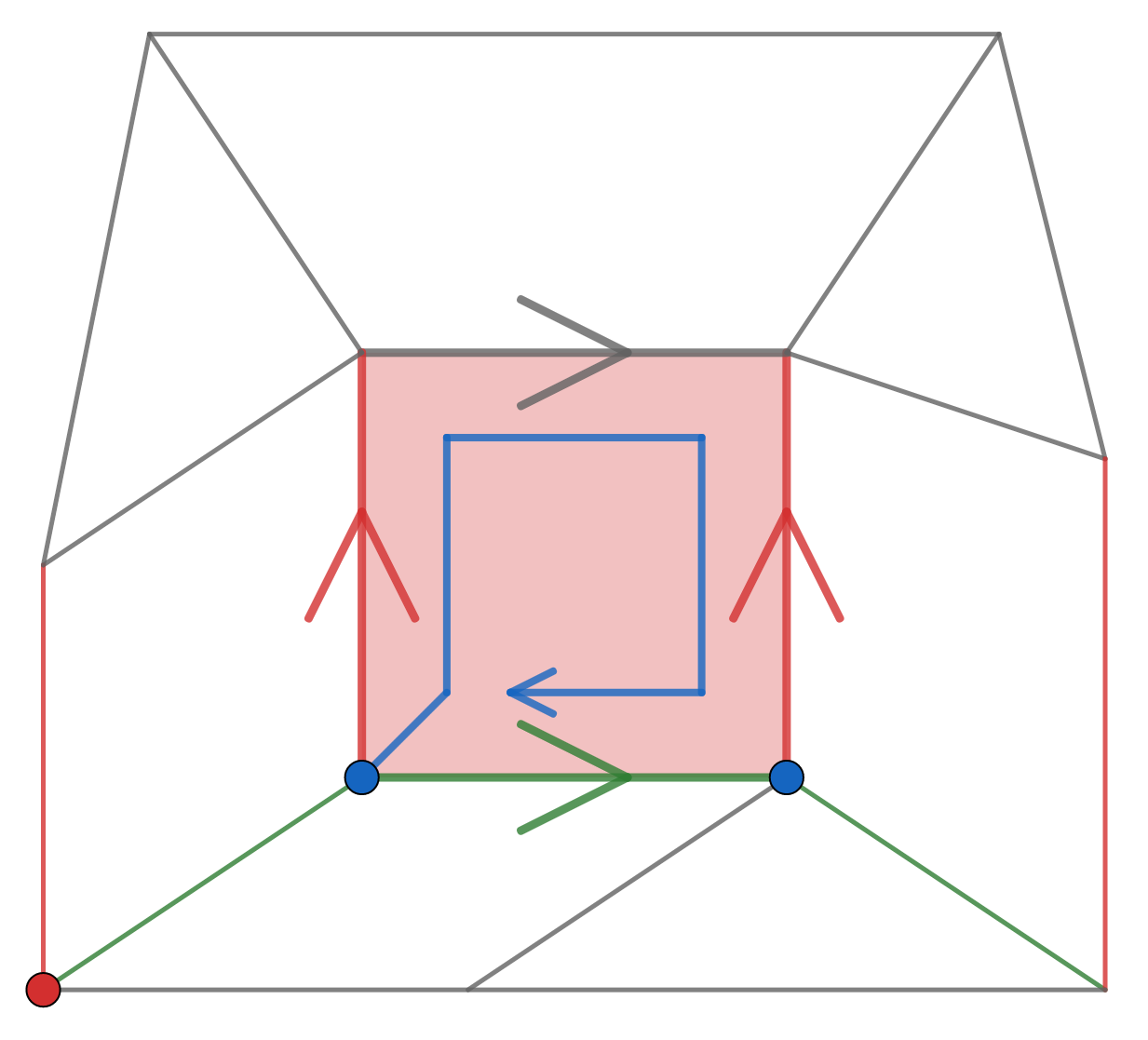}
				\put(0,0){start-point $s.p$}
				\put(30,18){$v_i$}
				\put(64,18){$v_{i+1}$}
				\put(16,8){$g(s.p-v_i)$}
				\put(42,14){$g(1-2)$}
				\put(50,64){$g_x$}
				\put(27,34){$g_1$}
				\put(70,34){$g_2$}
				\put(50,38){$e_p$}
			\end{overpic}
			\caption{Because we can freely change the base-point and orientation of a plaquette without affecting the plaquette energy term, we can choose the base-point of the plaquette from Figure \ref{plaquette_on_magnetic_ribbon} to be at $v_i$ and fix an orientation for the plaquette. This means that we have to check the commutation relation between the energy term and the ribbon operator for fewer cases.}
			\label{based_plaquette_on_magnetic_ribbon}
		\end{center}
	\end{figure}

	We first consider such internal plaquettes, such as the example in Figure \ref{plaquette_on_magnetic_ribbon}. The fact that the dual paths enters and exits the plaquette means that the labels of two edges on the boundary of the plaquette are changed by the magnetic ribbon operator (or more generally, an even number of transformations are applied on the edges if the ribbon self-intersects). As we described in Section \ref{Section_Recap_Paper_2} of the main text, the plaquette term checks that the plaquette holonomy (or flux) is the identity element $1_G$, and energetically punishes the state if this is not satisfied. The plaquette holonomy $H_1(p)$ of a plaquette $p$ is $\partial(e_p)g_p$, where $e_p$ is the label of the plaquette and $g_p$ is the path element for the boundary of the plaquette (starting at the base-point). When considering how this plaquette holonomy is changed by the magnetic ribbon operator there are several cases to consider, depending on the position of the base-point of the plaquette, the orientation of the plaquette and the orientations of the edges. However, the base-point that we choose does not actually affect the energy term, because using a different base-point only leads to conjugation of the plaquette holonomy, as we proved in the Appendix of Ref. \cite{HuxfordPaper1}, and the plaquette term checks that the plaquette holonomy is equal to the identity. Similarly, swapping the orientation of the plaquette (and changing the label from $e_p$ to $e_p^{-1}$) will invert the plaquette holonomy and so leave the identity invariant. Equally the base-point and orientation of the plaquette do not affect the action of the ribbon operator. Therefore, we can freely choose the base-point and orientation of the plaquette, and check if the action of the plaquette energy term and magnetic ribbon operator commute. If so, they will do so for any orientation of the plaquette and any choice of base-point. We choose the base-point and orientation to match the situation indicated in Figure \ref{based_plaquette_on_magnetic_ribbon}. Note that in Figure \ref{based_plaquette_on_magnetic_ribbon}, the grey and green paths can contain any number of edges, so we have not made any assumptions about the shape of the plaquette.

	Considering Figure \ref{based_plaquette_on_magnetic_ribbon}, we see that
	\begin{align}
		C^h(t):H_1(p) &= C^h : \partial(e_p) g_1 g_x g_2^{-1} g(1 -2)^{-1} \notag\\
		&= \partial(e_p) g(s.p-v_i)^{-1} hg(s.p-v_i) g_1 g_x g_2^{-1} (g(s.p-v_i)g(1-2))^{-1}h^{-1}(g(s.p-v_i)g(1-2)) g(1-2)^{-1} \notag\\
		&= \partial(e_p) g(s.p-v_i)^{-1} hg(s.p-v_i) g_1 g_x g_2^{-1} g(1 -2)^{-1} g(s.p-v_i)^{-1} h^{-1} g(s.p-v_i) \notag\\
		&= g(s.p-v_i)^{-1}hg(s.p-v_i) \partial(e_p) g_1g_x g_2^{-1} g(1 -2)^{-1} g(s.p-v_i)^{-1} h^{-1}g(s.p-v_i) \notag\\
		&= g(s.p-v_i)^{-1}hg(s.p-v_i) H_1(p) g(s.p-v_i)^{-1} h^{-1}g(s.p-v_i). \label{Equation_magnetic_ribbon_internal_plaquette_1}
	\end{align}
	This preserves the identity, so the ribbon operator commutes with the plaquette term $B_p =\delta(H_1(p),1_G)$. Note that if one of the edges labelled by $g_1$ and $g_2$ has the opposite orientation, then we must replace the label with its inverse. However if we also remember that swapping the orientation of the edges will cause them to transform in the opposite way under the magnetic ribbon, with their labels being post-multiplied by an element $g(t)^{-1}h^{-1}g(t)$, instead of pre-multiplied by $g(t)^{-1}hg(t)$, the mathematics works out in a similar way and our conclusion will still hold. Note also that if the dual path self intersects, so that the plaquette is passed through multiple times, then we have one transformation of the form given in Equation \ref{Equation_magnetic_ribbon_internal_plaquette_1} for each time the dual path passes through the plaquette. Because one of these transformations preserves the plaquette holonomy, so will a sequence of them.

	We can use the argument above to consider the two plaquettes at the ends of the dual path. The internal plaquettes have one transformation from the dual path entering the plaquette and one from the dual path exiting the plaquette, each contributing $g(s.p-v_i)^{-1}hg(s.p-v_i)$ or $g(s.p-v_i)^{-1}h^{-1}g(s.p-v_i)$. These factors cancel when acting on a fake-flat plaquette, as indicated in Equation \ref{Equation_magnetic_ribbon_internal_plaquette_1}. However if the dual path only exits or enters the plaquette then we only gain one such factor and so there is no cancellation. This means that the plaquette holonomy is multiplied by a factor of the form $g(t)^{-1}hg(t)$ (or the inverse), which is only trivial when $h=1_G$ (i.e., when the magnetic ribbon operator itself is trivial). Therefore, any non-trivial open magnetic ribbon operator excites the plaquettes at the two ends of the ribbon operator. Combining this with our previous results, we have therefore shown that the magnetic ribbon operator commutes with each energy term, except the vertex term at the start-point of the ribbon and the plaquette terms at the start and end of the ribbon.
	
	\subsection{$E$-valued membrane operators}
	\label{Section_E_Membrane_Commutation_Proof}

	We now want to derive the commutation relations between the $E$-valued membrane operators and the energy terms. The $E$-valued membrane operators measure the total value of a surface and assigns a weight for each value measured. Therefore, to find the commutation relations we need to understand how a surface built from multiple plaquettes interacts with each of the energy operators. The discussion will hold for both the 2+1d case considered in this paper and the 3+1d case that we examine in a future paper \cite{HuxfordPaper3}. Firstly, note that the $E$-valued membrane operators are diagonal in the configuration basis, so they commute with the plaquette energy terms (and the blob energy terms, when we consider the 3+1d case), which are similarly diagonal in this basis. This means that we only need to consider the vertex and edge transforms.

\begin{figure}[h]
		\begin{center}
			\begin{overpic}[width=0.8\linewidth]{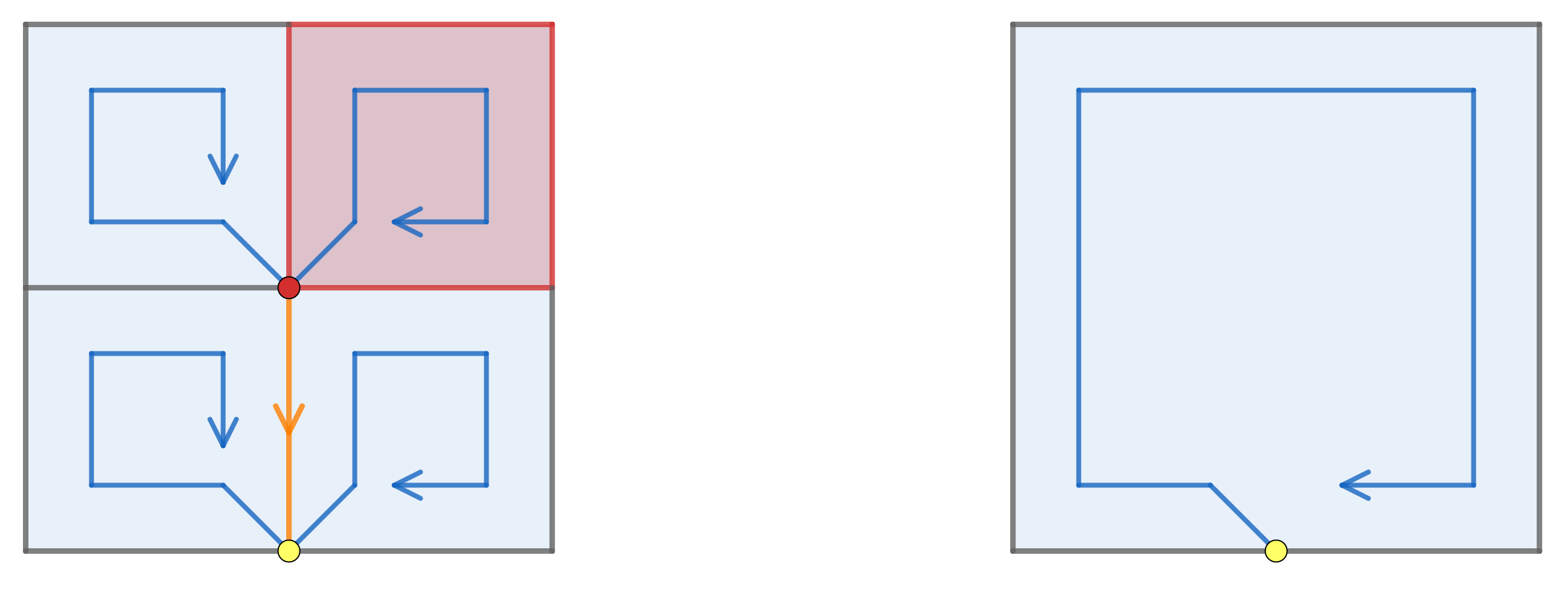}
				\put(47,18){\Huge $\rightarrow$}
				\put(45,14.5){combine}
				\put(20,30){$e_p \rightarrow$}
				\put(21,27){$g(t)^{-1} \rhd e_p$}
				\put(19,10){$g(t)$}
			\end{overpic}
			\caption{When we combine a plaquette (such as the red one in the top-right of the left side of the figure) with other plaquettes to form a larger surface, we must match the base-point of this plaquette (originally the red vertex, which is darker in grayscale) with the base-point of the combined surface (the yellow vertex, which is lighter in grayscale) by whiskering the plaquette to move the base-point of the plaquette. This changes the label from $e_p$ to $g(t)^{-1} \rhd e_p$ where $t$ is the path from the original base-point of the plaquette to the base-point of the combined surface.}
			\label{whiskered_plaquette_combination}
		\end{center}
	\end{figure}

	Recall that a surface is built from one or more plaquettes. In order to combine the individual plaquettes, we need to follow the rules for combining surfaces laid out in Ref. \cite{Bullivant2017}. Namely, we need to ensure that when we combine two surfaces, the target of one of these surfaces matches the source of the other surface. Part of this condition is that the base-points of the two surfaces must be the same. We can consider the composition of multiple surfaces as a sequence of pairwise combinations of surfaces. This means that when we combine the plaquettes into a total surface, we must whisker each of these plaquettes so that they have the same base-point (this is a necessary but not sufficient condition for being able to combine the surfaces). This common base-point is called the start-point of the membrane operator. Consider such a plaquette $p$, whiskered along a path $t$ with label $g(t)$, as illustrated in Figure \ref{whiskered_plaquette_combination}. If the label of the plaquette before whiskering is $e_p$, the label after whiskering is $g(t)^{-1} \rhd e_p$. This means that if we consider a membrane $m$ made out of a combination of such plaquettes, all whiskered so that their base-point is at the same vertex $s.p(m)$ (the start-point of the surface), the total surface label of a membrane $m$ has the form
	\begin{equation}
		\hat{e}(m) = \prod_{\text{plaquettes }p \in m} g(v_0(p) -s.p(m))^{-1} \rhd e_p^{\sigma_p}. \label{Equation_surface_label_product_appendix}
	\end{equation}
	In this expression $(v_0(p) -s.p(m))$ are paths from the base-point of the plaquette $p$ to the start-point of the membrane (i.e., the path that we must whisker $p$ along) and $\sigma_p$ is either $+1$ or $-1$ to account for the possibility that the plaquette orientation must be reversed to combine $p$ into the surface.

	We then wish to consider how the surface label transforms under the vertex and edge transforms, starting with the vertex transforms. Because the total surface is then made of a product of terms of the form $g(t)^{-1} \rhd e_p$, we wish to know how this transforms under vertex transforms. First, note that the vertex transforms that may change this quantity are the ones that change $g(t)$ and the ones that change $e_p$. $g(t)$ is changed by the vertex transforms at the start and end of the path $t$, while $e_p$ is changed by the transform at the original base-point of that plaquette (the red dot in Figure \ref{whiskered_plaquette_combination}), which is also the start of the path $t$. We therefore need to consider the vertex transforms at these two vertices. First consider the transform at the original base-point. The transform $A_v^g$ at the base-point takes $e_p$ to $g \rhd e_p$. Because this vertex is also the start of the path $t$, the transform at this vertex takes $g(t)$ to $gg(t)$, as described in Section \ref{Section_Electric_Ribbon_Operator_Proof}. This means that $A_v^g : g(t)^{-1}\rhd e_p = (gg(t))^{-1} \rhd (g \rhd e_p) = g(t)^{-1} \rhd e_p$. That is, this transform does not change the label $g(t)^{-1} \rhd e_p$ that appears in the total surface operator. Now consider the transform at the end of the path $t$ (the yellow dot in Figure \ref{whiskered_plaquette_combination}). This does not affect the label $e_p$, but the transform $A_v^g$ at this vertex takes $g(t)$ to $g(t)g^{-1}$. Therefore, the transform takes $g(t)^{-1} \rhd e_p$ to $ (g(t)g^{-1})^{-1} \rhd e_p = g \rhd (g(t)^{-1} \rhd e_p)$. That is, only the transform at the new base-point (after whiskering) changes the label and it changes the label by a $g \rhd$ action. Note that this action is the same as the action when we apply a vertex transform on the base-point of an unwhiskered plaquette, as we expect from the fact that the vertex transform is consistent with the procedure for changing the base-point of a plaquette (as we showed in the Appendix of Ref. \cite{HuxfordPaper1}).

	For the terms $g(t)^{-1} \rhd e_p$ in the surface element, the end of path $t$ is always the start-point of the membrane operator (which is the base-point of the combined surface). This tells us that the transform $A_v^g$ at the start-point of the membrane operator is the only vertex transform that affects the combined surface. Then the action of the vertex transform at the start-point on the total surface label is
	\begin{align*}
		A_v^g : e(m) &= \prod_{\text{plaquettes $p$ in } m} (gg(v_0(p)-s.p(m))^{-1}) \rhd e_p^{\sigma_p} \\
		&= \prod_{\text{plaquettes $p$ in } m} g \rhd (g(v_0(p)-s.p(m))^{-1} \rhd e_p^{\sigma_p})\\
		&= g \rhd e(m).
	\end{align*}
	This means that 
	$$\hat{e}(m) A_v^g = A_v^g \: [g \rhd \hat{e}(m)].$$
	Therefore
	\begin{align}
		\sum_{e \in E} \alpha_e \delta( \hat{e}(m),e) A_v^g &= A_v^g \sum_{e \in E} \alpha_e \delta( g \rhd \hat{e}(m),e) \notag \\
		&= A_v^g \sum_{e \in E} \alpha_e \delta( \hat{e}(m), g^{-1} \rhd e) \notag\\
		&= A_v^g \sum_{e' = g^{-1} \rhd e} \alpha_{ g \rhd e'} \delta(\hat{e}(m), e'). \label{Equation_E_valued_start_point_commutation_appendix}
	\end{align}
	
	Whether this leads to the start-point vertex being excited depends on the coefficients $\alpha$. If $\alpha_{g \rhd e} = \alpha_e$ for all $g \in G$ and $e \in E$ then the membrane operator commutes with the vertex transforms and the vertex is not excited. However if this does not hold, then the vertex may be excited, as described in Section \ref{Section_2D_RO_Fake_Flat} of the main text.

	An interesting thing to note about the action of the vertex transform on the membrane operator is that it has the same form as the transformation induced by moving the start-point of the membrane. Consider moving the start-point of $m$ to a new position $s.p(m')$, along a path $(s.p(m)-s.p(m'))$. We denote the membrane with this new start-point (but otherwise unchanged) by $m'$. Then the paths from the start-point of this new membrane to the plaquettes are the same as the paths on the old membrane, except that they start with the path section $(s.p(m')-s.p(m))$. Therefore, the surface element for the new membrane is given by
	\begin{align*}
		\hat{e}(m')&= \prod_{\text{plaquettes }p \in m} g(s.p(m')-v_0(p)) \rhd e_p^{\sigma_p}\\
		&= \prod_{\text{plaquettes }p \in m} (g(s.p(m')-s.p(m))g(s.p(m)-v_0(p))) \rhd e_p^{\sigma_p}\\
		&= g(s.p(m')-s.p(m)) \rhd \big(\prod_{\text{plaquettes }p \in m} g(s.p(m)-v_0(p)) \rhd e_p^{\sigma_p}\big)\\
		&= g(s.p(m')-s.p(m)) \rhd \hat{e}(m).
	\end{align*}
	
	Then when we change the start-point of a membrane operator $\sum_{e \in E} \alpha_e \delta( \hat{e}(m),e)$, it becomes
	\begin{align*}
		\sum_{e \in E} \alpha_e \delta( \hat{e}(m'),e) &= \sum_{e \in E} \alpha_e \delta( g(s.p(m')-s.p(m)) \rhd \hat{e}(m),e)\\
		&=\sum_{e \in E} \alpha_e \delta( \hat{e}(m), g(s.p(m')-s.p(m))^{-1} \rhd e)\\
		&=\sum_{e'= g(s.p(m')-s.p(m))^{-1} \rhd e \in E} \alpha_{g(s.p(m')-s.p(m)) \rhd e'} \delta( \hat{e}(m), e').
	\end{align*}
	This is the same transformation that would be induced by commutation with a vertex transform $A_v^{g(s.p(m')-s.p(m))}$ at the start-point of $m$ (see Equation \ref{Equation_E_valued_start_point_commutation_appendix}). This indicates that, if the membrane operator commutes with the vertex transforms (i.e., it does not excite the start-point) then it is also invariant under changes to the start-point position. We may expect this from the general idea that the vertex transform acts like parallel transport of that vertex along an edge, as we explained in Ref. \cite{HuxfordPaper1}.
	
\begin{figure}[h]
		\begin{center}
			\begin{overpic}[width=0.5\linewidth]{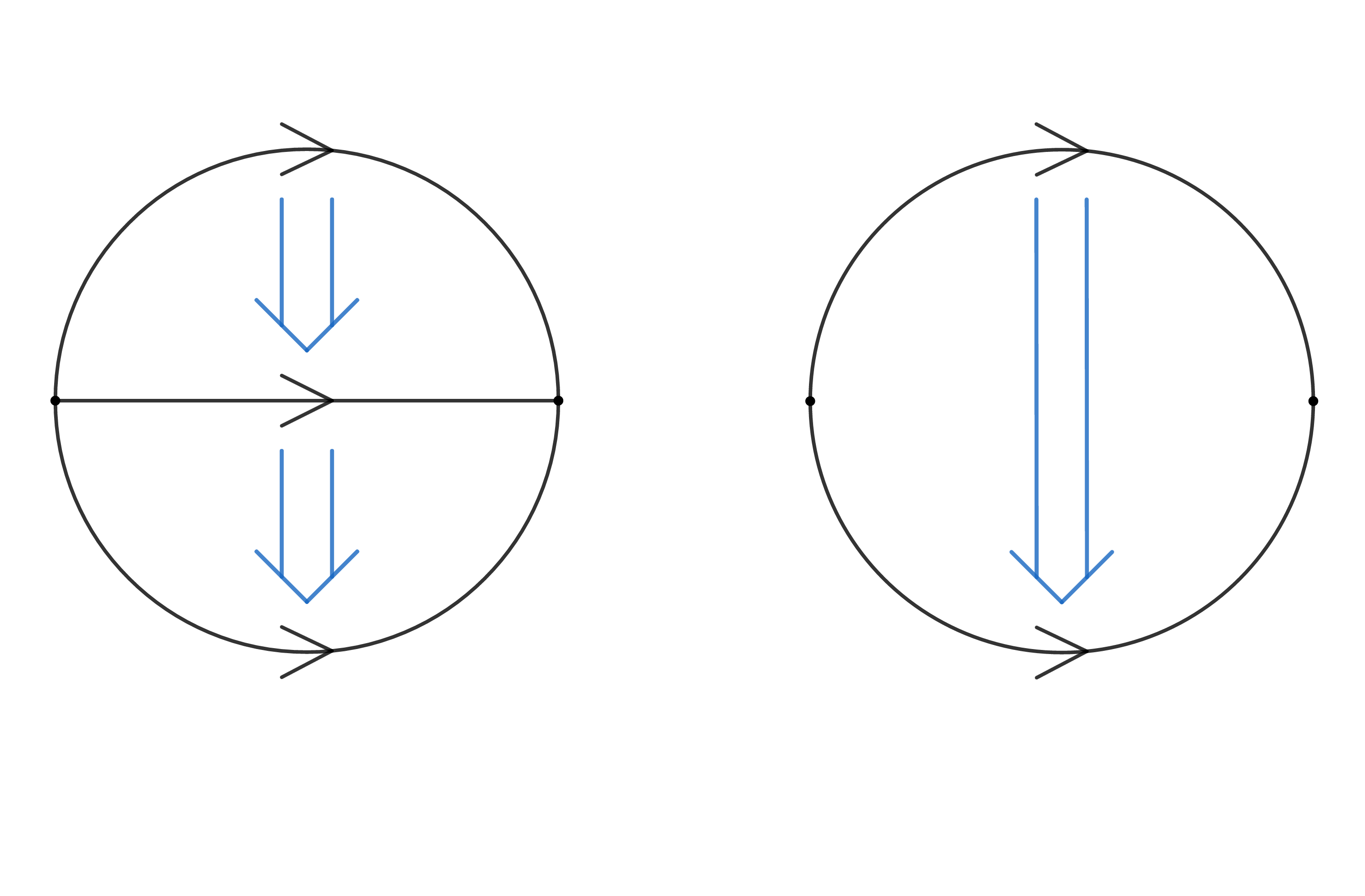}
				
				\put(46,32){\Huge $\rightarrow$}
				\put(43,38){combine}
				
				\put(20,58){$s_1$}
				\put(17,38.5){$t_1=s_2$}
				\put(20,13){$t_2$}
				
				\put(68,58){$s_1=s_{\text{combined}}$}
				\put(68,13){$t_2=t_{\text{combined}}$}
				
				\put(27,45){$e_1$}
				\put(27,27){$e_2$}
				\put(82,35){$e_2 e_1$}
				
			\end{overpic}
			\caption{Vertical composition of two 2-holonomies. $s$ refers to the source of a given surface and $t$ to its target. To combine we need the source of the second surface to be the target of the first. Then the source of the first surface becomes the source of the combined surface and the target of the second surface becomes the target of the combined surface.}
			\label{vertical_composition_source_target}
		\end{center}
	\end{figure}

	Next we must consider how the edge transforms commute with the total surface operator. In order to do this, we must consider the combination of surface elements in more detail. We consider the method laid out in Ref. \cite{Bullivant2017} for combining surfaces. Surfaces have a source path and a target path. To combine surfaces we match the target of one to the source of the other, as indicated in Figure \ref{vertical_composition_source_target}.
	
\begin{figure}[h]
		\begin{center}
			\begin{overpic}[width=0.4\linewidth]{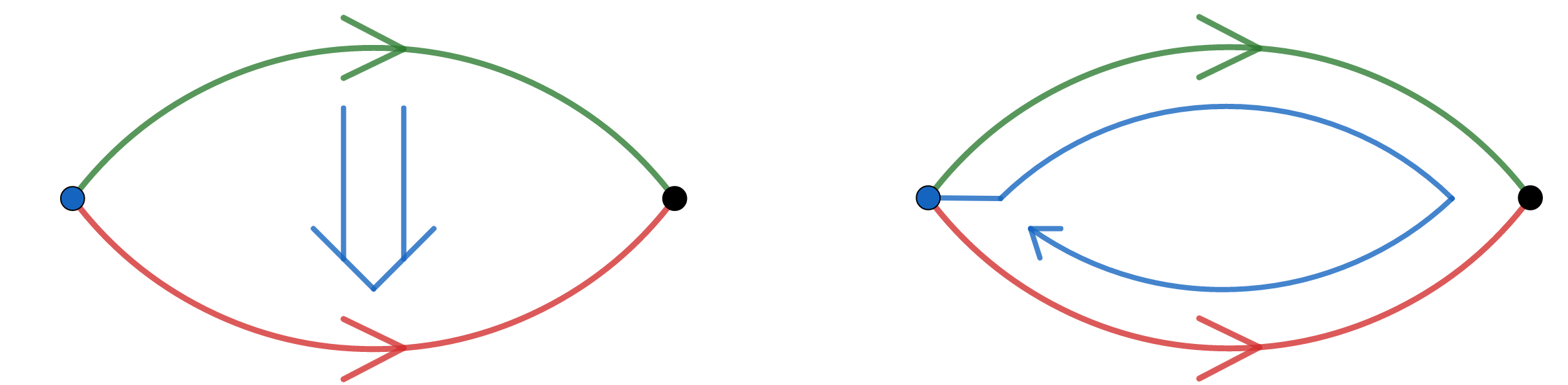}
				\put(46,9.5){\Huge $\rightarrow$}
			\end{overpic}
			
			\caption{(Copy of Figure 26 from Ref. \cite{HuxfordPaper1}.) Instead of illustrating the orientation of the surface as an arrow between the source and target, we can draw it as an arrow that passes clockwise or anticlockwise around the surface, starting and ending at the base-point of the surface. The direction of this arrow matches the direction of the source.}
			\label{surface_switch_notation_appendix}
		\end{center}
	\end{figure}

	As described in Ref. \cite{Bullivant2017}, we can move edges between source and target without changing the surface label, by moving the end-point of the surface (the end of the source and target paths). Because of this redundancy, in this series of papers we do not generally specify the end-point of each surface, and present this visually by combining the source and target of each surface, as indicated in Figure \ref{surface_switch_notation_appendix}, to draw the surface as a circulation. The rules for combining surfaces in our new presentation follow from the original rules for combining surfaces. Because we do not specify the end-point, which divides the source and target, instead we require that when combining two surfaces, it must be possible to choose the source and target of each surface (by choosing an end-point) so that the target of the first matches the source of the next. In other words, while we do not specify the end-points of surfaces, there must be some choice of end-points such that the composition of the surface is valid in the original presentation. This means that the two surfaces must have the same base-point (or we must change the base-point of one of the surfaces before combining them, with the accompanying change to the surface label) and the same orientation (or we must swap the orientations, again changing the surface label).
	
\begin{figure}[h]
		\begin{center}
			\begin{overpic}[width=0.7\linewidth]{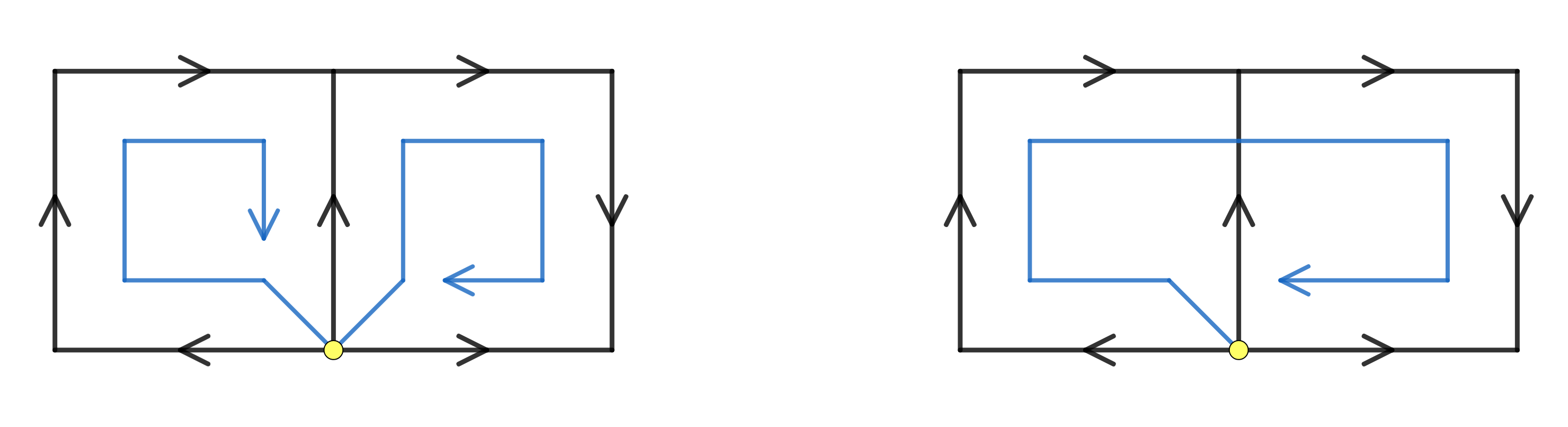}
				\put(45,12){\Huge $\rightarrow$}
				\put(43,10){Combine}

				\put(10,2){$t_1$}
				\put(0,15){$t_2$}
				\put(10,25){$t_3$}
				\put(18,16){$t_4$}
				\put(31,25){$t_5$}
				\put(41,15){$t_6$}
				\put(31,2){$t_7$}
				\put(11,13){$A$}
				\put(29,13){$B$}
				
				\put(68,2){$t_1$}
				\put(58,15){$t_2$}
				\put(68,25){$t_3$}
				\put(76,16){$t_4$}
				\put(89,25){$t_5$}
				\put(99,15){$t_6$}
				\put(89,2){$t_7$}
				\put(81,13){$AB$}
				
			\end{overpic}
			\caption{When we combine two surfaces $A$ and $B$, we combine their boundaries. In this case the boundary of $A$ is $t_1t_2t_3 t_4^{-1}$ and the boundary of $B$ is $t_4 t_5 t_6 t_7^{-1}$. Therefore, the boundary of $AB$ is $t_1t_2t_3 t_4^{-1}t_4 t_5 t_6 t_7^{-1}$. Note that the section $t_4^{-1} t_4$ can be removed from the boundary without affecting the surface label or the label of the boundary. This gives us the path $t_1t_2t_3t_5t_6t_7^{-1}$, which is the more natural boundary of the surface $AB$.}
			\label{combine_two_surfaces}
		\end{center}
	\end{figure}

	To describe the composition of surfaces in more detail, we define the \textit{boundary} of a surface as the source path of the surface, followed by the inverse of its target. That is, the boundary of the surface is the path all the way around the surface, starting at the base-point and following the orientation of the surface. The boundary is a composition of multiple edges, with inverses if the edge orientation is anti-aligned with the boundary's orientation. We start by taking the naive approach to composing surfaces, where the boundary path of the combined surface is the product of the paths of the two individual surfaces. This corresponds to the case where we manipulate the sources and targets of the surfaces so that the target of the first surface is empty (the source is the whole boundary) and the source of the second surface is similarly empty (the inverse of the target is the whole boundary). For example, consider Figure \ref{combine_two_surfaces}, where we have two surfaces $A$ and $B$, with boundary paths $\text{bd}(A)$ and $\text{bd}(B)$, and where the surfaces have labels $e_A$ and $e_B$. This figure illustrates the point that our naive way of combining surfaces will leave sections of the boundary that can be removed by manipulating the source and target. Generally, whenever we have a section of path that appears twice consecutively in the boundary, with opposite orientation (i.e., the path includes a section such as $t_1 t_1^{-1}$), the two appearances of the section can be removed without affecting the surface label or the label of the boundary.
	
	Then the combined surface $AB$ has a surface label $e_{AB}=e_Be_A$ and a boundary path $\text{bd}(A)\text{bd}(B)$. We can use these composition rules to show that the combined surface satisfies a fake-flatness rule, as long as the constituent surfaces do as well. We can write the fake-flatness conditions on the individual surfaces as
	\begin{align*}
		\partial(e_A)g(\text{bd}(A))&=1_G \iff g(\text{bd}(B))^{-1}\partial(e_A)g(\text{bd}(A))g(\text{bd}(B))=1_G \intertext{and}\\
		\partial(e_B)g(\text{bd}(B))&=1_G.
	\end{align*}
	Then we can put the two conditions together to obtain
	\begin{align*}
		[\partial(e_B)g(\text{bd}(B))]&[g(\text{bd}(B))^{-1}\partial(e_A)g(\text{bd}(A))g(\text{bd}(B))]=1_G \cdot 1_G\\
		&\implies \partial(e_B)\partial(e_A)g(\text{bd}(A))g(\text{bd}(B))=1_G\\
		&\implies \partial(e_Be_A)g(\text{bd}(A)\text{bd}(B))=1_G\\
		& \implies \partial(e_{AB})g(\text{bd}(AB))=1_G,
	\end{align*}
	where $e_{AB}$ is the surface label of surface $AB$ and $g(\text{bd}(AB))$ is the label of its boundary (note that surface labels are combined right-to-left whereas path labels are combined left-to-right)). This is the fake-flatness condition on the combined surface, so we see that fake-flatness on the constituent surfaces implies fake-flatness on the combined surface. Because any surface can be built from a series of pairwise combinations of surfaces and the building blocks of these surfaces (the plaquettes) satisfy fake-flatness in the ground-state (even if we whisker the plaquettes or invert their orientation, as illustrated in Section \ref{Section_Magnetic_Ribbon_Proof}), every surface has this fake-flatness condition in the ground-state. This in turn means that every contractible closed path on the lattice must have a path label in $\partial(E)$ when the surface enclosed by the path contains no plaquettes that violate fake-flatness.

	Now we want to consider the transformation properties of combined surfaces under the edge transforms. First we will give an ansatz for the transformation of surfaces under edge transforms. Then we will prove that this ansatz is correct by induction. To do this we first show that, given two surfaces that transform according to our ansatz, the combination of the two surface also transforms according to this ansatz. Finally we will check that these rules hold for individual plaquettes, from which we can build the larger surfaces.

	Consider applying an edge transform $\mathcal{A}_i^e$ onto an edge which belongs to the boundary of a surface $A$ (an edge transform on an edge not on the boundary will not affect the surface). If the edge $i$ on which we apply the edge transform $\mathcal{A}_i^e$ appears once in the set of edges associated to the boundary of the surface (the boundary passes through the edge precisely once), we claim that the surface transforms in the same way as an individual plaquette does under the edge transform:
	\begin{equation}
		e_A \rightarrow \begin{cases} e_A [g(v_0-v_i|A) \rhd e^{-1}] & \text{ edge `right-way' on A}\\
			[g(\overline{v_0-v_{i+1}}|A) \rhd e] e_A & \text{ edge `wrong-way' on A.} \end{cases}
		\label{Equation_edge_transform_appendix}
	\end{equation}

	In Equation \ref{Equation_edge_transform_appendix}, `right-way' indicates that the edge's orientation matches the orientation of the boundary and `wrong-way' indicates that it is anti-aligned with the boundary. $|A$ after a path indicates that the path lies on the boundary of $A$. We also used the convention from Ref. \cite{Bullivant2017} that an overlined path travels against the orientation of a surface and the non-overlined path travels with the orientation of the surface. Note that the vertex $v_i$ or $v_{i+1}$ which appears in the transform is the source vertex of the edge, $s(i)$, where the source vertex is the start of the edge and the target vertex is the end of the edge (the edge points from its source to its target). From now on we will use $s(i)$ instead of $v_i$ or $v_{i+1}$, because this notation is less ambiguous for other scenarios we consider in this section.

	While useful for many surfaces, the transformation formula given in Equation \ref{Equation_edge_transform_appendix} is not sufficient to describe the transformation of a general surface. Whereas for an individual plaquette, each edge only appears once in the boundary of the plaquette, for a more general surface the same edge can appear multiple times. For example, if a surface is whiskered along a path, then the edges in that path appear twice in the boundary, with opposite orientation each time. If the edge on which we applied the transform appears multiple times in the boundary, the surface label gains the transformation in Equation \ref{Equation_edge_transform_appendix} once for each appearance of the edge in the boundary (though the label of the edge itself is only transformed once, gaining a factor $\partial(e)$ from the transform $\mathcal{A}_i^e$). In this case the expression $(v_0-s(i)|A)$ (and the equivalent overlined expression) in Equation \ref{Equation_edge_transform_appendix} is ambiguous because there are multiple paths on the surface that end at $s(i)$. We must instead specify that the path $(v_0-s(i)|A)$ refers to the path up to that particular occurrence of the edge in the boundary. We refer to a particular occurrence of the edge $i$ in the boundary by ordering the occurrences according to which appearance occurs earliest in the boundary path. We then assign each occurrence a numerical index and use squared brackets to refer to it. For example, to refer to the first occurrence we use `$[1]$'. Consider a surface $A$ with boundary $t_1it_2it_3$, where $i$ is the edge on which we apply the transform and the $t_x$ are the other sections of the path. Then there are two right-way appearances of the edge $i$ (we would indicate a ``wrong-way" appearance with an inverse to indicate that the boundary includes $i$ in the opposite orientation). For the first appearance (the $i$ just after $t_1$), the relevant path element is $g(v_0-s(i)[1]|A)=g(t_1)$. Therefore, we gain a factor of $g(t_1) \rhd e^{-1}$. For the second (the $i$ after $t_2$), the relevant path element $(v_0-s(i)[2]|A)$ is $t_1 i t_2$ and so we gain a factor of $g(t_1it_2) \rhd e^{-1}$. There are, however, some issues with this. Firstly, these added factors do not commute in general, and it is not clear in which order we should add these factors. Secondly, when we apply an edge transform the edge label itself gains a factor of $\partial(e)$, and it is not obvious whether this factor should be included in paths elements such as $g(t_1it_2)$. When $E$ is Abelian neither of these concerns matter, but if we want to consider the case of a general crossed module, these must be addressed.

	The simplest way to sort this is to slightly alter the way in which we state the action of the transform. We will deal only with the cases where fake-flatness is satisfied in the region on which we apply our membrane operator, though our argument also holds with minimal alteration when we instead restrict our crossed module so that $E$ is Abelian and $\partial$ maps in to the centre of $G$ (i.e., if we take Case 2 from Table \ref{Table_Cases_2d} in the main text) but only enforce that fake-flatness is satisfied up to an element of $\partial(E)$ in the region of our operator. When considering this latter case, we say that two elements of $G$, $g_1$ and $g_2$, which differ only by a factor in $\partial(E)$ are similar, and write this as $g_1 \sim g_2$. Because we care only about the action of the group elements on surface labels (i.e., terms like $g_1 \rhd e$), the factor of $\partial(e)$ does not affect any of our results. Factors of $\partial(f)$ will not affect $g_1 \rhd e$, because $(\partial(f) g_1) \rhd e = f [g_1 \rhd e] f^{-1} = g_1 \rhd e$ if $E$ is Abelian, so that $g_1 \rhd e =g_2 \rhd e$ for all $e \in E$ if $g_1 \sim g_2$. In order to modify the argument given in the rest of this section to apply to this case, some equalities should be replaced by $\sim$ instead, but the final results hold. For the rest of the section we will assume fake-flatness, and not make any restrictions to the crossed module.

	The transformation on a surface element $e_A$ from the edge transform given in Equation \ref{Equation_edge_transform_appendix} can be rewritten into a more useful form. We use the fact that the two paths $(v_0 -s(i)|A)$ and $(\overline{v_0-s(i)}|A)^{-1}$, which both pass from the base-point of surface $A$ to a particular vertex on that surface (here $s(i)$), combine to give the boundary of $A$, $\text{bd}(A)$. This is because the two paths $(v_0 -s(i)|A)$ and $(\overline{v_0-s(i)}|A)$ travel in the opposite directions around the boundary of the surface and meet at the vertex. We use this to write that $g(v_0-s(i)|A) g(\overline{v_0-s(i)}|A)^{-1}= g(\text{bd}(A))= \partial(e_A)^{-1}$, utilizing fake-flatness in the last step. Therefore
	\begin{align}
		[g(\overline{v_0-s(i)}|A)\rhd e] e_A &= [(g(\text{bd}(A)^{-1})g(v_0-s(i)|A))\rhd e] e_A \notag\\
		&= [(\partial(e_A) g(v_0-s(i)|A))\rhd e] e_A \notag\\
		&=e_A [g(v_0-s(i)|A)\rhd e] e_A^{-1} e_A \text{ (using the second Peiffer condition, Equation \ref{Peiffer_2} in the main text)} \notag \\
		&=e_A [g(v_0-s(i)|A)\rhd e], \label{Equation_plaquette_transform_move_left_right}
	\end{align}
	where $g(\text{bd}(A)^{-1})$ is the path element for the path all the way around the surface $A$ in the wrong direction and where $|A$ after a path indicates that it is defined on the surface $A$. In addition, all paths take the value that they have before we apply the edge transform, so there are no additional $\partial(e)$ factors anywhere. The factor $g(v_0-s(i)|A)\rhd e$ is not quite the same as the factor which appears in a right-way edge transform because the factor $e$ rather than $e^{-1}$ appears, as indicated in Figure \ref{Equation_edge_transform_appendix} (and also $s(i)$ is now after the edge $i$ on the path $v_0-s(i)|A$).

\begin{figure}[h]
		\begin{center}
			\begin{overpic}[width=0.5\linewidth]{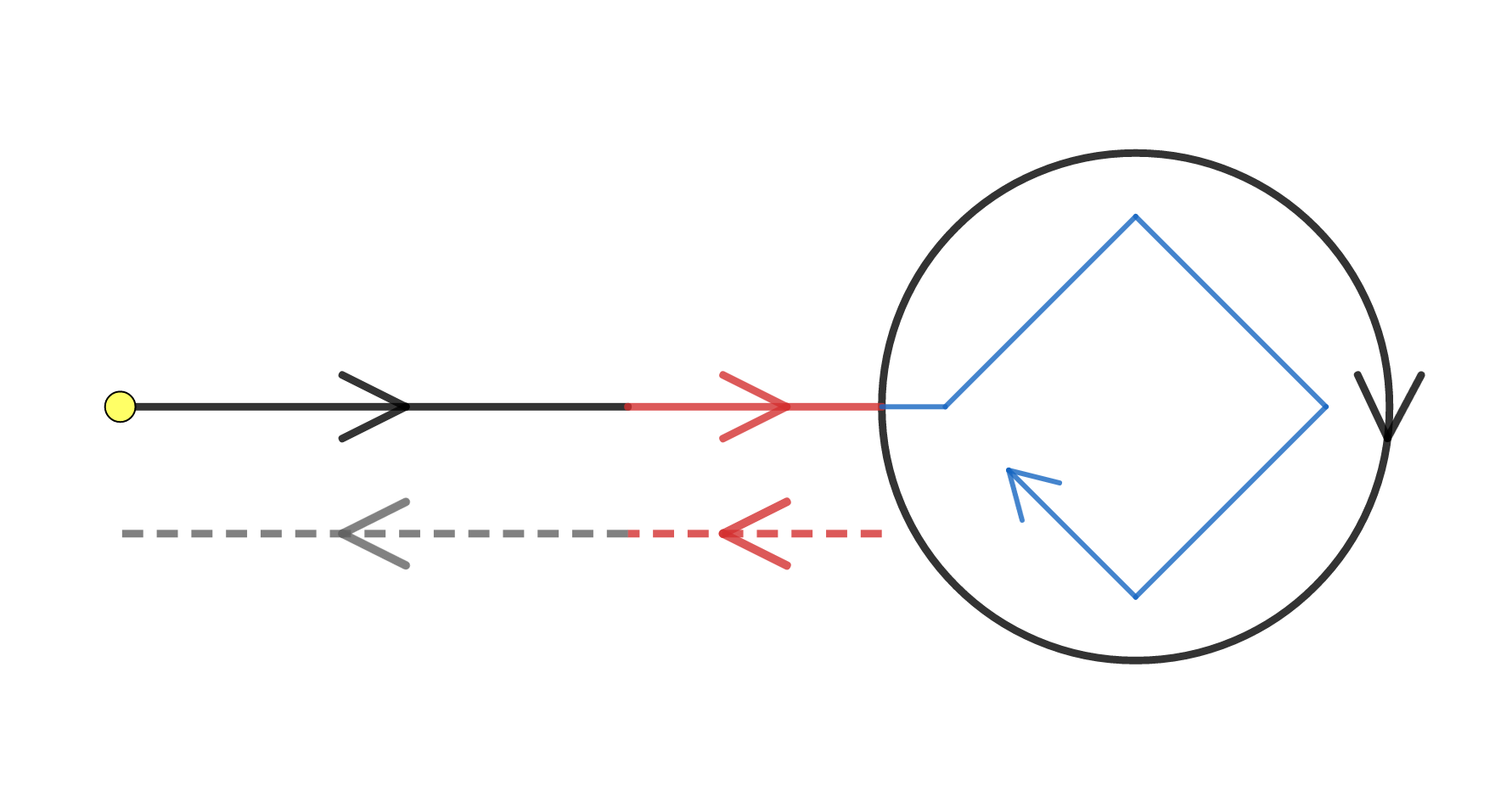}
				\put(25,30){$t_1$}
				\put(48,30){$i$}
				\put(75,43){$t_2$}
				\put(48,10){$i^{-1}$}
				\put(25,10){$t_1^{-1}$}
				\put(74,25){$A$}
				\put(0,30){\parbox{1.6cm}{base-point of $A$}}

			\end{overpic}
			\caption{If the beginning section of the boundary of a surface A (here the section is $t_1 i$) is repeated in reverse at the end of the boundary, this corresponds to the case where the surface is whiskered along that section. The result of an edge transform obtained from Equation \ref{Equation_surface_edge_transformation_1} matches the transformation we expect from the rules for whiskering a surface.}
			\label{edge_on_whisker_path}
		\end{center}
	\end{figure}

	Now that all of the paths that appear in our expressions are right-way, we can order each appearance of the edge on which we apply the transform from its first appearance in the boundary to its last. We apply the transformation factors in reverse order, applying the transform from the last appearance of the edge first. This means that the paths to each appearance do not cross over the appearances of the edge that we have already applied the transformation for. Then the surface label $e_A$ transforms under an edge transform $\mathcal{A}_i^e$ on the boundary as 
	\begin{equation}
		e_A \rightarrow e_A \prod_{\text{appearance }k} \big(g(v_0-s(i)[k]|A) \rhd e^{\gamma_k} \big),
		\label{Equation_surface_edge_transformation_1}
	\end{equation}
	where the product is taken from the last appearance of the edge in the boundary to the first (it counts down in $k$), and where $\gamma_k=-1$ if the edge points along the orientation of the boundary for appearance $k$ and is $+1$ if the edge points against the boundary for that appearance. Recall that $s(i)$ is the source vertex for the edge $i$. Now let's look at a couple of examples to see how this transformation works. First suppose that the boundary path of surface $A$ is $t_1it_2i^{-1}t_3$. Then the first appearance of the edge (which is ``right-way", or aligned with the boundary of the surface) is after $t_1$ and so $g(v_0-s(i)[1]|A)= g(t_1)$. On the other hand, the second appearance of the edge $i$ is ``wrong-way" (i.e., anti-aligned with the boundary). The path element $g(\overline{v_0-s(i)}[2]|A)$ corresponding to the path to this edge around the reversed boundary is simply $g(t_3)$. However the group element corresponding to the path aligned with the boundary that runs to this edge is $g(t_1it_2i^{-1})$. Therefore, under the edge transform the surface label $e_A$ transforms as
	\begin{align*}
		e_A \rightarrow& [g(t_3)^{-1} \rhd e] e_A [g(t_1) \rhd e^{-1}]\\
		=& e_A [g(t_1it_2i^{-1}) \rhd e] [g(t_1) \rhd e^{-1}],
	\end{align*}
	where the first line gives the transformation using the same form as Equation \ref{Equation_edge_transform_definition} (in the main text) for the transformation of an individual plaquette (albeit applying both transformations due to the two appearances of the edge in the boundary), and the second line gives the transformation in the form of Equation \ref{Equation_surface_edge_transformation_1}. Now consider the case where the boundary is $t_1it_2i^{-1}t_1^{-1}$ (i.e., we take $t_3=t_1^{-1}$ in the above case). This corresponds to whiskering, as indicated in Figure \ref{edge_on_whisker_path}.
	
	Then under the edge transform, Equation \ref{Equation_surface_edge_transformation_1} tells us that the surface transforms according to
	\begin{align*}
		e_A &\rightarrow e_A[(g(t_1)g_ig(t_2)g_i^{-1}) \rhd e] [g(t_1) \rhd e^{-1}].
	\end{align*}
	Using fake-flatness, we can write $\partial(e_A)g(t_1)g_ig(t_2)g_i^{-1}g(t_1)^{-1}=1_G$ and so
	$$(g(t_1)g_ig(t_2)g_i^{-1}) = \partial(e_A)^{-1} g(t_1).$$
	This allows us to use our standard trick to move the factor corresponding to the wrong-way appearance of edge $i$ to the left of the expression:
	\begin{align*}
		e_A &\rightarrow e_A[(g(t_1)g_ig(t_2)g_i^{-1}) \rhd e] [g(t_1) \rhd e^{-1}]\\
		&=e_A[(\partial(e_A)^{-1}g(t_1)) \rhd e] [g(t_1) \rhd e^{-1}]\\
		&=e_Ae_A^{-1}[g(t_1) \rhd e]e_A [g(t_1) \rhd e^{-1}]\\
		&=[g(t_1) \rhd e]e_A [g(t_1) \rhd e^{-1}],
	\end{align*}
	where we used the second Peiffer condition (Equation \ref{Peiffer_2} in the main text) to obtain the third line. In the case of this whiskered surface, we can further simplify this result:
	\begin{align*}
		e_A &\rightarrow [g(t_1) \rhd e] e_A [g(t_1) \rhd e^{-1}]\\
		&=g(t_1) \rhd (e [g(t_1)^{-1} \rhd e_A] e^{-1}) \text{ (using $g \rhd (e_1 e_2)= (g \rhd e_1)(g \rhd e_2))$}\\
		&=g(t_1) \rhd (\partial(e) \rhd [g(t_1)^{-1} \rhd e_A]) \text{ (using the second Peiffer condition, Equation \ref{Peiffer_2} in the main text)}\\
		&=(g(t_1)\partial(e)g(t_1)^{-1}) \rhd e_A \text{ (using $g_1 \rhd (g_2 \rhd e)=(g_1g_2) \rhd e$)}.
	\end{align*}
	
	The result given in the last line is exactly the same result that we expect from our rules for whiskering a surface. Consider the surface $A'$ that is obtained by moving the base-point of $A$ to the end of $i$ (unwhiskering it). Then the label of the original (whiskered) surface $A$ can be written in terms of this new surface as
	$$e_{A}=(g(t_1)g_i) \rhd e_{A'},$$
	from the rules for whiskering. If we apply the edge transform $\mathcal{A}_i^e$, then the surface $A'$ is unaffected (because $i$ is not on its boundary), but the term $g_i$ is altered by the transform (to $\partial(e)g_i$). We therefore obtain
	\begin{align*}
		e_A &\rightarrow (g(t_1) \partial(e)g_i) \rhd e_{A'}\\
		&=(g(t_1)\partial(e)g(t_1)^{-1} g(t_1)g_i) \rhd e_{A'},
	\end{align*}
	where we have inserted the identity in the form of $g(t_1)^{-1} g(t_1)$. This tells us
	\begin{align*}
		e_A &\rightarrow (g(t_1)\partial(e)g(t_1)^{-1} g(t_1)g_i) \rhd e_{A'}\\
		&= (g(t_1)\partial(e)g(t_1)^{-1}) \rhd (g(t_1)g_i) \rhd e_{A'}\\
		&= (g(t_1)\partial(e)g(t_1)^{-1}) \rhd e_A,
	\end{align*}
	which is the same result we obtained from applying Equation \ref{Equation_surface_edge_transformation_1}. This example illustrates a general idea, that the modified form of the edge transform we present here is consistent with the procedure for changing the base-point of a plaquette.

\begin{figure}[h]
		\begin{center}
			\begin{overpic}[width=0.7\linewidth]{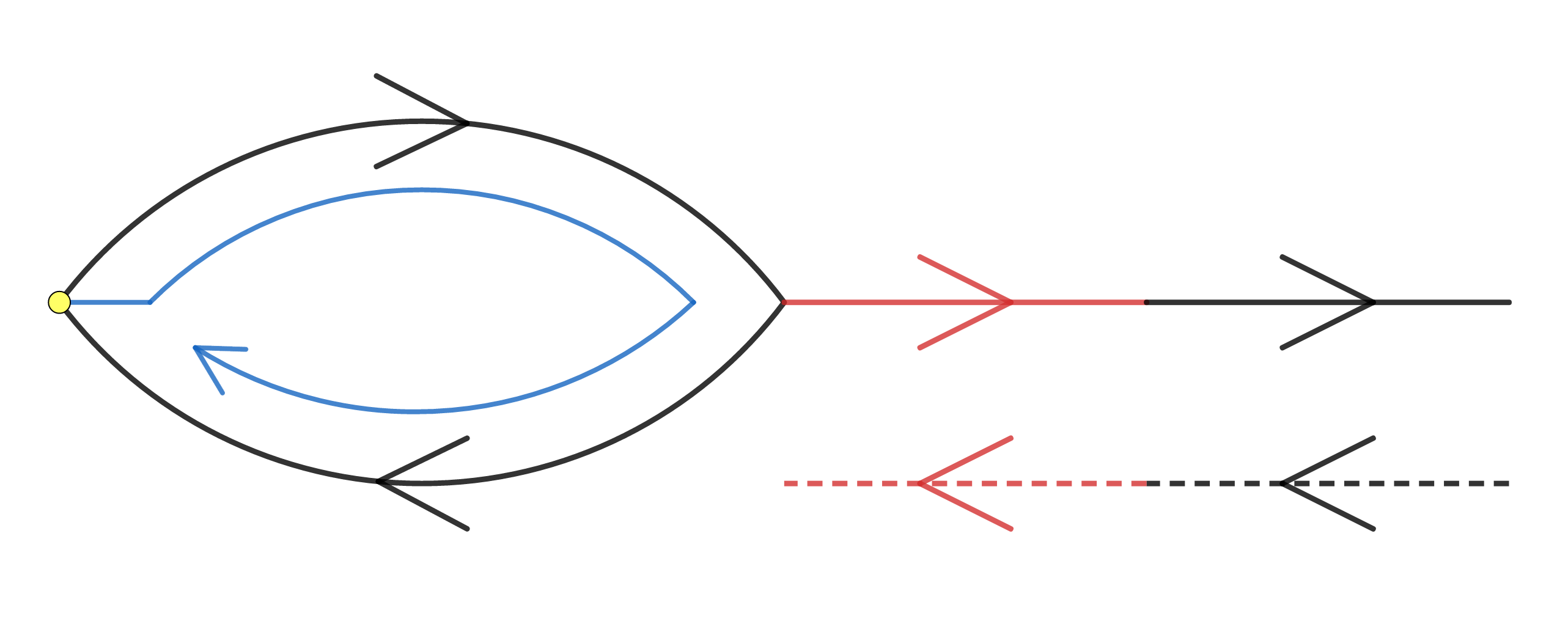}
				\put(30,34){$t_1$}
				\put(62,23){$i$}
				\put(86,23){$t_2$}
				\put(82,4){$t_2^{-1}$}
				\put(58,4){$i^{-1}$}
				\put(25,4){$t_3$}
				\put(26,19){$A$}

			\end{overpic}
			\caption{If a surface's boundary (here $t_1it_2t_2^{-1} i^{-1}t_3$) includes a section (here $i t_2$) which is travelled consecutively in opposite directions, then this section can be removed from the boundary without affecting the surface label. Therefore, we expect that an edge transform on the edge $i$ should not affect the surface label, to be consistent with the ability to remove the edge from the boundary.}
			\label{edge_on_end_point_whisker_path}
		\end{center}
	\end{figure}

	Next consider the case where the boundary is instead $\text{bd}(A)=t_1 i t_2 t_2^{-1} i^{-1} t_3$. This corresponds to the case shown in Figure \ref{edge_on_end_point_whisker_path} where we have included edges on the boundary that need not be on the boundary (we can remove them by switching edges between the source and target), and $i$ is one such edge (we can remove it by first eliminating the section $t_2t_2^{-1}$, then removing $ii^{-1}$). In this case Equation \ref{Equation_surface_edge_transformation_1} becomes
	\begin{align*}
		e_A &\rightarrow e_A (g(t_1 i t_2 t_2^{-1} i^{-1}) \rhd e) \: (g(t_1) \rhd e^{-1})\\
		&= e_A (g(t_1) \rhd e) \: (g(t_1) \rhd e^{-1})\\
		&=e_A,
	\end{align*}
	so that the surface label is unchanged. We expect this result, because the edge on which we apply the transform doesn't need to be included in the list of boundary edges at all, and edge transforms away from the boundary should leave the surface label invariant. This tells us that our transformation in Equation \ref{Equation_surface_edge_transformation_1} is sensible, because it does not depend on the parts of the boundary which can be arbitrarily chosen.
	
	Now we finally consider combining two surfaces. We consider a surface $A$ with label $e_A$ and a surface $B$ with label $e_B$ that combine into $AB$, with label $e_{AB} = e_B e_A$ (note that we assume that the surfaces have appropriate base-points and orientations to allow composition, or that we have already performed the necessary manipulations to ensure this). The boundary of $AB$ is $\text{bd}(AB)=\text{bd}(A)\text{bd}(B)$. Consider an edge transform on an edge on the boundary of $AB$. We assume that the surface labels $e_A$ and $e_B$ transform under the edge transform $\mathcal{A}_i^e$ according to Equation \ref{Equation_surface_edge_transformation_1}. This means that the total surface transforms as
	\begin{align}
		e_Be_A &\rightarrow e_B \bigg(\prod_{\substack{\text{appearance $k$} \\ \text{in }\text{bd}(B)}} \hspace{-0.5cm}g(v_0-s(i)[k]|B) \rhd e^{\gamma_k} \bigg) \: e_A \bigg(\prod_{\substack{\text{appearance $j$} \\ \text{in }\text{bd}(A)}} \hspace{-0.5cm}g(v_0-s(i)[j]|A) \rhd e^{\gamma_j}\bigg). \label{Equation_combined_surface_transformation_1}
	\end{align}
	
\begin{figure}[h]
		\begin{center}
			\begin{overpic}[width=0.8\linewidth]{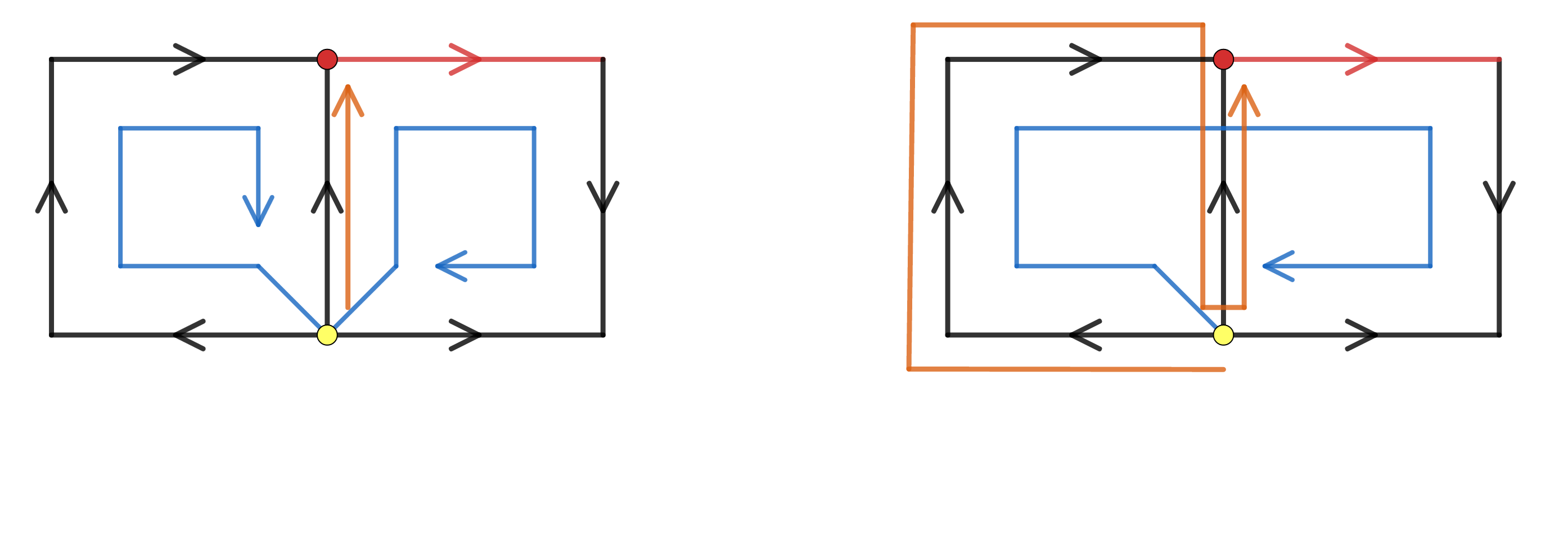}

				\put(11,22){$A$}
				\put(29,22){$B$}
				\put(16,33){$v_i=s(i)$}
				\put(28,33){$i$}
				
				\put(81,22){$AB$}
				
				\put(78,33){$s(i)$}
				\put(86,33){$i$}
				
				\put(15,5){\textcolor{orange}{$(v_0-s(i)|B)$}}
				\put(57,5){\textcolor{orange}{$(v_0-s(i)|AB)=\text{bd}(A)\cdot (v_0-s(i)|B)$}}
			\end{overpic}
			\caption{The path $v_0-s(i)$ (orange) that appears in the edge transform (on the red edge) when considering the effect of the transform on the surface $B$ is different from the path that appears when considering the effect on $AB$. This is because the boundary of $AB$ is the boundary of $A$ followed by the boundary of $B$, so the path to the vertex in $AB$ is the boundary of $A$ followed by the path to the vertex in $B$}
			\label{paths_on_combined_surface}
		\end{center}
	\end{figure}

	We want to show that this transformation is equivalent to applying Equation \ref{Equation_surface_edge_transformation_1} for the combined surface. In order to put the expression above into the same form as Equation \ref{Equation_surface_edge_transformation_1} for the combined surface, we want to commute the factors sandwiched between $e_B$ and $e_A$ over to the right of $e_A$, so that they are together with the other factors gained from the edge transform. We also need to replace the paths on $A$ and $B$ with paths on $AB$. We will start by replacing the paths. The boundary of $AB$ is the boundary of $A$ followed by the boundary of $B$. This means that any path that follows the boundary of $AB$ (with the correct orientation) will first travel along the boundary of $A$ and then (if the end-point of the path is on the boundary of $B$) will travel along $B$. This means the right-way paths on $A$ are also right-way paths on $AB$. However the right-way paths on $B$ are missing a factor of $g(\text{bd}(A))$ compared to the paths on $AB$, because a path on $AB$ that ends at a point on $B$ must first travel along the entire boundary of $A$ first. This is illustrated in Figure \ref{paths_on_combined_surface}.
	
	The fact that a path on $AB$ to an edge on $B$ must traverse the entirety of $A$ means that $g(v_0-s(i)[k] |AB)=g(\text{bd}(A))g(v_0-s(i)[k]|B)$, which using fake flatness gives us $g(v_0-s(i)[k]|B)= \partial(e_A) g(v_0-s(i)[k]|AB)$. This is a slight abuse of notation, because the $k$th appearance of the edge $i$ in $B$ does not correspond to the $k$th appearance in $AB$, but will instead be appearance $k+j_{max}$, where $j_{max}$ is the last appearance of edge $i$ in $A$. However $k$ is a dummy index, and we will soon be combining the two sets of appearances anyway, so we will not worry too much about this. Putting the relation $g(v_0-s(i)[k]|B)= \partial(e_A) g(v_0-s(i)[k]|AB)$ into Equation \ref{Equation_combined_surface_transformation_1} gives us
	\begin{align*}
		e_Be_A &\rightarrow e_B \bigg( \prod_{\substack{\text{appearance $k$} \\ \text{in }\text{bd}(B)}} \big(\partial(e_A) g(v_0-s(i)[k]|AB)\big) \rhd e^{\gamma_k} \bigg) e_A \bigg( \prod_{\substack{\text{appearance $j$} \\ \text{in }\text{bd}(A)}} g(v_0-s(i)[j]|AB) \rhd e^{\gamma_j}\bigg).
	\end{align*}
	
	We can then use the second Peiffer condition (Equation \ref{Peiffer_2} in the main text), which tells us that $\partial(e_A) \rhd x =e_A xe_A^{-1}$ for any $x \in E$, to write this transformation as
	\begin{align*}
		e_Be_A &\rightarrow e_Be_A \bigg( \prod_{\substack{\text{appearance $k$} \\ \text{in }\text{bd}(B)}}g(v_0-s(i)[k]|AB) \rhd e^{\gamma_k} \bigg) e_A^{-1}e_A \bigg( \prod_{\substack{\text{appearance $j$} \\ \text{in }\text{bd}(A)}}g(v_0-s(i)[j]|AB) \rhd e^{\gamma_j}\bigg)\\
		&=e_Be_A \bigg( \prod_{\substack{\text{appearance $k$} \\ \text{in }\text{bd}(B)}} g(v_0-s(i)[k]|AB) \rhd e^{\gamma_k} \bigg) \bigg( \prod_{\substack{\text{appearance $j$} \\ \text{in }\text{bd}(A)}} g(v_0-s(i)[j]|AB) \rhd e^{\gamma_j} \bigg).
	\end{align*}
	
	Then, because $\text{bd}(B)$ is after $\text{bd}(A)$ in $\text{bd}(AB)$, any appearance of the edge in $\text{bd}(B)$ is after any appearance of the edge in $\text{bd}(A)$, and so combining the two products as written will order the appearances from last to first appearance in $\text{bd}(AB)$. This gives us the result
	\begin{align*}
		e_Be_A &\rightarrow e_Be_A \bigg( \prod_{\text{appearance } k \text{ in } \text{bd}(AB)} g(v_0-s(i)[k]|AB) \rhd e^{\gamma_k} \bigg).
	\end{align*}
	
	This expression is in the same form as Equation \ref{Equation_surface_edge_transformation_1} for the transformation of a single surface. Therefore, if two surfaces transform according to our Equation \ref{Equation_surface_edge_transformation_1} and can be combined, then the combination of these two surfaces will also transform according to Equation \ref{Equation_surface_edge_transformation_1}. In order to complete the induction process, we need only show that the formula we gave is true for the building blocks that we can build all other surfaces out of, namely plaquettes, including plaquettes that have been whiskered or whose orientations have been flipped. We chose our ansatz for the transformation (Equation \ref{Equation_surface_edge_transformation_1}) to agree with the transformation of ordinary plaquettes, so we need only check the whiskered and inverted plaquettes.

	\begin{figure}[h]
		\begin{center}
			\begin{overpic}[width=0.7\linewidth]{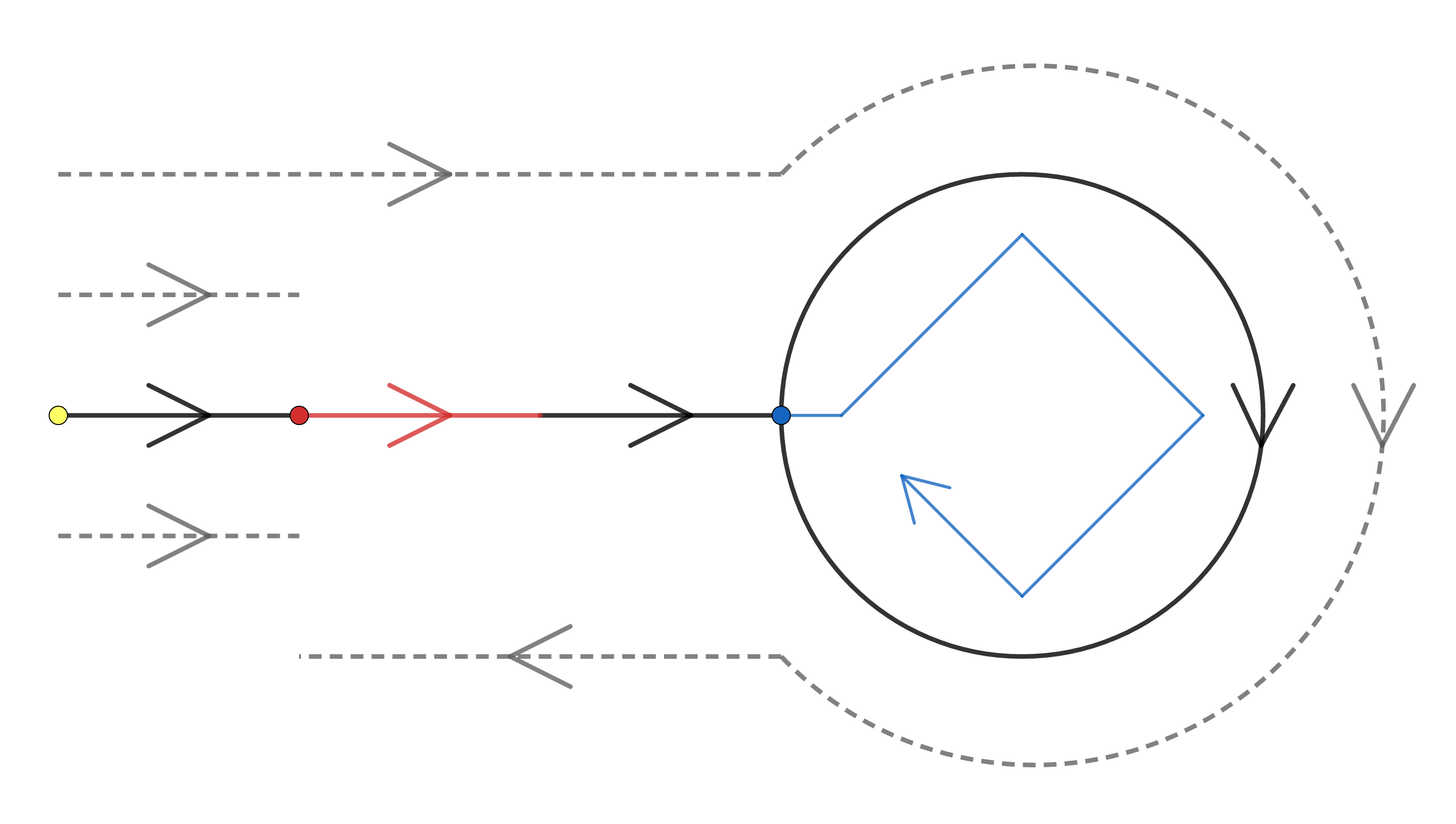}
				\put(69,28){$p$}
				\put(5,39){$s.p-s(i)[1]$}
				\put(30,46){$s.p-s(i)[2]$}
				\put(5,14){$\overline{s.p-s(i)[2]}$}
				\put(51,25){$v_0$}
				\put(32,29){$i$}
				\put(21,25){$s(i)$}
				\put(2,25){$s.p$}
				\put(27,22){$g_i \rightarrow \partial(e)g_i$}

			\end{overpic}
			\caption{We consider the edge transform applied on an edge along the whiskering path for a whiskered plaquette. First we consider the case where the edge points along the path $(s.p-v_0)$ that runs from the whiskered base-point to the original base-point of the plaquette. We find that Equation \ref{Equation_surface_edge_transformation_1} matches the explicit calculation for the transformation of the label of the whiskered plaquette in this case. This figure shows the relevant paths from the start-point to the source of edge $i$ that appear in the calculation. }
			\label{whiskered_edge_transform_1}
		\end{center}
	\end{figure}

	First we check how the whiskered plaquettes transform under an edge transform on one of the edges along which we whisker the plaquette, as illustrated in Figure \ref{whiskered_edge_transform_1}. While Ref. \cite{Bullivant2017} did not explicitly define the transformation for a whiskered plaquette, it is sensible to choose this transformation to be consistent with the rules for changing the base-point of a surface and the transformation of an unwhiskered plaquette, both of which are given in Ref. \cite{Bullivant2017}. The label for the whiskered plaquette is $e_p=g(s.p-v_0) \rhd e_0$, where $e_0$ is the label of the corresponding unwhiskered plaquette, $s.p$ is the base-point of the plaquette after whiskering and $v_0$ is the base-point of the original, unwhiskered plaquette. The edge on which we apply the transform is part of this path $s.p-v_0$. The edge transform therefore affects the path label and the transformation of the plaquette $p$ under the edge transform should be obtained from the transformation of this whiskering path. The edge can either point along $(s.p-v_0)$ or against it. If the edge points along $s.p-v_0$ (as in Figure \ref{whiskered_edge_transform_1}), we have that under the transform $\mathcal{A}_i^e$,
	\begin{align*}
		e_p &\rightarrow (g(s.p-s(i))\partial(e)g(s(i)-v_0))\rhd e_0\\
		&= g(s.p-s(i)) \rhd (\partial (e) \rhd [g(s(i)-v_0) \rhd e_0]),
	\end{align*}
	where $(s.p-s(i))$ is the part of $(s.p-v_0)$ up to the source of $i$ (which is $(s.p-s(i)[1])$ in Figure \ref{whiskered_edge_transform_1}) and $(s(i)-v_0)$ is the rest of $(s.p-v_0)$, so that $g(s.p-v_0)=g(s.p-s(i))g(s(i)-v_0)$. Then using the second Peiffer condition (Equation \ref{Peiffer_2} in the main text), this transformation for the plaquette label becomes
	\begin{align*}
		e_p &\rightarrow g(s.p-s(i)) \rhd (e [g(s(i)-v_0) \rhd e_0] e^{-1})\\
		&=[g(s.p-s(i)) \rhd e] [(g(s.p-s(i))g(s(i)-v_0)) \rhd e_0] [g(s.p-s(i)) \rhd e^{-1}],
	\end{align*}
	where we used the fact that $g \rhd (ef) = [g \rhd e] [g \rhd f]$ for all $g \in G$, $e,f \in E$. Then we can use $g(s.p-s(i))g(s(i)-v_0)= g(s.p-v_0)$ to obtain
	\begin{align}
		e_p &\rightarrow [g(s.p-s(i)) \rhd e] [g(s.p-v_0) \rhd e_0] [g(s.p-s(i)) \rhd e^{-1}]. \label{Equation_whiskered_plaquette_transform}
	\end{align}
	
	Then looking at Figure \ref{whiskered_edge_transform_1}, we can see that $(s.p-s(i))$ (which we should write as $(s.p-s(i)[1])$ to be unambiguous) is the same as the path $g(\overline{s.p-s(i)[2]})$. Writing Equation \ref{Equation_whiskered_plaquette_transform} in terms of this notation we have
	\begin{align*}
		e_p&\rightarrow [g(\overline{s.p-s(i)[2]}) \rhd e] [g(s.p-v_0) \rhd e_0] [g(s.p-s(i)[1]) \rhd e^{-1}].
	\end{align*}
	
	Then using Equation \ref{Equation_plaquette_transform_move_left_right} to move the factor from the left of $e_p$ to the immediate right by swapping the wrong-way path for the right-way one, we have
	\begin{align*}
		e_p&\rightarrow [g(\overline{s.p-s(i)[2]}) \rhd e] [g(s.p-v_0) \rhd e_0] [g(s.p-s(i)[1]) \rhd e^{-1}]\\
		&=[(\partial(e_p)g(s.p-s(i)[2])) \rhd e] e_p [g(s.p-s(i)[1])\rhd e^{-1}]\\
		&=e_p [g(s.p-s(i)[2]) \rhd e] [g(s.p-s(i)[1]) \rhd e^{-1}]\\
		&=e_p \bigg(\prod_{\text{appearance }k} g(s.p-s(i)[k]) \rhd e^{\gamma_k}\bigg),
	\end{align*}
	which agrees with Equation \ref{Equation_surface_edge_transformation_1}. Similarly, if the edge $i$ instead points in the other direction, we have from a direct calculation that the transformation of the surface label under the edge transform $\mathcal{A}_i^e$ is
	\begin{align*}
		e_p &\rightarrow (g(s.p-s(i)[1]) \partial(e)^{-1}g(s(i)[1]-v_0)) \rhd e_0\\
		&=[g(s.p-s(i)[1]) \rhd e^{-1}] e_p [g(s.p-s(i)[1]) \rhd e].
	\end{align*}
	
\begin{figure}[h]
		\begin{center}
			\begin{overpic}[width=0.7\linewidth]{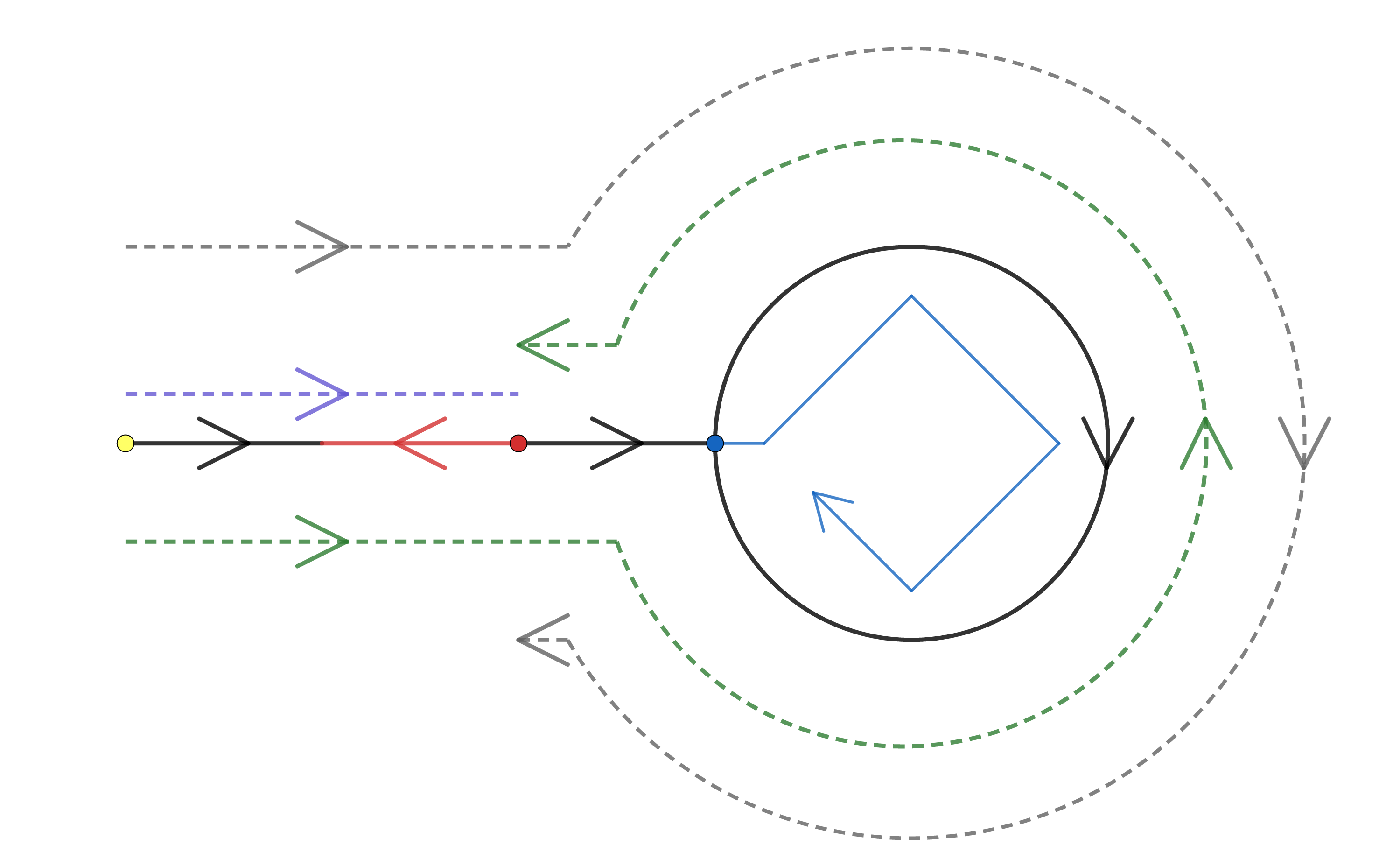}
				\put(65,29){$p$}
				\put(17,36){$(s.p-s(i)[1])$}
				\put(26,46){$(s.p-s(i)[2])$}
				\put(12,18){$(\overline{s.p-s(i)[1]})$}
				\put(48,27){$v_0$}
				\put(25,27){$i$}
				\put(36,26){$s(i)$}
				\put(8,27){$s.p$}

			\end{overpic}
			\caption{We consider the edge transform applied on an edge along the whiskering path for a whiskered plaquette. We consider the case where the edge points against the path $(s.p-v_0)$ that runs from the whiskered base-point to the original base-point of the plaquette. We find that our form for the transformation of the label of the whiskered plaquette, as given in Equation \ref{Equation_surface_edge_transformation_1}, matches the explicit calculation in this case. This figure shows the relevant paths from the start-point to the source of edge $i$ that appear in the calculation. }
			\label{whiskered_edge_transform_wrong_way}
		\end{center}
	\end{figure}

	Looking at Figure \ref{whiskered_edge_transform_wrong_way}, we see that $g(s.p-s(i)[2])=g_pg(s.p-s(i)[1])=\partial(e_p^{-1})g(s.p-s(i)[1])$, where $g_p$ is the label of the boundary of the plaquette and is equal to $\partial(e_p^{-1})$ by fake-flatness. Substituting this into our previous expression for the transformation gives
	\begin{align*}
		e_p &\rightarrow [(\partial(e_p) g(s.p-s(i)[2])) \rhd e^{-1}] e_p [g(s.p-s(i)[1]) \rhd e]\\
		&=e_p [g(s.p-s(i)[2]) \rhd e^{-1}] [g(s.p-s(i)[1]) \rhd e]\\
		&=e_p \bigg(\prod_{\text{appearance }k} g(s.p-s(i)[k]) \rhd e^{\gamma_k}\bigg),
	\end{align*}
	which also agrees with Equation \ref{Equation_surface_edge_transformation_1}.

	Next we need to check how the whiskered surface transforms under edge transforms on the unwhiskered edges. Firstly, consider ``right-way" edges on the plaquette (edges that point with the orientation of the plaquette). Applying the transform $\mathcal{A}_i^e$ changes the surface label $e_p =g(s.p-v_0) \rhd e_0$ as follows:
	\begin{align*}
		g(s.p-v_0) \rhd e_0 & \rightarrow g(s.p-v_0) \rhd (e_0 [g(v_0-s(i)) \rhd e^{-1}])\\
		&= e_p [(g(s.p-v_0)g(v_0-s(i))) \rhd e^{-1}]\\
		&= e_p [g(s.p-s(i)) \rhd e^{-1}],
	\end{align*}
	where $g(s.p-v_0)g(v_0-s(i))=g(s.p-s(i))$ is the right-way path on the whiskered plaquette, because this path is the path from the whiskered base-point to the original base-point, followed by the right-way path on the unwhiskered plaquette.

	Now consider a transform applied on a wrong-way edge. In this case we have that
	\begin{align*}
		g(s.p-v_0 )\rhd e_0 &\rightarrow g(s.p-v_0) \rhd ([g(\overline{v_0-s(i)}) \rhd e] e_0)\\
		&=[(g(s.p-v_0) g(\overline{v_0-s(i)}))\rhd e] e_p\\
		&=[g(\overline{s.p-s(i)}) \rhd e] e_p,
	\end{align*}
	where $g(\overline{s.p-s(i)}) =g(s.p-v_0)g(\overline{v_0-s(i)})$ is the appropriate wrong-way path to the edge on the whiskered plaquette, because it travels along the whiskered path and then along the wrong-way path on the unwhiskered plaquette. We can then use Equation \ref{Equation_plaquette_transform_move_left_right} to move the factor of $[g(\overline{s.p-s(i)}) \rhd e]$ to the right of $e_p$, by swapping the wrong-way path for the corresponding right-way path. This gives us
	\begin{align*}
		e_p&\rightarrow e_p [g(s.p-s(i))\rhd e],
	\end{align*}
	
	matching Equation \ref{Equation_surface_edge_transformation_1}.

	In addition to moving the base-points of plaquettes in order to combine them into a surface, we may need to invert the orientation of a plaquette before combining it with another surface. Inverting the orientation of a plaquette $p$ takes $e_p$ to $e_p^{-1}$, so we need to check that $e_p^{-1}$ transforms as an ordinary plaquette. In the Appendix of Ref. \cite{HuxfordPaper1}, we showed that the energy terms, including the edge transforms, are consistent with this inversion procedure. The inverted plaquettes therefore also transform according to Equation \ref{Equation_edge_transform_appendix}, and so match Equation \ref{Equation_surface_edge_transformation_1}. We can also whisker the inverted plaquette if necessary to combine it with a surface, and it will behave just like the whiskered plaquettes described earlier.

	We have therefore shown that plaquettes (including inverted and whiskered plaquettes) do indeed transform according to Equation \ref{Equation_surface_edge_transformation_1}. We also showed that, given two surfaces that transform according to Equation \ref{Equation_surface_edge_transformation_1}, the composition of these surfaces will also transform in the same way. Therefore, general surfaces do indeed transform as we proposed. However, as defined here, the boundary path of a combined surface contains the entirety of both boundary of the individual surfaces. That includes many extraneous edges and does not allow us to easily see that edge transforms on edges in the interior of the surface (i.e., edges within the surface, that do not need to be in the boundary path) commute with the membrane operator. Therefore, we need to show that we can get rid of edges that should be in the interior from the boundary path, without affecting the transform, as we have alluded to throughout this section. The sections of the path that can be removed are those that appear twice consecutively in the boundary, with opposite orientations (i.e., appear as ...$t t^{-1}$... in the boundary). The transformation of the surface label under the edge transform must be preserved when we remove these sections from the boundary.

\begin{figure}[h]
		\begin{center}
			\begin{overpic}[width=0.7\linewidth]{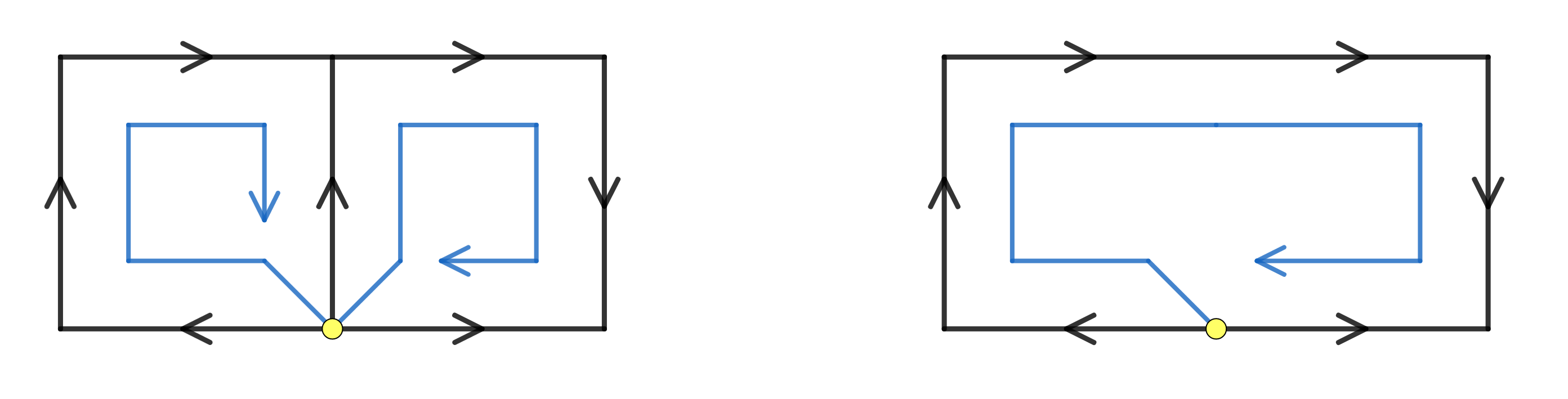}
				\put(45,12){\Huge $\rightarrow$}
				\put(43,10){Combine}

				\put(10,2){$t_1$}
				\put(0,15){$t_2$}
				\put(10,24){$t_3$}
				\put(18,16){$t_4$}
				\put(31,24){$t_5$}
				\put(41,15){$t_6$}
				\put(31,2){$t_7$}
				\put(11,12){$A$}
				\put(29,12){$B$}
				
				\put(67,2){$t_1$}
				\put(57,15){$t_2$}
				\put(68,24){$t_3$}
				
				\put(89,24){$t_5$}
				\put(97,15){$t_6$}
				\put(89,2){$t_7$}
				\put(75,12){$AB$}
				
			\end{overpic}
			\caption{We can consider surfaces, such as $AB$, where the boundary is not the naive product of the boundaries of the constituent surfaces. Compared to Figure \ref{combine_two_surfaces} we have removed the section of the boundary involving $t_4$.}
			\label{combine_two_surfaces_simplify}
		\end{center}
	\end{figure}

	We can see this property explicitly from our formula Equation \ref{Equation_surface_edge_transformation_1}. Suppose an edge in the boundary path appears twice consecutively, once in the orientation of the boundary (right-way) and once in the opposite orientation (wrong-way). Then the edge transform on this edge gives
	$$e_A \rightarrow e_A [g(v_0-s(i)) \rhd e] [g(v_0-s(i)) \rhd e^{-1}]=e_A.$$
	Then we can remove any such pairs of edges from the boundary paths, and iterate this process to remove sections like $itt^{-1}i^{-1}$ by sequentially removing edges from the middle. This will remove any sections of path that are travelled, then immediately re-travelled in the opposite direction. This means that surfaces obtained by removing these sections of path transform in the same way as the ones that have all such edges included. Therefore, when we combine two surfaces, we can consider the intuitive boundary of the total surface, such as that shown in Figure \ref{combine_two_surfaces_simplify}, as long as we can construct this intuitive boundary by using the procedure for removing edges from the product of the individual boundaries.

\begin{figure}[h]
		\begin{center}
			\begin{overpic}[width=0.7\linewidth]{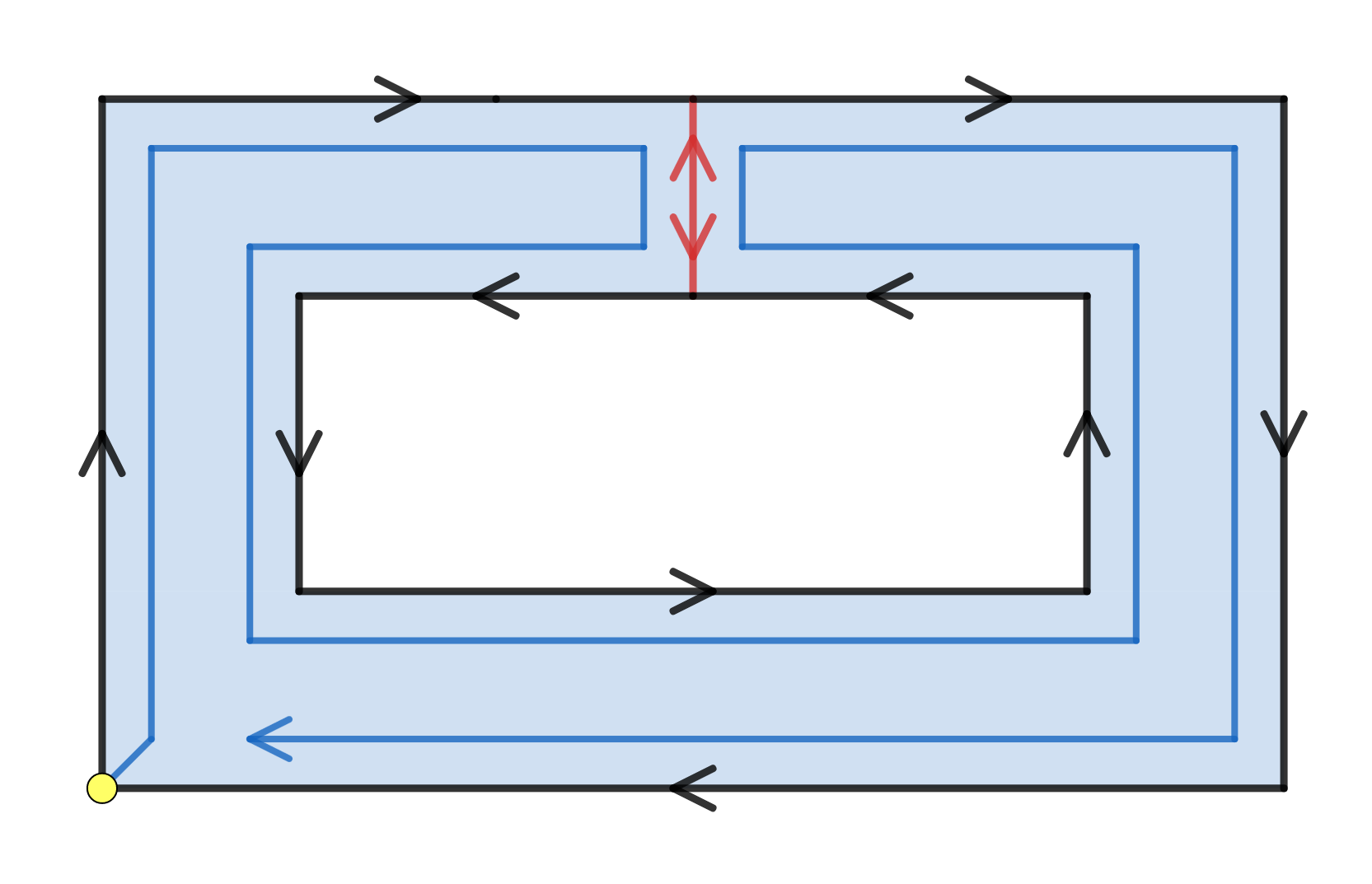}
				
			\end{overpic}
			\caption{The surface (shaded area) in this figure is an example of a surface where part of the boundary (the red edge with two arrows) appears twice in the boundary with opposite orientation, but cannot be removed from the boundary because the inner cycle may not be trivial.}
			\label{surface_boundary_touching}
		\end{center}
	\end{figure}

	We need to be more careful with more complicated surfaces, where the boundary may fold back on itself, such as in Figure \ref{surface_boundary_touching}, because then even though the same edge appears twice with opposite orientation in the path, we cannot remove the edges from the boundary with our procedure for removing edges because there is a non-trivial path element between the two appearances of the edge (e.g., the inner cycle in Figure \ref{surface_boundary_touching}). This is reflected in the transformation of the surface element under the edge transform applied on such an edge with multiple appearances. In such a case Equation \ref{Equation_surface_edge_transformation_1} becomes
	$$e_A \rightarrow e_A [g(v_0-s(i)) \rhd e] [(g(v_0-s(i)) g(c)) \rhd e^{-1}],$$
	where $c$ is a closed path. The factors $[g(v_0-s(i)) \rhd e]$ and $ [(g(v_0-s(i)) g(c)) \rhd e^{-1}]$ do not cancel if $g(c)$ acts non-trivially on $e$ via $\rhd$. In some cases, these factors will cancel. For example, if $E$ is Abelian and the closed path $c$ encloses a region satisfying fake-flatness, then $g(c)= \partial(f)$ for some $f \in E$ and then the second Peiffer condition (Equation \ref{Peiffer_2} in the main text) tells us that $g(c) \rhd e = \partial(f) \rhd e = fef^{-1}=e$. However, even if $c$ encloses a surface satisfying fake-flatness, if $E$ is non-Abelian then $g(c)$ need not act trivially on $e$. Furthermore, if the path does not enclose a fake-flat surface, then even if $E$ is Abelian $g(c)\rhd e$ need not be $e$. This case will be particularly relevant in the 3+1d model that we consider in more detail in Ref. \cite{HuxfordPaper3}. We therefore see that the edge transformation on such edges can act non-trivially on the surface element, so it makes sense that our procedure for removing edges from the boundary would not allow us to remove these ones.

	It is worth mentioning that the factor of $[g(v_0-s(i)) \rhd e] [(g(v_0-s(i)) g(c)) \rhd e^{-1}]$ obtained from the transform on this type of edge is not a general element of $E$ (i.e., this expression may be restricted to a subset of $E$, no matter what value of $e$ we take). For example, if $\rhd$ is trivial this element becomes the identity, and if $\rhd$ is non-trivial but $E$ is Abelian and $\partial$ maps to the centre of $G$ then this element belongs to the kernel of $\partial$. Therefore, even if this factor can be non-trivial, it does not describe the same transformation as for edge transforms elsewhere on the boundary of the surface, which means that the corresponding edge may or may not be excited, depending on whether the $E$-valued membrane is labelled by an irrep that is sensitive to such factors or not. For example, when $E$ is Abelian and $\partial$ maps to the centre of $G$, only $E$-valued membrane operators labelled by irreps which are sensitive to the kernel of $\partial$ may excite this edge.

	\begin{figure}[h]
		\begin{center}
			\begin{overpic}[width=0.8\linewidth]{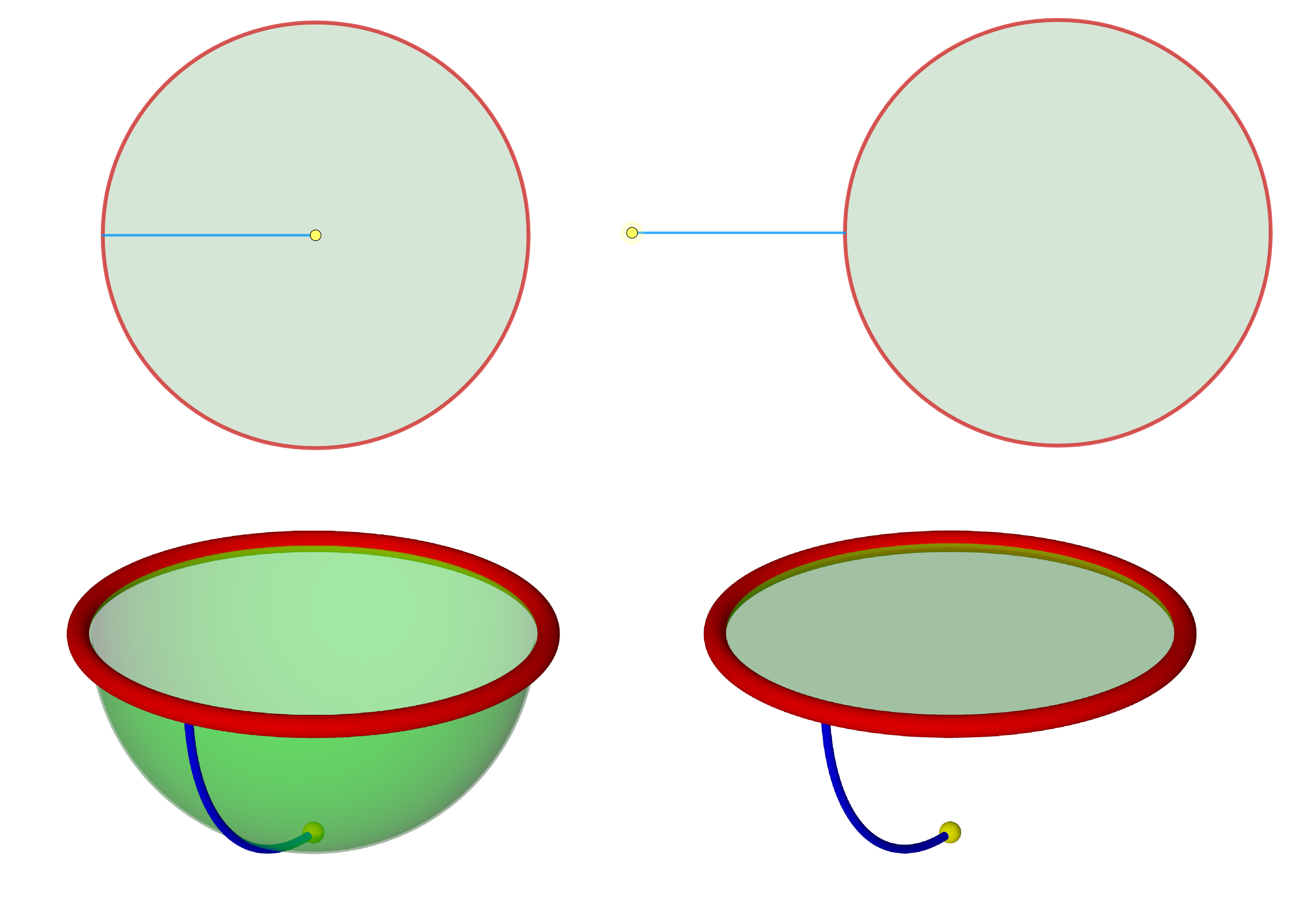}
				
			\end{overpic}
			\caption{Here we consider some examples of surfaces where the start-point (yellow dot or sphere) is located away from the naive boundary of the surface (represented in red). In each case, the thinner (blue) line attached to the start-point represents the ``seam" of edges that appear twice in the boundary, which may be excited depending on the membrane operator. In the first line, we show some examples in 2+1d, for which it is clear that the surface can be treated as a whiskered version of the circular surface. In the second line we show two examples in the 3+1d model. These two measurement operators in the 3+1d case are equivalent when acting on the ground state, because they can be deformed into one-another smoothly, but it is important to keep the boundary (including the seam) fixed during this process (at least if it is excited). Again these surfaces can also be thought of as whiskered versions of a surface where the start-point is attached to the loop-like excitation proper.}
			\label{surface_boundary_whiskered}
		\end{center}
	\end{figure}

	One particular example of this type of boundary edge that we want to consider in more detail is when the start-point of the membrane operator is located away from the naive boundary of the membrane, as shown in the examples in Figure \ref{surface_boundary_whiskered}. Because the boundary always starts and ends at the start-point, there is a ``seam" of edges that runs from the start-point to the expected location of the loop-like excitation, consisting of edges that appear twice in the boundary with opposite directions, but which do not appear consecutively in the boundary. This is equivalent to considering whiskering the start-point away from the loop-like excitation, as in the situation shown in Figure \ref{whiskered_edge_transform_1}. We therefore call this section of boundary the whiskered section of the boundary. If we then consider the transformation of the surface element under an edge transform $A_i^e$ on an edge $i$ along the whiskered section of boundary, then from Equation \ref{Equation_surface_edge_transformation_1} we get a transformation of the form
	\begin{align*}
		e_A \rightarrow& e_A [g(s.p-s(i)[2]) \rhd e] [g(s.p-s(i))[1] \rhd e^{-1}],
	\end{align*}
	where we assumed that the edge $i$ points away from the start-point as in Figure \ref{whiskered_edge_transform_1} (we would get a similar result for the opposite orientation of $i$, except that we would have $e^{-1}$ instead of $e$). As we may expect, this is the same form as that for a whiskered plaquette, regardless of whether the base-point is within the surface or away from it. Using the same manipulations that we did for the individual plaquettes (i.e., noting that the path element $(s.p-s(i)[2]) \cdot (s.p-s(i)[1])^{-1}$ is the entire boundary of the surface, and applying fake-flatness), we can write this transformation as
	\begin{align*}
		e_A \rightarrow& e_A [g(\text{bd}(A)) g(s.p-s(i)[1]) \rhd e] [g(s.p-s(i))[1] \rhd e^{-1}]\\
		&= e_A [\partial(e_A)^{-1} \rhd (g(s.p-s(i)[1]) \rhd e)] [g(s.p-s(i))[1] \rhd e^{-1}].
	\end{align*}
	Then using the second Peiffer condition (Equation \ref{Peiffer_2} in the main text), we can write this as
	\begin{align*}
		e_A \rightarrow&e_A e_A^{-1} [g(s.p-s(i)[1]) \rhd e] e_A [g(s.p-s(i))[1] \rhd e^{-1}]\\
		&=[g(s.p-s(i)[1]) \rhd e] e_A [g(s.p-s(i))[1] \rhd e^{-1}],
	\end{align*}
	just like Equation \ref{Equation_whiskered_plaquette_transform} for an individual whiskered plaquette.

	We note that this transformation has the form of conjugation of the surface label, and so is trivial when $E$ is Abelian. In particular, we note that $E$ is Abelian in Cases 1 and 2 from Table \ref{Table_Cases_2d} in the main text (where $\rhd$ is trivial for Case 1, implying $E$ is Abelian, and we explicitly assume $E$ is Abelian for Case 2) and so this transformation is trivial in those cases. That is, the edge transforms on these seams are trivial in those cases and so the edges are not excited (at least, if the surface satisfies fake-flatness). On the other hand, in Case 3, where the crossed module is general and so $E$ need not be Abelian, this is not the case. In that case, whether the edge transform acts trivially on an $E$-valued membrane operator depends on the coefficients for that operator. Consider an $E$-valued membrane operator
	$$L^{\vec{\alpha}}(m) = \sum_{f \in E} \alpha_f \delta(f, \hat{e}(m)),$$
	applied on a membrane such as the one in Figure \ref{whiskered_edge_transform_1}. Then applying an edge transform on one of the edges on the seam gives a transformation of the form
	\begin{align*}
		L^{\vec{\alpha}}(m) &= \sum_{f \in E} \alpha_f \delta(f, \hat{e}(m))\\
		& \rightarrow \sum_{f \in E} \alpha_f \delta(f, [g(s.p-s(i)[1]) \rhd e] \hat{e}(m) [g(s.p-s(i))[1] \rhd e^{-1}]).
	\end{align*}
	Denoting $g(s.p-s(i)[1]) \rhd e$ by $x$, we can write this as
	\begin{align}
		L^{\vec{\alpha}}(m) &= \sum_{f \in E} \alpha_f \delta(f, \hat{e}(m)) \notag \\
		&\hspace{-0.2cm} \rightarrow \sum_{f \in E} \alpha_f \delta(f, x \hat{e}(m) x^{-1}) \notag\\
		&=\sum_{f \in E} \alpha_f \delta(x^{-1}fx, \hat{e}(m)) \notag \\
		&= \sum_{f' =x^{-1}fx \in E} \alpha_{xf'x^{-1}} \delta(f', \hat{e}(m)). \label{Equation_E_membrane_seam_edge_transform_2}
	\end{align}
	
	We therefore see that the membrane operator will be invariant under any such edge transform if the coefficient $\alpha$ is a function of conjugacy class only. On the other hand, generally the membrane operator is not invariant under the edge transform and so the edge may be excited (this will occur if the coefficients for the elements within each conjugacy class sum to zero). This means that the line of edges on the whiskered section of boundary from the start-point to the loop-like excitation may be excited. In this case, there is an energy cost associated to moving the loop-like excitation away from the start-point, and so the loop-like excitation is in a sense confined. However, this energy cost does not grow with the size of the loop (apart from the energy cost associated to the length of the loop which we always get for loop-like excitations in this model). We note that a necessary condition to have this additional string of excited edges connecting to the start-point is for the start-point itself to be excited. To see this, first note that we can use the second Peiffer condition (Equation \ref{Peiffer_2} in the main text) to write $xf'x^{-1} = \partial(x) \rhd f'$. Inserting this relation into Equation \ref{Equation_E_membrane_seam_edge_transform_2}, we see that
	\begin{align*}
		L^{\vec{\alpha}}(m) \rightarrow \sum_{f' =x^{-1}fx \in E} \alpha_{\partial(x) \rhd f'} \delta(f', \hat{e}(m)). 
	\end{align*}

	If the coefficient is the same for $f'$ and $\partial(x) \rhd f'$ for all $x \in E$ and $f' \in E$, then the membrane operator is invariant under any such edge transform (and so the edge is not excited). However as we showed at the start of this section (Section \ref{Section_E_Membrane_Commutation_Proof}), the start-point of the membrane operator is not excited if the coefficient for $f'$ and $ g \rhd f'$ is the same for all $g \in G$ and $f' \in E$. Because $\partial(E)$ is a subgroup of $G$, if the coefficient is invariant under this $g \rhd$ action then it is also invariant under the $\partial(x) \rhd$ action. Therefore, if the start-point is not excited then we never get this line of excitations from the start-point to the loop-excitation proper. We expect this, because if the start-point is not excited then its location becomes arbitrary, and so we would not expect a line of excitations terminating at the start-point. This extra string of excitations connecting to the start-point is particularly interesting in the 3+1d model that we consider in a future paper (Ref. \cite{HuxfordPaper3}), because it suggests that the point-like charge associated to the loop (and balanced by the point-like charge of the start-point) is confined by this string (rather than the loop-like charge being confined, which would instead be reflected by an energy cost for the area enclosed by the loop).

\begin{figure}[h]
		\begin{center}
			\begin{overpic}[width=0.6\linewidth]{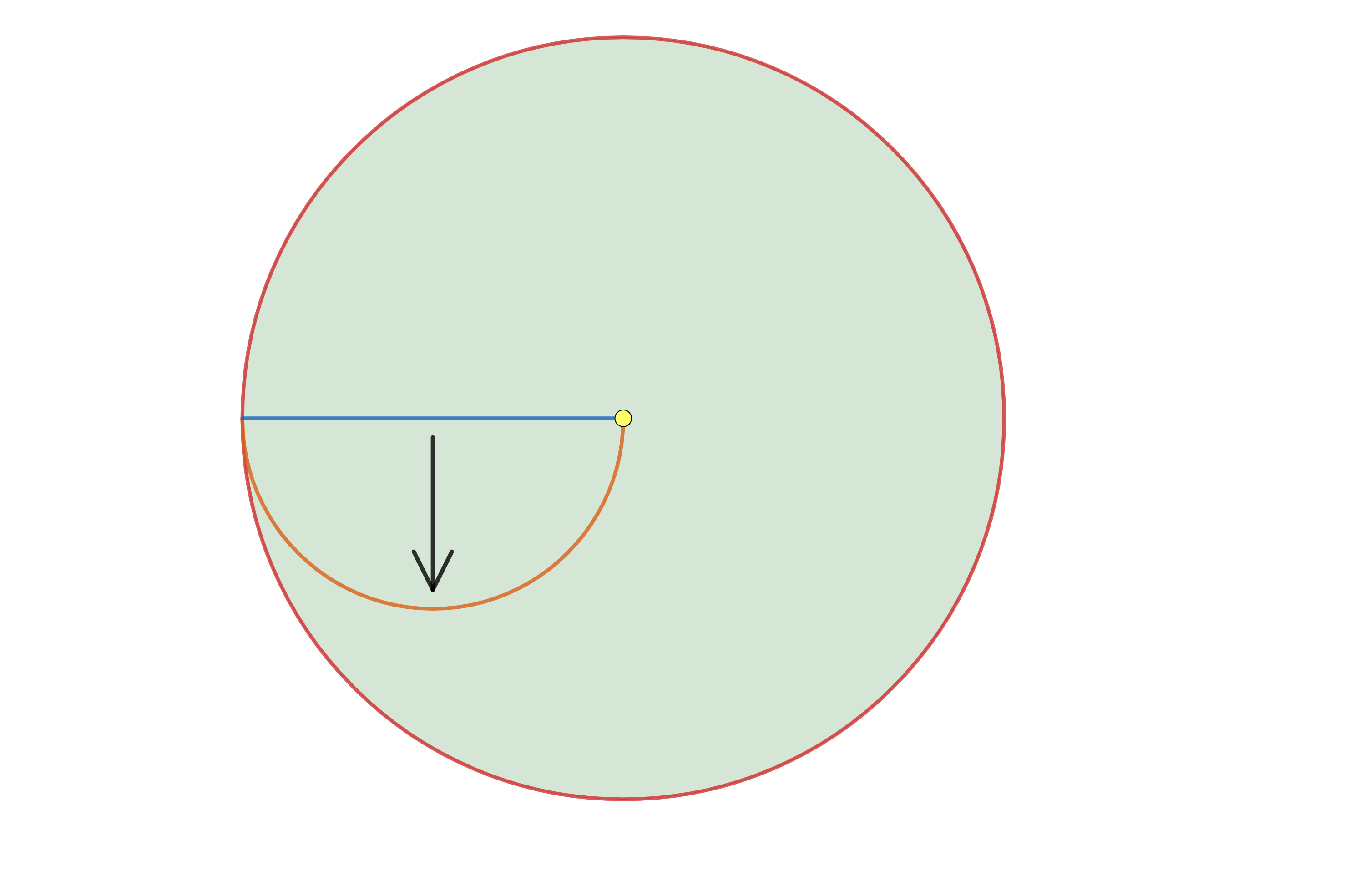}
				
			\end{overpic}
			\caption{If the whiskered section of the boundary (shown in blue) is not excited, then we can deform it over a fake-flat region (for example to the orange position) without affecting the action of the membrane operator. On the other hand, if it is excited then we cannot move this section of the boundary without changing the action of the membrane operator.}
			\label{surface_boundary_whiskered_deform}
		\end{center}
	\end{figure}

	It is also worth noting that if the line of edges from the start-point to the loop-like excitation proper is not excited, then we can freely deform this section of the boundary, as shown in Figure \ref{surface_boundary_whiskered_deform}, without affecting the action of the membrane operator. One way to see this is to note that applying an edge transform is equivalent to parallel transport of the edge over a surface, so if the membrane operator is invariant under an edge transform on this section of boundary, it is also invariant under transport of the boundary. Another way to see this is to write the surface element in terms of a membrane without the whiskered section of boundary. The surface label of the original membrane $\hat{e}(m)$ can be written as $g(t) \rhd \hat{e}(m')$, where $m'$ is the membrane obtained by moving the start-point of $m$ back along the whiskered section $t$ of boundary to remove that section from the boundary, from the rules for whiskering explained in Ref. \cite{Bullivant2017}. Then deforming $t$ over a fake-flat region to a new path $t'$ changes the path label $g(t)$ to $g(t')=g(t)\partial(e)$ for some $e \in E$, so that the new surface label is $(g(t)\partial(e)) \rhd \hat{e}(m') = g(t) \rhd (\partial(e) \rhd \hat{e}(m'))$. However the second Peiffer condition (Equation \ref{Peiffer_2} in the main text) states that $\partial(e) \rhd \hat{e}(m') = e \hat{e}(m') e^{-1}$, so the surface label is only changed by conjugation. If the membrane operator is not sensitive to conjugation (i.e., the whiskered section of boundary is not excited), then the membrane operator is also not sensitive to this deformation of the whiskered section of boundary.
	
	We have therefore established the effect of the edge transforms on a general surface element. Only transforms applied on edges along the boundary of the surface have an effect on the surface label. Because our $E$-valued membrane operator measures the surface label and applies a weight for each possible value, this means that the $E$-valued membrane operator commutes with all edge transforms that are not on the boundary of the surface. Generally the membrane operator will not commute with the edge energy terms (made of an average of edge transforms on each edge) on the boundary, unless the membrane operator is trivial, with all of the weights being equal. If the coefficients are orthogonal to this case (they sum to zero, or equivalently correspond to non-trivial irreps of $E$) the edges on the boundary are excited (except when the same edge appears multiple times, such as the case shown in Figure \ref{surface_boundary_touching}, in which the edge may or may not be excited, as described earlier). We also saw earlier that the membrane operator commutes with all of the other energy terms except potentially the vertex transform at the start-point of the membrane. This establishes the results stated in Sections \ref{Section_2D_Loop} and \ref{Section_2D_RO_Fake_Flat}. 
	
	\subsection{Single plaquette multiplication operators}
	\label{Section_single_plaquette_multiplication_commutation_relations}
	While not precisely a ribbon or membrane operator, the single plaquette multiplication operators may also produce excitations (albeit local ones) and are necessary to give a complete set of operators on the Hilbert space. While the electric ribbon operators measure path labels and the magnetic ribbon operators change path labels, the $E$-valued membrane operators measure surface labels and it is the single plaquette multiplication operators that change the surface labels. A single plaquette multiplication operator $M^e(p)$ multiplies the label of plaquette $p$ by $e$. We now wish to find the commutation relations between this operator and the various energy terms. First we consider the plaquette term, which has the form $\delta(\partial(e_p)g_p,1_G)$, where $e_p$ is the label of the plaquette $p$ on which we apply the energy term and $g_p$ is the path label of the boundary of that plaquette. We see that this depends on the label of plaquette $p$ through $\partial(e_p)$. The single plaquette multiplication operator $M^e(q)$ on a plaquette $q$, which only affects the plaquette $q$, will commute with the plaquette term unless $p=q$. When $p=q$ the multiplication operator will still commute with the plaquette term if $e$ is in the kernel of $\partial$, because multiplication of $e_p$ by an element of the kernel does not affect $\partial(e_p)$. On the other hand, if $e$ is not in the kernel then the multiplication operator will not commute with the plaquette term.

	The commutation relations with the other energy terms depend on which special case from Table \ref{Table_Cases_2d} in the main text we are considering. First we examine the case where $\rhd$ is trivial. In this case the vertex term only acts on the edges, whereas the plaquette multiplication term only acts on plaquettes, and so these operators commute. In addition, when $\rhd$ is trivial $E$ is Abelian (see Section \ref{Section_Recap_Paper_2} of the main text) and so the edge transforms (which multiply plaquette labels by elements of $E$) also commute with the single plaquette multiplication terms (because the order of multiplication does not matter).

	On the other hand, consider the case where $\rhd$ is non-trivial, but we restrict to the fake-flat configurations of our lattice. In this case, we must throw out the single plaquette multiplication operators with label $e$ outside the kernel of $\partial$, because these cause plaquette excitations as we discussed earlier. This leaves us only with multiplication operators labelled by elements of the kernel, which are also in the centre of $E$ (from the second Peiffer condition, Equation \ref{Peiffer_2} in the main text). This means that once again the single plaquette multiplication operators commute with the edge transforms. This is because the edge transform $A_i^x$ multiplies the labels of the adjacent plaquettes by some label $g(t) \rhd x^{\pm 1}$, but even if the single plaquette multiplication operator acts on one of these plaquettes, it does so by multiplication of the label by an element $e$ in the centre of $E$, and so these two multiplications commute. Finally, we wish to find the commutation between a single plaquette multiplication operator $M^e(p)$ and a vertex transform $A_v^g$. The vertex transform primarily acts on the edges, but does affect the label of any plaquette whose base-point is the vertex $v$. Therefore, the two operators may fail to commute only if $v$ is the base-point of $p$. In this case, the vertex transform acts on the plaquette label $e_p$ according to $A_v^g :e_p = g \rhd e_p$. This means that if we act first with the vertex transform $A_v^g$ and then with $M^e(p)$ we obtain 
	$$M^e(p) A_v^g: e_p = e [g \rhd e_p].$$
	On the other hand, if we act first with $M^e(p)$ and then $A_v^g$, we obtain
	$$A_v^g M^e(p) : e_p = A_v^g: (e e_p) = g \rhd (ee_p)= [g \rhd e] [g \rhd e_p]= M^{g \rhd e}(p) A_v^g:e_p.$$
	We therefore see that a vertex transform applied on the base-point of $p$ does not generally commute with the single plaquette multiplication operator applied on $p$.
	
	\section{The topological nature of ribbon operators}
	\label{Section_topological_membrane_operators}
	
	An important property of the unconfined ribbon operators (and membrane operators in the 3+1d case) is that they are topological. By topological we mean that we can deform the ribbon or membrane on which we apply these operators without changing their action on the ground state, provided that we keep the location of any excitations produced by the ribbon or membrane operator fixed. More generally, this holds for states other than the ground state provided that the region over which we deform the ribbon or membrane has no excitations. In this section we will prove that the ribbon operators in the 2+1d model that we have considered so far are topological in this sense, except for the ribbon operators corresponding to confined excitations. The ribbon operators corresponding to the confined excitations are not topological, as can be seen from the fact that they produce excitations along the length of the ribbon, the location of which can then be detected through the energy terms. 
	
	\subsection{Electric ribbon operators}
	
	First we consider the electric ribbon operators. Recall from Section \ref{Section_2D_electric} of the main text that the electric ribbon operator measures the label $g(t)$ of a path $t$ and assigns a weight to each value. If we deform this path, while keeping the end-points of the path fixed, then we obtain a new path $t'$. If we pull the path over a fake-flat surface then the group label of the final path $t'$ is related to that of the initial path $t$ by an element in $\partial(E)$, due to fake-flatness. That is, $g(t')=\partial(e)g(t)$ for some $e \in E$. We can now consider how this affects the electric ribbon operator. Consider an electric ribbon operator applied on the path $t$:
	$$S^{\vec{\alpha}}(t) = \sum_{g \in G} \alpha_g \delta(g, g(t)).$$
	Then when we deform the path $t$ into $t'$ the ribbon operator becomes
	\begin{align*}
		\sum_{g \in G} \alpha_g \delta(g,g(t'))&= \sum_{g \in G} \alpha_g \delta(g,\partial(e)g(t))\\
		&= \sum_{g \in G} \alpha_g \delta(\partial(e)^{-1} g,g(t))\\
		&= \sum_{g' \in G} \alpha_{\partial(e)g'} \delta(g',g(t)).
	\end{align*}
	
	As described in Section \ref{Section_Electric_Ribbon_Operator_Proof}, the electric ribbon operators corresponding to unconfined excitations satisfy $\alpha_g = \alpha_{\partial(e)g}$. Therefore, for such an operator the ribbon operator on the deformed path satisfies
	\begin{align*}
		\sum_{g \in G} \alpha_g \delta(g,g(t'))&= \sum_{g' \in G} \alpha_{\partial(e)g'} \delta(g',g(t))\\
		&= \sum_{g' \in G} \alpha_{g'} \delta(g',g(t)),
	\end{align*}
	which is the same as the original (un-deformed) ribbon operator. This indicates that the ribbon operators for unconfined electric excitations are topological (this also holds for the 3+1d case).
	
	\subsection{Magnetic ribbon operators}
	\label{Section_Topological_Magnetic_Ribbons}
	We next consider the magnetic ribbon operators. We only considered the magnetic excitations in 2+1d when we restricted the crossed module so that $\rhd$ is trivial (apart from some special cases in Section \ref{Section_Z2_Z3_Magnetic} of the main text and Section \ref{Section_confined_magnetic} of the Supplemental Material), and so we only use this special case in this discussion. We wish to prove that the magnetic ribbon operators are topological, in that we can deform the ribbon without changing the action of the ribbon operator, provided that we keep any excitations produced by the operator fixed and do not deform the ribbon over any other excitations. We will do this in several steps. First, we will demonstrate that a closed magnetic ribbon operator acts trivially, provided that the ribbon is contractible and encloses no excitations. Then we will show that we can deform an open magnetic ribbon operator by multiplying the open ribbon operator by a contractible closed ribbon operator of appropriate label and with a correctly chosen ribbon. Then, because the action of the closed ribbon operator is trivial, this multiplication does not change the action of the open ribbon operator, and therefore neither does the deformation.

	For the first step, consider a magnetic ribbon operator $C^h(c)$ applied on a contractible closed ribbon $c$. Let $\ket{\psi}$ be a state which has no excitations enclosed by $c$. We wish to show that the ribbon operator acts trivially on the state:
	$$C^{h}(c) \ket{\psi}=\ket{\psi}.$$
	To do this we will show that the closed ribbon operator is equivalent to a product of vertex transforms. If $S$ is the region enclosed by the ribbon, then we claim that the operator $C^h(c)$ is equivalent to the product of vertex transforms
	$$C^h(c) = \prod_{ v \in S} A_v^{g(s.p-v)^{-1}hg(s.p-v)},$$
	where $v$ runs over each vertex in $S$, including the vertices on the direct path of $c$, and $s.p$ is the start-point of the ribbon $c$. The reason that we wish to show this is that vertex transforms can be absorbed into a state for which the vertex is unexcited, which means that the vertex transforms have no effect on a state for which the corresponding vertex is unexcited. Recall from Section \ref{Section_Recap_Paper_2} that a vertex transform applied on vertex $v$ acts trivially on a state $\ket{\psi}$ for which the vertex $v$ is unexcited:
	$$A_v^x \ket{\psi} = A_v^x A_v \ket{\psi} =A_v \ket{\psi}= \ket{\psi}.$$
	
	By showing that the closed magnetic ribbon operator is equivalent to a series of vertex transforms, we can therefore show that the ribbon operator has trivial action on states for which the vertices are unexcited. There is a slight subtlety, however, in that the labels $g(s.p-v)^{-1}hg(s.p-v)$ of the vertex transforms are configuration dependent (i.e., the labels are operators). We therefore need to prove that we can absorb these types of vertex transforms into $A_v$, rather than just constant-labelled transforms. We can write a configuration-dependent $A_v^{\hat{g}_\text{config}}$ transform acting on a state $\ket{\psi}$ as follows:
	\begin{align*}
		A_v^{\hat{g}_\text{config}} \ket{\psi}&=\sum_{x \in G} A_v^x \delta(x,\hat{g}_{\text{config}}) \ket{\psi}.
	\end{align*}
	Then if $\ket{\psi}$ is an eigenstate of $A_v$ with eigenvalue +1 ( i.e., the vertex $v$ is unexcited in $\ket{\psi}$), we can write $\ket{\psi}=A_v \ket{\psi}$ to obtain
	\begin{align*}
		A_v^{\hat{g}_\text{config}} \ket{\psi}&=\sum_{x \in G} A_v^x \delta(x,\hat{g}_\text{config}) A_v \ket{\psi}\\
		&=\sum_{x \in G} \delta(x, ({A_v^x})^{-1}:\hat{g}_\text{config})A_v^gA_v \ket{\psi}\\
		&=\sum_{x \in G} \delta(x, {A_v^{x^{-1}}}:\hat{g}_\text{config}) \ket{\psi}.
	\end{align*}
	
	The configuration-dependent label that we need in order to reconstruct the magnetic ribbon operator has the form $\hat{g}_\text{config}=g(s.p-v)^{-1}hg(s.p-v)$, where $h$ is the label of the magnetic ribbon operator and $g(s.p-v)$ is the configuration-dependent path label to the vertex $v$ on which we apply the vertex transform. In Section \ref{Section_Electric_Ribbon_Operator_Proof} we showed that a vertex transform affects a path element that terminates at that vertex according to Equation \ref{Equation_vertex_transform_end_path_1}:
	$$A_v^x: g(s.p-v)= g(s.p-v)x^{-1}.$$
	This means that the vertex transform acts on $g(s.p-v)^{-1}hg(s.p-v)$ as
	$$A_v^{x^{-1}}:g(s.p-v)^{-1}hg(s.p-v)=x^{-1}g(s.p-v)^{-1}hg(s.p-v)x,$$
	unless $v$ is the start-point itself. If $v$ is the start-point, then the path $s.p-s.p$ is trivial, or the path is closed. If the path is closed, then as long as it encloses a fake-flat surface, the path has label in the image of $\partial$ and so in the centre of $G$, which means that $g(s.p-v)^{-1}hg(s.p-v)=h$. In either case, the label of the vertex transform is $h$, which is a constant group element. This means that the vertex transform is just an ordinary transform, so we already know that the vertex transform can be absorbed. For the other cases we can write the action of the configuration-dependent vertex transform on the state as
	\begin{align*}
		A_v^{g(s.p-v)^{-1}hg(s.p-v)} \ket{\psi}&= A_v^{g(s.p-v)^{-1}hg(s.p-v)} A_v \ket{\psi}\\
		&= \sum_{x \in G} A_v^x \delta(x, g(s.p-v)^{-1}hg(s.p-v) ) A_v \ket{\psi}\\
		&= \sum_{x \in G} \delta(x, x^{-1}g(s.p-v)^{-1}hg(s.p-v)x ) A_v^x A_v \ket{\psi}\\
		&= \sum_{x \in G} \delta(x, g(s.p-v)^{-1}hg(s.p-v) ) A_v \ket{\psi}\\
		&= (\sum_{x \in G} \delta(x, g(s.p-v)^{-1}hg(s.p-v) )) \ket{\psi}\\
		&= \ket{\psi}.
	\end{align*}
	Therefore
	$$A_v^{g(s.p-v)^{-1}hg(s.p-v)} \ket{\psi}= \ket{\psi},$$
	which indicates that these configuration-dependent vertex transforms do not change states that satisfy the vertex energy term, such as the ground state, and so any operator that we build out of such transforms is equivalent to the identity operator when acting on these states.

\begin{figure}[h]
		\begin{center}
			\begin{overpic}[width=0.75\linewidth]{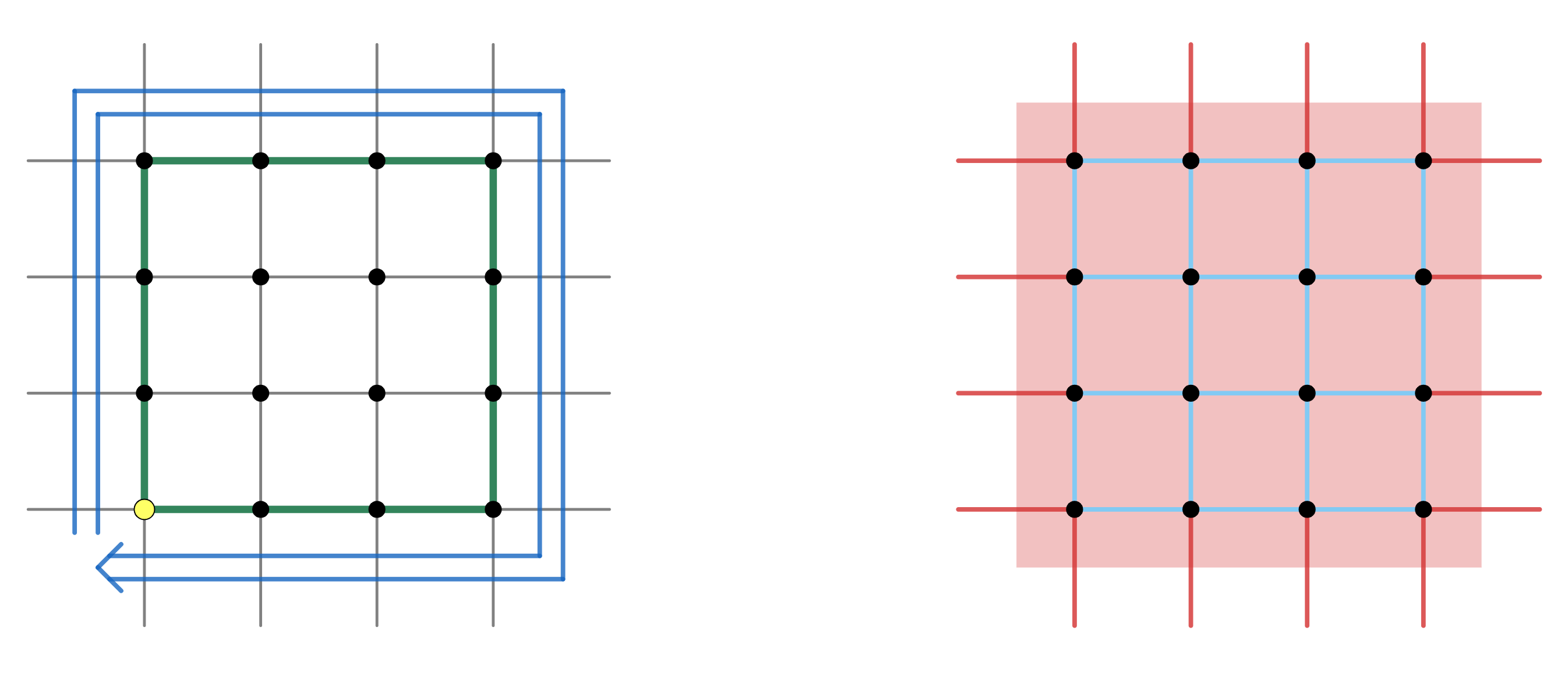}
				\put(45,20){\Huge $\rightarrow$}
				\put(16.5,0){ribbon $c$}
				\put(75,0){region $S$}
			\end{overpic}
			\caption{In order to reproduce the closed magnetic ribbon operator indicated on the left, where the blue arrow indicates the dual path of the ribbon and the dark (green) path is the direct path, we apply a series of vertex transforms in the shaded region indicated on the right. We apply a vertex transform on each of the vertices in this region, so that an internal edge (light blue) is affected by a vertex transform at each end of the edge, while a boundary edge (dark red) is only affected by a vertex transform at one end of the edge.}
			\label{closed_magnetic_ribbon_region}
		\end{center}
	\end{figure}

	Having shown that these vertex transforms do not affect the state, we now need to prove that the closed ribbon operator can indeed be written as a product of such vertex transforms. We claimed that
	$$C^h(c) = \prod_{v\text{ in }S} A_v^{\hat{g}(t)^{-1}h \hat{g}(t)},$$
	where $S$ is the region enclosed by the ribbon $c$ (including vertices on the direct ribbon itself) and $t$ is a path from the start-point of the ribbon operator to the vertex in question (the precise path is not important because we are looking at the case where the region is initially fake-flat). An example of the ribbon and the corresponding vertices on which we apply the transforms is shown in Figure \ref{closed_magnetic_ribbon_region}.
	
	To show that these vertex transforms reproduce the action of the ribbon operator we first consider the action of the transforms on edges internal to the region $S$. That is, consider edges for which both of the vertices that are attached to the edge are within the region $S$ (the blue edges in the right side of Figure \ref{closed_magnetic_ribbon_region}). These edges are not cut by the dual ribbon and so are not affected by the ribbon operator. We wish to see if this is also true for the vertex transforms. Consider such an edge $i$, with source $s(i)$ and target $t(i)$. Under the action of the two vertex transforms at the source and target of $i$, the label $g_i$ becomes
	\begin{equation}
		g_i \rightarrow g(s.p-s(i))^{-1} h g(s.p-s(i))g_i g(s.p-t(i))^{-1} h^{-1} g(s.p-t(i)). \label{Equation_magnetic_topological_edge_2D}
	\end{equation}

	However the path $g(s.p-s(i)) g_i g(s.p-t(i))^{-1}$ (which is the path from the start-point to the edge, along the edge and back to the start-point) is a closed path. Provided that there are no excitations in the region, this closed path will enclose a fake-flat surface. Therefore, for some element $e \in E$ we have $g(s.p-s(i))g_ig(s.p-t(i))^{-1} = \partial(e)=h^{-1}\partial(e)h$ (using the fact that $\partial(e)$ is in the centre of $G$ for that last equality). Inserting this relationship between the path elements into the previous expression for the transformation of the edge label, we obtain
	\begin{align*}
		g_i \rightarrow& g(s.p-s(i))^{-1} h [g(s.p-s(i))g_i g(s.p-t(i))^{-1}] h^{-1} g(s.p-t(i))\\
		=& g(s.p-s(i))^{-1} h [h^{-1}\partial(e)h] h^{-1} g(s.p-t(i))\\
		&= g(s.p-s(i))^{-1} \partial(e)g(s.p-t(i))\\
		&= g(s.p-s(i))^{-1} [g(s.p-s(i))g_ig(s.p-t(i))^{-1}] g(s.p-t(i))\\
		&=g_i.
	\end{align*}

	This means that the edge label is unchanged by the action of the series of vertex transforms, just as it is unaffected by the action of the ribbon operator. Next we consider the edges that are cut by the dual ribbon of the magnetic ribbon operator and so undergo a non-trivial transformation (the red edges in the right of Figure \ref{closed_magnetic_ribbon_region}). For such edges, only one of the vertices attached to the edge is within the region $S$ and so such an edge is affected by a single vertex transform. The action of the vertex transform on such an edge $i$, with label $g_i$, is $$g_i \rightarrow g(s.p-s(i))^{-1} h g(s.p-s(i)) g_i$$ if the vertex is the source of the edge, and $$g_i \rightarrow g_i g(s.p-t(i))^{-1}h^{-1}g(s.p-t(i))$$ if it is the target. Note that if the vertex in region $S$ is the source of the edge, then the edge points away from the direct path, and if the vertex is the target then the edge points towards the direct path. This means that the action of the vertex transform is the same as the action of the ribbon operator on the edge (compare to Equation \ref{Equation_magnetic_ribbon_appendix}), so we have reproduced the ribbon operator with our sequence of vertex transforms.

	One subtlety is that we have assumed that none of the vertex transforms affect the labels of the other vertex transforms (i.e., that the vertex transforms all commute). Ordinarily, vertex transforms on different vertices commute. However we are considering vertex transforms with configuration-dependent labels and so we need to consider how the labels change under the action of other vertex transforms. Recall that the vertex transform labels depend on the path element for the path from the start-point to the vertex in question. We therefore need to consider how these path elements are affected. We need to worry about this because all of the paths in the magnetic ribbon operator are determined simultaneously before the action of the ribbon operator, whereas when we apply a sequence of vertex transforms it is possible that a vertex transform affects the path label for a subsequent vertex transform. Only the vertex transform at the start-point affects the path label $g(s.p-v)$ (a vertex transform at $v$ would also change this label, but we only apply one transform at $v$, so we do not need to worry about two vertex transforms at $v$ failing to commute with each-other). However we can always apply the transform at the start-point last (so that it doesn't affect any path labels) and even if we applied it in a different order, it doesn't affect the label of any transforms because we have the relation $A^h_{s.p}: g(s.p-v)^{-1}hg(s.p-v) = g(s.p-v)^{-1}h^{-1}hhg(s.p-v)=g(s.p-v)^{-1}hg(s.p-v)$.

\begin{figure}[h]
		\begin{center}
			\begin{overpic}[width=0.95\linewidth]{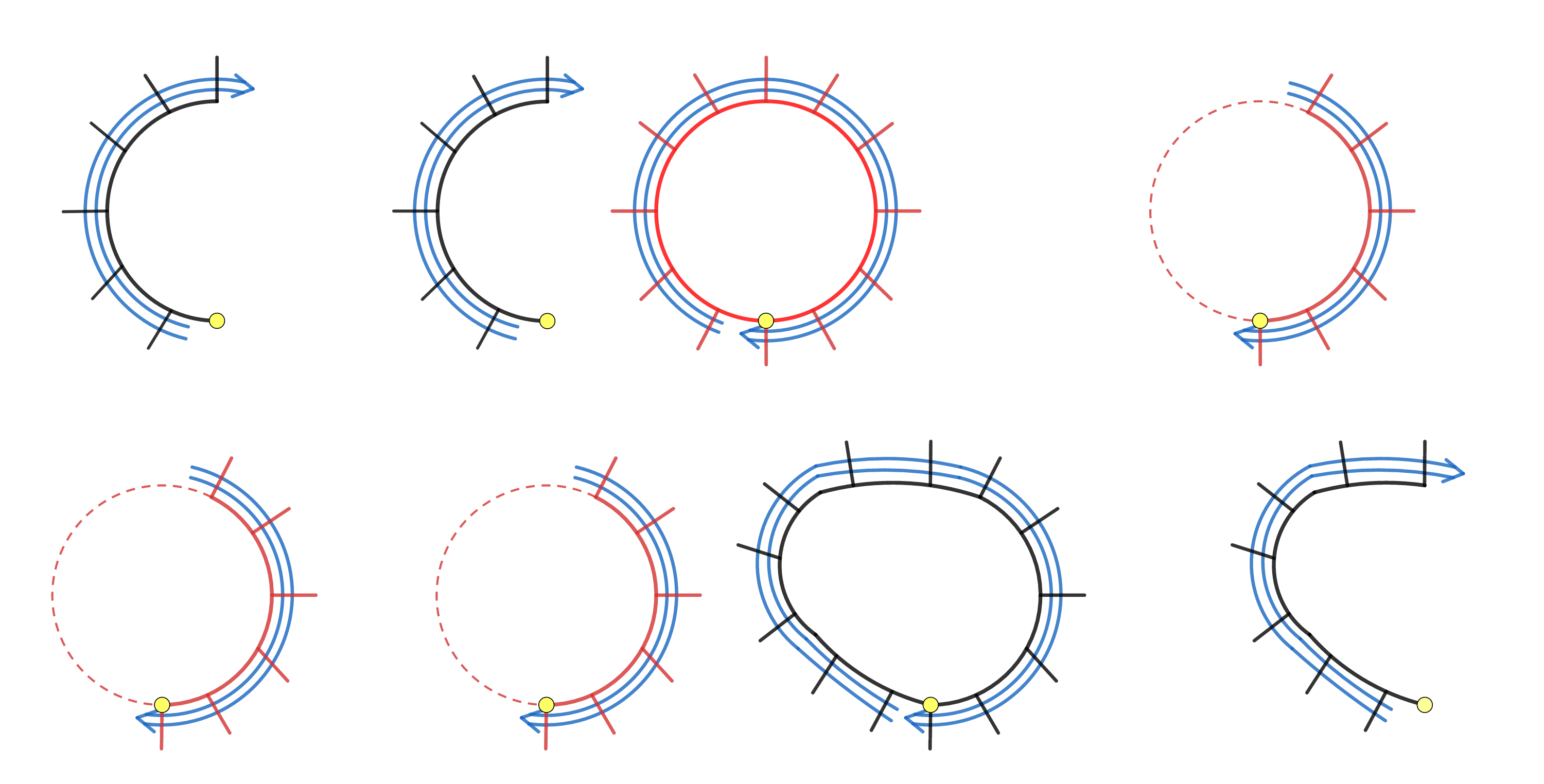}
				\put(16,36){\large $\cdot \ket{\psi}=$}
				\put(36,36){\large$\cdot$}
				\put(62,36){\large$\cdot \ket{\psi} \hspace{0.2cm} =$}
				\put(92,36){\large $\cdot \ket{\psi}$}
				\put(20,11){\large $\cdot \ket{\psi}=$}
				\put(47,11){\large $\cdot$}
				\put(71,11){\large $\cdot \ket{\psi}=$}
				\put(93,11){\large $\cdot \ket{\psi}$}
				\put(5,48){$C^h(t)$}
				\put(26,48){$C^h(t)$}
				\put(43,48){$C^{h^{-1}}(c)$}
				\put(84,48){$C^{h^{-1}}(s)$}
				\put(15,2){$C^{h^{-1}}(s)$}
				\put(40,2){$C^{h^{-1}}(s)$}
				\put(65,2){$C^h(c')$}
				\put(80,2){$C^h(t')$}
			\end{overpic}
			\caption{We consider a magnetic ribbon operator $C^h(t)$ acting on a state $\ket{\psi}$ which has no excitations in the region depicted, as shown in the top left image. The black curve represents the direct path (with the edges pointing outwards from the curve representing the edges pierced by the dual ribbon), while the blue arrow represents the dual path of the ribbon. We can extract a closed ribbon operator $C^{h^{-1}}(c)$ (the red closed ribbon in the second image) from the state, because such an operator acts trivially on the state if it encloses no excitations. We choose this closed ribbon operator so that the ribbon includes $t$ (the left part of $c$ in this example is the same as $t$). Then we can combine the open and closed ribbon operators into a single ribbon operator $C^{h^{-1}}(s)$, because the action of the closed ribbon operator $C^{h^{-1}}(c)$ cancels with the action of $C^h(t)$ on the shared edges (those cut by the dual paths of both ribbon), leaving just the action of the closed ribbon operator on the edges unique to $c$. This is shown in the third image on the top line. The resulting ribbon operator has the opposite orientation and inverse label to $C^h(t)$, but repeating the process (as shown in the lower line) gives us a new ribbon operator $C^h(t')$ which has the same label and orientation as $C^h(t)$, but on a deformed ribbon. Therefore, the ribbon operator is topological.}
			\label{deform_magnetic_ribbon}
		\end{center}
	\end{figure}

	We have therefore shown that we can write the contractible closed magnetic ribbon operator as a product of configuration-dependent gauge transforms, which can be absorbed into $\ket{\psi}$. Therefore, such ribbon operators act trivially on states for which the ribbon encloses no excitations. Next we will show that combining the closed ribbon operators with an open one allows us to deform the open ribbon operator. In order to do so, consider a magnetic ribbon operator $C^h(t)$ applied on a ribbon $t$, such as the one shown in Figure \ref{deform_magnetic_ribbon}. Then we multiply this operator from the right with a closed magnetic ribbon operator $C^{h^{-1}}(c)$. The closed ribbon $c$ is a concatenation of the original open ribbon $t$ with another open ribbon $s$ (so that $c=t \cdot s$ includes $t$ as a part of itself, and the start-point of $c$ is the start-point of $t$). Then we wish to consider the action of the product $C^h(t)C^{h^{-1}}(c)$ on the various edges cut by the dual path ribbon $c$. There are two types of edge to consider. First, we have the edges cut by $t$, which are also cut by $c$ because $c$ includes $t$. Such edges are acted on by both ribbon operators. We consider such an edge $i$, with initial label $g_i$. Using Equation \ref{Equation_magnetic_ribbon_appendix}, which describes the action of a magnetic ribbon operator, the combined action of the two ribbon operators on the edge is given by
	\begin{align*}
		C^h(t)C^{h^{-1}}(c):g_i &=\begin{cases} C^h(t): g(s.p-v_i)^{-1}h^{-1}g(s.p-v_i)g_i & \text{if $i$ points away from the shared direct path of $t$ and $c$}\\
			C^h(t): g_i g(s.p-v_i)^{-1}hg(s.p-v_i) & \text{if $i$ points towards the shared direct path of $t$ and $c$} \end{cases}\\
		&= \begin{cases} g(s.p-v_i)^{-1}hg(s.p-v_i)g(s.p-v_i)^{-1}h^{-1}g(s.p-v_i)g_i & \text{if $i$ points away from the shared} \\ & \text{ direct path of $t$ and $c$}\\
			g_i g(s.p-v_i)^{-1}hg(s.p-v_i) g(s.p-v_i)^{-1}h^{-1}g(s.p-v_i) & \text{if $i$ points towards the shared} \\ & \text{ direct path of $t$ and $c$,} \end{cases}
	\end{align*}
	where the action of each magnetic ribbon operator depends on the same path $s.p-v_i$ because the two ribbon operators share a direct path up to this edge. Then we can cancel the factors from each ribbon operator to obtain
	\begin{align*}
		C^h(t)C^{h^{-1}}(c):g_i &=g_i,
	\end{align*}
	which indicates that the product of the ribbon operators acts trivially on the edges cut by both ribbon operators.

	Next we consider the edges on the ribbon $s$, which are cut by $c$, but which are not cut by $t$. These edges transform according to the action of $C^{h^{-1}}(c)$, so that the label $g_j$ of such an edge $j$ becomes
	$$C^{h^{-1}}(c):g_j= \begin{cases} g(s.p-v_j)^{-1}h^{-1}g(s.p-v_j)g_j & \text{if $j$ points away from the direct path of $c$}\\
		g_j g(s.p-v_j)^{-1}hg(s.p-v_j) & \text{if $j$ points towards the direct path of $c$.} \end{cases}$$
	The product of the two ribbon operators therefore acts trivially on the edges cut by both ribbon operators and acts like a single ribbon operator on the edges cut only by the dual path of $c$. This action is equivalent to that of a single open ribbon operator $C^{h^{-1}}(s)$ on the ribbon $s$, which is made from the part of the dual path of $c$ not in the dual path of $t$, with the direct path of $c$, as shown in Figure \ref{deform_magnetic_ribbon}. That is, by combining a closed and open magnetic ribbon operator, we produce a new open ribbon operator. However this new ribbon operator is not simply a deformed version of the original, because it is oriented in the opposite direction and has the inverse label. However we can then repeat the process, right-multiplying the resulting ribbon $C^{h^{-1}}(s)$ by a new closed ribbon operator $C^h(c')$, whose ribbon is a concatenation of $s$ and the desired final position of the ribbon. This will invert the label of the ribbon operator back to $h$ and put the ribbon in the desired final position (note that we can deform the direct path of the ribbon freely in the unexcited region, as long as it still passes through the vertex attached to each cut edge). This indicates that we can indeed deform the ribbon by applying closed ribbon operators. However we showed earlier that these closed ribbon operators act trivially on states for which they enclose an unexcited region, which means that the deformation also does not affect the action of the ribbon operators as long as we deform them through unexcited regions. That is, the magnetic ribbon operators are indeed topological.
	
	\section{Condensation of magnetic excitations from the ribbon operator picture}
	\label{Section_Condensation_Magnetic_2D}
	
	In this section we will prove the claims we made in Section \ref{Section_2D_Condensation_Confinement} of the main text, that some of the magnetic excitations are condensed. To do this, we will show that the ribbon operators (in 2+1d) corresponding to the condensed excitations act like local operators on the ground state. Note that some of the $E$-valued loop excitations are also condensed, but we gave an argument to show this in Section \ref{Section_2D_Loop} of the main text, so we will not repeat the argument here.

\begin{figure}[h]
		\begin{center}
			\begin{overpic}[width=0.6\linewidth]{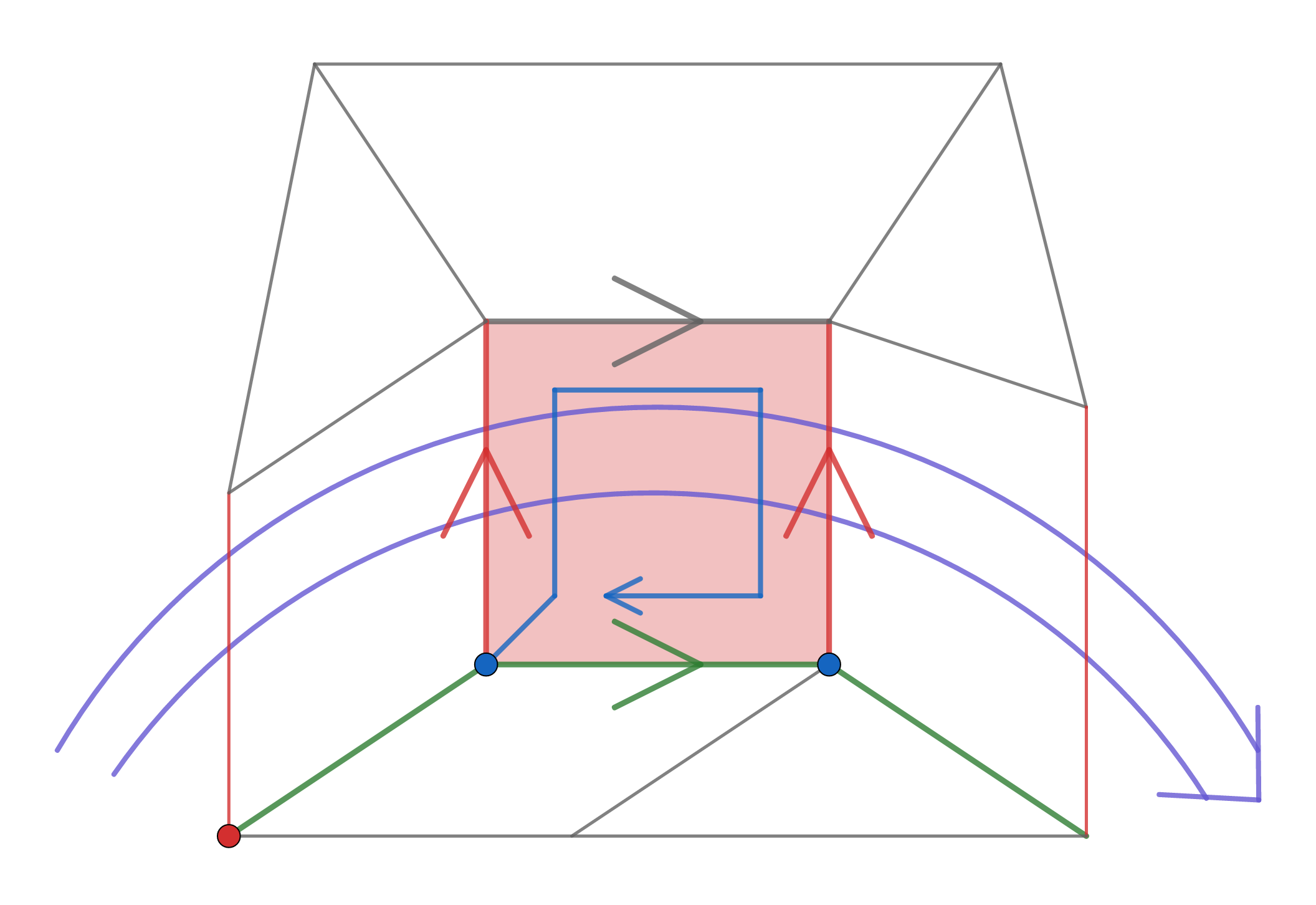}
				\put(5,2){start-point $s.p$}
				\put(0,21){dual path}
				
				\put(40,11){direct path}
				
				\put(32,30){$i$}
				\put(65,30){$j$}
				\put(48,28){$p$}
			\end{overpic}
			\caption{An internal plaquette on a ribbon is one that is passed through by the dual path. This means that two of the edges on the boundary of the plaquette are cut by the dual path. When we attempt to replicate the action of a ribbon operator $C^{\partial(e)}(t)$, we apply an edge transform on each edge cut by the dual path. The label of plaquette $p$ in the figure is therefore affected by an edge transform on each edge $i$ and $j$. The effects of these two edge transforms on the plaquette label cancel, so that the plaquette label is left unaffected. This matches the action of the magnetic ribbon operator, which does not affect any plaquette labels.}
			\label{condensed_magnetic_ribbon_internal_plaquette}
		\end{center}
	\end{figure}

	In the 2+1d model, we only look at the magnetic excitations in the case where $\rhd$ is trivial. We claim that the magnetic excitations labelled by elements of $\partial(E)$ are condensed. To show this, we will prove that the ribbon operators that produce such excitations can be reproduced by local operators, when acting on the ground state (or any state which has no excitations in the region of the ribbon). Consider the magnetic ribbon operator labelled by $\partial(e)$, for arbitrary $e \in E$. The action of such a ribbon operator on an edge $i$ cut by the dual path, initially labelled by $g_i$, is
	$$C^{\partial(e)}(t):g_i = \begin{cases} \partial(e)g_i & i \text{ points away from the direct path} \\ g_i \partial(e)^{-1} & i \text{ points towards the direct path,} \end{cases}$$
	where we have used the fact that $\partial(e)$ is in the centre of $G$ to simplify the expression $g(t)^{-1}hg(t)$ that would usually appear in the action of the ribbon operator (with $h=\partial(e)$ in this case). This action on the edges is very similar to that of a series of edge transforms on each edge $i$ cut by the ribbon. Indeed, consider applying an edge transform on each of the affected edges, with label $e$ if the edge points away from the direct path and $e^{-1}$ if the edge points towards the direct path. This reproduces the action of the magnetic operator on the edges, because the action on an affected edge $i$ is to take the label $g_i$ to $\partial(e)g_i$ if the edge points away from the direct path and $\partial(e)^{-1}g_i=g_i\partial(e)^{-1}$ if the edge points towards the direct path. This suggests that the action of the ribbon operator can at least partially be reproduced by edge transforms, which act trivially on the ground state. However, unlike the magnetic ribbon operator, the edge transforms also affect the plaquette elements. Consider the action of the series of edge transforms on a plaquette internal to the ribbon, that is a plaquette that is not at either end of the dual path, as shown in Figure \ref{condensed_magnetic_ribbon_internal_plaquette}. Two of the edges on the boundary of such a plaquette are cut by the dual path, and so the plaquette label is affected by two of the edge transforms. First we consider the case where both edges point away from the direct path. In this case, one of the edges (edge $i$) points along the boundary of the plaquette and the other (edge $j$) points against the boundary. Therefore, the net result of the two edge transforms on the plaquette $p$, initially labelled by $e_p$, is:
	\begin{align*}
		\mathcal{A}_i^e \mathcal{A}_j^e :e_p &=\mathcal{A}_i^e : e e_p\\
		&= e e_p e^{-1} =e_p,
	\end{align*}
	so the plaquette label is unaffected by the edge transforms.

	Now consider the case where one of the edges has the opposite orientation, so that it points towards the direct path. For instance, suppose that $i$ points towards the direct path. In this case, the edge now points against the boundary of the plaquette, so an edge transform $\mathcal{A}_i^f$ acts on the plaquette as $\mathcal{A}_i^f: e_p = fe_p$. However, we also change the label of the edge transform from $e$ to $e^{-1}$. Therefore, we apply the operator $\mathcal{A}_i^{e^{-1}}\mathcal{A}_j^e$. This means that the effect on the plaquette $p$ is
	\begin{align*}
		\mathcal{A}_i^{e^{-1}} \mathcal{A}_j^e :e_p &=\mathcal{A}_i^{e^{-1}} : \:e e_p\\
		&= e^{-1} e e_p =e_p,
	\end{align*}
	so that the plaquette is again unaffected by the edge transforms. A similar situation occurs if we change the orientation of edge $j$: the change in action of the edge transform from swapping the orientation of the edge is countered by the change in the label of the edge transform that we apply. This means that, regardless of the orientation of the edges, the label of an internal plaquette is unchanged by the sequence of edge transforms that we apply. The magnetic ribbon operator also does not affect the label of plaquettes. Therefore, we have shown that the sequence of edge transforms reproduces the action of the magnetic ribbon on the edges and on internal plaquettes. That is, the edge transforms act the same as the magnetic ribbon operator everywhere except at the two plaquettes at the ends of the ribbon. These plaquettes are only adjacent to one of the edges on which we apply the edge transforms. This means that there is no cancellation of factors from different edge transforms and so the plaquettes are changed by the edge transforms. The plaquette label $e_p$ of such a plaquette becomes either $e e_p$ or $e_p e^{-1}$, depending on the orientation of the plaquette. However this extra factor of $e$ or $e^{-1}$ on the two plaquettes at the ends of the ribbon can be removed by local operators (one at each end of the ribbon) which simply multiply the label of the affected plaquette by either $e^{-1}$ or $e$.

\begin{figure}[h]
		\begin{center}
			\begin{overpic}[width=0.8\linewidth]{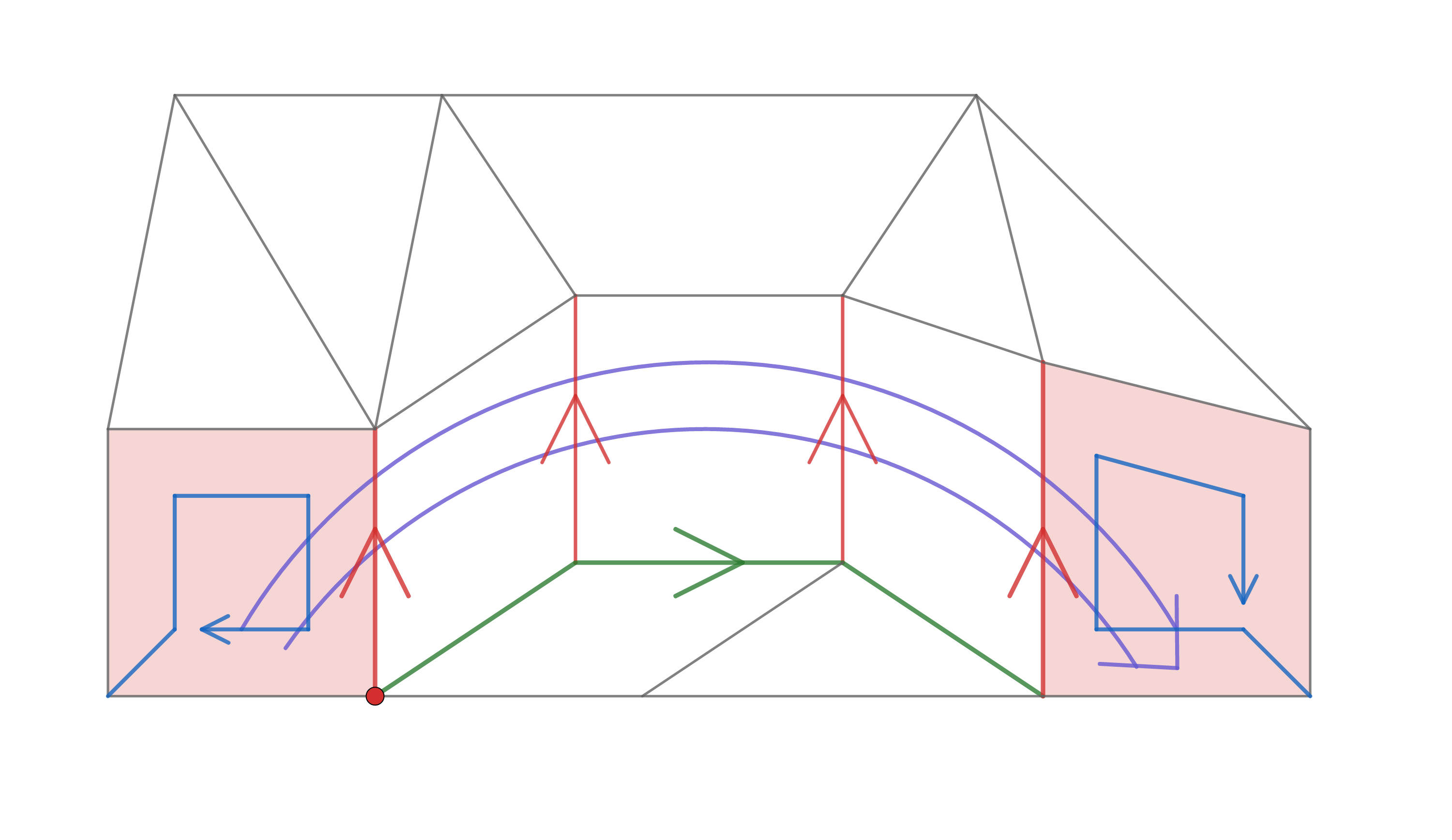}
				\put(15,7){start-point $s.p$}
				\put(43,34){dual path}
				
				\put(40,14){direct path}
				
				\put(12,26){$e_1 \rightarrow e e_1$}
				\put(78,26){$e_5 \rightarrow e_5 e^{-1}$}
				\put(30,21){$e_2 \rightarrow e_2$}
				\put(45,24){$e_3 \rightarrow e_3$}
				\put(60,21){$e_4 \rightarrow e_4$}
			\end{overpic}
			\caption{While we can reproduce the action of the magnetic ribbon operator $C^{\partial(e)}(t)$ on the edges cut by the dual path with a series of edge transforms on these edges, these edge transforms also act on the plaquettes. The edge transforms leave the internal plaquettes unaffected, but change the labels of the plaquettes at the two ends of the dual ribbon. In order to reproduce the action of the magnetic ribbon operator, we must correct this action on the plaquettes at the two ends with local operators. In this example, the label of the plaquette at the start of the ribbon changes from $e_1$ to $ee_1$ (which we can correct by multiplying it by $e^{-1}$), while the label of the plaquette at the end goes from $e_5$ to $e_5e^{-1}$ (which we can correct by multiplying it by $e$). We note that whether the plaquette label is multiplied by $e$ or $e^{-1}$ depends on the orientation of both the plaquette and the ribbon operator.}
			\label{condensed_magnetic_ribbon_end_plaquettes}
		\end{center}
	\end{figure}

	For example, consider Figure \ref{condensed_magnetic_ribbon_end_plaquettes} which shows the action of the series of edge transforms on a set of plaquettes. In this example, we have oriented all of the edges to point away from the direct path, so we apply the edge transform $A_i^e$ on each edge $i$. Considering the two boundary plaquettes (shaded in Figure \ref{condensed_magnetic_ribbon_end_plaquettes}), which are only affected by a single edge transform, we know that the plaquette label is right-multiplied by $e^{-1}$ under the action of $A_i^e$ if the plaquette orientation matches the direction of the edge (i.e., the circulation of the plaquette through that edge lines up with the edge). Therefore, the plaquette at the end of the ribbon in Figure \ref{condensed_magnetic_ribbon_end_plaquettes} is right-multiplied by $e^{-1}$, which can be corrected to match the action of the magnetic ribbon operator by multiplying the plaquette label by $e$ (i.e., applying a single plaquette multiplication operator $M^e(\text{end})$, as defined in Section \ref{Section_Single_Plaquette_1} of the main text, on the end plaquette). On the other hand, if the plaquette circulation is anti-aligned with the edge, as is the case for the plaquette at the start of the ribbon, the label of the plaquette is left-multiplied by $e$ (and so is corrected by applying a single plaquette multiplication operator $M^{e^{-1}}(\text{start})$). The factor that determines whether we use $e$ or $e^{-1}$ is the relative orientation of the plaquette with the affected edges on those plaquettes, once the edges are chosen to point away from the direct path. For example, if we made the plaquettes clockwise, the plaquette at the start would instead have its label multiplied by $e^{-1}$ and be corrected by multiplying the label by $e$. Similarly, if we had put the direct path above the dual path in the figure, rather than below it, we would have to reverse the direction of the edges in order to have them point away from the direct path. This would reverse the relative orientation of the plaquettes and edges and so would again invert the effect of the edge transforms on the two plaquettes.
	
	Apart from these factors on the plaquettes at the two ends of the ribbon operator, the series of edge transforms perfectly replicates the action of the magnetic ribbon operator $C^{\partial(e)}(t)$. This means that the action of the magnetic ribbon operator differs from a series of edge transforms only by local operators (the operators which we apply on the two ends of the ribbon to correct the action on the plaquettes). In the ground state, or any state where the region around the ribbon is not excited, the edge transforms have no effect on the state, because a state $\ket{\psi}$ for which an edge $i$ is unexcited satisfies $\mathcal{A}_i^e\ket{\psi}=\ket{\psi}$. Therefore, the action of the magnetic ribbon operator on such a state is the same as the action of the local operators which change the labels of the boundary plaquettes. This indicates that a magnetic ribbon operator with label in the image of $\partial$ is equivalent to operators that are local to the excitations produced by the ribbon, and so the excitations are condensed.
	
	\section{Calculation of braiding statistics}
	\label{Section_braiding_supplemental}
	
	In Section \ref{Section_2D_Braiding} of the main text, we discussed the braiding statistics of the various excitations in the 2+1d case. In this Section we will verify the results that we claimed in Section \ref{Section_2D_Braiding}. As explained in that section, in the case where $\rhd$ is non-trivial we were unable to define magnetic excitations and so found no non-trivial braiding. Therefore, in this section we will only consider the case where $\rhd$ is trivial.

	\subsection{Braiding between magnetic and electric excitations}
	\label{Section_2D_braiding_electric_magnetic}
\begin{figure}[h]
		\begin{center}
			\begin{overpic}[width=0.75\linewidth]{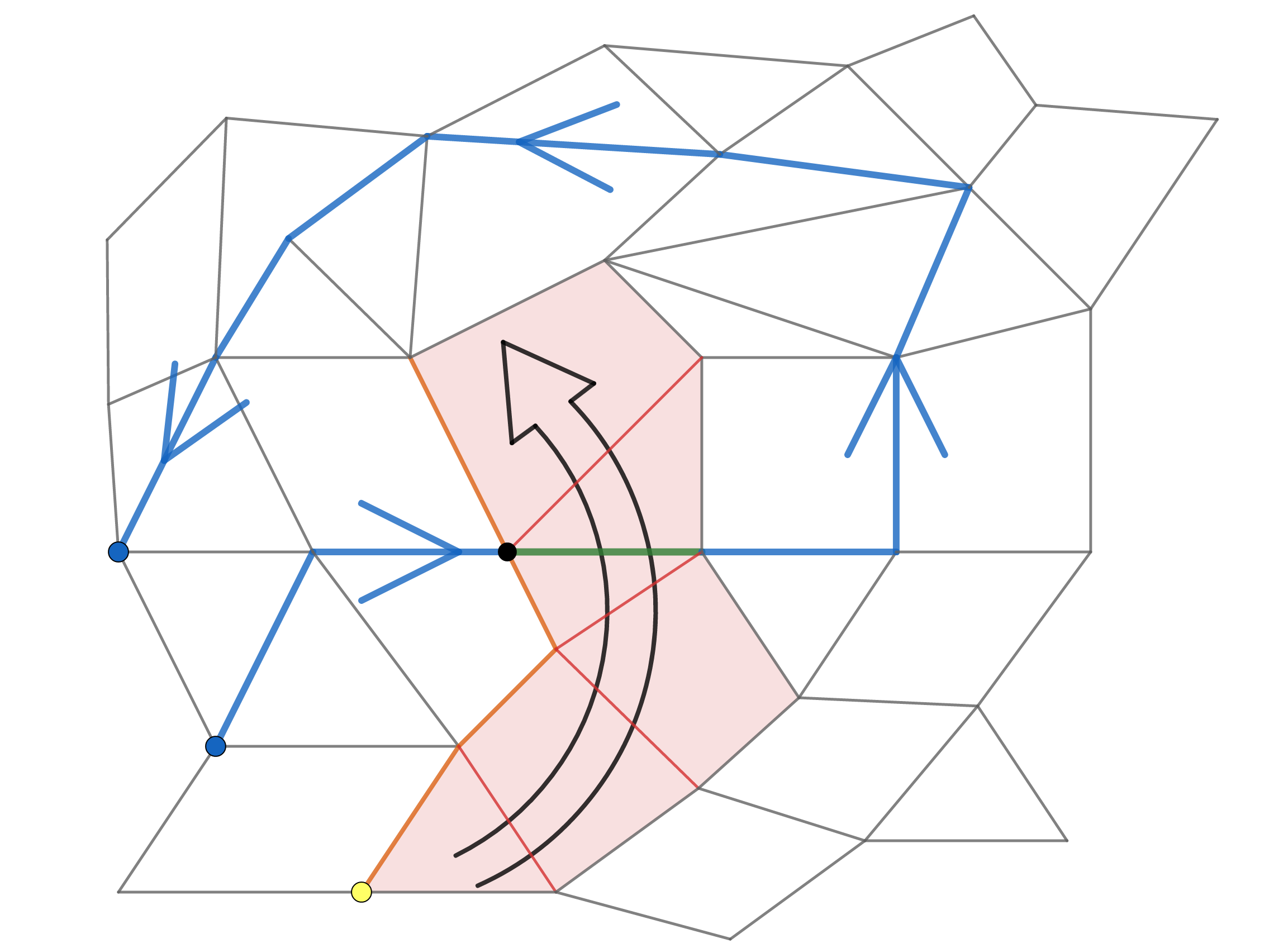}
				\put(16,13){$s.p(t)$}
				\put(25,1){$s.p(r)$}
				\put(28,57){$t$}
				\put(53,20){$r$}
				\put(38,28){$v_i$}
				\put(52,31){$i$}
			\end{overpic}
			\caption{We consider braiding an electric excitation anticlockwise around a magnetic excitation. In order to do so, we apply a magnetic ribbon operator on the red ribbon $r$, followed by an electric ribbon operator on the blue path $t$, which intersects the red ribbon on the green edge, edge $i$. Comparing this to the case where we apply the operators in the opposite order gives us the braiding relation. The start-point of the ribbon, $s.p(r)$ is shown as a yellow dot. When this start-point is in the same position as the start-point of the electric ribbon, $s.p(t)$, the braiding relation becomes simpler as described later in this section. Note that this figure shows one particular crossing orientation. If the path $t$ wound clockwise around the flux, or the orientation of the flux were flipped, then we would obtain a different braiding result, corresponding to inverting the label of the flux.}
			\label{magnetic_electric_braid_2D_appendix}
		\end{center}
	\end{figure}

	In Section \ref{Section_2D_Braiding_Tri_Trivial} of the main text, we argued that the only non-trivial braiding in the $\rhd$ trivial case is between electric and magnetic excitations or two magnetic excitations. The braiding relations for the electric and magnetic excitations are the same as those for the corresponding excitations in Kitaev's Quantum Double model \cite{Kitaev2003}. This is because the ribbon operators that produce these excitations act on the $G$-valued labels in the same way as the corresponding ribbon operators in the Quantum Double model and the difference in the ground state between the models does not interfere with this for the non-confined excitations. We will now verify these braiding relations with explicit calculations. First consider the case of an electric excitation braiding with a magnetic one. In order to do this, we consider a crossing of the ribbon operators that produce and move these excitations, as shown in Figure \ref{magnetic_electric_braid_2D_appendix}. By applying the magnetic ribbon operator on the red ribbon we produce a pair of magnetic excitations and separate them. Following this by applying an electric ribbon operator on the blue path moves an electric excitation anticlockwise around the magnetic excitation at the end of the red ribbon. On the other hand, if we first applied the electric ribbon operator, then the magnetic ribbon operator, we would move the electric excitation through the vacuum before producing the magnetic excitations. This indicates that we are interested in the commutation relation between the two ribbon operators.

	We consider a magnetic ribbon operator $C^h(r)$ applied on the ribbon $r$ and an electric ribbon operator $\delta(g,\hat{g}(t))$ on the path $t$. The commutation relation between the two operators is non-trivial because the magnetic ribbon operator affects the label of the edge where the path of the electric operator crosses the ribbon of the magnetic operator. Taking the example in Figure \ref{magnetic_electric_braid_2D_appendix}, this edge is the green edge, edge $i$. The effect of the magnetic membrane operator on this edge is
	$$C^h(r):g_i = g(s.p(r)-v_i)^{-1}hg(s.p(r)-v_i)g_i$$
	if the edge points away from the direct path of the ribbon (the orange path in Figure \ref{magnetic_electric_braid_2D_appendix}) and 
	$$C^h(r):g_i = g_i g(s.p(r)-v_i)^{-1}h^{-1}g(s.p(r)-v_i)$$
	otherwise. In order to determine how this affects the path label of the electric ribbon, $\hat{g}(t)$, we split the path $t$ into three parts. Firstly, we have the path up to $v_i$, which we label by $t_1$. Then we have the edge $i$ itself and finally we have the path after the edge, which we label $t_2$. The path elements $g(t_1)$ and $g(t_2)$ corresponding to the paths $t_1$ and $t_2$ are unaffected by the magnetic ribbon operator. Therefore, writing that $g(t)=g(t_1)g_i^{\sigma_i}g(t_2)$, where $\sigma_i$ is $+1$ if the edge $i$ points along $t$ and $-1$ if it points against $t$, we see that the action of the magnetic operator on the path element $g(t)$ measured by $\hat{g}(t)$ is
	\begin{align}
		C^h(r): g(t) &= C^h(r): g(t_1)g_i^{\sigma_i}g(t_2) \notag\\
		&=g(t_1) g(s.p(r)-v_i)^{-1}hg(s.p(r)-v_i)g_i^{\sigma_i} g(t_2) \notag\\
		&= g(t_1) g(s.p(r)-v_i)^{-1}hg(s.p(r)-v_i)g(t_1)^{-1} g(t_1) g_i^{\sigma_i} g(t_2) \notag\\
		&=g(s.p(t)-s.p(r))hg(s.p(t)-s.p(r))^{-1} g(t), \label{Magnetic_electric_braid_2D_1_appendix}
	\end{align}
	where $g(s.p(t)-s.p(r))= g(t_1) g(s.p(r)-v_i)^{-1}$ corresponds to a path between the start-points of the two ribbon operators. Therefore, the commutation relation between the two ribbon operators is
	\begin{align}
		\delta(g,\hat{g}(t))C^h(r)\ket{GS}&=C^h(r) \delta(g, C^h(r):\hat{g}(t))\ket{GS} \notag \\
		&= C^h(r) \delta( g, g(s.p(t)-s.p(r))hg(s.p(t)-s.p(r))^{-1} \hat{g}(t))\ket{GS} \notag \\
		&= C^h(r) \delta(g(s.p(t)-s.p(r))h^{-1}g(s.p(t)-s.p(r))^{-1}g, \hat{g}(t)\ket{GS}. \label{Magnetic_electric_braid_2D_2_appendix}
	\end{align}
	
	It is helpful to rewrite this result using the irrep basis for the electric ribbon operators described in Section \ref{Section_2D_electric} of the main text. That is, we consider the commutation relation between the magnetic ribbon operator and an electric ribbon operator of the form $S^{R,a,b}(t)=\sum_{g \in G} [D^R(g)]_{ab} \delta(g, \hat{g}(t))$, where $R$ is an irrep of $G$ and $D^R(g)$ is the matrix representation of $g$ in the irrep $R$. In this case, we have
	\begin{align*}
		S^{R,a,b}(t)C^h(r)\ket{GS}&= \sum_{g \in G} [D^R(g)]_{ab} \delta(g,\hat{g}(t))C^h(r)\ket{GS}\\
		&=C^h(r) \sum_{g \in G} [D^R(g)]_{ab} \delta(g(s.p(t)-s.p(r))h^{-1}g(s.p(t)-s.p(r))^{-1}g, \hat{g}(t)\ket{GS}\\
		&= C^h(r) \sum_{\substack{g'= g(s.p(t)-s.p(r))h^{-1}\\ \hspace{0.2cm} \times g(s.p(t)-s.p(r))^{-1}g}} \hspace{-1cm} [D^R(g(s.p(t)-s.p(r))hg(s.p(t)-s.p(r))^{-1}g')]_{ab} \delta(g', \hat{g}(t))\ket{GS}\\
		&= C^h(r) \sum_{g' \in G} \sum_{c=1}^{|R|} [D^R(g(s.p(t)-s.p(r))hg(s.p(t)-s.p(r))^{-1})]_{ac} [D^R(g')]_{cb} \delta(g', \hat{g}(t))\ket{GS}\\
		&= C^h(r) \sum_{c=1}^{|R|} [D^R(g(s.p(t)-s.p(r))hg(s.p(t)-s.p(r))^{-1})]_{ac} S^{R,c,b}(t) \ket{GS}.
	\end{align*}
	
	We see that the commutation of the two ribbon operators mixes different electric ribbon operators labelled by the same irrep but different matrix indices. Furthermore, we see that the coefficients for each of these operators depend on the operator $g(s.p(t)-s.p(r))$. Certain special cases eliminate this operator dependence. If $R$ is a 1D irrep, as is the case when $G$ is Abelian, the result of the commutation becomes 
	\begin{align*}
		S^{R}(t)C^h(r)\ket{GS}&=R(h) C^h(r) S^R(t) \ket{GS},
	\end{align*}
	so that the result of the braiding is just a phase $R(h)$. Another interesting case is where the start-points of the two ribbon operators are taken to be the same. In this case the path $s.p(t)-s.p(r)$ is a closed path. Provided that this path is contractible, this means that the label of this path in the ground state belongs to the image of $\partial$, due to fake-flatness constraints. When $\rhd$ is trivial, elements of $\partial(E)$ are in the centre of $G$. Therefore, $g(s.p(t)-s.p(r))hg(s.p(t)-s.p(r))^{-1})=h$. In this case, the commutation relation becomes
	\begin{equation}
		S^{R,a,b}(t)C^h(r)\ket{GS}=C^h(r) \sum_{c=1}^{|R|} [D^R(h)]_{ac} S^{R,c,b}(t) \ket{GS},
		\label{Equation_magnetic_electric_braiding_2D_same_sp_appendix}
	\end{equation}
	which still exhibits the mixing of different electric ribbon operators, but this time with constant coefficients for the different ribbon operators. As we explained in Section \ref{Section_2D_Braiding_Tri_Trivial} of the main text, we should only expect to obtain such constant coefficients in general when the start-points of the ribbon operators are the same, because in this case the flux created by the magnetic ribbon operator is defined with respect the same point as the charge created by the electric ribbon operator.

	We note that we only considered one particular orientation of the crossing. We can easily consider the case where the electric ribbon operator passes through the magnetic ribbon in the opposite direction, and so braids clockwise around the magnetic excitation, by considering the inverse path $s=t^{-1}$. Then $g(s)=g(t)^{-1}$ and the start-point of $t$ is the end-point of $s$, which we denote by $e.p(s)$. We see that 
	\begin{align}
		C^h(r):g(s)&=C^h(r):g(t)^{-1} \notag \\
		&= [g(s.p(t)-s.p(r))hg(s.p(t)-s.p(r))^{-1} g(t)]^{-1}\notag \\
		&=g(t)^{-1}g(s.p(t)-s.p(r))h^{-1}g(s.p(t)-s.p(r))^{-1}.\notag
	\end{align}
	We then note that $g(t)^{-1}=g(s)=g(s.p(s)-e.p(s))$ and $s.p(t)=e.p(s)$, so that
	\begin{align}
		C^h(r):g(s)&=g(s)g(e.p(s)-s.p(r))h^{-1} g(e.p(s)-s.p(r))^{-1} \notag\\
		&=g(s.p(s)-e.p(s))g(e.p(s)-s.p(r))h^{-1} g(e.p(s)-s.p(r))^{-1} \notag\\
		&= g(s.p(s)-s.p(r))h^{-1}g(s.p(s)-s.p(r))^{-1} g(s.p(s)-s.p(r))g(e.p(s)-s.p(r))^{-1}\notag \\
		&=g(s.p(s)-s.p(r))h^{-1}g(s.p(s)-s.p(r))^{-1} g(s). \label{Magnetic_electric_braid_reverse}
	\end{align}
	
	From this we see that the only change that we need to make compared to our results for the path $t$ is to replace $h$ with $h^{-1}$, so we have
	\begin{align}
		\delta(g,\hat{g}(s))C^h(r)\ket{GS}&=C^h(r) \delta(g, C^h(r):\hat{g}(s))\ket{GS} \notag \\
		&= C^h(r) \delta( g, g(s.p(s)-s.p(r))h^{-1}g(s.p(s)-s.p(r))^{-1} \hat{g}(s))\ket{GS} \notag \\
		&= C^h(r) \delta(g(s.p(s)-s.p(r))hg(s.p(s)-s.p(r))^{-1}g, \hat{g}(s)\ket{GS}. \label{Magnetic_electric_braid_reverse_2} 
	\end{align}
	This reflects the fact that the label of a flux is associated with a particular orientation, and moving the charge around the flux in the opposite direction is like moving the charge around a flux of the inverse label. In a similar way, we could have braided the electric excitation around the magnetic excitation at the start of the ribbon $r$. Braiding the electric excitation clockwise around that excitation is equivalent to braiding the electric excitation anticlockwise around the excitation at the end of $r$, because the region where the ribbon operators cross is the same in either case (in Figure \ref{magnetic_electric_braid_2D_appendix}, the electric ribbon operator could pass down and around the magnetic ribbon operator to perform the clockwise braiding around the start of the ribbon $r$, without changing the region where the two ribbons cross, and so without changing the commutation relation).

	\subsection{Braiding between two magnetic excitations}
	\label{Section_2D_braiding_two_magnetic_appendix}

\begin{figure}[h]
		\begin{center}
			\begin{overpic}[width=0.75\linewidth]{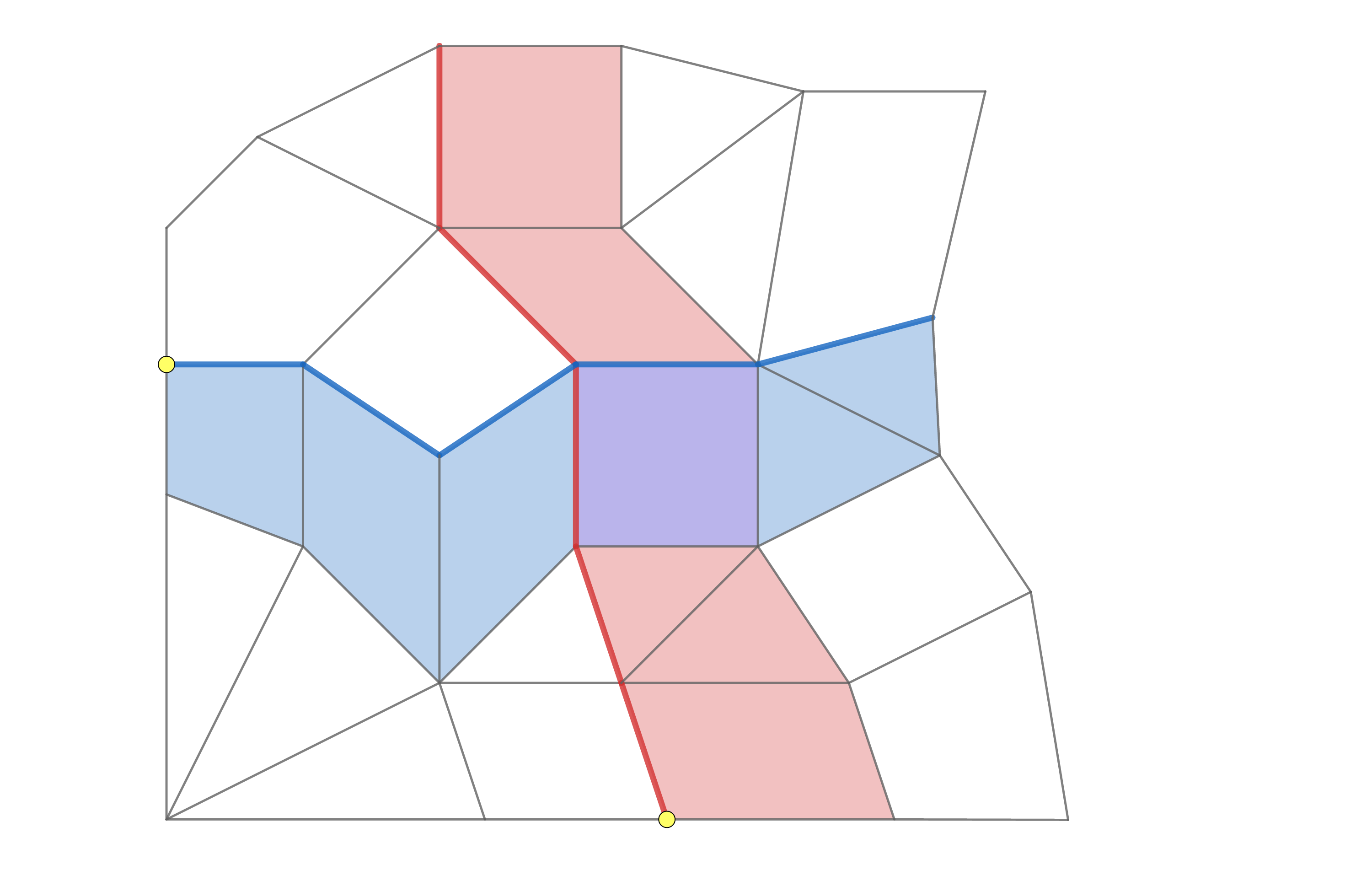}
				\put(5,38){$s.p(s)$}
				\put(42,1){$s.p(t)$}
				\put(60,32){$C^h(s)$}
				\put(35,55){$C^k(t)$}
			\end{overpic}
			\caption{We consider a crossing between two magnetic ribbon operators, $C^k(t)$ and $C^h(s)$. Such a crossing corresponds to a braiding event. If we first apply $C^h(s)$ and then $C^k(t)$, then we are considering the case where we first produce a pair of excitations labelled by $h$ at the start and end of the ribbon $s$, before producing a pair of excitations labelled by $k$ and moving one of them in the presence of the other pair of excitations. Comparing this to the case where we first produce the excitations at the end of $t$ and then produce the excitations at the end of $s$ gives us a braiding relation.}
			\label{Magnetic_magnetic_crossing}
		\end{center}
	\end{figure}

	Next, we consider the braiding between two magnetic excitations. These can braid non-trivially when the group $G$ is non-Abelian. Just as we did with the electric and magnetic excitations, we consider a crossing of their respective ribbon operators, as shown in Figure \ref{Magnetic_magnetic_crossing}. Then we consider the commutation relation between two such crossing magnetic ribbon operators.

\begin{figure}[h]
		\begin{center}
			\begin{overpic}[width=0.75\linewidth]{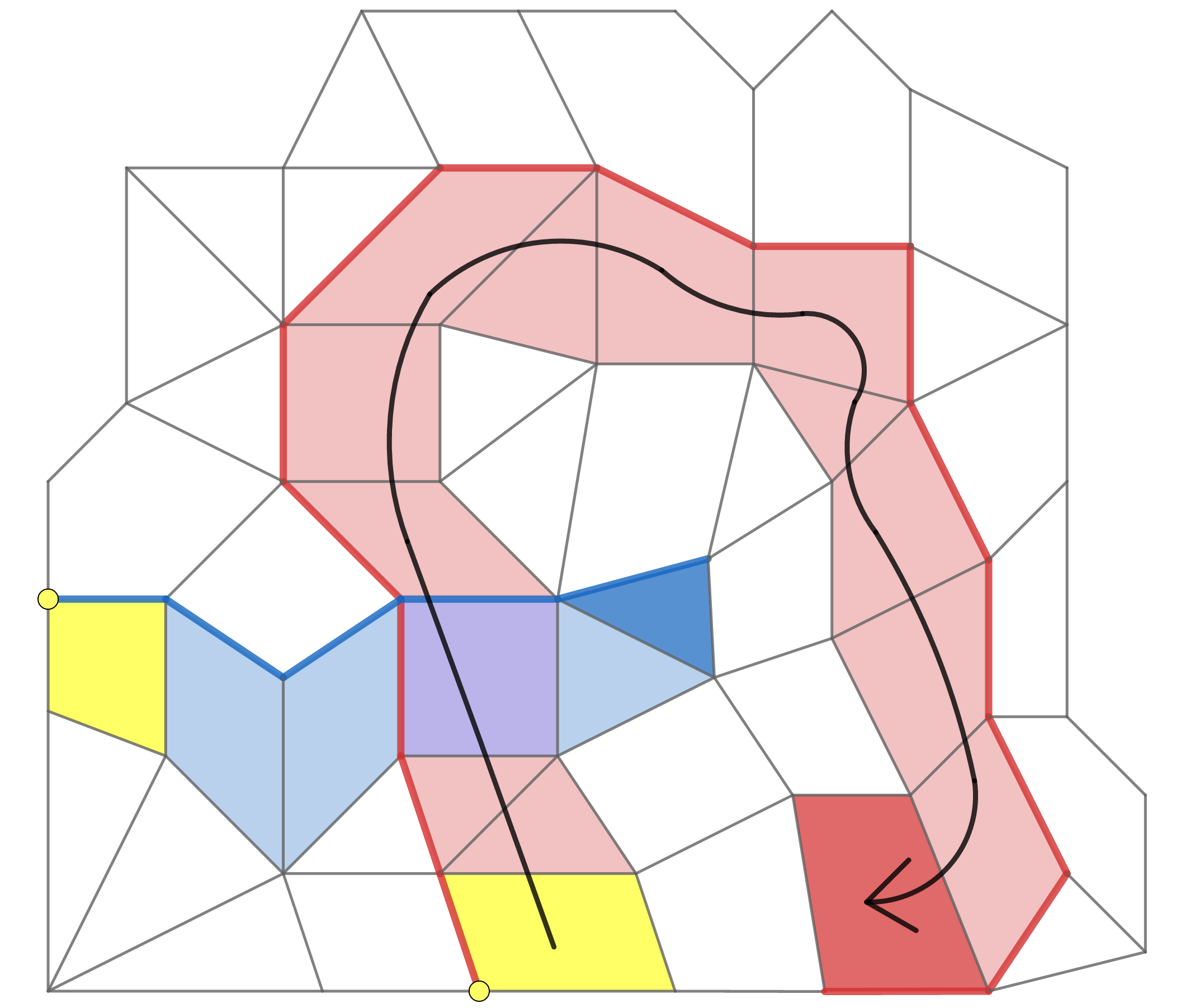}
				\put(2,36){$s.p(s)$}
				\put(32,2){$s.p(t)$}
				\put(16,25){$C^h(s)$}
				\put(48,59){$C^k(t)$}
			\end{overpic}
			\caption{By extending the red ribbon $t$ from Figure \ref{Magnetic_magnetic_crossing}, we can see that the situation in that figure corresponds to the case where we braid the magnetic excitation at the end of the red ribbon $t$ (the excitation shown in darker red) clockwise around the excitation at the end of $s$ (the excitation shown in darker blue). The excitations at the start of the ribbons (shown in yellow) do not undergo the braiding.}
			\label{Magnetic_magnetic_crossing_extended}
		\end{center}
	\end{figure}

	Taking the crossing shown in Figure \ref{Magnetic_magnetic_crossing}, we consider applying the magnetic ribbon operator $C^h(s)$ on the blue ribbon before applying the ribbon operator $C^k(t)$ on the red ribbon. That is, we consider the state $C^k(t) C^h(s) \ket{GS}$, where $t$ is the red ribbon in Figure \ref{Magnetic_magnetic_crossing} and $s$ is the blue ribbon. This crossing can be used to calculate the result of various related braiding moves. We can imagine extending the ribbon $t$, so that the red ribbon passes all the way around the excitation at the end of ribbon $s$, as shown in Figure \ref{Magnetic_magnetic_crossing_extended}. From this, we see that applying $C^h(s)$ then $C^k(t)$ corresponds to the case where we produce and separate a pair of magnetic excitations along the ribbon $s$, before braiding an excitation labelled by $k$ clockwise around the excitation at the end of $s$. By commuting the operator $C^k(t)$ to the right of $C^h(s)$, we can compare this situation to one in which we instead first produce the excitations labelled by $k$ and move them through the vacuum, before producing the excitations labelled by $h$. This means that we must take the state $C^k(t)C^h(s)\ket{GS}$ and move $C^h(s)$ to the left of this expression. In order to do so, we first consider why these operators do not necessarily commute in the case where $G$ is non-Abelian.

\begin{figure}[h]
		\begin{center}
			\begin{overpic}[width=0.75\linewidth]{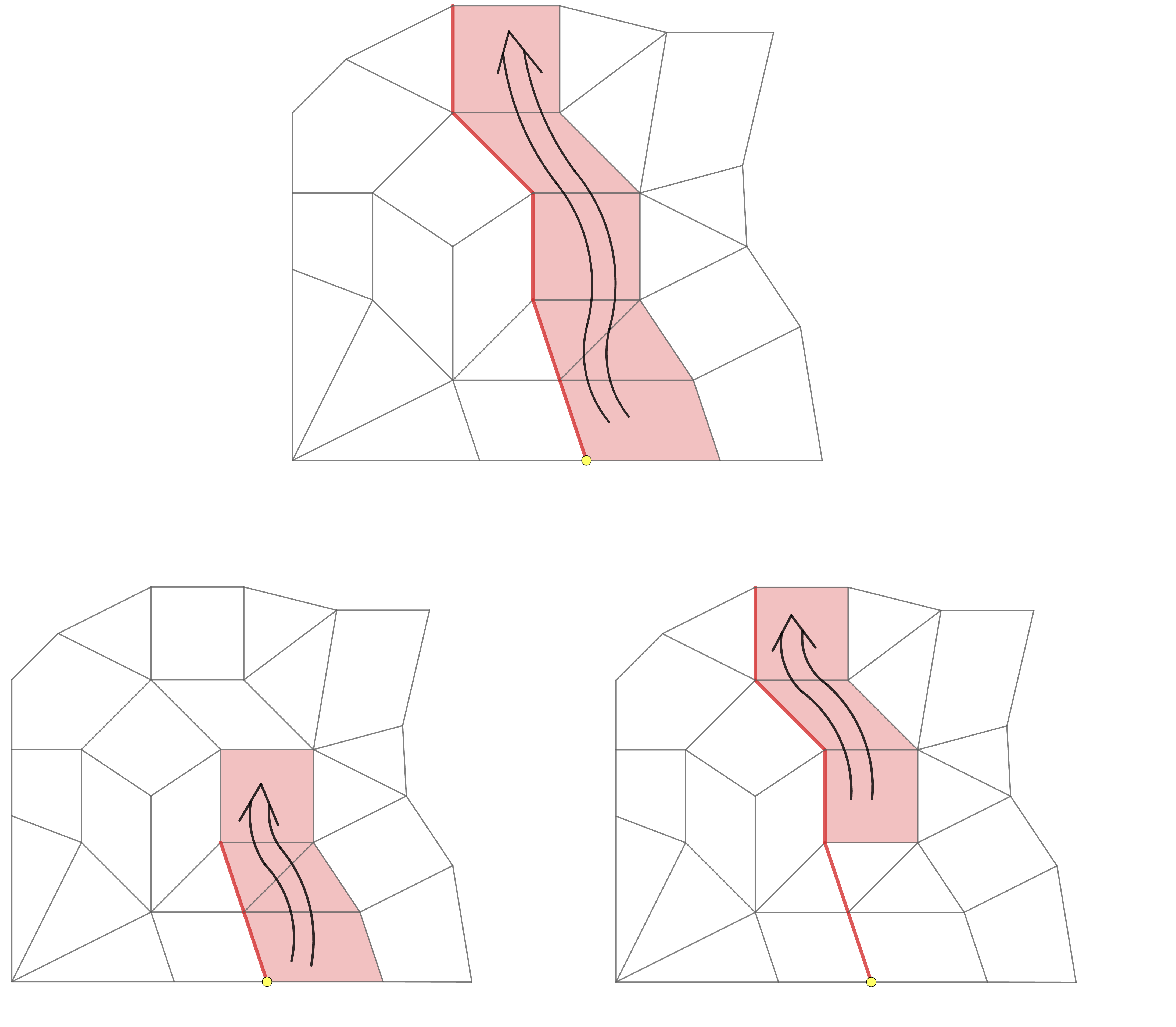}
				\put(38,64){$C^k(t)$}
				\put(45,38){\Huge $\downarrow$}
				\put(45,20){\Huge $\cdot$}
				\put(11,20){$C^k(t_1)$}
				\put(62,20){$C^k(t_2)$}
			\end{overpic}
			\caption{In order to commute the ribbon operator $C^k(t)$ past $C^h(s)$, we split it into two parts, corresponding to the sections of ribbon $t_1$ and $t_2$ (bottom-left and bottom-right respectively). Then $C^k(t)=C^k(t_1)C^k(t_2)$. Note that the dual paths (indicated by the double arrows) of the two fragments concatenate, whereas the direct path of $t_2$ (solid red line) includes the direct path of $t_1$. Each ribbon ($t$, $t_1$ and $t_2$) has the same start-point (yellow dot).}
			\label{Magnetic_ribbon_red_split}
		\end{center}
	\end{figure}

	As we discussed in Section \ref{Section_2D_Magnetic} of the main text, the action of a magnetic ribbon operator $C^x(r)$ on an edge $g_i$ depends on the two paths that make up the ribbon. The dual path determines which edges are affected: if an edge is cut by the dual path then it is acted on by the magnetic ribbon operator. The direct path (the solid red or blue lines in Figure \ref{Magnetic_magnetic_crossing}) determines the precise action on such an edge. For an edge $i$ of label $g_i$ cut by the dual path, the magnetic ribbon operator acts on that edge as
	$$C^x(r):g_i = \begin{cases} g(s.p-v_i)^{-1}xg(s.p-v_i)g_i & \text{ if edge $i$ points away from the direct path} \\ g_i g(s.p-v_i)^{-1}x^{-1}g(s.p-v_i) & \text{ if edge $i$ points towards the direct path},\end{cases}$$
	where $g(s.p-v_i)$ is the label of the section of direct path from the start-point of the ribbon operator to the edge being affected. If two magnetic ribbon operators cross, as in Figure \ref{Magnetic_magnetic_crossing}, the action of the first ribbon operator to be applied affects the label of the direct path of the second ribbon operator, by changing the labels of an edge along that direct path. For instance, $C^h(s)$ affects the label of the red edge making up the left side of the purple square in Figure \ref{Magnetic_magnetic_crossing}, and so affects the label of the direct path of $C^g(t)$. However, because the action of a magnetic ribbon operator on an edge only depends on the section of the direct path up to that edge, changing an edge on the direct path only affects the action of the ribbon operator on edges beyond the affected edge. In Figure \ref{Magnetic_magnetic_crossing}, the blue ribbon only affects the action of the red ribbon on edges past the purple square (including the blue edge that forms the top side of the square). This indicates that the commutation of the two magnetic ribbon operators acts differently on different parts of the ribbon. Therefore, we split each ribbon operator into two parts, corresponding to the ribbon before and after the crossing. The part of each ribbon operator before the crossing is unaffected by the presence of the other ribbon operator, whereas the part afterwards is affected. In Figure \ref{Magnetic_ribbon_red_split}, we indicate how we split the ribbon operator $C^k(t)$ applied on the red ribbon from Figure \ref{Magnetic_magnetic_crossing}. Denoting the first part of the ribbon by $t_1$ and the second part by $t_2$, we have $C^k(t)=C^k(t_1)C^k(t_2)$. Note that the dual paths of the two sections, indicated by the arrows in Figure \ref{Magnetic_ribbon_red_split}, concatenate to form the dual path of the ribbon $t$. By contrast, the direct path of $t_2$ includes the entire direct path of $t_1$, because the two ribbons have the same start-point.

\begin{figure}[h]
		\begin{center}
			\begin{overpic}[width=0.75\linewidth]{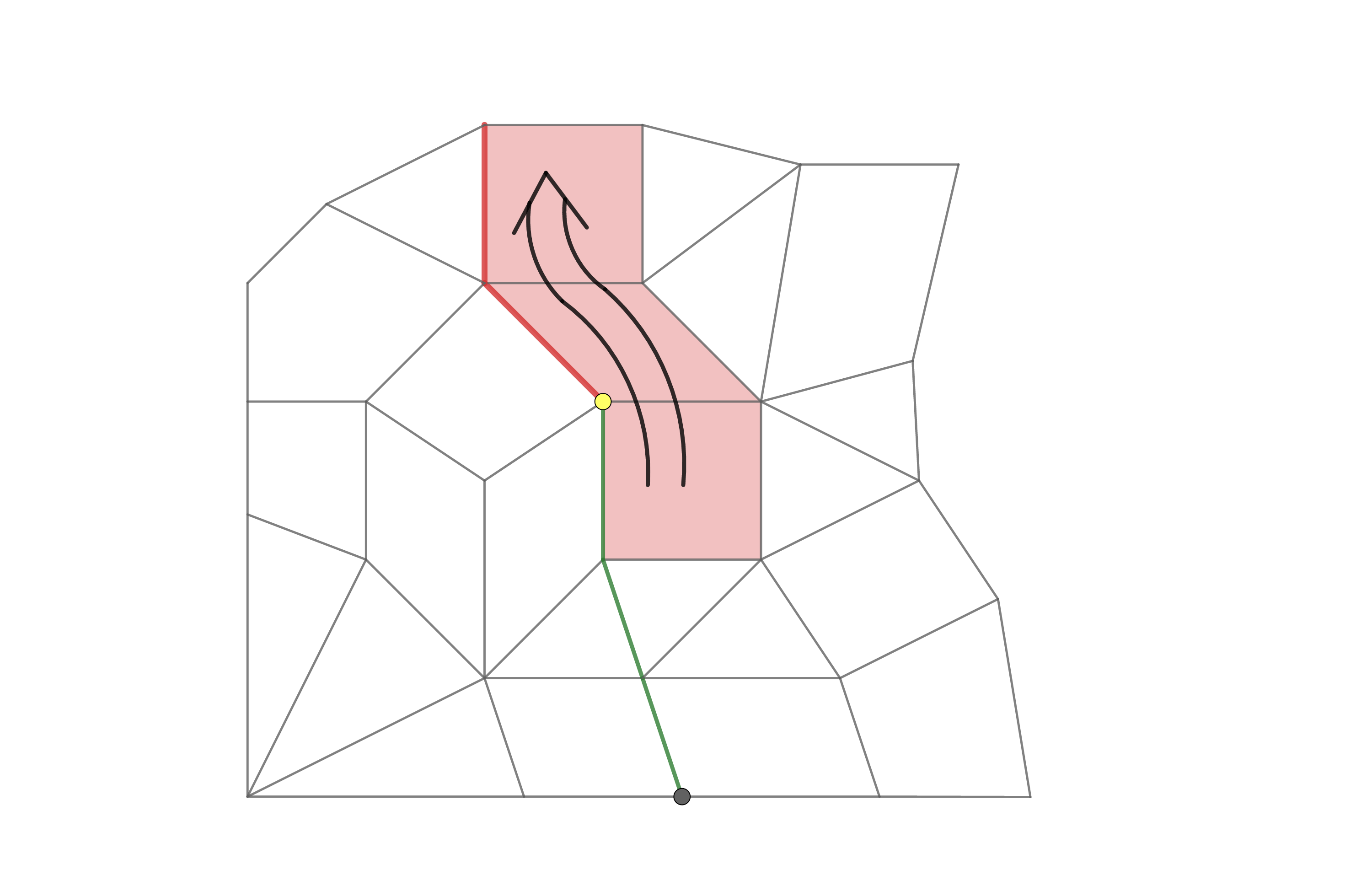}
				\put(51,6){$s.p(t)$}
				\put(35,33){$s.p(t_2')$}
				\put(45,42){$t_2'$}
				\put(29,12){$(s.p(t)-s.p(t_2'))$}
			\end{overpic}
			
			\caption{We can move the start-point of the ribbon $t_2$, from the (grey) vertex $s.p(t)$ to the (yellow) vertex $s.p(t_2')$. We denote the ribbon with this new start-point by $t_2'$. Performing this change of start-point necessitates changing the label of $C^k(t_2)$ from $k$ to $g(s.p(t)-s.p(t_2'))^{-1} k g(s.p(t)-s.p(t_2'))$, where $(s.p(t)-s.p(t_2'))$ is the (green) path between the start-points.}
			\label{Magnetic_ribbon_red_move_sp}
		\end{center}
	\end{figure}

	Because the direct path of the ribbon $t_2$ passes through the ribbon $s$, the operator $C^k(t_2)$ is affected by the operator $C^h(s)$. In order to formalise this, it is convenient to change the start-point of the ribbon $t_2$, to give a new ribbon $t_2'$ as shown in Figure \ref{Magnetic_ribbon_red_move_sp}. The two ribbons $t_2$ and $t_2'$ have the same dual path, so they act on the same edges. However the action of the magnetic operator $C^k(t_2)$ on an edge $g_i$ cut by the dual path is $C^k(t_2):g_i = g(s.p(t)-v_i)^{-1}kg(s.p(t)-v_i)g_i$, assuming that the edge points away from the direct path (and recalling that $s.p(t_2)=s.p(t)$). By contrast, the action of a ribbon operator $C^x(t_2')$ on the same edge is 
	\begin{align*}
		C^x(t_2'):g_i &=g(s.p(t_2')-v_i)^{-1}xg(s.p(t_2')-v_i)g_i\\
		&= g(s.p(t)-v_i)^{-1}g(s.p(t)-s.p(t_2'))xg(s.p(t)-s.p(t_2'))^{-1}g(s.p(t)-v_i) g_i.
	\end{align*}
	We see that this agrees with the action of $C^k(t_2)$ provided that the label $x$ satisfies $g(s.p(t)-s.p(t_2'))xg(s.p(t)-s.p(t_2'))^{-1}=k$, or equivalently $x= g(s.p(t)-s.p(t_2'))^{-1} k g(s.p(t)-s.p(t_2'))$. This is an example of a general feature of magnetic ribbon operators: we can move the start-point of a magnetic ribbon operator, without changing the dual path, by conjugating the label by the path element along which we move the start-point. As long as we do not change the dual path, this leaves the action of the ribbon operator invariant. By moving the start-point of the ribbon $t_2$ beyond the intersection, we remove the overlap between the ribbons $t_2$ and $s$. However, note that the label $x=g(s.p(t)-s.p(t_2'))^{-1} k g(s.p(t)-s.p(t_2'))$ includes the operator elements $g(s.p(t)-s.p(t_2'))$. This means that the non-commutativity of the blue and red ribbon operators is now encoded in the labels of the operators. The blue ribbon operator $C^h(s)$ will change the label $g(s.p(t)-s.p(t_2'))^{-1} k g(s.p(t)-s.p(t_2'))$ of the ribbon operator applied on $t_2'$ when we commute the two ribbon operators past each-other. However we know how the path label $g(s.p(t)-s.p(t_2'))$ is affected by such commutation from our results concerning the braiding of the electric and magnetic excitations earlier in this section. 
	
\begin{figure}[h]
		\begin{center}
			\begin{overpic}[width=0.75\linewidth]{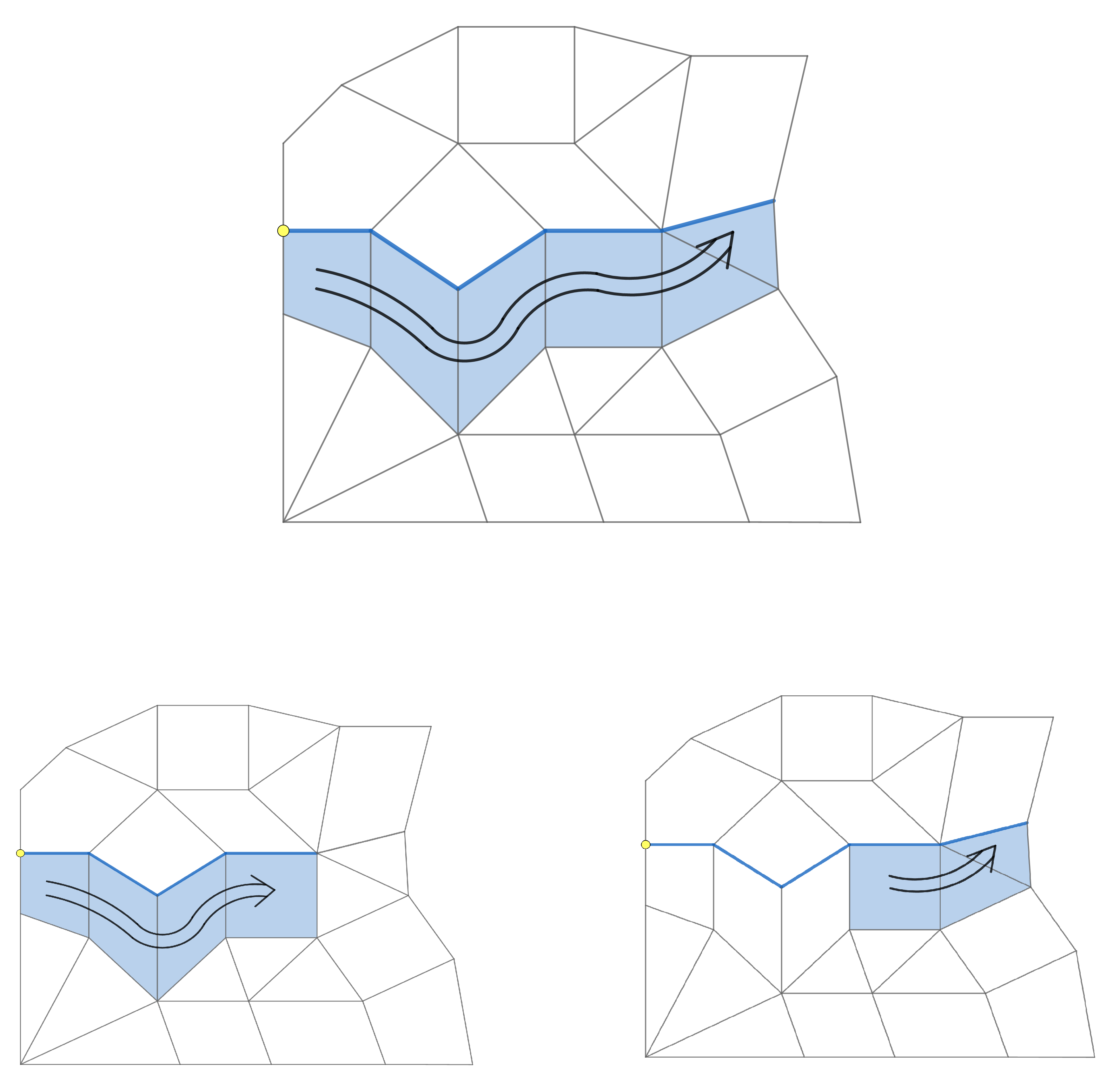}
				\put(48,62){$C^h(s)$}
				\put(45,42){\Huge $\downarrow$}
				\put(45,18){\Huge $\cdot$}
				\put(11,22){$C^h(s_1)$}
				\put(74,23){$C^h(s_2)$}	
			\end{overpic}
			\caption{Just as we split the ribbon $t$ from Figure \ref{Magnetic_magnetic_crossing} into two parts, we must split the ribbon $s$ into a part before and a part after the intersection of the two ribbons. We denote these sections of ribbon by $s_1$ and $s_2$. Then $C^h(s)=C^h(s_1) C^h(s_2)$.}
			\label{Magnetic_ribbon_blue_split}
		\end{center}
	\end{figure}

\begin{figure}[h]
		\begin{center}
			\begin{overpic}[width=0.75\linewidth]{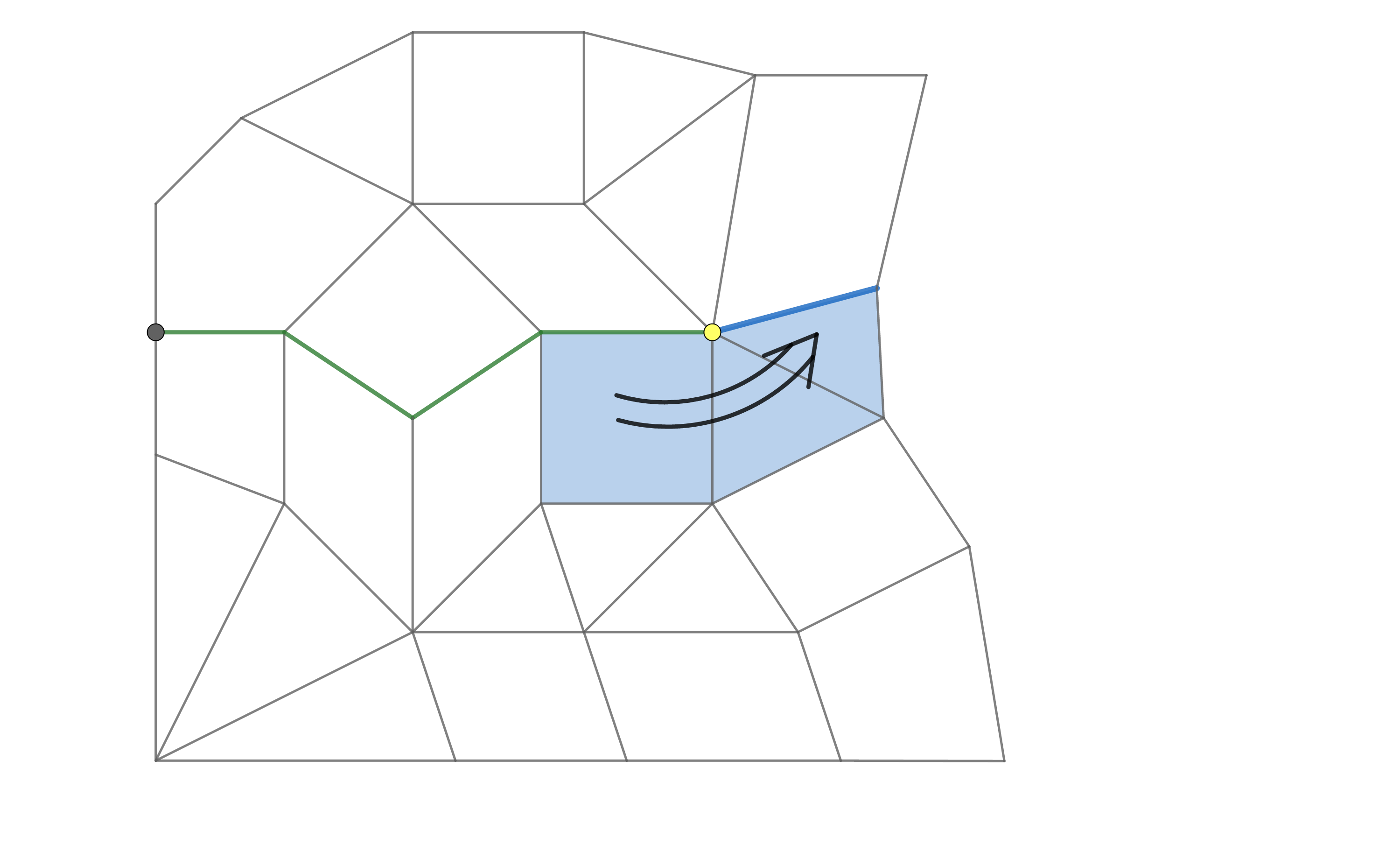}
				\put(4,38){$s.p(s)$}
				\put(20,40){$(s.p(s)-s.p(s_2'))$}
				\put(46,40){$s.p(s_2')$}
				\put(49,28){$s_2'$}
			\end{overpic}
			
			\caption{We move the start-point of the ribbon $s_2$ from $s.p(s_2)=s.p(s)$ (grey vertex) to $s.p(s_2')$ (yellow vertex), thereby changing the label of $C^h(s_2)$ from $h$ to $g(s.p(s)-s.p(s_2'))^{-1} h g(s.p(s)-s.p(s_2'))$. Here $(s.p(s)-s.p(s_2'))$ is the (green) path between the two start-points. We call the ribbon obtained by changing the start-point in this way $s_2'$.}
			\label{Magnetic_ribbon_blue_move_sp}
		\end{center}
	\end{figure}

	In addition to splitting the ribbon $t$ into two parts, before and after the intersection, we split the ribbon $s$ in a similar way, as indicated in Figure \ref{Magnetic_ribbon_blue_split}, giving two ribbon sections $s_1$ and $s_2$, both of which have the start-point $s.p(s)$ for their direct paths. We also change the start-point of the ribbon $s_2$ so that the start-point lies after the intersection, to give us a ribbon $s_2'$. This is indicated in Figure \ref{Magnetic_ribbon_blue_move_sp}. Just as with the ribbon operator applied on $t_2$, changing the start-point of the ribbon results in conjugation of the operator label by the path element along which we move the start-point.

	Having described how we can split the ribbons, we are now able to consider the expression $C^k(t)C^h(s) \ket{GS}$. We write $C^k(t)=C^k(t_1)C^k(t_2)$ where $t_1$ and $t_2$ are the sections of ribbon $t$ shown in Figure \ref{Magnetic_ribbon_red_split}. Similarly, we write $C^h(s)=C^h(s_1)C^h(s_2)$, where $s_1$ and $s_2$ are shown in Figure \ref{Magnetic_ribbon_blue_split}. Then moving the start-points of the ribbons $s_2$ and $t_2$ past the intersection point changes the label of the operators to give us
	\begin{equation}
		C^k(t_2)=C^{g(s.p(t)-s.p(t_2'))^{-1} k g(s.p(t)-s.p(t_2'))}(t_2') \label{Equation_magnetic_ribbon_t2_change_sp}
	\end{equation}
	and
	\begin{equation}
		C^h(s_2)=C^{g(s.p(s)-s.p(s_2'))^{-1}hg(s.p(s)-s.p(s_2'))}(s_2').
		\label{Equation_magnetic_ribbon_s2_change_sp}
	\end{equation}
	We can then write the state $C^k(t) C^h(s) \ket{GS}$ as 
	\begin{align}
		C^k(t)&C^h(s)\ket{GS}=C^k(t_1)C^k(t_2) C^h(s_1) C^h(s_2) \ket{GS} \notag\\
		&=C^k(t_1) C^{g(s.p(t)-s.p(t_2'))^{-1} k g(s.p(t)-s.p(t_2'))}(t_2') C^h(s_1) C^{g(s.p(s)-s.p(s_2'))^{-1}hg(s.p(s)-s.p(s_2'))}(s_2')\ket{GS}. \label{Magnetic_magnetic_commute_2D_0.5}
	\end{align}

	Using these new ribbons, the direct paths of the various ribbons do not intersect with other ribbons. However, as we explained earlier, this does not mean that the various ribbon operators commute, because the labels of the ribbon operators on $t_2'$ and $s_2'$ are now operators: they include the path elements $g(s.p(t)-s.p(t_2'))$ and $g(s.p(s)-s.p(s_2'))$. This means that we must consider how these path elements are affected by the commutation with other magnetic ribbon operators. As we showed in Section \ref{Section_2D_braiding_electric_magnetic}, a path element is affected by a magnetic ribbon operator if the path passes through that ribbon. Looking at Figures \ref{Magnetic_magnetic_crossing}, \ref{Magnetic_ribbon_red_move_sp} and \ref{Magnetic_ribbon_blue_move_sp}, we see that the path $s.p(t)-s.p(t_2')$ passes through the ribbon $s_1$, whereas the path $s.p(s)-s.p(s_2')$ passes through the ribbon $t_2$. We will first consider the commutation relation between $C^h(s_1)$ and 
	$C^{g(s.p(t)-s.p(t_2'))^{-1} k g(s.p(t)-s.p(t_2'))}(t_2')$. We have
	\begin{align*}
		C^{g(s.p(t)-s.p(t_2'))^{-1} k g(s.p(t)-s.p(t_2'))}(t_2')C^h(s_1) &= C^h(s_1) C^{(C^h(s_1):g(s.p(t)-s.p(t_2')))^{-1} k (C^h(s_1):g(s.p(t)-s.p(t_2')))}(t_2').
	\end{align*}
	
	Using our results from the crossing of an electric and magnetic ribbon operator (see Equation \ref{Magnetic_electric_braid_reverse}), we know that 
	$$C^h(s_1):g(s.p(t)-s.p(t_2'))= g(s.p(t)-s.p(s))h^{-1}g(s.p(t)-s.p(s))^{-1}g(s.p(t)-s.p(t_2')),$$
	where we used the fact that the start-point of $s_1$ is the start-point of $s$. Therefore
	\begin{align*}
		[C^h(s_1):g(s.p(t)-s.p(t_2'))]^{-1} k &[C^h(s_1):g(s.p(t)-s.p(t_2'))]\\
		&= \big[g(s.p(t)-s.p(s))h^{-1}g(s.p(t)-s.p(s))^{-1}g(s.p(t)-s.p(t_2'))\big]^{-1} k\\ & \hspace{1cm}g(s.p(t)-s.p(s))h^{-1}g(s.p(t)-s.p(s))^{-1}g(s.p(t)-s.p(t_2')).
	\end{align*}
	In order to simplify this expression, we define $h_{[t-s]}^{\phantom{-1}}= g(s.p(t)-s.p(s))hg(s.p(t)-s.p(s))^{-1}$. Then we can write
	\begin{align*}
		[C^h(s_1):g(s.p(t)-s.p(t_2'))]^{-1} k [C^h(s_1):g(s.p(t)-s.p(t_2'))]&= [h_{[t-s]}^{-1}g(s.p(t)-s.p(t_2'))]^{-1} k [h_{[t-s]}^{-1}g(s.p(t)-s.p(t_2'))]\\
		&=g(s.p(t)-s.p(t_2'))^{-1} [h_{[t-s]}^{\phantom{-1}}k h_{[t-s]}^{-1}] g(s.p(t)-s.p(t_2')).
	\end{align*}
	This tells us that
	\begin{align*}
		C^{g(s.p(t)-s.p(t_2'))^{-1} k g(s.p(t)-s.p(t_2'))}(t_2')C^h(s_1) &= C^h(s_1) C^{[C^h(s_1):g(s.p(t)-s.p(t_2'))]^{-1} k [C^h(s_1):g(s.p(t)-s.p(t_2'))]}(t_2')\\
		&= C^h(s_1) C^{g(s.p(t)-s.p(t_2'))^{-1} [h_{[t-s]}^{\phantom{-1}}k h_{[t-s]}^{-1}] g(s.p(t)-s.p(t_2'))}(t_2').
	\end{align*}
	Substituting this into Equation \ref{Magnetic_magnetic_commute_2D_0.5}, we obtain
	\begin{align}
		&C^k(t)C^h(s)\ket{GS}\notag \\
		&= C^k(t_1) C^h(s_1) C^{g(s.p(t)-s.p(t_2'))^{-1} [h_{[t-s]}^{\phantom{-1}}k h_{[t-s]}^{-1}] g(s.p(t)-s.p(t_2'))}(t_2') C^{g(s.p(s)-s.p(s_2'))^{-1}hg(s.p(s)-s.p(s_2'))}(s_2')\ket{GS}.
	\end{align}
	Then, using the fact that $C^k(t_1)$ and $C^h(s_1)$ commute (there is no intersection of these ribbons, and their operators have constant labels), this becomes
	\begin{align}
		&C^k(t)C^h(s)\ket{GS}\notag \\
		&= C^h(s_1) C^k(t_1) C^{g(s.p(t)-s.p(t_2'))^{-1} [h_{[t-s]}^{\phantom{-1}}k h_{[t-s]}^{-1}] g(s.p(t)-s.p(t_2'))}(t_2') C^{g(s.p(s)-s.p(s_2'))^{-1}hg(s.p(s)-s.p(s_2'))}(s_2')\ket{GS}.
		\label{Magnetic_magnetic_commute_2D_3}
	\end{align}
	
	The next step is to commute $C^{g(s.p(s)-s.p(s_2'))^{-1}hg(s.p(s)-s.p(s_2'))}(s_2')$ past $C^{g(s.p(t)-s.p(t_2'))^{-1} [h_{[t-s]}^{\phantom{-1}}k h_{[t-s]}^{-1}] g(s.p(t)-s.p(t_2'))}(t_2')$. As we mentioned before, the path $s.p(s)-s.p(s_2')$ passes through the ribbon $t_2'$. We therefore need to consider how the magnetic ribbon operator applied on ribbon $t_2'$ affects the path element $g(s.p(s)-s.p(s_2'))$. Denoting the label of the ribbon operator applied on $t_2'$ by $x$, we have 
	\begin{align*}
		C^x(t_2') C^{g(s.p(s)-s.p(s_2'))^{-1}hg(s.p(s)-s.p(s_2'))}(s_2') &=C^{([C^x(t_2')]^{-1}:g(s.p(s)-s.p(s_2')))^{-1}h([C^x(t_2')]^{-1}:g(s.p(s)-s.p(s_2')))}(s_2') C^x(t_2'),
	\end{align*}
	where $x= g(s.p(t)-s.p(t_2'))^{-1} [h_{[t-s]}^{\phantom{-1}}k h_{[t-s]}^{-1}] g(s.p(t)-s.p(t_2'))$. Using Equation \ref{Magnetic_electric_braid_2D_1_appendix}, we see that
	$$C^x(t_2'): g(s.p(s)-s.p(s_2'))= g(s.p(s)-s.p(t_2'))xg(s.p(s)-s.p(t_2'))^{-1} g(s.p(s)-s.p(s_2')),$$ 
	which means that 
	\begin{align*}
		C^x(t_2')^{-1}: g(s.p(s)-s.p(s_2')) &= g(s.p(s)-s.p(t_2'))x^{-1}g(s.p(s)-s.p(t_2'))^{-1} g(s.p(s)-s.p(s_2')).
	\end{align*}
	
	Using this result, we have
	\begin{align*}
		[C^x(t_2')^{-1}:g(s.p(s)-s.p(s_2'))]^{-1}h&[C^x(t_2')^{-1}:g(s.p(s)-s.p(s_2'))]\\
		&= \big[g(s.p(s)-s.p(t_2'))x^{-1}g(s.p(s)-s.p(t_2'))^{-1} g(s.p(s)-s.p(s_2'))\big]^{-1} h\\ & \hspace{1cm}\big[g(s.p(s)-s.p(t_2'))x^{-1}g(s.p(s)-s.p(t_2'))^{-1} g(s.p(s)-s.p(s_2'))\big]\\
		&=\big[g(s.p(s)-s.p(s_2'))^{-1}g(s.p(s)-s.p(t_2'))xg(s.p(s)-s.p(t_2'))^{-1}\big] h\\& \hspace{1cm} \big[g(s.p(s)-s.p(t_2'))x^{-1}g(s.p(s)-s.p(t_2'))^{-1} g(s.p(s)-s.p(s_2'))\big]\\
		&=g(s.p(s)-s.p(s_2'))^{-1}\big[g(s.p(s)-s.p(t_2'))xg(s.p(s)-s.p(t_2'))^{-1} h \\& \hspace{1cm} g(s.p(s)-s.p(t_2'))x^{-1}g(s.p(s)-s.p(t_2'))^{-1} \big] g(s.p(s)-s.p(s_2')),
	\end{align*}
	so that 
	\begin{align*}
		&C^x(t_2') C^{g(s.p(s)-s.p(s_2'))^{-1}hg(s.p(s)-s.p(s_2'))}(s_2')\\
		&=C^{g(s.p(s)-s.p(s_2'))^{-1}\big[g(s.p(s)-s.p(t_2'))xg(s.p(s)-s.p(t_2'))^{-1} h g(s.p(s)-s.p(t_2'))x^{-1}g(s.p(s)-s.p(t_2'))^{-1} \big] g(s.p(s)-s.p(s_2'))}(s_2') C^x(t_2').\\
	\end{align*}
	Substituting this into our commutation relation Equation \ref{Magnetic_magnetic_commute_2D_3}, we see that
	\begin{align}
		&C^k(t)C^h(s)\ket{GS} \notag\\
		&= C^h(s_1) C^{g(s.p(s)-s.p(s_2'))^{-1}\big[g(s.p(s)-s.p(t_2'))xg(s.p(s)-s.p(t_2'))^{-1}hg(s.p(s)-s.p(t_2'))x^{-1}g(s.p(s)-s.p(t_2'))^{-1}\big]g(s.p(s)-s.p(s_2'))}(s_2') \notag\\
		& \hspace{1cm} C^k(t_1) C^{g(s.p(t)-s.p(t_2'))^{-1} [h_{[t-s]}^{\phantom{-1}}k h_{[t-s]}^{-1}] g(s.p(t)-s.p(t_2')) }(t_2') \ket{GS}. \label{Magnetic_magnetic_commute_2D_4}
	\end{align}
	
	Next, we wish to move the start-points of $t_2'$ and $s_2'$ back to the original start-points so that we again consider ribbon operators on the ribbons $t_2$ and $s_2$. This will allow us to compare the labels of these sections of ribbon operators to the original labels. In order to move the start-point, we undo the conjugation by $g(s.p(t)-s.p(t_2'))$ or $g(s.p(s)-s.p(s_2'))$ that we introduced in Equations \ref{Equation_magnetic_ribbon_t2_change_sp} and \ref{Equation_magnetic_ribbon_s2_change_sp}. This gives us
	\begin{align}
		&C^k(t)C^h(s)\ket{GS}\notag \\
		&=C^h(s_1)C^{[g(s.p(s)-s.p(t_2'))xg(s.p(s)-s.p(t_2'))^{-1}hg(s.p(s)-s.p(t_2'))x^{-1}g(s.p(s)-s.p(t_2'))^{-1}]}(s_2) C^k(t_1) C^{h_{[t-s]}^{\phantom{-1}}k h_{[t-s]}^{-1}}(t_2)\ket{GS}. \label{Magnetic_magnetic_commute_2D_5}
	\end{align}
	
	Substituting $x=g(s.p(t)-s.p(t_2'))^{-1} [h_{[t-s]}^{\phantom{-1}}k h_{[t-s]}^{-1}] g(s.p(t)-s.p(t_2'))$, the label for the magnetic operator applied on $s_2$ is
	\begin{align*}
		g(s.p(s)-s.p(t_2'))\big[g(s.p(t)-s.p(t_2')&)^{-1} [h_{[t-s]}^{\phantom{-1}}k h_{[t-s]}^{-1}] g(s.p(t)-s.p(t_2'))\big]g(s.p(s)-s.p(t_2'))^{-1}hg(s.p(s)-s.p(t_2'))\\
		&\big[g(s.p(t)-s.p(t_2'))^{-1} [h_{[t-s]}^{\phantom{-1}}k h_{[t-s]}^{-1}] g(s.p(t)-s.p(t_2'))\big]^{-1}g(s.p(s)-s.p(t_2'))^{-1}.
	\end{align*}
	
	We then write $g(s.p(s)-s.p(t_2'))g(s.p(t)-s.p(t_2'))^{-1}=g(s.p(s)-s.p(t))$, in order to obtain
	\begin{align}
		&C^k(t)C^h(s)\ket{GS} \notag \\
		&=C^h(s_1) C^{[g(s.p(s)-s.p(t)) [h_{[t-s]}^{\phantom{-1}}k h_{[t-s]}^{-1}] g(s.p(s)-s.p(t))^{-1}hg(s.p(s)-s.p(t)) [h_{[t-s]}^{\phantom{-1}}k^{-1} h_{[t-s]}^{-1}] g(s.p(s)-s.p(t))^{-1}]}(s_2') \notag \\
		& \hspace{2cm} C^k(t_1) C^{h_{[t-s]}^{\phantom{-1}}k h_{[t-s]}^{-1}}(t_2')\ket{GS}\notag\\
		&=C^h(s_1) C^{ hg(s.p(s)-s.p(t))k g(s.p(s)-s.p(t))^{-1} h^{-1} h h g(s.p(s)-s.p(t))k^{-1}g(s.p(s)-s.p(t))^{-1} h^{-1}}(s_2) C^k(t_1) C^{h_{[t-s]}^{\phantom{-1}}k h_{[t-s]}^{-1}}(t_2)\ket{GS}\notag \\
		&= C^h(s_1) C^{ [hg(s.p(s)-s.p(t))k g(s.p(s)-s.p(t))^{-1} h^{-1}] h [h g(s.p(s)-s.p(t))kg(s.p(s)-s.p(t))^{-1} h^{-1}]^{-1}}(s_2) C^k(t_1) C^{h_{[t-s]}^{\phantom{-1}}k h_{[t-s]}^{-1}}(t_2)\ket{GS}. \label{Magnetic_magnetic_commute_2D_6}
	\end{align}

\begin{figure}[h]
		\begin{center}
			\begin{overpic}[width=0.6\linewidth]{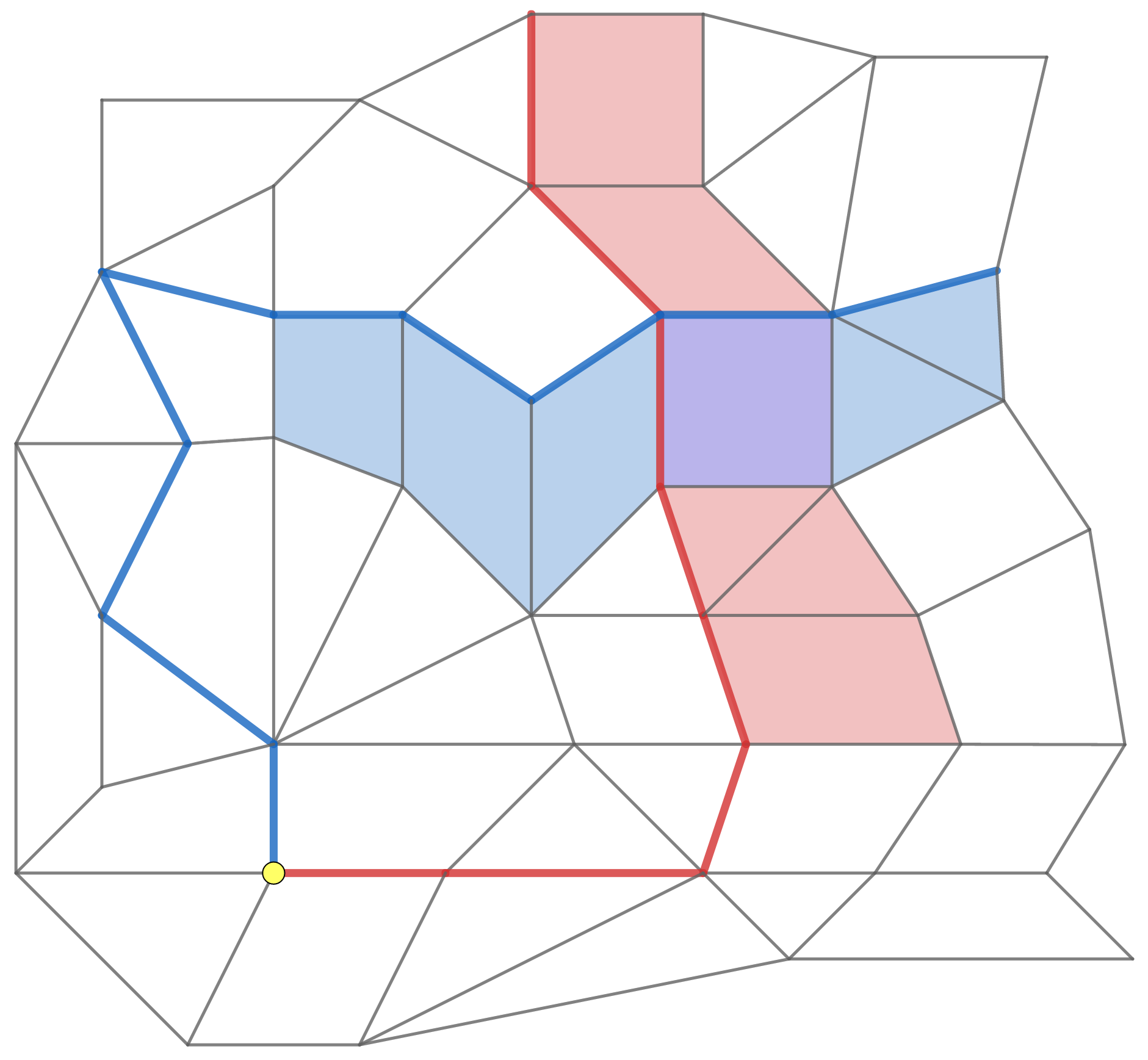}
				\put(12,13){$s.p(s)=s.p(t)$}
				\put(42,50){$C^h(s)$}
				\put(52,80){$C^k(t)$}
			\end{overpic}
			\caption{The braiding relation between the two magnetic excitations is simplified when their corresponding ribbon operators have the same start-point. In this figure, we see an example of such a situation, where the direct paths of the two ribbons meet at the same start-point, $s.p(s)$, even though the dual paths are far from each-other (this prevents one of the magnetic ribbons from acting on the direct path of the other before their intersection at the purple square). The path between the start-points that appears in our calculations of the braiding, $(s.p(s)-s.p(t))$, is then the closed path that passes from the mutual start-point along the direct path of $s$ (shown in blue) to the intersection of the direct paths of $s$ and $t$ at the top-left of the purple (darker gray in grayscale) square, before passing back to the start-point along the direct path in $t$ (shown in red). In the ground state, this closed path must have an element in $\partial(E)$ due to fake-flatness, though if we evaluate this path after the action of $C^h(s)$ then this is not so, because the path then encloses an excitation.}
			\label{Magnetic_ribbon_braid_same_sp}
		\end{center}
	\end{figure}

	We see that the labels of the ribbons $t_2$ and $s_2$, which correspond to the parts of the ribbon after braiding, are conjugated by the braiding. The label $k$ of $t_2$ is conjugated by $h_{[t-s]}=g(s.p(t)-s.p(s))hg(s.p(t)-s.p(s)^{-1})$, whereas the label $h$ of $s_2$ is conjugated by $[hg(s.p(s)-s.p(t))k g(s.p(s)-s.p(t))^{-1} h^{-1}]$. These relationships are simplified in the case where the two ribbons have the same start-point, such as in the case shown in Figure \ref{Magnetic_ribbon_braid_same_sp}. In this case $g(s.p(s)-s.p(t))$ is in the image of $\partial$ and so is in the centre of $G$. Note that $g(s.p(s)-s.p(t))$ is an operator and we must be sure that the path does not enclose any excitations when we evaluate it. This path could enclose the excitation at the start of $s$ that is produced by $C^h(s_1)$, but all evaluations of this path are to the right of $C^h(s_1)$, so this is not a concern. Therefore, $g(s.p(s)-s.p(t))$ is in the image of $\partial$ whenever we evaluate it and so has no effect when it conjugates another label. As a result, $h_{[t-s]}^{\phantom{-1}}=h$ and $g(s.p(s)-s.p(t))k g(s.p(s)-s.p(t))^{-1}=k$. This means that the labels of $s_2$ and $t_2$ simplify to $(hkh^{-1})h (hkh^{-1})^{-1} = hkh k^{-1}h^{-1} = (hk) h (hk)^{-1}$ and $hkh^{-1} = (hk)k(hk)^{-1}$ respectively. That is, the commutation relation Equation \ref{Magnetic_magnetic_commute_2D_6} becomes
	\begin{align}
		C^k(t)C^h(s)\ket{GS} &= C^h(s_1) C^{ [hkh^{-1}] h [h k h^{-1}]^{-1}}(s_2) C^k(t_1) C^{hk h^{-1}}(t_2)\ket{GS} \notag\\
		&=C^h(s_1) C^{ hkh k^{-1} h^{-1}}(s_2) C^k(t_1) C^{hk h^{-1}}(t_2)\ket{GS} \notag\\
		&= C^h(s_1) C^{ (hk)h (hk)^{-1}}(s_2) C^k(t_1) C^{(hk)k (hk)^{-1}}(t_2)\ket{GS}. \label{Magnetic_magnetic_commute_same_sp}
	\end{align}
	We therefore see that in this case, the labels of both excitations are conjugated by the product of the two labels. This is equivalent to the braiding of two fluxes in the Non-Abelian Superconductor model \cite{Preskill}, once we account for the orientation of the fluxes we have defined here. We can see that this braiding relation preserves the product $hk$ of the two fluxes, which is the total flux of the pair of excitations that we braided.

	Recall that we have been considering the braiding result for moving the excitation at the end of the ribbon labelled by $k$ clockwise around the end of the ribbon labelled by $h$. We worked out the clockwise braiding because the calculation is easier to present, but we can also deduce the anticlockwise braiding without redoing the calculation. The anticlockwise braiding of the flux labelled by $k$ around the one labelled by $h$ will give the inverse result. Keeping in mind that the product $hk$ is preserved by the braiding, we therefore deduce that the label $h$ of the flux associated with $s$ will become $(hk)^{-1} h (hk)=k^{-1}hk$ and the label $k$ will become $(hk)^{-1} k(hk)$ under the braiding.

	\subsection{Braiding an electric excitation around a non-trivial cycle in the fake-flat case}
	\label{Section_2D_braid_fake_flat_non_trivial_cycle}
	
	In Section \ref{Section_braiding_2D_fake_flat} of the main text, we discussed how moving an electric excitation around a non-contractible cycle could induce a non-trivial transformation, even in the ground state. In that section, we considered applying an electric ribbon operator $S^{R,a,b}(s \cdot t)$ on a path $s \cdot t$, where $t$ is a non-contractible cycle and $s$ is an arbitrary path. We found that, in terms of a ribbon operator only applied on the path $s$, the original ribbon operator could be written as
	\begin{align*}
		S^{R,a,b}(s \cdot t) &= \sum_{c=1}^{|R|} [D^R(\hat{g}(t))]_{cb} S^{R,a,c}(s),
	\end{align*}
	which indicates that there is a non-trivial transformation from moving the excitation around $t$ in addition to $s$, with this transformation described by the matrix $D^R(\hat{g}(t))$. However, $\hat{g}(t)$ is an operator and so this transformation is neither simple nor well-defined for generic ground-states (ground states are not usually eigenstates of $\hat{g}(t)$). However we can simplify this transformation in certain circumstances. Suppose that we apply the electric ribbon operator just on the closed path $t$, so that the ribbon operator is given by
	$$S^{R,a,b}(t)= \sum_{g \in G} [D^R(g)]_{ab} \delta( \hat{g}(t),g) = [D^R(\hat{g}(t))]_{ab}.$$
	
	Now consider a ground state where the path label $\hat{g}(t)$ is minimally mixed. That is, consider starting from a state which is an eigenstate of $\hat{g}(t)$, with some eigenvalue $h$, and generating a ground state from this by applying all of the energy terms to project to the ground state sector. Of the energy terms, the ones that affect the path label are the vertex term at the start (which is also the end) of path $t$ and the edge terms along the path. The vertex term mixes the eigenvalue $h$ with the other elements of its conjugacy class. This is because (just as for the case of ordinary lattice gauge theory), a vertex transform $A_v^x$ applied at the start of a closed loop pre-multiplies the path element by $x$ (because the vertex is as the start of the path) and also post-multiplies it by $x^{-1}$ (because the vertex is also at the end of the path), so that the vertex transform conjugates the path element by $x$. Then the vertex term is an average over the vertex transforms labelled by each element of $G$, so the vertex term results in an equal mix of path labels in the conjugacy class of $h$. On the other hand, as we showed in Section \ref{Section_Electric_Ribbon_Operator_Proof}, the edge terms mix the path label with other elements of the same coset of $\partial(E)$ in $G$. Putting these together, the energy terms mix the path label, which is initially $h$, with other elements of the form $\partial(e) xhx^{-1}$. Then we can write the path element $\hat{g}(t)$ acting on this state as $\hat{g}(t) \ket{GS} = \partial(\hat{e}) \hat{x}h\hat{x}^{-1} \ket{GS}$, where we have used $\hat{e}$ and $\hat{x}$ to remind the reader that these are generally operators, because the ground-state is still not an eigenstate of $\hat{g}(t)$. Now consider acting with the electric ribbon operator on a ground state where the path element belongs to this class of elements. We have
	$$S^{R,a,b}(t) \ket{GS} = [D^R(\partial(\hat{e}) \hat{x}h\hat{x}^{-1})]_{ab} \ket{GS}.$$
	
	If the electric ribbon operator is not confined then, as we discussed in Section \ref{Section_2D_electric} of the main text (where we described confinement when $\rhd$ is non-trivial at the end of the discussion of confinement in the $\rhd$ trivial case), the irrep $R$ is not sensitive to elements in $\partial(E)$. This means that the action of the electric ribbon operator simplifies to
	\begin{equation}
		S^{R,a,b}(t) \ket{GS} = [D^R(\hat{x}h\hat{x}^{-1})]_{ab} \ket{GS}. \label{Equation_electric_non_contractible_1}
	\end{equation}
	This still involves the operator $\hat{x}$, and so the result is still not well-defined in general (unless $h$ is in the centre of $G$). One case in which this is well-defined however is when the electric ribbon operator produces a pair of excitations and annihilates them after moving one around the cycle. That is, if the electric ribbon operator leaves no excitations afterwards. In this case, the ribbon operator must commute with the vertex transform at the start (which is also the end) of the path $t$. As we mentioned previously in this section, the action of a vertex transform on the path element starting and ending at that vertex is to conjugate it:
	$$\hat{g}(t)A_v^x = A_v^x x\hat{g}(t)x^{-1}.$$
	This means that the action of the vertex transform on an electric ribbon operator with general coefficients $\alpha_g$ is
	\begin{align*}
		\sum_{g \in G} \alpha_g \delta(g, \hat{g}(t))A_v^x &= A_v^x \sum_{g \in G} \alpha_g \delta(g, x\hat{g}(t)x^{-1})\\
		&=A_v^x \sum_{g \in G} \alpha_g \delta(x^{-1}gx, \hat{g}(t))\\
		&= A_v^x \sum_{g'= x^{-1}gx \in G} \alpha_{xg'x^{-1}} \delta(g', \hat{g}(t)).
	\end{align*}
	
	Therefore, the electric ribbon operator commutes with the vertex transforms at the start-point, and so does not produce an excitation, when the coefficient $\alpha$ is invariant under conjugation. For example, we can use the irrep basis to define an electric ribbon operator whose coefficients are given by the character of an irrep (rather than a general matrix element):
	\begin{align*}
		\sum_{c=1}^{|R|} S^{R,c,c}(t) & = \sum_{g \in G}\sum_{c=1}^{|R|} [D^R(g)]_{cc} \delta( \hat{g}(t),g)\\
		& = \sum_{g \in G} \chi_R(g) \delta( \hat{g}(t),g).
	\end{align*}
	Then when acting on our minimally-mixed ground state, this electric ribbon operator gives us (from Equation \ref{Equation_electric_non_contractible_1})
	\begin{align*}
		\sum_{c=1}^{|R|} S^{R,c,c}(t) \ket{GS} &= \chi_R(\hat{x}h\hat{x}^{-1})\ket{GS}\\
		&= \chi_R(h) \ket{GS},
	\end{align*}
	where in the last line we used the fact that the character of an irrep is a function of conjugacy class (i.e., is invariant under conjugation of the group element).
	
	\section{Topological sectors when $\rhd$ is trivial}
	\label{Section_topological_charge_supplemental}
	\subsection{The space of measurement operators}
	\label{Section_topological_sectors_2D_space_of_operators}
\begin{figure}[h]
		\begin{center}
			\begin{overpic}[width=0.75\linewidth]{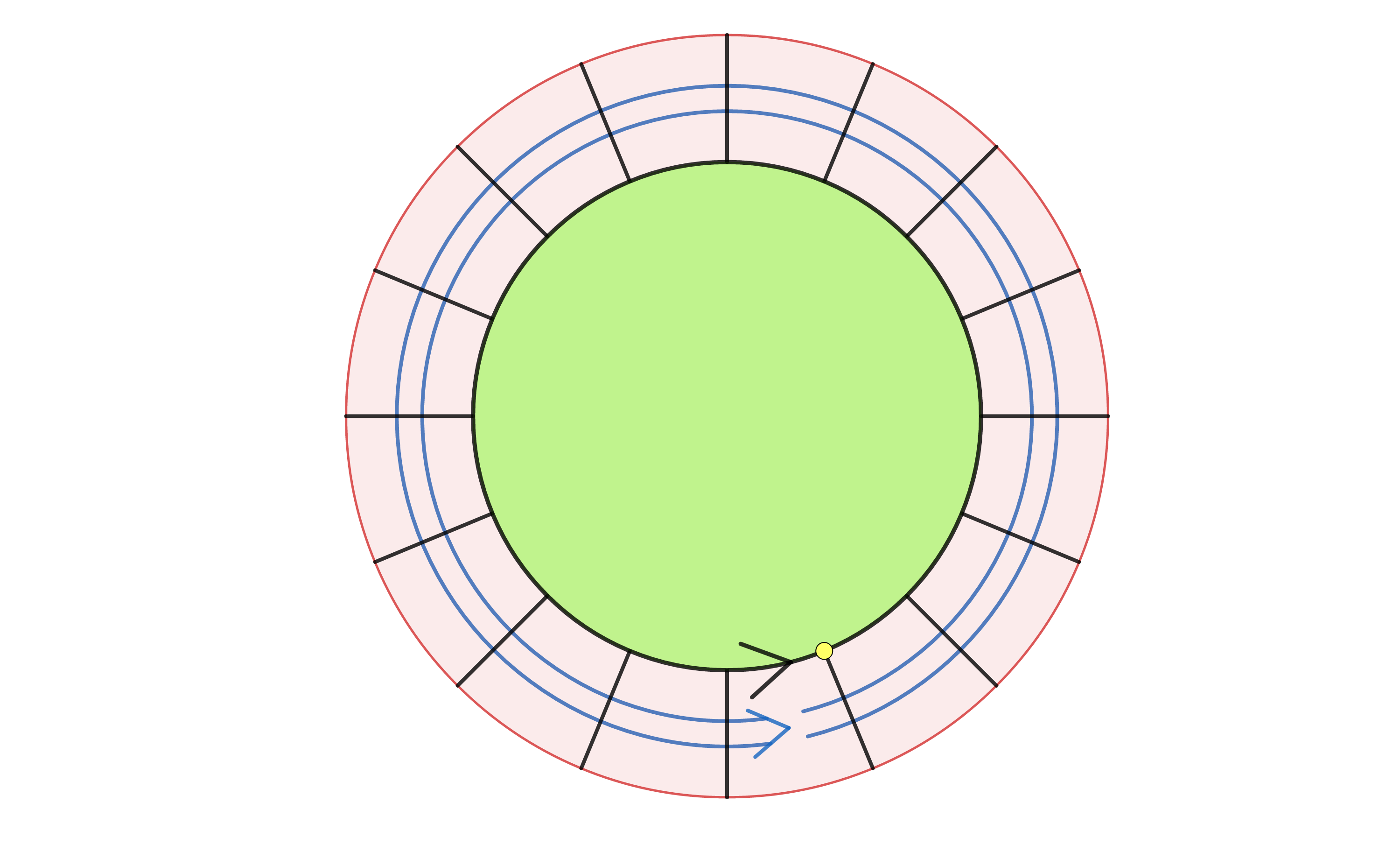}
				\put(47,32){Region $S$}
				\put(70,10){Ribbon operator $F^{h,g}(\sigma)$}
			\end{overpic}
			
			\caption{In order to measure the topological charge held within the green inner region $S$, we apply a ribbon operator enclosing that region. The ribbon operator includes both an electric and magnetic component. For example, here we apply the operator $F^{h,g}(\sigma)$, which could be part of a measurement operator for the topological charge. The black inner circle of the ribbon is the path of the electric ribbon operator and is also the direct path for the magnetic ribbon, while the arrow in the middle of the ribbon illustrates the dual path of the magnetic ribbon. The (yellow) dot is the start-point for both parts of the ribbon operator.}
			\label{2D_charge_measurement_ribbon}
			
		\end{center}
	\end{figure}

	In this section we will construct the measurement operators for topological charge in 2+1d in the $\rhd$ trivial case, which we presented in Section \ref{Section_2D_topological_Charge} of the main text. As described in Section \ref{Section_2D_topological_Charge}, the measurement operators can be constructed from closed ribbon operators and single plaquette multiplication operators (which change the surface label of a single plaquette). Therefore, to measure the topological charge within a region of our lattice, we consider applying a closed ribbon operator around the boundary of that region, as illustrated in Figure \ref{2D_charge_measurement_ribbon}. Setting aside the single plaquette multiplication operators for now, the space of ribbon operators that we can apply on a ribbon $\sigma$ is spanned by the set of operators $\set{F^{h,g}(\sigma)=C^h(\sigma)\delta(g(\sigma),g)|g,h \in G}$. We note that in this expression we have implicitly set the start-points of the electric and magnetic ribbon parts of the ribbon operator to be the same. We will explain why we can do this later and also consider what happens when we change this common start-point.

\begin{figure}[h]
		\begin{center}
			\begin{overpic}[width=0.75\linewidth]{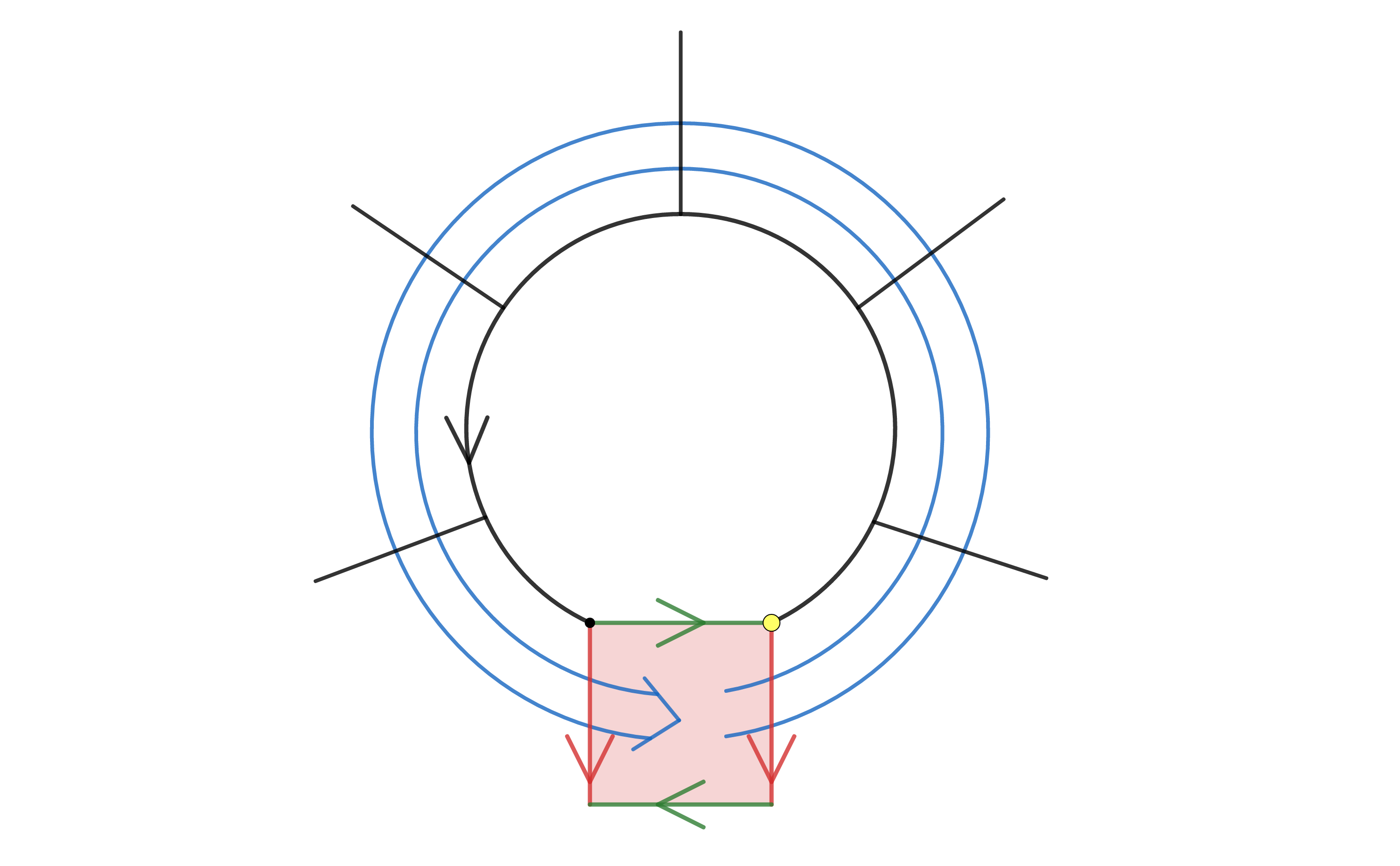}
				\put(51,18){$b$}
				\put(51,2){$u$}
				\put(57,5){$i_1$}
				\put(39,5){$i_2$}
			\end{overpic}
			
			\caption{We consider the plaquette (shaded red) at which the dual path of the closed ribbon operator starts and ends. If the labels of the magnetic and electric parts of the ribbon operator do not commute, then this plaquette will be excited by the action of the ribbon operator. Throughout this section, we will take this plaquette to be oriented clockwise, while the ribbon is oriented anticlockwise.}
			\label{2D_charge_measurement_first_plaquette}
			
		\end{center}
	\end{figure}

	We now wish to consider which operators in this space are valid measurement operators, i.e., which do not produce any excitations. We therefore consider the various energy terms which may be excited by a general ribbon operator $\sum_{g,h \in G} \alpha_{g,h} F^{h,g}(\sigma)$. The magnetic ribbon operator $C^h(\sigma)$ may excite the plaquette which is at the common start and end of the ribbon, and it may also excite the start-point vertex of the ribbon. The electric ribbon operator $\delta(g(\sigma),g)$ may also excite this start-point, and may excite the edge energy terms along the direct path of the ribbon. No other energy terms can be excited by the ribbon operator (see Section \ref{Section_2D_Ribbon_Operators_Appendix} for the properties of ribbon operators). We start by considering the plaquette at the start and end of the ribbon, as shown in Figure \ref{2D_charge_measurement_first_plaquette}. Just like the other plaquettes pierced by the dual path of the magnetic ribbon, two of the edges on the plaquette are affected by the dual path. However one of these edges is near the beginning of the ribbon, while the other is near the end. The magnetic ribbon operator $C^h(\sigma)$ acts on an edge $i$ cut by the dual path as
	\begin{equation*}
		C^h(\sigma):g_i = g(s.p-v_i)^{-1}hg(s.p-v_i)g_i
	\end{equation*}
	if the edge $i$ points away from the direct path, where $(s.p-v_i)$ is the path along the direct path from the start-point of the ribbon to the edge $i$. If we set up the branching structure of the lattice as indicated in Figure \ref{2D_charge_measurement_first_plaquette}, then the first edge, $i_1$, with initial label $g_1$, transforms as
	$$g_1 \rightarrow hg_1$$
	under the action of $F^{h,g}(\sigma)$, because $v_{i_1}$ is the start-point, and the dual ribbon pierces the edge immediately (rather than wrapping around the entire circle first). On the other hand the second edge, $i_2$, transforms as
	$$g_2 \rightarrow (g(\sigma)g(b)^{-1})^{-1}hg(\sigma)g(b)^{-1}g_2,$$
	because the direct path from the start-point of the ribbon to the edge is labelled by $g(\sigma)g(b)^{-1}$, where $g(\sigma)$ is the label of the entire path around the closed loop and $b$ is the base of the plaquette $p$. The electric part of $F^{h,g}(\sigma)$ ensures that $g(\sigma)=g$, so we can write the action of the magnetic ribbon operator on $g_2$ as
	$$g_2 \rightarrow (gg(b)^{-1})^{-1}hgg(b)^{-1} g_2,$$

	The plaquette holonomy is initially given by $H_1(p)=\partial(e_p)g_1g(u)g_2^{-1}g(b)$, which is the identity if the plaquette is originally unexcited. After the action of the ribbon operator on the two edges, this becomes
	\begin{align*}
		H_1(p) \rightarrow& \partial(e_p)[hg_1] g(u)[g_2^{-1} (gg(b)^{-1})^{-1}h^{-1}gg(b)^{-1}]g(b)\\
		=& \partial(e_p) hg_1g(u) g_2^{-1}g(b)g^{-1}h^{-1}g.
	\end{align*}
	
	When $\rhd$ is trivial, $\partial(E)$ is a subgroup in the centre of $G$ as discussed in Section \ref{Section_Recap_Paper_2} of the main text. Therefore, we can commute $\partial(e_p)$ past $h$ to obtain
	\begin{align*}
		H_1(p) \rightarrow& h [\partial(e_p)g_1g(u) g_2^{-1}g(b)]g^{-1}h^{-1}g.
	\end{align*}
	Then, using the fact that the original plaquette holonomy is $\partial(e_p)g_1g(u)g_2^{-1}g(b)=1_G$, this becomes
	\begin{align*}
		H_1(p) \rightarrow&hg^{-1}h^{-1}g.
	\end{align*}
	This indicates that if $g$ and $h$ do not commute then the plaquette is excited by the action of the ribbon operator (if it is not originally excited). If $hg^{-1}h^{-1}g$ is outside of $\partial(E)$ then there is nothing we can do about this and the ribbon operator $F^{h,g}(\sigma)$ does not contribute to a valid measurement operator. On the other hand, consider the case where $hg^{-1}h^{-1}g$ is equal to $\partial(f)$ for some $f \in E$. In this case we can apply a single plaquette multiplication operator $M^{f^{-1}}(p)$ which will take the plaquette label $e_p$ to $f^{-1}e_p$. Then the total action on the plaquette holonomy is
	\begin{align*}
		F^{h,g}(\sigma)M^{f^{-1}}(p):H_1(p) =&\partial(f^{-1}e_p)[hg_1] g(u)[g_2^{-1} (gg(b)^{-1})^{-1}h^{-1}gg(b)^{-1}]g(b)\\
		=& \partial(f^{-1}) \partial(e_p) hg_1g(u) g_2^{-1}g(b)g^{-1}h^{-1}g\\
		=& \partial(f^{-1})hg^{-1}h^{-1}g\\
		=&1_G.
	\end{align*}
	
	That is, by applying this single plaquette operator in addition to the ribbon operator, we can ensure that the plaquette remains unexcited, provided that the commutator of $g$ and $h$ lies within the image of $\partial$. Note that, because $\partial(f)=hg^{-1}h^{-1}g$ is in the centre of $G$, we have $\partial(f)=g\partial(f)g^{-1}$, which implies that $hg^{-1}h^{-1}g=ghg^{-1}h^{-1}=[g,h]$. This means that our measurement operator must actually be
	\begin{equation}
		\sum_{g\in G} \sum_{\substack{h \in G |\\ [g,h] \in \partial(E)}} \alpha_{g,h}F^{h,g}(\sigma) M^{f(g,h)^{-1}}(p), \label{Equation_2D_charge_introduce_M}
	\end{equation}
	where 
	\begin{equation}
		\partial(f(g,h))=[g,h]. \label{Equation_2D_charge_define_f}
	\end{equation}
	Note that we only sum over elements $g$ and $h$ which commute up to an element of $\partial(E)$, or alternatively the coefficient $\alpha_{g,h}$ must satisfy
	\begin{equation}
		\alpha_{g,h} \neq 0 \implies [g,h] \in \partial(E). \label{Equation_2D_charge_condition_plaquette}
	\end{equation}
	When $\rhd$ is trivial, the single plaquette multiplication operator commutes with all energy terms (except the plaquette term that we just considered), as discussed in Section \ref{Section_single_plaquette_multiplication_commutation_relations}, so introducing this single plaquette multiplication operator to the measurement operator will not affect the other commutation relations between the measurement operator and the energy terms.

	Considering Equation \ref{Equation_2D_charge_introduce_M}, we note that, depending on the properties of the map $\partial$, there may be many elements $f \in E$ which satisfy $\partial(f)=[g,h]$. The possible elements differ by elements of the kernel of $\partial$. However, as described in Section \ref{Section_2D_irrep_basis} of the main text, the single plaquette multiplication operators with label in the kernel of $\partial$ are related to projectors to the different symmetry-related ground-states. These projectors are given by
	$$P^{\alpha}=\frac{1}{|\ker(\partial)|} \alpha(e_k) \sum_{e_k \in \ker (\partial)}M^{e_k}(p),$$
	where $\alpha$ is an irrep of the kernel of $\partial$. This means that including all of the different elements $f$ satisfying $\partial(f)=[g,h]$ will introduce these projectors. For example, we can obtain one particular measurement operator by averaging over all of these elements by replacing $M^{f(g,h)^{-1}}(p)$ with $M^{f(g,h)^{-1}}(p) \frac{1}{|\ker(\partial)|} \sum_{e_k \in \ker (\partial)}M^{e_k}(p)$, where $f(g,h)$ is an arbitrarily chosen representative of the elements $f$ in $E$ satisfying $\partial(f)=[g,h]$. This will project to the state where the plaquette at the start of the measurement ribbon $\sigma$ is in the trivial symmetry state (labelled by a trivial irrep of the kernel). On the other hand, we could have chosen any other linear combination of the different $M^{e_k}(p)$, and any such linear combination can be decomposed into irreps of the kernel. In principle these give us different measurement operators. However because the single plaquette multiplication operators relate to the symmetry rather than the topological content of the theory, we will just consider using the trivial irrep for now and later obtain the operators corresponding to the other irreps of the kernel. We also note that this is the same reason why we do not apply single plaquette multiplication operators on other plaquettes, even though they do not excite any energy terms when they have a label in the kernel of $\partial$. That is, even though this would result in a valid measurement operator, the difference is related to the symmetry. For now we will use the averaging procedure and will explain later how we may adapt the measurement operator to the case where the plaquette corresponds to a different symmetry-related ground state.

	Having ensured that the plaquette condition is satisfied by the ribbon operator, we now consider the edge energy terms. The magnetic part of the electromagnetic ribbon operator automatically satisfies these energy terms (see Section \ref{Section_Magnetic_Ribbon_Proof}) and we know that the condition for the electric ribbon operator to satisfy the edge energy terms along its length (i.e., for the electric excitations to be unconfined) is given by Equation \ref{Equation_electric_unconfined} in Section \ref{Section_Electric_Ribbon_Operator_Proof}. Therefore, our measurement operator will satisfy the edge terms provided that the coefficients satisfy
	\begin{equation}
		\alpha_{g,h}=\alpha_{\partial(e)g,h} \label{Equation_2D_charge_condition_edge}
	\end{equation}
	for all $e \in E$ and $h \in G$.

	Finally we must consider the vertex term at the start-point of the ribbon. We first consider how the electric part of the electromagnetic ribbon is affected by a vertex transform at the start-point. From Section \ref{Section_Electric_Ribbon_Operator_Proof}, we know that a path element $g(t)$ transforms according to $g(t)A_v^x=A_v^x xg(t)$ under a vertex transform at the start of the path, and according to $g(t)A_v^x = A_v^x g(t)x^{-1}$ under a transform at the end of the path. When the same vertex is the start and end of the path, this means that the path element transforms as $g(t)A_v^x = A_v^x xg(t)x^{-1}$. Therefore, the path element $g(\sigma)$ appearing in our ribbon operator transforms as $g(\sigma)A_{s.p}^x= A_{s.p}^x x g(\sigma)x^{-1}$. Consequently, the commutation relation between $\delta(g(\sigma),g)$ and $A_{s.p}^x$ is
	\begin{align*}
		A_{s.p}^x\delta(g(\sigma),g) &= \delta(x^{-1}g(\sigma)x,g)A_{s.p}^x\\
		&= \delta(g(\sigma),xgx^{-1})A_{s.p}^x.
	\end{align*}
	
	Now consider the magnetic part of the ribbon operator. From Equation \ref{Equation_magnetic_ribbon_vertex_commutation_appendix} in Section \ref{Section_Magnetic_Ribbon_Proof}, we have
	\begin{equation}
		A_{s.p}^x C^h(\sigma)=C^{xhx^{-1}}(\sigma)A_{s.p}^x. \label{Equation_2D_charge_vertex_magnetic_commutation_1}
	\end{equation}
	
	We also note that the vertex transform commutes with the single plaquette multiplication operators when $\rhd$ is trivial, as described in Section \ref{Section_single_plaquette_multiplication_commutation_relations}. Putting this together, we have
	\begin{align*}
		A_{s.p}^x&\sum_{g\in G} \sum_{\substack{h \in G |\\ [g,h] \in \partial(E)}} \alpha_{g,h}F^{h,g}(\sigma) M^{f(g,h)^{-1}}(p) \frac{1}{|\ker(\partial)|} \sum_{e_k \in \ker (\partial)}M^{e_k}(p) \\
		&=\sum_{g\in G} \sum_{\substack{h \in G |\\ [g,h] \in \partial(E)}} \alpha_{g,h} A_{s.p}^x C^h(\sigma) \delta(g(\sigma),g) M^{f(g,h)^{-1}}(p) \frac{1}{|\ker(\partial)|} \sum_{e_k \in \ker (\partial)}M^{e_k}(p) \\
		&=\sum_{g\in G} \sum_{\substack{h \in G |\\ [g,h] \in \partial(E)}} \alpha_{g,h}C^{xhx^{-1}}(\sigma) A_{s.p}^x \delta(g(\sigma),g) M^{f(g,h)^{-1}}(p) \frac{1}{|\ker(\partial)|} \sum_{e_k \in \ker (\partial)}M^{e_k}(p)\\
		&= \sum_{g\in G} \sum_{\substack{h \in G |\\ [g,h] \in \partial(E)}} \alpha_{g,h}C^{xhx^{-1}}(\sigma) \delta(g(\sigma),xgx^{-1}) M^{f(g,h)^{-1}}(p) \frac{1}{|\ker(\partial)|} \sum_{e_k \in \ker (\partial)}M^{e_k}(p) A_{s.p}^x.
	\end{align*}
	
	We now introduce new dummy variables $h' = xhx^{-1}$ and $g'=xgx^{-1}$, to rewrite the relation above as 
	\begin{align}
		A_{s.p}^x&\sum_{g\in G} \sum_{\substack{h \in G |\\ [g,h] \in \partial(E)}} \alpha_{g,h}F^{h,g}(\sigma) M^{f(g,h)^{-1}}(p) \frac{1}{|\ker(\partial)|} \sum_{e_k \in \ker (\partial)}M^{e_k}(p) \notag \\
		&= \sum_{g' \in G} \sum_{\substack{h' \in G |\\ [x^{-1}g'x,x^{-1}h'x] \in \partial(E)}} \alpha_{x^{-1}g'x,x^{-1}h'x}C^{h'}(\sigma)\delta(g(\sigma),g') M^{f(g,h)^{-1}}(p) \frac{1}{|\ker(\partial)|} \sum_{e_k \in \ker (\partial)}M^{e_k}(p)A_{s.p}^x. \label{Equation_2D_charge_vertex_condition_1}
	\end{align}
	
	Consider the condition 
	$$[x^{-1}g'x,x^{-1}h'x] \in \partial(E),$$
	from the sum over $h'$. If we write the commutator out explicitly, we have
	\begin{align*}
		[x^{-1}g'x,x^{-1}h'x]&= (x^{-1}g'x) (x^{-1}h'x) (x^{-1}g'x)^{-1} (x^{-1}h'x)^{-1} \\
		&= x^{-1} g'h'{g'}^{-1} {h'}^{-1} x.
	\end{align*}
	
	Therefore, requiring $[x^{-1}g'x,x^{-1}h'x]$ to be in $\partial(E)$ is equivalent to requiring $x^{-1} g'h'{g'}^{-1} {h'}^{-1} x$ to be in $\partial(E)$. This is in turn equivalent to requiring $g'h'{g'}^{-1} {h'}^{-1}$ to be in $\partial(E)$ (because $\partial(E)$ is a normal subgroup of $G$). Furthermore, because $\partial(E)$ is in the centre of $G$ when $\rhd$ is trivial, 
	$$[g,h]=x^{-1} g'h'{g'}^{-1} {h'}^{-1} x =g'h'{g'}^{-1} {h'}^{-1} =[g',h'].$$ 
	Therefore, we can rewrite Equation \ref{Equation_2D_charge_vertex_condition_1} as
	\begin{align*}
		A_{s.p}^x&\sum_{g\in G} \sum_{h \in G | [g,h] \in \partial(E)} \alpha_{g,h}F^{h,g}(\sigma) M^{f(g,h)^{-1}}(p) \frac{1}{|\ker(\partial)|} \sum_{e_k \in \ker (\partial)}M^{e_k}(p)\\
		&= \sum_{g' \in G} \sum_{h' \in G | [g',h'] \in \partial(E)} \alpha_{x^{-1}g'x,x^{-1}h'x}C^{h'}(\sigma)\delta(g(\sigma),g') M^{f(g,h)^{-1}}(p) \frac{1}{|\ker(\partial)|} \sum_{e_k \in \ker (\partial)}M^{e_k}(p) A_{s.p}^x.
	\end{align*}
	
	The only place that the original $g$ and $h$ are present in this expression is in $f(g,h)$. Recall that $f(g,h)$ is an element of $E$ satisfying $\partial(f(g,h))=[g,h]$. However we have established that $[g,h] = [g',h']$ for $g$, $h$ satisfying the conditions in the sum. We can therefore choose $f(g',h')=f(g,h)$ (even if we chose a different $f(g',h')$, it must differ at most by an element of the kernel and this difference can be absorbed into the sum $\sum_{e_k \in \ker (\partial)}M^{e_k}(p)$). We can therefore write
	\begin{align*}
		A_{s.p}^x&\sum_{g\in G} \sum_{h \in G | [g,h] \in \partial(E)} \alpha_{g,h}F^{h,g}(\sigma) M^{f(g,h)^{-1}}(p) \frac{1}{|\ker(\partial)|} \sum_{e_k \in \ker (\partial)}M^{e_k}(p)\\
		&= \sum_{g' \in G} \sum_{h' \in G | [g',h'] \in \partial(E)} \alpha_{x^{-1}g'x,x^{-1}h'x}C^{h'}(\sigma)\delta(g(\sigma),g') M^{f(g',h')^{-1}}(p) \frac{1}{|\ker(\partial)|} \sum_{e_k \in \ker (\partial)}M^{e_k}(p) A_{s.p}^x.
	\end{align*}
	For simplicity, we then relabel the dummy indices $g'$ and $h'$ to $g$ and $h$ respectively, to obtain
	\begin{align*}
		A_{s.p}^x&\sum_{g\in G} \sum_{h \in G | [g,h] \in \partial(E)} \alpha_{g,h}F^{h,g}(\sigma) M^{f(g,h)^{-1}}(p) \frac{1}{|\ker(\partial)|} \sum_{e_k \in \ker (\partial)}M^{e_k}(p) \\
		&= \sum_{g \in G} \sum_{h \in G | [g,h] \in \partial(E)} \alpha_{x^{-1}gx,x^{-1}hx}C^{h}(\sigma)\delta(g(\sigma),g') M^{f(g,h)^{-1}}(p) \frac{1}{|\ker(\partial)|} \sum_{e_k \in \ker (\partial)}M^{e_k}(p)A_{s.p}^x.
	\end{align*}

	If we act with the measurement operator on a state where the start-point vertex is initially unexcited, then the vertex transform on the right can be absorbed into the state (see Section \ref{Section_Recap_Paper_2} of the main text), giving us
	\begin{align*}
		A_{s.p}^x&\sum_{g\in G} \sum_{h \in G | [g,h] \in \partial(E)} \alpha_{g,h}F^{h,g}(\sigma) M^{f(g,h)^{-1}}(p) \frac{1}{|\ker(\partial)|} \sum_{e_k \in \ker (\partial)}M^{e_k}(p) \ket{\psi} \\
		&= \frac{1}{|G|} \sum_{g \in G} \sum_{h \in G | [g,h] \in \partial(E)} \alpha_{x^{-1}gx,x^{-1}hx}C^{h}(\sigma)\delta(g(\sigma),g') M^{f(g,h)^{-1}}(p) \frac{1}{|\ker(\partial)|} \sum_{e_k \in \ker (\partial)}M^{e_k}(p)\ket{\psi} .
	\end{align*}
	
	Requiring the vertex to remain unexcited after the action of the measurement operator means that the vertex transform must also act trivially after the measurement operator is applied:
	\begin{align*}
		A_{s.p}^x&\sum_{g\in G} \sum_{h \in G | [g,h] \in \partial(E)} \alpha_{g,h}F^{h,g}(\sigma) M^{f(g,h)^{-1}}(p) \frac{1}{|\ker(\partial)|} \sum_{e_k \in \ker (\partial)}M^{e_k}(p) \ket{\psi} \\
		&=\sum_{g\in G} \sum_{h \in G | [g,h] \in \partial(E)} \alpha_{g,h}F^{h,g}(\sigma) M^{f(g,h)^{-1}}(p) \frac{1}{|\ker(\partial)|} \sum_{e_k \in \ker (\partial)}M^{e_k}(p) \ket{\psi} .
	\end{align*}
	This gives us the requirement
	\begin{align*}
		\sum_{g\in G}& \sum_{h \in G | [g,h] \in \partial(E)} \alpha_{g,h}F^{h,g}(\sigma) M^{f(g,h)^{-1}}(p) \frac{1}{|\ker(\partial)|} \sum_{e_k \in \ker (\partial)}M^{e_k}(p) \ket{\psi} \\
		&= \frac{1}{|G|} \sum_{g \in G} \sum_{h \in G | [g,h] \in \partial(E)} \alpha_{x^{-1}gx,x^{-1}hx}C^{h}(\sigma)\delta(g(\sigma),g') M^{f(g,h)^{-1}}(p) \frac{1}{|\ker(\partial)|} \sum_{e_k \in \ker (\partial)}M^{e_k}(p)\ket{\psi} .
	\end{align*}
	We therefore see that the coefficients for the measurement operator must satisfy
	\begin{equation}
		\alpha_{x^{-1}gx,x^{-1}hx}=\alpha_{g,h}. \label{Equation_2D_charge_condition_vertex}
	\end{equation}
	This must hold for each $x \in G$ (from the different vertex transforms), so this gives us $|G|$ restrictions on our coefficients $\alpha_{g,h}$.

	At this point, we can come back to a point we mentioned at the beginning of Section \ref{Section_topological_sectors_2D_space_of_operators}. We have assumed that we can take the start-points of the magnetic and electric ribbon operators to be the same without loss of generality. We will now prove that this is indeed the case. Suppose that the start-points of the magnetic and electric ribbon operators were different. Then each operator would individually have to commute with the vertex transforms at its start-point. Let the start-point of the magnetic ribbon operator be $s.p(1)$. Then, because this is not the start-point of the electric ribbon operator, vertex transforms at $s.p(1)$ commute with the electric part of the ribbon operator $F^{h,g}(\sigma)$. On the other hand, the magnetic ribbon operator again satisfies the commutation relation $$A_{s.p(1)}^x C^h(\sigma)=C^{xhx^{-1}}(\sigma)A_{s.p(1)}^x.$$ Therefore, we have
	\begin{align*}
		A_{s.p(1)}^x \sum_{g,h} \alpha_{g,h} F^{h,g}(\sigma) &= \sum_{g,h} \alpha_{g,h} F^{xhx^{-1},g}(\sigma) A_{s.p(1)}^x\\
		&=\sum_{g,h' =xhx^{-1}} \alpha_{g,x^{-1}hx} F^{h',g}(\sigma) A_{s.p(1)}^x.
	\end{align*}
	
	Then to commute with the vertex term $\frac{1}{|G|} \sum_{x \in G}A_{s.p(1)}^x$, the magnetic part of the ribbon operator must include an equal sum of each magnetic ribbon operator with a label of the form $x^{-1}hx$. That is, the magnetic part of the ribbon operator must be made of terms of the form $\sum_{x \in G} C^{xhx^{-1}}(\sigma)$. However, if we change the start-point of the magnetic ribbon operator then this is equivalent to conjugating the label of the ribbon operator by the path element along which we would move the start-point. That is, if we consider changing the start-point from $s.p(1)$ to a new start-point, $s.p(2)$, then this is equivalent to changing the label to $g(s.p(1)-s.p(2))hg(s.p(1)-s.p(2))^{-1}$. To see this, consider the action of the magnetic ribbon operator on an edge $g_i$ cut by the dual ribbon, when the start-point is $s.p(2)$. If the edge points away from the direct path, this action is given by 
	\begin{align*}
		g_i \rightarrow& g(s.p(2)-v_i)^{-1}hg(s.p(2)-v_i)g_i\\
		=& (g(s.p(2)-s.p(1))g(s.p(1)-v_i))^{-1}h(g(s.p(2)-s.p(1))g(s.p(1)-v_i))g_i\\
		=& g(s.p(1)-v_i)^{-1}(g(s.p(2)-s.p(1))^{-1}hg(s.p(2)-s.p(1)))g(s.p(1)-v_i)g_i,
	\end{align*}
	which is the same as the action of a ribbon operator $C^{g(s.p(2)-s.p(1))^{-1}hg(s.p(2)-s.p(1))}(\sigma)$ with the original start-point, $s.p(1)$, on the edge. Comparing this transformation of the label of the magnetic ribbon operator under moving the start-point with Equation \ref{Equation_2D_charge_vertex_magnetic_commutation_1}, we note that the effect of moving the start-point on the magnetic ribbon operator is equivalent to the effect of commuting the ribbon operator with a vertex transform of label $g(s.p(1)-s.p(2))$ at the start-point.

	We expect this equivalence of changing the label and moving the start-point from the idea that a vertex transform acts like parallel transport of that vertex. In the case of this ribbon operator, in order to satisfy the vertex transform, the magnetic ribbon operator is an equal sum over all labels in a given conjugacy class. This means that the conjugation of the label from moving the start-point has no effect on the magnetic ribbon operator. That is, moving the start-point does not induce any transformation on the magnetic ribbon operator. We can therefore freely move the start-point of the magnetic ribbon operator and in particular can move it to the same location as the start-point of the electric ribbon operator. Therefore, the cases where the two start-points of the magnetic and electric ribbon operators are distinct is equivalent to (a subset of) the cases where the start-points are the same. In a similar way, if we start with the start-points of the ribbon operators being the same, then we can consider moving the common start-point of the electric and magnetic ribbon operator. This will induce a transformation equivalent to applying a vertex transform on the start-point. However, because the ribbon operator is required to commute with the vertex transform at the start-point, this has no effect and so the choice of start-point is truly arbitrary.

	\subsection{Constructing projectors to definite topological charge}
	\label{Section_2D_Topological_Charge_Projectors}
	
	So far, we have discussed the conditions that our ribbon operator must satisfy to be a valid measurement operator. Combining Equations \ref{Equation_2D_charge_condition_plaquette}, \ref{Equation_2D_charge_condition_edge} and \ref{Equation_2D_charge_condition_vertex}, we see that the coefficients $\alpha_{g,h}$ for a valid measurement operator must satisfy the following conditions:
	\begin{align*}
		\alpha_{g,h} \neq 0 &\implies [g,h] \in \partial(E)\\
		\alpha_{\partial(e)g,h}&=\alpha_{g,h} \: \forall e \in E\\
		\alpha_{x^{-1}gx,x^{-1}hx}&=\alpha_{g,h} \: \forall x \in G
	\end{align*}
	for all $g,h \in G$. These conditions describe the space of measurement operators. We now wish to construct basis vectors for this space and then construct projectors to definite topological charge, following a similar approach to that used by Ref. \cite{Bombin2008} for a family of models related to Kitaev's Quantum Double model \cite{Kitaev2003}. The simplest way to construct a set of operators that span our space of measurement operators is to start with the set of operators $F^{h,g}(\sigma)M^{f(g,h)^{-1}}(p) \frac{1}{|\ker(\partial)|} \sum_{e_k \in \ker (\partial)}M^{e_k}(p)$ and then apply our constraints. First, we require $g$ and $h$ to commute up to an element in $\partial(E)$. This just restricts which pairs $(g,h)$ are allowed. We denote the set of elements in $G$ that commute with $g$ up to an element of $\partial(E)$ by $N'_g$. Then our operators are labelled by a pair $(g,h)$, where $g \in G$ and $h \in N'_g$.

	Next, consider the other two conditions. These mean that, if $F^{h,g}(\sigma)$ appears in our measurement operator, then so must $F^{h,\partial(e)g}(\sigma)$ and $F^{x^{-1}hx, x^{-1}gx}(\sigma)$, with equal coefficient. We denote this with an equivalence relation between the labels of these $F$ operators:
	\begin{align*}
		(g,h) &\sim (\partial(e)g,h)\\
		&\sim(xgx^{-1},xhx^{-1}),
	\end{align*}
	where in the latter relation we swapped $x$ for $x^{-1}$ for simplicity (the relation holds for all $x \in G$, so this is just a relabelling). Then we can make our operator, 
	$$F^{h,g}(\sigma)M^{f(g,h)^{-1}}(p) \frac{1}{|\ker(\partial)|} \sum_{e_k \in \ker (\partial)}M^{e_k}(p),$$
	satisfy this condition by summing over all labels related to $(g,h)$ by these equivalence relations. This gives us the new operator
	$$K^{h,g}_{\sigma}= \sum_{x \in G} \sum_{e \in E} F^{xhx^{-1}, x\partial(e)gx^{-1}}(\sigma)M^{f(x\partial(e)gx^{-1},xhx^{-1})^{-1}}(p) \frac{1}{|\ker(\partial)|} \sum_{e_k \in \ker (\partial)}M^{e_k}(p).$$
	
	Now $f(x\partial(e)gx^{-1},xhx^{-1})$ is an element of $E$ satisfying (see Equation \ref{Equation_2D_charge_define_f})
	\begin{align*}
		\partial(f(x\partial(e)gx^{-1},xhx^{-1})) &= [x\partial(e)gx^{-1},xhx^{-1}]\\
		&= x\partial(e)gx^{-1} xhx^{-1} (x\partial(e)gx^{-1})^{-1} (xhx^{-1})^{-1}\\
		&= x\partial(e)gh g^{-1} \partial(e)^{-1} h^{-1}x^{-1}.
	\end{align*}
	Because $\partial(e)$ is in the centre of $G$, we can move the elements $\partial(e)$ and $\partial(e)^{-1}$ next to each-other and cancel them. This gives us
	\begin{align*}
		\partial(f(x\partial(e)gx^{-1},xhx^{-1})) &= xghg^{-1}h^{-1}x^{-1}\\
		&= x[g,h]x^{-1}.
	\end{align*}
	However because $\partial(f)$ is also in the centre of $G$, this means that
	\begin{align*}
		\partial(f(x\partial(e)gx^{-1},xhx^{-1})) &= [g,h].
	\end{align*}
	We can therefore choose $f(x\partial(e)gx^{-1},xhx^{-1})=f(g,h)$. This means that we can write our $K^{h,g}_{\sigma}$ operators more simply as
	$$K^{h,g}_{\sigma}= \sum_{x \in G} \sum_{e \in E} F^{xhx^{-1}, x\partial(e)gx^{-1}}(\sigma)M^{f(g,h)^{-1}}(p) \frac{1}{|\ker(\partial)|} \sum_{e_k \in \ker (\partial)}M^{e_k}(p).$$
	
	At the moment, we have one such operator for each pair $(g,h)$, where $g \in G$ and $g=h \in N'_h=g$. However this includes many redundant operators. For example, $K^{h,g}_{\sigma}=K^{h,\partial(r)g}_{\sigma}$, because
	\begin{align*}
		K^{h,\partial(r)g}_{\sigma}&= \sum_{x \in G} \sum_{e \in E} F^{xhx^{-1}, x\partial(e)\partial(r)gx^{-1}}(\sigma)M^{f(\partial(r)g,h)^{-1}}(p) \frac{1}{|\ker(\partial)|} \sum_{e_k \in \ker (\partial)}M^{e_k}(p).
	\end{align*}
	Then, noting that $f(\partial(r)g,h)=f(g,h)$, as we showed previously, we have
	\begin{align}
		K^{h,\partial(r)g}_{\sigma}&=\sum_{x \in G} \sum_{e \in E} F^{xhx^{-1}, x\partial(er)gx^{-1}}(\sigma)M^{f(g,h)^{-1}}(p) \frac{1}{|\ker(\partial)|} \sum_{e_k \in \ker (\partial)}M^{e_k}(p) \notag \\
		&=\sum_{x \in G} \sum_{e' = er \in E} F^{xhx^{-1}, x\partial(e')gx^{-1}}(\sigma)M^{f(g,h)^{-1}}(p) \frac{1}{|\ker(\partial)|} \sum_{e_k \in \ker (\partial)}M^{e_k}(p) \notag \\
		&=K^{h,g}_{\sigma}. \label{Equation_2d_redundant_measurement_1}
	\end{align}
	
	Similarly we have
	\begin{align}
		K^{yhy^{-1},ygy^{-1}}_{\sigma}&= \sum_{x \in G} \sum_{e \in E} F^{xyhy^{-1}x^{-1}, x\partial(e)ygy^{-1}x^{-1}}(\sigma)M^{f(ygy^{-1},yhy^{-1})^{-1}}(p) \frac{1}{|\ker(\partial)|} \sum_{e_k \in \ker (\partial)}M^{e_k}(p)\notag \\
		&= \sum_{x \in G} \sum_{e \in E} F^{xyh(xy)^{-1}, xy\partial(e)g(xy)^{-1}}(\sigma)M^{f(g,h)^{-1}}(p) \frac{1}{|\ker(\partial)|} \sum_{e_k \in \ker (\partial)}M^{e_k}(p) \notag\\
		&= \sum_{z=xy \in G} \sum_{e \in E} F^{zhz^{-1}, z\partial(e)gz^{-1}}(\sigma)M^{f(g,h)^{-1}}(p) \frac{1}{|\ker(\partial)|} \sum_{e_k \in \ker (\partial)}M^{e_k}(p) \notag \\
		&= K^{h,g}_{\sigma} \label{Equation_2d_redundant_measurement_2}.
	\end{align}

	We need to remove such redundant operators. We start by defining a set of equivalence classes with the equivalence relation
	\begin{equation}
		g \overset{G}{\sim} g' \text{ if and only if there exists a $x\in G$ and $e \in E$ such that }g'= x\partial(e)gx^{-1}. \label{Equation_union_coset_class_definition}
	\end{equation}
	
	Then for each such class $C$, we choose a representative $r_C$. For each such class, we also define $N'_C=N'_{r_C}$, where $N'_{r_C}$ is the set of all elements of $G$ that commute with $r_C$ up to an element in $\partial(E)$. We then define a new set of operators by $K^{h,C}_{\sigma}=K^{h,r_C}_{\sigma}$, where we have one such operator for each equivalence class $C$ defined by Equation \ref{Equation_union_coset_class_definition} and $h$ in $N'_C$. We wish to show that this set of operators includes all of the previously defined operators $K^{h,g}_{\sigma}$. To show this, consider an arbitrary $K^{h,g}_{\sigma}$. The element $g$ belongs to some class $C$ and therefore can be written as $g=y\partial(r)r_Cy^{-1}$ for some $r \in E$ and $y \in G$. Then
	\begin{align*}
		K^{h,g}_{\sigma}&=K^{h, y\partial(r)r_Cy^{-1}}\\
		&= K^{y^{-1}hy, \partial(r)r_C},
	\end{align*}
	as we showed in Equation \ref{Equation_2d_redundant_measurement_2}. Using Equation \ref{Equation_2d_redundant_measurement_1}, we then have
	\begin{align*}
		K^{h,g}_{\sigma}&= K^{y^{-1}hy, \partial(r)r_C}\\
		&=K^{y^{-1}hy, r_C}\\
		&= K^{y^{-1}hy,C},
	\end{align*}
	which belongs to our new set of operators. Therefore, our new set of operators includes all of our old ones, but removes some of the redundant operators. We are still not done excluding repeated operators yet however. Consider the operator $K^{yhy^{-1},C}_{\sigma}$, where $y$ commutes with $r_C$ up to an element $\partial(s^{-1})$ in $\partial(E)$, so that $y\partial(s)r_Cy^{-1}=r_C$ (i.e., $y$ is in $N'_C$). We can write this operator as
	\begin{align*}
		K^{yhy^{-1},C}_{\sigma}&= \sum_{x \in G} \sum_{e \in E} F^{xyhy^{-1}x^{-1}, x\partial(e)r_Cx^{-1}}(\sigma)M^{f(r_C,yhy^{-1})^{-1}}(p) \frac{1}{|\ker(\partial)|} \sum_{e_k \in \ker (\partial)}M^{e_k}(p)\\
		&= \sum_{x \in G} \sum_{e \in E} F^{xyhy^{-1}x^{-1}, x\partial(e)[y\partial(s)r_Cy^{-1}]x^{-1}}(\sigma)M^{f(r_C,h)^{-1}}(p) \frac{1}{|\ker(\partial)|} \sum_{e_k \in \ker (\partial)}M^{e_k}(p)\\
		&= \sum_{x \in G} \sum_{e \in E} F^{xyhy^{-1}x^{-1}, xy\partial(es)r_Cy^{-1}x^{-1}}(\sigma)M^{f(r_C,h)^{-1}}(p) \frac{1}{|\ker(\partial)|} \sum_{e_k \in \ker (\partial)}M^{e_k}(p),
	\end{align*}
	where we used the fact that $\partial(E)$ is in the centre of $G$ to collect $\partial(e)$ and $\partial(s)$. Then switching the dummy variables $x$ and $e$ for $z=xy$ and $e'=es$ respectively, we obtain
	\begin{align*}
		K^{yhy^{-1},C}_{\sigma}&=\sum_{z=xy \in G} \sum_{e'=es \in E} F^{zhz^{-1},z\partial(e')r_Cz^{-1}}(\sigma)M^{f(r_C,h)^{-1}}(p) \frac{1}{|\ker(\partial)|} \sum_{e_k \in \ker (\partial)}M^{e_k}(p)\\
		&=K^{h,C}_{\sigma},
	\end{align*}
	which is the same as another operator in our set. To remove this redundancy, we split the sum over all elements $x\in G$ into a sum over elements in $N'_C$ and a sum over elements representing the quotient group $Q_C=G/N'_C$, so that each $q \in Q_C$ is an element of a distinct coset $qN'_C$ and we have one such element for each unique coset. Then we can write the $K^{h,C}$ operator as 
	\begin{align*}
		K^{h,C}_{\sigma}&= \sum_{x \in G} \sum_{e \in E} F^{xhx^{-1}, x\partial(e)r_Cx^{-1}}(\sigma)M^{f(r_C,h)^{-1}}(p) \frac{1}{|\ker(\partial)|} \sum_{e_k \in \ker (\partial)}M^{e_k}(p)\\
		&= \sum_{q \in Q_C} \sum_{a \in N'_C} \sum_{e \in E} F^{qaha^{-1}q^{-1}, qa\partial(e)r_Ca^{-1}q^{-1}}(\sigma)M^{f(r_C,h)^{-1}}(p) \frac{1}{|\ker(\partial)|} \sum_{e_k \in \ker (\partial)}M^{e_k}(p).
	\end{align*}
	
	As established previously, for $a \in N'_C$ we have $a\partial(e)r_Ca^{-1} = \partial(es)r_C$ for some $s$ in $E$, and the factor of $s$ can be absorbed into the sum over $e \in E$. Therefore
	\begin{align*}
		K^{h,C}_{\sigma}&=\sum_{q \in Q_C} \sum_{a \in N'_C} \sum_{e \in E} F^{qaha^{-1}q^{-1}, q\partial(e)r_Cq^{-1}}(\sigma)M^{f(r_C,h)^{-1}}(p) \frac{1}{|\ker(\partial)|} \sum_{e_k \in \ker (\partial)}M^{e_k}(p).
	\end{align*}
	
	This means that the label $a$ only appears in the expression $qaha^{-1}q^{-1}$. Because $a$ and $h$ are both in $N'_C$, this sum over $a\in N'_C$ is equivalent (up to a normalisation factor) to a sum over the elements of the conjugacy class $D$ in $N'_C$ which contains $h$. That is
	\begin{align*}
		K^{h,C}_{\sigma}&= N \sum_{q \in Q_C} \sum_{d \in D} \sum_{e \in E} F^{qdq^{-1}, q\partial(e)r_Cq^{-1}}(\sigma)M^{f(r_C,d)^{-1}}(p) \frac{1}{|\ker(\partial)|} \sum_{e_k \in \ker (\partial)}M^{e_k}(p),
	\end{align*}
	where $D$ contains $h$ and $N$ is a constant, which we do not care about and will drop (we will worry about normalization later when we construct projectors from these operators). Because this operator then depends only on the conjugacy class $D$ containing $h$, rather than on $h$ itself, it is sensible to label this operator with the conjugacy class $D$ instead of $h$. In addition, the label $e$ only appears in the form $\partial(e)$, so we replace the sum over $E$ with a sum over $\partial(E)$ (and again drop the resulting multiplicative factor that arises from the sum over $E$ covering each element of $\partial(E)$ multiple times). Then we define the set of operators
	\begin{align*}
		K^{D,C}_{\sigma}=\sum_{q \in Q_C} \sum_{d \in D} \sum_{k \in \partial(E)} F^{qdq^{-1}, qkr_Cq^{-1}}(\sigma)M^{f(r_C,d)^{-1}}(p) \frac{1}{|\ker(\partial)|} \sum_{e_k \in \ker (\partial)}M^{e_k}(p).
	\end{align*}

	These operators form a basis for our space of measurement operators, but they are not projectors. We next construct a set of operators using the natural map from conjugacy classes to irreps of the group $N'_C$:
	\begin{align*}
		K^{R,C}_{\sigma}=\frac{|R|}{|N'_C|} \sum_{D \in (N'_C)_{cj}} \overline{\chi}_R(D) K^{DC}_{\sigma},
	\end{align*}
	where $R$ is an irrep of dimension $|R|$, $\chi_R$ is the character for that irrep, and $(N'_C)_{cj}$ is the set of conjugacy classes of $N'_C$, so that the index $D$ runs over conjugacy classes of $N'_C$. Then inserting our expression for $K^{DC}$, we have
	\begin{align*}
		K^{R,C}_{\sigma}&=\frac{|R|}{|N'_C|} \sum_{D \in (N'_C)_{cj}} \overline{\chi}_R(D) \sum_{q \in Q_C} \sum_{d \in D} \sum_{k \in \partial(E)} F^{qdq^{-1}, qkr_Cq^{-1}}(\sigma)M^{f(r_C,d)^{-1}}(p) \frac{1}{|\ker(\partial)|} \sum_{e_k \in \ker (\partial)}M^{e_k}(p)\\
		&=\frac{|R|}{|N'_C|} \sum_{d \in N'_C} \sum_{q \in Q_C} \sum_{k \in \partial(E)} \overline{\chi}_R(d) F^{qdq^{-1}, qkr_Cq^{-1}}(\sigma)M^{f(r_C,d)^{-1}}(p) \frac{1}{|\ker(\partial)|} \sum_{e_k \in \ker (\partial)}M^{e_k}(p).
	\end{align*}
	
	We then wish to check that these are in fact projectors. We therefore consider multiplying two such operators, $K^{R,C}_{\sigma}$ and $K^{R',C'}_{\sigma}$. This gives us
	\begin{align}
		K^{R,C}_{\sigma}K^{R',C'}_{\sigma}&= \frac{|R|}{|N'_C|} \sum_{d \in N'_C} \sum_{q \in Q_C} \sum_{k \in \partial(E)} \overline{\chi}_R(D) F^{qdq^{-1}, qkr_Cq^{-1}}(\sigma)M^{f(r_C,d)^{-1}}(p) \frac{1}{|\ker(\partial)|} \sum_{e_k \in \ker (\partial)}M^{e_k}(p) \notag \\
		& \hspace{0.5cm} \frac{|R'|}{|N'_{C'}|} \sum_{d' \in N'_{C'}} \sum_{q' \in Q_{C'}} \sum_{k' \in \partial(E)} \overline{\chi}_{R'}(D') F^{q'd'{q'}^{-1}, q'k'r_{C'}{q'}^{-1}}(\sigma)M^{f(r_{C'},d')^{-1}}(p)\frac{1}{|\ker(\partial)|} \sum_{e'_k \in \ker (\partial)}M^{e'_k}(p) \notag\\
		&= \frac{|R|}{|N'_C|} \frac{|R'|}{|N'_{C'}|} \sum_{d \in N'_C} \sum_{d' \in N'_{C'}} \sum_{q \in Q_C} \sum_{q' \in Q_{C'}} \sum_{k \in \partial(E)} \sum_{k' \in \partial(E)} \overline{\chi}_R(d) \overline{\chi}_{R'}(d') F^{qdq^{-1}, qkr_Cq^{-1}}(\sigma) F^{q'd'{q'}^{-1}, q'k'r_{C'}{q'}^{-1}}(\sigma) \notag \\
		& \hspace{0.5cm} M^{f(r_C,d)^{-1}}(p) M^{f(r_{C'},d')^{-1}}(p) \frac{1}{|\ker(\partial)|} \sum_{e_k \in \ker (\partial)}M^{e_k}(p) \frac{1}{|\ker(\partial)|} \sum_{e'_k \in \ker (\partial)}M^{e'_k}(p). \label{Equation_2D_charge_projectors_product_1}
	\end{align}
	
	We can then use Equation \ref{Equation_fusion_EM_ribbons_1} from Section \ref{Section_2D_Magnetic} of the main text, which describes the algebra of the $F$ operators, to combine the two $F$ operators. We have $F^{h_1,g_1}(\sigma)F^{h_2,g_2}(\sigma)=\delta(g_1,g_2)F^{h_1h_2,g_1}(\sigma)$, which gives us
	\begin{align}
		K^{R,C}_{\sigma}K^{R',C'}_{\sigma}&= \frac{|R|}{|N'_C|} \frac{|R'|}{|N'_{C'}|} \sum_{d \in N'_C} \sum_{d' \in N'_{C'}} \sum_{q \in Q_C} \sum_{q' \in Q_{C'}} \sum_{k \in \partial(E)} \sum_{k' \in \partial(E)} \overline{\chi}_R(d) \overline{\chi}_{R'}(d') \delta(qkr_Cq^{-1},q'k'r_{C'}{q'}^{-1} ) \notag \\
		& \hspace{0.4cm}F^{qdq^{-1}q'd'{q'}^{-1}, qkr_Cq^{-1}}(\sigma)M^{f(r_C,d)^{-1}}(p) M^{f(r_{C'},d')^{-1}}(p) \frac{1}{|\ker(\partial)|} \sum_{e_k \in \ker (\partial)} \hspace{-0.1cm}M^{e_k}(p) \frac{1}{|\ker(\partial)|} \sum_{e'_k \in \ker (\partial)} \hspace{-0.1cm}M^{e'_k}(p). \label{Equation_2D_charge_projectors_product_2}
	\end{align}
	
	We wish to take a closer look at the expression $\delta(qkr_Cq^{-1},q'k'r_{C'}{q'}^{-1} )$. Because $qkr_Cq^{-1}$ belongs to class $C$, while $q'k'r_{C'}{q'}^{-1}$ belongs to class $C'$, and these equivalence classes partition the group $G$, this Kronecker delta can only be one if $C=C'$. Each class has a unique representative, so this also implies that $r_C=r_{C'}$ when the Kronecker delta is non-zero. We can therefore write the Kronecker delta as $\delta(C,C') \delta(qkr_Cq^{-1},q'k'r_C,{q'}^{-1})$. From the second Kronecker delta in this product, we can also show that we must have $q=q'$ and $k=k'$ for the delta to be non-zero. Recall that each $q$ labels a coset $qN'_C$. We will show that $q$ and $q'$ must belong to the same coset, and therefore we must have $q=q'$ because each coset only has one representative. If the Kronecker delta is satisfied, then we have
	\begin{align*}
		qkr_Cq^{-1}=q'k'r_C{q'}^{-1}, 
	\end{align*}
	which is true if and only if
	\begin{align*}
		r_C &= q^{-1}q' k^{-1}k'r_C {q'}^{-1} q\\
		&= (q^{-1}q')(k^{-1}k')r_C (q^{-1}q')^{-1}.
	\end{align*}
	
	However this implies that $q^{-1}q'$ is in $N'_C$, because it commutes with $r_C$ up to an element ($k^{-1}k'$) of $\partial(E)$. This means that $q$ and $q'$ are in the same coset, and so $q=q'$. This then implies that $r_C=k^{-1}k' r_C$, so we also have $k=k'$. This means that we can write
	$$\delta(qkr_Cq^{-1},q'k'r_{C'}{q'}^{-1} )=\delta(C,C') \delta(q,q') \delta(k,k').$$
	Inserting this into our product of $K$ operators from Equation \ref{Equation_2D_charge_projectors_product_2}, we have
	\begin{align}
		K^{R,C}_{\sigma}&K^{R',C'}_{\sigma} \notag\\
		&= \frac{|R|}{|N'_C|} \frac{|R'|}{|N'_{C'}|} \sum_{d \in N'_C} \sum_{d' \in N'_{C'}} \sum_{q \in Q_C} \sum_{q' \in Q_{C'}} \sum_{k \in \partial(E)} \sum_{k' \in \partial(E)} \overline{\chi}_R(d) \overline{\chi}_{R'}(d') \delta(C,C') \delta(q,q') \delta(k,k') \notag \\
		& \hspace{0.5cm}F^{qdq^{-1}q'd'{q'}^{-1}, qkr_Cq^{-1}}(\sigma) M^{f(r_C,d)^{-1}}(p) M^{f(r_{C'},d')^{-1}}(p) \frac{1}{|\ker(\partial)|} \sum_{e_k \in \ker (\partial)}M^{e_k}(p) \frac{1}{|\ker(\partial)|} \sum_{e'_k \in \ker (\partial)}M^{e'_k}(p) \notag\\
		&= \delta(C,C') \frac{|R||R'|}{|N'_C|^2} \sum_{d \in N'_C} \sum_{d' \in N'_{C}} \sum_{q \in Q_C} \sum_{k \in \partial(E)} \overline{\chi}_R(d) \overline{\chi}_{R'}(d') F^{qdd'{q}^{-1}, qkr_Cq^{-1}}(\sigma) M^{f(r_C,d)^{-1}}(p) M^{f(r_{C},d')^{-1}}(p) \notag\\
		& \hspace{0.5cm} \frac{1}{|\ker(\partial)|} \sum_{e_k \in \ker (\partial)}M^{e_k}(p) \frac{1}{|\ker(\partial)|} \sum_{e'_k \in \ker (\partial)}M^{e'_k}(p) \label{Equation_2D_charge_projectors_product_3}.
	\end{align}
	
	We can remove the sum over $e_k'$ in the kernel of $\partial$ by combining the terms involving $e_k$ and $e_k'$. This is because these terms are projectors:
	\begin{align}
		\frac{1}{|\ker(\partial)|} \sum_{e_k \in \ker (\partial)}M^{e_k}(p) \frac{1}{|\ker(\partial)|} \sum_{e'_k \in \ker (\partial)}M^{e'_k}(p) &=\frac{1}{|\ker(\partial)|} \sum_{e_k \in \ker (\partial)}\frac{1}{|\ker(\partial)|} \sum_{e'_k \in \ker (\partial)}M^{e_k}(p) M^{e'_k}(p) \notag\\
		&=\frac{1}{|\ker(\partial)|} \sum_{e_k \in \ker (\partial)}\frac{1}{|\ker(\partial)|} \sum_{e'_k \in \ker (\partial)}M^{e_k e'_k} (p)\notag\\
		&=\frac{1}{|\ker(\partial)|} \sum_{e_k''=e_k e'_k \in \ker (\partial)}\frac{1}{|\ker(\partial)|} \sum_{e'_k \in \ker (\partial)}M^{e''_k } (p)\notag\\ 
		&=\big(\frac{1}{|\ker(\partial)|} \sum_{e'_k \in \ker (\partial)} 1) \frac{1}{|\ker(\partial)|} \sum_{e_k'' \in \ker (\partial)}M^{e''_k }(p) \notag\\ 
		&= \frac{1}{|\ker(\partial)|} \sum_{e_k'' \in \ker (\partial)}M^{e''_k }(p). \label{Equation_2D_charge_symmetry_projector_product}
	\end{align}
	
	We can then relabel $e''_k$ to $e_k$ for simplicity. Next we must consider the product of single plaquette multiplication operators $M^{f(r_C,d)^{-1}}(p)M^{f(r_{C},d')^{-1}}(p)$. We can combine these into a single operator $M^{f(r_C,d)^{-1}f(r_C,d')^{-1}}(p)$, and because $E$ is Abelian we can write this as $M^{(f(r_C,d)f(r_C,d'))^{-1}}(p)$. We wish to know whether we can write this in terms of a single $f$ element $f(r_C,dd')$. Recall that $f(r_C,d)$ is an element of $E$ satisfying $\partial(f(r_C,d))=r_Cdr_C^{-1}d^{-1}$. Because $\partial(E)$ is in the centre of $G$, this means that $\partial(f(r_C,d))= d^{-1} (r_C d r_C^{-1} d^{-1}) d =d^{-1}r_C d r_C^{-1}$. Then we can write 
	\begin{align*}
		\partial(f(r_C,d)f(r_C,d')) &= \partial(f(r_C,d)) \partial(f(r_C,d'))\\
		&= (d^{-1}r_C d r_C^{-1}) (r_C d' r_C^{-1} {d'}^{-1})\\
		&=d^{-1}r_C d d' r_C^{-1} {d'}^{-1}.
	\end{align*}
	Then, because this lies in the centre of $G$, we can conjugate the right-hand side by $d$, with no effect on the left side of the equation, to obtain
	\begin{align*}
		\partial(f(r_C,d)f(r_C,d')) &= r_C d d' r_C^{-1} {d'}^{-1}d^{-1}\\
		&= r_C (dd')r_C^{-1} (dd')^{-1}.
	\end{align*}
	This implies that $f(r_C,d)f(r_C,d')$ is an element of $E$ satisfying $\partial(f(r_C,d)f(r_C,d'))= [r_C,dd']$, just as $f(r_C,dd')$ satisfies the same condition. However it does not guarantee that $f(r_C,d)f(r_C,d')$ is the same as the chosen representative $f(r_C,dd')$. That is, it is not guaranteed that we can choose our representatives $f$ such that $f(r_C,d)f(r_C,d')=f(r_C,dd')$ for all $d$ and $d'$. $f(r_C,d)f(r_C,d')$ may differ from $f(r_C,dd')$ by at most an element $a$ of the kernel of $\partial$. This is why it was necessary to average over the elements of the kernel, because we can absorb this extra factor into that average. We have 
	\begin{align}
		M^{f(r_C,d)^{-1}f(r_C,d')^{-1}}(p) \sum_{e_k \in \ker(\partial)}M^{e_k}(p)&=M^{f(r_C,dd')^{-1}}(p)M^{a}(p)\sum_{e_k \in \ker(\partial)}M^{e_k}(p) \notag\\
		&=M^{f(r_C,dd')^{-1}}(p)\sum_{e_k \in \ker(\partial)}M^{ae_k}(p) \notag\\
		&=M^{f(r_C,dd')^{-1}}(p)\sum_{e_k'=ae_k \in \ker(\partial)}M^{e_k'}(p) \label{Equation_2D_charge_single_plaquette_product}.
	\end{align}
	
	We note that, depending on the groups $G$ and $E$ and the map $\partial$, it may be possible to choose the representatives $f(r_C,x)$ to form a group for the different values of $x$, in which case this averaging procedure is unnecessary and we do not need to worry about the symmetry state of the plaquette at all. For example, if $G$ is Abelian then we always have $\partial(f(r_C,d))=[r_C,d]=1_G$, so we can always take $f(r_C,x)=1_E$. In this case the $f$ elements form a (trivial) group satisfying $f(r_C,d)f(r_C,d')=f(r_C,dd')$ and so we can just combine the two $M^{f(r_C,d)^{-1}}(p)$ operators directly, without any additional factors to absorb.

	Substituting the results from Equations \ref{Equation_2D_charge_symmetry_projector_product} and \ref{Equation_2D_charge_single_plaquette_product} into Equation \ref{Equation_2D_charge_projectors_product_3} gives us
	\begin{align*}
		K^{R,C}_{\sigma}K^{R',C'}_{\sigma}&= \delta(C,C') \frac{|R||R'|}{|N'_C|^2} \sum_{d \in N'_C} \sum_{d' \in N'_{C}} \sum_{q \in Q_C} \sum_{k \in \partial(E)} \overline{\chi}_R(d) \overline{\chi}_{R'}(d') F^{qdd'{q}^{-1}, qkr_Cq^{-1}}(\sigma) M^{f(r_C,dd')^{-1}}(p)\\ & \hspace{0.5cm}\frac{1}{|\ker(\partial)|}\sum_{e_k' \in \ker(\partial)}M^{e_k'}(p). 
	\end{align*}
	
	Now the only place where $d$ and $d'$ appear individually is in the characters $\chi_R$ and $\chi_{R'}$. We now make the change of variable from $d'$ to $ \tilde{d}=dd'$ to obtain
	\begin{align}
		K^{R,C}_{\sigma}K^{R',C'}_{\sigma} &= \delta(C,C') \frac{|R||R'|}{|N'_C|^2} \sum_{d \in N'_C} \sum_{\tilde{d}=dd' \in N'_{C}} \sum_{q \in Q_C} \sum_{k \in \partial(E)}\overline{\chi}_R(d) \overline{\chi}_{R'}(d^{-1} \tilde{d}) F^{q\tilde{d}{q}^{-1}, qkr_Cq^{-1}}(\sigma) \notag\\
		& \hspace{0.5cm} M^{f(r_C,\tilde{d})^{-1}}(p) \frac{1}{|\ker(\partial)|} \sum_{e_k' \in \ker(\partial)}M^{e_k'}(p). \label{Equation_2D_charge_projectors_product_4} 
	\end{align}
	
	We can separate the expression $\overline{\chi}_{R'}(d^{-1} \tilde{d})$ into contributions from $d$ and $\tilde{d}$, by writing
	\begin{align*}
		\chi_{R'}(d^{-1} \tilde{d})&= \sum_{c=1}^{|R'|} [D^{R'}(d^{-1}\tilde{d})]_{cc}\\
		&= \sum_{c=1}^{|R'|} \sum_{b=1}^{|R'|} [D^{R'}(d^{-1})]_{cb} [D^{R'}(\tilde{d})]_{bc}.
	\end{align*}
	Substituting this into Equation \ref{Equation_2D_charge_projectors_product_4}, we have
	\begin{align}
		K^{R,C}_{\sigma}K^{R',C'}_{\sigma}&= \delta(C,C') \frac{|R||R'|}{|N'_C|^2} \sum_{d \in N'_C} \sum_{\tilde{d}=dd' \in N'_{C}} \sum_{q \in Q_C} \sum_{k \in \partial(E)} \bigg(\sum_{a=1}^{R} \sum_{b=1}^{|R'|} \sum_{c=1}^{|R'|} [D^R(d)]_{aa} [D^{R'}(d^{-1})]_{cb} [D^{R'}(\tilde{d})]_{bc}\bigg)^* \notag \\
		& \hspace{0.5cm} F^{q\tilde{d}{q}^{-1}, qkr_Cq^{-1}}(\sigma) M^{f(r_C,\tilde{d})^{-1}}(p) \frac{1}{|\ker(\partial)|} \sum_{e_k' \in \ker(\partial)}M^{e_k'}(p). \label{Equation_2D_charge_projectors_product_5} 
	\end{align}
	
	Then because $d$ only appears in the expression $ [D^R(d)]_{aa} [D^{R'}(d^{-1})]_{cb}$, we can use the Grand Orthogonality Theorem to write
	$$\sum_{d' \in N'_{C}} [D^R(d)]_{aa} [D^{R'}(d^{-1})]_{cb} = \frac{|N'_C|}{|R|} \delta_{ab} \delta_{ac} \delta(R,R').$$
	Substituting this into Equation \ref{Equation_2D_charge_projectors_product_5} gives us
	\begin{align*}
		&K^{R,C}_{\sigma}K^{R',C'}_{\sigma}\\
		&= \delta(C,C') \frac{|R||R'|}{|N'_C|^2} \sum_{d \in N'_C} \sum_{\tilde{d}=dd' \in N'_{C}} \sum_{q \in Q_C} \sum_{k \in \partial(E)} \bigg(\sum_{a=1}^{R} \sum_{b=1}^{|R'|} \sum_{c=1}^{|R'|} \frac{|N'_C|}{|R|} \delta_{ab} \delta_{ac} \delta(R,R') [D^{R'}(\tilde{d})]_{bc}\bigg)^*F^{q\tilde{d}{q}^{-1}, qkr_Cq^{-1}}(\sigma) \\
		& \hspace{0.5cm} M^{f(r_C,\tilde{d})^{-1}}(p) \frac{1}{|\ker(\partial)|} \sum_{e_k' \in \ker(\partial)}M^{e_k'}(p)\\
		&=\delta(C,C') \delta(R,R') \frac{|R|}{|N'_C|} \sum_{d \in N'_C} \sum_{\tilde{d}=dd' \in N'_{C}} \sum_{q \in Q_C} \sum_{k \in \partial(E)} \big( \sum_{c=1}^{|R|} [D^{R}(\tilde{d})]_{cc}\big)^* F^{q\tilde{d}{q}^{-1}, qkr_Cq^{-1}}(\sigma) M^{f(r_C,\tilde{d})^{-1}}(p) \\
		& \hspace{0.5cm} \frac{1}{|\ker(\partial)|} \sum_{e_k' \in \ker(\partial)}M^{e_k'}(p)\\
		&=\delta(C,C') \delta(R,R') \frac{|R|}{|N'_C|}\sum_{d \in N'_C} \sum_{\tilde{d}=dd' \in N'_{C}} \sum_{q \in Q_C} \sum_{k \in \partial(E)} \overline{\chi}_R(\tilde{d}) F^{q\tilde{d}{q}^{-1}, qkr_Cq^{-1}}(\sigma) M^{f(r_C,\tilde{d})^{-1}}(p) \frac{1}{|\ker(\partial)|} \sum_{e_k' \in \ker(\partial)}M^{e_k'}(p)\\
		&= \delta(C,C') \delta(R,R') K^{R,C}_{\sigma},
	\end{align*}
	so that the $K$ operators are indeed orthogonal projectors.

	Next, we wish to show that these projectors are complete, at least in the space where the plaquette $p$ affected by $M^{f}(p)$ is in the space described by the identity irrep of the kernel of $\partial$. To see this, consider summing all of the projectors. We have
	\begin{align}
		\sum_{R}\sum_C K^{R,C}_{\sigma} &= \sum_R \sum_C \frac{|R|}{|N'_C|} \sum_{d \in N'_C} \overline{\chi}_R(d) \sum_{q \in Q_C} \sum_{k \in \partial(E)} C^{qdq^{-1}}(\sigma) \delta(g(\sigma), qkr_Cq^{-1}) \frac{1}{|\ker(\partial)|} \sum_{e \in E |[r_C,d]= \partial(e)} M^e(p) \notag \\
		&= \sum_C \sum_{d \in N'_C} \big[ \sum_R \frac{|R|}{|N'_C|} \overline{\chi}_R(d) \big] \sum_{q \in Q_C} \sum_{k \in \partial(E)} C^{qdq^{-1}}(\sigma) \delta(g(\sigma), qkr_Cq^{-1}) \frac{1}{|\ker(\partial)|}\sum_{e \in E |[r_C,d]= \partial(e)} M^e(p), \label{Equation_2D_charge_projectors_sum_1}
	\end{align}
	where we have written $M^{f(r_C,\tilde{d})^{-1}}(p) \frac{1}{|\ker(\partial)|} \sum_{e_k' \in \ker(\partial)}M^{e_k'}(p)$ as $\frac{1}{|\ker(\partial)|}\sum_{e \in E|[r_C,d]= \partial(e)} M^e(p)$. We now wish to consider the expression $\big[\sum_R \frac{|R|}{|N'_C|} \overline{\chi}_R(d) \big]$. We can write $|R| =\chi^R(1_{N'_C})$. Then we have
	$$( \sum_R \frac{1}{|N'_C|} \chi_R(1_{N'_C})\overline{\chi}_R(d) ),$$
	which, from the column orthogonality relations for irreps, gives $\delta(d,1_{N'_C})$, where $1_{N'_C}$ is also the identity of $G$. Then substituting this into Equation \ref{Equation_2D_charge_projectors_sum_1} gives us
	\begin{align*}
		\sum_{R}\sum_C K^{R,C}_{\sigma}&= \sum_C \sum_{d \in N'_C} ( \delta(d,1_{N'_C})) \sum_{q \in Q_C} \sum_{k \in \partial(E)} C^{qdq^{-1}}(\sigma) \delta(g(\sigma), qkr_Cq^{-1}) \frac{1}{|\ker(\partial)|} \sum_{e \in E|[r_C,d]= \partial(e)} M^e(p)\\
		&=\sum_C \sum_{q \in Q_C} \sum_{k \in \partial(E)} C^{1_G}(\sigma) \delta(g(\sigma), qkr_Cq^{-1}) \frac{1}{|\ker(\partial)|} \sum_{e \in E|1_G= \partial(e)} M^e(p).
	\end{align*}
	$C^{1_G}(\sigma)$ is just the identity operator (it multiplies edges cut by the dual path of $\sigma$ by the identity element, which is a trivial action). Therefore
	\begin{align*}
		\sum_{R}\sum_C K^{R,C}_{\sigma}&=\sum_C \sum_{q \in Q_C} \sum_{k \in \partial(E)} \delta(g(\sigma), qkr_Cq^{-1}) \frac{1}{|\ker(\partial)|} \sum_{e \in \ker(\partial)} M^e(p).
	\end{align*}

	Now consider $\sum_{q \in Q_C} \sum_{k \in \partial(E)}\delta(g(\sigma), qkr_Cq^{-1})$. Recall that we defined $Q_C$ as representatives of the cosets of $N'_C$, where $N'_C$ is the subgroup of $G$ consisting of elements which commute with $r_C$ up to elements in $\partial(E)$. This means that under the sum $\sum_{q \in Q_C} \sum_{k \in \partial(E)}$, the expression $qkr_Cq^{-1}$ reaches every element in $C$ precisely once, as we will now prove. First, we will show that each element in $C$ is reached by this sum. By definition, an element $g$ in the class $C$ can be written as $x y r_Cx^{-1}$, where $x$ is some element of $G$ and $y$ is some element of $\partial(E)$. Now $x$ belongs to some coset $xN'_C$ because these partition the group $G$, and so $x$ can be written as $q b$, for some element $q$ in $Q_C$ and some element $b$ in $N'_C$. Then we have
	$$g= qb yr_C b^{-1}q^{-1} = q y br_Cb^{-1} q^{-1},$$
	where we used the fact that $y$ is in the centre of $G$ (because it is in $\partial(E)$) to move it outside of $br_Cb^{-1}$ in the second equality. However because $b$ is in $N'_C$, it commutes with $r_C$ up to some element in $\partial(E)$, which we will denote by $z$. That is $b r_Cb^{-1} =z r_C$. Then we have 
	$$g= q y br_Cb^{-1} q^{-1}=q y z r_C q^{-1}= q k r_C q^{-1},$$
	where $k=yz$. We have therefore shown that any element $g \in C$ can be written as $q k r_C q^{-1}$ for some $q \in Q_C$ and $k \in \partial(E)$ and so our sum reaches every element of the class. Next we will show that each element is visited precisely once by the sum. Suppose to the contrary that there were two pairs, $(q_1,k_1)$ and $(q_2,k_2)$, such that 
	$$q_1k_1r_Cq_1^{-1} =q_2k_2r_Cq_2^{-1}.$$
	
	Then we would have
	\begin{align*}
		r_C &= k_1^{-1} q_1^{-1} q_2 k_2 r_Cq_2^{-1} q_1\\
		&=q_1^{-1} q_2 (k_1^{-1}k_2) r_C (q_1^{-1}q_2)^{-1},
	\end{align*}
	which implies that $q_1^{-1}q_2$ commutes with $r_C$ up to an element $k_1^{-1}k_2$ of $\partial(E)$ and so is in $N'_C$. However this implies that $q_1$ and $q_2$ are in the same coset of $N'_C$, and so cannot be distinct representatives of $G/N'_C$. Therefore, $q_1$ and $q_2$ must be the same. This implies that
	\begin{align*}
		r_C &= (k_1^{-1}k_2) r_C, 
	\end{align*} 
	which implies that $k_1=k_2$. Therefore, the pairs $(q_1,k_1)$ and $(q_2,k_2)$ must in fact be the same. Therefore, in the sum $\sum_{q \in Q_C} \sum_{k \in \partial(E)}$, the expression $qkr_Cq^{-1}$ reaches every element in $C$ precisely once, and so 
	$$\sum_{q \in Q_C} \sum_{k \in \partial(E)} \delta(g(\sigma), qkr_Cq^{-1}) = \sum_{g \in C} \delta(g(\sigma),g) .$$
	
	Therefore, we have
	\begin{align}
		\sum_{R}\sum_C K^{R,C}_{\sigma}&=\sum_C \sum_{q \in Q_C} \sum_{k \in \partial(E)} \delta(g(\sigma), qkr_Cq^{-1}) \frac{1}{|\ker(\partial)|} \sum_{e \in \ker(\partial)} M^e(p)\notag \\
		&= \sum_C \sum_{g \in C} \delta(g(\sigma), g) \frac{1}{|\ker(\partial)|} \sum_{e \in \ker(\partial)} M^e(p) \notag \\
		&= \big(\sum_{g \in G}\delta(g(\sigma), g) \big) \frac{1}{|\ker(\partial)|} \sum_{e \in \ker(\partial)} M^e(p) \notag \\
		&= \frac{1}{|\ker(\partial)|}\sum_{e \in \ker(\partial)} M^e(p), \label{Equation_2D_charge_projectors_sum_2}
	\end{align}
	which we can recognise as the projector which projects onto the space where the plaquette at the start of the ribbon is labelled by the trivial irrep of the kernel of $\partial$. Therefore, the set of projectors $K^{R,C}$ is complete in this space.

	So far we have restricted to the case where the plaquette at the start of the ribbon is in a state labelled by the trivial irrep of the kernel of $\partial$. However this state is linked to the ones where the plaquette is labelled by a non-trivial irrep through the symmetry operators $U^{\nu}$ that we introduced in Section \ref{Section_2D_irrep_basis} of the main text (where $\nu$ is an irrep of $E$ with an appropriate restriction to the kernel of $\partial$). In order to generate the projectors corresponding to a different plaquette state, we can first apply the symmetry operator ${U^{\nu}}^{-1}$ that takes the plaquette to a state labelled by the trivial irrep of the kernel, then apply our projector for the trivial irrep state, then apply the inverse symmetry operation $U^{\nu}$ to return to the original state of the plaquette. Let us consider the measurement operator $U^{\nu}K^{R,C}_{\sigma}{U^{\nu}}^{-1}$ generated by this procedure. Recall from Section \ref{Section_2D_irrep_basis} of the main text that the symmetry operator $U^{\nu}$ applied on our lattice $L$ can be written as
	\begin{align*}
		U^{\nu} &= \sum_{e \in E} \nu(e) \delta(\hat{e}(L),e)\\
		&=\sum_{e \in E} \nu(e) \delta(\prod_{\text{plaquette } p \in L} \hat{e}_p ,e),
	\end{align*}
	where $\hat{e}_p$ is the label of plaquette $p$ and we have assumed for simplicity that each plaquette has the same orientation. This can also be written as
	\begin{align*}
		U^{\nu} &= \nu(\prod_{\text{plaquette } p \in L} \hat{e}_p)\\
		&= \prod_{\text{plaquette } p \in L} \nu(\hat{e}_p).
	\end{align*}
	Then the inverse operator can be written as
	\begin{align*}
		{U^{\nu}}^{-1} &= \nu(\prod_{\text{plaquette } p \in L} \hat{e}_p)^{-1}\\
		&= \prod_{\text{plaquette } p \in L} \nu(\hat{e}_p)^{-1}.
	\end{align*}
	
	Now because our measurement operator $K^{R,C}_{\sigma}$ only affects the label of one plaquette, the plaquette at the start and end of the dual path of $\sigma$, the factors $\nu(\hat{e}_p)$ corresponding to other plaquettes will commute with $K^{R,C}_{\sigma}$ and so the factors from $U^{\nu}$ and its inverse will cancel. Therefore, we can replace $U^{\nu}$ with the operator $U^{\nu}_p= \nu(\hat{e}_p)$, which only acts on the plaquette $p$ at the start and end of the dual path. Then
	\begin{align*}
		U^{\nu}K^{R,C}_{\sigma}{U^{\nu}}^{-1}&= U^{\nu}_p K^{R,C}_{\sigma}{U^{\nu}_p}^{-1}\\
		&= \nu(\hat{e}_p) K^{R,C}_{\sigma} \nu(\hat{e}_p)^{-1}\\
		&= \nu(\hat{e}_p) \frac{|R|}{|N'_C|} \sum_{d \in N'_C} \sum_{q \in Q_C} \sum_{k \in \partial(E)} \overline{\chi}_R(d) F^{qdq^{-1}, qkr_Cq^{-1}}(\sigma)M^{f(r_C,d)^{-1}}(p) \frac{1}{|\ker(\partial)|} \sum_{e_k \in \ker (\partial)}M^{e_k}(p) \nu(\hat{e}_p)^{-1}.
	\end{align*}

	Now only the single plaquette multiplication operators $M^{f(r_C,d)^{-1}}(p)$ and $M^{e_k}(p)$ affect the plaquette label $\hat{e}_p$. We have $\hat{e_p}M^e(p) = M^e(p) (e \hat{e}_p)$ and so
	\begin{align*}
		U^{\nu}&K^{R,C}_{\sigma}{U^{\nu}}^{-1}\\
		&= \frac{|R|}{|N'_C|} \sum_{d \in N'_C} \sum_{q \in Q_C} \sum_{k \in \partial(E)} \overline{\chi}_R(d) F^{qdq^{-1}, qkr_Cq^{-1}}(\sigma)\nu(\hat{e}_p)M^{f(r_C,d)^{-1}}(p) \frac{1}{|\ker(\partial)|} \sum_{e_k \in \ker (\partial)}M^{e_k}(p) \nu(\hat{e}_p)^{-1}\\
		&= \frac{|R|}{|N'_C|} \sum_{d \in N'_C} \sum_{q \in Q_C} \sum_{k \in \partial(E)} \overline{\chi}_R(d) F^{qdq^{-1}, qkr_Cq^{-1}}(\sigma)M^{f(r_C,d)^{-1}}(p) \nu(f(r_C,d)^{-1}\hat{e}_p)\frac{1}{|\ker(\partial)|} \sum_{e_k \in \ker (\partial)}M^{e_k}(p) \nu(\hat{e}_p)^{-1}\\
		&= \frac{|R|}{|N'_C|} \sum_{d \in N'_C} \sum_{q \in Q_C} \sum_{k \in \partial(E)} \overline{\chi}_R(d) F^{qdq^{-1}, qkr_Cq^{-1}}(\sigma)M^{f(r_C,d)^{-1}}(p) \frac{1}{|\ker(\partial)|} \sum_{e_k \in \ker (\partial)}M^{e_k}(p)\nu(e_kf(r_C,d)^{-1}\hat{e}_p)\nu(\hat{e}_p)^{-1}\\ 
		&=\frac{|R|}{|N'_C|} \sum_{d \in N'_C} \sum_{q \in Q_C} \sum_{k \in \partial(E)} \overline{\chi}_R(d) F^{qdq^{-1}, qkr_Cq^{-1}}(\sigma)M^{f(r_C,d)^{-1}}(p) \frac{1}{|\ker(\partial)|} \sum_{e_k \in \ker (\partial)}M^{e_k}(p) \nu(e_k) \nu(f(r_C,d)^{-1})\\
		& \hspace{0.5cm} \nu(\hat{e}_p)\nu(\hat{e}_p)^{-1}\\
		&=\frac{|R|}{|N'_C|} \sum_{d \in N'_C} \sum_{q \in Q_C} \sum_{k \in \partial(E)} \overline{\chi}_R(d) F^{qdq^{-1}, qkr_Cq^{-1}}(\sigma)\nu(f(r_C,d)^{-1}) M^{f(r_C,d)^{-1}}(p) \frac{1}{|\ker(\partial)|} \sum_{e_k \in \ker (\partial)} \nu(e_k)M^{e_k}(p).
	\end{align*}
	
	We see that this is largely the same as our original operator, except that we have the phase $\nu(e_k)$ in the averaging over $M^{e_k}(p)$ for $e_k$ in the kernel of $\partial$, and we have the phase $\nu(f(r_C,d)^{-1})$ before the single plaquette multiplication operator with label outside the kernel. The former difference means that our operator now includes $\sum_{e_k \in \ker (\partial)} \nu(e_k)M^{e_k}(p)$, which is the projector to the space where the plaquette $p$ is labelled by an irrep whose restriction to the kernel of $\partial$ is the same as that of $\nu$. This highlights the fact that this projector should be applied to a different space to our original projector $K^{R,C}_{\sigma}$. We also note that we can guarantee that these new symmetry related operators do indeed form a new set of orthogonal projectors, because taking the product of two such operators gives us
	\begin{align*}
		(U^{\nu}K^{R,C}_{\sigma}{U^{\nu}}^{-1}) (U^{\nu}K^{R',C'}_{\sigma}{U^{\nu}}^{-1})&= U^{\nu}K^{R,C}_{\sigma}K^{R',C'}_{\sigma}{U^{\nu}}^{-1}\\
		&=U^{\nu} \delta(R,R') \delta(C,C')K^{R,C}_{\sigma}{U^{\nu}}^{-1}\\
		&= \delta(R,R') \delta(C,C') U^{\nu} K^{R,C}_{\sigma}{U^{\nu}}^{-1}.
	\end{align*}

	Having defined our projectors to definite topological charge, we now wish to see how these projectors change in a particular special case for our crossed module. In Section \ref{Section_2D_Condensation_Confinement} of the main text, for each pair of groups $G$ and $E$ we defined an ``uncondensed" model, for which $\partial$ maps to the identity element of $G$. In that section we presented projectors to definite topological charge. We will now show how these arise from our general (uncondensed) projectors. Because $\partial$ maps to the identity of $G$, the class $C$ of elements that have the form $xyr_C x^{-1}$, where $x$ is an element of $G$ and $y$ an element of $\partial(E)$, is just the conjugacy class of $r_C$ ($y$ can only be the identity element). In addition, the group $N'_C$ of elements that commute with $r_C$ up to elements in $\partial(E)$ just becomes the centralizer $N_C$ of $r_C$. We can also drop the sum over elements in $\partial(E)$. Finally, because $\partial(E)=\set{1_G}$, the single plaquette multiplication operator $M^{f(r_C,d)}(p)$ has a label satisfying $\partial(f(r_C,d))=1$. We can just take $f(r_C,d)=1_E$ for all $r_C$ and $d$. This in turn means that we do not need to project to a particular symmetry state for the plaquette at the start of the ribbon (i.e., we do not need $\frac{1}{|\ker(\partial)|} \sum_{e \in \ker(\partial)} M^e(p)$ or similar in the projector), but if we wish to project to a given symmetry state we can just multiply by the relevant projector. Making these various simplifications, we can see that the projector to definite topological charge is just given by
	\begin{align*}
		K^{R,C}_{\sigma} &= \frac{|R|}{|N_C|} \sum_{d \in N_C} \overline{\chi}_R(d) \sum_{q \in Q_C} F^{qdq^{-1}, qr_Cq^{-1}}(\sigma)\\
		&=\frac{|R|}{|N_C|} \sum_{D \in (N_C)_{cj}} \overline{\chi}_R(D)\sum_{q \in Q_C} \sum_{d \in D} F^{qdq^{-1},qr_Cq^{-1}}(\sigma),
	\end{align*}
	as we claimed in Section \ref{Section_2D_topological_Charge} of the main text.
	
	\subsection{Examples of topological charge}
	\label{Section_2D_Topological_Charge_Examples}

	\subsubsection{Charge of an electric excitation at the end of an electric ribbon operator}
	\label{Section_2D_charge_electric_excitation}
\begin{figure}[h]
		\begin{center}
			\begin{overpic}[width=0.75\linewidth]{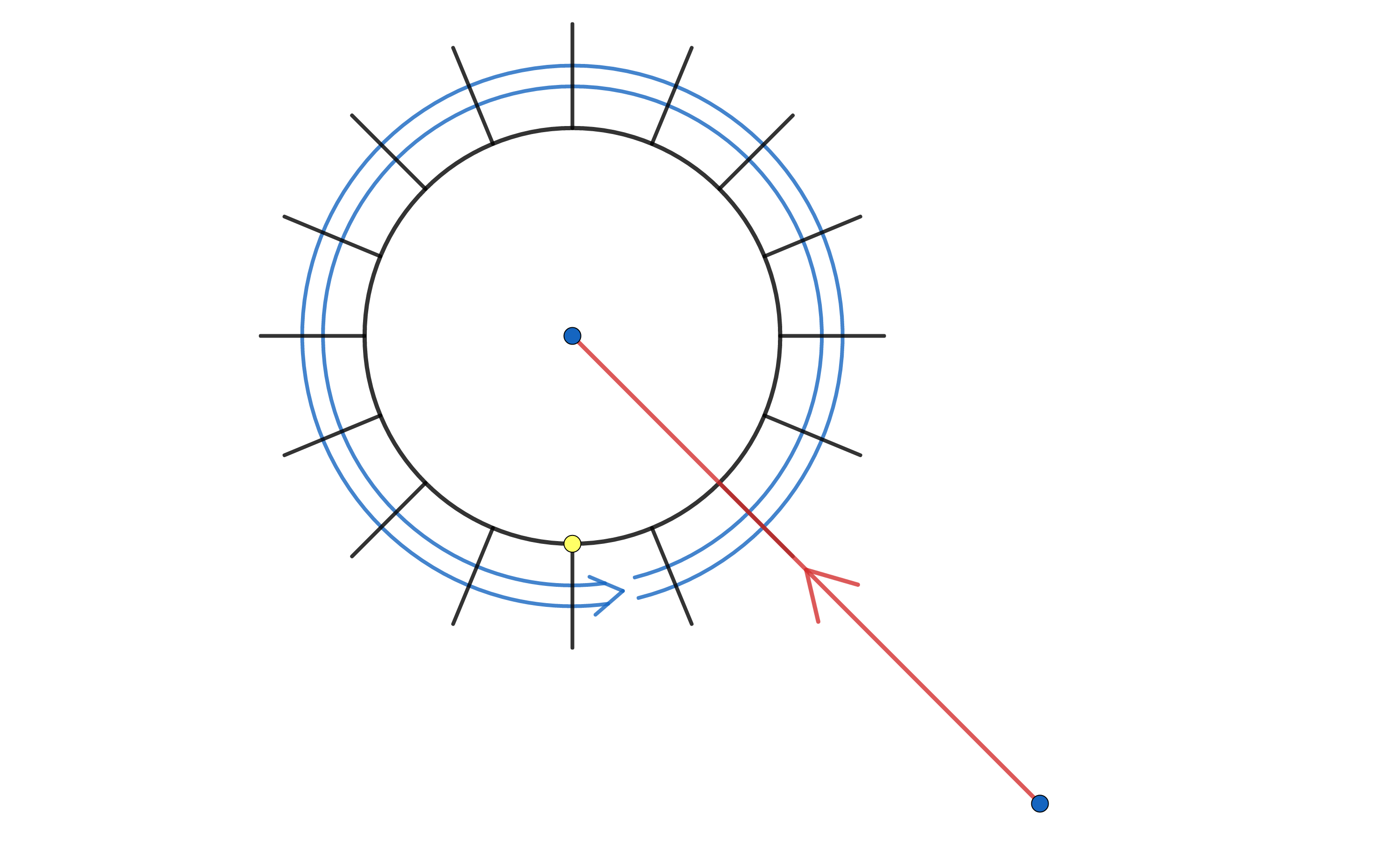}
				\put(15,30){$K^{R,C}_{\sigma}$}
				\put(65,15){$S^{R_1,a,b}(t)$}
			\end{overpic}
			
			\caption{In order to measure the charge of an electric excitation produced by an electric ribbon operator $S^{R_1,a,b}(t)$, we apply one of our projectors $K^{R,C}_{\sigma}$ on a ribbon enclosing that excitation. }
			\label{2D_charge_measurement_electric}
			
		\end{center}
	\end{figure}

	Next we wish to give some simple examples of the use of the projectors to definite topological charge, by applying these projectors to states with simple excitations. For example, we first consider measuring the charge of a pure electric excitation produced at the end of an electric ribbon operator $S^{R_1,a,b}(t)= \sum_{g \in G} [D^{R_1}(g)]_{ab} \delta(g,g(t))$. To do so, we consider the situation shown in Figure \ref{2D_charge_measurement_electric}, where we apply a measurement operator $K^{R,C}_{\sigma}$ on a ribbon enclosing the excitation whose charge we wish to measure. We therefore must consider the state where we apply the electric ribbon operator on a ground state, followed by the measurement operator:
	\begin{align}
		K_{\sigma}^{R,C} S^{R_1,a,b}(t) \ket{GS}&= \frac{|R|}{|N'_C|} \sum_{d \in N'_C} \overline{\chi}_R(d) \sum_{q \in Q_C} \sum_{k \in \partial(E)} C^{qdq^{-1}}(\sigma) \delta(g(\sigma), qkr_Cq^{-1}) M^{f(r_C,d)^{-1}}(p) \frac{1}{|\ker(\partial)|} \sum_{e_k \in \ker (\partial)} M^{e_k}(p) \notag\\
		& \hspace{0.5cm} \sum_{g \in G} [D^{R_1}(g)]_{ab} \delta(g,g(t)) \ket{GS}. \label{Equation_2D_topological_charge_electric_1}
	\end{align}

	The electric ribbon operator commutes with everything in $K_{\sigma}^{RC}$ except for the closed magnetic ribbon operator $C^{qdq^{-1}}(\sigma)$. Therefore,
	\begin{align*}
		K_{\sigma}^{R,C} S^{R_1,a,b}(t) \ket{GS}&= \frac{|R|}{|N'_C|} \sum_{d \in N'_C} \overline{\chi}_R(d) \sum_{q \in Q_C} \sum_{k \in \partial(E)} C^{qdq^{-1}}(\sigma) \sum_{g \in G} [D^{R_1}(g)]_{ab} \delta(g,g(t)) \delta(g(\sigma), qkr_Cq^{-1}) \notag \\ & \hspace{0.5cm}M^{f(r_C,d)^{-1}}(p) \frac{1}{|\ker(\partial)|} \sum_{e_k \in \ker (\partial)} M^{e_k}(p) \ket{GS}. 
	\end{align*}
	
	We can also move the electric ribbon operator $\delta(g(\sigma), qkr_Cq^{-1})$ from the projector to the right, so that it acts directly on the ground state. Then we have
	\begin{align*}
		K_{\sigma}^{R,C} S^{R_1,a,b}(t) \ket{GS}&= \frac{|R|}{|N'_C|} \sum_{d \in N'_C} \overline{\chi}_R(d) \sum_{q \in Q_C} \sum_{k \in \partial(E)} C^{qdq^{-1}}(\sigma) \sum_{g \in G} [D^{R_1}(g)]_{ab} \delta(g,g(t)) \notag\\
		& \hspace{0.5cm}M^{f(r_C,d)^{-1}}(p) \frac{1}{|\ker(\partial)|} \sum_{e_k \in \ker (\partial)} M^{e_k}(p) \delta(g(\sigma), qkr_Cq^{-1}) \ket{GS}. 
	\end{align*}
	
	We do this because $\delta(g(\sigma), qkr_Cq^{-1})$ acting on the ground state only gives a non-zero result in certain circumstances. Because $g(\sigma)$ is the label of a contractible closed path, and the closed path must satisfy fake-flatness in the ground state, this label must be equal to $\partial(e_m)^{-1}$, where $e_m$ is the label of the surface enclosed by the closed ribbon. This means that $\delta(g(\sigma), qkr_Cq^{-1})$ is only non-zero if $qkr_Cq^{-1}=\partial(e_m)^{-1}$. We can also write this condition as $qk \partial(e_m)r_Cq^{-1}=1_G$, which implies that $k\partial(e_m)r_C=1_G$ and so $r_C=\partial(e_m)^{-1}k^{-1}$. Because $k$ is in $\partial(E)$, we see that we only obtain a non-zero result when $r_C$ is in $\partial(E)$. From the definition of the equivalence classes in Equation \ref{Equation_union_coset_class_definition}, this indicates that the class $C$ must be $\partial(E)$. Therefore, we have
	\begin{align*}
		K_{\sigma}^{R,C} S^{R_1,a,b}(t) \ket{GS}&= \delta(C, \partial(E))\frac{|R|}{|N'_C|} \sum_{d \in N'_C} \overline{\chi}_R(d) \sum_{q \in Q_C} \sum_{k \in \partial(E)} C^{qdq^{-1}}(\sigma) \sum_{g \in G} [D^{R_1}(g)]_{ab} \delta(g,g(t)) \notag \\
		& \hspace{0.5cm} M^{f(r_C,d)^{-1}}(p) \frac{1}{|\ker(\partial)|} \sum_{e_k \in \ker (\partial)} M^{e_k}(p) \delta(r_Ck, \partial(e_m)^{-1}) \ket{GS}. 
	\end{align*}
	
	This expression is only non-zero when $C=\partial(E)$. This means that $r_C \in \partial(E)$ is in the centre of $G$, so we can replace $N'_C$, the subgroup of $G$ consisting of elements that commute with $r_C$ up to elements of $\partial(E)$, by the group $G$. This in turn means that the quotient group $Q_C=G/N'_C$ is the trivial group and we can take $q=1_G$. Then
	\begin{align}
		K_{\sigma}^{R,C} S^{R_1,a,b}(t) \ket{GS}&= \delta(C, \partial(E))\frac{|R|}{|G|} \sum_{d \in G} \overline{\chi}_R(d) \sum_{k \in \partial(E)} C^{d}(\sigma) \sum_{g \in G} [D^{R_1}(g)]_{ab} \delta(g,g(t)) \notag \\
		& \hspace{0.5cm} M^{f(r_C,d)^{-1}}(p) \frac{1}{|\ker(\partial)|} \sum_{e_k \in \ker (\partial)} M^{e_k}(p) \delta(r_Ck, \partial(e_m)^{-1}) \ket{GS}. \label{Equation_2D_topological_charge_electric_5}
	\end{align}
	Then we can rewrite $\delta(r_Ck, \partial(e_m)^{-1}) $ as $\delta(k, \partial(e_m)^{-1} r_C^{-1} )$ and use it to remove the sum over $k \in \partial(E)$. In addition, because $[r_C,d]=1_G$, we can take $f(r_C,d)=1_E$ and so we only have the single plaquette multiplication operators $\sum_{e \in \ker(\partial)}M^{e_k}(p)$. Assuming that the plaquette on which we apply the single plaquette operator is in the correct symmetry state (carrying a trivial irrep of the kernel of $\partial$), $\frac{1}{|\ker(\partial)|} \sum_{e \in \ker(\partial)}M^{e_k}(p)$ will act trivially on the ground state as described in Section \ref{Section_2D_irrep_basis} of the main text, so we can also remove this term. Then Equation \ref{Equation_2D_topological_charge_electric_5} becomes
	\begin{align}
		K_{\sigma}^{R,C} S^{R_1,a,b}(t) \ket{GS}&= \delta(C, \partial(E))\frac{|R|}{|G|} \sum_{d \in G} \overline{\chi}_R(d) C^{d}(\sigma) \sum_{g \in G} [D^{R_1}(g)]_{ab} \delta(g,g(t)) \ket{GS}. \label{Equation_2D_topological_charge_electric_6}
	\end{align}

	We can use the commutation relation given in Equation \ref{Magnetic_electric_braid_reverse_2} (note that we have the reversed orientation of ribbons compared to that indicated in Figure \ref{magnetic_electric_braid_2D_appendix}, so we must use Equation \ref{Magnetic_electric_braid_reverse_2} rather than Equation \ref{Magnetic_electric_braid_2D_2_appendix}) to commute $\delta(g,g(t))$ past $C^d(\sigma)$. Noting that Equation \ref{Magnetic_electric_braid_reverse_2} must be inverted to give this result, we have
	\begin{align*}
		C^{d}(\sigma) \delta(g,g(t)) = \delta(g(s.p(t)-s.p(\sigma))d^{-1}g(s.p(t)-s.p(\sigma))^{-1}g,g(t)).
	\end{align*}
	
	When inverting Equation \ref{Magnetic_electric_braid_reverse_2}, we may worry about whether the operator $g(s.p(t)-s.p(\sigma))$ is evaluated before or after the action of $C^d(\sigma)$. However if the path element before the action of $C^d(\sigma)$ is $g(s.p(t)-s.p(\sigma))$ then the path element after $C^d(\sigma) $ acts is $g(s.p(t)-s.p(\sigma))d^{-1}$, which does not change the expression $g(s.p(t)-s.p(\sigma))d^{-1}g(s.p(t)-s.p(\sigma))^{-1}$. Inserting the commutation relation into Equation \ref{Equation_2D_topological_charge_electric_6}, we have
	\begin{align*}
		& K_{\sigma}^{R,C} S^{R_1,a,b}(t) \ket{GS} \notag\\
		&= \delta(C, \partial(E))\frac{|R|}{|G|} \sum_{d \in G} \overline{\chi}_R(d) \sum_{g \in G} [D^{R_1}(g)]_{ab} \delta(g(s.p(t)-s.p(\sigma))d^{-1}g(s.p(t)-s.p(\sigma))^{-1}g,g(t)) C^{d}(\sigma) \ket{GS}.
	\end{align*}
	
	Next we change variables from $g$ to $g' = g(s.p(t)-s.p(\sigma))d^{-1}g(s.p(t)-s.p(\sigma))^{-1}g$ to obtain
	\begin{align}
		&K_{\sigma}^{R,C} S^{R_1,a,b}(t) \ket{GS} \notag\\
		&= \delta(C, \partial(E))\frac{|R|}{|G|} \sum_{d \in G} \overline{\chi}_R(d) \sum_{g' \in G} [D^{R_1}(g(s.p(t)-s.p(\sigma))dg(s.p(t)-s.p(\sigma))^{-1}g')]_{ab} \delta(g',g(t)) C^{d}(\sigma) \ket{GS}. \label{Equation_2D_topological_charge_electric_8}
	\end{align}
	
	Because $C^d(\sigma)$ is a contractible closed magnetic ribbon operator acting directly on the ground state, it acts trivially (as shown in Section \ref{Section_Topological_Magnetic_Ribbons}) and so we can remove it. We then write 
	\begin{align*}
		[D^{R_1}(g(s.p(t)-s.p(\sigma))dg(s.p(t)-s.p(\sigma))^{-1}g')]_{ab}&= \sum_{c=1}^{|R_1|} [D^{R_1}(g(s.p(t)-s.p(\sigma))dg(s.p(t)-s.p(\sigma))^{-1})]_{ac} [D^{R_1}(g')]_{cb}
	\end{align*}
	and substitute this into Equation \ref{Equation_2D_topological_charge_electric_8} to find
	\begin{align}
		K_{\sigma}^{R,C} S^{R_1,a,b}(t) \ket{GS}&= \delta(C, \partial(E))\frac{|R|}{|G|} \sum_{d \in G} \overline{\chi}_R(d) \sum_{g' \in G} \sum_{c=1}^{|R_1|} [D^{R_1}(g(s.p(t)-s.p(\sigma))dg(s.p(t)-s.p(\sigma))^{-1})]_{ac} \notag\\
		& \hspace{0.5cm} [D^{R_1}(g')]_{cb} \delta(g',g(t)) \ket{GS}. \label{Equation_2D_topological_charge_electric_9}
	\end{align} 
	
	Now $d$ only appears in the character $\overline{\chi}_R(d)$ and the matrix element $[D^{R_1}(g(s.p(t)-s.p(\sigma))dg(s.p(t)-s.p(\sigma))^{-1})]_{ac}$. This will allow us to use the orthogonality of irreps. First we use the fact that the character is a property of conjugacy class to write 
	$$\overline{\chi}_R(d) = \overline{\chi}_R(g(s.p(t)-s.p(\sigma))dg(s.p(t)-s.p(\sigma))^{-1}).$$
	Then, introducing
	$$d'=g(s.p(t)-s.p(\sigma))dg(s.p(t)-s.p(\sigma))^{-1},$$
	we have
	\begin{align*}
		\sum_{d \in G}& \overline{\chi}_R(g(s.p(t)-s.p(\sigma))dg(s.p(t)-s.p(\sigma))^{-1}) [D^{R_1}(g(s.p(t)-s.p(\sigma))dg(s.p(t)-s.p(\sigma))^{-1})]_{ac}\\	
		&=\sum_{d' \in G} \overline{\chi}_R(d') [D^{R_1}(d')]_{ac}\\
		&= \sum_{d' \in G}\sum_{n=1}^{|R|} [D^R(d')]_{nn}^* [D^{R_1}(d')]_{ac}.
	\end{align*}
	
	From the Grand Orthogonality Theorem, this gives us
	\begin{align*}
		\sum_{d' \in G}\sum_{n=1}^{|R|} [D^R(d')]_{nn}^* [D^{R_1}(d')^{-1})]_{ac}
		&= \frac{|G|}{|R|} \sum_{n=1}^{|R|} \delta(R_1,R) \delta(n,c) \delta(n,a)\\
		&= \frac{|G|}{|R_1|} \delta(R_1,R) \delta(c,a).
	\end{align*}
	Then substituting this into Equation \ref{Equation_2D_topological_charge_electric_9} gives us
	\begin{align}
		K_{\sigma}^{R,C} S^{R_1,a,b}(t) \ket{GS}&= \delta(C, \partial(E))\frac{|R|}{|G|} \sum_{g' \in G} \sum_{c=1}^{|R_1|} \frac{|G|}{|R_1|} \delta(R_1,R) \delta(c,a) [D^{R_1}(g')]_{cb} \delta(g',g(t)) \ket{GS} \notag \\
		&= \delta(C, \partial(E)) \delta(R_1,R) \sum_{g' \in G} [D^{R_1}(g')]_{ab} \delta(g',g(t)) \ket{GS}. \label{Equation_2D_topological_charge_electric_10}
	\end{align} 
	
	Note that $ \sum_{g' \in G} [D^{R_1}(g')]_{ab} \delta(g',g(t))$ is the original electric ribbon operator. We therefore see that if $C=\partial(E)$ and $R=R_1$, we obtain our original electric ribbon operator acting on the ground state and otherwise we get zero. Recall that the measurement operator projects onto states of definite charge in the region enclosed by the measurement operator. The fact that the projector either leaves the state produced by the electric ribbon operator invariant or returns zero depending on the labels $R$ and $C$ indicates that the electric ribbon operator $S^{R_1,a,b}(t)$ produces a state with definite topological charge in the region enclosed by the closed ribbon. We see that this charge is labelled by $C=\partial(E)$ and $R=R_1$, independently of the matrix indices $a$ and $b$ in the electric ribbon operator. This supports the claim made in Section \ref{Section_2D_electric} of the main text that the irrep $R_1$ labels the charge of the electric excitation and not the matrix indices (which correspond to some internal space).
	
	\subsubsection{Charge of a magnetic excitation at the start of a magnetic ribbon operator}
	\label{Section_2D_charge_magnetic_excitation}
\begin{figure}[h]
		\begin{center}
			\begin{overpic}[width=0.75\linewidth]{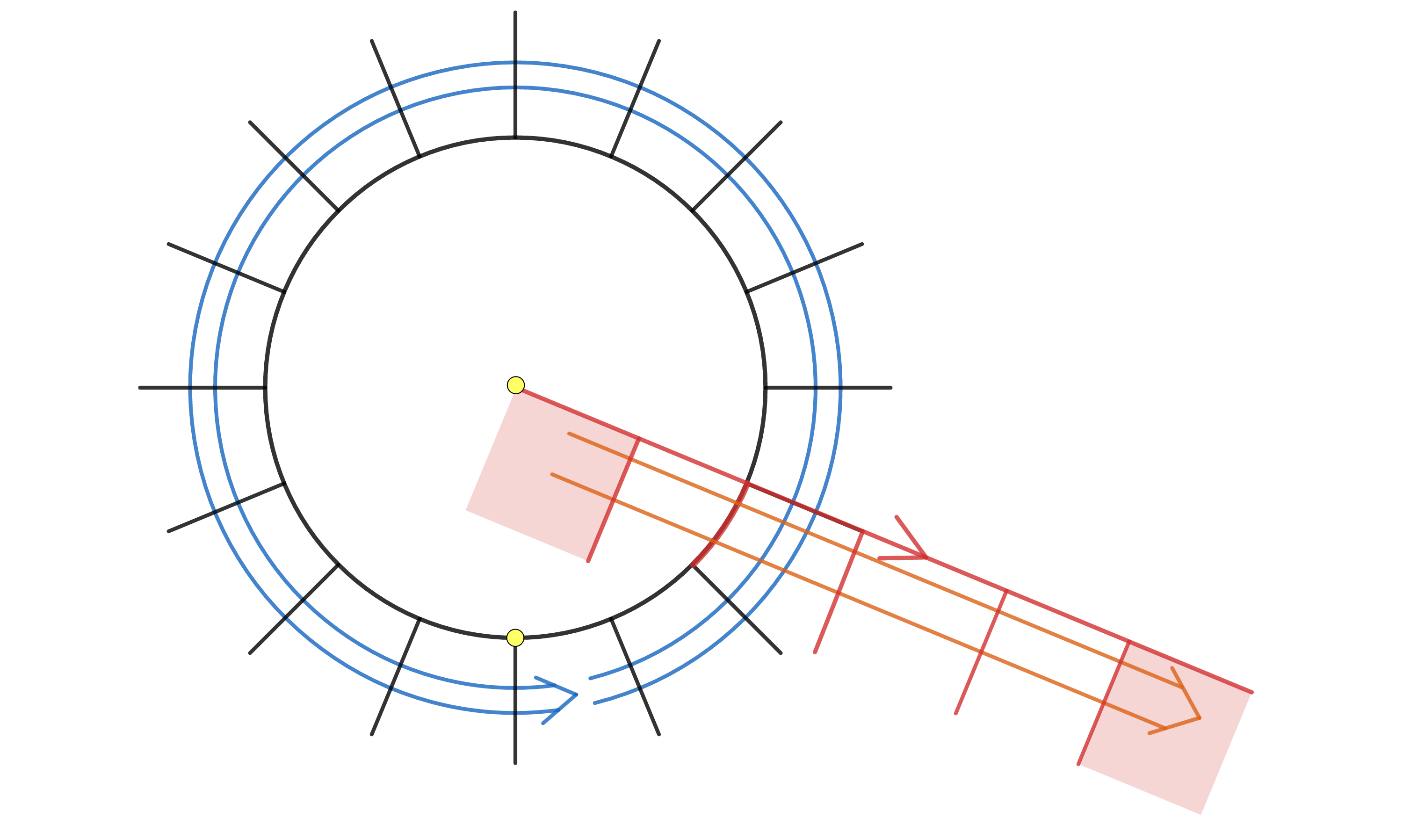}
				\put(12,15){$K^{R,C}_{\sigma}$}
				\put(80,15){$C^h(t)$}
				
			\end{overpic}
			
			\caption{We consider measuring the topological charge of the excitation produced at the start of a magnetic ribbon operator $C^h(t)$, by applying a measurement operator $K^{R,C}_{\sigma}$ on a ribbon $\sigma$ enclosing that excitation.}
			\label{2D_charge_measurement_magnetic}
			
		\end{center}
	\end{figure}

	Next we wish to give an example of the measurement of the topological charge of a magnetic excitation. We choose to measure the charge of the magnetic excitation produced at the start of the magnetic ribbon operator $C^h(t)$, rather than the end. This is because the magnetic ribbon operator may also excite the start-point of the ribbon $t$, if we take an appropriate sum of magnetic ribbon operators with label in the conjugacy class of $h$. We wish to highlight the point that the presence of this vertex excitation does not affect the topological charge of the magnetic excitation. That is, the topological charge of the magnetic excitation is the same whether we just have $C^h(t)$ (which would not necessarily produce an energy eigenstate for the start-point vertex), an equal linear combination of ribbon operators labelled by elements in the conjugacy class of $h$ (which would leave the start-point unexcited) or a linear combination that leaves the start-point excited. Therefore, we consider the situation shown in Figure \ref{2D_charge_measurement_magnetic}, where we apply the projector $K^{R,C}_{\sigma}$ around the start of the magnetic excitation. Then we consider the state
	$$K^{R,C}_{\sigma} C^h(t)\ket{GS}.$$

	We can then unpack the projector to write the state $K^{R,C}_{\sigma} C^h(t)\ket{GS}$ as 
	\begin{align}
		K_{\sigma}^{R,C}& C^h(t) \ket{GS} \notag \\
		&= \frac{|R|}{|N'_C|} \sum_{d \in N'_C} \overline{\chi}_R(d) \sum_{q \in Q_C} \sum_{k \in \partial(E)} C^{qdq^{-1}}(\sigma) \delta(g(\sigma), qkr_Cq^{-1}) M^{f(r_C,d)^{-1}}(p) \frac{1}{|\ker(\partial)|} \sum_{e_k \in \ker (\partial)} M^{e_k}(p) C^h(t) \ket{GS}. \label{Equation_2D_measurement_magnetic_1}
	\end{align}
	
	Assuming that the ground state is in the appropriate symmetry sector, the symmetry projector $\frac{1}{|\ker(\partial)|} \sum_{e_k \in \ker (\partial)} M^{e_k}(p)$ acts trivially on the ground state and so can be removed. We then wish to commute the magnetic ribbon operator $C^h(t)$ to the left of this expression, so that the measurement operator acts directly on the ground state. To do so, we first note that we can commute $C^h(t)$ past the single plaquette multiplication operators, because the former only acts on the edges and the latter only acts on a plaquette. Next we need to commute the magnetic ribbon operator past $C^{qdq^{-1}}(\sigma) \delta(g(\sigma), qkr_Cq^{-1})$. The first step is to move the magnetic ribbon $C^h(t)$ past the electric ribbon operator $\delta(g(\sigma), qkr_Cq^{-1})$. We performed an equivalent calculation in Section \ref{Section_2D_braiding_electric_magnetic}, when considering the braiding of a magnetic and an electric excitation. Given the orientations of the ribbons that we chose in Figure \ref{2D_charge_measurement_magnetic}, the relevant equation is Equation \ref{Magnetic_electric_braid_reverse_2}. This equation tells us that
	\begin{align*}
		\delta(qkr_Cq^{-1},g(\sigma))C^h(t)\ket{GS}&= C^h(t) \delta(g(s.p(\sigma)-s.p(t))hg(s.p(\sigma)-s.p(t))^{-1}qkr_Cq^{-1}, g(\sigma))\ket{GS},
	\end{align*}
	so that
	\begin{align}
		K_{\sigma}^{R,C}& C^h(t) \ket{GS} \notag \\
		&= \frac{|R|}{|N'_C|} \sum_{d \in N'_C} \overline{\chi}_R(d) \sum_{q \in Q_C} \sum_{k \in \partial(E)} C^{qdq^{-1}}(\sigma) C^h(t) M^{f(r_C,d)^{-1}}(p)\notag \\ & \hspace{0.5cm}\delta(g(s.p(\sigma)-s.p(t))hg(s.p(\sigma)-s.p(t))^{-1}qkr_Cq^{-1}, g(\sigma)) \ket{GS}. \label{Equation_2D_measurement_magnetic_1.5}
	\end{align}
	
	Now because $\delta(g(s.p(\sigma)-s.p(t))hg(s.p(\sigma)-s.p(t))^{-1}qkr_Cq^{-1}, g(\sigma))$ acts on the ground state and $g(\sigma)$ is a contractible closed path, we must have $g(\sigma) =\partial(e_m)^{-1}$ where $e_m \in E$ is the surface element of the region enclosed by the ribbon, due to fake-flatness. This implies that the Kronecker delta only gives a non-zero result if 
	\begin{align}
		g(s.p(\sigma)-s.p(t))&hg(s.p(\sigma)-s.p(t))^{-1}qkr_Cq^{-1} = \partial(e_m)^{-1} \notag\\
		& \implies hg(s.p(\sigma)-s.p(t))^{-1}qkr_Cq^{-1} = g(s.p(\sigma)-s.p(t))^{-1}\partial(e_m)^{-1} \notag\\
		& \implies h= g(s.p(\sigma)-s.p(t))^{-1}qr_C^{-1}k^{-1}q^{-1}\partial(e_m)^{-1}g(s.p(\sigma)-s.p(t)) \notag\\
		& \implies h= g(s.p(\sigma)-s.p(t))^{-1}qk^{-1}\partial(e_m)^{-1}r_C^{-1}q^{-1}g(s.p(\sigma)-s.p(t)). \label{Equation_2D_measurement_magnetic_2}
	\end{align}
	
	This implies that $h$ has the form $h= xy r_C^{-1} x^{-1}$, where $x=g(s.p(\sigma)-s.p(t))^{-1}q$ is an element of $G$ and $y=k^{-1}\partial(e_m)^{-1}$ is an element of $\partial(E)$. Therefore, $h$ must be in the same equivalence class as $r_C^{-1}$ for the projector to give a non-zero result. Denoting that class by $C^{-1}$, we have that $K_{\sigma}^{R,C} C^h(t) \ket{GS}$ is zero unless $h$ is in $C^{-1}$ ( if we had measured the excitation at the end of the ribbon, the result would have been zero unless $h$ was in $C$). We can then use Equation \ref{Equation_2D_measurement_magnetic_2} to write Equation \ref{Equation_2D_measurement_magnetic_1.5} as
	\begin{align}
		K_{\sigma}^{R,C} C^h(t) \ket{GS} =& \delta(h \in C^{-1}) \frac{|R|}{|N'_C|} \sum_{d \in N'_C} \overline{\chi}_R(d) \sum_{q \in Q_C} \sum_{k \in \partial(E)} C^{qdq^{-1}}(\sigma) C^h(t) \notag\\
		& \delta(h, g(s.p(\sigma)-s.p(t))^{-1}qk^{-1}\partial(e_m)^{-1}r_C^{-1}q^{-1}g(s.p(\sigma)-s.p(t))) M^{f(r_C,d)^{-1}}(p) \ket{GS}. \label{Equation_2D_measurement_magnetic_3}
	\end{align}
	
\begin{figure}[h]
		\begin{center}
			\begin{overpic}[width=0.75\linewidth]{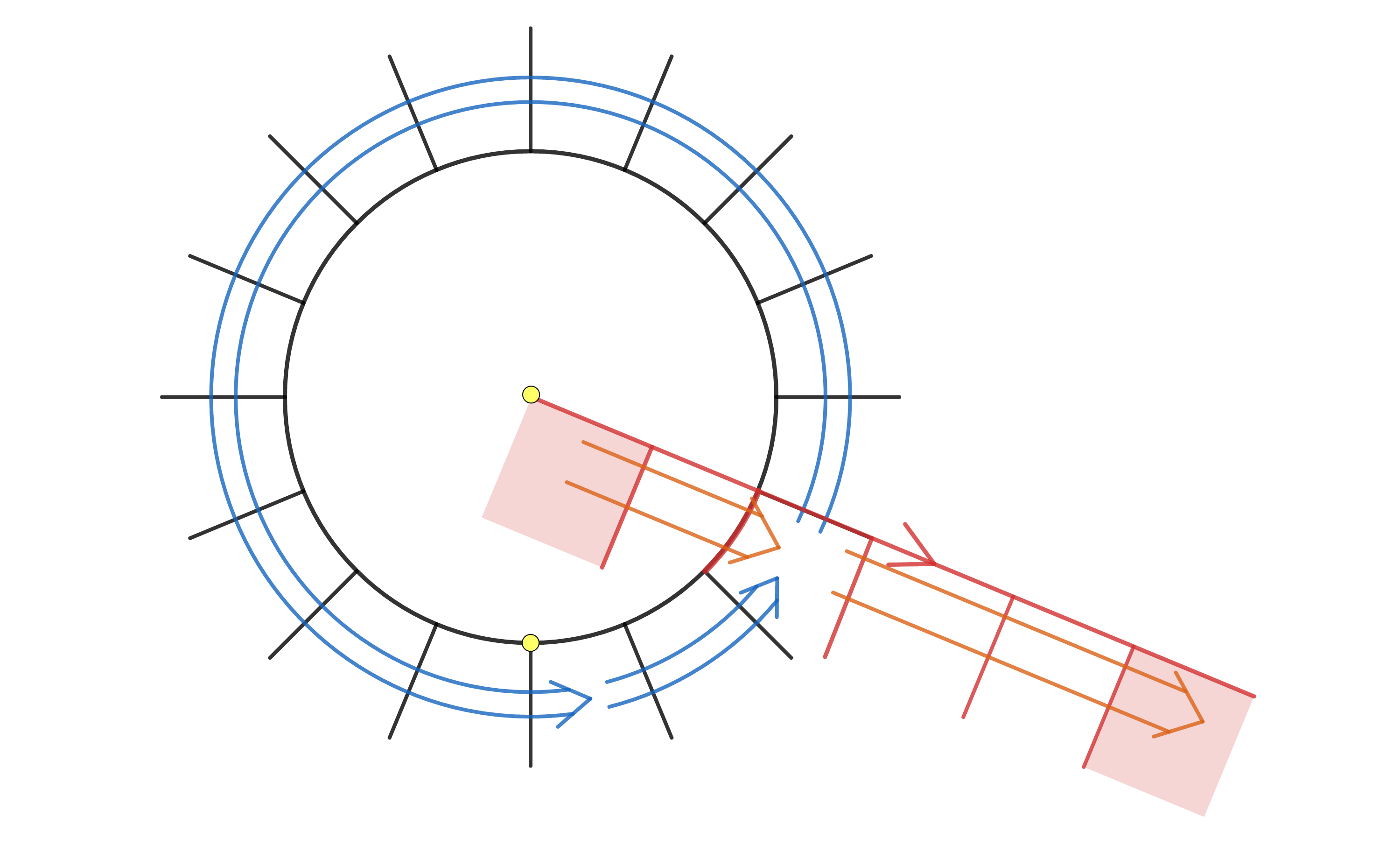}
				\put(50,12){$\sigma_1$}
				\put(62,36){$\sigma_2$}
				\put(45,22){$t_1$}
				\put(72,11){$t_2$}
				\put(56,21){$A$}
				\put(82,7){$B$}
				
			\end{overpic}
			
			\caption{In order to commute the magnetic ribbon operator in the measurement operator past the magnetic ribbon operator that produces the excitation we are measuring, we must split the ribbons $t$ and $\sigma$ into two parts, corresponding to the parts of each ribbon before and after their intersection at plaquette A (with the parts before and after intersection indicated by subscripts 1 and 2 respectively). }
			\label{2D_charge_measurement_magnetic_split_ribbons}
			
		\end{center}
	\end{figure}

	Then we must commute $C^h(t)$ past the closed ribbon operator $C^{qdq^{-1}}(\sigma)$. We found this commutation relation in Section \ref{Section_2D_braiding_two_magnetic_appendix}, when we considered braiding two magnetic excitations. Using Equation \ref{Magnetic_magnetic_commute_2D_6} from that section, we have
	\begin{align}
		C^{qdq^{-1}}(\sigma) C^h(t) &=C^h(t_1) C^{ [hg(s.p(t)-s.p(\sigma))qdq^{-1} g(s.p(t)-s.p(\sigma))^{-1} ] h [h g(s.p(t)-s.p(\sigma))qdq^{-1}g(s.p(t)-s.p(\sigma))^{-1}]^{-1}}(t_2) \notag \\
		& \hspace{0.5cm} C^{qdq^{-1}}(\sigma_1) C^{h_{[\sigma-t]}^{\phantom{-1}}qdq^{-1} h_{[\sigma-t]}^{-1}}(\sigma_2), \label{Equation_2D_measurement_magnetic_commutation}
	\end{align}
	where $\sigma_1$ and $t_1$ are the parts of $\sigma$ and $t$ before their intersection, while $\sigma_2$ and $t_2$ are the parts after the intersection, as shown in Figure \ref{2D_charge_measurement_magnetic_split_ribbons}. In this expression we have also defined $h_{[\sigma-t]}= g(s.p(\sigma)-s.p(t))hg(s.p(\sigma)-s.p(t))^{-1}$. We note that this is an operator, because it depends on the path element $g(s.p(\sigma)-s.p(t))$, so its value will depend on the state on which it acts. Substituting Equation \ref{Equation_2D_measurement_magnetic_commutation} into Equation \ref{Equation_2D_measurement_magnetic_3}, the total state is
	\begin{align}
		K_{\sigma}^{R,C} C^h(t) \ket{GS} &=\delta(h \in C^{-1}) \frac{|R|}{|N'_C|} \sum_{d \in N'_C} \overline{\chi}_R(d) \sum_{q \in Q_C} \sum_{k \in \partial(E)} C^h(t_1) \notag \\
		& \hspace{0.5cm} C^{ [hg(s.p(t)-s.p(\sigma))qdq^{-1} g(s.p(t)-s.p(\sigma))^{-1} ] h [h g(s.p(t)-s.p(\sigma))qdq^{-1}g(s.p(t)-s.p(\sigma))^{-1}]^{-1}}(t_2) C^{qdq^{-1}}(\sigma_1) \notag \\ & \hspace{0.5cm} C^{h_{[\sigma-t]}^{\phantom{-1}}qdq^{-1} h_{[\sigma-t]}^{-1}}(\sigma_2) \delta(h, g(s.p(\sigma)-s.p(t))^{-1}qk^{-1}\partial(e_m)^{-1}r_C^{-1}q^{-1}g(s.p(\sigma)-s.p(t))) \notag \\ &\hspace{0.5cm} M^{f(r_C,d)^{-1}}(p) \ket{GS}. \label{Equation_2D_measurement_magnetic_4}
	\end{align}

	Now that the ribbon operators from the measurement operator are to the right of those from the open magnetic ribbon operator, we can begin to simplify this expression. We start by considering the label $h_{[\sigma-t]}^{\phantom{-1}}qdq^{-1} h_{[\sigma-t]}^{-1}$ of the right-most ribbon operator in Equation \ref{Equation_2D_measurement_magnetic_4}. This operator is adjacent to the Kronecker delta in Equation \ref{Equation_2D_measurement_magnetic_4}. That means that we can use the condition enforced by the delta to write $h=g(s.p(\sigma)-s.p(t))^{-1}qk^{-1}\partial(e_m)^{-1}r_C^{-1}q^{-1}g(s.p(\sigma)-s.p(t))$ and so
	$$h_{[\sigma-t]}^{\phantom{-1}}=qk^{-1}\partial(e_m)^{-1}r_C^{-1}q^{-1}.$$
	
	This means that
	\begin{align*}
		h_{[\sigma-t]}^{\phantom{-1}}qdq^{-1} h_{[\sigma-t]}^{-1}&= (qk^{-1}\partial(e_m)^{-1}r_C^{-1}q^{-1}) qdq^{-1} (qk^{-1}\partial(e_m)^{-1}r_C^{-1}q^{-1})^{-1}\\
		&=(qk^{-1}\partial(e_m)^{-1}r_C^{-1}q^{-1}) qdq^{-1} (qr_C \partial(e_m)kq^{-1}) \\
		&= qk^{-1}\partial(e_m)^{-1}r_C^{-1} d r_C \partial(e_m)kq^{-1}.
	\end{align*}
	
	Now $\partial(e_m)$ and $k$ are both in $\partial(E)$ and so are both in the centre of $G$. This means that we can commute $k$ and $k^{-1}$ together and cancel them, then do the same with $\partial(e_m)$ and its inverse, to obtain
	\begin{align}
		h_{[\sigma-t]}^{\phantom{-1}}qdq^{-1} h_{[\sigma-t]}^{-1}&= qr_C^{-1} d r_C q^{-1}. \label{Equation_2D_measurement_magnetic_h_sigma_t_2}
	\end{align}
	
	Now $d$ is in $N'_C$ and so $d$ and $r_C$ commute up to an element of $\partial(E)$. We have $[r_C,d]=\partial(f(r_C,d))$. This tells us that
	\begin{align*}
		r_Cdr_C^{-1}d^{-1} & = \partial(f(r_C,d))\\
		& \implies d r_C d^{-1} r_C^{-1} = \partial(f(r_C,d))^{-1}\\
		& \implies r_C^{-1}d r_C d^{-1} = r_C^{-1} \partial(f(r_C,d))^{-1} r_C = \partial(f(r_C,d))^{-1}\\
		& \implies r_C^{-1}d r_C = \partial(f(r_C,d))^{-1}d.
	\end{align*}
	
	Substituting this into Equation \ref{Equation_2D_measurement_magnetic_h_sigma_t_2}, we have
	\begin{align}
		h_{[\sigma-t]}^{\phantom{-1}}qdq^{-1} h_{[\sigma-t]}^{-1}&= \partial(f(r_C,d))^{-1}q d q^{-1}. \label{Equation_2D_measurement_magnetic_h_sigma_t_3}
	\end{align}
	Substituting this into Equation \ref{Equation_2D_measurement_magnetic_4}, we see that the total state can be written as
	\begin{align}
		K_{\sigma}^{R,C} C^h(t) \ket{GS} &=\delta(h \in C^{-1}) \frac{|R|}{|N'_C|} \sum_{d \in N'_C} \overline{\chi}_R(d) \sum_{q \in Q_C} \sum_{k \in \partial(E)} C^h(t_1) \notag \\
		& \hspace{0.5cm} C^{ [hg(s.p(t)-s.p(\sigma))qdq^{-1} g(s.p(t)-s.p(\sigma))^{-1} ] h [h g(s.p(t)-s.p(\sigma))qdq^{-1}g(s.p(t)-s.p(\sigma))^{-1}]^{-1}}(t_2) C^{qdq^{-1}}(\sigma_1)\notag \\ & \hspace{0.5cm} C^{\partial(f(r_C,d))^{-1}q d q^{-1}}(\sigma_2) \delta(h, g(s.p(\sigma)-s.p(t))^{-1}qk^{-1}\partial(e_m)^{-1}r_C^{-1}q^{-1}g(s.p(\sigma)-s.p(t))) \notag \\ & \hspace{0.5cm} M^{f(r_C,d)^{-1}}(p) \ket{GS}. \label{Equation_2D_measurement_magnetic_5}
	\end{align}
	
	Next we look at the label of the ribbon operator on $t_2$, 
	$$[hg(s.p(t)-s.p(\sigma))qdq^{-1} g(s.p(t)-s.p(\sigma))^{-1} ] h [h g(s.p(t)-s.p(\sigma))qdq^{-1}g(s.p(t)-s.p(\sigma))^{-1}]^{-1}.$$
	The ribbon operator on $t_2$ is not adjacent to the Kronecker delta, but is instead separated from it by the ribbon operators applied on $\sigma_1$ and $\sigma_2$ in Equation \ref{Equation_2D_measurement_magnetic_5}. However looking at Figure \ref{2D_charge_measurement_magnetic_split_ribbons}, we see that the path element $g(s.p(t)-s.p(\sigma))$ will not be affected by the ribbon operators on $\sigma$, because both $s.p(t)$ and $s.p(\sigma)$ are inside the ribbon $\sigma$. We can therefore use the Kronecker delta to fix the label of the ribbon operator on $t_2$. In order to do this, we first rewrite the label of the ribbon operator,
	$$[hg(s.p(t)-s.p(\sigma))qdq^{-1} g(s.p(t)-s.p(\sigma))^{-1} ] h [h g(s.p(t)-s.p(\sigma))qdq^{-1}g(s.p(t)-s.p(\sigma))^{-1}]^{-1},$$
	as
	\begin{align*}
		g(s.p(t)-s.p(\sigma))&h_{[\sigma-t]}^{\phantom{-1}}qdq^{-1} h_{[\sigma-t]}^{\phantom{-1}} qd^{-1}q^{-1} h_{[\sigma-t]}^{-1} g(s.p(t)-s.p(\sigma))^{-1}\\
		&= g(s.p(t)-s.p(\sigma))h_{[\sigma-t]}^{\phantom{-1}}qdq^{-1} (h_{[\sigma-t]}^{\phantom{-1}} qd^{-1}q^{-1} h_{[\sigma-t]}^{-1}) g(s.p(t)-s.p(\sigma))^{-1}.
	\end{align*}
	
	Then using Equation \ref{Equation_2D_measurement_magnetic_h_sigma_t_3}, we can replace $(h_{[\sigma-t]}^{\phantom{-1}}qd^{-1}q^{-1} h_{[\sigma-t]}^{\phantom{-1}})$ with $(\partial(f(r_C,d))^{-1}q d q^{-1})^{-1}$ to write this as
	\begin{align*}
		g(s.p(t)-s.p(\sigma))&h_{[\sigma-t]}^{\phantom{-1}}qdq^{-1} (\partial(f(r_C,d))^{-1}q d q^{-1})^{-1} g(s.p(t)-s.p(\sigma))^{-1}\\
		&=g(s.p(t)-s.p(\sigma))h_{[\sigma-t]}^{\phantom{-1}}\partial(f(r_C,d)) g(s.p(t)-s.p(\sigma))^{-1}\\
		&=g(s.p(t)-s.p(\sigma))g(s.p(\sigma)-s.p(t))h g(s.p(\sigma)-s.p(t))^{-1}\partial(f(r_C,d)) g(s.p(t)-s.p(\sigma))^{-1}\\
		&= \partial(f(r_C,d))h,
	\end{align*}
	where we used the fact that $\partial(f(r_C,d))$ is in the centre of $G$ in the last step. Therefore, we have
	\begin{align*}
		C^{qdq^{-1}}(\sigma) C^h(t)&=C^h(t_1) C^{ \partial(f(r_C,d))h}(t_2) C^{qdq^{-1}}(\sigma_1) C^{\partial(f(r_C,d))^{-1}qdq^{-1} }(\sigma_2). 
	\end{align*}
	
	We can then write $C^{ \partial(f(r_C,d))h}(t_2)= C^{\partial(f(r_C,d))}(t_2)C^h(t_2)$ (and similarly for $C^{\partial(f(r_C,d))^{-1}qdq^{-1} }(\sigma_2)$). We can then join $C^h(t_1)$ to $C^h(t_2)$ to obtain $C^h(t)$ (and similarly for $C^{qdq^{-1} }(\sigma)$). This gives us
	\begin{align}
		C^{qdq^{-1}}(\sigma) C^h(t)&=C^h(t) C^{\partial(f(r_C,d))}(t_2) C^{qdq^{-1}}(\sigma) C^{\partial(f(r_C,d))^{-1}}(\sigma_2), \label{Equation_2D_measurement_magnetic_commute_3}
	\end{align}
	where we note that $C^{\partial(f(r_C,d))}(t_2)$ and $C^{\partial(f(r_C,d))^{-1}}(\sigma_2)$ commute with the other magnetic ribbon operators because they have label in the centre of $G$. Substituting Equation \ref{Equation_2D_measurement_magnetic_commute_3} into Equation \ref{Equation_2D_measurement_magnetic_5}, we see that the total state resulting from our measurement is then
	\begin{align*}
		K_{\sigma}^{R,C} C^h(t) \ket{GS} &=\delta(h \in C^{-1}) \frac{|R|}{|N'_C|} \sum_{d \in N'_C} \overline{\chi}_R(d) \sum_{q \in Q_C} \sum_{k \in \partial(E)} C^h(t) C^{\partial(f(r_C,d))}(t_2) C^{qdq^{-1}}(\sigma) C^{\partial(f(r_C,d))^{-1}}(\sigma_2) \notag \\
		& \hspace{0.5cm} \delta(h, g(s.p(\sigma)-s.p(t))^{-1}qk^{-1}\partial(e_m)^{-1}r_C^{-1}q^{-1}g(s.p(\sigma)-s.p(t))) M^{f(r_C,d)^{-1}}(p) \ket{GS}. 
	\end{align*}
	
	We then commute $C^{qdq^{-1}}(\sigma)$ to the right, so it acts on the ground state directly. As we showed in Section \ref{Section_Topological_Magnetic_Ribbons}, a magnetic ribbon operator acting on a closed ribbon (like $\sigma$) acts trivially on a ground state. Therefore, we can remove this ribbon operator to write
	\begin{align}
		K_{\sigma}^{R,C} C^h(t) \ket{GS} &=\delta(h \in C^{-1}) \frac{|R|}{|N'_C|} \sum_{d \in N'_C} \overline{\chi}_R(d) \sum_{q \in Q_C} \sum_{k \in \partial(E)} C^h(t) C^{\partial(f(r_C,d))}(t_2) C^{\partial(f(r_C,d))^{-1}}(\sigma_2) \notag \\
		& \hspace{0.5cm} \delta(h, g(s.p(\sigma)-s.p(t))^{-1}qk^{-1}\partial(e_m)^{-1}r_C^{-1}q^{-1}g(s.p(\sigma)-s.p(t)))M^{f(r_C,d)^{-1}}(p)\ket{GS}. \label{Equation_2D_measurement_magnetic_7}
	\end{align}
	
	Precisely one pair $(q,k)$ will satisfy the Kronecker delta condition for any given $e_m$ and $g(s.p(\sigma)-s.p(t))$, as we showed earlier (in the discussion before Equation \ref{Equation_2D_charge_projectors_sum_2}), and because $q$ and $k$ only appear in this Kronecker delta, we can remove the sums over $q$ and $k$ together with the delta, to obtain
	\begin{align}
		K_{\sigma}^{R,C} C^h(t) \ket{GS} 
		&=\delta(h \in C^{-1}) \frac{|R|}{|N'_C|} \sum_{d \in N'_C} \overline{\chi}_R(d) C^h(t) C^{\partial(f(r_C,d))}(t_2) C^{\partial(f(r_C,d))^{-1}}(\sigma_2) M^{f(r_C,d)^{-1}}(p) \ket{GS}. \label{Equation_2D_measurement_magnetic_8}
	\end{align}
	
	Next, we examine the ribbon operators on $t_2$ and $\sigma_2$, where the ribbons $t_2$ and $\sigma_2$ are illustrated in Figure \ref{2D_charge_measurement_magnetic_split_ribbons}. We note that because the labels of these magnetic ribbon operators are in $\partial(E)$, the ribbon operators are condensed. This means that, as shown in Section \ref{Section_Condensation_Magnetic_2D}, they are equivalent to a series of edge transforms and a single plaquette multiplication operator at each end of the ribbon. However the edge transforms act trivially on the ground state, so when acting on the ground state these two ribbon operators are equivalent to a series of single plaquette multiplication operators. First consider the ribbon operator on $t_2$. The plaquette at the start of the ribbon is the intersection between $t$ and $\sigma$, which we will denote by $A$, while the plaquette at the end of the ribbon is the plaquette at the end of $t$ (one of the two excited by the ribbon operator $C^h(t)$), which we will denote by $B$. The ribbon operator $C^{\partial(f(r_C,d))}(t_2)$ is equivalent to a product of single plaquette multiplication operators on these two plaquettes. As discussed in Section \ref{Section_Condensation_Magnetic_2D}, these plaquette multiplication operators have label $f(r_C,d)$ or $f(r_C,d)^{-1}$, depending on the orientation of the plaquettes. We can deduce whether the label should be $f(r_C,d)$ or $f(r_C,d)^{-1}$ in each case by comparing to the case described in Section \ref{Section_Condensation_Magnetic_2D}. Comparing Figure \ref{2D_charge_measurement_magnetic_split_ribbons} to Figure \ref{condensed_magnetic_ribbon_end_plaquettes} from Section \ref{Section_Condensation_Magnetic_2D} we see that the dual path is on the opposite side of the direct path in Figure \ref{2D_charge_measurement_magnetic_split_ribbons}. This means that, as discussed in Section \ref{Section_Condensation_Magnetic_2D}, we need the product $M^{f(r_C,d)}(A) M^{f(r_C,d)^{-1}}(B)$ if the two plaquettes are clockwise with respect the normal of the lattice. Then considering $C^{\partial(f(r_C,d))^{-1}}(\sigma_2)$, the start of $\sigma_2$ is again the intersection plaquette $A$, while the end is the plaquette $p$ at the start and end of $\sigma$. Therefore, this ribbon operator acting on the ground state is equivalent to $M^{f(r_C,d)^{-1}}(A) M^{f(r_C,d)}(p)$. We note that the single plaquette multiplication operators acting on the intersection plaquette $A$ from the two ribbon operators cancel, because the ribbon operator on $t_2$ contributes $M^{f(r_C,d)}(A)$, while the ribbon operator on $\sigma_2$ contributes the inverse. Then we can write our state as
	\begin{align*}
		K_{\sigma}^{R,C} C^h(t) \ket{GS} 
		&=\delta(h \in C^{-1}) \frac{|R|}{|N'_C|} \sum_{d \in N'_C} \overline{\chi}_R(d) C^h(t) M^{f(r_C,d)^{-1}}(B) M^{f(r_C,d)}(p) M^{f(r_C,d)^{-1}}(p)\ket{GS}. 
	\end{align*}

	The single plaquette multiplication operator $M^{f(r_C,d)^{-1}}(p)$ already present in the projector acts on the plaquette $p$ at the start of the dual path of $\sigma$, so we can cancel this with the operator $M^{f(r_C,d)}(p)$. This leaves us with 
	\begin{align}
		K_{\sigma}^{R,C} C^h(t) \ket{GS} &=\delta(h \in C^{-1}) \frac{|R|}{|N'_C|} \sum_{d \in N'_C} \overline{\chi}_R(d) C^h(t) M^{f(r_C,d)^{-1}}(B) \ket{GS}. \label{Equation_2D_measurement_magnetic_10}
	\end{align}
	
	In order to see which charges give non-zero values, we now consider calculating the expectation value of the charge in the state $C^h(t) \ket{GS}$:
	\begin{align}
		\braket{K_{\sigma}^{R,C}}&= \bra{GS}C^h(t)^\dagger K_{\sigma}^{R,C} C^h(t) \ket{GS}. \label{Equation_2D_measurement_magnetic_10.5}
	\end{align}
	$C^h(t)$ is unitary, so that $C^h(t)^\dagger= (C^h(t))^{-1}=C^{h^{-1}}(t)$. To see this, consider the matrix element $$\bra{ \set{g'_i}, \set{e'_p}} C^h(t) \ket{\set{g_i}, \set{e_p}},$$ where $\ket{\set{g_i}, \set{e_p}}$ is a basis state where each edge $i$ is labelled by a group element $g_i \in G$ and each plaquette $p$ by a group element $e_p \in E$ (and similar for $\bra{ \set{g'_i}, \set{e'_p}}$). The action of $C^h(t)$ on an edge $j$ cut by the dual path is
	$$C^h(t) :g_j = \begin{cases}g(s.p-v_j)^{-1}hg(s.p-v_j)g_j & \text{ if $j$ points away from the direct path} \\ g_j g(s.p-v_j)^{-1}h^{-1}g(s.p-v_j) & \text{ if $j$ points towards the direct path.} \end{cases}$$
	
	On the other hand, if an edge is not cut by the dual path, it is left invariant. We therefore split the edges of the lattice into two groups: those cut by the dual path and those not. That is, we write
	$$\ket{\set{g_i}, \set{e_p}}= \ket{\set{g_i}_{\text{uncut}}, \set{g_j}_{\text{cut}}, \set{e_p}}.$$
	
	Then the matrix element $\bra{ \set{g'_i}, \set{e'_p}} C^h(t) \ket{\set{g_i}, \set{e_p}}$ is either one or zero. It is one if all of the plaquette labels agree ($\set{e'_p}= \set{e_p}$), the edges not cut by the dual path agree ($i$ not cut by dual path implies that $g_i'=g_i$) and each edge $j$ cut by the dual path satisfies $g_j' =C^h(t) :g_j$. It is zero in all other cases. That is
	\begin{align*}
		\bra{ \set{g'_i}, \set{e'_p}} C^h(t) \ket{\set{g_i}, \set{e_p}} &= \bra{ \set{g'_i}_{\text{uncut}}, \set{g'_j}_{\text{cut}}, \set{e'_p}} C^h(t) \ket{\set{g_i}_{\text{uncut}}, \set{g_j}_{\text{cut}}, \set{e_p}}\\
		&= \delta(\set{e_p},\set{e'_p}) \delta(\set{g_i}_{\text{uncut}}, \set{g'_i}_{\text{uncut}}) \delta(\set{C^h(t):g_j}_{\text{cut}}, \set{g'_j}_{\text{cut}}). 
	\end{align*}
	
	On the other hand, consider $$\bra{\set{g_i}, \set{e_p}} C^{h^{-1}}(t) \ket{ \set{g'_i}, \set{e'_p}}.$$ This is given by
	\begin{align*}
		\bra{\set{g_i}, \set{e_p}} C^{h^{-1}}(t) \ket{ \set{g'_i}, \set{e'_p}} &= \bra{ \set{g_i}_{\text{uncut}}, \set{g_j}_{\text{cut}}, \set{e_p}} C^{h^{-1}}(t) \ket{\set{g'_i}_{\text{uncut}}, \set{g'_j}_{\text{cut}}, \set{e'_p}}\\
		&= \delta(\set{e'_p},\set{e_p}) \delta(\set{g'_i}_{\text{uncut}}, \set{g_i}_{\text{uncut}}) \delta(\set{C^{h^{-1}}(t):g'_j}_{\text{cut}}, \set{g_j}_{\text{cut}}),
	\end{align*}
	using the same logic as before. However because $C^{h^{-1}}(t)$ is the inverse of $C^h(t)$, we have 
	$$\delta(\set{C^{h^{-1}}(t):g'_j}_{\text{cut}}, \set{g_j}_{\text{cut}}) = \delta(\set{C^h(t):g_j}_{\text{cut}}, \set{g'_j}_{\text{cut}}),$$
	and so 
	$$\bra{\set{g_i}, \set{e_p}} C^{h^{-1}}(t) \ket{ \set{g'_i}, \set{e'_p}} = \bra{ \set{g'_i}, \set{e'_p}} C^h(t) \ket{\set{g_i}, \set{e_p}}.$$
	
	Furthermore, this matrix element is real, so
	$$(\bra{\set{g_i}, \set{e_p}} C^{h^{-1}}(t) \ket{ \set{g'_i}, \set{e'_p}})^*= \bra{ \set{g'_i}, \set{e'_p}} C^h(t) \ket{\set{g_i}, \set{e_p}}.$$
	This is true for all basis states and so $C^{h^{-1}}(t)$ is the Hermitian conjugate of $C^h(t)$. Substituting this into Equation \ref{Equation_2D_measurement_magnetic_10.5}, we see that the expectation value for the topological charge measurement is given by
	\begin{align}
		\braket{K_{\sigma}^{R,C}}&= \bra{GS}C^{h^{-1}}(t) \delta(h \in C^{-1}) \frac{|R|}{|N'_C|} \sum_{d \in N'_C} \overline{\chi}_R(d) C^h(t) M^{f(r_C,d)^{-1}}(B) \ket{GS} \notag\\
		&= \delta(h \in C^{-1}) \frac{|R|}{|N'_C|} \sum_{d \in N'_C} \overline{\chi}_R(d) \bra{GS}C^{h^{-1}}(t) C^h(t) M^{f(r_C,d)^{-1}}(B) \ket{GS} \notag\\
		&=\delta(h \in C^{-1}) \frac{|R|}{|N'_C|} \sum_{d \in N'_C} \overline{\chi}_R(d) \bra{GS} M^{f(r_C,d)^{-1}}(B) \ket{GS}. \label{Equation_2D_measurement_magnetic_11}
	\end{align}

	Now $\bra{GS} M^{f(r_C,d)^{-1}}(B) \ket{GS}$ is zero unless $f(r_C,d)$ is in the kernel of $\partial$, because otherwise $M^{f(r_C,d)^{-1}}(B)$ excites the plaquette $B$ and so $M^{f(r_C,d)^{-1}}(B) \ket{GS}$ is orthogonal to the ground state. On the other hand, $f(r_C,d)$ is a representative element satisfying $\partial(f(r_C,d))=[r_C,d]$ and we choose the representative for the kernel to be the identity element $1_E$. So if $f(r_C,d)$ is in the kernel it is the identity element and $M^{f(r_C,d)^{-1}}(B)$ is the identity operator. From $\partial(f(r_C,d))=[r_C,d]$, we can tell that this occurs when $r_C$ and $d$ commute. Denoting the set of elements of $G$ that commute with $r_C$ by $N_C$ (as opposed to $N'_C$), the expectation value in Equation \ref{Equation_2D_measurement_magnetic_11} can be written as
	\begin{align}
		\braket{K_{\sigma}^{R,C}}&= \delta(h \in C^{-1}) \frac{|R|}{|N'_C|} \sum_{d \in N'_C} \delta(d \in N_C) \overline{\chi}_R(d) \braket{GS|GS} \notag \\
		&=\delta(h \in C^{-1}) \frac{|R|}{|N'_C|} \sum_{d \in N_C} \overline{\chi}_R(d).\label{Equation_2D_measurement_magnetic_12}
	\end{align}
	
	We note that $R$ is an irrep of $N'_C$, the subgroup of $G$ consisting of elements which commute with $r_C$ up to elements in $\partial(E)$. $N_C$, the subgroup of $G$ consisting of elements which commute completely with $r_C$, is a normal subgroup of $N'_C$. To see this, consider an element $y \in N_C$ and an element $x \in N'_C$. Then $N_C$ is a normal subgroup of $N'_C$ if $xyx^{-1}$ is also in $N_C$ for all such $x$ and $y$, i.e., if $xyx^{-1}$ commutes with $r_C$. We have
	\begin{align*}
		xyx^{-1}r_C =xyr_C x^{-1} \partial(e),
	\end{align*}
	for some $e \in E$, from the fact that $y$ is in $N'_C$. Then
	\begin{align*}
		xyx^{-1}r_C &=xyr_C x^{-1} \partial(e)\\
		&= xr_Cyx^{-1} \partial(e)\\
		&=r_C xyx^{-1} \partial(e)^{-1} \partial(e)\\
		&= r_C xyx^{-1},
	\end{align*}
	where we used the fact that $y$ is in $N_C$ to commute it past $r_C$, and the fact that $xr_C= r_C x \partial(e)^{-1}$, with $\partial(e)$ in the centre of $G$. Therefore $xyx^{-1}$ commutes with $r_C$ and so $N_C$ is a normal subgroup of $N'_C$. This means that, from Clifford's theorem \cite{Clifford1937}, the irrep $R$ branches to conjugacy related irreps of $N_C$. If this irrep is non-trivial then the sum in Equation \ref{Equation_2D_measurement_magnetic_12} is zero from the Grand Orthogonality Theorem. The sum is non-zero only if $R$ has a trivial restriction to the subgroup $N_C$.

	Now, recall from Section \ref{Section_2D_Magnetic} of the main text that the magnetic ribbon operator may excite the start-point vertex of the ribbon, if we take an appropriate sum of terms $C^h(t)$ with labels $h$ in a given conjugacy class. The measurement ribbon $\sigma$ considered in Figure \ref{2D_charge_measurement_magnetic} encloses the start-point as well as the starting plaquette. It is therefore interesting to see whether the start-point vertex excitation changes which topological charge the magnetic excitation can carry. To do this, we replace the magnetic ribbon operator $C^h(t)$ with a sum $\sum_{x \in [h]} \alpha_x C^x(t)$, where $[h]$ is the conjugacy class containing $h$. This will cause a vertex excitation if $\sum_{x \in [h]} \alpha_x=0$ and will not if $\alpha_x$ is the same for all $x$. Any other case will result in the state not being an eigenstate of the vertex energy term. We denote the expectation value of the topological charge measurement operator in the state $\sum_{x \in [h]} \alpha_x C^x(t)\ket{GS}$ by $\braket{K_{\sigma}^{R,C}}_2$. Then we have
	\begin{align}
		\braket{K_{\sigma}^{R,C}}_2&= \mathcal{N} \bra{GS}(\sum_{y \in [h]} \alpha_{y} C^{y}(t))^\dagger K_{\sigma}^{R,C} \sum_{x \in [h]} \alpha_x C^x(t) \ket{GS},
	\end{align}
	where $$\mathcal{N}= \frac{1}{\bra{GS}(\sum_{y \in [h]} \alpha_{y} C^{y}(t))^\dagger \sum_{x \in [h]} \alpha_x C^x(t) \ket{GS}}.$$
	
	Then 
	\begin{align}
		\braket{K_{\sigma}^{R,C}}_2
		&= \mathcal{N} \bra{GS}\sum_{y \in [h]} \alpha_y^* C^{y^{-1}}(t) \sum_{x \in [h]} \alpha_x \delta(x \in C^{-1}) \frac{|R|}{|N'_C|} \sum_{d \in N'_C} \overline{\chi}_R(d) C^x(t) M^{f(r_C,d)^{-1}}(B)\ket{GS}.
	\end{align}
	
	The class $C^{-1}$ includes the conjugacy class of each of its elements, so $x$ is in $C^{-1}$ if and only if $h$ is in $C^{-1}$. Therefore, the expectation value is
	\begin{align}
		\braket{K_{\sigma}^{R,C}}_2 &= \mathcal{N}\sum_{y \in [h]} \sum_{x \in [h]} \alpha_y^* \alpha_x \delta(h \in C^{-1}) \frac{|R|}{|N'_C|} \sum_{d \in N'_C} \overline{\chi}_R(d) \bra{GS} C^{y^{-1}}(t) C^x(t) M^{f(r_C,d)^{-1}}(B)\ket{GS} \notag\\
		&=\mathcal{N} \sum_{y \in [h]} \sum_{x \in [h]} \alpha_y^* \alpha_x \delta(h \in C^{-1}) \frac{|R|}{|N'_C|} \sum_{d \in N'_C} \overline{\chi}_R(d) \bra{GS} C^{y^{-1}x}(t) M^{f(r_C,d)^{-1}}(B)\ket{GS}.
	\end{align}
	
	Now we note that $C^{y^{-1}x}(t)$ excites the plaquette at the start of $t$ if $y \neq x$, which would make the overlap with the ground state zero, so we only get contributions from the $y=x$ cases. Then when $y=x$, $M^{f(r_C,d)^{-1}}(B)$ excites the plaquette at the end of $t$ unless $f(r_C,d)$ is $1_E$, so this condition is required for a non-zero contribution. This means that the expectation value is
	\begin{align}
		\braket{K_{\sigma}^{R,C}}_2 &= \mathcal{N} \sum_{x \in [h]} |\alpha_x|^2 \delta(h \in C^{-1}) \frac{|R|}{|N'_C|} \sum_{d \in N_C} \overline{\chi}_R(d) \notag\\
		&= \mathcal{N}\sum_{x \in [h]} |\alpha_x|^2 \braket{K_{\sigma}^{R,C}},
	\end{align}
	where $\braket{K_{\sigma}^{R,C}}$ is the expectation value of the charge measurement operator in the state $C^h(t)\ket{GS}$, given in Equation \ref{Equation_2D_measurement_magnetic_12}. We have
	\begin{align}
		\frac{1}{\mathcal{N}} &= \bra{GS}(\sum_{y \in [h]} \alpha_{y} C^{y}(t))^\dagger \sum_{x \in [h]} \alpha_x C^x(t) \ket{GS} \notag \\
		&=\bra{GS}\sum_{y \in [h]} \alpha_{y}^* C^{y^{-1}}(t)\sum_{x \in [h]} \alpha_x C^x(t) \ket{GS} \notag \\
		&= \sum_{x \in [h]} |\alpha_x|^2,
	\end{align}
	where we again used the fact that the overlap is zero unless $y=x$. We therefore see that the expectation value $\braket{K_{\sigma}^{R,C}}_2$ is 
	\begin{align}
		\braket{K_{\sigma}^{R,C}}_2 &= \frac{1}{ \sum_{x \in [h]} |\alpha_x|^2}\sum_{x \in [h]} |\alpha_x|^2 \braket{K_{\sigma}^{R,C}} \notag\\
		&=\braket{K_{\sigma}^{R,C}}.
	\end{align}
	
	That is, the expectation value is independent of the coefficients and so the allowed charges are the same regardless of the presence of the start-point vertex excitation.
	
	\subsubsection{Charge of the start-point of a magnetic ribbon operator}
	
\begin{figure}[h]
		\begin{center}
			\begin{overpic}[width=0.75\linewidth]{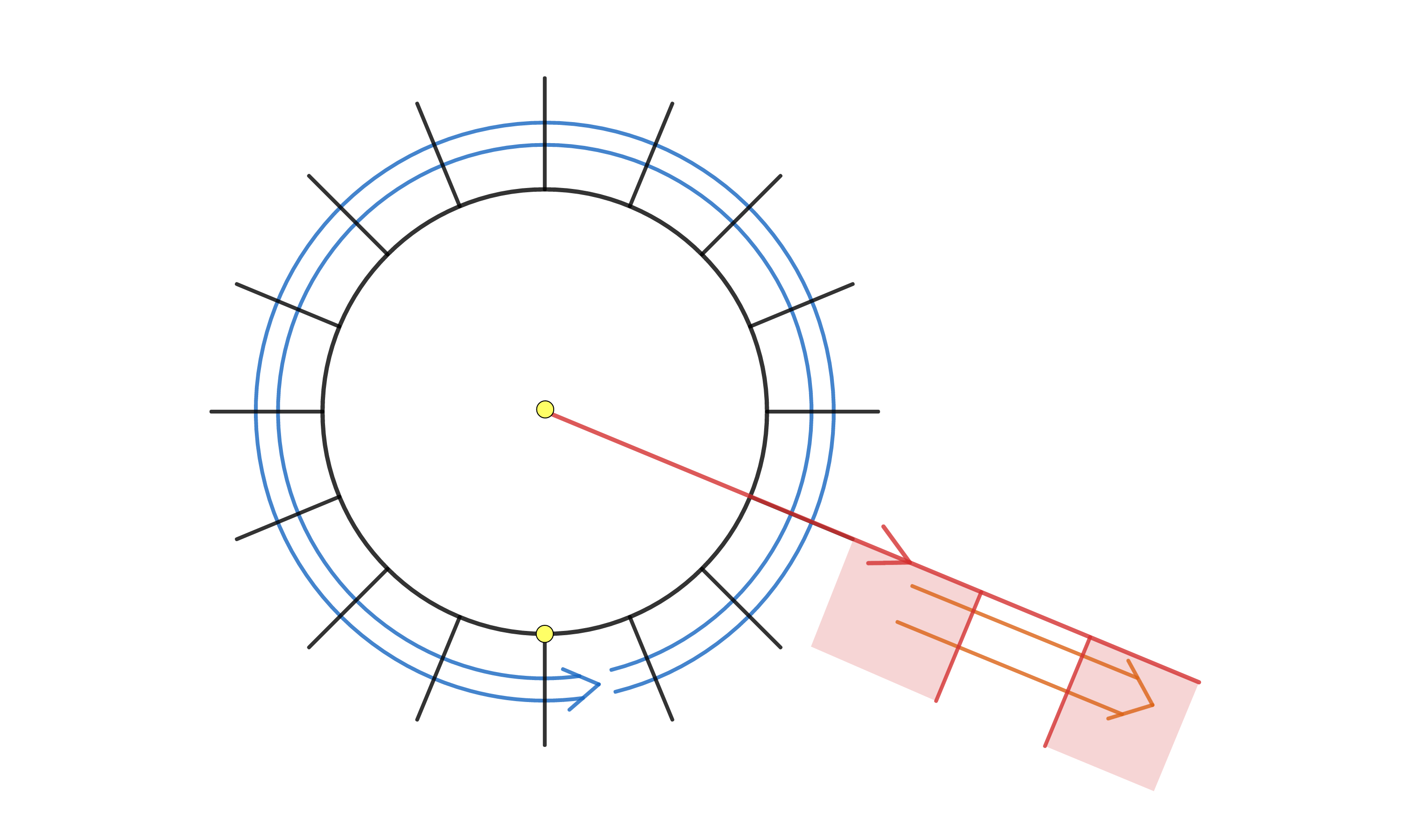}
				\put(22,10){$K^{R,C}_{\sigma}$}
				\put(68,9){$C^h(s)$}
				\put(35,28){$s.p(s)$}
			\end{overpic}
			
			\caption{We now wish to consider measuring the topological charge of the start-point of the magnetic ribbon, without the plaquette excitation. To do so, we consider a ribbon operator $C^h(s)$ for which the start-point of the direct path (the direct path is illustrated by the red arrow) is separated from the plaquette at the start of the dual path (the dual path is represented by the orange double arrow). We then apply a measurement operator $K^{R,C}_{\sigma}$ that encloses the start-point, but not the plaquette excitations. }
			\label{2D_charge_measurement_magnetic_start_point}
			
		\end{center}
	\end{figure}

	From the calculation in Section \ref{Section_2D_charge_magnetic_excitation}, we see that even though the start-point vertex may be excited, this does not change the topological charge of the magnetic excitation. This does not mean that the excited vertex does not carry a topological charge of its own however. To see this, we must isolate the start-point vertex from the plaquette excitations. To do so, we consider ribbon operators for which the start-point is not adjacent to either of the plaquette excitations. We can measure the charge of the vertex excitation on its own, by applying a measurement operator around the vertex that does not enclose the plaquette excitations, as shown in Figure \ref{2D_charge_measurement_magnetic_start_point}. We then consider the state
	\begin{align}
		K^{R,C}_{\sigma}&C^h(s) \ket{GS} \notag\\
		&= \frac{|R|}{|N'_C|} \sum_{d \in N'_C} \overline{\chi}_R(d) \sum_{q \in Q_C} \sum_{k \in \partial(E)} C^{qdq^{-1}}(\sigma) \delta(g(\sigma), qkr_Cq^{-1}) M^{f(r_C,d)^{-1}}(p) \frac{1}{|\ker(\partial)|} \sum_{e_k \in \ker (\partial)} M^{e_k}(p) C^h(s) \ket{GS}. \label{Equation_2D_charge_start_point_magnetic_1}
	\end{align}

	In this case, the magnetic ribbon operator $C^h(s)$ does not affect any of the edges in the path element $g(\sigma)$ (because the dual path does not cross $\sigma$), and so commutes with the electric part of the measurement operator. Therefore, we have
	\begin{align}
		K^{R,C}_{\sigma}&C^h(s) \ket{GS} \notag \\
		&= \frac{|R|}{|N'_C|} \sum_{d \in N'_C} \overline{\chi}_R(d) \sum_{q \in Q_C} \sum_{k \in \partial(E)} C^{qdq^{-1}}(\sigma) C^h(s) \delta(g(\sigma), qkr_Cq^{-1}) M^{f(r_C,d)^{-1}}(p) \frac{1}{|\ker(\partial)|} \sum_{e_k \in \ker (\partial)} M^{e_k}(p) \ket{GS}. \label{Equation_2D_charge_start_point_magnetic_2}
	\end{align}
	Then, just as we discussed when we measured the charge of an electric excitation in Section \ref{Section_2D_charge_electric_excitation}, the Kronecker delta $\delta(g(\sigma), qkr_Cq^{-1}) $ acting on the ground state only gives one if $qkr_Cq^{-1}= \partial(e_m)^{-1}$, where $e_m$ is the label of the surface enclosed by $\sigma$. This means that $qkr_Cq^{-1}$ must be in the image of $\partial$ for the relevant term to be non-zero. Because $\partial(E)$ is in the centre of $G$, this in turn means that $qkr_Cq^{-1}=kr_C$. Then we see that $r_C$ must be in $\partial(E)$ and indeed the class $C$ must just contain the elements of $\partial(E)$. In fact, we can take the representative $r_C$ to be the identity element $1_G$. Then substituting these results into Equation \ref{Equation_2D_charge_start_point_magnetic_2} we find that
	\begin{align}
		&K^{R,C}_{\sigma}C^h(s) \ket{GS} \notag \\
		&= \delta(C, \partial(E)) \frac{|R|}{|N'_C|} \sum_{d \in N'_C} \overline{\chi}_R(d) \sum_{q \in Q_C} \sum_{k \in \partial(E)} C^{qdq^{-1}}(\sigma) C^h(s) \delta(g(\sigma), k) M^{f(r_C,d)^{-1}}(p) \frac{1}{|\ker(\partial)|} \sum_{e_k \in \ker (\partial)} M^{e_k}(p) \ket{GS}. \label{Equation_2D_charge_start_point_magnetic_3}
	\end{align}
	
	Then because $r_C$ is the identity element when this expression is non-zero, $r_C$ commutes with all of the elements of $G$. Therefore, the subgroup $N'_C$, which consists of the elements which commute with $r_C$ up to an element of $\partial(E)$, is the entire group $G$. Then $Q_C=G/N'_C$ is the trivial group and so we can drop the sum over $q$. Furthermore, $[r_C,d]=1_G$, which implies that $\partial(f(r_C,d))=[r_C,d]=1_G$ and so we can take $M^{f(r_C,d)^{-1}}(p)$ to be trivial. Equation \ref{Equation_2D_charge_start_point_magnetic_3} then becomes
	\begin{align}
		K^{R,C}_{\sigma}C^h(s) \ket{GS}&= \delta(C, \partial(E)) \frac{|R|}{|G|} \sum_{d \in G} \overline{\chi}_R(d) \sum_{k \in \partial(E)} C^{d}(\sigma) C^h(s) \delta(g(\sigma), k) \frac{1}{|\ker(\partial)|}\sum_{e_k\in \ker{\partial}} M^{e_k}(p) \ket{GS}. \label{Equation_2D_charge_start_point_magnetic_4}
	\end{align}
	
	We have assumed that the ground state is the one for which the plaquettes carry a trivial irrep of the kernel of $\partial$ and so $\frac{1}{|\ker(\partial)|} \sum_{e_k \in \ker{\partial}} M^{e_k}(p) \ket{GS}=\ket{GS}$. In addition, $k$ only appears in $\sum_{k \in \partial(E)} \delta(g(\sigma), k)$ and this expression acts as the identity on the ground state because it sums over all possible values for $g(\sigma)$ (which must be in $\partial(E)$ due to fake-flatness). This means that we can write Equation \ref{Equation_2D_charge_start_point_magnetic_4} as
	\begin{align}
		K^{R,C}_{\sigma}C^h(s) \ket{GS}&= \delta(C, \partial(E)) \frac{|R|}{|G|} \sum_{d \in G} \overline{\chi}_R(d) C^{d}(\sigma) C^h(s) \ket{GS}. \label{Equation_2D_charge_start_point_magnetic_5}
	\end{align}
	
	Now we wish to commute $C^d(\sigma)$ past $C^h(s)$ so that it may act directly on the ground state. In order to do this, we must first establish why these operators do not commute. The action of the operator $C^h(s)$ on an edge $i$ cut by the dual path is (assuming that the edge points away from the direct path of $s$)
	$$C^h(s):g_i = g(s.p(s)-v_i)^{-1}hg(s.p(s)-v_i)g_i,$$
	where $v_i$ is the vertex attached to edge $i$ and lying on the direct path. In the case of the ribbon $s$ indicated in Figure \ref{2D_charge_measurement_magnetic_start_point}, the start-point $s.p(s)$ is inside the region enclosed by the ribbon $\sigma$ while the entire dual path is outside the region. This means that the path $(s.p(s)-v_i)$ intersects the ribbon $\sigma$ for every edge $i$ affected by $C^h(s)$. The magnetic ribbon operator $C^d(\sigma)$ will affect the label of each such path, according to
	$$C^d(\sigma):g(s.p(s)-v_i) =g(s.p(s)-s.p(\sigma))dg(s.p(s)-s.p(\sigma))^{-1}g(s.p(s)-v_i),$$
	as we showed in Section \ref{Section_2D_braiding_electric_magnetic} when considering the braiding between a magnetic and electric excitation (see Equation \ref{Magnetic_electric_braid_2D_1_appendix}). Defining $d_{[s-\sigma]}= g(s.p(s)-s.p(\sigma))dg(s.p(s)-s.p(\sigma))^{-1}$, we can write this as
	$$C^d(\sigma):g(s.p(s)-v_i) = d_{[s-\sigma]} g(s.p(s)-v_i).$$
	
	Then if we consider first acting with $C^d(\sigma)$ and then $C^h(s)$, the magnetic ribbon operator on $\sigma$ will change the path labels before $C^h(s)$ acts. This means that the action on an edge $i$ cut by the dual path of $C^h(s)$ is given by
	\begin{align*}
		C^h(s) C^d(\sigma):g_i &= (C^d(\sigma):g(s.p(s)-v_i))^{-1}h(C^d(\sigma):g(s.p(s)-v_i))g_i\\
		&= (d_{[s-\sigma]} g(s.p(s)-v_i))^{-1}h(d_{[s-\sigma]} g(s.p(s)-v_i))g_i\\
		&=g(s.p(s)-v_i)^{-1}d_{[s-\sigma]}^{-1}hd_{[s-\sigma]}g(s.p(s)-v_i)g_i\\
		&=C^{d_{[s-\sigma]}^{-1}hd_{[s-\sigma]}}(s):g_i.
	\end{align*}
	
	This holds for the action on all edges $i$ cut by the dual path of $s$, and the two ribbon operators do not interfere in any other way, so we can write
	\begin{align*}
		C^h(s) C^d(\sigma)= C^d(\sigma)C^{d_{[s-\sigma]}^{-1}hd_{[s-\sigma]}}(s).
	\end{align*}
	Then this implies that
	\begin{align}
		C^d(\sigma) C^h(s) = C^{d_{[s-\sigma]}hd_{[s-\sigma]}^{-1}}(s)C^d(\sigma), \label{Equation_2D_charge_start_point_magnetic_magnetic_commutation}
	\end{align}
	where, as we discussed previously when considering the charge of a plaquette excitation, $d_{[s-\sigma]}$ is the same before and after the action of $C^d(\sigma)$, which allows us to invert the commutation relation in this simple way. Inserting Equation \ref{Equation_2D_charge_start_point_magnetic_magnetic_commutation} into Equation \ref{Equation_2D_charge_start_point_magnetic_5}, we obtain
	\begin{align}
		K^{R,C}_{\sigma}C^h(s) \ket{GS}&= \delta(C, \partial(E)) \frac{|R|}{|G|} \sum_{d \in G} \overline{\chi}_R(d) C^{d_{[s-\sigma]}hd_{[s-\sigma]}^{-1}}(s) C^{d}(\sigma) \ket{GS}. \label{Equation_2D_charge_start_point_magnetic_6}
	\end{align}

	Now, because $C^d(\sigma)$ is a closed magnetic ribbon operator acting directly on the ground state, it acts trivially (as we showed in Section \ref{Section_Topological_Magnetic_Ribbons}). Then we have
	\begin{align}
		K^{R,C}_{\sigma}C^h(s) \ket{GS}&= \delta(C, \partial(E)) \frac{|R|}{|G|} \sum_{d \in G} \overline{\chi}_R(d) C^{d_{[s-\sigma]}hd_{[s-\sigma]}^{-1}}(s) \ket{GS}. \label{Equation_2D_charge_start_point_magnetic_7}
	\end{align}

	Then we can change the sum over $d$ to a sum over $d_{[s-\sigma]}$, using the fact that $\overline{\chi}_R(d)$ is a function of conjugacy class and so is the same for $d$ and $d_{[s-\sigma]}$:
	\begin{align}
		K^{R,C}_{\sigma}C^h(s) \ket{GS}&= \delta(C, \partial(E)) \frac{|R|}{|G|} \sum_{d_{[s-\sigma]} \in G} \overline{\chi}_R(d_{[s-\sigma]}) C^{d_{[s-\sigma]}hd_{[s-\sigma]}^{-1}}(s) \ket{GS}. \label{Equation_2D_charge_start_point_magnetic_8}
	\end{align}
	
	Now we consider the case where, instead of a magnetic ribbon operator $C^h(s)$ labelled by a single group element $h$, we have a sum of ribbon operators with labels in a conjugacy class $[h]$ of $G$, $\sum_{g \in [h]} \alpha_g C^g(s)$. Using Equation \ref{Equation_2D_charge_start_point_magnetic_8}, the result of applying a topological measurement operator on the state created by this operator acting on the ground state is
	\begin{align}
		K^{R,C}_{\sigma}\sum_{g \in [h]} \alpha_g C^g(s) \ket{GS}&= \delta(C, \partial(E)) \frac{|R|}{|G|} \sum_{d_{[s-\sigma]} \in G} \overline{\chi}_R(d_{[s-\sigma]}) \sum_{g \in [h]} \alpha_g C^{d_{[s-\sigma]}gd_{[s-\sigma]}^{-1}}(s) \ket{GS}. \label{Equation_2D_charge_start_point_magnetic_9}
	\end{align}
	
	We can change the sum over $g$ on the right to a sum over $g'=d_{[s-\sigma]}gd_{[s-\sigma]}^{-1}$, to obtain
	\begin{align}
		K^{R,C}_{\sigma}\sum_{g \in [h]} \alpha_g C^g(s) \ket{GS}&= \delta(C, \partial(E)) \frac{|R|}{|G|} \sum_{d_{[s-\sigma]} \in G} \overline{\chi}_R(d_{[s-\sigma]}) \sum_{g'= d_{[s-\sigma]}gd_{[s-\sigma]}^{-1} \in [h]} \alpha_{d_{[s-\sigma]}^{-1}g'd_{[s-\sigma]}} C^{g'}(s) \ket{GS}. \label{Equation_2D_charge_start_point_magnetic_10}
	\end{align}
	
	The result now depends on the coefficients $\alpha$. First, we consider the case where $\alpha_g$ is the same for all elements of the conjugacy class (and so there is no vertex excitation). Denoting this common coefficient by $\alpha$, Equation \ref{Equation_2D_charge_start_point_magnetic_10} becomes
	\begin{align}
		K^{R,C}_{\sigma}\sum_{g \in [h]} \alpha C^g(s) \ket{GS}&= \delta(C, \partial(E)) \frac{|R|}{|G|} \sum_{d_{[s-\sigma]} \in G} \overline{\chi}_R(d_{[s-\sigma]}) \sum_{g' \in [h]} \alpha C^{g'}(s) \ket{GS}, \label{Equation_2D_charge_start_point_magnetic_11}
	\end{align}
	from which we see that the magnetic ribbon operator is left invariant. Then $d_{[s-\sigma]}$ only appears in $ \sum_{d_{[s-\sigma]} \in G} \overline{\chi}_R(d_{[s-\sigma]})$ and so we can apply the Grand Orthogonality Theorem to obtain
	\begin{align}
		K^{R,C}_{\sigma}\sum_{g \in [h]} \alpha C^g(s) \ket{GS}&= \delta(C, \partial(E)) \delta(R,1_{\text{Rep}}) \sum_{g' \in [h]} \alpha C^{g'}(s) \ket{GS}, \label{Equation_2D_charge_start_point_magnetic_12}
	\end{align}	
	where $1_{\text{Rep}}$ is the trivial irrep of $G$. This tells us that the charge enclosed by the ribbon $\sigma$ is the trivial charge, as we may expect from the fact that there is no vertex excitation and so no excitation within the closed ribbon on which we apply the measurement operator.

	However, now consider the case where $\alpha_g$ is not the same for each element $g$ of the conjugacy class $[h]$. In order to do so, we first construct the centralizer of $[h]$, $N_{[h]}$, which we define as the subgroup of $G$ consisting of elements which commute with $h$. Note that we have chosen an element $h$ as the representative of $[h]$ and choosing a different representative would give a different, but isomorphic, centralizer. We can then construct the quotient group $Q_{[h]}= G/N_{[h]}$, which consists of the cosets of $N_{[h]}$ in $G$. Then these cosets are in one-to-one correspondence with the elements of the conjugacy class $h$. That is, for any element $g' \in [h]$, all of the elements $x \in G$ such that $g' = xhx^{-1}$ belong to a single coset $xN_{[h]}$. This is very similar to our construction of $N'_C$ and $Q_C$ when we defined the projectors to definite topological charge. We can then write the ribbon operator $ \sum_{g' \in [h]} \alpha_{g'} C^{g'}(s)$ as 
	$$\sum_{g' \in [h]} \alpha_{g'} C^{g'}(s)= \sum_{x \in G} \beta_{x} C^{xhx^{-1}}(s)$$
	for some coefficients $\beta_x$, where $\beta_x$ is the same for each element $x$ of a given coset. Then Equation \ref{Equation_2D_charge_start_point_magnetic_10} becomes
	\begin{align*}
		K^{R,C}_{\sigma}\sum_{x \in G} \beta_x C^{xhx^{-1}}(s) \ket{GS}&= \delta(C, \partial(E)) \frac{|R|}{|G|} \sum_{d_{[s-\sigma]} \in G} \overline{\chi}_R(d_{[s-\sigma]}) \sum_{x \in G} \beta_{d_{[s-\sigma]}^{-1}x} C^{xhx^{-1}}(s) \ket{GS} .
	\end{align*}
	
	Now $d_{[s-\sigma]}$ only appears in the coefficient
	$$ \sum_{d_{[s-\sigma]} \in G} \overline{\chi}_R(d_{[s-\sigma]}) \beta_{d_{[s-\sigma]}^{-1}x}.$$
	We can decompose $\beta_{d_{[s-\sigma]}^{-1}x}$ in terms of irreps of $G$ and then apply the Grand Orthogonality Theorem. This guarantees that only the irreps $R$ for which the conjugate representation $\overline{R}$ is in $\beta$ will contribute (the conjugate because $d_{[s-\sigma]}$ appears with an inverse in the subscript of $\beta$). However, because $\beta$ is a function of the cosets of $N_{[h]}$, $\beta$ can only contain irreps that are trivial in $N_{[h]}$ and so the irrep $R$ must be trivial in $N_{[h]}$ for it to contribute. Because the irrep $R$ (together with the class $C =\partial(E)$) labels the topological charge of the start-point vertex excitation, this indicates that the vertex carries a charge labelled by the trivial class $C=\partial(E)$ and an irrep that is generally non-trivial, but has trivial restriction to the subgroup $N_{[h]}$. 
	
	\section{Algebraic proofs for Section VII}
	\label{Section_2D_irrep_basis_Appendix}
	
	In Section \ref{Section_2D_irrep_basis} of the main text, we performed a change of basis in order to uncover more information about the higher lattice gauge theory model with $\rhd$ trivial. We considered labelling the edges and plaquettes by irreps of the groups $G$ and $E$ respectively, rather than group elements. In that section, we made some claims about the action of the various energy terms, skipping over some straightforward, but lengthy, algebraic manipulations. In this section, we present those manipulations for the sake of completeness. These calculations are also relevant for Section \ref{Section_Z4_Z4_irrep_basis}, where we considered a specific model for which $\rhd$ is trivial and skipped over the same manipulations. The first relation to prove is Equation \ref{Equation_irrep_basis_vertex_term_1}, which describes the action of the vertex energy term in the irrep basis. As described in Section \ref{Section_2D_irrep_basis}, we define a basis state in the irrep basis for the edges around a vertex $v$, in terms of basis states in the group element basis, by 
	\begin{align}
		&\ket{ \set{R_i,a_i,b_i}_{\text{out}}, \set{R_j,a_j,b_j}_{\text{in}}} \notag\\
		& \hspace{2cm} = \sum_{\set{g_i}_{\text{out}}} \sum_{\set{g_j}_{\text{in}}} \bigg(\prod_{\substack{\text{outgoing } \\ \text{edges }i}} \sqrt{\frac{|R|}{|G|}} [D^{R_i}(g_i)]_{a_i b_i} \bigg) \bigg(\prod_{\substack{\text{incoming } \\ \text{edges }j}} \sqrt{\frac{|R|}{|G|}} [D^{R_j}(g_j)]_{a_j b_j} \bigg)\ket{\set{g_i}_{\text{out}}, \set{g_j}_{\text{in}}}, \label{Equation_appendix_vertex_irrep}
	\end{align}
	where $\sum_{\set{g_i}_{\text{out}}}$ sums over each element of $G$ for each outgoing edge and $\sum_{\set{g_j}_{\text{in}}}$ does the same for each incoming edge. Then we can apply the vertex transform to this state:
	\begin{align*}
		A_v^x &\ket{ \set{R_i,a_i,b_i}_{\text{out}}, \set{R_j,a_j,b_j}_{\text{in}}}\\
		&=\sum_{\set{g_i}_{\text{out}}} \sum_{\set{g_j}_{\text{in}}} \bigg(\prod_{\substack{\text{outgoing } \\ \text{edges }i}} \sqrt{\frac{|R|}{|G|}} [D^{R_i}(g_i)]_{a_i b_i} \bigg) \bigg(\prod_{\substack{\text{incoming } \\ \text{edges }j}} \sqrt{\frac{|R|}{|G|}} [D^{R_j}(g_j)]_{a_j b_j} \bigg) A_v^x\ket{\set{g_i}_{\text{out}}, \set{g_j}_{\text{in}}} \\
		&= \sum_{\set{g_i}_{\text{out}}} \sum_{\set{g_j}_{\text{in}}} \bigg(\prod_{\substack{\text{outgoing } \\ \text{edges }i}} \sqrt{\frac{|R|}{|G|}} [D^{R_i}(g_i)]_{a_i b_i} \bigg) \bigg(\prod_{\substack{\text{incoming } \\ \text{edges }j}} \sqrt{\frac{|R|}{|G|}} [D^{R_j}(g_j)]_{a_j b_j} \bigg) \ket{\set{xg_i}_{\text{out}}, \set{g_jx^{-1}}_{\text{in}}}.
	\end{align*}
	
	In order to write the resulting state in the irrep basis, we start by making a change of variable for the dummy edge variables by defining $g'_i=xg_i$ for the outgoing edges and $g'_j=g_jx^{-1}$ for the incoming edges. This gives us
	\begin{align}
		A_v^x &\ket{ \set{R_i,a_i,b_i}_{\text{out}}, \set{R_j,a_j,b_j}_{\text{in}}} \notag \\
		& \hspace{1cm}=\sum_{\set{g'_i}_{\text{out}}} \sum_{\set{g'_j}_{\text{in}}} \bigg(\prod_{\substack{\text{outgoing } \\ \text{edges }i}} \sqrt{\frac{|R|}{|G|}} [D^{R_i}(x^{-1}g'_i)]_{a_i b_i} \bigg) \bigg(\prod_{\substack{\text{incoming } \\ \text{edges }j}} \sqrt{\frac{|R|}{|G|}} [D^{R_j}(g'_jx)]_{a_j b_j} \bigg) \ket{\set{g'_i}_{\text{out}}, \set{g'_j}_{\text{in}}}. \label{Equation_vertex_transform_irrep_basis_2_appendix}
	\end{align}
	
	We can then use the fact that the matrices are part of a representation to split off the part of each matrix corresponding to $x$. For example, 
	$$[D^{R_i}(x^{-1}g'_i)]_{a_i b_i} = \sum_{c_i=1}^{|R_i|} [D^{R_i}(x^{-1})]_{a_i c_i} [D^{R_i}(g'_i)]_{c_i b_i}.$$
	Substituting this into Equation \ref{Equation_vertex_transform_irrep_basis_2_appendix}, we have
	\begin{align}
		A_v^x &\ket{ \set{R_i,a_i,b_i}_{\text{out}}, \set{R_j,a_j,b_j}_{\text{in}}} \notag \\
		&=\sum_{\set{g'_i}_{\text{out}}} \sum_{\set{g'_j}_{\text{in}}} \bigg( \prod_{\substack{\text{outgoing } \\ \text{edges }i}} \sqrt{\frac{|R|}{|G|}} \sum_{c_i=1}^{|R_i|} [D^{R_i}(x^{-1})]_{a_i c_i} [D^{R_i}(g'_i)]_{c_i b_i} \bigg) \notag \\
		& \hspace{1cm} \bigg(\prod_{\substack{\text{incoming } \\ \text{edges }j}} \sqrt{\frac{|R|}{|G|}} \sum_{c_j=1}^{|R_j|} [D^{R_j}(g'_j)]_{a_j c_j} [D^{R_j}(x)]_{c_j b_j} \bigg)\ket{\set{g'_i}_{\text{out}}, \set{g'_j}_{\text{in}}} \notag \\
		&= \sum_{\set{c_i}} \sum_{\set{c_j}} \bigg(\prod_{\substack{\text{outgoing } \\ \text{edges }i}} [D^{R_i}(x^{-1})]_{a_i c_i} \bigg) \bigg(\prod_{\substack{\text{incoming } \\ \text{edges }j}} [D^{R_j}(x)]_{c_j b_j} \bigg) \sum_{\set{g'_i}_{\text{out}}} \sum_{\set{g'_j}_{\text{in}}} \bigg( \prod_{\substack{\text{outgoing } \\ \text{edges }i}} \sqrt{\frac{|R|}{|G|}} [D^{R_i}(g'_i)]_{c_i b_i} \bigg) \notag \\
		& \hspace{1cm} \bigg(\prod_{\substack{\text{incoming } \\ \text{edges }j}} \sqrt{\frac{|R|}{|G|}} [D^{R_j}(g'_j)]_{a_j c_j} \bigg) \ket{\set{g'_i}_{\text{out}}, \set{g'_j}_{\text{in}}}. \label{Equation_vertex_transform_irrep_basis_3_appendix}
	\end{align} 
	
	We can then recognise the expression 
	\begin{align*}
		&\sum_{\set{g'_i}_{\text{out}}} \sum_{\set{g'_j}_{\text{in}}} \bigg( \prod_{\substack{\text{outgoing } \\ \text{edges }i}} \sqrt{\frac{|R|}{|G|}} [D^{R_i}(g'_i)]_{c_i b_i} \bigg) \bigg(\prod_{\substack{\text{incoming } \\ \text{edges }j}} \sqrt{\frac{|R|}{|G|}} [D^{R_j}(g'_j)]_{a_j c_j} \bigg)\ket{\set{g'_i}_{\text{out}}, \set{g'_j}_{\text{in}}}
	\end{align*}
	as the state $\ket{\set{R_i,c_i,b_i}_{\text{out}}, \set{R_j,a_j,c_j}_{\text{in}}}$ (see Equation \ref{Equation_appendix_vertex_irrep}). This means that Equation \ref{Equation_vertex_transform_irrep_basis_3_appendix} can be written as 
	\begin{align*}
		A_v^x &\ket{ \set{R_i,a_i,b_i}_{\text{out}}, \set{R_j,a_j,b_j}_{\text{in}}}\\
		&= \sum_{\set{c_i}} \sum_{\set{c_j}} \bigg(\prod_{\substack{\text{outgoing } \\ \text{edges }i}} [D^{R_i}(x^{-1})]_{a_i c_i} \bigg) \bigg(\prod_{\substack{\text{incoming } \\ \text{edges }j}} [D^{R_j}(x)]_{c_j b_j} \bigg)\ket{\set{R_i,c_i,b_i}_{\text{out}}, \set{R_j,a_j,c_j}_{\text{in}}}.
	\end{align*}
	
	We then wish to consider the vertex energy term, which is the average of the vertex transforms with each label $x \in G$. That is $A_v = \frac{1}{|G|} \sum_{x \in G} A_v^x$. We have
	\begin{align}
		A_v &\ket{ \set{R_i,a_i,b_i}_{\text{out}}, \set{R_j,a_j,b_j}_{\text{in}}} \notag \\
		&= \frac{1}{|G|} \sum_{x \in G} \sum_{\set{c_i}} \sum_{\set{c_j}} \bigg(\prod_{\substack{\text{outgoing } \\ \text{edges }i}} [D^{R_i}(x^{-1})]_{a_i c_i} \bigg) \bigg(\prod_{\substack{\text{incoming } \\ \text{edges }j}} [D^{R_j}(x)]_{c_j b_j} \bigg) \ket{\set{R_i,c_i,b_i}_{\text{out}}, \set{R_j,a_j,c_j}_{\text{in}}}. \label{Equation_vertex_term_irrep_basis_1_appendix}
	\end{align}
	
	We note that, because we are using unitary representations, 
	$$[D^{R_i}(x^{-1})]_{a_i c_i}= [D^{R_i}(x)^{-1}]_{a_i c_i}= [D^{R_i}(x)]_{c_i a_i}^*,$$
	which we can write in terms of the conjugate representation $\overline{R}_i$, whose matrix elements satisfy
	$$ [D^{\overline{R}_i}(x)]_{c_i a_i}=[D^{R_i}(x)]_{c_i a_i}^*=[D^{R_i}(x^{-1})]_{a_i c_i}.$$
	
	Then Equation \ref{Equation_vertex_term_irrep_basis_1_appendix} can be written as
	\begin{align*}
		A_v &\ket{ \set{R_i,a_i,b_i}_{\text{out}}, \set{R_j,a_j,b_j}_{\text{in}}}\\
		&= \frac{1}{|G|} \sum_{x \in G} \sum_{\set{c_i}} \sum_{\set{c_j}} \bigg(\prod_{\substack{\text{outgoing } \\ \text{edges }i}} [D^{\overline{R}_i}(x)]_{c_i a_i} \bigg) \bigg(\prod_{\substack{\text{incoming } \\ \text{edges }j}} [D^{R_j}(x)]_{c_j b_j} \bigg) \ket{\set{R_i,c_i,b_i}_{\text{out}}, \set{R_j,a_j,c_j}_{\text{in}}}.
	\end{align*}
	This is the same as Equation \ref{Equation_irrep_basis_vertex_term_1}, and so we have proven the result claimed in that equation. When $G$ is Abelian, so that the irreps $R$ are all one-dimensional, this equation simplifies greatly. We can drop the matrix indices to obtain
	\begin{align*}
		A_v \ket{ \set{R_i}_{\text{out}}, \set{R_j}_{\text{in}}}
		&= \frac{1}{|G|} \sum_{x \in G} \big(\prod_{\substack{\text{outgoing } \\ \text{edges }i}} R^*_i(x) \big) \big(\prod_{\substack{\text{incoming } \\ \text{edges }j}} R_j(x)\big) \ket{\set{R_i}_{\text{out}}, \set{R_j}_{\text{in}}},
	\end{align*}
	where $R_i(x)$ is the phase representing element $x$ in irrep $R_i$ (equivalently this is the character, because the irrep is 1D) and $R^*_i(x)$ is the conjugate of this, which is equivalently the inverse (because we use unitary irreps). $R^*_i(x)$ is also the phase representing element $x$ in the conjugate representation $\overline{R}_i$ of $G$. Using the orthogonality condition for characters, we then have
	\begin{align*}
		A_v \ket{ \set{R_i}_{\text{out}}, \set{R_j}_{\text{in}}}&= \delta\bigg( \big(\prod_{\substack{\text{outgoing } \\ \text{edges }i}} \hspace{-0.3cm} \overline{R}_i\big) \cdot \big(\prod_{\substack{\text{incoming } \\ \text{edges }j}} \hspace{-0.3cm} R_j\big) ,1_{\text{Rep}}\bigg) \ket{\set{R_i}_{\text{out}}, \set{R_j}_{\text{in}}},
	\end{align*}
	as stated in Equation \ref{Equation_vertex_transform_irrep_basis_Abelian} in Section \ref{Section_2D_irrep_basis} of the main text. We note that this is also equivalent to Equation \ref{Equation_Z4_Z4_irrep_vertex} in Section \ref{Section_Z4_Z4_irrep_basis} of the main text, which is the analogous result for the example model $(\mathbb{Z}_4,\mathbb{Z}_4, \partial \rightarrow \mathbb{Z}_2, \rhd \text{ trivial})$ considered in that section.

	The next result from the main text to prove is Equation \ref{Equation_edge_transform_irrep_basis_2}, which describes the action of the edge transform in the new basis. We define the state of the degrees of freedom around an edge $i$ in the irrep basis by
	\begin{align*}
		\ket{ \set{R,a,b}, \mu_1, \mu_2} = \sqrt{\frac{|R|}{|G|}}& \sum_{g \in G} \frac{1}{|E|} \sum_{e_1,e_2 \in E} [D^R(g)]_{ab} \mu_1(e_1) \mu_2(e_2) \ket{g,e_1,e_2},
	\end{align*}
	where $g$ is the label of the edge in the group element basis and $e_1$ and $e_2$ are the labels of the two plaquettes on the left and right of the edge (if the edge points upwards), as described in Figure \ref{edge_support_rhd_trivial} in Section \ref{Section_2D_irrep_basis} of the main text. The action of the edge transform on an irrep basis state is given by
	\begin{align}
		\mathcal{A}_i^e \ket{ \set{R,a,b}, \mu_1, \mu_2} 
		&=\sqrt{\frac{|R|}{|G|}} \sum_{g \in G} \frac{1}{|E|} \sum_{e_1,e_2 \in E} [D^R(g)]_{ab} \mu_1(e_1) \mu_2(e_2) \mathcal{A}_i^e\ket{g,e_1,e_2} \notag\\
		&=\sqrt{\frac{|R|}{|G|}} \sum_{g \in G} \frac{1}{|E|} \sum_{e_1,e_2 \in E} [D^R(g)]_{ab} \mu_1(e_1) \mu_2(e_2)\ket{\partial(e)g,e_1e^{\sigma_1},e_2e^{\sigma_2}}, \label{Equation_edge_transform_irrep_basis_1_appendix}
	\end{align}
	where $\sigma_x$ is $-1$ if plaquette $x$ is aligned with the edge and $1$ otherwise. Just as we did with the vertex transform, we relabel the dummy indices to remove the factors of $e$ in the state $\ket{\partial(e)g,e_1e^{\sigma_1},e_2e^{\sigma_2}}$. To do this, we define $g' = \partial(e)g$, $e_1'=e_1e^{\sigma_1}$ and $e_2'=e_2e^{\sigma_2}$. Inserting these into Equation \ref{Equation_edge_transform_irrep_basis_1_appendix} gives us
	\begin{align}
		\mathcal{A}_i^e \ket{ \set{R,a,b}, \mu_1, \mu_2} &=\sqrt{\frac{|R|}{|G|}} \sum_{g' \in G} \frac{1}{|E|} \sum_{e_1',e_2' \in E} [D^R(\partial(e)^{-1}g')]_{ab} \mu_1(e_1' e^{- \sigma_1}) \mu_2(e_2' e^{- \sigma_2})\ket{g',e_1',e_2'} \notag \\
		&= \sqrt{\frac{|R|}{|G|}} \sum_{g' \in G} \frac{1}{|E|} \sum_{e_1',e_2' \in E} \sum_{c=1}^{|R|} [D^R(\partial(e)^{-1})]_{ac} [D^R(g')]_{cb} \mu_1(e_1') \mu_1(e^{- \sigma_1}) \mu_2(e_2') \mu(e^{- \sigma_2})\ket{g',e_1',e_2'} \notag \\
		&=\sum_{c=1}^{|R|} [D^R(\partial(e)^{-1})]_{ac} \mu_1(e^{- \sigma_1}) \mu_2(e^{- \sigma_2})
		\sqrt{\frac{|R|}{|G|}} \sum_{g' \in G} \frac{1}{|E|} \sum_{e_1',e_2' \in E} [D^R(g')]_{cb} \mu_1(e_1') \mu_2(e_2') \ket{g',e_1',e_2'} \notag\\
		&= \sum_{c=1}^{|R|} [D^R(\partial(e)^{-1})]_{ac} \mu_1(e^{- \sigma_1}) \mu_2(e^{- \sigma_2}) \ket{\set{R,c,b}, \mu_1, \mu_2}. \label{Equation_edge_transform_irrep_basis_2_appendix}
	\end{align}
	
	This establishes the result claimed in Equation \ref{Equation_edge_transform_irrep_basis_2} for the action of the edge transform in the irrep basis. As we discussed previously in Section \ref{Section_2D_irrep_basis} of the main text, due to Clifford's theorem we can write $[D^R( \partial(e)^{-1})]_{ac} =R_{\partial}(\partial(e^{-1})) \delta_{ac}$, where $R_{\partial}$ is the (1D) irrep of the subgroup $\partial(E)$ that $R$ decomposes into. Furthermore, we can define an irrep $\mu^R$ of $E$ by $\mu^R(e)=R_{\partial}(\partial(e))$. This means that
	\begin{align*}
		\mathcal{A}_i^e &\ket{ \set{R,a,b}, \mu_1, \mu_2}= \mu^R(e^{-1}) \mu_1(e^{- \sigma_1}) \mu_2(e^{- \sigma_2}) \ket{\set{R,a,b}, \mu_1, \mu_2}.
	\end{align*}
	
	Then the edge energy term $\mathcal{A}_i = \frac{1}{|E|} \mathcal{A}_i^e$ acts on this basis as
	\begin{align*}
		\mathcal{A}_i \ket{\set{R,a,b}, \mu_1, \mu_2} &= \frac{1}{|E|} \sum_{e \in E} \mu^R(e^{-1}) \mu_1(e^{- \sigma_1}) \mu_2(e^{- \sigma_2}) \ket{\set{R,a,b}, \mu_1, \mu_2}.
	\end{align*}
	Using orthogonality of characters, this becomes
	\begin{align*}
		\mathcal{A}_i \ket{\set{R,a,b}, \mu_1, \mu_2}&= \delta( \mu^R \cdot \mu^{\sigma_1} \cdot \mu^{\sigma_2}, 1_{\text{Rep}(E)})\ket{\set{R,a,b}, \mu_1, \mu_2},
	\end{align*}
	which is the result claimed in Equation \ref{Equation_edge_transform_irrep} of Section \ref{Section_2D_irrep_basis} of the main text (and is equivalent to Equation \ref{Equation_Z4_Z4_irrep_edge} in Section \ref{Section_Z4_Z4_irrep_basis}).

	Finally, we must prove the result given in Equation \ref{Equation_plaquette_term_irrep_basis_3} of the main text for the action of the plaquette term in the irrep basis. We defined a basis state of a plaquette $p$ and the edges $\set{i}$ on its boundary, in the irrep basis, by
	\begin{align*}
		\ket{ \mu , \set{R_i,a_i,b_i}}
		& = \sum_{\set{g_i \in G}} \bigg(\prod_{\substack{\text{edge }i \text{ in } \\ \text{boundary}(p)}} \sqrt{\frac{|R|}{|G|}} [D^{R_i}(g_i)]_{a_i,b_i} \bigg)\sum_{e_p \in E} \sqrt{\frac{1}{|E|}} \mu(e_p)\ket{e_p, \set{g_i}}.
	\end{align*}
	Here $\ket{e_p, \set{g_i}}$ is a basis state in the group element basis, which is acted on by the plaquette term according to
	\begin{align*}
		B_p\ket{e_p, \set{g_i}} & = \delta\big(\partial(e_p)\prod_{\substack{\text{edge }i \text{ in } \\ \text{boundary}(p)}} g_i^{\sigma_i}, 1_G \big)\ket{e_p, \set{g_i}},
	\end{align*}
	where $\sigma_i$ is $+1$ if edge $i$ is aligned with the plaquette $p$ and $-1$ if it is anti-aligned. Then the action of the plaquette operator on the irrep basis state is
	\begin{align}
		B_p&\ket{ \mu, \set{R_i,a_i,b_i}} \notag \\
		&= \sum_{\set{g_i \in G}} \bigg( \prod_{\substack{\text{edge }i \text{ in } \\ \text{boundary}(p)}} \sqrt{\frac{|R|}{|G|}} [D^{R_i}(g_i)]_{a_i,b_i} \bigg) \sum_{e_p \in E} \sqrt{\frac{1}{|E|}} \mu(e_p) B_p\ket{e_p, \set{g_i}} \notag \\
		&= \sum_{\set{g_i \in G}} \bigg( \prod_{\substack{\text{edge }i \text{ in } \\ \text{boundary}(p)}} \sqrt{\frac{|R|}{|G|}} [D^{R_i}(g_i)]_{a_i,b_i} \bigg) \sum_{e_p \in E} \sqrt{\frac{1}{|E|}} \mu(e_p) \delta\bigg(\partial(e_p)\prod_{\substack{\text{edge }i \text{ in } \\ \text{boundary}(p)}} \hspace{-0.4cm}g_i^{\sigma_i}, 1_G \bigg)\ket{e_p, \set{g_i}}. \label{Equation_plaquette_term_irrep_basis_1_appendix}
	\end{align} 
	
	We can then use the Grand Orthogonality Theorem to write
	\begin{align*}
		\delta\bigg(\partial(e_p)\prod_{\substack{\text{edge }i \text{ in } \\ \text{boundary}(p)}} g_i^{\sigma_i}, 1_G \bigg) &= \sum_{\text{irreps }R \text{ of }G} \frac{|R|}{|G|} \chi_R \big( \partial(e_p) \hspace{-0.3cm}\prod_{\substack{\text{edge }i \text{ in } \\ \text{boundary}(p)}} \hspace{-0.4cm} g_i^{\sigma_i}\big).
	\end{align*}
	
	Substituting this into Equation \ref{Equation_plaquette_term_irrep_basis_1_appendix}, we have
	\begin{align}
		B_p&\ket{ \mu , \set{R_i,a_i,b_i}} \notag \\
		&= \sum_{\set{g_i \in G}} \bigg( \prod_{\substack{\text{edge }i \text{ in } \\ \text{boundary}(p)}} \sqrt{\frac{|R|}{|G|}} [D^{R_i}(g_i)]_{a_i,b_i} \bigg) \sum_{e_p \in E} \sqrt{\frac{1}{|E|}} \mu(e_p) \sum_{\text{irreps }R \text{ of }G} \frac{|R|}{|G|} \chi_R\big( \partial(e_p)\prod_{\substack{\text{edge }i \text{ in } \\ \text{boundary}(p)}} \hspace{-0.4cm} g_i^{\sigma_i}\big) \ket{e_p, \set{g_i}} \notag \\
		&= \sum_{\set{g_i \in G}} \bigg( \prod_{\substack{\text{edge }i \text{ in } \\ \text{boundary}(p)}} \sqrt{\frac{|R|}{|G|}} [D^{R_i}(g_i)]_{a_i,b_i} \bigg) \sum_{e_p \in E} \sqrt{\frac{1}{|E|}} \mu(e_p) \sum_{\text{irreps }R \text{ of }G} \frac{|R|}{|G|} \sum_{c=1}^{|R|} \big[D^R( \partial(e_p) \hspace{-0.2cm}\prod_{\substack{\text{edge }i \text{ in } \\ \text{boundary}(p)}} \hspace{-0.4cm} g_i^{\sigma_i})\big]_{cc}\ket{e_p, \set{g_i}}. \label{Equation_plaquette_term_irrep_basis_2_appendix}
	\end{align}
	
	We can then split the expression 
	$$\sum_{c=1}^{|R|} [D^R( \partial(e_p)\prod_{\substack{\text{edge }i \text{ in } \\ \text{boundary}(p)}}g_i^{\sigma_i})]_{cc}$$
	into a product of matrices, with one matrix for each degree of freedom. We start by separating off the part corresponding to $e_p$:
	\begin{align*}
		\sum_{c=1}^{|R|} [D^R( \partial(e_p)\prod_{\substack{\text{edge }i \text{ in } \\ \text{boundary}(p)}} \hspace{-0.4cm}g_i^{\sigma_i})]_{cc}
		&= \sum_{c=1}^{|R|} \sum_{c_0=1}^{|R|} [D^R( \partial(e_p))]_{cc_0} [D^R(\prod_{\substack{\text{edge }i \text{ in } \\ \text{boundary}(p)}} \hspace{-0.4cm}g_i^{\sigma_i})]_{c_0 c}.
	\end{align*}

	Using Clifford's theorem to write $[D^R( \partial(e_p))]_{cc_0} = R_{\partial}(\partial(e_p)) \delta_{cc_0} = \mu^R(e_p) \delta_{cc_0} $ (as described earlier in this section), we have
	\begin{align*}
		\sum_{c=1}^{|R|} [D^R( \partial(e_p)\prod_{\substack{\text{edge }i \text{ in } \\ \text{boundary}(p)}}\hspace{-0.4cm}g_i^{\sigma_i})]_{cc}
		&= \sum_{c=1}^{|R|} \sum_{c_0=1}^{|R|} R_{\partial}(\partial(e_p)) \delta_{cc_0} [D^R(\prod_{\substack{\text{edge }i \text{ in } \\ \text{boundary}(p)}}\hspace{-0.4cm}g_i^{\sigma_i})]_{c_0 c}\\
		&=\sum_{c=1}^{|R|} \mu^R(e_p)[D^R(\prod_{\substack{\text{edge }i \text{ in } \\ \text{boundary}(p)}}\hspace{-0.4cm}g_i^{\sigma_i})]_{c c}.
	\end{align*}
	
	Then we split off the contributions from the individual $g_i$:
	\begin{align*}
		\sum_{c=1}^{|R|} [D^R( \partial(e_p)\prod_{\substack{\text{edge }i \text{ in } \\ \text{boundary}(p)}} \hspace{-0.4cm}g_i^{\sigma_i})]_{cc}
		&= \mu^R(e_p) \sum_{\set{c_i}} \prod_{\substack{\text{edge }i \text{ in } \\ \text{boundary}(p)}} \hspace{-0.4cm}[D^R(g_i^{\sigma_i})]_{c_i c_{i+1}},
	\end{align*}
	where $c_1=c$ and $c_{N+1}=c$, given that $N$ is the total number of edges in the boundary. Substituting this into Equation \ref{Equation_plaquette_term_irrep_basis_2_appendix} gives us
	\begin{align*}
		B_p\ket{ \mu , \set{R_i,a_i,b_i}}
		&= \sum_{\set{g_i \in G}} \bigg( \prod_{\substack{\text{edge }i \text{ in } \\ \text{boundary}(p)}} \sqrt{\frac{|R|}{|G|}} [D^{R_i}(g_i)]_{a_i,b_i} \bigg) \sum_{e_p \in E} \sqrt{\frac{1}{|E|}} \mu(e_p) \\
		&\hspace{0.5cm}\sum_{\text{irreps }R \text{ of }G} \frac{|R|}{|G|} \mu^R(e_p) \sum_{\set{c_i}} \bigg( \prod_{\substack{\text{edge }i \text{ in } \\ \text{boundary}(p)}} [D^R(g_i^{\sigma_i})]_{c_i c_{i+1}}\bigg) \ket{e_p, \set{g_i}}.
	\end{align*}
	
	We then group the terms corresponding to each group element $g_i$, to obtain
	\begin{align}
		B_p&\ket{ \mu, \set{R_i,a_i,b_i}} \notag\\
		&= \sum_{\text{irreps }R \text{ of }G} \sum_{\set{c_i}} \sum_{\set{g_i \in G}} \frac{|R|}{|G|} \bigg(\prod_{\substack{\text{edge }i \text{ in } \\ \text{boundary}(p)}} \sqrt{\frac{|R|}{|G|}} [D^{R_i}(g_i)]_{a_i,b_i} [D^R(g_i^{\sigma_i})]_{c_i c_{i+1}} \bigg) \sum_{e_p \in E} \sqrt{\frac{1}{|E|}} \mu(e_p) \mu^R(e_p) \ket{e_p, \set{g_i}}, \label{Equation_plaquette_term_irrep_basis_3_appendix}
	\end{align}
	as we claimed in Equation \ref{Equation_plaquette_term_irrep_basis_3}. When $G$ is Abelian, this becomes
	\begin{align}
		B_p&\ket{ \mu, \set{R_i,a_i,b_i}} \notag\\
		&= \sum_{\text{irreps }R \text{ of }G} \sum_{\set{g_i \in G}} \frac{1}{|G|} \bigg( \prod_{\substack{\text{edge }i \text{ in } \\ \text{boundary}(p)}} \sqrt{\frac{1}{|G|}} {R_i}(g_i) R(g_i^{\sigma_i}) \bigg) \sum_{e_p \in E} \sqrt{\frac{1}{|E|}} \mu(e_p) \mu^R(e_p) \ket{e_p, \set{g_i}} \notag\\
		&= \frac{1}{|G|} \sum_{\text{irreps }R \text{ of }G} \ket{\mu \cdot \mu^R, \set{R_i\cdot R^{\sigma_i}}}, \label{Equation_plaquette_term_irrep_basis_3_appendix_Abelian}
	\end{align}
	as we showed in Section \ref{Section_2D_irrep_basis} of the main text. We note that Equation \ref{Equation_plaquette_term_irrep_basis_3_appendix_Abelian} is equivalent to Equation \ref{Equation_Z4_Z4_irrep_plaquette} in Section \ref{Section_Z4_Z4_irrep_basis}.
	
	\section{Condensation and confinement when $G$ is Abelian but $\rhd$ is non-trivial}
	\label{Section_rhd_non-trivial_condense_confine}
	
	\subsection{Confinement of magnetic excitations}
	\label{Section_confined_magnetic}
	In the case of the $\mathbb{Z}_2$, $\mathbb{Z}_3$ model (with a simple lattice), we were able to construct the magnetic excitation (there is only one non-trivial one), in spite of the potential problems associated with taking $\rhd$ to be non-trivial. We found that the magnetic excitation is confined or unconfined depending on which ground state we raise these excitations from. Here we aim to generalize this to various types of crossed module. Rather than start with the most general case we can manage, we will first consider a simple case to introduce the ideas that we will need later.

	\subsubsection{The case where both groups are Abelian and $\partial \rightarrow 1_G$}
	\label{Section_magnetic_confinement_partial_trivial}
	A rather simple generalization of the $\mathbb{Z}_2$, $\mathbb{Z}_3$ model is one defined by a crossed module $(G, E, \partial \rightarrow 1_G, \rhd)$, where $G$ and $E$ are Abelian groups. We will use the same square lattice from the $\mathbb{Z}_2$, $\mathbb{Z}_3$ case, so we have four types of terms in the Hamiltonian to consider (from the vertices, the plaquettes, the horizontal edges and the vertical edges). We wish to use the basis where the plaquettes are labelled by irreps of $E$, rather than group elements, but the edges are still labelled by elements of $G$. In order to do this, we must use the usual change of basis
	$$\ket{\mu} =\frac{1}{\sqrt{|E|}} \sum_{e \in E} \mu(e) \ket{e}.$$

\begin{figure}[h]
		\begin{center}
			\hspace{-1cm}
			\begin{overpic}[width=0.5\linewidth]{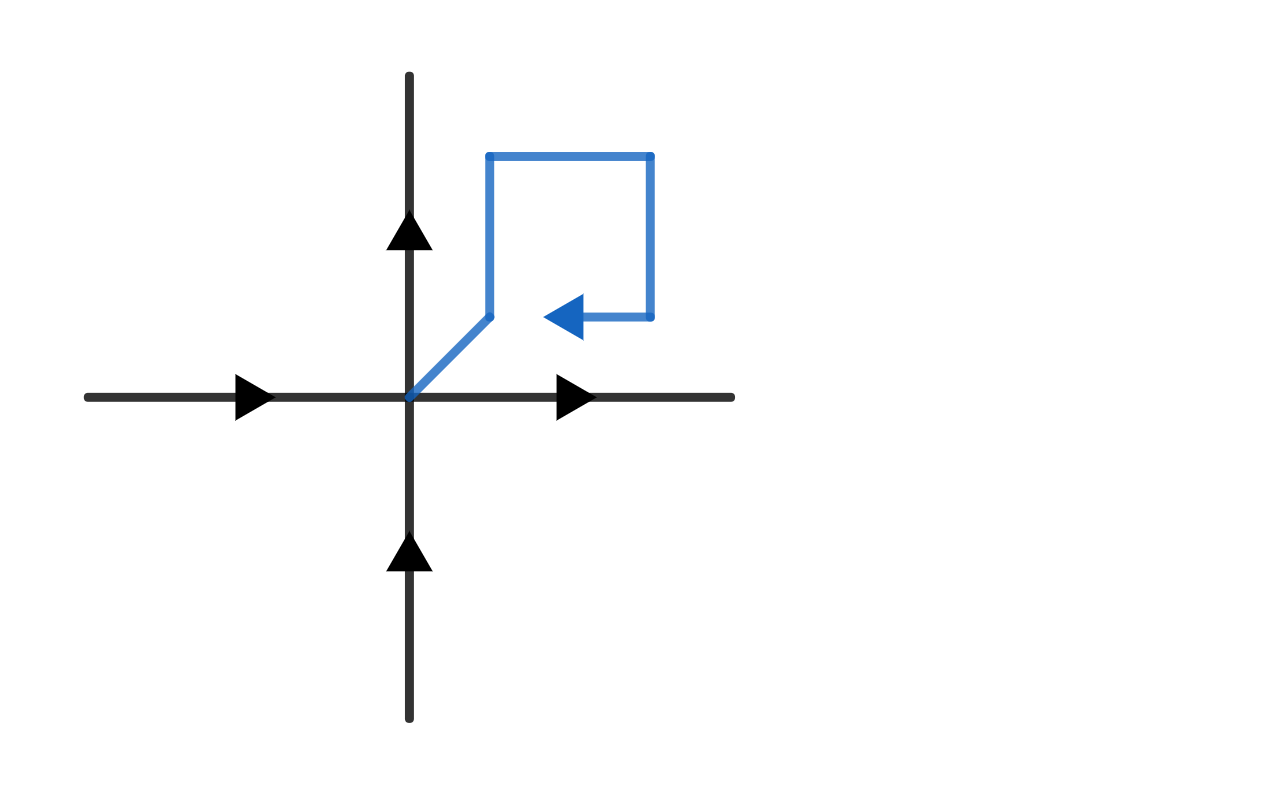}
				\put(71,33){\large $:=\ket{\mu,g_1,g_2,g_3,g_4}$}	
				\put(35,10){\large $g_4$}
				\put(15,36){\large $g_1$}
				\put(30,60){\large $g_2$}
				\put(43,43){\large $\mu$}
				\put(60,31){\large $g_3$}
			\end{overpic}
			\caption{We use this shorthand for the degrees of freedom affected by the vertex transform.}
			\label{support_vertex_1_E_Abelian}
		\end{center}
	\end{figure}

	We can use this expression to evaluate the different energy terms in the new basis. First consider the vertex transforms. We will use the short-hand given in Figure \ref{support_vertex_1_E_Abelian} to express the state of the degrees of freedom affected by the transform. Then we have
	\begin{align*}
		A_v^g \ket{\mu,g_1,g_2,g_3,g_4} &= \frac{1}{\sqrt{|E|}} \sum_{e \in E} \mu(e) A_v^g \ket{e,g_1,g_2,g_3,g_4}\\
		&=\frac{1}{\sqrt{|E|}} \sum_{e \in E} \mu(e) \ket{g \rhd e,g_1g^{-1},gg_2,gg_3,g_4g^{-1}}.
	\end{align*}
	
	We can then switch the dummy variable $e$ for $e' = g \rhd e$, using the fact that the map $g \rhd$ is one-to-one, to obtain
	\begin{align*}
		A_v^g \ket{\mu,g_1,g_2,g_3,g_4} &=\frac{1}{\sqrt{|E|}} \sum_{e' = g \rhd e \in E} \mu(g^{-1} \rhd e) \ket{e',g_1g^{-1},gg_2,gg_3,g_4g^{-1}}.
	\end{align*}
	
	Recalling from Section \ref{Section_2D_RO_Fake_Flat} of the main text that $\mu(g^{-1} \rhd e)$ can be written as $g^{-1} \rhd \mu(e)$ for some irrep $g^{-1} \rhd \mu$ of $E$, we can write the action of the vertex transform as
	\begin{align*}
		A_v^g \ket{\mu,g_1,g_2,g_3,g_4} &=\frac{1}{\sqrt{|E|}} \sum_{e' = g \rhd e \in E} g^{-1} \rhd \mu( e) \ket{e',g_1g^{-1},gg_2,gg_3,g_4g^{-1}}\\
		&= \ket{g^{-1} \rhd \mu,g_1g^{-1},gg_2,gg_3,g_4g^{-1}}.
	\end{align*}
	We note that this fluctuates the irrep label $\mu$, but only within a $\rhd$-Rep class (the class of irreps related by the $\rhd$ action of $G$).

	\begin{figure}[h]
		\begin{center}
			\hspace{-1cm}
			\begin{overpic}[width=0.5\linewidth]{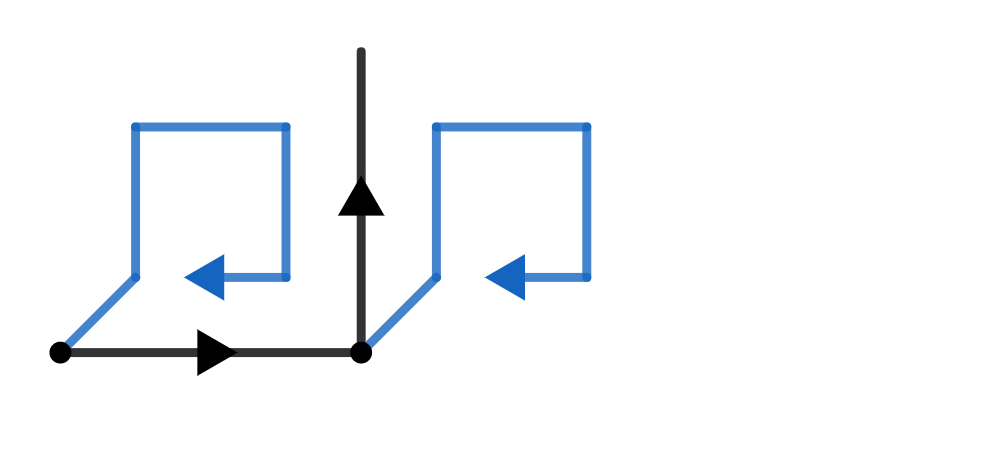}
				\put(70,25){\large $:=\ket{\mu_1,\mu_2,g,h}$}
				\put(35,46){\large $g$}
				\put(20,2){\large $h$}
				\put(19,27){\large $\mu_1$}
				\put(51,27){\large $\mu_2$}	
				
			\end{overpic}
			\caption{Shorthand for the degrees of freedom affected by the vertical edge transform}
			\label{vert_edge_op_E_Abelian}
		\end{center}
	\end{figure}

	Next we consider the vertical edge transform $\mathcal{A}_{\uparrow}$. We use the shorthand for the degrees of freedom around the edge illustrated in Figure \ref{vert_edge_op_E_Abelian}. Then we have
	\begin{align*}
		\mathcal{A}_{\uparrow} \ket{\mu_1, \mu_2, g, h } &= (\frac{1}{|E|} \sum_{f \in E} \mathcal{A}_{\uparrow}^f) \frac{1}{|E|} \sum_{e_1, e_2 \in E} \mu_1(e_1) \mu_2(e_2) \ket{e_1, e_2, g,h}\\
		&= \frac{1}{|E|} \sum_{f \in E} \frac{1}{|E|} \sum_{e_1, e_2 \in E} \mu_1(e_1) \mu_2(e_2) \ket{[h \rhd f] e_1, e_2f^{-1}, g,h},
	\end{align*}
	where we note that, because $\partial$ maps to the identity, the label of the edge on which we apply the edge transform is unchanged by the transform. Then we change the sums over dummy indices $e_1$ and $e_2$ in $E$ for sums over $e_1'= [h \rhd f] e_1$ and $e_2' = e_2 f^{-1}$ in $E$, to obtain
	\begin{align*}
		\mathcal{A}_{\uparrow} \ket{\mu_1, \mu_2, g, h } &= \frac{1}{|E|} \sum_{f \in E} \frac{1}{|E|} \sum_{e_1', e_2' \in E} \mu_1([h \rhd f^{-1}] e_1') \mu_2(e_2' f) \ket{e_1', e_2', g,h}\\
		&= \frac{1}{|E|} \sum_{f \in E} \frac{1}{|E|} \sum_{e_1', e_2' \in E} \mu_1([h \rhd f^{-1}]) \mu_1(e_1') \mu_2(e_2') \mu_2( f) \ket{e_1', e_2', g,h}\\
		&= \big(\frac{1}{|E|} \sum_{f \in E} \mu_1([h \rhd f^{-1}]) \mu_2( f) \big) \frac{1}{|E|} \sum_{e_1', e_2' \in E} \mu_1(e_1') \mu_2(e_2') \ket{e_1', e_2', g,h}\\
		&=\big(\frac{1}{|E|} \sum_{f \in E} h \rhd \mu_1(f^{-1}) \mu_2( f) \big) \ket{\mu_1, \mu_2, g, h }.
	\end{align*} 
	
	From the Grand Orthogonality Theorem for irreps, we can write
	$$\big(\frac{1}{|E|} \sum_{f \in E} h \rhd \mu_1(f^{-1}) \mu_2( f) \big) =\delta( h \rhd \mu_1, \mu_2),$$
	and so we finally have
	\begin{align*}
		\mathcal{A}_{\uparrow} \ket{\mu_1, \mu_2, g, h } &= \delta( h \rhd \mu_1, \mu_2) \ket{\mu_1, \mu_2, g, h },
	\end{align*}
	similar to the result from the $\mathbb{Z}_2, \mathbb{Z}_3$ case. Note in particular that this enforces that the adjacent plaquettes have irrep labels in the same $\rhd$-Rep class. We obtain an analogous result for the horizontal edge transform (the result for the horizontal edges has the same form as Equation \ref{Equation_Z2_Z3_irrep_basis_horizontal_edge_transform_1} from the $\mathbb{Z}_2$, $\mathbb{Z}_3$ case discussed in the main text). This just leaves us with the plaquette term, which does not act on the plaquette labels when $\partial$ maps to the identity.

	To recap, the vertex term fluctuates irrep labels, but only within a $\rhd$-Rep class, while the edge term forces the plaquettes adjacent to that edge to have labels in the same $\rhd$-Rep class in the ground state. Applying this condition sequentially on all of the edges fixes all plaquettes that are connected by paths in the dual lattice to be in the same $\rhd$-Rep class, as we discussed in the $\mathbb{Z}_2$, $\mathbb{Z}_3$ case in Section \ref{Section_Z2_Z3_Ground_States} in the main text (although in order to guarantee that this can be done in a consistent way we require the lattice to be simply connected). This means that we can construct ground states which are labelled by these $\rhd$-Rep classes: the fact that no energy terms fluctuate outside the class means that the class of any particular plaquette label is fixed, and the edge terms enforce that every plaquette in the (simply connected) lattice has label in the same $\rhd$-Rep class. The $\rhd$-Rep class labelling a ground state is the ground state property that we will use to determine whether magnetic excitations are confined or not in that particular ground state.

	\begin{figure}[h]
		\begin{center}
			\begin{overpic}[width=0.5\linewidth]{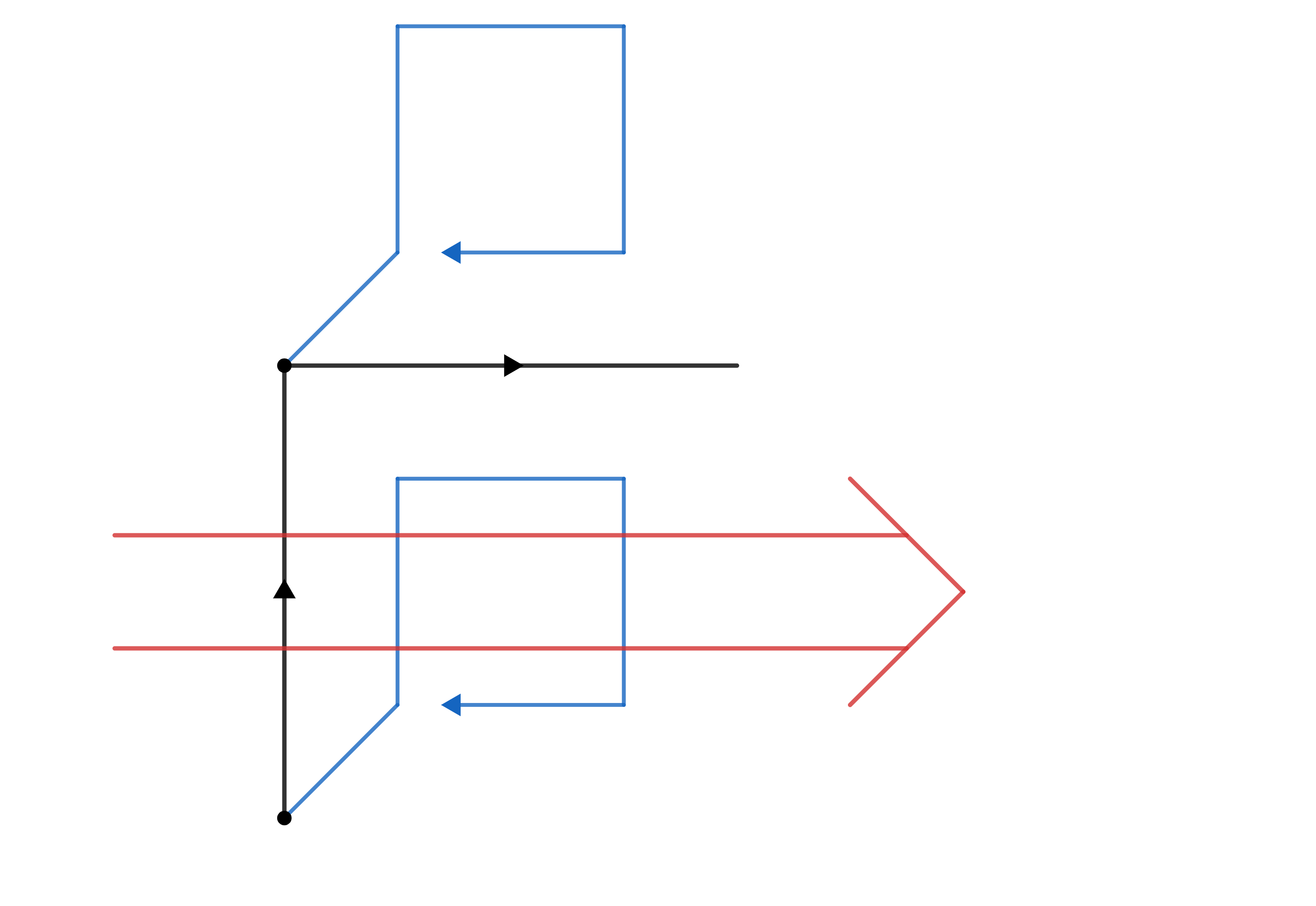}
				\put(37,57){$\mu_2$}
				\put(37,24){$\mu_1$}
				\put(18,30){$h$}
				\put(38,43){$g$}
				\put(70,18){\textcolor{red}{$C^x(t)$}}
				\put(70,40){$= \ket{ \mu_1, \mu_2, h, g}$}
			\end{overpic}
			\caption{The edge transforms that may fail to commute with the magnetic ribbon operator are those above the ribbon. Here we introduce a shorthand for the degrees of freedom affected by the transform.}
			\label{E_Abelian_magnetic_shorthand}
		\end{center}
	\end{figure}

\begin{figure}[h]
		\begin{center}
			\begin{overpic}[width=0.7\linewidth]{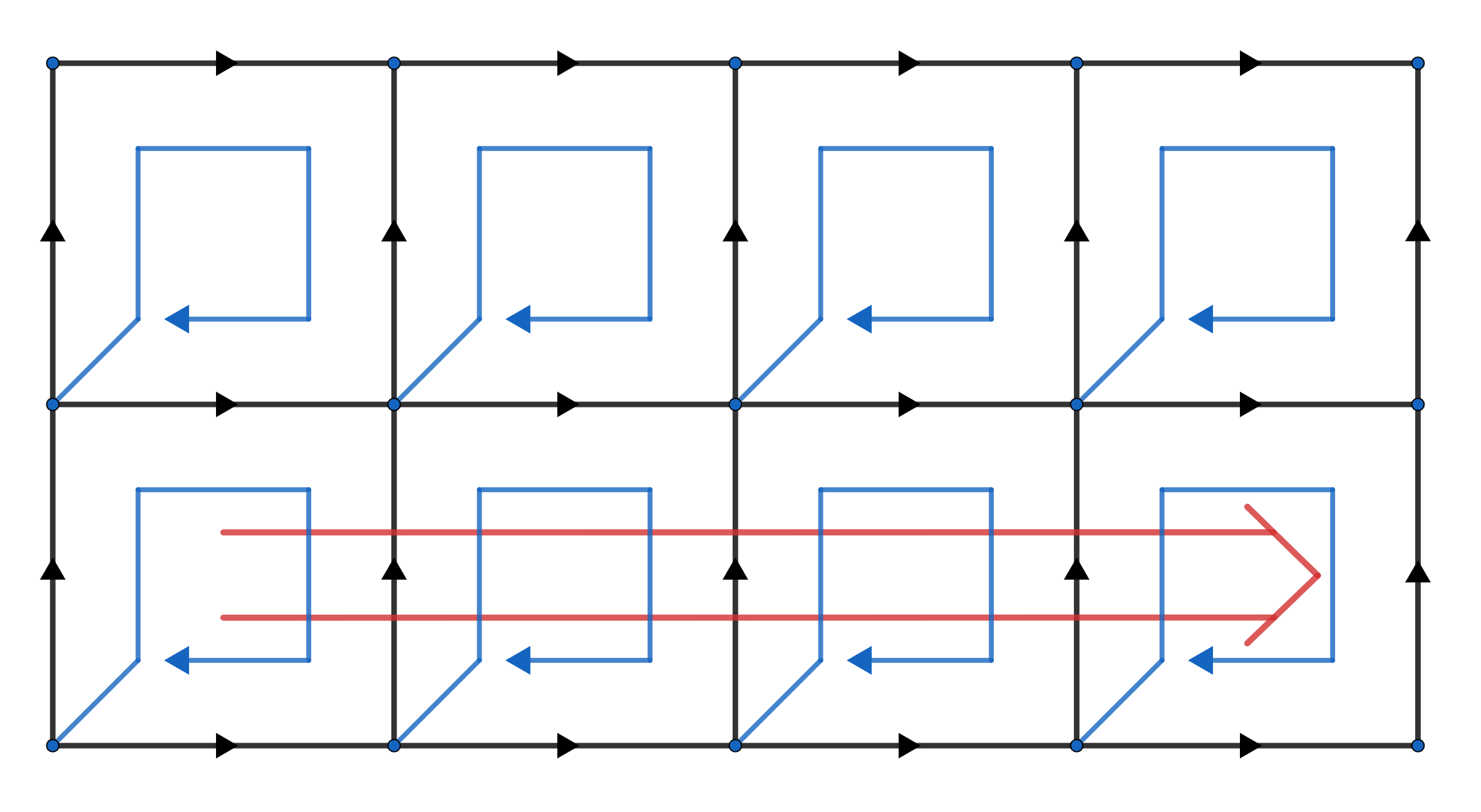}
				
				\put(27,22){$g_1 \rightarrow xg_1$}
				\put(50,22){$g_2 \rightarrow xg_2$}
				\put(73,22){$g_3 \rightarrow xg_3$}
				\put(87,12){\textcolor{red}{$C^x(t)$}}
			\end{overpic}
			\caption{For simplicity, we consider a magnetic ribbon operator $C^x(t)$ on a horizontal ribbon. Because $G$ is Abelian, the action of the ribbon operator only depends on the orientation of the edges cut by the ribbon (rather than on any direct path elements). In this case all of the edges point in the same way relative to the ribbon (the orientation of each edge can be obtained by rotating the local ribbon orientation ninety degrees anti-clockwise), so all of the edge labels are multiplied by $x$.}
			\label{Horizontal_ribbon_1}
		\end{center}
	\end{figure}

	In order to see which magnetic excitations are confined however, we must first examine the commutation relation between the magnetic ribbon operator (defined in the same way as in the $\rhd$ trivial case, as presented in Section \ref{Section_2D_Magnetic} of the main text) and the different energy terms. The magnetic ribbon operator only acts on the edges of the lattice, and will commute (or not) in the same way with the vertex transforms and plaquette terms as in the $\rhd$ trivial case (just as we discussed for the $\mathbb{Z}_2, \mathbb{Z}_3$ model). For simplicity, we consider a ribbon passing horizontally through the lattice, as shown in Figure \ref{Horizontal_ribbon_1}. 
	
	The terms we need to consider are the edge terms for which the path element $g(v_0(p)-s(i)$ (or $g(\overline{v_0(p)-s(i)})$) appearing in the action of the edge transform is affected by the magnetic ribbon operator. As we discussed in Section \ref{Section_Z2_Z3_Magnetic} of the main text for the $\mathbb{Z}_2$, $\mathbb{Z}_3$ case, this means that we must consider the edges above the ribbon (above and to the right of the edges cut by the ribbon operator). In Figure \ref{E_Abelian_magnetic_shorthand} we show the degrees of freedom affected by such an edge transform, and introduce a shorthand for their state. The action of the horizontal edge term on these degrees of freedom is
	$$\mathcal{A}_{\rightarrow} \ket{\mu_1, \mu_2,h, g} = \delta(h \rhd \mu_1, \mu_2) \ket{\mu_1, \mu_2,h, g},$$
	while the magnetic ribbon operator acts as
	$$C^x(t)\ket{\mu_1, \mu_2,h, g} =\ket{\mu_1, \mu_2,xh, g}.$$

	We want to consider the state
	$$\mathcal{A}_{\rightarrow} C^x(t) \mathcal{A}_{\rightarrow} \ket{\mu_1, \mu_2,h, g},$$
	which will give zero if $C^x(t)$ produces an edge excitation, and will give
	$$C^x(t) \mathcal{A}_{\rightarrow} \ket{\mu_1, \mu_2,h, g}$$
	if it does not. We have
	\begin{align*}
		\mathcal{A}_{\rightarrow}C^x(t) \mathcal{A}_{\rightarrow} \ket{\mu_1, \mu_2,h, g}&= 	\mathcal{A}_{\rightarrow}C^x(t) \delta(h \rhd \mu_1, \mu_2) \ket{\mu_1, \mu_2,h, g}\\
		&= \mathcal{A}_{\rightarrow} \delta(h \rhd \mu_1, \mu_2) \ket{\mu_1, \mu_2,xh, g}\\
		&= \delta((xh) \rhd \mu_1, \mu_2)\delta(h \rhd \mu_1, \mu_2) \ket{\mu_1, \mu_2,xh, g}.
	\end{align*}
	
	We can write this as
	\begin{align}
		\mathcal{A}_{\rightarrow}C^x(t) \mathcal{A}_{\rightarrow} \ket{\mu_1, \mu_2,h, g}&= \delta((xh) \rhd \mu_1, h \rhd \mu_1)\delta(h \rhd \mu_1, \mu_2) \ket{\mu_1, \mu_2,xh, g} \notag\\
		&= \delta((xh) \rhd \mu_1, h \rhd \mu_1) C^x(t) \mathcal{A}_{\rightarrow} \ket{\mu_1, \mu_2,h, g}. \label{Equation_Abelian_confined_magnetic}
	\end{align}
	We can simplify this slightly by noting that $(xh) \rhd \mu_1 = h \rhd (x \rhd \mu_1)$ (see Equation \ref{rhd_irrep_composition} in Section \ref{Section_Example_Z_2_Z_3} of the main text), and because the $\rhd$ map is invertible
	$$\delta((xh) \rhd \mu_1, h \rhd \mu_1) = \delta(h \rhd (x \rhd \mu_1), h \rhd \mu_1) = \delta(x \rhd \mu_1, \mu_1).$$
	The edge is therefore excited if $\delta(x \rhd \mu_1, \mu_1) =0$ and unexcited if $\delta(x \rhd \mu_1, \mu_1) =1$. We aim to show that the ground states labelled by $\rhd$-Rep class are eigenstates of the Kronecker delta, which in turn implies that the states produced by acting on these ground states with the ribbon operator are eigenstates of the edge term from Equation \ref{Equation_Abelian_confined_magnetic}. To demonstrate this, we will show that $\delta((xh) \rhd \mu_1, h \rhd \mu_1)$ depends on which $\rhd$-Rep class $\mu_1$ is in, but not on which irrep within that class $\mu_1$ is. Because the $\rhd$-Rep class is well defined for each basis ground state, this will imply that the Kronecker delta is also well defined for the ground state. Suppose that, for some irrep $\mu_I$ in the same class as $\mu_1$, that we do indeed have $x \rhd \mu_I =\mu_I$. Then any other irrep in the class can be written as $\mu_g= g \rhd \mu_I$ for some $g \in G$. We then have 
	$$x \rhd \mu_g =x \rhd (g \rhd \mu_I) = (gx) \rhd \mu_I.$$
	
	Because $G$ is Abelian we then have
	$$x \rhd \mu_g =(xg) \rhd \mu_I= g \rhd (x \rhd \mu_I).$$
	Then because $\mu_I$ does satisfy $x \rhd \mu_I =\mu_I$, we have
	$$x \rhd \mu_g = g \rhd \mu_I= \mu_g,$$
	and so $\mu_g$ satisfies this condition. Therefore, if any irrep in the class satisfies the condition, then they all do. We say that a $\rhd$-Rep class satisfying $x \rhd \mu =\mu$ for all its members is stabilised by $x$. Conversely, if any irrep does not satisfy the condition then none of them do. This means that whether the edge is excited or not does not depend on which irrep within the class the plaquettes adjacent to the edge are labelled by. Whether the edge is excited or not only depends on the $\rhd$-Rep class, and so only on which ground state we act on with the ribbon operator. This holds for all of the (horizontal) edges just above the ribbon (like the one labelled by $g$ in Figure \ref{E_Abelian_magnetic_shorthand}), and so we either get a confining string or not depending on the ground state. Note however that not all of the magnetic excitations are simultaneously confined: a particular $\rhd$-Rep class could be stabilised by some labels $x$ of magnetic ribbon operator but not by others, leading to interesting patterns of confinement.

	Before we generalize to the next case, we'd like to briefly describe what happens when we take $G$ to be non-Abelian (even if we restrict $\partial$ to map to $1_G$). In this case, even if one irrep $\mu_I$ in a $\rhd$-Rep class does satisfy the condition $x \rhd \mu_I =\mu_I$, it does not guarantee that another will. This is because the $\rhd$ actions of different elements of $G$ no longer commute. For an irrep $\mu_g =g \rhd \mu_I$, we have
	$$x \rhd \mu_g = x \rhd (g \rhd \mu_I) = (gx) \rhd \mu_I,$$
	which does not in general equal $g \rhd (x \rhd \mu_I)$. Instead we have
	$$x \rhd \mu_g = (xx^{-1}gx) \rhd \mu_I = (x^{-1}gx) \rhd (x \rhd \mu_I) = (x^{-1}gx) \rhd \mu_I = (g^{-1}x^{-1}gx) \rhd \mu_g.$$
	Some $\rhd$-Rep classes may have the property that this always gives $\mu_g$ (it holds if any irrep $\mu_I$ in the class is stabilised by all elements in the conjugacy class of $x$), but this does not seem to be generic. Because the vertex terms fluctuate the irrep labels within the entire $\rhd$-Rep class, this means that generally some irrep labels in the ground state give an excited edge and others do not, leading to indefinite confinement. 
	
	\subsubsection{The case where $G$ is Abelian but $E$ is non-Abelian and $\partial$ is general}
	
	While taking $G$ to be non-Abelian may present problems for determining the confinement of the magnetic excitations, taking $E$ to be non-Abelian (which also necessitates generalizing $\partial$) does not. When $E$ is Abelian, the plaquettes are labelled by the (now generally higher-dimensional) irreps of $E$, along with two matrix indices. The change of basis from the group element basis to the irrep basis is now given by
	$$\ket{\mu,a,b} = \sqrt{\frac{|\mu|}{|E|}} \sum_{e \in E} [D^{\mu}(e)]_{ab} \ket{e}.$$
	
	\begin{figure}[h]
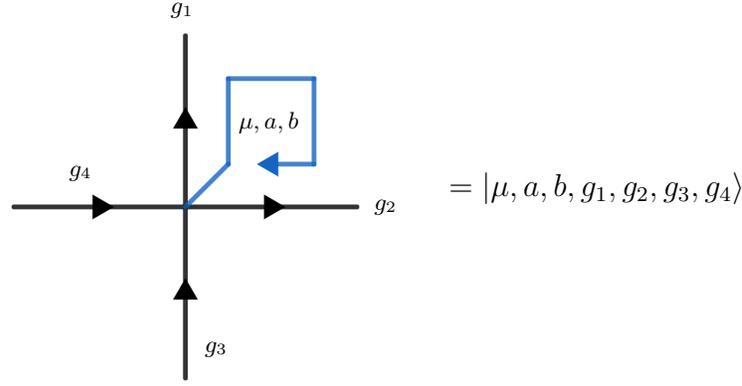

		\begin{center}
			\hspace{-1cm}
			\begin{overpic}[width=0.5\linewidth]{2D_support_vertex_2_image}
				\put(71,33){\large $=\ket{\mu,a,b,g_1,g_2,g_3,g_4}$}	
				\put(35,10){$g_3$}
				\put(15,36){$g_4$}
				\put(30,60){$g_1$}
				\put(40,43){$\mu,a,b$}
				\put(60,31){$g_2$}
			\end{overpic}
			\caption{Again we introduce a shorthand for the degrees of freedom affected by the vertex transform. Now that $E$ is non-Abelian, the plaquette is labelled by a (generally higher dimensional) irrep and two matrix indices.}
			\label{support_vertex_1_E_non_Abelian}
		\end{center}
	\end{figure}
	
	Now we want to consider how the various terms in the Hamiltonian act on this basis, starting with the vertex transforms. Denoting the degrees of freedom around the vertex by $\ket{\mu_1,a,b, g_1,g_2,g_3,g_4}$, as shown in Figure \ref{support_vertex_1_E_non_Abelian}, we have
	\begin{align*}
		A_v^x \ket{\mu_1,a,b, g_1,g_2,g_3,g_4} &= \sqrt{\frac{|\mu|}{|E|}} \sum_{e \in E} [D^{\mu}(e)]_{ab} A_v^x \ket{e,g_1,g_2,g_3,g_4}\\
		&=\sqrt{\frac{|\mu|}{|E|}} \sum_{e \in E} [D^{\mu}(e)]_{ab} \ket{x \rhd e,xg_1,xg_2,g_3x^{-1},g_4x^{-1}}.
	\end{align*}
	
	Replacing the sum over $e \in E$ with a sum over $e' = x \rhd e$ gives us
	\begin{align*}
		A_v^x \ket{\mu_1,a,b, g_1,g_2,g_3,g_4}&=\sqrt{\frac{|\mu|}{|E|}} \sum_{e' \in E} [D^{\mu}(x^{-1} \rhd e')]_{ab} \ket{e',xg_1,xg_2,g_3x^{-1},g_4x^{-1}}\\
		&= \ket{x^{-1} \rhd \mu,a,b,xg_1,xg_2,g_3x^{-1},g_4x^{-1}},
	\end{align*}
	where we used the fact that $[D^{\mu}(x^{-1} \rhd e')]_{ab}$ can be written as $[D^{x^{-1} \rhd \mu}( e')]_{ab}$, where $x^{-1} \rhd \mu$ is also an irrep of the same dimension (it is guaranteed to be irreducible because the new irrep $x^{-1} \rhd \mu$ gives us the same set of matrices as $\mu$, although which matrix is assigned to which group element may change). It is worth noting here that, because the irreps are generally higher-dimensional, when choosing our basis we must choose a representative for each set of irreps that are related by conjugation. This means that $x^{-1} \rhd \mu$ may not be one of the chosen representatives and so $\ket{x^{-1} \rhd \mu,a,b}$ may be some linear combination of our basis vectors, rather than a basis state itself. Thankfully we are not interested in the full irrep, but instead how the irrep restricts to the kernel of $\partial$. Because the kernel is always in the centre of $E$ (from the second Peiffer condition, given in Equation \ref{Peiffer_2} in the main text), we know from Schur's lemma that the full irrep will give a scalar multiple of the identity when restricted to the kernel. That is, for an element $e_k$ in the kernel of $\partial$,
	$$[D^{\mu}(x^{-1} \rhd e_k)]_{ab} = \delta_{ab} [D^{\mu}(x^{-1} \rhd e_k)]_{11},$$
	and furthermore $[D^{\mu}(x^{-1} \rhd e_k)]_{11}$ will be a 1D irrep of the kernel of $\partial$, which we call $\mu^{\text{ker}}$. Then
	$$[D^{\mu}(x^{-1} \rhd e_k)]_{ab} = \delta_{ab} \: \mu^{\text{ker}}(x^{-1} \rhd e_k) = \delta_{ab} \: x^{-1} \rhd \mu^{\text{ker}}(e_k),$$
	from which we see that the irrep of the kernel is only changed within a $\rhd$-Rep class. Even if $\ket{x^{-1} \rhd \mu,a,b}$ must be written in terms of another (equivalent) irrep that belongs our original basis, that labelling irrep will still restrict to the same irrep of the kernel of $\partial$.

	\begin{figure}[h]
		\begin{center}
			\hspace{-1cm}
			\begin{overpic}[width=0.3\linewidth]{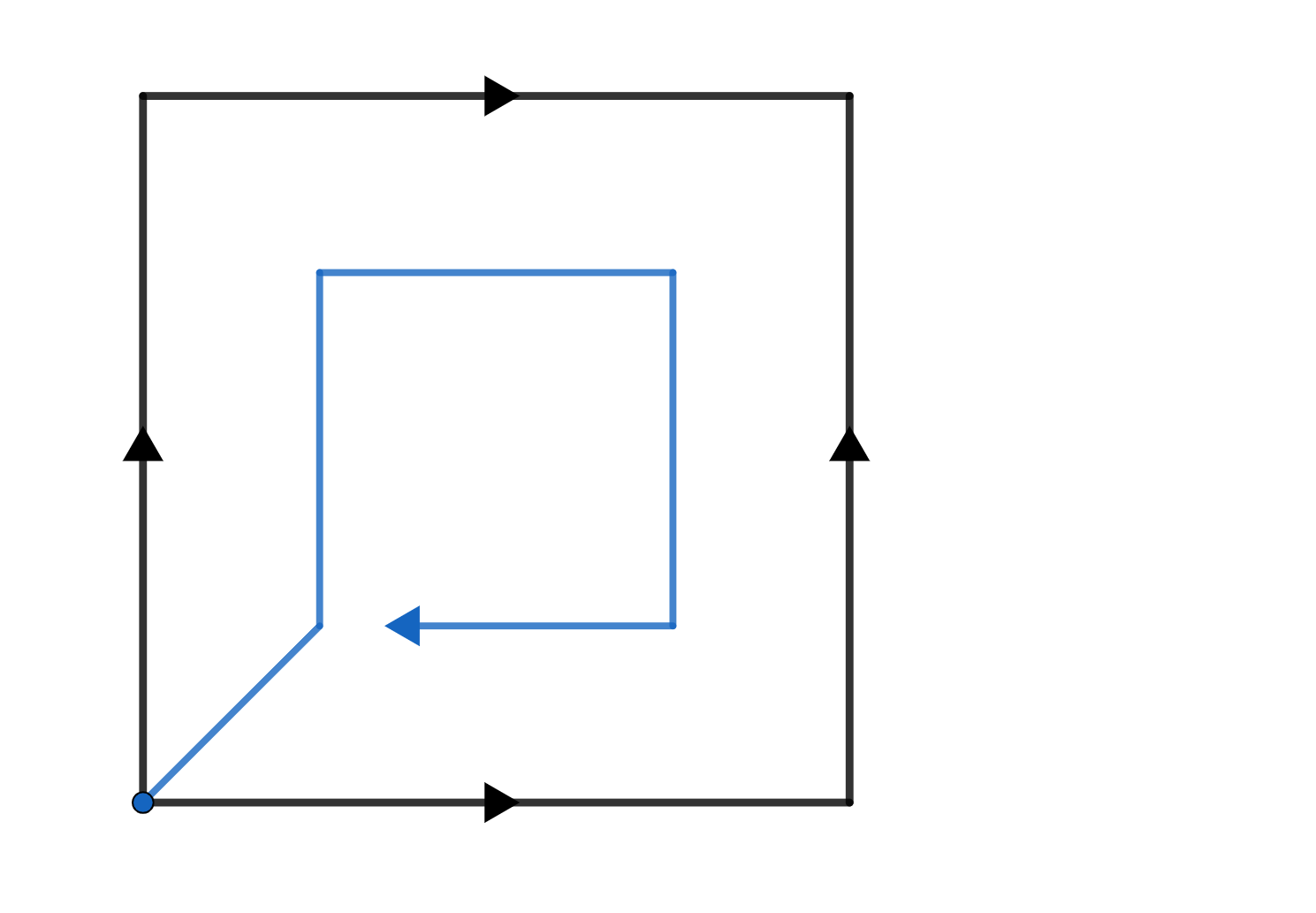}
				\put(31,35){$\mu,a,b$}
				\put(4,35){$g_1$}
				\put(35,66){$g_2$}
				\put(68,35){$g_3$}
				\put(35,4){$g_4$}
				\put(80,35){\large$= \ket{\mu,a,b, g_1,g_2,g_3,g_4}$}
			\end{overpic}
			\caption{Because $\partial$ does not just map to the identity, the plaquette term acts non-trivially on the plaquette labels. Here we introduce notation to use in our discussion of this term.}
			\label{support_plaquette_E_non_Abelian}
		\end{center}
	\end{figure}
	
	Next we consider the plaquette terms. Now that $\partial$ does not map only to the identity of $G$, the plaquette term acts non-trivially on the plaquette label itself. Using the notation for the degrees of freedom affected by the plaquette term shown in Figure \ref{support_plaquette_E_non_Abelian}, we have
	\begin{align*}
		B_p \ket{\mu,a,b, g_1,g_2,g_3,g_4} & = \sqrt{\frac{|\mu|}{|E|}} \sum_{e \in E} [D^{\mu}(e)]_{ab} B_p \ket{e, g_1,g_2,g_3,g_4}\\
		&= \sqrt{\frac{|\mu|}{|E|}} \sum_{e \in E} [D^{\mu}(e)]_{ab} \delta( \partial(e)g_1g_2g_3^{-1}g_4^{-1},1_G)\ket{e, g_1,g_2,g_3,g_4}.
	\end{align*}
	
	In order to separate off the action on the irrep label, we need to look at $\delta( \partial(e)g_1g_2g_3^{-1}g_4^{-1},1_G)$. We can manipulate this term by introducing a resolution of the identity:
	$$\delta( \partial(e)g_1g_2g_3^{-1}g_4^{-1},1_G) = \big( \sum_{f \in \partial(E)} \delta(\partial(e),f^{-1}) \big) \delta( \partial(e)g_1g_2g_3^{-1}g_4^{-1},1_G).$$
	Then we have
	\begin{align*}
		\delta( \partial(e)g_1g_2g_3^{-1}g_4^{-1},1_G) = \sum_{f \in \partial(E)} \delta(\partial(e)f,1_G) \delta(g_1g_2g_3^{-1}g_4^{-1},f).
	\end{align*}
	
	We can then use the column orthogonality relation for irreps, using specifically the irreps of $\partial(E)$ (which are all 1D because this group is Abelian), to write
	$$\delta(\partial(e)f,1_G) = \frac{1}{|\partial(E)|} \sum_{\substack{ \text{irreps } R^{\partial} \\ \text{ of }\partial(E)}} R^{\partial}(\partial(e)f).$$
	This then gives us
	\begin{align*}
		\delta( \partial(e)g_1g_2g_3^{-1}g_4^{-1},1_G) &= \sum_{f \in \partial(E)} \frac{1}{|\partial(E)|} \sum_{\substack{ \text{irreps } R^{\partial} \\ \text{ of }\partial(E)}} R^{\partial}(\partial(e)f) \delta(g_1g_2g_3^{-1}g_4^{-1},f)\\
		&= \sum_{f \in \partial(E)} \frac{1}{|\partial(E)|} \sum_{\substack{ \text{irreps } R^{\partial} \\ \text{ of }\partial(E)}} R^{\partial}(\partial(e)) R^{\partial}(f) \delta(g_1g_2g_3^{-1}g_4^{-1},f).
	\end{align*}
	The action of the plaquette term can therefore be written as
	\begin{align*}
		B_p \ket{\mu,a,b, g_1,g_2,g_3,g_4} & = \sqrt{\frac{|\mu|}{|E|}} \sum_{e \in E} [D^{\mu}(e)]_{ab} \sum_{f \in \partial(E)} \frac{1}{|\partial(E)|} \sum_{\substack{ \text{irreps } R^{\partial} \\ \text{ of }\partial(E)}} R^{\partial}(\partial(e)) R^{\partial}(f) \delta(g_1g_2g_3^{-1}g_4^{-1},f) \ket{e, g_1,g_2,g_3,g_4}\\
		&= \sum_{f \in \partial(E)} \frac{1}{|\partial(E)|} \sum_{\substack{ \text{irreps } R^{\partial} \\ \text{ of }\partial(E)}} R^{\partial}(f) \delta(g_1g_2g_3^{-1}g_4^{-1},f) \sqrt{\frac{|\mu|}{|E|}} \sum_{e \in E} [D^{\mu}(e)]_{ab} R^{\partial}(\partial(e)) \ket{e, g_1,g_2,g_3,g_4}.
	\end{align*}
	
	We can use $R^{\partial}$ to define an irrep $\mu^R$ of $E$ by $\mu^R(e) = R^{\partial}(\partial(e))$ for all $e \in E$. $\mu^R$ is guaranteed to be an irrep of $E$ because it is one-dimensional (and is a representation due to the fact that $\partial$ is a group homomorphism). Then 
	$$[D^{\mu}(e)]_{ab} R^{\partial}(\partial(e))=[D^{\mu}(e)]_{ab} \mu^R(e)$$ 
	is the fusion of the irreps $\mu$ and $\mu^R$ (or rather a particular matrix element of the fusion), and this fused irrep $\mu \cdot \mu^R$ will have the same dimension as $\mu$ (because $\mu^R$ only introduces a phase, due to the fact that it is a 1D irrep). We can therefore write the action of the plaquette term as
	\begin{align*}
		B_p \ket{\mu,a,b, g_1,g_2,g_3,g_4} &= \sum_{f \in \partial(E)} \frac{1}{|\partial(E)|} \sum_{\substack{ \text{irreps } R^{\partial} \\ \text{ of }\partial(E)}} R^{\partial}(f) \delta(g_1g_2g_3^{-1}g_4^{-1},f) \sqrt{\frac{|\mu|}{|E|}} \sum_{e \in E} [D^{\mu \cdot \mu^R}(e)]_{ab} \ket{e, g_1,g_2,g_3,g_4}\\
		&=\sum_{f \in \partial(E)} \frac{1}{|\partial(E)|} \sum_{\substack{ \text{irreps } R^{\partial} \\ \text{ of }\partial(E)}} R^{\partial}(f) \delta(g_1g_2g_3^{-1}g_4^{-1},f) \ket{\mu \cdot \mu^R,a,b, g_1,g_2,g_3,g_4}.
	\end{align*}
	
	We see that the plaquette term fluctuates the plaquette label by irreps $\mu^R$, but because these irreps were derived from $\partial(E)$ they do not affect the kernel of $\partial$, and so do not change the restriction of the irrep label to the kernel. That is, for an element $e_k$ of the kernel,
	\begin{align*}
		[D^{\mu \cdot \mu^R}(e_k)]_{ab} &= [D^{\mu}(e_k)]_{ab} R^{\partial}(\partial(e_k))\\
		&= [D^{\mu}(e_k)]_{ab} R^{\partial}(1_G)\\
		&=[D^{\mu}(e_k)]_{ab} .
	\end{align*}
	This means that none of the terms we have seen so far fluctuate the part of the plaquette label corresponding to an irrep of the kernel, except within a $\rhd$-Rep class.

	\begin{figure}[h]
		\begin{center}
			\begin{overpic}[width=0.4\linewidth]{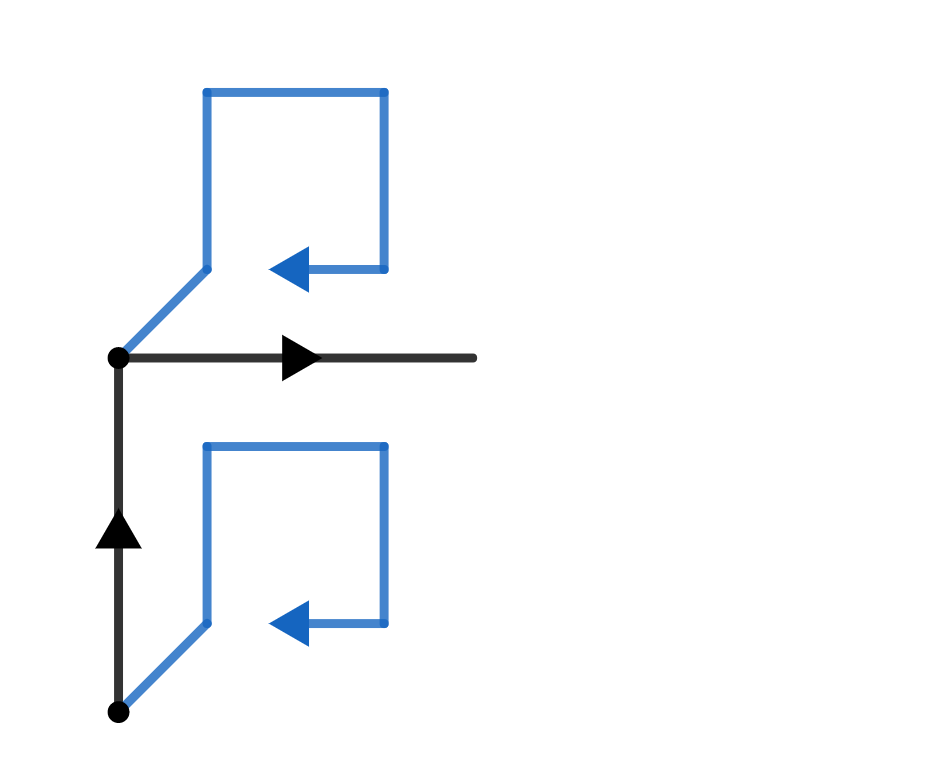}
				\put(23,25){$\mu_1,a_1,b_1$}
				\put(23,62){$\mu_2, a_2, b_2$}
				\put(8,30){$h$}
				\put(38,41){$g$}
				
				\put(63,44){\large $= \ket{ \mu_1, a_1, b_1, \mu_2, a_2, b_2, g, h}$}
			\end{overpic}
			\caption{We use this shorthand for the degrees of freedom affected by the horizontal edge transform}
			\label{E_non_Abelian_horizontal_edge_shorthand}
		\end{center}
	\end{figure}

\begin{figure}[h]
		\begin{center}
			\begin{overpic}[width=0.5\linewidth]{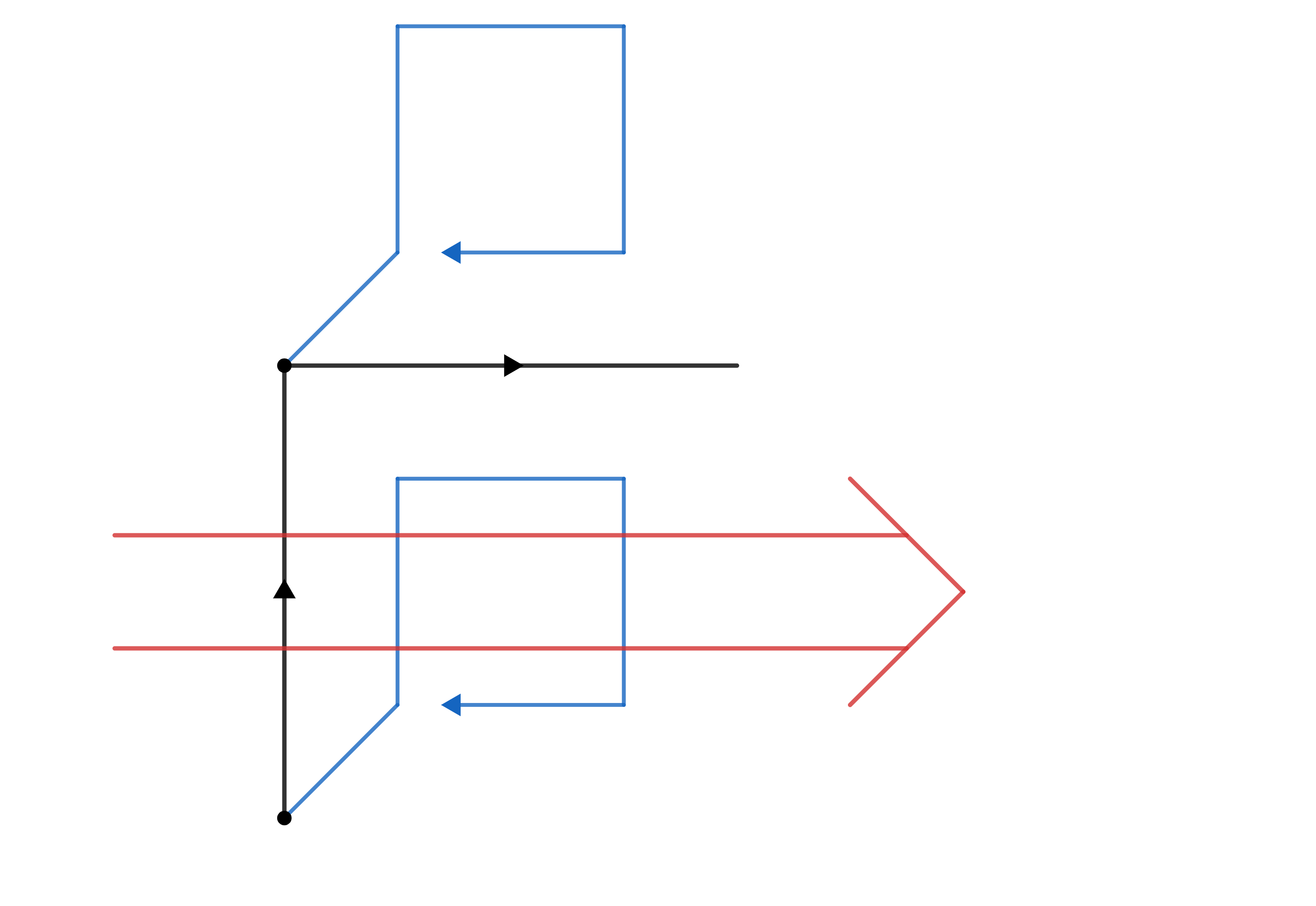}
				\put(31,24){$\mu_1,a_1,b_1$}
				\put(31,57){$\mu_2, a_2, b_2$}
				\put(18,30){$h$}
				\put(38,43){$g$}
				\put(70,18){\textcolor{red}{$C^x(t)$}}
				\put(63,40){\large $= \ket{ \mu_1, a_1, b_1, \mu_2, a_2, b_2, g, h}$}
			\end{overpic}
			\caption{The edge transforms that may fail to commute with the magnetic ribbon operator are those above the ribbon. Here we introduce a shorthand for the degrees of freedom affected by the transform.}
			\label{E_non_Abelian_magnetic_shorthand}
		\end{center}
	\end{figure}

	Finally, consider the edge terms, in particular the ones associated to the horizontal edges. As usual we will introduce a shorthand for the degrees of freedom affected by the energy term, shown in Figure \ref{E_non_Abelian_magnetic_shorthand}. Then we have
	\begin{align*}
		\mathcal{A}_i^f \ket{\mu_1,a_1,b_1, \mu_2, a_2, b_2, g, h} &= \frac{\sqrt{|\mu_1| | \mu_2|}}{|E|} \sum_{e_1, e_2 \in E} [D^{\mu_1}(e_1)]_{a_1b_1} [D^{\mu_2}(e_2)]_{a_2b_2} \mathcal{A}_i^f \ket{e_1, e_2, g, h}\\
		&=\frac{\sqrt{|\mu_1| | \mu_2|}}{|E|} \sum_{e_1, e_2 \in E} [D^{\mu_1}(e_1)]_{a_1b_1} [D^{\mu_2}(e_2)]_{a_2b_2} \ket{e_1 [h \rhd f^{-1}], fe_2, \partial(f)g, h}.
	\end{align*}
	
	We then use the standard trick of changing variables, from $e_1$ and $e_2$ to $e_1'=e_1 [h \rhd f^{-1}]$ and $e_2' = fe_2$. This gives us
	\begin{align*}
		\mathcal{A}_i^f \ket{\mu_1,a_1,b_1, \mu_2, a_2, b_2, g, h} &= \frac{\sqrt{|\mu_1| | \mu_2|}}{|E|} \sum_{e_1', e_2' \in E} [D^{\mu_1}(e_1' [h \rhd f])]_{a_1b_1} [D^{\mu_2}(f^{-1}e_2')]_{a_2b_2} \ket{e_1', e_2', \partial(f)g, h}.
	\end{align*}
	We can then separate the matrices into the contributions from $e_1'$ and $e_2'$ and the contributions from $f$, to obtain
	\begin{align}
		\mathcal{A}_i^f &\ket{\mu_1,a_1,b_1, \mu_2, a_2, b_2, g, h}\notag \\
		&= \frac{\sqrt{|\mu_1| | \mu_2|}}{|E|} \sum_{e_1', e_2' \in E} \sum_{c_1=1}^{|\mu_1|} [D^{\mu_1}(e_1')]_{a_1c_1} [D^{\mu_1}([h \rhd f])]_{c_1b_1} \sum_{c_2=1}^{|\mu_2|}[D^{\mu_2}(f^{-1})]_{a_2c_2}[D^{\mu_2}(e_2')]_{c_2b_2} \ket{e_1', e_2', \partial(f)g, h} \notag \\
		&=\sum_{c_1=1}^{|\mu_1|} \sum_{c_2=1}^{|\mu_2|} [D^{\mu_1}([h \rhd f])]_{c_1b_1} [D^{\mu_2}(f^{-1})]_{a_2c_2} \frac{\sqrt{|\mu_1| | \mu_2|}}{|E|} \sum_{e_1', e_2' \in E} [D^{\mu_1}(e_1')]_{a_1c_1} [D^{\mu_2}(e_2')]_{c_2b_2} \ket{e_1', e_2', \partial(f)g, h} \notag \\
		&=\sum_{c_1=1}^{|\mu_1|} \sum_{c_2=1}^{|\mu_2|} [D^{\mu_1}([h \rhd f])]_{c_1b_1} [D^{\mu_2}(f^{-1})]_{a_2c_2} \ket{\mu_1,a_1,c_1, \mu_2, c_2, b_2, \partial(f)g, h} \label{E_non_Abelian_edge_transform}.
	\end{align}
	
	If we then consider the edge energy term $\mathcal{A}_i = \frac{1}{|E|} \sum_{f \in E} \mathcal{A}_i^f$ we have
	\begin{align}
		\mathcal{A}_i \ket{\mu_1,a_1,b_1, \mu_2, a_2, b_2, g, h} &= \frac{1}{|E|} \sum_{f \in E} \sum_{c_1=1}^{|\mu_1|} \sum_{c_2=1}^{|\mu_2|} [D^{\mu_1}([h \rhd f])]_{c_1b_1} [D^{\mu_2}(f^{-1})]_{a_2c_2} \ket{\mu_1,a_1,c_1, \mu_2, c_2, b_2, \partial(f)g, h} \notag \\
		&= \sum_{c_1=1}^{|\mu_1|} \sum_{c_2=1}^{|\mu_2|} \big(\frac{1}{|E|} \sum_{f \in E} [D^{\mu_1}([h \rhd f])]_{c_1b_1} [D^{\mu_2}(f^{-1})]_{a_2c_2} \big) \ket{\mu_1,a_1,c_1, \mu_2, c_2, b_2, \partial(f)g, h}. \label{E_non_Abelian_edge_term}
	\end{align}
	
	We see from this that the edge term does not mix different irrep labels. However, unlike in the $\partial \rightarrow 1_G$ case, we cannot simply use orthogonality of irreps to relate $\mu_1$ to $\mu_2$, due to the change to the edge label $g$. However, we can use another property of the edge terms to extract some additional information about the ground states. The edge term can absorb individual edge transforms: $\mathcal{A}_i = \mathcal{A}_i \mathcal{A}_i^e$. We can therefore write 
	$$\mathcal{A}_i = \mathcal{A}_i \frac{1}{| \ker(\partial)|} \sum_{e_k \in \ker(\partial)} \mathcal{A}_i^{e_k}.$$
	Therefore
	\begin{align*}
		\mathcal{A}_i \ket{\mu_1,a_1,b_1, \mu_2, a_2, b_2, g, h} &= \mathcal{A}_i \frac{1}{| \ker(\partial)|} \sum_{e_k \in \ker(\partial)} \mathcal{A}_i^{e_k}\ket{\mu_1,a_1,b_1, \mu_2, a_2, b_2, g, h}\\
		&=\mathcal{A}_i \frac{1}{| \ker(\partial)|} \sum_{e_k \in \ker(\partial)} \sum_{c_1=1}^{|\mu_1|} \sum_{c_2=1}^{|\mu_2|} [D^{\mu_1}([h \rhd e_k])]_{c_1b_1} [D^{\mu_2}(e_k^{-1})]_{a_2c_2} \ket{\mu_1,c_1,b_1, \mu_2, c_2, b_2, g, h},
	\end{align*}
	where $\partial(e_k)=1_G$ for elements of the kernel by definition. We then use our standard approach of restricting the irreps $\mu_1$ and $\mu_2$ to the kernel of $\partial$ to obtain
	$$[D^{\mu_1}([h \rhd e_k])]_{c_1b_1}= \delta_{c_1 b_1} \mu_1^{\ker}(h \rhd e_k)= \delta_{c_1 b_1} h \rhd \mu_1^{\ker}(e_k),$$
	and
	$$[D^{\mu_2}(e_k^{-1})]_{a_2c_2} = \delta_{a_2c_2} \mu_2^{\ker}(e_k^{-1}),$$
	where $\mu_1^{\ker}(e_k) =[D^{\mu_1}([h \rhd e_k])]_{11}$ is an irrep of the kernel of $\partial$ (and similar for $\mu_2^{\ker}$). We then have
	\begin{align*}
		\mathcal{A}_i \ket{\mu_1,a_1,b_1, \mu_2, a_2, b_2, g, h} &= \mathcal{A}_i \frac{1}{| \ker(\partial)|} \sum_{e_k \in \ker(\partial)} h \rhd \mu_1^{\ker}(e_k) \mu_2^{\ker}(e_k^{-1}) \ket{\mu_1,a_1,b_1, \mu_2, a_2, b_2, g, h},
	\end{align*}
	which, using standard orthogonality of irreps, gives us
	\begin{align*}
		\mathcal{A}_i \ket{\mu_1,a_1,b_1, \mu_2, a_2, b_2, g, h} &= \mathcal{A}_i \delta(h \rhd \mu_1^{\ker}, \mu_2^{\ker})\ket{\mu_1,a_1,b_1, \mu_2, a_2, b_2, g, h}.
	\end{align*}
	
	In other words, a necessary condition for the edge term to be satisfied is that the two irreps $\mu_1$ and $\mu_2$ satisfy $h \rhd \mu_1^{\ker} = \mu_2^{\ker}$. This means that two plaquettes connected by an edge must have irreps whose restriction to the kernel lie in the same $\rhd$-Rep class. In a path connected lattice, each plaquette must, therefore, be associated to the same $\rhd$-Rep class of the kernel. We have already shown that no energy terms fluctuate the parts of the labels of the plaquettes corresponding to irreps of the kernel outside of the $\rhd$-Rep class of the kernel. Now we have shown that spatially separated plaquettes must also be labelled by the same $\rhd$-Rep classes in the ground state. Combining these two facts, we see that the $\rhd$-Rep class is a good label for the ground states.

	With this in mind, let us now consider the commutation relation between the magnetic ribbon operator and the energy terms, taking a horizontal ribbon operator as shown in Figure \ref{Horizontal_ribbon_1}. Once again the relevant energy terms to consider are the edge terms above the ribbon. This is because the ribbon operator only acts on the edges, so its commutation relations with the plaquette and vertex terms are the same as in the $\rhd$ trivial case, and $G$ is Abelian so we still don't need to worry about the factor of $\partial(e)$ on an edge label from the edge transform affecting the action of the ribbon operator. We consider the edge term shown in Figure \ref{E_non_Abelian_magnetic_shorthand}, where this figure also defines our shorthand for the degrees of freedom affected by the edge term. We wish to evaluate $\mathcal{A}_iC^x(t)\mathcal{A}_i$, which will give us zero if the magnetic ribbon operator excites the edge and will give $C^x(t)\mathcal{A}_i$ if the operator does not excite the edge. Then, using Equation \ref{E_non_Abelian_edge_term} (which describes the action of the edge term), and the standard action of the magnetic ribbon operator (which multiplies the edges it cuts by $x$), we have
	\begin{align}
		\mathcal{A}_i&C^x(t)\mathcal{A}_i \ket{\mu_1,a_1,b_1, \mu_2, a_2, b_2, g, h} \notag\\
		&= \mathcal{A}_iC^x(t) \sum_{c_1=1}^{|\mu_1|} \sum_{c_2=1}^{|\mu_2|} \big(\frac{1}{|E|} \sum_{f \in E} [D^{\mu_1}([h \rhd f])]_{c_1b_1} [D^{\mu_2}(f^{-1})]_{a_2c_2} \big) \ket{\mu_1,a_1,c_1, \mu_2, c_2, b_2, \partial(f)g, h} \notag \\
		&=\mathcal{A}_i \sum_{c_1=1}^{|\mu_1|} \sum_{c_2=1}^{|\mu_2|} \big(\frac{1}{|E|} \sum_{f \in E} [D^{\mu_1}([h \rhd f])]_{c_1b_1} [D^{\mu_2}(f^{-1})]_{a_2c_2} \big) \ket{\mu_1,a_1,c_1, \mu_2, c_2, b_2, \partial(f)g, xh} \notag \\
		&=\sum_{c_1=1}^{|\mu_1|} \sum_{c_2=1}^{|\mu_2|} \big(\frac{1}{|E|} \sum_{f \in E} [D^{\mu_1}([h \rhd f])]_{c_1b_1} [D^{\mu_2}(f^{-1})]_{a_2c_2} \big) \sum_{d_1=1}^{|\mu_1|} \sum_{d_2=1}^{|\mu_2|} \big(\frac{1}{|E|} \sum_{e \in E} [D^{\mu_1}([(xh) \rhd e])]_{d_1c_1} [D^{\mu_2}(e^{-1})]_{c_2d_2} \big) \notag \\
		& \hspace{1cm}\ket{\mu_1,a_1,d_1, \mu_2, d_2, b_2, \partial(e)\partial(f)g, xh}. \label{Equation_E_non_abelian_magnetic_edge_energy_1}
	\end{align}
	
	We can then contract the matrices associated to the irreps $\mu_1$ and $\mu_2$ on the indices $c_1$ and $c_2$ to obtain
	\begin{align*}
		\mathcal{A}_i&C^x(t)\mathcal{A}_i \ket{\mu_1,a_1,b_1, \mu_2, a_2, b_2, g, h}\\
		&= \frac{1}{|E|^2} \sum_{f \in E} \sum_{e \in E} \sum_{d_2=1}^{|\mu_2|} \sum_{d_1=1}^{|\mu_1|} [D^{\mu_2}(f^{-1}e^{-1})]_{a_2d_2} [D^{\mu_1}([(xh) \rhd e] [h \rhd f])]_{d_1b_1} \ket{\mu_1,a_1,d_1, \mu_2, d_2, b_2, \partial(e)\partial(f)g, xh}.
	\end{align*}
	
	We want to replace the variable $f$ with the product $f'=ef$ so that the change to the edge label $g$ is parametrized by only one variable (leaving us able to sum over the remaining variable). In order to do this, we must first perform some simple manipulations. Firstly, we can write
	$$[D^{\mu_2}(f^{-1}e^{-1})]_{a_2d_2}= [D^{\mu_2}((ef)^{-1})]_{a_2d_2}.$$
	Secondly, we can insert the identity, in the form of $[h \rhd e^{-1}] [h \rhd e]$, into the argument of the other matrix, to obtain
	\begin{align*}
		[D^{\mu_1}([(xh) \rhd e] [h \rhd f])]_{d_1b_1} &= [D^{\mu_1}([(xh) \rhd e] ([h \rhd e^{-1}] [h \rhd e]) [h \rhd f])]_{d_1b_1}\\
		&=[D^{\mu_1}([(xh) \rhd e] [h \rhd e^{-1}] [h \rhd (ef)])]_{d_1b_1}.
	\end{align*}
	Then we can make the change of variables from $f$ to $f'=ef$ to obtain
	\begin{align*}
		&\mathcal{A}_iC^x(t)\mathcal{A}_i \ket{\mu_1,a_1,b_1, \mu_2, a_2, b_2, g, h}\\
		&= \frac{1}{|E|^2} \sum_{f \in E} \sum_{e \in E} \sum_{d_2=1}^{|\mu_2|} \sum_{d_1=1}^{|\mu_1|} [D^{\mu_2}((ef)^{-1})]_{a_2d_2} [D^{\mu_1}([(xh) \rhd e] [h \rhd e^{-1}] [h \rhd (ef)])]_{d_1b_1} \ket{\mu_1,a_1,d_1, \mu_2, d_2, b_2, \partial(e)\partial(f)g, xh}\\
		&=\frac{1}{|E|^2} \sum_{e \in E} \sum_{f'=ef \in E} \sum_{d_2=1}^{|\mu_2|} \sum_{d_1=1}^{|\mu_1|} [D^{\mu_2}((f')^{-1})]_{a_2d_2} [D^{\mu_1}([(xh) \rhd e] [h \rhd e^{-1}] [h \rhd f'])]_{d_1b_1} \ket{\mu_1,a_1,d_1, \mu_2, d_2, b_2, \partial(f')g, xh}.
	\end{align*}

	We then split the matrix which still involves $e$:
	$$[D^{\mu_1}([(xh) \rhd e] [h \rhd e^{-1}] [h \rhd f'])]_{d_1b_1} = \sum_{j=1}^{|\mu_1|}[D^{\mu_1}([(xh) \rhd e] [h \rhd e^{-1}])]_{d_1 j} [D^{\mu_1}([h \rhd f'])]_{jb_1},$$
	to find
	\begin{align}
		\mathcal{A}_i&C^x(t)\mathcal{A}_i \ket{\mu_1,a_1,b_1, \mu_2, a_2, b_2, g, h} \notag \\
		&= \frac{1}{|E|^2} \sum_{e \in E} \sum_{f'=ef \in E} \sum_{d_2=1}^{|\mu_2|} \sum_{d_1=1}^{|\mu_1|} [D^{\mu_2}((f')^{-1})]_{a_2d_2} \sum_{j=1}^{|\mu_1|}[D^{\mu_1}([(xh) \rhd e] [h \rhd e^{-1}])]_{d_1 j} [D^{\mu_1}([h \rhd f'])]_{jb_1} \notag \\
		& \hspace{1cm} \ket{\mu_1,a_1,d_1, \mu_2, d_2, b_2, \partial(f')g, xh} \notag \\
		&= \sum_{d_2=1}^{|\mu_2|} \sum_{d_1=1}^{|\mu_1|} \frac{1}{|E|} \sum_{f'\in E} [D^{\mu_2}((f')^{-1})]_{a_2d_2} \sum_{j=1}^{|\mu_1|} [D^{\mu_1}([h \rhd f'])]_{jb_1} \frac{1}{|E|} \sum_{e \in E} [D^{\mu_1}([(xh) \rhd e] [h \rhd e^{-1}])]_{d_1 j} \notag \\
		& \hspace{1cm} \ket{\mu_1,a_1,d_1, \mu_2, d_2, b_2, \partial(f')g, xh}. \label{Equation_E_non_abelian_magnetic_edge_energy_2}
	\end{align}
	
	The only term that now depends on the variable $e$ is $[D^{\mu_1}([(xh) \rhd e] [h \rhd e^{-1}])]_{d_1 j}$, so we can evaluate the sum 
	$$\sum_{e \in E} [D^{\mu_1}([(xh) \rhd e] [h \rhd e^{-1}])]_{d_1 j} $$
	algebraically. We can simplify this expression slightly by writing 
	$$[D^{\mu_1}([(xh) \rhd e] [h \rhd e^{-1}])]_{d_1 j} = [D^{\mu_1}([x \rhd (h \rhd e)] [h \rhd e^{-1}])]_{d_1 j},$$
	using the fact that $G$ is Abelian (so the order of $\rhd$ actions does not matter). Then we can replace the sum over $e$ with a sum over $k = h \rhd e$ to give
	$$\sum_{k \in E} [D^{\mu_1}([x \rhd k] k^{-1})]_{d_1 j}. $$

	Next, we note that $[x \rhd k] k^{-1}$ is always in the kernel of $\partial$. This can be seen directly from
	$$\partial([x \rhd k] k^{-1})=\partial([x \rhd k]) \partial(k^{-1}),$$
	which (using the first Peiffer condition Equation \ref{Peiffer_1} from the main text) gives us
	$$\partial([x \rhd k] k^{-1})=x\partial(k)x^{-1} \partial(k^{-1})=1_G,$$
	where the final equality follows from the fact that $G$ is Abelian. However, while all elements of the form $[x \rhd k] k^{-1}$ are in the kernel of $\partial$, it is not necessarily true that all elements of the kernel can be written in the form $[x \rhd k] k^{-1}$ for some $k \in E$. Instead, we claim that elements of this form are a subgroup inside the kernel. To prove that this is so, note that, given two elements $[x \rhd k] k^{-1}$ and $[x \rhd z] z^{-1}$ of the proposed subgroup, we have
	\begin{align*}
		[x \rhd k] k^{-1} [x \rhd z] z^{-1}&= [x \rhd z] 	[x \rhd k] k^{-1} z^{-1} ,
	\end{align*}
	because, as we just showed, $[x \rhd k] k^{-1} $ is in the kernel of $\partial$ and therefore in the centre of $E$. This means that
	\begin{align*}
		[x \rhd k] k^{-1} [x \rhd z] z^{-1}&= [x \rhd (zk)] (zk)^{-1},
	\end{align*}
	so the product of the elements can also be written in the same form (i.e., the proposed subgroup is indeed closed under multiplication). With this in mind, we can begin to evaluate the sum
	$$\sum_{k \in E} [D^{\mu_1}([x \rhd k] k^{-1})]_{d_1 j}. $$
	
	The subgroup proposed above, which we will call $\text{GC}_x$ (``GC" for generalized commutator) is central in $E$ (because it is contained within the kernel), and so the irrep $\mu_1$ will branch to identical 1D irreps of this subgroup when we restrict it to the subgroup (just as we have done before when restricting to the kernel). Calling the resulting 1D irrep $\mu_1^{\text{restr.}}$ (``restr." for restricted), we have
	$$\sum_{k \in E} [D^{\mu_1}([x \rhd k] k^{-1})]_{d_1 j}= \sum_{k \in E} \mu_1^{\text{restr.}} ([x \rhd k] k^{-1}) \delta_{d_1 j}. $$
	
	The $\delta_{j_1 k}$ here is useful, because we can put it back into our expression for $\mathcal{A}_i C^x(t) \mathcal{A}_i$ to obtain something proportional to $C^x(t) \mathcal{A}_i$. Before we do this however, we would like to evaluate $\sum_{k \in E} \mu_1^{\text{restr.}} ([x \rhd k] k^{-1})$. This looks very much like one side of the standard orthogonality relation for irreps, but we are summing over elements $k \in E$ rather than summing directly over the subgroup $\text{GC}_x$ of $E$. However we can write
	\begin{align*}
		\sum_{k \in E} \mu_1^{\text{restr.}} ([x \rhd k] k^{-1})&= \sum_{k \in E} \sum_{y \in \text{GC}_x} \delta([x \rhd k] k^{-1}, y ) \mu_1^{\text{restr.}} (y)\\
		&= \sum_{y \in \text{GC}_x} \mu_1^{\text{restr.}} (y) (\sum_{k \in E} \delta([x \rhd k] k^{-1}, y )).
	\end{align*}
	
	If we can show that
	$$\sum_{k \in E} \delta([x \rhd k] k^{-1}, y )$$
	is independent of the element $y$ in the subgroup, we will then be able to apply the orthogonality relations on $\sum_{y \in \text{GC}_x} \mu_1^{\text{restr.}} (y)$. It is relatively simple to demonstrate this independence. Because $y$ is an element of the subgroup $\text{GC}_x$, it can be written as $[x \rhd e] e^{-1}$ for some element $e \in E$. Then
	\begin{align*}
		\sum_{k \in E} \delta([x \rhd k] k^{-1}, y )&= \sum_{k \in E} \delta([x \rhd k] k^{-1},[x \rhd e] e^{-1} )\\
		&= \sum_{k \in E} \delta([x \rhd e^{-1}][x \rhd k] k^{-1} e,1_E )\\
		&=\sum_{k \in E} \delta([x \rhd (e^{-1}k)] (e^{-1}k)^{-1},1_E ).
	\end{align*}
	
	We then replace the sum over $k \in E$ with a sum over $k'=e^{-1}k$ to obtain
	\begin{align*}
		\sum_{k \in E} \delta([x \rhd k] k^{-1}, y )&= \sum_{k' \in E} \delta([x \rhd k'] k^{ \prime -1}, 1_E ),
	\end{align*}
	which is now manifestly independent of $y$. Knowing this, we can also evaluate the expression $\sum_{k \in E} \delta([x \rhd k] k^{-1}, y )$. We have
	\begin{align*}
		| \text{GC}_x | \sum_{k \in E} \delta([x \rhd k] k^{-1}, y ) &= \sum_{y \in \text{GC}_x}	
		\sum_{k \in E} \delta([x \rhd k] k^{-1}, y ),
	\end{align*}
	from the fact that the expression is independent of $y$. Then because $[x \rhd k] k^{-1}$ is a member of the subgroup $\text{GC}_x$ by definition, 
	$$\sum_{y \in \text{GC}_x}	\delta([x \rhd k] k^{-1}, y )=1,$$
	and so
	\begin{align*}
		| \text{GC}_x | \sum_{k \in E} \delta([x \rhd k] k^{-1}, y ) &= \sum_{y \in \text{GC}_x}	
		\sum_{k \in E} \delta([x \rhd k] k^{-1}, y )= \sum_{k \in E} 1 = |E|.
	\end{align*}
	Therefore
	$$\sum_{k \in E} \delta([x \rhd k] k^{-1}, y ) = \frac{|E|}{|\text{GC}_x|}.$$
	
	Then
	\begin{align*}
		\sum_{k \in E} \mu_1^{\text{restr.}} ([x \rhd k] k^{-1})&= \sum_{y \in \text{GC}_x} \mu_1^{\text{restr.}} (y) (\sum_{k \in E} \delta([x \rhd k] k^{-1}, y ))\\
		&= \frac{|E|}{|\text{GC}_x|} \sum_{y \in \text{GC}_x} \mu_1^{\text{restr.}} (y)\\
		&= \frac{|E|}{|\text{GC}_x|} |\text{GC}_x| \delta(\mu_1^{\text{restr.}},1_R)\\
		&= |E|\delta(\mu_1^{\text{restr.}},1_R),
	\end{align*}
	where $1_R$ is the trivial irrep of $\text{GC}_x$, using the usual orthogonality rules. Therefore
	\begin{align}
		\sum_{k \in E} [D^{\mu_1}([x \rhd k] k^{-1})]_{d_1 j} &=\sum_{k \in E} \mu_1^{\text{restr.}} ([x \rhd k] k^{-1}) \delta_{d_1j} \notag\\
		&= |E|\delta(\mu_1^{\text{restr.}},1_R) \delta_{d_1j}. \label{Equation_subgroup_decomposition_1}
	\end{align}
	
	Now we can finally return to evaluating the energy of the edge using the expression $\mathcal{A}_i C^x(t) \mathcal{A}_i$. Substituting Equation \ref{Equation_subgroup_decomposition_1} into Equation \ref{Equation_E_non_abelian_magnetic_edge_energy_2}, we now have
	\begin{align*}
		\mathcal{A}_i&C^x(t)\mathcal{A}_i \ket{\mu_1,a_1,b_1, \mu_2, a_2, b_2, g, h}\\
		&= \sum_{d_2=1}^{|\mu_2|} \sum_{d_1=1}^{|\mu_1|} \frac{1}{|E|} \sum_{f'\in E} [D^{\mu_2}((f')^{-1})]_{a_2d_2} \sum_{j=1}^{|\mu_1|} [D^{\mu_1}([h \rhd f'])]_{jb_1} \frac{1}{|E|} \sum_{e \in E} [D^{\mu_1}([(xh) \rhd e] [h \rhd e^{-1}])]_{d_1 j} \\
		& \hspace{1cm} \ket{\mu_1,a_1,d_1, \mu_2, d_2, b_2, \partial(f')g, xh}\\
		&= \sum_{d_2=1}^{|\mu_2|} \sum_{d_1=1}^{|\mu_1|} \frac{1}{|E|} \sum_{f'\in E} [D^{\mu_2}((f')^{-1})]_{a_2d_2} \sum_{j=1}^{|\mu_1|} [D^{\mu_1}([h \rhd f'])]_{jb_1} \frac{1}{|E|} |E| \delta(\mu_1^{\text{restr.}},1_R) \delta_{d_1 j} \ket{\mu_1,a_1,d_1, \mu_2, d_2, b_2, \partial(f')g, xh}\\
		&= \sum_{d_2=1}^{|\mu_2|} \sum_{d_1=1}^{|\mu_1|} \frac{1}{|E|} \sum_{f'\in E} [D^{\mu_2}((f')^{-1})]_{a_2d_2} [D^{\mu_1}([h \rhd f'])]_{d_1b_1} \delta(\mu_1^{\text{restr.}},1_R) \ket{\mu_1,a_1,d_1, \mu_2, d_2, b_2, \partial(f')g, xh}\\
		&= \delta(\mu_1^{\text{restr.}},1_R) \sum_{d_2=1}^{|\mu_2|} \sum_{d_1=1}^{|\mu_1|} \frac{1}{|E|} \sum_{f'\in E} [D^{\mu_2}((f')^{-1})]_{a_2d_2} [D^{\mu_1}([h \rhd (f')])]_{d_1b_1} \ket{\mu_1,a_1,d_1, \mu_2, d_2, b_2, \partial(f')g, xh}.
	\end{align*}
	
	We can recognise the expression
	$$\sum_{d_2=1}^{|\mu_2|} \sum_{d_1=1}^{|\mu_1|} \frac{1}{|E|} \sum_{f'\in E} [D^{\mu_2}((f')^{-1})]_{a_2d_2} [D^{\mu_1}([h \rhd f'])]_{d_1b_1} \ket{\mu_1,a_1,d_1, \mu_2, d_2, b_2, \partial(f')g, xh}$$
	as the state
	$$C^x(t)\mathcal{A}_i\ket{\mu_1,a_1,b_1, \mu_2, a_2, b_2, g, h},$$
	by examining the second equality in Equation \ref{Equation_E_non_abelian_magnetic_edge_energy_1}. Therefore
	\begin{align*}
		\mathcal{A}_iC^x(t)\mathcal{A}_i \ket{\mu_1,a_1,b_1, \mu_2, a_2, b_2, g, h} &= \delta(\mu_1^{\text{restr.}},1_R) C^x(t) \mathcal{A}_i\ket{\mu_1,a_1,b_1, \mu_2, a_2, b_2, g, h}.
	\end{align*}
	
	This indicates that whether the edge is excited or not depends entirely on the function $\delta(\mu_1^{\text{restr.}},1_R)$. The question is whether this a ground state property, so that it is well defined for each plaquette in a ground state. Recall that we showed that the $\rhd$-Rep class of the restriction of the irrep plaquette label to the kernel is such a ground state property. Furthermore, because $\text{GC}_x$ is a (normal) subgroup of the kernel, we can obtain $\mu_1^{\text{restr.}}$ from the irrep of the kernel by making a further restriction. If the irrep $\delta(\mu_1^{\text{restr.}},1_R)$ is the same for each irrep of the kernel in each given $\rhd$-Rep class, then we will have shown that $\delta(\mu_1^{\text{restr.}},1_R)$ is also a ground state property. Suppose that one irrep, $\alpha$, in a particular $\rhd$-Rep class does satisfy the Kronecker delta. Then
	$$\alpha( [x \rhd e] e^{-1})=1$$
	for all $e$ by definition. Any other irrep in the $\rhd$-Rep class can be written as $g \rhd \alpha$ for some $g \in G$. Then such an irrep $\beta$ will satisfy
	\begin{align*}
		\beta([x \rhd e] e^{-1}) &= g \rhd \alpha([x \rhd e] e^{-1})\\
		& = \alpha( g \rhd ([x \rhd e] e^{-1})) \\
		&= \alpha ( [x \rhd (g \rhd e)] [g \rhd e^{-1}]),
	\end{align*}
	using the fact that $G$ is Abelian to swap the order of $\rhd$ actions on $e$. However this is just 
	\begin{align*}
		\beta([x \rhd e] e^{-1}) &= \alpha ( [x \rhd e'] e^{\prime -1}),
	\end{align*}
	for $e' = g \rhd e$, which gives
	\begin{align*}
		\beta([x \rhd e] e^{-1}) &=1,
	\end{align*}
	from our assumption about $\alpha$. This indicates that if any irrep of the kernel in the $\rhd$-Rep class restricts to the trivial irrep of the subgroup $\text{GC}_x$, then all of the irreps in the $\rhd$-Rep class do. This also means that if any of the irreps in the class do not restrict to the trivial subgroup, then none of them will. Therefore, $\delta(\mu_1^{\text{restr.}},1_R)$ is a property of the $\rhd$-Rep class and therefore of the ground states. This means that whether the magnetic excitation is confined or not depends only on the ground state (and on its own label $x$).

	\subsection{Condensation}
	\label{Section_condensed_magnetic_electric}
	
	In addition to finding the magnetic excitations in the example $\mathbb{Z}_2$, $\mathbb{Z}_3$ model that we considered in Section \ref{Section_Z2_Z3_Condensation} of the main text, and finding that their pattern of confinement depends on the ground state, we were able to show that some of the electric excitations were condensed (with this pattern again depending on the ground state). This contrasts with the other cases (those in Table \ref{Table_Cases_2d} in the main text) that we have been considering so far, where only the magnetic excitations (and $E$-valued loop-like excitations) could be condensed. In the previous section, we found that some generalizations of the $\mathbb{Z}_2$, $\mathbb{Z}_3$ model also supported an interesting pattern of confinement for the magnetic excitations, and so we may expect them to similarly support condensation that depends on the ground states. In this section, we will show that this is true, and will see cases for which both magnetic and electric excitations may be condensed.
	
	\subsubsection{The case where both groups are Abelian and $\partial \rightarrow 1_G$}
	
	One simple way to generalize the $\mathbb{Z}_2$, $\mathbb{Z}_3$ model is to allow each group to be any Abelian group, while still fixing $\partial$ to map only to the identity element of $G$. We discussed this case when looking at the confined magnetic excitations, where we found that magnetic excitations labelled by group elements which did not stabilise the irreps found in a particular ground state were confined in that ground state. In this case, because $\partial$ maps to $1_G$ there are no confined electric excitations and we expect the condensed excitations to be purely electric. We can generate these condensed excitations by using local operators which are a slight generalization of the ones used for the $\mathbb{Z}_2,\mathbb{Z}_3$ model. In the ground state, two plaquettes 1 and 2 separated by a path $t$ satisfy the relationship $\mu_2 = g(t) \rhd \mu_1$, just like in the $\mathbb{Z}_2, \mathbb{Z}_3$ case. We therefore consider the local operator
	$$\delta(\hat{\mu}_2, q \rhd \hat{\mu}_1),$$
	where $q$ is any element of $G$. Then
	\begin{align*}
		\delta(\hat{\mu}_2, q \rhd \hat{\mu}_1) \ket{GS}&= \delta(\hat{g}(t) \rhd \hat{\mu}_1, q \rhd \hat{\mu}_1)\ket{GS}\\
		&= \delta((q\hat{g}(t)^{-1}) \rhd \hat{\mu}_1, \hat{\mu}_1) \ket{GS}.
	\end{align*}
	
	That is, we get the state back only if $q\hat{g}(t)^{-1}$ \textit{stabilises} the irrep $\mu_1$. The elements that stabilise $\mu_1$ form a subgroup $S$ of $G$, as can be verified simply. If $s_1$ and $s_2$ are elements of $S$ then
	$$(s_1s_2) \rhd \mu_1 = s_2 \rhd (s_1 \rhd \mu_1)= s_2 \rhd \mu_1 = \mu_1,$$
	and so the product element $s_1s_2$ is also an element of $S$, so the subgroup is indeed closed under multiplication. This stabiliser group is the same for all irreps that appear in a particular basis ground state (using the same basis as for the discussion of the magnetic excitations), so the stabiliser group is a good ground state property. To see this, note that all irreps in the ground state belong to the same $\rhd$-Rep class and so can be written in the form $\mu_x = g \rhd \mu_1$ for some $g \in G$ (as shown in Section \ref{Section_magnetic_confinement_partial_trivial}). Then if $s$ is in $S$ we have
	$$s \rhd \mu_x = s \rhd (g \rhd \mu_1) = (gs) \rhd \mu_1.$$
	Because $G$ is Abelian, we then have
	$$(gs) \rhd \mu_1 = (sg) \rhd \mu_1 = g \rhd (s \rhd \mu_1)= g \rhd \mu_1 = \mu_x.$$
	
	This also holds in the reverse direction by the same logic (i.e., if $s$ stabilises $\mu_x$ it stabilises $\mu_1$), so the stabiliser group is the same for all irreps in the same $\rhd$-Rep class (and so for all irreps in a particular ground state). Then we can write our local operator acting on the ground state in terms of this subgroup as
	\begin{align}
		\delta(\hat{\mu}_2, q \rhd \hat{\mu}_1) \ket{GS}&= \delta((q\hat{g}(t)^{-1}) \rhd \hat{\mu}_1, \hat{\mu}_1) \ket{GS} \notag\\
		&= \delta(q\hat{g}(t)^{-1} \in S) \ket{GS}\notag \\
		&= \sum_{s \in S} \delta(q \hat{g}(t)^{-1},s^{-1})\ket{GS} \notag \\
		&= \sum_{s \in S} \delta(\hat{g}(t),qs)\ket{GS}, \label{Equation_partial_trivial_condensed_magnetic}
	\end{align}
	which is an electric ribbon operator acting on the ground state. This electric ribbon operator acts in the same way on the ground state as the local operator and so must be condensed. A general condensed electric ribbon operator can be built out of these elements by forming linear combinations of these elements with different $q \in G$. This means that any electric ribbon operator labelled by an irrep with trivial restriction to the subgroup $S$ is condensed. To see this explicitly, consider such an electric ribbon operator
	$$\sum_{g \in G} R(g) \delta(\hat{g}(t),g).$$
	We can split the sum over $g \in G$ into a sum over elements $s \in S$ and cosets of $S$ in $G$. Denoting representatives for these cosets by $q$ we have
	\begin{align*}
		\sum_{g \in G} R(g) \delta(\hat{g}(t),g) = \sum_{s \in S} \sum_{\text{cosets } qS} R(qs) \delta(\hat{g}(t),qs).
	\end{align*}
	
	If $R$ has trivial restriction to the stabiliser then $R(s)=1$ and so $R(qs)=R(q)$, giving us
	\begin{align*}
		\sum_{g \in G} R(g) \delta(\hat{g}(t),g) = \sum_{\text{cosets } qS} R(q) \sum_{s \in S} \delta(\hat{g}(t),qs),
	\end{align*}
	which is just a linear combination of condensed elements of the form given in Equation \ref{Equation_partial_trivial_condensed_magnetic}, implying that the electric excitations produced by this operator are condensed. This pattern of condensation makes sense when viewed from the perspective of the confined magnetic excitations, because the confined excitations are those labelled by elements outside the stabiliser subgroup, and so can have non-trivial braiding with these condensed electric excitations. On the other hand any non-confined magnetic excitation is labelled by an element of the stabiliser and so has trivial braiding with the condensed excitations, due to the fact that the irreps labelling the condensed excitations restrict trivially to the stabiliser group.
	
	\subsubsection{The case where both groups are Abelian but $\partial$ is general}
	
	Next we consider the case where the two groups are still Abelian, but $\partial$ can be general (while still satisfying the Peiffer conditions). When considering the magnetic excitations, we passed over this case to look at the more general situation where $E$ can be non-Abelian, but due to the jump in complexity here from the Abelian to non-Abelian case it will be useful to look at the Abelian case first.

\begin{figure}[h]
		\begin{center}
			\begin{overpic}[width=0.5\linewidth]{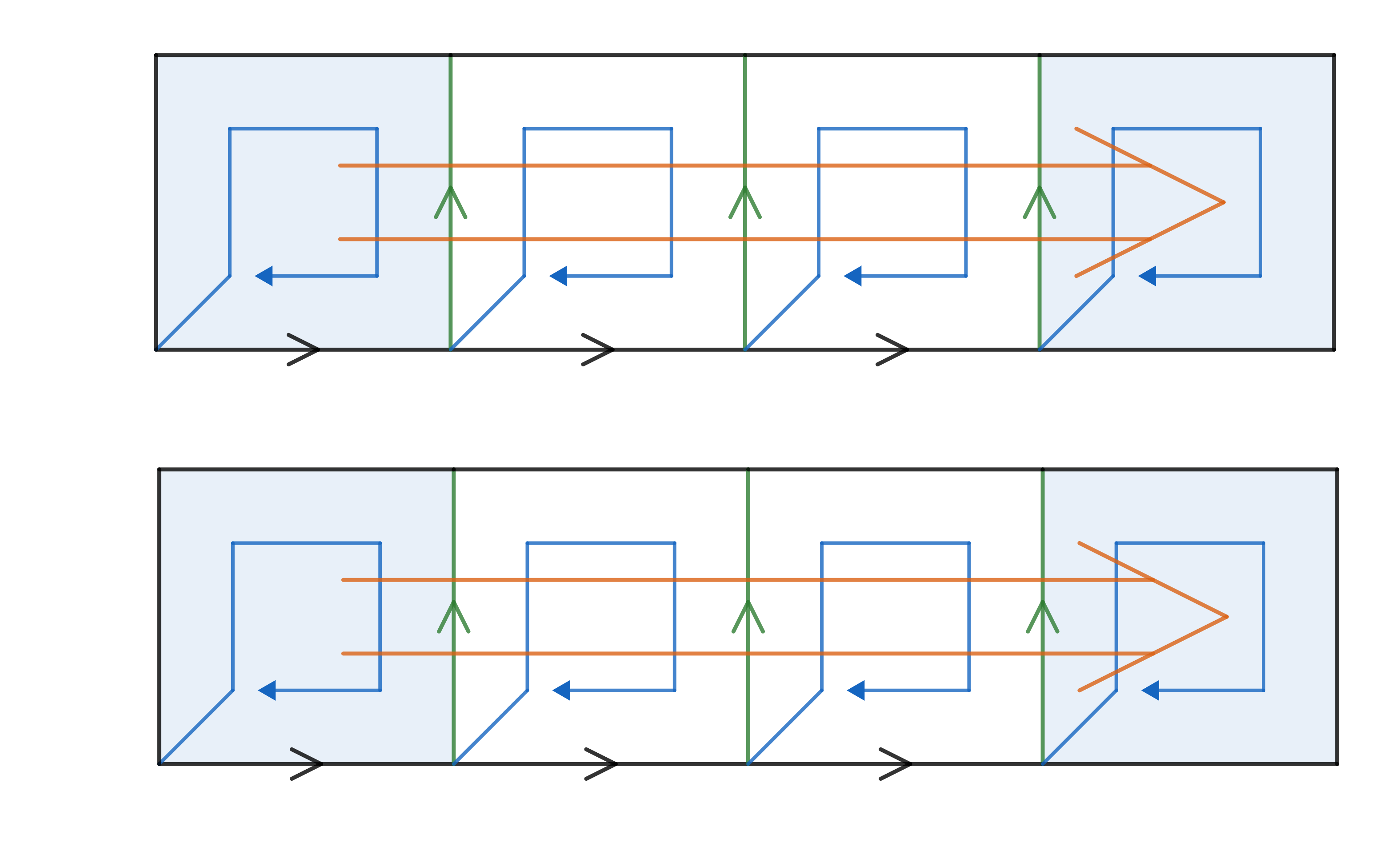}
				\put(21,45){$e_1$}
				\put(42,45){$e_2$}
				\put(63,45){$e_3$}
				\put(83,45){$e_4$}
				
				\put(19,15){$e e_1$}
				\put(39,14){\parbox{1cm}{$ e e_2 e^{-1}$\\ $=e_2$}}
				\put(60,14){\parbox{1cm}{$e e_3 e^{-1}$\\ $=e_3$}}
				\put(81,15){$e_4 e^{-1}$}
				
				\put(31,58){$g_1$}
				\put(52,58){$g_2$}
				\put(73,58){$g_3$}
				
				\put(28,29){$\partial(e)g_1$}
				\put(49,29){$\partial(e)g_2$}
				\put(70,29){$\partial(e)g_3$}
				
				\put(-7,45){$\mathcal{A}_{i_1}^e \mathcal{A}_{i_2}^e \mathcal{A}_{i_3}^e$}
				\put(5,15){$=$}
				
				\put(18,37){$h_1$}
				\put(39,37){$h_2$}
				\put(60,37){$h_3$}
				
				\put(18,7){$h_1$}
				\put(39,7){$h_2$}
				\put(60,7){$h_3$}
				
			\end{overpic}
			\caption{When $\rhd$ is trivial (which also implies that $E$ is Abelian), we can reproduce the action of a ribbon operator $C^{\partial(e)}(t)$ by applying edge transforms (which can be absorbed into the ground state) and local operators on the two ends of the ribbon. This is not generally possible when $\rhd$ is non-trivial however.}
			\label{Condensed_magnetic_1}
		\end{center}
	\end{figure}

	In the previous case, we saw that some electric excitations are condensed because we can deduce some information about path elements by measuring the irreps labelling the plaquettes at either end of the path, without measuring the path element directly by applying an electric ribbon operator. This occurs when the map $\rhd$ is non-trivial, because this map couples the plaquette labels with the path element. However we also saw that there is some condensation when $\rhd$ is trivial in Section \ref{Section_Condensation_Magnetic_2D}. Specifically, in the $\rhd$ trivial case there is condensation of some of the magnetic excitations (those with label in $\partial(E)$), resulting from their corresponding ribbon operators acting in a similar way to edge transforms. In this new case, where both $\partial$ and $\rhd$ are non-trivial, we expect some combination of these effects, but it is not as simple as just including both patterns simultaneously (after all, the mechanism for both forms of condensation is the edge transform, so there may be some interference). Indeed, our argument for the condensation of the magnetic excitations breaks down in this new case due to $\rhd$ being non-trivial. First let us briefly recall that previous argument, given in Section \ref{Section_Condensation_Magnetic_2D}, by taking a simple case where the magnetic ribbon operator acts on just a few edges, as shown in Figure \ref{Condensed_magnetic_1}. We can see that the magnetic ribbon operator multiplies each edge label cut by the ribbon by an element $\partial(e)$. This is similar to the action of edge transforms $\mathcal{A}_i^e$ applied on each edge, so we can try to reproduce the action of the ribbon operator with a series of edge transforms, as shown in Figure \ref{Condensed_magnetic_1}. We see that, when $\rhd$ is trivial, the action of the edge transforms on the intermediate plaquettes (the plaquettes cut by the ribbon operator, but not at either end of the ribbon) cancels, while the action on the edges reproduces the action of the magnetic ribbon operator. The only difference is that the two plaquettes at the ends of the ribbon operator are affected by the edge transforms (but not by the magnetic ribbon operator), but this can be corrected by local operators which multiply these two plaquette labels by $e$ or $e^{-1}$. The edge transforms act trivially on the ground state, so the action of the magnetic ribbon operator is then just equivalent to the local operators on the two end plaquettes, implying that these particular magnetic excitations are condensed.

	However, now consider the case where $\rhd$ is non-trivial. If we naively applied the same edge transform on each edge, the action on the intermediate plaquettes would not cancel out. Each plaquette would gain a factor of $e^{-1}$ from the edge to the left, but a factor of the form $h_{x} \rhd e$ from the edge on the right (where $h_{x}$ is the label of the edge connecting the base-point of plaquettes $x$ and $x+1$, as shown in Figure \ref{Condensed_magnetic_1}). This is due to the edge transform depending on the path from the base-point of the plaquette to the edge on which we apply the transform, as indicated by Equation \ref{Equation_edge_transform_definition} in Section \ref{Section_Recap_Paper_2} of the main text. Instead, we must apply edge transforms whose labels evolve along the ribbon with a $\rhd$ action in order to allow the factors from the edge transforms on different edges to cancel. Specifically, for each edge $i_x$ (these are the vertical edges in the example shown in Figure \ref{Condensed_magnetic_1}, where $x=1$, 2 or 3), we must apply an edge transform $\mathcal{A}_{i_x}^{g(s.p-s(i_x))^{-1} \rhd e}$, where $g(s.p-s(i_x)) = \prod_{j=1}^{x} h_j$ is the path from the leftmost vertex (the base-point of the first plaquette, which we call the start-point) up to the edge in question. These transforms have the same action on the edge labels as the original series of transforms, because $\partial(g \rhd e)= \partial(e)$ when $G$ is Abelian (due to the first Peiffer condition, Equation \ref{Peiffer_1} in the main text), but the action on the internal plaquettes (i.e., not the plaquettes on the ends of the ribbon) now cancels out as required. This is because the label of plaquette $p_x$, which is adjacent to edges $i_{x-1}$ and $i_x$, gains a factor of 
	$$g(s.p-s(i_{x-1}))^{-1} \rhd e^{-1} = (\prod_{j=1}^{x-1} h_j)^{-1} \rhd e^{-1}$$ 
	from the edge to the left, and a factor of
	\begin{align*}
		h_{x} \rhd ( g(s.p-s(i_{x}))^{-1} \rhd e)&= h_{x} \rhd (\prod_{j=1}^{x} h_j)^{-1} \rhd e^{-1}\\
		&=(\prod_{j=1}^{x-1} h_j)^{-1} \rhd e\\
		&= g(s.p-s(i_{x-1}))^{-1} \rhd e
	\end{align*}
	from the edge to the right, which cancel. However in order to correct the labels of the two end plaquettes (which would not be changed by the ribbon operator we wish to reproduce), we now need to multiply the left plaquette (plaquette 1) by $e^{-1}$, but the right plaquette (plaquette 4) by $g(s.p-s(i_3))^{-1} \rhd e$. This is now a non-local operation, because it depends on the path element between the two end plaquettes. This indicates that the magnetic ribbon operators labelled by elements of $\partial(E)$ may not act as local operators on the ground state (at the least, our old argument no longer holds), and so the associated excitations might not be condensed.

\begin{figure}[h]
		\begin{center}
			\begin{overpic}[width=0.5\linewidth]{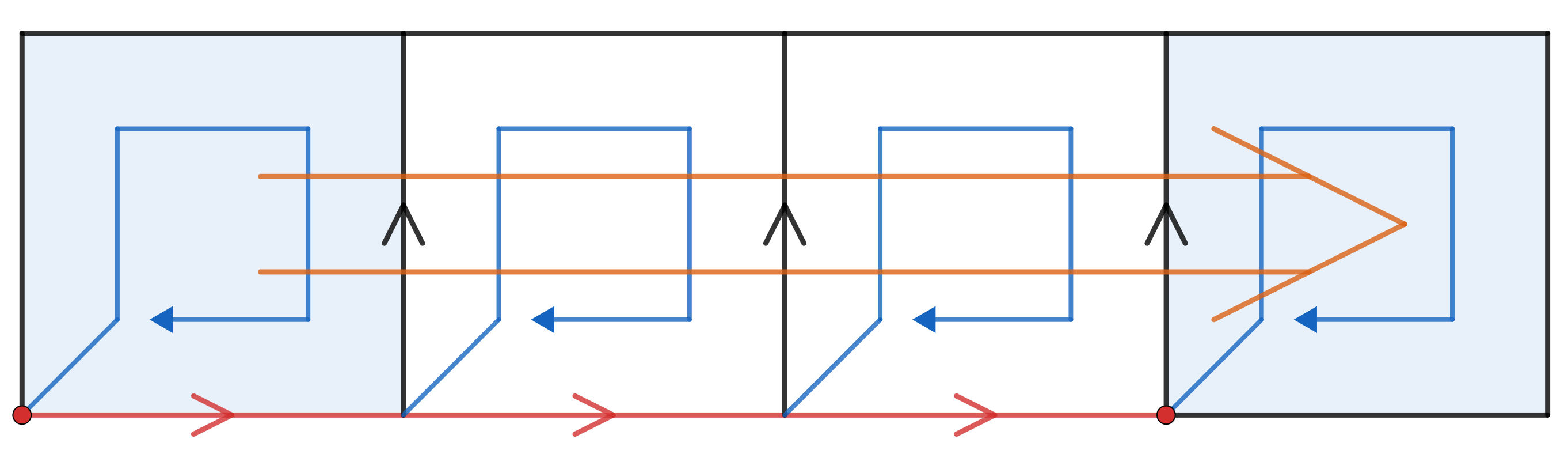}
				\put(10,0){$h_1$}
				\put(36,0){$h_2$}
				\put(58,0){$h_3$}
				\put(26,20){$R_1$}
				\put(51,20){$R_2$}
				\put(74.5,20){$R_3$}
				\put(12,15){$\mu_1$}
				\put(90,15){$\mu_n$}
			\end{overpic}
			\caption{We wish to reproduce the action of a ribbon operator by applying local operators on the two ends of the ribbon, together with edge transforms which can be absorbed into the ground state. We consider an electromagnetic ribbon operator, with an electric ribbon operator acting on the direct path (here shown by red horizontal edges) and magnetic ribbon operator on the dual path (here indicated by the orange double arrow). We consider the action of this ribbon operator in a basis where the edges along the direct path of the ribbon (the red horizontal edges) are labelled by group elements of $G$, while the vertical edges cut by the dual path are labelled by irreps of $G$ and the plaquettes by irreps of $E$.}
			\label{Condensed_ribbon_1}
		\end{center}
	\end{figure}

\begin{figure}[h]
		\begin{center}
			
			\begin{overpic}[width=0.5\linewidth]{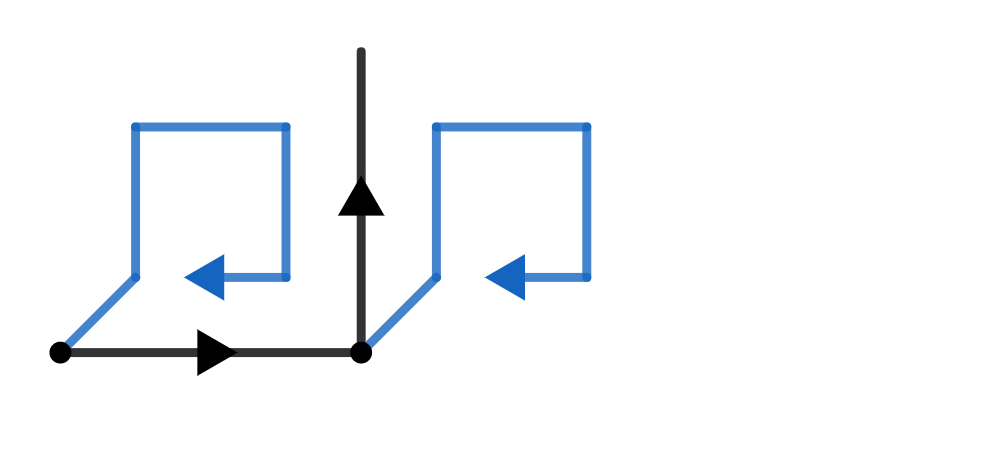}
				\put(63,25){ $:=\ket{\mu_1,\mu_2, R_i,h}$}
				\put(35,46){$R_i$}
				\put(20,6){ $h$}
				\put(19,27){$\mu_1$}
				\put(49,27){ $\mu_2$}	
				
			\end{overpic}
			\caption{Shorthand for the degrees of freedom affected by the vertical edge transform}
			\label{vertical_edge_transform_hybrid_basis_1}
		\end{center}
	\end{figure}

	Having seen why this approach doesn't work, we now want to find the actual pattern of condensation, of both electric and magnetic excitations (and crucially of dyonic excitations, which are combinations of the two). To do so, we follow a similar approach to that used in the previous section for the case where $\partial$ maps to $1_G$. We want to use the irrep basis to show that a measurement performed on the two ends of the ribbon can be related to the action of certain ribbon operators. In order to motivate the local operators we will use, we first want to look at the edge energy terms in a new basis. We start by considering the ribbon operator that we would like to reproduce, taking it to be an electromagnetic ribbon operator passing horizontally through our lattice, as shown in Figure \ref{Condensed_ribbon_1}. Then we consider a basis where the edges cut by the ribbon are labelled by irreps of $G$, while the edges on the direct path (equivalently those connecting the base-points of the plaquettes on the ribbon) are labelled by elements of $G$ (and the plaquettes are labelled by irreps of $E$), as shown in Figure \ref{vertical_edge_transform_hybrid_basis_1}. Then the edge transforms on the vertical edges act as (similar to Equation \ref{E_non_Abelian_edge_transform} for the horizontal edge transforms)
	\begin{align*}
		\mathcal{A}_i^e \ket{\mu_1, \mu_2, R_i, h}&= \frac{1}{\sqrt{|G|}}\sum_{g \in G} R_i(g) \mathcal{A}_i^e \ket{\mu_1, \mu_2, g, h} \\
		&= \frac{1}{\sqrt{|G|}}\sum_{g \in G} R_i(g) \mu_1 (h \rhd e^{-1}) \mu_2(e) \ket{\mu_1, \mu_2, \partial(e)g, h}\\
		&= \frac{1}{\sqrt{|G|}} \sum_{g' = \partial(e)g} R_i( \partial(e)^{-1}g') \mu_1 (h \rhd e^{-1}) \mu_2(e) \ket{\mu_1, \mu_2, g', h}\\
		&= R_i(\partial(e)^{-1}) \mu_1 (h \rhd e^{-1}) \mu_2(e) \ket{\mu_1, \mu_2, R_i, h}.
	\end{align*}

	Now consider the edge term $\mathcal{A}_i = \frac{1}{|E|} \sum_{e \in E} \mathcal{A}_i^e$, which gives us
	\begin{align*}
		\mathcal{A}_i \ket{\mu_1, \mu_2, R_i, h} &= \frac{1}{|E|} \sum_{e \in E} R_i(\partial(e)^{-1}) \mu_1 (h \rhd e^{-1}) \mu_2(e) \ket{\mu_1, \mu_2, R_i, h} \\
		&= \delta( \mu_2, \mu^{R_i} [h \rhd \mu_1])\ket{\mu_1, \mu_2, R_i, h},
	\end{align*}
	where $\mu^{R_i}$ is the irrep of $E$ defined by $ \mu^{R_i}(e)= R_i(\partial(e))$. This tells us how neighbouring plaquette labels should be related in the ground state. If we iterate this condition along the ribbon in Figure \ref{Condensed_ribbon_1}, we get
	$$\mu_n = [\hat{g}(t) \rhd \mu_1] \mu^{R_t},$$
	where $R_t$ is obtained by fusing the irreps labelling the edges cut by the dual path. Note that the irreps $\mu^{R_i}$ are unaffected by the $\rhd$ action when $G$ is Abelian, because 
	\begin{equation}
		\mu^{R_i}(g \rhd e)= R_i(\partial(g \rhd e))=R_i(g\partial(e)g^{-1})=R_i(\partial(e))=\mu^{R_i}(e). \label{Equation_irrep_partial_rhd}
	\end{equation}
	
	With this ground state condition in mind, we now propose the (bi-)local operator
	$$\delta(\mu_n, [q \rhd \mu_1] \mu^X),$$
	where $q$ is some element of $G$ and $\mu^X$ is an irrep of $E$ of the form $\mu^X(e)=X(\partial(e))$ for some irrep $X$ of $\partial(E)$. We have
	\begin{align*}
		\delta(\mu_n, [q \rhd \mu_1] \mu^X) &= \delta([\hat{g}(t) \rhd \mu_1] \mu^{R_t}, [q \rhd \mu_1] \mu^X)\\
		&= \delta( [(q^{-1}\hat{g}(t)) \rhd \mu_1] \mu^{R_t}, \mu_1 \mu^X)\\
		&= \delta([(q^{-1} \hat{g}(t)) \rhd \mu_1] \mu_1^{-1}, \mu^X (\mu^{R_t})^{-1}),
	\end{align*}
	where we have used the fact that $\mu^{R_t}$ and $\mu^X$ are invariant under the $\rhd$ action. The irrep $\mu^X (\mu^{R_t})^{-1})$ is not a generic irrep of $E$, but instead can be written in terms of an irrep of $\partial(E)$, and so is not sensitive to elements in the kernel of $E$. This means that, when the Kronecker delta is non-zero, the element $q^{-1}\hat{g}(t)$ must only affect the irrep $\mu_1$ in a limited way via the $\rhd$ action (it doesn't affect the restriction of the irrep to the kernel). This motivates us to consider the subset $M$ of elements in $G$ which stabilise $\mu_1$ up to an irrep of the form $\mu^A(e)=A(\partial(e))$ for some irrep $A$ of $G$. This set forms a subgroup of $G$: for any two elements $m$ and $n$ of $M$, we have
	\begin{align*}
		m \rhd \mu_1 &= \mu_1 \mu^A\\
		n \rhd \mu_1 &= \mu_1 \mu^B\\
		& \implies (mn) \rhd \mu_1 = n \rhd (m \rhd \mu_1) = n \rhd (\mu_1 \mu^A) = (n \rhd \mu_1) \mu^A = \mu_1 \mu^B\mu^A= \mu_1 \mu^{B \cdot A},
	\end{align*}
	so that $mn$ is also in $M$ (i.e., the subset is indeed closed under multiplication). Then
	\begin{align*}
		\delta(\mu_n, [q \rhd \mu_1] \mu^X) &=\delta([(q^{-1} \hat{g}(t)) \rhd \mu_1] \mu_1^{-1}, \mu^X (\mu^{R_t})^{-1})\\
		&=\delta(q^{-1} \hat{g}(t) \in M)\delta([(q^{-1} \hat{g}(t)) \rhd \mu_1] \mu_1^{-1}, \mu^X (\mu^{R_t})^{-1})\\
		&= \sum_{ m \in M} \delta(q^{-1}\hat{g}(t), m) \delta([m \rhd \mu_1] \mu_1^{-1}, \mu^X (\mu^{R_t})^{-1})\\
		&= \sum_{m \in M} \delta(\hat{g}(t), qm) \frac{1}{|\partial(E)|} \sum_{h \in \partial(E)} R_t(h) X(h^{-1}) \big[(\mu_1 [m \rhd \mu_1^{-1}])(\partial^{-1}(h^{-1}))\big],
	\end{align*}
	where in the last line we used the Grand Orthogonality Theorem. We note that the $\partial^{-1}(h^{-1})$ in the argument for the irrep $\mu_1 [m \rhd \mu_1^{-1}]$ indicates any element $e$ satisfying $\partial(e)=h^{-1}$, and which particular element is taken does not matter for evaluating the expression (because we know that $\mu_1 [m \rhd \mu_1^{-1}]$ is equivalent to an irrep of $\partial(E)$ from the other Kronecker delta). We also note that the factor $R_t(h)$ is the same as the action of a magnetic ribbon operator $C^{h^{-1}}(t)$, because the ribbon operator multiplies each edge by $h^{-1}$, resulting in a phase $R_i(h)$ from each edge $i$:
	\begin{align}
		C^{h^{-1}}(t)\ket{R_i} &= \frac{1}{\sqrt{|G|}} \sum_{g_i \in G} R_i(g_i) C^{h^{-1}}(t) \ket{g_i} \notag \\
		&= \frac{1}{\sqrt{|G|}} \sum_{g_i \in G} R_i(g_i)\ket{h^{-1}g_i} \notag \\
		&=\frac{1}{\sqrt{|G|}} \sum_{g'=h^{-1}g_i \in G} R_i(hg')\ket{g'} \notag \\
		&= R_i(h)\frac{1}{\sqrt{|G|}} \sum_{g'=h^{-1}g_i \in G} R_i(g')\ket{g'} \notag \\
		&=R_i(h) \ket{R_i}. \label{Equation_magnetic_irrep_basis_1}
	\end{align}

	Therefore, we have
	\begin{align*}
		\delta(\mu_n, [q \rhd \mu_1] \mu^X)&= \sum_{m \in M} \delta(\hat{g}(t), qm) \frac{1}{|\partial(E)|} \sum_{h \in \partial(E)} C^{h^{-1}}(t) X(h^{-1}) \big[(\mu_1 [m \rhd \mu_1^{-1}])(\partial^{-1}(h^{-1}))\big]\\
		&= \sum_{m \in M} \frac{1}{|\partial(E)|} \sum_{h \in \partial(E)} X(h^{-1}) \big[(\mu_1 [m \rhd \mu_1^{-1}])(\partial^{-1}(h^{-1}))\big] \delta(\hat{g}(t), qm) C^{h^{-1}}(t),
	\end{align*}
	which can be recognised as an electromagnetic ribbon operator. The operators of this form are then a basis for the condensed ribbon operators (with different basis operators given by different values of $q$ and $\mu^X$, although we must pick $q$ only from a set of representatives to ensure that all of the operators are distinct). However we can make these operators easier to understand by taking appropriate linear combinations. In particular, we consider
	$$\frac{1}{|\partial(E)|} \sum_{X \in \text{Rep}(\partial(E))} \delta(\mu_n, [q \rhd \mu_1] \mu^X) X(x^{-1}),$$
	where we have one such operator for each $x \in \partial(E)$ (and each distinct choice of $q$). Then we have
	\begin{align*}
		\frac{1}{|\partial(E)|}& \sum_{X \in \text{Rep}(\partial(E))} \delta(\mu_n, [q \rhd \mu_1] \mu^X) X(x^{-1})\\
		&= \frac{1}{|\partial(E)|} \sum_{X \in \text{Rep}(\partial(E))} X(x^{-1}) \sum_{m \in M} \frac{1}{|\partial(E)|} \sum_{h \in \partial(E)} X(h^{-1}) \big[(\mu_1 [m \rhd \mu_1^{-1}])(\partial^{-1}(h^{-1}))\big] \delta(\hat{g}(t), qm) C^{h^{-1}}(t)\\
		&= \sum_{m \in M} \frac{1}{|\partial(E)|} \sum_{h \in \partial(E)} \bigg(\frac{1}{|\partial(E)|} \sum_{X \in \text{Rep}(\partial(E))} X(h^{-1}) X(x^{-1})\bigg) \big[(\mu_1 [m \rhd \mu_1^{-1}])(\partial^{-1}(h^{-1}))\big] \delta(\hat{g}(t), qm) C^{h^{-1}}(t)\\
		&= \sum_{m \in M} \frac{1}{|\partial(E)|} \sum_{h \in \partial(E)} \partial(x,h^{-1}) \big[(\mu_1 [m \rhd \mu_1^{-1}])(\partial^{-1}(h^{-1}))\big] \delta(\hat{g}(t), qm) C^{h^{-1}}(t)\\
		&= \sum_{m \in M} \frac{1}{|\partial(E)|} \big[(\mu_1 [m \rhd \mu_1^{-1}])(\partial^{-1}(x))\big] \delta(\hat{g}(t), qm) C^{x}(t),
	\end{align*}
	where we used the sum over irreps $X$ of $\partial(E)$ with the Grand Orthogonality Theorem to remove the sum over $h$ and thereby isolate a single label for the magnetic part of the ribbon operator.

	As an example, we can now see from this which of the pure electric and magnetic excitations are condensed. To see which of the electric ribbon operators are condensed, we take $x$ to be the identity element, so that $C^{x}(t)$ is just the identity operator. Then we have
	\begin{align*}
		\sum_{m \in M} \frac{1}{|\partial(E)|} \big[(\mu_1 [m \rhd \mu_1^{-1}])(1_E)\big] \delta(\hat{g}(t), qm) &= \frac{1}{|\partial(E)|} \sum_{m \in M} \delta(\hat{g}(t), qm),
	\end{align*}
	which means that any electric ribbon operator labelled by an irrep with trivial restriction to the subgroup $M$ is condensed. Here $M$ is the subgroup of $G$ consisting of elements that stabilise $\mu_1$ up to an irrep of the form $\mu^A(e)=A_{\partial}(\partial(e))$ for some irrep $A_{\partial}$ of $\partial(E)$.

	In order to see which of the pure magnetic excitations are condensed, we instead sum over the label $q$ (over all elements in $G$ for simplicity, although many values of $q$ may give the same ribbon operator). This gives us
	\begin{align*}
		\sum_{q \in G} \frac{1}{|\partial(E)|} \sum_{X \in \text{Rep}(\partial(E))} \delta(\mu_n, [q \rhd \mu_1] \mu^X) \mu^X(x^{-1}) &= \sum_{m \in M} \frac{1}{|\partial(E)|} \big[(\mu_1 [m \rhd \mu_1^{-1}])(\partial^{-1}(x))\big] (\sum_{q \in G} \delta(\hat{g}(t), qm)) C^{x}(t)\\
		&= \sum_{m \in M} \frac{1}{|\partial(E)|} \big[(\mu_1 [m \rhd \mu_1^{-1}])(\partial^{-1}(x))\big] C^x(t).
	\end{align*}
	
	Then any magnetic ribbon operator labelled by $x \in \partial(E)$ for which
	$$\sum_{m \in M} \big[(\mu_1 [m \rhd \mu_1^{-1}])(\partial^{-1}(x))\big]$$
	is non-zero will be condensed. We therefore want to evaluate this quantity to find when it is non-zero. To simplify the notation, we choose some element $f \in E$ such that $\partial(f)=x$ and replace $\partial^{-1}(x)$ with $f$ in the expression above. We then define $\mu_1^m =\mu_1 [m \rhd \mu_1^{-1}]$, which is itself an irrep of $E$. We can show that the irreps of this type form a group with multiplication defined by $\mu_1^{m_1} \mu_1^{m_2}= \mu_1^{m_1m_2}$:
	\begin{align*}
		\mu_1^{m_1m_2} &= \mu_1 [(m_1 m_2) \rhd \mu_1^{-1}]\\
		&= \mu_1 [m_1 \rhd \mu_1^{-1}] [m_1 \rhd \mu_1] [(m_1 m_2) \rhd \mu_1^{-1}]\\
		&= \mu_1^{m_1} [m_1 \rhd (\mu_1 [m_2 \rhd \mu_1^{-1}])]\\
		&= \mu_1^{m_1} [m_1 \rhd (\mu_1^{m_2})].
	\end{align*}

	By the definition of $M$, $m_2$ stabilises $\mu_1$ up to an irrep $\mu^Y$ which satisfies $\mu^Y(e)=Y(\partial(E))$ for some irrep $Y$ of $G$. Then $m_2$ also stabilises $\mu_1^{-1}$ in a similar way:
	\begin{align*}
		m_2 \rhd \mu_1^{-1}(e) &= \mu_1^{-1}(m_2 \rhd e)=(\mu_1(m_2 \rhd e))^{-1}\\
		& = (m_2 \rhd \mu_1(e))^{-1} = (\mu_1(e) \mu^Y(e))^{-1}\\
		& = \mu_1^{-1}(e) \mu^{Y^{-1}}(e).
	\end{align*}
	
	This means that $\mu_1^{m_2} = \mu_1 [m_2 \rhd \mu_1^{-1}]$ is equal to $\mu^A$ for some irrep $\mu^A$ satisfying $\mu^A(e)= A(\partial(e))$ for an irrep $A$ of $\partial(E)$. These irreps are invariant under the $\rhd$ action as shown by Equation \ref{Equation_irrep_partial_rhd}, and so $m_1 \rhd (\mu_1^{m_2}) = \mu_1^{m_2}$. Therefore,
	$$\mu_1^{m_1m_2}= \mu_1^{m_1} \mu_1^{m_2},$$
	as we claimed. We can then define an irrep $R_f$ of this subgroup by $R_f(\mu_1^{m_1})=\mu_1^{m_1}(f)$. This is a representation because $R_f( \mu_1^{m_1} \mu_1^{m_2})= \mu_1^{m_1 m_2}(f) = \mu_1^{m_1}(f) \mu_1^{m_2}(f) = R_f(\mu_1^{m_1}) R_f(\mu_1^{m_2}).$ Then 
	\begin{align*}
		\sum_{m \in M} \big[(\mu_1 [m \rhd \mu_1^{-1}])(f)\big] &= \sum_{m \in M} \mu_1^m(f)\\
		&= \sum_{m \in M} R_f(\mu_1^m).
	\end{align*}
	
	The group structure of the irreps $\mu_1^m$ guarantees that $\sum_m R_f(\mu_1^m) \propto \sum_{\mu_1^m} R_f(\mu_1^m)$. To see this, note that
	\begin{align*}
		\sum_{m \in M} R_f(\mu_1^m)&= \sum_{m \in M} \sum_{\mu^n} \delta(\mu_1^n, \mu_1^m ) R_f(\mu_1^n)\\
		&= \sum_{m \in M} \sum_{\mu^n} \delta(\mu_1^n (\mu_1^m)^{-1}, 1_{\text{Rep}} ) R_f(\mu_1^n)\\
		&= \sum_{m \in M} \sum_{\mu^n} \delta(\mu_1^{nm^{-1}}, 1_{\text{Rep}} ) R_f(\mu_1^n)\\
		&=\sum_{m' = nm^{-1}} \sum_{\mu^n} \delta(\mu_1^{m'}, 1_{\text{Rep}} ) R_f(\mu_1^n)\\
		&= (\sum_{m' \in M} \delta(\mu_1^{m'}, 1_{\text{Rep}} ) ) \sum_{\mu^n} R_f(\mu_1^n).
	\end{align*}
	
	Then $(\sum_{m' = nm^{-1}} \delta(\mu_1^{m'}, 1_{\text{Rep}} ) ) $ is some constant independent of $n$ (it depends only on the group of irreps), so $\sum_m R_f(\mu_1^m)$ is proportional to $ \sum_{\mu^n} R_f(\mu_1^n)$. Therefore, we obtain
	\begin{align*}
		\sum_{m \in M} \big[(\mu_1 [m \rhd \mu_1^{-1}])(f)\big] &\propto \sum_{\mu_1^m} R_f(\mu_1^m)\\
		&\propto \delta(R_f, 1_R),
	\end{align*}
	where $1_R$ is the trivial irrep and we used the Grand Orthogonality Theorem in the last line. If $R_f$ is the trivial irrep, then $\mu_1^m(f) =1$ for all $m \in M$, which means
	$$\mu_1(f) [m \rhd \mu_1^{-1}(f)]=1$$
	for all $m \in M$. If this is satisfied, the magnetic excitation labelled by $\partial(f)$ is condensed, but otherwise it is not.

	Throughout this section, we have been using $\mu_1$, the plaquette label of the first plaquette on the ribbon in Figure \ref{Condensed_ribbon_1}. We defined $M$ in terms of this label and found the condensation properties in terms of $M$. We therefore wish to show that $M$ does not depend on which value of $\mu_1$ in a particular ground state we have, only which ground state we are in (so that the pattern of condensation does not depend on any local properties). If an irrep $\mu$ is in the ground state, the other irreps appearing in the ground state have the form
	$$ \mu_x = [g \rhd \mu] \mu^A,$$
	where $\mu_A$ is an irrep of $E$ satisfying $\mu_A(e)=A(\partial(e))$ for some irrep $A$ of $\partial(E)$. We will now show that each such irrep gives rise to the same subgroup $M$. Consider an element $m$ that satisfies
	$$m \rhd \mu = \mu \mu^B$$
	(i.e., $m$ is in $M$ for $\mu$). Then 
	\begin{align*}
		m \rhd \mu_x &= m \rhd ([g \rhd \mu] \mu^A)\\
		&= [(gm) \rhd \mu] \: [m \rhd \mu^A]\\
		&=[g \rhd (m \rhd \mu)] \mu^A\\
		&= [g \rhd (\mu \mu^B)] \mu^A\\
		&= [g \rhd \mu] [g \rhd \mu^B] \: \mu^A. 
	\end{align*}
	
	Then, using the fact that $g \rhd \mu^B= \mu^B$ (because $g \rhd \mu^B(e)= B(\partial(g \rhd e))=B(\partial(e))=\mu^B(e)$), we have
	\begin{align*}
		m \rhd \mu_x &= [g \rhd \mu] \: \mu^A \mu^B\\
		&= \mu_x \mu^B,
	\end{align*}
	so $\mu_X$ is also stabilised up to an irrep of the form $\mu^R$, so $m$ is also in $M$ for $\mu_x$. Therefore, all $\mu_x$ have the same group $M$. Furthermore, the irrep $\mu^B$ in $	m \rhd \mu_x= \mu_x \mu^B$ is the same for any irrep in the ground state, which means that $\mu_1(f) [m \rhd \mu_1^{-1}(f)]=1$ (which is equal to $\mu^B(f^{-1})$) is also the same for any irrep $\mu_1$ in the ground state. Therefore, none of the properties we have discussed in this section depend on the arbitrary choice of $\mu_1$ within the ground state.
	
	\subsubsection{The case where $G$ is Abelian but $E$ is non-Abelian and $\partial$ is general}
	\label{Section_condensation_rhd_nontrivial_E_non_Abelian}
	
\begin{figure}[h]
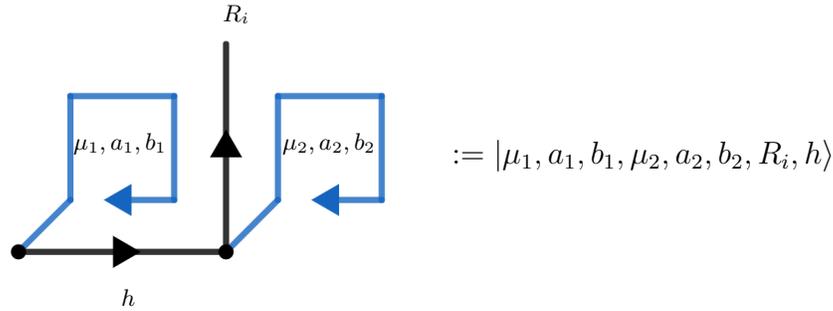

		\begin{center}
			\hspace{-1cm}
			\begin{overpic}[width=0.5\linewidth]{2D_upwards_edge_transform_2_image}
				\put(70,25){\large $:=\ket{\mu_1,a_1,b_1,\mu_2,a_2,b_2, R_i,h}$}
				\put(35,46){ $R_i$}
				\put(20,4){ $h$}
				\put(13,27){ $\mu_1,a_1,b_1$}
				\put(45,27){$\mu_2,a_2,b_2$}	
				
			\end{overpic}
			\caption{Shorthand for the degrees of freedom affected by the vertical edge transform}
			\label{vertical_edge_transform_hybrid_basis_2}
		\end{center}
	\end{figure}

	The case where we allow $E$ to be non-Abelian is somewhat similar to the previous one, except that there is some added complexity with matrix indices for the irreps of $E$. Let us start by considering the edge term in more detail, in the hybrid basis shown in Figure \ref{vertical_edge_transform_hybrid_basis_2}. In terms of the group element basis, this hybrid basis state is given by
	\begin{align*}
		\ket{\mu_1,a_1, b_1, \mu_2, a_2, b_2, R_i, h} &= \frac{\sqrt{|\mu_1| | \mu_2|}}{|E|} \sum_{e_1, e_2 \in E} [D^{\mu_1}(e_1)]_{a_1b_1} [D^{\mu_2}(e_2)]_{a_2b_2} \frac{1}{\sqrt{|G}|} \sum_{g \in G} R_i(g) \ket{e_1,e_2,g,h}.
	\end{align*}

	From the definition of the edge transform given in Equation \ref{Equation_edge_transform_definition} of the main text (noting that plaquette 1 is anti-aligned with the vertical edge, while plaquette 2 is aligned with it), the action of the edge term on the basis state is given by
	\begin{align*}
		\mathcal{A}_i &\ket{\mu_1,a_1, b_1, \mu_2, a_2, b_2, R_i, h} \\
		&= 	\mathcal{A}_i \frac{\sqrt{|\mu_1| | \mu_2|}}{|E|} \sum_{e_1, e_2 \in E} [D^{\mu_1}(e_1)]_{a_1b_1} [D^{\mu_2}(e_2)]_{a_2b_2} \frac{1}{\sqrt{|G}|} \sum_{g \in G} R_i(g) \ket{e_1,e_2,g,h}\\
		&= \frac{1}{|E|} \sum_{e \in E} \mathcal{A}_i^e\frac{\sqrt{|\mu_1| | \mu_2|}}{|E|} \sum_{e_1, e_2 \in E} [D^{\mu_1}(e_1)]_{a_1b_1} [D^{\mu_2}(e_2)]_{a_2b_2} \frac{1}{\sqrt{|G}|} \sum_{g \in G} R_i(g) \ket{e_1,e_2,g,h}\\
		&= \frac{1}{|E|} \sum_{e \in E}\frac{\sqrt{|\mu_1| | \mu_2|}}{|E|} \sum_{e_1, e_2 \in E} [D^{\mu_1}(e_1)]_{a_1b_1} [D^{\mu_2}(e_2)]_{a_2b_2} \frac{1}{\sqrt{|G}|} \sum_{g \in G} R_i(g) \ket{ [ h \rhd e]e_1,e_2e^{-1}, \partial(e)g,h}.
	\end{align*}
	
	Changing variables from $e_1$, $e_2$ and $g$ to $e_1' =[h \rhd e]e_1$, $e_2' = e_2 e^{-1}$ and $g' = \partial(e)g_i$, we then have
	\begin{align*}
		\mathcal{A}_i & \ket{\mu_1,a_1, b_1, \mu_2, a_2, b_2, R_i, h}\\
		&=\frac{1}{|E|} \sum_{e \in E}\frac{\sqrt{|\mu_1| | \mu_2|}}{|E|} \sum_{e_1' =[h \rhd e]e_1 } \sum_{ e_2'=e_2e^{-1}} [D^{\mu_1}([h \rhd e^{-1}]e_1' )]_{a_1b_1} [D^{\mu_2}(e_2'e)]_{a_2b_2} \frac{1}{\sqrt{|G}|} \sum_{g' = \partial(e)g} R_i(\partial(e)^{-1}g') \ket{e_1',e_2',g',h}\\
		&=\frac{1}{|E|} \sum_{e \in E}\frac{\sqrt{|\mu_1| | \mu_2|}}{|E|} \sum_{e_1', e_2' \in E} \sum_{c_1=1}^{|\mu_1|}[D^{\mu_1}(h \rhd e^{-1})]_{a_1c_1} [D^{\mu_1}(e_1')]_{c_1b_1} \sum_{c_2=1}^{|\mu_2|} [D^{\mu_2}(e_2')]_{a_2c_2} [D^{\mu_2}(e)]_{c_2b_2}\\
		& \hspace{0.5cm} \frac{1}{\sqrt{|G}|} \sum_{g' \in G} R_i(\partial(e)^{-1}) R_i(g') \ket{e_1',e_2',g',h}\\
		&=\sum_{c_1=1}^{|\mu_1|} \sum_{c_2=1}^{|\mu_2|} \frac{1}{|E|} \sum_{e \in E} [D^{\mu_1}(h \rhd e^{-1})]_{a_1c_1} [D^{\mu_2}(e)]_{c_2b_2} R_i(\partial(e)^{-1}) \frac{\sqrt{|\mu_1| | \mu_2|}}{|E|} \sum_{e_1', e_2' \in E} [D^{\mu_1}(e_1')]_{c_1b_1} [D^{\mu_2}(e_2')]_{a_2c_2} \\ 
		&	 \hspace{0.5cm} \frac{1}{\sqrt{|G}|} \sum_{g' \in G}R_i(g') \ket{e_1',e_2',g',h}\\
		&= \sum_{c_1=1}^{|\mu_1|} \sum_{c_2=1}^{|\mu_2|} \frac{1}{|E|} \sum_{e \in E} [D^{\mu_1}(h \rhd e^{-1})]_{a_1c_1} [D^{\mu_2}(e)]_{c_2b_2} R_i(\partial(e)^{-1}) \ket{\mu_1,c_1,b_1, \mu_2,a_2,c_2,R_i,h}\\
		&= \sum_{c_1=1}^{|\mu_1|} \sum_{c_2=1}^{|\mu_2|} \frac{1}{|E|} \sum_{e \in E} [D^{\mu_1}(h \rhd e^{-1})]_{a_1c_1} [D^{\mu_2}(e)]_{c_2b_2} \mu^{R_i}(e^{-1}) \ket{\mu_1,c_1,b_1, \mu_2,a_2,c_2,R_i,h}.
	\end{align*}

	Now, while we can use orthogonality of irreps to evaluate the sum over $e \in E$, we have to account for the fact that $\mu^{R_i} \cdot [h \rhd \mu_1]$ may be equivalent to $\mu_2$, but not identical (i.e., the two irreps can be related by conjugation by a constant matrix). This gives us
	\begin{align*}
		\mathcal{A}_i \ket{\mu_1,a_1, b_1, \mu_2, a_2, b_2, R_i, h} &= \sum_{c_1, c_2=1}^{|\mu_1|} \frac{1}{|\mu_2|} \delta(\mu^{R_i} \cdot h \rhd \mu_1 \sim \mu_2) S_{a_1b_2}S^{-1}_{c_2 c_1}\ket{\mu_1,c_1,b_1, \mu_2, a_2, c_2, R_i, h},
	\end{align*}
	where $\sim$ indicates that the two irreps are related by conjugation and $S$ is the relevant conjugating matrix (note that, when the expression is non-zero, $|\mu_1|= |\mu_2|$, because the two irreps differ only by conjugation, multiplication by a 1D irrep, and the $\rhd$ action, none of which effect the dimension of the irreps). If we then apply another edge term on the edge separating plaquette 2 from another plaquette, plaquette 3, we get something proportional to
	$$ \delta(\mu^{R_i} \cdot h \rhd \mu_1 \sim \mu_2) \delta(\mu^{R_{i_2}} \cdot h_2 \rhd \mu_2 \sim \mu_3).$$
	
	If these two Kronecker deltas are satisfied (e.g., in the ground state), then we can show that the Kronecker delta
	$$\delta(\mu_3 \sim \mu^{R_1} \cdot \mu^{R_2} \cdot (hh_2) \rhd \mu_1)$$
	relating $\mu_1$ to $\mu_3$ is also satisfied. To show this, assume that the two original Kronecker deltas are satisfied. Then we can write
	$$D^{\mu_2}(e) =S_2 \mu^{R_1}(e)D^{\mu_1}(h \rhd e)S_2^{-1}$$
	and
	$$D^{\mu_3}(e) = S_3 \mu^{R_2}(e) D^{\mu_2}(h_2 \rhd e)S_3^{-1}.$$
	
	Noting that the $S$ matrices are independent of $e$, we can substitute the first relation into the second to obtain
	\begin{align*}
		D^{\mu_3}(e) =S_3 \mu^{R_2}(e) S_2 \mu^{R_1}(h_2 \rhd e)D^{\mu_1}(h \rhd (h_2 \rhd e))S_2^{-1} S_3^{-1}.
	\end{align*}
	
	Now $\mu^{R_2}(e)$ is a 1D irrep and so is just a phase. This means that it commutes with the matrix $S_2$. Furthermore, $\mu^{R_1}(h_2 \rhd e)=\mu^{R_1}(e)$, and so we have
	\begin{align*}
		D^{\mu_3}(e) &=S_3 S_2 \mu^{R_2}(e) \mu^{R_1}(e)D^{\mu_1}((h_2h) \rhd e)S_2^{-1} S_3^{-1}\\
		&= (S_3 S_2) (\mu^{R_2} \cdot \mu^{R_1})(e) D^{\mu_1}((hh_2) \rhd e)(S_3 S_2)^{-1},
	\end{align*}
	where in the last line we used the fact that $G$ is Abelian to switch the order of $h$ and $h_2$. This means that $\mu_3$ is indeed equivalent to $\mu^{R_1} \cdot \mu^{R_2} \cdot (hh_2) \rhd \mu_1$. Iterating this condition across a ribbon $t$, such as the one shown in Figure \ref{Condensed_ribbon_1}, gives us
	\begin{equation}
		\mu_n \sim \mu^{R_t} [\hat{g}(t) \rhd \mu_1], \label{Equation_condensation_non_Abelian_plaquette_relation}
	\end{equation}
	where $\mu^{R_t}$ is the defect label for the dual path (obtained by fusing the irreps associated to the edges cut by the dual path) and $\hat{g}(t)$ is the direct path element between the base-points of plaquettes 1 and $n$. We claim that, just as for the case where $E$ is Abelian, this relation between distant plaquettes in the ground state means that some of the ribbon operators condense (act equivalently to local operators on the ground state). We consider a bilocal operator of the form
	$$\delta(\mu_n \sim [q \rhd \mu_1] \mu^X).$$
	
	Let $U$ be the subset of elements in $G$ which stabilise $\mu_1$ up to conjugation by a constant matrix and multiplication by an irrep $\mu^A$ of the form $\mu^A(e)=A(\partial(e))$ for some irrep $A$ of $\partial(E)$. That is, for any $u \in U$
	$$D^{u \rhd \mu_1}(e)= SD^{\mu_1}(e) S^{-1} \mu^A(e) \ \forall e \in E.$$ 
	
	This subset forms a subgroup of $G$. To see this, consider two elements $u$ and $r$ in $U$. We have
	$$D^{u \rhd \mu_1}(e)=SD^{\mu_1}(e)S^{-1}\mu^A(e)$$
	and
	$$D^{r \rhd \mu_1}(e)=PD^{\mu_1}(e)P^{-1}\mu^B(e).$$
	Then for $ur$, we have
	\begin{align*}
		D^{(ur) \rhd \mu_1}(e)&= D^{r \rhd (u \rhd \mu_1)}(e)\\
		&=D^{u \rhd \mu_1}(r \rhd e)\\
		&=SD^{\mu_1}(r \rhd e)S^{-1}\mu^A(r \rhd e).
	\end{align*}
	$\mu^A$ satisfies 
	$$\mu^A(r \rhd e)=A(\partial(r \rhd e))=A(r\partial(e)r^{-1})=A(\partial(e))=\mu^A(e),$$
	using the first Peiffer condition (see Equation \ref{Peiffer_1} in the main text) and the fact that $G$ is Abelian. This means that
	\begin{align*}
		D^{(ur) \rhd \mu_1}(e)&=SD^{r \rhd \mu_1}(e)S^{-1}\mu^A(e)\\
		&=SPD^{\mu_1}(e)P^{-1}\mu^B(e)S^{-1}\mu^A(e)\\
		&=SP D^{\mu_1}(e) P^{-1} S^{-1} \mu^B(e)\mu^A(e)\\
		&=(SP) D^{\mu_1}(e)(SP)^{-1} (\mu^B \cdot \mu^A)(e),
	\end{align*}
	so $ur$ is also in $U$. Furthermore, this subgroup $U$ is the same for any irrep $\mu_x$ appearing in the same ground state as $\mu_1$ (i.e., it is a ground state property). To see this, note that, as we showed earlier in Section \ref{Section_condensation_rhd_nontrivial_E_non_Abelian}, an irrep $\mu_2$ in the same ground state as $\mu_1$ satisfies 
	$$D^{\mu_2}(e)=PD^{g \rhd \mu_1}(e) \mu^A(e)P^{-1},$$
	for some matrix $P$, element $g \in G$ and irrep $\mu^A$ (with this expression holding for all $e \in E$). Then for $u \in U_{\mu_1}$, where $U_{\mu_1}$ denotes the subgroup $U$ for irrep $\mu_1$, we have
	\begin{align*}
		D^{u \rhd \mu_2}(e)&= D^{\mu_2}( u \rhd e)\\
		&= P D^{g \rhd \mu_1}( u \rhd e) \mu^A(e)P^{-1}\\
		&= P D^{\mu_1}( (ug) \rhd e) \mu^A(e)P^{-1}\\
		&= P D^{ \mu_1}( u \rhd (g\rhd e)) \mu^A(e)P^{-1}\\
		&= P D^{ u \rhd \mu_1}(g\rhd e) \mu^A(e)P^{-1}\\
		&= P S D^{ \mu_1}(g\rhd e) \mu^B(e)S^{-1} \mu^A(e)P^{-1},
	\end{align*}
	where the last equality follows from the fact that $u$ is in $U_{\mu_1}$. Then we can write this as
	\begin{align*}
		D^{u \rhd \mu_2}(e)&= P S D^{ g\rhd \mu_1}( e) \mu^A(e)S^{-1} P^{-1} \mu^B(e),
	\end{align*}
	where we used the fact that $\mu^A(e)$ and $\mu^B(e)$ are just phases and so commute with the other terms. Inserting the identity in the form of $P^{-1}P$ gives us
	\begin{align*}
		D^{u \rhd \mu_2}(e)&= PSP^{-1} P D^{ g\rhd \mu_1}( e) \mu^A(e)\mu^B(e)P^{-1} PS^{-1} P^{-1}\\
		&= (PSP^{-1}) D^{ \mu_2}( e) \mu^B(e)(PSP^{-1})^{-1},
	\end{align*}
	where the last identity follows from the relationship between $\mu_2$ and $\mu_1$. We therefore see that $u$ stabilises $\mu_2$ up to conjugation and an irrep sensitive only to $\partial(E)$. That is $u$ is in the group $U_{\mu_2}$, as well as the group $U_{\mu_1}$. We can show that the converse holds by exactly analogous arguments, and so the two groups $U_{\mu_1}$ and $U_{\mu_2}$ are the same, and we will continue to call them $U$.

	Now we can consider the operator
	$$\delta(\mu_n \sim [q \rhd \mu_1] \mu^X)$$
	acting on a particular ground state. Using the fact that $\mu_n$ is equivalent to $\mu^{R_t} [\hat{g}(t) \rhd \mu_1]$ as stated in Equation \ref{Equation_condensation_non_Abelian_plaquette_relation}, we can write
	\begin{align*}
		\delta(\mu_n \sim [q \rhd \mu_1] \mu^X)\ket{GS} = \delta(\mu^{R_t} [\hat{g}(t) \rhd \mu_1] \sim [q \rhd \mu_1] \mu^X ) \ket{GS},
	\end{align*}
	using the transitive property of the equivalence relation. Now we note that
	$$\delta(\mu^{R_t} [\hat{g}(t) \rhd \mu_1] \sim [q \rhd \mu_1] \mu^X ) \ket{GS} = \delta( [(\hat{g}(t)q^{-1}) \rhd \mu_1] \mu^{R_t} \sim \mu_1 \mu^X)\ket{GS}.$$
	
	To see this, note that
	$$\mu^{R_t} [\hat{g}(t) \rhd \mu_1] \sim [q \rhd \mu_1] \mu^X$$
	means that
	\begin{align*}
		D^{\hat{g}(t) \rhd \mu_1}(e) \mu^{R_t}(e) = S D^{q \rhd \mu_1}(e)\mu^X(e)S^{-1},
	\end{align*}
	for all $e \in E$ and for $S$ independent of $e$. Then
	\begin{align*}
		D^{ \mu_1}(\hat{g}(t) \rhd e)& \mu^{R_t}(e) = S D^{ \mu_1}(q \rhd e)\mu^X(e)S^{-1}\\
		& \implies D^{ \mu_1}((\hat{g}(t)q^{-1}) \rhd (q \rhd e)) \mu^{R_t}(q \rhd e) = S D^{ \mu_1}(q \rhd e)\mu^X(q \rhd e)S^{-1},
	\end{align*}
	using the fact that $\mu^{R_t}(q \rhd e)=\mu^{R_t}(e)$ (and similar for $\mu^X$). Then because this is true for all $e \in E$ it must be true for all $e' =q^{-1} \rhd e \in E$ and so
	\begin{align*}
		D^{ \mu_1}((\hat{g}(t)q^{-1}) \rhd e') \mu^{R_t}(e') &= S D^{ \mu_1}(e')\mu^X(e')S^{-1}.
	\end{align*}
	This means that
	$$[(\hat{g}(t)q^{-1}) \rhd \mu_1] \mu^{R_t} \sim \mu_1 \mu^X,$$
	and so $\mu^{R_t} [\hat{g}(t) \rhd \mu_1] \sim [q \rhd \mu_1] \mu^X$ implies that $[(\hat{g}(t)q^{-1}) \rhd \mu_1] \mu^{R_t} \sim \mu_1 \mu^X$. The reverse direction is also true by similar reasoning (i.e., $[(\hat{g}(t)q^{-1}) \rhd \mu_1] \mu^{R_t} \sim \mu_1 \mu^X$ implies $\mu^{R_t} [\hat{g}(t) \rhd \mu_1] \sim [q \rhd \mu_1] \mu^X$) and so the two Kronecker delta are the same. That is
	\begin{align*}
		\delta([\hat{g}(t) \rhd \mu_1] \mu^{R_t} \sim [q \rhd \mu_1] \mu^X) &=\delta([(\hat{g}(t)q^{-1}) \rhd \mu_1] \mu^{R_t} \sim \mu_1 \mu^X)\\
		&= \delta((\hat{g}(t)q^{-1}) \rhd \mu_1 \sim \mu_1 \mu^X{\mu^{R_t}}^{-1}),
	\end{align*}
	where the last line follows from the fact that $\mu^{R_t}$ is just a phase (and so commutes with conjugation). This indicates that $\hat{g}(t)q^{-1}$ stabilises $\mu_1$ up to $\mu^X{\mu^{R_t}}^{-1}$ and conjugation, meaning that $\hat{g}(t)q^{-1}$ belongs to the subgroup $U$. Therefore
	\begin{align*}
		\delta((\hat{g}(t)q^{-1}) \rhd \mu_1 \sim \mu_1 \mu^X{\mu^{R_t}}^{-1})&= \delta((\hat{g}(t)q^{-1}) \rhd \mu_1 \sim \mu_1 \mu^X{\mu^{R_t}}^{-1})\delta(\hat{g}(t)q^{-1} \in U)\\
		&=\sum_{u \in U} \delta(\hat{g}(t)q^{-1},u) \delta(u \rhd \mu_1 \sim \mu_1 \mu^X{\mu^{R_t}}^{-1})\\
		&= \sum_{u \in U} \delta(\hat{g}(t),uq) \delta(u \rhd \mu_1 \sim \mu_1 \mu^X{\mu^{R_t}}^{-1}).
	\end{align*}
	We can then use orthogonality of characters to write this as
	\begin{align*}
		\sum_{u \in U} \delta(\hat{g}(t),uq) \delta(u \rhd \mu_1 \sim \mu_1 \mu^X{\mu^{R_t}}^{-1}) &= \sum_{u \in U} \delta(\hat{g}(t),uq) \frac{1}{|E|} \sum_{e \in E} \chi_{\mu_1}(e)\mu^X(e) \mu^{R_t}(e)^{-1} \overline{\chi}_{u \rhd \mu_1}(e).	
	\end{align*}
	
	Noting that $\mu^{R_t}(e)^{-1}=R_t(\partial(e)^{-1})$ is equivalent to the action of a magnetic ribbon operator $C^{\partial(e)}(t)$ on the ground state (as shown in Equation \ref{Equation_magnetic_irrep_basis_1}), we have
	\begin{align*}
		\sum_{u \in U} \delta(\hat{g}(t),uq) \delta(u \rhd \mu_1 \sim \mu_1 \mu^X{\mu^{R_t}}^{-1}) &= \sum_{u \in U} \delta(\hat{g}(t),uq) \frac{1}{|E|} \sum_{e \in E} \chi_{\mu_1}(e)\mu^X(e) \overline{\chi}_{u \rhd \mu_1}(e) C^{\partial(e)}(t)\\
		&= \sum_{u \in U} \frac{1}{|E|} \sum_{e \in E} \chi_{\mu_1}(e)\mu^X(e) \overline{\chi}_{u \rhd \mu_1}(e) \delta(\hat{g}(t),uq) C^{\partial(e)}(t),
	\end{align*}
	which can be recognised as an electromagnetic ribbon operator. Just as we did in the case where $E$ was Abelian, we want to take linear combinations of these in order to obtain simpler expressions. We therefore consider the operators
	$$\frac{1}{|\partial(E)|} \sum_{X \in \text{Rep}(\partial(E))} \mu^X(f^{-1})\delta(\mu_n \sim [q \rhd \mu_1] \mu^X),$$
	of which we have one operator for each representative $f$ of the quotient group $E/ \ker(E)$ and each distinct representative $q$ of the quotient group $G/ U$. For such an operator, we have
	\begin{align*}
		\frac{1}{|\partial(E)|}& \sum_{X \in \text{Rep}(\partial(E))} \mu^X(f^{-1})\delta(\mu_n \sim [q \rhd \mu_1] \mu^X)\\
		&= \frac{1}{|\partial(E)|} \sum_{X \in \text{Rep}(\partial(E))} \mu^X(f^{-1}) \sum_{u \in U} \frac{1}{|E|} \sum_{e \in E} \chi_{\mu_1}(e)\mu^X(e) \overline{\chi}_{u \rhd \mu_1}(e) \delta(\hat{g}(t),uq) C^{\partial(e)}(t)\\
		&= \sum_{u \in U} \frac{1}{|E|} \sum_{e \in E} \chi_{\mu_1}(e) \bigg[\frac{1}{|\partial(E)|} \sum_{X \in \text{Rep}(\partial(E))} \mu^X(f^{-1})\mu^X(e)\bigg] \overline{\chi}_{u \rhd \mu_1}(e) \delta(\hat{g}(t),uq) C^{\partial(e)}(t).
	\end{align*}
	
	We can then use the Grand Orthogonality Theorem to evaluate the term in squared brackets, giving
	$$\bigg[\frac{1}{|\partial(E)|} \sum_{X \in \text{Rep}(\partial(E))} \mu^X(f^{-1})\mu^X(e)\bigg]= \delta(\partial(e),\partial(f)).$$
	Therefore,
	\begin{align*}
		\frac{1}{|\partial(E)|}& \sum_{X \in \text{Rep}(\partial(E))} \mu^X(f^{-1})\delta(\mu_n \sim [q \rhd \mu_1] \mu^X)\\
		&=\sum_{u \in U} \frac{1}{|E|} \sum_{e \in E} \chi_{\mu_1}(e) \delta(\partial(e),\partial(f)) \overline{\chi}_{u \rhd \mu_1}(e) \delta(\hat{g}(t),uq) C^{\partial(e)}(t).
	\end{align*}
	
	While the Kronecker delta fixes $\partial(e)$, it does not fix the part of $e$ in the kernel, so we are left with a sum over elements $e_k$ in the kernel, where $e=fe_k$:
	\begin{align*}
		\frac{1}{|\partial(E)|}& \sum_{X \in \text{Rep}(\partial(E))} \mu^X(f^{-1})\delta(\mu_n \sim [q \rhd \mu_1] \mu^X)\\
		&=\sum_{u \in U} \frac{1}{|E|} \big[\sum_{e_k \in \ker{\partial}} \chi_{\mu_1}(fe_k) \overline{\chi}_{u \rhd \mu_1}(fe_k)\big] \delta(\hat{g}(t),uq) C^{\partial(f)}(t).
	\end{align*}
	
	We can then evaluate the expression $\big[\sum_{e_k \in \ker{\partial}} \chi_{\mu_1}(fe_k) \overline{\chi}_{u \rhd \mu_1}(fe_k)\big]$ by splitting it into parts corresponding to $e_k$ and parts corresponding to $f$. We have
	\begin{align*}
		\sum_{e_k \in \ker{\partial}} \chi_{\mu_1}(fe_k) \overline{\chi}_{u \rhd \mu_1}(fe_k) &= \sum_{a=1}^{|\mu_1|} \sum_{b=1}^{|u \rhd \mu_1|} \sum_{e_k \in \ker{\partial}} [D^{\mu_1}(fe_k)]_{aa} [D^{u \rhd \mu_1}(e_k^{-1}f^{-1})]_{bb}\\
		&= \sum_{a=1}^{|\mu_1|} \sum_{b=1}^{|u \rhd \mu_1|} \sum_{c=1}^{|\mu_1|} \sum_{c=1}^{|u \rhd\mu_1|} \sum_{e_k \in \ker{\partial}} [D^{\mu_1}(f)]_{ac} [D^{\mu_1}(e_k)]_{ca} [D^{u \rhd \mu_1}(e_k^{-1})]_{bd} [D^{u \rhd \mu_1}(f^{-1})]_{db}.
	\end{align*}
	Now by Schur's lemma, using the fact that the kernel of $\partial$ is in the centre of $E$, we have $$[D^{\mu_1}(e_k)]_{ca}= [D^{\mu_1}(e_k)]_{11} \delta_{ac} = \mu_1^{\ker}(e_k) \delta_{ca}.$$ Therefore,
	\begin{align*}
		\sum_{e_k \in \ker{\partial}} \chi_{\mu_1}(fe_k) \overline{\chi}_{u \rhd \mu_1}(fe_k) &= \sum_{a=1}^{|\mu_1|} \sum_{b=1}^{|u \rhd \mu_1|} \sum_{c=1}^{|\mu_1|} \sum_{c=1}^{|u \rhd\mu_1|} \sum_{e_k \in \ker{\partial}} [D^{\mu_1}(f)]_{ac} \mu_1^{\ker}(e_k) \delta_{ca} u \rhd \mu_1^{\ker}(e_k^{-1}) \delta_{bd} [D^{u \rhd \mu_1}(f^{-1})]_{db}\\
		&= \sum_{a=1}^{|\mu_1|} \sum_{b=1}^{|u \rhd \mu_1|} \sum_{e_k \in \ker{\partial}} [D^{\mu_1}(f)]_{aa} \mu_1^{\ker}(e_k) u \rhd \mu_1^{\ker}(e_k^{-1}) [D^{u \rhd \mu_1}(f^{-1})]_{bb}\\
		&= \chi_{\mu_1}(f)\overline{\chi}_{u \rhd \mu_1}(f) \big[\sum_{e_k \in \ker{\partial}} \mu_1^{\ker}(e_k) u \rhd \mu_1^{\ker}(e_k^{-1})\big]\\
		&= \chi_{\mu_1}(f)\overline{\chi}_{u \rhd \mu_1}(f) |\ker(\partial)| \delta(\mu_1^{\ker}, u \rhd \mu_1^{\ker}),
	\end{align*}
	where in the last step we use orthogonality of irreps of the kernel of $\partial$. The Kronecker delta $\delta(\mu_1^{\ker}, u \rhd \mu_1^{\ker})$ is automatically satisfied by the definition of $u$. To see this, note that the definition of $u \in U$ means that $\mu_1 \sim [u \rhd \mu_1] \mu^A$ for some $\mu^A$ derived from a representation of $\partial(E)$. When we restrict $\mu_1$ to the kernel of $\partial$, the conjugation implied by $\sim$ becomes trivial (because the restricted representation is one-dimensional) and $\mu^A$ also becomes trivial. This means that
	\begin{align*}
		\sum_{e_k \in \ker{\partial}} \chi_{\mu_1}(fe_k) \overline{\chi}_{u \rhd \mu_1}(fe_k) &= \chi_{\mu_1}(f)\overline{\chi}_{u \rhd \mu_1}(f) |\ker(\partial)|,
	\end{align*}
	and our final expression for the condensed ribbon operators becomes
	$$\sum_{u \in U} \frac{1}{|\partial(E)|} \chi_{\mu_1}(f) \overline{\chi}_{u \rhd \mu_1}(f) \delta(\hat{g}(t),uq) C^{\partial(f)}(t).$$

\end{document}